\PassOptionsToPackage{dvipsnames,table}{xcolor}
\pdfoutput=1
\documentclass[11pt]{article}
\usepackage[utf8]{inputenc}
\usepackage{changepage}
\usepackage{enumitem}
\usepackage{nips10submit_e,times}
\nipsfinalcopy  % <-- This disables the line numbers
\usepackage{latexsym}
\usepackage[T1]{fontenc}
\usepackage[utf8]{inputenc}
\usepackage{placeins}
\usepackage{microtype}
\usepackage{inconsolata}
\usepackage{changepage}
\usepackage[margin=1in]{geometry}
\usepackage{caption}
\usepackage[normalem]{ulem}
\usepackage{booktabs}
\usepackage{siunitx}
\usepackage{pifont}

\usepackage[belowskip=-1.0pt,aboveskip=5pt]{caption}
\usepackage{float}

\usepackage{tikz}
\usepackage{tikz-3dplot}
\usetikzlibrary{arrows.meta, positioning, calc, shadows.blur, shapes.gates.logic.US}

\usepackage{threeparttable}

\RequirePackage{etex}
\usepackage{textcomp,upquote}

\usepackage{mdframed}
\usepackage[numbers]{natbib}
\setcitestyle{authoryear,aysep={},maxcitenames=1}

\usepackage[table]{xcolor}
\usepackage{comment}
\usepackage[linguistics]{forest}
\usepackage{bigfoot}
\usepackage{pgfplots,pgfplotstable}
\usepackage{sansmath} % optional: nicer number rendering
\usepackage{neuralnetwork}
\pgfplotsset{compat=1.18}
\usepgfplotslibrary{groupplots}
\usepgfplotslibrary{statistics}
\usepackage{times}
\usepackage{helvet}
\usepackage{courier}
\usepackage{setspace}
\usepackage{amsmath}
\usepackage{caption}
\usepackage{bigfoot}
\usepackage{latexsym}
\usepackage{amssymb}
\usepackage{wrapfig}
\usepackage{url}
\usepackage{lipsum}
\usepackage{minted}
\usepackage[amsmath,hyperref,thmmarks]{ntheorem}
\usepackage{alltt}
\usepackage[bottom]{footmisc}
\usepackage{fancyheadings} 
\usepackage{supertabular}              
\usepackage{fancybox}             
\usepackage{float}
\usepackage{afterpage}
\usepackage{longtable}
\usepackage{graphicx,scalerel}
\usepackage[percent]{overpic}
\usepackage{relsize}
\usepackage{curves}
\usepackage{epsfig}
\usepackage{lingmacros}
\usepackage{tree-dvips}
\usepackage{amsfonts}
\usepackage{diagbox}
\usepackage{booktabs}
\usepackage{mathtools}
\usepackage{thmtools}
\usepackage{footnote}
\usepackage{etoolbox}
\usepackage{multirow}
\usepackage{subcaption}
\usepackage[skins,most,hooks,xparse,breakable]{tcolorbox}

\usepackage{chngcntr}
\usepackage{blindtext}
\definecolor{greentitle}{RGB}{61,170,61}
\definecolor{greentitleback}{RGB}{216,233,213}

\newtcolorbox{llmquote}[1]{
  colback=gray!5!white,
  colframe=red!45!black,
  fonttitle=\bfseries,
  coltitle=black,           % title text color
  colbacktitle=blue!20,
  title={#1},
  rounded corners,
  boxrule=1pt,
  left=6pt,
  right=6pt,
  top=6pt,
  bottom=6pt,
 fontupper=\footnotesize%\sffamily\itshape
}

\usepackage{expl3}
\usepackage{xparse}

\def\mathunderline#1#2{\color{#1}\underline{{\color{black}#2}}\color{black}}

\usepackage{hyperref} 
\usepackage{cleveref}

\DeclareMathOperator*{\argmax}{argmax}

\definecolor{deepblue}{rgb}{0,0,0.5}
\definecolor{deepred}{rgb}{0.6,0,0}
\definecolor{deepgreen}{rgb}{0,0.5,0}

\definecolor{codegreen}{rgb}{0,0.6,0}
\definecolor{codegray}{rgb}{0.5,0.5,0.5}
\definecolor{codepurple}{rgb}{0.58,0,0.82}
\definecolor{backcolour}{rgb}{0.95,0.95,0.92}

\usepackage{listings-fix}
\usepackage{textcomp}
\definecolor{backgroundColor}{HTML}{FDF5E6}
\definecolor{kwdColor}{HTML}{0071BC}
\definecolor{Sienna}{HTML}{A0522D}
\definecolor{Maroon}{HTML}{800000}

\definecolor{theoremColor}{HTML}{800080}

\lstdefinelanguage{Athena}
 {alsoletter={-*'~%,=>,:=,!,(,),\|,
    },
   morekeywords={dlet,,method,primitive-method, lambda,domain,domains,with-witness,for,datatype-cases,on,%\!,| , 
     by, BY, for, OP, apply-method,generalize-over,datatypes,clear-assumption-base,assume,from,=>,suppose-absurd,let,begin,end,by-cases,set-flag,%=>,\{,\},
letrec,cell,ref,dseq,seq,set!,set,vector-set!,vector-set,vector-sub,make-vector,pick-any,pick-witness,pick-witnesses,by,assume-let,match,dmatch,conclude,%_,
retract,suppose-absurd-let,assert,assert*,define,try,dtry,dcheck,check,else,while,split, where, set-precedence,expand-input,transform-output,define-sort, 
left-assoc,right-assoc,list-of,bind,as,val-of,some-list,some-term,some-sentence,some-sent,some-symbol,some-method,
some-quant,some-sent-con,some-prop-con,some-atom,some-proc,some-var,some-cell,some-vector,some-sub,some-char,
define-symbol,module,open,extend-module,private,quit,foobar,
overload,secure-file,load-file,load,declare,datatype,structure,by-induction,datatype-cases},
 otherkeywords={%!,|,
                },
keywordstyle=\color{kwdColor}\bfseries,
commentstyle=\itshape\color{green!40!black}\itshape,
stringstyle=\ttfamily,
morecomment=[l]{\#},
morestring=[b]",
}

\lstdefinelanguage{AthenaNC}
 {alsoletter={-*'~%,=>,:=,!,(,),\|,
    },
   morekeywords={dlet,,method,primitive-method, lambda,domain,domains,with-witness,for,datatype-cases,on,%\!,| , 
     by, BY, for, OP, apply-method,generalize-over,datatypes,clear-assumption-base,assume,from,=>,suppose-absurd,let,begin,end,by-cases,set-flag,%=>,\{,\},
letrec,cell,ref,dseq,seq,set!,set,vector-set!,vector-set,vector-sub,make-vector,pick-any,pick-witness,pick-witnesses,by,assume-let,match,dmatch,conclude,%_,
retract,suppose-absurd-let,assert,assert*,define,try,dtry,dcheck,check,else,while,split, where, set-precedence,expand-input,transform-output,define-sort, 
left-assoc,right-assoc,list-of,bind,as,val-of,some-list,some-term,some-sentence,some-sent,some-symbol,some-method,
some-quant,some-sent-con,some-prop-con,some-atom,some-proc,some-var,some-cell,some-vector,some-sub,some-char,
define-symbol,module,open,extend-module,private,quit,foobar,
overload,secure-file,load-file,load,declare,datatype,structure,by-induction,datatype-cases},
 otherkeywords={%!,|,
                },
keywordstyle=\color{kwdColor}\bfseries,
stringstyle=\ttfamily,
morestring=[b]",
}

\definecolor{cinereous}{rgb}{0.6, 0.51, 0.48}
\definecolor{amber}{rgb}{1.0, 0.75, 0.0}
\definecolor{almond}{rgb}{0.94, 0.87, 0.8}
\definecolor{amethyst}{rgb}{0.6, 0.4, 0.8}
\definecolor{arylideyellow}{rgb}{0.91, 0.84, 0.42}
\definecolor{ballblue}{rgb}{0.13, 0.67, 0.8}

\newcommand{\customsize}{\fontsize{9.7}{11.6}}

\renewcommand{\small}{\customsize}

\colorlet{backgroundColor}{red!50!yellow!4.5!white}

\definecolor{brightmaroon}{rgb}{0.76, 0.13, 0.28}
\definecolor{amber}{rgb}{1.0, 0.49, 0.0}
\definecolor{alizarin}{rgb}{0.82, 0.1, 0.26}
\definecolor{cadmiumred}{rgb}{0.89, 0.0, 0.13}

\newcommand{\bangforce}[1]{{!{\bf\color{alizarin}force}}}  % Ignore the argument

\newcommand{\textize}[2]{\texttt{#1}}
\newcommand{\pairop}[1]{\symbol{64}} % Ignore the argument
\newcommand{\betternot}[1]{$\sim$} % Ignore the argument

\lstnewenvironment{Athena}[1][]
{\lstset{numbers=none,#1}}
{}

\lstnewenvironment{AthenaNN}[1][]
{\lstset{numbers=none,#1}}
{}

\lstnewenvironment{AthenaWN}[1][]
{\lstset{numbers=left,#1}}
{}

\lstnewenvironment{AthenaIOttfor}[1][]
{\lstset{frameround=ffff,
        stringstyle=\color{black},  % or remove this line entirely
        morestring=[b]{"},          % disable special handling for double quotes
        morestring=[b]{'},
        basicstyle=\scriptsize\ttfamily,
        frame=trBL,
        numbers=none,emph={[2]for,where,else,description,set,premise,complement,while,by,structure,dlet,method,primitive-method, lambda,domain,domains,with-witness,for,datatype-cases,on,by, BY, for, OP, apply-method,generalize-over,datatypes,clear-assumption-base,assume,FROM,from,=>,suppose-absurd,let,begin,end,by-cases,set-flag,=>, letrec,cell,ref,dseq,seq,set!,set,vector-set!,vector-set,vector-sub,make-vector,pick-any,pick-witness,pick-witnesses,by,assume-let,match,dmatch,conclude,retract,suppose-absurd-let,assert,assert*,define,try,dtry,dcheck,check,else,while,split, where, set-precedence,expand-input,transform-output,define-sort, left-assoc,right-assoc,list-of,bind,as,val-of,some-list,some-term,some-sentence,some-sent,some-symbol,some-method,some-quant,some-sent-con,some-prop-con,some-atom,some-proc,some-var,some-cell,some-vector,some-sub,some-char,define-symbol,module,open,extend-module,private,quit,overload,secure-file,load-file,load,declare,datatype,structure,by-induction,datatype-cases},emphstyle={[2]\tt},#1}}
{}

\definecolor{foocolor}{HTML}{006795}
\definecolor{bcolor}{gray}{0.92}
\lstnewenvironment{tcAthena}[1][]
 {\lstset{language=Athena,          % <-- put the language back
           backgroundcolor=\color{bcolor},
           basicstyle=\scriptsize\ttfamily,
           keywordstyle=\color{foocolor}\bfseries,
           commentstyle=\color{green!40!black},
           frame=lines,
           framerule=0.8pt,
           framesep=5pt,
           numbers=none,#1}}{}

\lstnewenvironment{tcAthenaWithNumbers}[1][]
 {\lstset{language=Athena,          % <-- put the language back
           backgroundcolor=\color{bcolor},
           basicstyle=\scriptsize\ttfamily,
           keywordstyle=\color{foocolor}\bfseries,
           frame=lines,
           framerule=0.8pt,
           framesep=5pt,
           commentstyle=\color{green!40!black},
           numbers=left,#1}}
{}

\lstnewenvironment{tcAthenaWithNoComments}[1][]
 {\lstset{language=AthenaNC,          % <-- put the language back
           backgroundcolor=\color{bcolor},
           basicstyle=\scriptsize\ttfamily,
           keywordstyle=\color{foocolor}\bfseries,
           frame=lines,
           framerule=0.8pt,
           framesep=5pt,
           numbers=left,#1}}
{}

\newcommand{\kwd}[1]{\mbox{\small{\color{kwdColor}\bf{\ttfamily #1}}}}
\newcommand{\skwd}[1]{\mbox{\small{\color{kwdColor}\bf{\ttfamily #1}}}}
\newcommand{\fkwd}[1]{\mbox{\footnotesize{\color{kwdColor}\bf{\ttfamily #1}}}}

\lstdefinelanguage{Bathena}
 {alsoletter={-*'~%,=>,:=,!,(,),\|,
  _,!\[,
    \{,\}
    },
keywords={foobarchoo},
keywordstyle=\color{kwdColor}\bfseries,
commentstyle={\rm},
commentstyle=\itshape\color{purple!40!black},
commentstyle=\itshape\color{green!40!black}\itshape,
stringstyle=\ttfamily 
}

\lstnewenvironment{tcBathena}[1][]
{\lstset{backgroundcolor=\color{bcolor},
belowskip=0pt,
frame=lines,
framerule=0.8pt,
framesep=5pt,
basicstyle=\scriptsize\ttfamily,
keywordstyle=\color{foocolor}\bfseries,numbers=none,#1}}
{}

\DeclareFixedFont{\ttb}{T1}{txtt}{bx}{n}{9} % for bold
\DeclareFixedFont{\ttm}{T1}{txtt}{m}{n}{9}  % for normal

\lstdefinestyle{mystyle}{
    backgroundcolor=\color{backcolour},   
    commentstyle=\color{codegreen},
    keywordstyle=\color{magenta},
    numberstyle=\tiny\color{codegray},
    stringstyle=\color{codepurple},
    basicstyle=\ttfamily\footnotesize,
    breakatwhitespace=false,         
    breaklines=true,                 
    captionpos=b,                    
    keepspaces=true,                 
    numbers=left,                    
    numbersep=5pt,                  
    showspaces=false,                
    showstringspaces=false,
    showtabs=false,                  
    tabsize=2
}

\newtcolorbox{maskingpromptbox}{
  colback=gray!5,
  colframe=black,
  title=Main Prompt for Proof Mask-Filling,
  breakable
}

\newtcolorbox{restrictedNDLbox}{
  colback=gray!5,
  colframe=black,
  title=Proof for NDL proof generation under a restricted set of rules,
  breakable
}

\newtcolorbox{expandedNDLbox}{
  colback=gray!5,
  colframe=black,
  title=Proof for NDL proof generation under an expanded set of rules,
  breakable
}

\newtcolorbox{ndlFROMbox}{
  colback=gray!5,
  colframe=black,
  title=Proof for NDL proof generation under an expanded set of rules plus the FROM rule,
  breakable
}

\newtcolorbox{equationalpcpromptbox}{
  colback=gray!5,
  colframe=black,
  title=Main Prompt for Equational Proof Checking,
  breakable
} 

\newtcolorbox{maskingpromptboxsequel}{
  colback=gray!5,
  colframe=black,
  title=Prompt Sequel for Proof Mask-Filling,
  breakable
} 

\newtcolorbox{hilbertpromptbox}{
  colback=gray!5,
  colframe=black,
  title=Prompt for Hilbert-Style Proof Generation with Expanded Axiom List,
  breakable
}

\newcommand\YAMLcolonstyle{\color{red}\mdseries}
\newcommand\YAMLkeystyle{\color{black}\bfseries}
\newcommand\YAMLvaluestyle{\color{blue}\mdseries}
\makeatletter
\newcommand\language@yaml{yaml}
\expandafter\expandafter\expandafter\lstdefinelanguage
\expandafter{\language@yaml}
{
  keywords={true,false,null,y,n},
  keywordstyle=\color{darkgray}\bfseries,
  basicstyle=\YAMLkeystyle,                                 % assuming a key comes first
  sensitive=false,
  comment=[l]{\#},
  morecomment=[s]{/*}{*/},
  commentstyle=\color{purple}\ttfamily,
  stringstyle=\YAMLvaluestyle\ttfamily,
  moredelim=[l][\color{orange}]{\&},
  moredelim=[l][\color{magenta}]{*},
  moredelim=**[il][\YAMLcolonstyle{:}\YAMLvaluestyle]{:},   % switch to value style at :
  morestring=[b]',
  morestring=[b]",
  literate =    {---}{{\ProcessThreeDashes}}3
                {>}{{\textcolor{red}\textgreater}}1     
                {|}{{\textcolor{red}\textbar}}1 
                {\ -\ }{{\mdseries\ -\ }}3,
}
\lst@AddToHook{EveryLine}{\ifx\lst@language\language@yaml\YAMLkeystyle\fi}
\makeatother
\newcommand\ProcessThreeDashes{\llap{\color{cyan}\mdseries-{-}-}}

\newcommand{\boxsparkcolored}[8]{%
  \begingroup
  \pgfmathsetmacro{\W}{3.2}
  \pgfmathsetmacro{\a}{#1}
  \pgfmathsetmacro{\b}{#7}
  \pgfmathsetmacro{\denom}{max(\b-\a,1e-9)}
  \pgfmathsetmacro{\xMin}{0}
  \pgfmathsetmacro{\xMax}{\W}
  \pgfmathsetmacro{\xPfive}{\W*(#2-\a)/\denom}
  \pgfmathsetmacro{\xQone}{\W*(#3-\a)/\denom}
  \pgfmathsetmacro{\xMed}{\W*(#4-\a)/\denom}
  \pgfmathsetmacro{\xQthree}{\W*(#5-\a)/\denom}
  \pgfmathsetmacro{\xPninetyFive}{\W*(#6-\a)/\denom}
  \pgfmathsetmacro{\xMean}{\W*(#8-\a)/\denom}
  \begin{tikzpicture}[x=1cm,y=1cm,baseline=1.2ex]
    \draw[gray!30, line width=0.2pt] (\xMin,0) -- (\xMax,0);
    \draw[gray!70, line width=0.4pt] (\xPfive,0) -- (\xPninetyFive,0);
    \draw[gray!70, line width=0.4pt] (\xPfive,-0.15) -- (\xPfive,0.15);
    \draw[gray!70, line width=0.4pt] (\xPninetyFive,-0.15) -- (\xPninetyFive,0.15);
    \draw[fill=gray!20, draw=gray!70, line width=0.4pt] (\xQone,-0.25) rectangle (\xQthree,0.25);
    \draw[red, line width=1.0pt] (\xMed,-0.25) -- (\xMed,0.25);
    \fill[black] (\xMean,0) circle (0.045);
    \node[font=\tiny, anchor=north east] at (\xMin,-0.38) {#1};
    \node[font=\tiny, anchor=north west] at (\xMax,-0.38) {#7};
    \node[font=\tiny, anchor=south]      at (\xMed, 0.34) {#4};
  \end{tikzpicture}%
  \endgroup
}

\title{Stress-Testing the Reasoning Competence of LLMs With Proofs \\  Under Minimal Formalism \\[0.12in] 
{\rm\fontsize{12}{14}\selectfont (Extended Technical Report\thanks{This article 
is an extended version of the 2025 EMNLP Findings paper 
{\em Stress-Testing the Reasoning Competence of LLMs With Formal Proofs}.
It contains substantial new material beyond the conference version, including new tasks, results, methods, and analyses.
}\hspace*{0.05in})
}}

\author{Konstantine Arkoudas\thanks{Corresponding author} \\
  \texttt{konstantine.arkoudas@gmail.com} \\\And
  Serafim Batzoglou \\
  \texttt{serafim.batzoglou@gmail.com} \\}

\tikzset{
  >=Latex,
  condlabel/.style = {midway, font=\footnotesize\itshape, text=black}, % horizontal edge labels
}
\forestset{
  mytree/.style={
    for tree = {
      draw,
      edge = {->, very thick, draw=gray!70, /tikz/shorten <= 0pt},
      rounded corners,
      align = center,
      inner sep = 3pt,
      s sep = 10mm,      % base horizontal spacing
      l sep = 12mm,      % vertical spacing
      anchor = north,
      blur shadow,
    },
    where level=0{for children={s sep=26mm}}{},
    where level=1{for children={s sep=24mm}}{},
  },
  decision/.style   = {circle, draw=teal!60!black, fill=teal!10, very thick, minimum size=10mm},
  leaf/.style       = {rectangle, rounded corners=3pt, very thick, inner sep=3pt},
}

\forestset{
  decision/.style = {circle, draw=teal!60!black, fill=teal!10, very thick, minimum size=7mm},
}
\begin{document}

\maketitle

\newcounter{algorithm}[subsection]
\renewcommand{\thealgorithm}{\thesection.\arabic{algorithm}}

\newcommand{\mendmarker}{\rule{1.2ex}{1.2ex}}
%% Theorems: 

%   \theoremstyle{marginbreak}
 \theoremstyle{plain}
   \theoremheaderfont{\normalfont\bfseries}\theorembodyfont{\slshape}
%   \theoremsymbol{\ensuremath{\diamondsuit}}
   \theoremseparator{:}
   \newtheorem{theorem}{Theorem}

 \theoremstyle{plain}
   \theoremheaderfont{\normalfont\bfseries}\theorembodyfont{\slshape}
%   \theoremsymbol{\ensuremath{\diamondsuit}}
   \theoremseparator{:}
   \newtheorem{proposition}{Proposition}

 \theoremstyle{plain}
   \theoremheaderfont{\normalfont\bfseries}\theorembodyfont{\slshape}
   \theoremseparator{:}
   \newtheorem{problem}{Problem}

%% Lemmas:

   \theoremstyle{plain}
%   \theoremsymbol{\ensuremath{\heartsuit}}
 %  \theoremindent0.5cm
%   \theoremnumbering{greek}
   \newtheorem{lemma}[theorem]{Lemma}
%% Corollaries:
  
   \theoremindent0cm
%   \theoremsymbol{\ensuremath{\spadesuit}}
   \theoremnumbering{arabic}
   \newtheorem{corollary}[theorem]{Corollary}
%% Examples:
  
   \theoremstyle{plain}
   \theorembodyfont{\upshape}
   \theoremsymbol{\ensuremath{\mendmarker}}
   \theoremseparator{:}
   \newtheorem{example}{Example}
%% Definitions:
  
   \theoremstyle{plain}
   \theoremsymbol{\ensuremath{\mendmarker}}
   \theoremseparator{:}
   \newtheorem{definition}{Definition}
%% Proofs:   
   \theoremheaderfont{\sc}\theorembodyfont{\upshape}
   \theoremstyle{nonumberplain}
   \theoremseparator{:}
   \theoremsymbol{\mendmarker}
   \newtheorem{proof}{Proof}

\newcounter{exc}
\newcounter{inexc}
\newcounter{Romanc}
\newcounter{romanc}
                
\newenvironment{exlist} {\begin{list} {{\bf \arabic{exc}}.} {\usecounter{exc}} 
\addtolength{\itemsep}{-0.6\itemsep}}{\end{list}}

\newenvironment{inexlist} {\begin{list} {(\alph{inexc})} {\usecounter{inexc}}}{\end{list}}
               
\newenvironment{romaninexlist} {\begin{list} {(\roman{inexc})} {\usecounter{inexc}}}{\end{list}}

\newenvironment{Romanlist} {\begin{list} {(\Roman{Romanc})} {\usecounter{Romanc}}}{\end{list}}

\newenvironment{romanlist} {\begin{list} {(\roman{romanc})} {\usecounter{romanc}}}{\end{list}}

\newcommand{\restrict}{\mbox{$\restriction$}}
\newcommand{\restrictsp}{\mbox{$\restriction$} }
\newcommand{\ar}{\mbox{\em r}}
\newcommand{\arsp}{\mbox{\em r} }
\newcommand{\arity}{\mbox{\em r}}
\newcommand{\aritysp}{\mbox{\em r} }
\newcommand{\dom}[1]{\mbox{\rm $\temls{Dom}(#1)$}}
\newcommand{\domsp}{\mbox{\em Dom\/}}
\newcommand{\ran}{\mbox{\em Ran}}
\newcommand{\ransp}{\mbox{\em Ran }}
\newcommand{\Bool}{\mbox{\bf Bool}}
\newcommand{\Boolsp}{\mbox{\bf Bool} }
\newcommand{\eps}{\mbox{$\epsilon$}}
\newcommand{\epsp}{\mbox{$\epsilon$} }
\newcommand{\sig}{\mbox{$\sigma$}}
\newcommand{\sigsp}{\mbox{$\sigma$} }
\newcommand{\Sig}{\mbox{$\Sigma$}}
\newcommand{\Sigsp}{\mbox{$\Sigma$} }
\newcommand{\Gam}{\mbox{$\Gamma$}}
\newcommand{\Gamsp}{\mbox{$\Gamma$} }
\newcommand{\gam}{\mbox{$\gamma$}}
\newcommand{\gamsp}{\mbox{$\gamma$} }
\newcommand{\mrho}{\mbox{$\rho$}}
\newcommand{\mrhosp}{\mbox{$\rho$ }}
\newcommand{\msigma}{\mbox{$\sigma$}}
\newcommand{\msigmasp}{\mbox{$\sigma$} }
\newcommand{\mtheta}{\mbox{$\theta$}}
\newcommand{\mthetasp}{\mbox{$\theta$ }}
\newcommand{\al}{\mbox{$\alpha$}}
\newcommand{\alt}{\mbox{$\alpha$ }}
\newcommand{\bet}{\mbox{$\beta$}}
\newcommand{\bett}{\mbox{$\beta$ }}
\newcommand{\mlam}{\mbox{$\lambda$}}
\newcommand{\mlamsp}{\mbox{$\lambda$} }
\newcommand{\Lam}{\mbox{$\Lambda$}}
\newcommand{\Lamt}{\mbox{$\Lambda$ }}
\newcommand{\negsp}{\mbox{$\neg$ }}
\newcommand{\negnsp}{\mbox{$\neg$}}
\newcommand{\mand}{\mbox{$\:\wedge \:$}}
\newcommand{\mandsp}{\mbox{$\wedge$ }}
\newcommand{\mor}{\mbox{$\:\vee \:$}}
\newcommand{\morsp}{\mbox{$\vee$ }}
\newcommand{\mif}{\mbox{$\:\Rightarrow\,$}}
\newcommand{\mlif}{\mbox{$\:\Leftarrow\,$}}
\newcommand{\mifsp}{\mbox{$\:\Rightarrow\,$ }}
\newcommand{\miff}{\mbox{$\:\Leftrightarrow \,$}}
\newcommand{\miffsp}{\mbox{$\Leftrightarrow$ }}
\newcommand{\mnegnsp}{\mbox{$\neg$}}
\newcommand{\mandnsp}{\mbox{$\wedge$}}
\newcommand{\mornsp}{\mbox{$\vee$}}
\newcommand{\mifnsp}{\mbox{$\Rightarrow$}}
\newcommand{\miffnsp}{\mbox{$\Leftrightarrow$}}
\newcommand{\at}{\mbox{\em ATOM}}
\newcommand{\tvs}{\mbox{$\{${\bf T, F}$\}\:$}}
\newcommand{\tv}{\mbox{${\mathcal TV}$}}
\newcommand{\True}{\mbox{\bf T}}
\newcommand{\False}{\mbox{\bf F}}
\newcommand{\true}{\mbox{\bf true}}
\newcommand{\false}{\mbox{\bf false}}
\newcommand{\strue}{\mbox{\bf t}}
\newcommand{\sfalse}{\mbox{\bf f}}
\newcommand{\struesp}{\mbox{\bf t} }
\newcommand{\sfalsesp}{\mbox{\bf f} }
\newcommand{\Truesp}{\mbox{\bf T }}
\newcommand{\Falsesp}{\mbox{\bf F }}
\newcommand{\truesp}{\mbox{\bf true }}
\newcommand{\falsesp}{\mbox{\bf false }}
\newcommand{\sposet}{\mbox{$(A,\prec)$ }}
\newcommand{\wposet}{\mbox{$(A,\preceq)$ }}
\newcommand{\ssqposet}{\mbox{$(A,\sqpsub)$ }}
\newcommand{\wsqposet}{\mbox{$(A,\sqsub)$ }}
\newcommand{\rarrs}{\mbox{$\;\rightarrow\;$}}
\newcommand{\rarrls}{\mbox{$\,\rightarrow\:$}}
\newcommand{\dhrarr}{\mbox{$\twoheadrightarrow$}}
\newcommand{\dhrarrs}{\mbox{$\;\twoheadrightarrow\;$}}
\newcommand{\dhrarrls}{\mbox{$\,\twoheadrightarrow\,$}}
\newcommand{\larrls}{\mbox{$\,\leftarrow\;$}}
\newcommand{\rarrstarls}{\mbox{$\,\rightarrow^*\,$}}
\newcommand{\rarrcrossls}{\mbox{$\,\rightarrow+*\,$}}
\newcommand{\rarr}{\mbox{$\rightarrow$}}
\newcommand{\rarrt}{\mbox{$\rightarrow$} }
\newcommand{\Rarr}{\mbox{$\Rightarrow$}}
\newcommand{\Rarrls}{\mbox{$\,\Rightarrow\,$}}
\newcommand{\Rarrtc}{\mbox{$\,\Rightarrow^+ \,$}}
\newcommand{\Rarrtrc}{\mbox{$\,\Rightarrow^* \,$}}
\newcommand{\larr}{\mbox{$\leftarrow$}}
\newcommand{\Larr}{\mbox{$\Leftarrow$}}
\newcommand{\hrarr}{\mbox{$\hookrightarrow$}}
\newcommand{\hrarrsp}{\mbox{$\hookrightarrow$ }}
\newcommand{\hrarrls}{\mbox{$\:\hookrightarrow\:$}}
\newcommand{\sub}{\mbox{$\; \subseteq \;$}}
\newcommand{\psub}{\mbox{$\; \subset \;$}}
\newcommand{\super}{\mbox{$\; \supseteq \;$}}
\newcommand{\psuper}{\mbox{$\; \supset \;$}}
\newcommand{\lrarr}{\mbox{$\leftrightarrow$}}
\newcommand{\lrarrls}{\mbox{$\:\leftrightarrow\:$}}
\newcommand{\Lrarr}{\mbox{$\Leftrightarrow$}}
\newcommand{\lRarr}{\mbox{$\Longrightarrow$}}
\newcommand{\lRarrsp}{\mbox{$\Longrightarrow$ }}
\newcommand{\lRarrls}{\mbox{$\,\Longrightarrow\,$}}
\newcommand{\sep}{\mbox{$\; | \;$}}
\newcommand{\sepls}{\mbox{$\:\, | \:$}}
\newcommand{\sepms}{\mbox{$\; | \;$}}
\newcommand{\sats}{\mbox{$\, \models \,$}}
\newcommand{\nsats}{\mbox{$\, \not\models \,$}}
\newcommand{\nullset}{\mbox{$\emptyset\,\:$}}
\newcommand{\eset}{\mbox{$\emptyset$}}
\newcommand{\esetsp}{\mbox{$\emptyset$ }}
\newcommand{\sqsub}{\mbox{$\; \sqsubseteq \;$}}
\newcommand{\sqpsub}{\mbox{$\; \sqsubset \;$}}
\newcommand{\sqsup}{\mbox{$\; \sqsupseteq \;$}}
\newcommand{\sqpsup}{\mbox{$\; \sqsupset \;$}}
\newcommand{\bsquare}{\rule{2.2mm}{2.2mm}}
\newcommand{\bsig}{\mbox{$\;\bsquare$}}
\newcommand{\inc}{\mbox{\em s}}
\newcommand{\incsp}{\mbox{\em s} }
\newcommand{\tar}{\mbox{${\mathcal T}\hspace*{-0.017in}{\mathcal A}{\mathcal R}$}}
\newcommand{\tarsp}{\mbox{${\mathcal T}\hspace*{-0.017in}{\mathcal A}{\mathcal R}$} }
\newcommand{\zero}{\mbox{\em 0}}
\newcommand{\zerosp}{\mbox{\em 0 }}
\newcommand{\plus}{\mbox{\em plus}}
\newcommand{\plusp}{\mbox{\em plus }}
\newcommand{\mult}{\mbox{\em times}}
\newcommand{\multsp}{\mbox{\em times }}
\newcommand{\pset}[1]{\mbox{$\mathcal{P}(#1)$}}
\newcommand{\psetsp}[1]{\mbox{$\mathcal{P}(#1)$} }
\newcommand{\finpset}[1]{\mbox{$\mathcal{P}_{\infty}(#1)$}}
\newcommand{\finpsetsp}[1]{\mbox{$\mathcal{P}_{\infty}(#1)$} }
\newcommand{\ded}{\mbox{$\mathcal{D}$}}
\newcommand{\dedsp}{\mbox{$\mathcal{D}$} }
\newcommand{\lttp}{\mbox{\tt (}}
\newcommand{\lttpsp}{\mbox{\tt (} }
\newcommand{\rttp}{\mbox{\tt )}}
\newcommand{\rttpsp}{\mbox{\tt )} }
\newcommand{\mc}[1]{\mbox{$\mathcal{#1}$}}
\newcommand{\beq}[2]{\begin{equation} #1 \label{#2} \end{equation}}
\newcommand{\irule}[3]{
\begin{tabular}{cl} 
$\;\;$#1$\;$ & \mbox{}  \\[-0.099in] \hrulefill & \hspace*{-0.08in} \mbox{$\;\:$#3} \\[-0.03in] 
$\;\;$#2$\;$ & \mbox{} 
\end{tabular}}

\newcommand{\lirule}[3]{
\begin{tabular}{rc} 
\mbox{} & $\;\;$#1$\;$  \\[-0.099in] 
#3 \hspace*{-0.08in} & \hrulefill \\[-0.03in] 
\mbox{} & $\;\;$#2$\;$ 
\end{tabular}}

\newcommand{\extrairule}[4]{
\begin{tabular}{cl} 
\begin{tabular}{c}
$\;\;$#1$\;$ \\[-0.099in] 
\hrulefill  \\[-0.03in] $\;\;$#2$\;$ \end{tabular} & \hspace*{-0.2in} #3 \\
#4 & \mbox{} 
\end{tabular}}

\newcommand{\extralirule}[6]{
\begin{tabular}{rc}  
#3 \hspace*{#4} & 
\begin{tabular}{c}
$\;\;$#1$\;$ \\[-0.099in] 
\hrulefill  \\[-0.03in] $\;\;$#2$\;$ \end{tabular} \\[#6]
\mbox{} & #5 
\end{tabular}}

\newcommand{\mtt}[1]{\mbox{\tt #1}}
\newcommand{\smtt}[1]{\mbox{\small\tt #1}}
\newcommand{\fmtt}[1]{\mbox{\footnotesize\tt #1}}
\newcommand{\scmtt}[1]{\mbox{\scriptsize\tt #1}}
\newcommand{\mbf}[1]{\mbox{\bf #1}}
\newcommand{\mforall}{\mbox{$\forall$}}
\newcommand{\mforallsp}{\mbox{$\forall$} }
\newcommand{\mexists}{\mbox{$\exists$}}
\newcommand{\mexistsp}{\mbox{$\exists$} }
\newcommand{\freevar}{\mbox{\em FV\/}}
\newcommand{\freevarls}{\mbox{\em FV}\hspace*{0.006in}}
\newcommand{\freevarsp}{\mbox{\em FV\/} }
\newcommand{\boundvar}{\mbox{\em BV}}
\newcommand{\boundvarls}{\mbox{\em BV}\hspace*{0.006in}}
\newcommand{\boundvarsp}{\mbox{\em BV\/} }
\newcommand{\be}{\begin{enumerate}}
\newcommand{\ee}{\end{enumerate}}
\newcommand{\bi}{\begin{itemize}}
\newcommand{\ei}{\end{itemize}}
\renewcommand{\bit}{\begin{itemize}}
\newcommand{\eit}{\end{itemize}}
\newcommand{\bdesc}{\begin{description}}
\newcommand{\edesc}{\end{description}}
\newcommand{\ben}{\begin{enumerate}}
\newcommand{\een}{\end{enumerate}}
\newcommand{\tem}[1]{\mbox{\em #1}}
\newcommand{\temv}[1]{\mbox{\em #1\/}}
\newcommand{\temls}[1]{\mbox{\em #1}\hspace*{0.006in}}
\newcommand{\temts}[1]{\mbox{\em #1}\hspace*{0.009in}}
\newcommand{\temTMode}[1]{\mbox{#1}}
\newcommand{\temvTMode}[1]{\mbox{#1\/}}
\newcommand{\temlsTMode}[1]{\mbox{#1}\hspace*{0.006in}}
\newcommand{\mcup}{\mbox{$\cup$}}
\newcommand{\mcupsp}{\mbox{$\cup$ }}
\newcommand{\mcupls}{\mbox{$\:\cup\:$}}

\newcommand{\terms}[2]{\mbox{$\mbf{Terms}(#1,#2)$}}

\newcommand{\tterms}[3]{\mbox{$\mbf{Terms}(#1,#2,#3)$}}

\newcommand{\grterms}[1]{\mbox{$\mbf{Terms}(#1)$}}

\newcommand{\nats}{\mbox{$N$}}
\newcommand{\natsp}{\mbox{$N$ }}
\newcommand{\pnats}{\mbox{$N_+$}}
\newcommand{\pnatsp}{\mbox{$N_+$ }}
\newcommand{\bools}{\mbox{$B$}}
\newcommand{\boolsp}{\mbox{$B$ }}
\newcommand{\nat}{\mbox{nat}}
\newcommand{\formsub}[4]{$\mbox{#1}\hspace*{-0.04in}\overset{\small \mbox{#2}/\mbox{#3}}
{\leadsto}\hspace*{-0.04in}\mbox{#4}$}
\newcommand{\altformsub}[4]{$\mbox{#1}\leadsto\mbox{#2}\{\mbox{#3}/\mbox{#4}\}$}
\newcommand{\propsent}{\mbox{\bf PropSent}}
\newcommand{\propsentsp}{\mbox{\bf PropSent }}
\newcommand{\propat}{\mbox{\bf PropAt}}
\newcommand{\propatsp}{\mbox{\bf PropAt }}
\newcommand{\var}[1]{\mbox{\rm $\mbox{\em Var\/}(#1)$}}
\newcommand{\varls}{\mbox{\em Var}\hspace*{0.006in}}
\newcommand{\varsp}{\mbox{\em Var\/} }
\newcommand{\typsub}{\mbox{$\{v_1\mapsto t_1,\ldots,v_n \mapsto t_n\}$}}
\newcommand{\typsubsp}{\mbox{$\{v_1\mapsto t_1,\ldots,v_n \mapsto t_n\}$} }
\newcommand{\id}[1]{$\mbox{id}_{#1}$}
\newcommand{\idsp}[1]{$\mbox{id}_{#1}$ }
\newcommand{\aka}{\mbox{a.k.a. }}
\newcommand{\ie}{\mbox{i.e.} }
\newcommand{\iensp}{\mbox{i.e.}}
\newcommand{\Iensp}{\mbox{I.e.}}
\newcommand{\iesp}{\mbox{i.e. }}
\newcommand{\Ie}{\mbox{I.e.} }
\newcommand{\Iesp}{\mbox{I.e. }}
\newcommand{\etc}{\mbox{etc.}}
\newcommand{\etcsp}{\mbox{etc.} }
\newcommand{\eg}{\mbox{e.g.} }
\newcommand{\egnsp}{\mbox{e.g.}}
\newcommand{\Egnsp}{\mbox{E.g.}}
\newcommand{\Eg}{\mbox{E.g.} }
\newcommand{\egsp}{\mbox{e.g. }}
\newcommand{\Egsp}{\mbox{E.g. }}
\newcommand{\successor}{\mbox{$\,^{\prime}$}}
\newcommand{\successorsp}{\mbox{$\,^{\prime}\,$ }}
\newcommand{\mtilde}{\mbox{$\tilde{}\:$}}
\newcommand{\mbar}{\mbox{$\:\bar{}\:$}}
\newcommand{\mbarsp}{\mbox{$\:\bar{}\:$ }}
\newcommand{\mbarls}{\mbox{$\,\bar{}\,$}}

\newcommand{\mqed}{\hfill \rule{5pt}{5pt}}
\newtheorem{solution}{Solution}

\newenvironment{solutionFollowedByNewParagraph}{{\bf Solution}. }{\vspace*{-0.03in} \\}

\newcommand{\bs}{\begin{solution}}
\newcommand{\es}{\end{solution}}
\newcommand{\bsnp}{\begin{solutionFollowedByNewParagraph}}
\newcommand{\esnp}{\end{solutionFollowedByNewParagraph}}

\newenvironment{alphaEnum}{\begin{list}{(\alph{enumi})}{\usecounter{enumi}
\setlength{\leftmargin}{0.44in}}}{\end{list}}
\newcommand{\mprime}[1]{\mbox{$#1'$}}

\newcommand{\fcomp}{\mbox{$\cdot$}}
\newcommand{\fcompls}{\mbox{$\:\cdot\:$}}
\newcommand{\fcompsp}{\mbox{$\cdot$} }

\newcommand{\lam}[2]{\mbox{$\lambda\,#1\,.\,#2$}}
\newcommand{\lamsp}[2]{\mbox{$\lambda\,#1\,.\,#2$} }
\newcommand{\plam}[2]{\mbox{$(\lambda\,#1\,.\,#2)$}}
\newcommand{\plamsp}[2]{\mbox{$(\lambda\,#1\,.\,#2$)} }

\newcommand{\fix}[2]{\mbox{$\mbf{fix}\,#1\,.\,#2$}}
\newcommand{\fixsp}[2]{\mbox{$\mbf{fix}\,#1\,.\,#2$} }

\newcommand{\wh}[1]{\mbox{$\widehat{#1}$}}
\renewcommand\wh[1]{\hstretch{2}{\hat{\hstretch{.5}{#1}}}}
\newcommand\whs[1]{\hstretch{2}{\hat{\hstretch{.5}{\hspace*{-0.02in}#1}}}}

\newcommand{\ol}[1]{\mbox{$\overline{#1}$}}

\newcommand{\mul}[1]{\mbox{$\underline{#1}$}}

\newcommand{\coloredul}[2]{\mbox{\coloreduline{#1}{#1}}}

\newcommand{\blueul}[1]{\mbox{$\mathunderline{blue}{#1}$}}
\newcommand{\redul}[1]{\mbox{$\mathunderline{red}{#1}$}}

\newcommand{\nth}[1]{\mbox{$#1^{\mbox{\scriptsize\em th}}$}}
\newcommand{\nthsp}[1]{\mbox{$#1^{\mbox{\scriptsize\em th}}$} }

\newcommand{\darr}{\mbox{$\downarrow$}}
\newcommand{\uarr}{\mbox{$\downarrow$}}

\newcommand{\mless}{\mbox{\symbol{60}}}
\newcommand{\mlesssp}{\mbox{\symbol{60}} }

\newcommand{\mgreater}{\mbox{\symbol{62}}}
\newcommand{\mgreatersp}{\mbox{\symbol{62}} }

\newcommand{\types}{\mbox{$\vdash$}}
\newcommand{\typessp}{\mbox{$\vdash$} }
\newcommand{\typesls}{\mbox{$\;\vdash\;$}}

\newcommand{\mmin}[2]{\mbox{$\mbox{min}_{#1}\,#2$}}
\newcommand{\mmax}[2]{\mbox{$\mbox{max}_{#1}\,#2$}}

\newcommand{\proves}{\mbox{$\vdash$}}
\newcommand{\provesp}{\mbox{$\vdash$} }
\newcommand{\provesls}{\mbox{$\;\vdash\:$}}

\newcommand{\tproves}[1]{\mbox{$\vdash_{#1}$}}
\newcommand{\tprovesp}[1]{\mbox{$\vdash_{#1}$} }
\newcommand{\tprovesls}[1]{\mbox{$\:\vdash_{#1}\:$}}

\newcommand{\tnproves}[1]{\mbox{$\not{\hspace*{-0.007in}\vdash_{#1}}$}}
\newcommand{\tnprovesp}[1]{\mbox{$\not{\hspace*{-0.007in}\vdash_{#1}}$} }
\newcommand{\tnprovesls}[1]{\mbox{$\not{\hspace*{-0.007in}\vdash_{#1}}\:$} }

\newcommand{\mdef}[1]{\mbox{\em #1\/}}

\newcommand{\mdash}{\mbox{\symbol{45}}}
\newcommand{\mdashsp}{\mbox{\symbol{45}} }

\newcommand{\singleton}[1]{\mbox{$\langle #1 \rangle$}}

\newcommand{\lefttuplemarker}{\mbox{$\langle$}}
\newcommand{\righttuplemarker}{\mbox{$\rangle$}}

\newcommand{\mpair}[2]{\mbox{$\langle #1, #2\rangle$}}
\newcommand{\mpairsp}[2]{\mbox{$\langle #1, #2\rangle$} }

\newcommand{\mtriple}[3]{\mbox{$\langle #1, #2, #3\rangle$}}
\newcommand{\mtriplesp}[3]{\mbox{$\langle #1, #2, #3\rangle$} }

\newcommand{\mquadruple}[4]{\mbox{$\langle #1, #2, #3, #4\rangle$}}
\newcommand{\mquadruplesp}[4]{\mbox{$\langle #1, #2, #3, #4\rangle$} }

\newcommand{\ttlp}{\mbox{\tt (}}
\newcommand{\ttlpsp}{\mbox{\tt ( }}
\newcommand{\ttrp}{\mbox{\tt )}}
\newcommand{\ttrpsp}{\mbox{\tt ) }}

\newcommand{\lamc}{\mbox{$\lambda$-calculus}}
\newcommand{\lamcsp}{\mbox{$\lambda$-calculus} }

\newcommand{\concat}[2]{\mbox{$#1\mbox{::}#2$}}
\newcommand{\append}[2]{\mbox{$#1\oplus#2$}}
\newcommand{\appendsymbol}{\mbox{$\oplus$}}
\newcommand{\appendsymbolsp}{\mbox{$\oplus$} }

\newcommand{\prefix}{\mbox{$\sqsubset$}}
\newcommand{\prefixeq}{\mbox{$\sqsubseteq$}}
\newcommand{\prefixls}{\mbox{$\:\sqsubset\:$}}
\newcommand{\prefixeqls}{\mbox{$\:\sqsubseteq\:$}}

\newcommand{\card}[1]{\mbox{$|#1|$}}

\newcommand{\longpage}{\enlargethispage{\baselineskip}}
\newcommand{\shortpage}{\enlargethispage{-\baselineskip}}
\newcommand{\clearemptydoublepage}{\newpage{\pagestyle{empty}\cleardoublepage}}

\newcommand{\mphi}{\mbox{$\Phi$}}
\newcommand{\mphisp}{\mbox{$\Phi$} }

\newcommand{\rsp}{\mbox{$\;\;\;\;\;$}}

\newcommand{\aquote}[1]{\mbox{$\lceil #1\rceil$}}

\newcommand{\elsesym}{\mbox{$\diamondsuit$}}
\newcommand{\elsesymls}{\mbox{$\,\diamondsuit\,$}}
\newcommand{\elsesymsp}{\mbox{$\diamondsuit$} }
\newcommand{\mcond}[3]{\mbox{$#1\,$?$\,\rarrls #2 \, \elsesym \,#3$}}

\newcommand{\derives}{\mbox{$\Vdash$}}
\newcommand{\derivesp}{\mbox{$\Vdash$} }
\newcommand{\derivesls}{\mbox{$\:\Vdash\:$}}

\newcommand{\subderives}[1]{\mbox{$\Vdash_{\mbox{\scriptsize #1}}$}}
\newcommand{\subderivesp}[1]{\mbox{$\Vdash_{\mbox{\scriptsize #1}}$} }
\newcommand{\subderivesls}[1]{\mbox{$\:\Vdash_{\mbox{\scriptsize #1}}\:$}}

\newcommand{\subprovesls}[1]{\mbox{$\:\vdash_{\mbox{\scriptsize #1}}\:$}}
\newcommand{\subproves}[1]{\mbox{$\vdash_{\mbox{\scriptsize #1}}$}}
\newcommand{\subprovesp}[1]{\mbox{$\vdash_{\mbox{\scriptsize #1}}$} }

\newcommand{\propsof}[1]{\mbox{$\mbf{Prop}[#1]$}}
\newcommand{\propsofsp}[1]{\mbox{$\mbf{Prop}[#1]$} }

\newcommand{\abof}[1]{\mbox{$\mbf{AB}[#1]$}}
\newcommand{\abofsp}[1]{\mbox{$\mbf{AB}[#1]$} }

\newcommand{\dedsof}[1]{\mbox{$\mbf{Ded}[#1]$}}
\newcommand{\dedsofsp}[1]{\mbox{$\mbf{Ded}[#1]$} }

\newcommand{\mlb}{\textnormal{[\kern-.15em[}}
\newcommand{\mrb}{\textnormal{]\kern-.15em]}}

\newcommand{\denot}[2]{\mbox{$#1\mlb#2\mrb$}}
\newcommand{\denotsp}[2]{\mbox{$#1\mlb#2\mrb$} }

\newcommand{\yields}{\mbox{$\hookrightarrow$}}
\newcommand{\yieldsls}{\mbox{$\:\hookrightarrow\:$}}
\newcommand{\yieldsp}{\mbox{$\hookrightarrow$} }

\newcommand{\msp}{\mbox{$\;\:$}}

\newcommand{\mssp}{\mbox{$\;$}}

\newcommand{\fappsp}{\mbox{$\;\,$}}

\newcommand{\ala}[1]{\mbox{\`{a} la #1}}

\newcommand{\lc}{\mbox{$\lambda$-calculus}}
\newcommand{\lcsp}{\mbox{$\lambda$-calculus} }

\newcommand{\funtype}[2]{\mbox{$#1 \rarrls #2$}}
\newcommand{\prodtype}[2]{\mbox{$#1 \times #2$}}
\newcommand{\sumtype}[2]{\mbox{$#1 + #2$}}

\newcommand{\fpslc}{\mbox{$\lambda_{\,\rarr,\,\times,\,+}$}}
\newcommand{\fpslcsp}{\mbox{$\lambda_{\,\rarr,\,\times,\,+}$} }

\newcommand{\lapp}[2]{\mbox{$#1\;#2$}}
\newcommand{\lappsp}[2]{\mbox{$#1\;#2$} }

\newcommand{\lproj}[1]{\mbox{$\pi_1(#1)$}}
\newcommand{\rproj}[1]{\mbox{$\pi_2(#1)$}}

\newcommand{\mcase}[3]{\mbox{$\mbf{case}(#1,#2,#3)$}}
\newcommand{\mcasesp}[3]{\mbox{$\mbf{case}(#1,#2,#3)$} }

\newcommand{\linj}[2]{\mbox{$\mbf{in-left}_{#1}(#2)$}}
\newcommand{\linjsp}[2]{\mbox{$\mbf{in-left}_{#1}(#2)$} }
\newcommand{\rinj}[2]{\mbox{$\mbf{in-right}_{#1}(#2)$}}
\newcommand{\rinjsp}[2]{\mbox{$\mbf{in-right}_{#1}(#2)$} }
\newcommand{\qdot}{\mbox{$\ . \ $}}

\newcommand{\wrt}{\mbox{w.r.t.} } 

\newcommand{\mrestriction}[2]{\mbox{$#1 \restriction #2$}}

\newcommand{\lessgen}{\mbox{$\preceq$}}
\newcommand{\lessgensp}{\mbox{$\preceq$ }}
\newcommand{\varment}{\mbox{$\rarr_{\text{M}}$}}
\newcommand{\varmentsp}{\mbox{$\rarr_{\text{M}}$} }
\newcommand{\varmentls}{\mbox{$\:\rarr_{\text{M}}\:$}}
\newcommand{\algmod}{\mbox{Mod}}
\newcommand{\algmodsp}{\mbox{Mod }}
\newcommand{\gterms}{\mbox{$\terms{\fsyms}{\variables}$}}
\newcommand{\gtermsp}{\mbox{$\terms{\fsyms}{\variables}$ }}
\newcommand{\natmap}{\mbox{nat}}
\newcommand{\natmapsp}{\mbox{nat }}
\newcommand{\height}{\mbox{hgt}}
\newcommand{\heightsp}{\mbox{hgt }}
\newcommand{\subcomp}{\mbox{$\circ$}}
\newcommand{\subeq}{\mbox{$\equiv$}}
\newcommand{\subeqls}{\mbox{$\:\equiv\:$}}
\newcommand{\subeqsp}{\text{$\equiv$ }}
\newcommand{\subcompsp}{\mbox{$\circ$} }
\newcommand{\subcompls}{\mbox{$\:\circ\:$}}
\newcommand{\compsub}[2]{\mbox{$#1 \subcompls #2$}}
\newcommand{\eqsfv}{\text{$\text{Eqn}(\fsyms,V)$}}
\newcommand{\eqsfvsp}{\text{$\text{Eqn}(\fsyms,V)$} }
\newcommand{\subalg}[1]{\text{\em Sub$\{\hspace*{-0.01in}#1\}$}}
\newcommand{\tmsubalg}[1]{\text{Sub$(\hspace*{-0.01in}#1)$}} %% (``Theorem-Mode'')
\newcommand{\simalg}[1]{$\text{Sim}\{#1\}$}
\newcommand{\con}{\tem{Con}}
\newcommand{\conls}{\temls{Con}}
\newcommand{\consp}{\temv{Con} }
\newcommand{\conTMode}{\temTMode{Con}}
\newcommand{\conlsTMode}{\temlsTMode{Con}}
\newcommand{\conspTMode}{\temvTMode{Con} }
\newcommand{\iso}{\text{$\cong$}}
\newcommand{\isosp}{\text{$\cong$ }}
\newcommand{\isols}{\text{$\:\cong\:$}}
\newcommand{\grar}{\text{\bf GA}}
\newcommand{\grarsp}{\text{\bf GA} }
\newcommand{\pa}{\text{\mc{P}\hspace*{-0.02in}\mc{A}}}
\newcommand{\pasp}{\text{\mc{P}\hspace*{-0.02in}\mc{A}} }
\newcommand{\quot}[1]{\text{$#1/\hspace*{-0.04in}\equiv$}}
\newcommand{\quotls}[1]{\text{$#1/\hspace*{-0.04in}\equiv\,$}}
\newcommand{\quotsp}[1]{\text{$#1/\hspace*{-0.04in}\equiv$ }}
\newcommand{\indexQuot}[2]{\text{$#1/\hspace*{-0.04in}\equiv_{#2}$}}
\newcommand{\indexQuotls}[2]{\text{$#1/\hspace*{-0.04in}\equiv_{#2}\,$}}
\newcommand{\indexQuotsp}[2]{\text{$#1/\hspace*{-0.04in}\equiv_{#2}$ }}
\newcommand{\alg}[1]{\text{$\text{Alg}(#1)$}}
%% Standard algebra:
\newcommand{\salg}{\text{$\mc{A}$}}
\newcommand{\salgsp}{\text{$\mc{A}$} }

\newcommand{\ranvar}[1]{\text{\rm $\text{\em RanVar\/}(#1)$}}
\newcommand{\ranvarsp}{\text{\em RanVar} }
\newcommand{\ranvarnsp}{\text{\em RanVar}}
\newcommand{\supp}[1]{\text{\rm $\text{\em Supp\/}(#1)$}}
\newcommand{\uppsp}{\text{\em Supp} }
\newcommand{\tvar}{\text{Var}}
\newcommand{\tvarsp}{\text{Var} }
\newcommand{\hex}[1]{\ol{#1}}
\newcommand{\hext}{\text{$\hex{\theta}$}}
\newcommand{\hextsp}{\text{$\hex{\theta}$} }

\newcommand{\unifset}[1]{\text{$U(#1)$}}
\newcommand{\unifsetsp}{\text{$U$}}

\newcommand{\termdom}[1]{\text{$\temls{TDom}(#1)$}}
\newcommand{\termdomsp}{\temv{TDom} }
\newcommand{\fdom}[1]{\text{$\temls{FunDom}(#1)$}}
\newcommand{\fdomsp}{\temv{FunDom} }
\newcommand{\vdom}[1]{\text{$\temls{VarDom}(#1)$}}
\newcommand{\vdomsp}{\temv{VarDom} }

\newcommand{\comp}{\mbox{$\cdot$}}
\newcommand{\compsp}{\mbox{$\cdot$} }
\newcommand{\compls}{\mbox{$\:\cdot\:$}}

\newcommand{\ssig}{\text{$\Sigma$}}
\newcommand{\ssigsp}{\text{$\Sigma$} }

\newcommand{\termlabel}[1]{\text{$\temls{TLabel}(#1)$}}
\newcommand{\termlabelsp}{\temv{TLabel} }
\newcommand{\termlabelnsp}{\tem{TLabel}}

\newcommand{\hil}{\text{$\mathcal{H}$}}
\newcommand{\hilb}{\text{\hil\hspace*{-0.02in}}}
\newcommand{\hilsp}{\text{$\mathcal{H}$ }}
\newcommand{\hilder}{\text{$\vdash_{\mathcal{H}}$}}
\newcommand{\hilderls}{\text{$\:\vdash_{\mathcal{H}}\:$}}
\newcommand{\hildersp}{\text{$\vdash_{\mathcal{H}}$ }}
\newcommand{\jder}{\text{$\vdash_{\mathcal{J}}$}}
\newcommand{\jderls}{\text{$\:\vdash_{\mathcal{J}}\:$}}
\newcommand{\jdersp}{\text{$\vdash_{\mathcal{J}}$ }}
\newcommand{\hilj}{\text{$\mathcal{J}$}}
\newcommand{\hiljb}{\text{$\mathcal{J}$\hspace*{-0.02in}}}
\newcommand{\hiljsp}{\text{$\mathcal{J}$ }}
\newcommand{\hiljder}{\text{$\vdash_{\mathcal{J}}$}}
\newcommand{\hiljderls}{\text{$\:\vdash_{\mathcal{J}}\:$}}
\newcommand{\hiljdersp}{\text{$\vdash_{\mathcal{J}}$ }}
\newcommand{\propform}{\text{\bf P}}
\newcommand{\propformsp}{\text{\bf P} }
\newcommand{\judge}{\text{$\Gamma \rarrls A$}}
\newcommand{\judgesp}{\text{$\Gamma \rarrls A$} }
\newcommand{\nhil}[1]{\text{$\hil_{#1}$}}
\newcommand{\nhilsp}[1]{\text{$\hil_{#1}$ }}
\newcommand{\nhilj}[1]{\text{$\hilj_{#1}$}}
\newcommand{\nhiljsp}[1]{\text{$\hilj_{#1}$ }}

\newcommand{\zhil}{\text{$\hilb_0$}}
\newcommand{\zhilsp}{\text{$\hilb_0$ }}
\newcommand{\zhilj}{\text{$\hiljb_0$}}
\newcommand{\zhiljsp}{\text{$\hiljb_0$ }}
\newcommand{\ohil}{\text{$\hilb_1$}}
\newcommand{\ohilsp}{\text{$\hilb_1$ }}
\newcommand{\ohilj}{\text{$\hiljb_1$}}
\newcommand{\ohiljsp}{\text{$\hiljb_1$ }}

\newcommand{\zhilder}{\text{$\vdash_{\mathcal{H}_0}$}}
\newcommand{\zhilderls}{\text{$\:\vdash_{\mathcal{H}_0}\:$}}
\newcommand{\zhildersp}{\text{$\vdash_{\mathcal{H}_0}$ }}

\newcommand{\ohilder}{\text{$\vdash_{\mathcal{H}_1}$}}
\newcommand{\ohilderls}{\text{$\:\vdash_{\mathcal{H}_1}\:$}}
\newcommand{\ohildersp}{\text{$\vdash_{\mathcal{H}_1}$ }}

\newcommand{\zhiljder}{\text{$\vdash_{\mathcal{J}_0}$}}
\newcommand{\zhiljderls}{\text{$\:\vdash_{\mathcal{J}_0}\:$}}
\newcommand{\zhiljdersp}{\text{$\vdash_{\mathcal{J}_0}$ }}
\newcommand{\ohiljder}{\text{$\vdash_{\mathcal{J}_1}$}}
\newcommand{\ohiljderls}{\text{$\:\vdash_{\mathcal{J}_1}\:$}}
\newcommand{\ohiljdersp}{\text{$\vdash_{\mathcal{J}_1}$ }}

%% NEW FOL COMMANDS:

\newcommand{\sfola}{\text{$\Sigma_{\text{\it fol}}$}}
\newcommand{\sfolasp}{\text{$\Sigma_{\text{\it fol}}$} }
\newcommand{\lvoc}{\text{$\mathcal{L}$}}
\newcommand{\lvocsp}{\text{$\mathcal{L}$} }
\newcommand{\variables}{\text{$\mathcal{V}$}}
\newcommand{\variablesp}{\text{$\mathcal{V}$} }
%% Standard variable set: 
\newcommand{\svars}{\text{$V$}}
\newcommand{\svarsp}{\text{$V$}}
\newcommand{\csyms}{\text{$\mathcal{C}$}}
\newcommand{\csymsp}{\text{$\mathcal{C}$} }
\newcommand{\fsyms}{\text{$\mathcal{F}$}}
\newcommand{\fsymsp}{\text{$\mathcal{F}$} }
\newcommand{\rsyms}{\text{$\mathcal{R}$}}
\newcommand{\rsymsp}{\text{$\mathcal{R}$} }
\newcommand{\psig}{\text{$\mathbf{\Omega}$}}
\newcommand{\psigvar}{\text{$(\mathbf{\Omega},\variables)$}}
\newcommand{\fsig}{\text{$\mathbf{\Sigma}$}}
%% ``Standard'' relational signature:
\newcommand{\spsig}{\text{$(\csyms,\fsyms,\rsyms)$}}
\newcommand{\spsigsp}{\text{$(\consyms,\fsyms,\rsyms)$} }

\newcommand{\svoc}{\text{$(\psig,\variables)$}}
\newcommand{\svocsp}{\text{$(\psig,\variables)$} }

\newcommand{\psigsp}{\text{$\mathbf{\Omega}$} }
\newcommand{\consymbols}{\text{$\mathcal{C}$}}
\newcommand{\consymbolsp}{\text{$\mathcal{C}$} }
\newcommand{\funsig}{\text{$\mathcal{F}$}}
\newcommand{\funsigsp}{\text{$\mathcal{F}$} }
\newcommand{\relsig}{\text{$\mathcal{R}$}}
\newcommand{\relsigsp}{\text{$\mathcal{R}$} }

\newcommand{\arlvoc}{\text{$\mathcal{L}_{\text{\footnotesize\em ar}}$}}
\newcommand{\arlvocsp}{\text{$\mathcal{L}_{\text{\footnotesize\em ar}}$ }}

\newcommand{\arsig}{\text{$\psig_{\text{\em Ar}}$}}
\newcommand{\arsigsp}{\text{$\psig_{\text{\em Ar}}$} }

\newcommand{\lfam}{\text{$\mc{L}_{\text{\em Fam}}$}}
\newcommand{\lfamsp}{\text{$\mc{L}_{\text{\em Fam}}$ }}

\newcommand{\lblock}{\text{$\mc{L}_{\text{\em Blocks}}$}}
\newcommand{\lblocksp}{\text{$\mc{L}_{\text{\em Blocks}}$ }}

\newcommand{\inter}{\mc{I}}
\newcommand{\intersp}{\mc{I} }
\newcommand{\struc}{\mc{D}}
\newcommand{\strucsp}{\mc{D} }
\newcommand{\interar}{\text{$\mathcal{I}_{\text{\em Ar}}$}}
\newcommand{\interarsp}{\text{$\mathcal{I}_{\text{\em Ar}}$ }}

\newcommand{\arstruc}{\text{$\struc_{\text{\em Ar}}$}}
\newcommand{\arstrucsp}{\text{$\struc_{\text{\em Ar}}$} }

\newcommand{\form}[1]{\text{$\text{\bf Form}_{#1}$}}
\newcommand{\tform}[2]{\text{$\text{\bf Form}(#1,#2)$}}
\newcommand{\forml}{\form{\lvoc}}
\newcommand{\formlsp}{\form{\lvoc} }

\newcommand{\alphaeq}{\text{$\approx_{\alpha}$}}
\newcommand{\alphaeqsp}{\text{$\approx_{\alpha}$} }
\newcommand{\alphaeqls}{\text{$\:\approx_{\alpha}\,$}}

\newcommand{\aleq}{\text{$\approx_{\alpha}$}}
\newcommand{\aleqsp}{\text{$\approx_{\alpha}$} }
\newcommand{\aleqls}{\text{$\:\approx_{\alpha}\,$}}

%% Apply a substitution to a variable:
\newcommand{\appv}[2]{\text{$#1(#2)$}}
%% Apply a substitution to a term: 
\newcommand{\appt}[2]{\text{$\hex{#1}(#2)$}}
%% Apply an extended substitution---in long hand---to a term: 
\newcommand{\appte}[2]{\text{$#1\,#2$}}
\newcommand{\appf}[2]{\text{$#1\,#2$}}

%% Apply a valuation to a term: 
\newcommand{\valappt}[2]{\text{$\hex{#1}(#2)$}}
%% Apply a valuation to a variable:
\newcommand{\valappv}[2]{\text{$#1(#2)$}}

%% Subscripted valappt: 
\newcommand{\svalappt}[3]{\text{$\hex{#1}_{#2}(#3)$}}

\newcommand{\quant}[3]{\text{$(#1\,#2)\,#3$}}

\newcommand{\fvar}[1]{\text{\rm $\text{\em FV\/}(#1)$}}
\newcommand{\fvarsp}{\text{\rm\text{\em FV\/}} }

\newcommand{\bvar}[1]{\text{\rm $\text{\em BV\/}(#1)$}}
\newcommand{\bvarsp}{\text{\rm\text{\em BV\/}} }

\newcommand{\valmod}[2]{\text{\rm $#1/#2$}}

%% Subscripted valmod:
\newcommand{\svalmod}[3]{\text{\rm $#1/_{#2}\,#3$}}

%% ``Standard'' logic vocabulary:
\newcommand{\slvoc}{\text{$(\consymbols,\funsig,\relsig,\variables)$}}
\newcommand{\slvocsp}{\text{$(\consymbols,\funsig,\relsig,\variables)$} }

\newcommand{\ssub}[2]{\text{$\{#1 \mapsto #2\}$}}
\newcommand{\extend}[3]{\text{$#1[#2 \mapsto #3]$}}
\newcommand{\hv}{\text{$\mathbf{V}$}}
\newcommand{\hvsp}{\text{$\mathbf{V}$} }

\newcommand{\allvars}{\text{$\variables \cup \mathbf{V}$}}
\newcommand{\allvarsp}{\text{$\variables \cup \mathbf{V}$} }

\newcommand{\hlvoc}{\text{$\lvoc^{\hv}$}}
\newcommand{\hlvocsp}{\text{$\lvoc^{\hv}$} }

\newcommand{\hvoc}{\text{$(\psig,\variables \cup \hv)$}}
\newcommand{\hvocsp}{\text{$(\psig,\variables \cup \hv)$} }

\newcommand{\jone}{\text{$\mc{J}_1$}}
\newcommand{\jonesp}{\text{$\mc{J}_1$} }

\newcommand{\jprovesls}{\tprovesls{\jone}}

\newcommand{\njproves}{\text{$\not{\hspace*{-0.007in}\vdash_{\jone}$}}}

\newcommand{\njprovesp}{\text{$\not{\hspace*{-0.007in}\vdash_{\jone}}$} }

\newcommand{\njprovesls}{\text{$\:\not{\hspace*{-0.007in}\vdash_{\jone}}\,$}}

\newcommand{\mequiv}{\text{$\equiv$}}
\newcommand{\mequivls}{\text{$\:\equiv\:$}}
\newcommand{\mequivsp}{\text{$\equiv$} }

\newcommand{\cnd}{\text{\mc{NDL}}}
\newcommand{\cndsp}{\text{\mc{NDL}} }

\newcommand{\ndl}{\text{\mc{NDL}}}
\newcommand{\ndlsp}{\text{\mc{NDL}} }

\newcommand{\cndo}{\text{$\cnd_1$}}
\newcommand{\cndosp}{\text{$\cnd_1$} }

\newcommand{\cndz}{\text{$\cnd_0$}}
\newcommand{\cndzsp}{\text{$\cnd_0$} }

\newcommand{\ndlz}{\text{$\cnd_0$}}
\newcommand{\ndlzsp}{\text{$\cnd_0$} }

\newcommand{\ndlo}{\text{$\cnd_1$}}
\newcommand{\ndlosp}{\text{$\cnd_1$} }

\newcommand{\cndzomega}{\text{$\cnd_0^{\omega}$}}
\newcommand{\cndzomegasp}{\text{$\cnd_0^{\omega}$} }

\newcommand{\ndlzomega}{\text{$\cnd_0^{\omega}$}}
\newcommand{\ndlzomegasp}{\text{$\cnd_0^{\omega}$} }

\newcommand{\ndloomega}{\text{$\cnd_1^{\omega}$}}
\newcommand{\ndloomegasp}{\text{$\cnd_1^{\omega}$} }

\newcommand{\cndoomega}{\text{$\cnd_1^{\omega}$}}
\newcommand{\cndoomegasp}{\text{$\cnd_1^{\omega}$} }

\newcommand{\rds}{\mbox{${\cal R}$}}
\newcommand{\rdsp}{\mbox{${\cal R}$ }}
\newcommand{\rder}{\mbox{$\vdash_{\cal R}$}}
\newcommand{\rderls}{\mbox{$\: \vdash_{\cal R}\:$}}
\newcommand{\rdersp}{\mbox{$\vdash_{\cal R}$ }}

\newcommand{\cndprovesls}{\text{$\:\vdash_{\text{\scriptsize \cnd}}\:$}}
\newcommand{\cndproves}{\text{$\vdash_{\text{\scriptsize \cnd}}$}}
\newcommand{\cndprovesp}{\text{$\vdash_{\text{\scriptsize \cnd}}$} }
\newcommand{\ncndprovesls}
{\text{$\:\not{\hspace*{-0.007in}\vdash_{\scriptsize \cnd}}\:$}}

\newcommand{\wf}{\text{$\vdash_{\text{\em\tiny W}}\;$}}
\newcommand{\wfnsp}{\text{$\vdash_{\text{\em\scriptsize W}}$}}
\newcommand{\wfsp}{\text{$\vdash_{\text{\em\scriptsize W}}$}}

\newcommand{\man}{\text{$\mathcal{M}$}}
\newcommand{\mansp}{\text{$\mathcal{M}$ }}
\newcommand{\mander}{\text{$\vdash_{\mathcal{M}}$}}
\newcommand{\manderls}{\text{$\:\vdash_{\mathcal{M}}\:$}}
\newcommand{\mandersp}{\text{$\vdash_{\mathcal{M}}$ }}

\newcommand{\myprop}{\mbox{\bf P}}
\newcommand{\mypropsp}{\mbox{\bf P} }

\newcommand{\props}{\text{Props}}

\newcommand{\ainter}{\text{$\mathcal{I}$}}  
\newcommand{\aintersp}{\text{$\mathcal{I}$} }  
\newcommand{\aprop}{\text{\bf Prop}}
\newcommand{\apropsp}{\text{\bf Prop} }
\newcommand{\aide}{\text{\bf Ide}}
\newcommand{\aidesp}{\text{\bf Ide} }
\newcommand{\freeid}{\text{\em FId}$\,$}
\newcommand{\freeidls}{\text{\em FId$\,$ }}
\newcommand{\freeidsp}{\text{\em FId\/} }
\newcommand{\freepatid}{\text{\em FPtId}$\,$}
\newcommand{\freepatidls}{\text{\em FPtId$\,$ }}
\newcommand{\freepatidsp}{\text{\em FPtId\/} }
\newcommand{\patvar}{\text{\em PtVar}$\,$}
\newcommand{\patvarsp}{\text{\em PtVar$\,$} }
\newcommand{\patidsp}{\text{\em PatId\/} }
\newcommand{\avar}{\text{\bf Var}}
\newcommand{\avarsp}{\text{\bf Var} }
\newcommand{\apat}{\text{\bf Pat}}
\newcommand{\apatsp}{\text{\bf Pat} }
\newcommand{\compat}{\text{\bf ComPat}}
\newcommand{\compatsp}{\text{\bf ComPat} }
\newcommand{\aexp}{\text{\bf Exp}}
\newcommand{\aexpsp}{\text{\bf Exp} }
\newcommand{\aded}{\text{\bf Ded}}
\newcommand{\adedsp}{\text{\bf Ded} }
\newcommand{\aphrase}{\text{\bf Phr}}
\newcommand{\aphrasesp}{\text{\bf Phr} }
\newcommand{\avexp}{\text{\bf ValExp}}
\newcommand{\avexpsp}{\text{\bf ValExp} }
\newcommand{\ttrue}{\text{\tt true}}
\newcommand{\ttruesp}{\text{\tt true} }
\newcommand{\ttfalse}{\text{\tt false}}
\newcommand{\ttfalsesp}{\text{\tt false} }
\newcommand{\aerror}{\text{\em error\/}}
\newcommand{\aerrorsp}{\text{\em error\/ }}
\newcommand{\idwo}{\text{$\prec$}}
\newcommand{\idwosp}{\text{$\prec$} }
\newcommand{\adom}{\text{$\mathcal{D}$}}
\newcommand{\adomsp}{\text{$\mathcal{D}$ }}
\newcommand{\aval}{\text{\bf Val}}
\newcommand{\avalsp}{\text{\bf Val }}
\newcommand{\propval}{\text{\bf PropVal}}
\newcommand{\propvalsp}{\text{\bf PropVal }}
\newcommand{\fval}{\text{\bf FVal}}
\newcommand{\fvalsp}{\text{\bf FVal }}
\newcommand{\tval}{\text{\bf MVal}}
\newcommand{\tvalsp}{\text{\bf MVal }}
\newcommand{\topval}{\text{\bf TopVal}}
\newcommand{\topvalsp}{\text{\bf TopVal }}
\newcommand{\propc}{\text{\em prop}}
\newcommand{\propcls}{\text{\em prop$\,$}}
\newcommand{\propcsp}{\text{\em prop\/} }
\newcommand{\fclos}{\text{\em fclos}}
\newcommand{\fclosls}{\text{\em fclos$\,$}}
\newcommand{\fclosp}{\text{\em fclos\/} }
\newcommand{\tclos}{\text{\em mclos}}
\newcommand{\tclosls}{\text{\em mclos$\,$}}
\newcommand{\tclosp}{\text{\em mclos\/} }
\newcommand{\aop}{\text{\em op}}
\newcommand{\aopls}{\text{\em op$\,$}}
\newcommand{\aopsp}{\text{\em op} }
\newcommand{\arule}{\text{\em rule}}
\newcommand{\arulels}{\text{\em rule$\,$}}
\newcommand{\arulesp}{\text{\em rule} }
\newcommand{\aenv}{\text{$\rho$}}
\newcommand{\aenvsp}{\text{$\rho$ }}
\newcommand{\ab}{\text{$\beta$}}
\newcommand{\absp}{\text{$\beta$ }}
\newcommand{\aenvdom}{\text{\em Env}}
\newcommand{\aenvdomsp}{\text{\em Env} }
\newcommand{\aprocdom}{\text{\em Proc}}
\newcommand{\aprocdomsp}{\text{\em Proc} }
\newcommand{\unbound}{\text{\em unbound}}
\newcommand{\unboundsp}{\text{\em unbound} }
\newcommand{\emean}{\text{$\mathcal{E}$}}
\newcommand{\emeansp}{\text{$\mathcal{E}$ }}
\newcommand{\cmean}{\text{$\mathcal{A}$}}
\newcommand{\cmeansp}{\text{$\mathcal{A}$ }}
\newcommand{\mmean}{\text{$\mathcal{M}$}}
\newcommand{\mmeansp}{\text{$\mathcal{M}$ }}
\newcommand{\smallsome}{\text{\small SOME}}
\newcommand{\smallsomesp}{\text{\small SOME} }
\newcommand{\smallnone}{\text{\small NONE}}
\newcommand{\smallnonesp}{\text{\small NONE} }
\newcommand{\enew}{\text{$E_{\text{new}}$}}
\newcommand{\enewsp}{\text{$E_{\text{new}}$} }
\newcommand{\isub}{\text{$I_{\text{sub}}$}}
\newcommand{\isubsp}{\text{$I_{\text{sub}}$} }
\newcommand{\sab}{\text{$\{\enew/\isub\}$}}
\newcommand{\sabsp}{\text{$\{\enew/\isub\}$} }
\newcommand{\apsem}{\text{$\mathcal{A}\mathcal{S}$}}
\newcommand{\apsemsp}{\text{$\mathcal{A}\mathcal{S}$} }
\newcommand{\tupn}{\text{\em tup$_n$}}
\newcommand{\tupnsp}{\text{\em tup$_n$} }
\newcommand{\proj}{\text{\em proj}}
\newcommand{\projsp}{\text{\em proj} }
\newcommand{\cmode}{\text{$\mathcal{C}$}}
\newcommand{\cmodesp}{\text{$\mathcal{C}$} }
\newcommand{\dmode}{\text{$\mathcal{D}$}}
\newcommand{\dmodesp}{\text{$\mathcal{D}$} }
\newcommand{\cder}{\text{$\vdash_{\mathcal C}$}}
\newcommand{\cderls}{\text{$\;\,\vdash_{\mathcal C}\:$}}
\newcommand{\cdersp}{\text{$\vdash_{\mathcal C}$} }
\newcommand{\dder}{\text{$\vdash_{\mathcal D}$}}
\newcommand{\dderls}{\text{$\:\,\vdash_{\mathcal D}\:$}}
\newcommand{\ddersp}{\text{$\vdash_{\mathcal D}$} }
\newcommand{\ttand}{\text{\tt and}}
\newcommand{\ttandsp}{\text{\tt and }}
\newcommand{\ttor}{\text{\tt or}}
\newcommand{\ttorsp}{\text{\tt or }}
\newcommand{\ttif}{\text{\tt if}}
\newcommand{\ttifsp}{\text{\tt if }}
\newcommand{\ttiff}{\text{\tt iff}}
\newcommand{\ttiffsp}{\text{\tt iff }}
\newcommand{\ttnot}{\text{\tt not}}
\newcommand{\ttnotsp}{\text{\tt not }}
\newcommand{\ttqm}{\text{\tt ?}}
\newcommand{\ttqmsp}{\text{\tt ?} }
\newcommand{\ttbq}{\text{\tt `}}
\newcommand{\ttbqsp}{\text{\tt `} }
\newcommand{\ttdollar}{\text{\tt \$}}
\newcommand{\ttdollarsp}{\text{\tt \$} }

\newcommand{\ttin}{\text{\tt in}}
\newcommand{\ttinsp}{\text{\tt in }}
\newcommand{\ttul}{\text{\tt \_}}
\newcommand{\ttulsp}{\text{\tt \_ }}
\newcommand{\topname}{\text{\em TopName}}
\newcommand{\topnamesp}{\text{\em TopName\/ }}
\newcommand{\logop}{\text{\em LogOp}}
\newcommand{\logopsp}{\text{\em LogOp\/ }}
\newcommand{\primrule}{\text{\em Prim-Rule}}
\newcommand{\primrulesp}{\text{\em Prim-Rule\/ }}
\newcommand{\mlceil}{\text{$\lceil$}}
\newcommand{\mlceilsp}{\text{$\lceil$ }}
\newcommand{\mrceil}{\text{$\rceil$}}
\newcommand{\mrceilsp}{\text{$\rceil$ }}
\newcommand{\mlfloor}{\text{$\lfloor$}}
\newcommand{\mlfloorsp}{\text{$\lfloor$ }}
\newcommand{\mrfloor}{\text{$\rfloor$}}
\newcommand{\mrfloorsp}{\text{$\rfloor$ }}
\newcommand{\eval}[1]{\text{$\temls{Eval}(#1)$}}
\newcommand{\evalsp}{\text{\em Eval\/} }
\newcommand{\evalnsp}{\temv{Eval}}
\newcommand{\sev}[1]{\text{$\temls{ev}(#1)$}}
\newcommand{\evals}{\text{$\:\leadsto\;$}}
\newcommand{\tevals}{\text{$\:\leadsto^*\:$}}
\newcommand{\eevals}{\text{$\:\hookrightarrow\:$}}
\newcommand{\error}{\text{\tem{error}}}
\newcommand{\errorsp}{\text{\tem{error}} }
\newcommand{\proptype}{\text{\bf prop}}
\newcommand{\proptypesp}{\text{\bf prop }}
\newcommand{\env}{\text{\bf Env}}
\newcommand{\envsp}{\text{\bf Env }}

\newcommand{\fdl}{\text{\mc{FND}}}
\newcommand{\fdlsp}{\text{\mc{FND}} }

\newcommand{\obeqproves}{\text{$\vdash_{\text{\scriptsize OE}}\,$}}

\newcommand{\obeq}{\text{$\approx$}}
\newcommand{\obeqsp}{\text{$\approx$} }
\newcommand{\obeqls}{\text{$\:\approx\,$}}
\newcommand{\nobeqls}{\text{$\:\not\approx\,$}}

\newcommand{\subobeq}[1]{\text{$\approx_{#1}$}}
\newcommand{\subobeqsp}[1]{\text{$\approx_{#1}$} }
\newcommand{\subobeqls}[1]{\text{$\:\approx_{#1}\,$}}
\newcommand{\subnobeqls}[1]{\text{$\:\not\approx_{#1}\,$}}

\newcommand{\give}{\text{$\rightarrowtail$}}
\newcommand{\givesp}{\text{$\rightarrowtail$} }
\newcommand{\gives}{\text{$\,\rightarrowtail\:$}}
\newcommand{\ngives}{\text{$\:\not\rightarrowtail\,$}}

\newcommand{\dep}{\text{$\twoheadrightarrow$}}
\newcommand{\depsp}{\text{$\twoheadrightarrow$} }
\newcommand{\depls}{\text{$\,\twoheadrightarrow\:$}}
\newcommand{\ndepls}{\text{$\,\not\twoheadrightarrow\:$}}

\newcommand{\mdep}{\text{$\Rrightarrow$}}
\newcommand{\mdepsp}{\text{$\Rrightarrow$} }
\newcommand{\mdepls}{\text{$\,\Rrightarrow\:$}}
\newcommand{\mndepls}{\text{$\,\not\Rrightarrow\:$}}

\newcommand{\tcdep}{\text{$\twoheadrightarrow^+$}}
\newcommand{\tcdepsp}{\text{$\twoheadrightarrow^+$} }
\newcommand{\tcdepls}{\text{$\,\twoheadrightarrow^+\:$}}
\newcommand{\tcndepls}{\text{$\,\not\twoheadrightarrow^+\:$}}

\newcommand{\ttcdep}[1]{\text{$\twoheadrightarrow_{#1}^+$}}
\newcommand{\ttcdepsp}[1]{\text{$\twoheadrightarrow_{#1}^+$} }
\newcommand{\ttcdepls}[1]{\text{$\,\twoheadrightarrow_{#1}^+\:$}}
\newcommand{\ttncdepls}[1]{\text{$\,\not\twoheadrightarrow_{#1}^+\:$}}

\newcommand{\strict}{\text{$\vdash_{\text{\em Strict}}$}}
\newcommand{\strictsp}{\text{$\vdash_{\text{\em Strict}}$} }
\newcommand{\strictls}{\text{$\vdash_{\text{\em Strict}}\,$}}

\newcommand{\el}[1]{\text{$\mc{E}(#1)$}}

\newcommand{\fass}[1]{\text{\rm $\text{\em FA\/}(#1)$}}
\newcommand{\fassp}{\text{\rm\text{\em FA\/}} }
\newcommand{\fassnsp}{\text{\rm\text{\em FA}}}

\newcommand{\st}[1]{\fass{#1}}

\newcommand{\fst}[1]{\text{$\tem{Strict}(#1)$}}

\newcommand{\concl}[1]{\text{$\mc{C}(#1)$}}

\newcommand{\conclnsp}{\mc{C}}

\newcommand{\conclsp}{\mc{C} }

\newcommand{\fconcl}[1]{\text{$\tem{Con}(#1)$}}

\newcommand{\mhex}[1]{\widehat{#1}}

\newcommand{\grab}[1]{\text{$\mless #1 \mgreater\:$}}
\newcommand{\grabsp}[1]{\text{$\mless #1 \mgreater$} }

\newcommand{\sz}[1]{\text{$\temls{SZ}(#1)$}}

\newcommand{\lsz}[1]{\text{$\temls{LSZ}(#1)$}}

\newcommand{\allprops}[2]{\text{$\text{\bf Prop}(#1,#2)$}}

\newcommand{\appdsub}[3]{\text{$#1\{#2/#3\}$}}

\newcommand{\dsub}[2]{\text{$\{#1/#2\}$}}

\newcommand{\mappt}[2]{\text{$\hex{#1}(#2)$}}

\newcommand{\fa}[1]{\text{$\tem{FA}(#1)$}}

\newcommand{\abdom}{\mc{B}}
\newcommand{\abdomsp}{\mc{B} }

\newcommand{\err}{\tem{error}}
\newcommand{\errsp}{\tem{error} }

\newcommand{\PrimDed}{\mbf{PrimDed}}
\newcommand{\PrimDedsp}{\mbf{PrimDed} }

\newcommand{\pdom}{\mbf{Prop}}
\newcommand{\pdomsp}{\mbf{Prop} }

\newcommand{\Luk}{\mbox{{\L}ukasiewicz}}
\newcommand{\Jask}{\mbox{J\'{a}skowski}}

\newcommand{\Luksp}{\mbox{{\L}ukasiewics} }
\newcommand{\Jasksp}{\mbox{J\'{a}skowski} }

\newcommand{\ass}[2]{\text{$\mbf{assume}\;#1\,.\,#2$}}
\newcommand{\assp}[2]{\text{$\mbf{assume}\;#1\,.\,#2$} }

\newcommand{\assin}[2]{\text{$\mbf{assume}\,\;#1\,\;\mbf{in}\;#2$}}
\newcommand{\assinsp}[2]{\text{$\mbf{assume}\;#1\,\mbf{in}\;#2$} }

\newcommand{\supab}[2]{\text{$\mbf{suppose-absurd}\,\;#1\,\;\mbf{in}\;#2$}}
\newcommand{\supabsp}[2]{\text{$\mbf{suppose-absurd}\;#1\,\mbf{in}\;#2$} }

\newcommand{\abst}[2]{\text{$\lambda\;#1\,.\,#2$}}
\newcommand{\abstsp}[2]{\text{$\lambda\;#1\,.\,#2$} }

\newcommand{\type}{\text{$\tau$}}
\newcommand{\typesp}{\text{$\tau$} }

\newcommand{\xt}{\text{$x_{\tau}$}}
\newcommand{\xtsp}{\text{$x_{\tau}$} }

\newcommand{\xs}{\text{$x_{\sigma}$}}
\newcommand{\xsp}{\text{$x_{\sigma}$} }

\newcommand{\modus}[2]{\text{$\mbf{mp}(#1,#2)$}}
\newcommand{\modusp}[2]{\text{$\mbf{mp}(#1,#2)$} }

\newcommand{\app}[2]{\text{$\mbf{app}(#1,#2)$}}
\newcommand{\appsp}[2]{\text{$\mbf{app}(#1,#2)$} }

\newcommand{\both}[2]{\text{$\mbf{both}(#1,#2)$}}
\newcommand{\bothsp}[2]{\text{$\mbf{both}(#1,#2)$} }

\newcommand{\pair}[2]{\text{$\mbf{pair}(#1,#2)$}}
\newcommand{\pairsp}[2]{\text{$\mbf{pair}(#1,#2)$} }

\newcommand{\leftand}[1]{\text{$\mbf{left}(#1)$}}
\newcommand{\leftandsp}[1]{\text{$\mbf{left}(#1)$} }

\newcommand{\first}[1]{\text{$\mbf{first}(#1)$}}
\newcommand{\firstsp}[1]{\text{$\mbf{first}(#1)$} }

\newcommand{\secondsp}[1]{\text{$\mbf{second}(#1)$} }

\newcommand{\rightand}[1]{\text{$\mbf{right}(#1)$}}
\newcommand{\rightandsp}[1]{\text{$\mbf{right}(#1)$} }

\newcommand{\fv}[1]{\text{$\tem{FV}(#1)$}}
\newcommand{\fvsp}[1]{\text{$\tem{FV}(#1)$} }

\newcommand{\ch}{\text{Curry-Howard}}
\newcommand{\chsp}{\text{Curry-Howard} }

\newcommand{\genproof}{\text{$\mathfrak{D}$}}

\newcommand{\genproofsp}{\text{$\mathfrak{D}$} }

%% Alphabetic equivalence for first-order CND proofs: 
\newcommand{\paleq}{\text{$\equiv_{\epsilon}$}}
\newcommand{\paleqsp}{\text{$\equiv_{\epsilon}$} }
\newcommand{\paleqls}{\text{$\:\equiv_{\epsilon}\,$}}

\newcommand{\eigenrename}[1]{\text{$\temls{EigenRename}(#1)$}}
\newcommand{\eigenrenamensp}{\tem{EigenRename}}
\newcommand{\eigenrenamesp}{\temv{EigenRename} }

\newcommand{\evar}[1]{\text{\rm $\text{\em EV\/}(#1)$}}
\newcommand{\evarsp}{\text{\rm\text{\em EV\/}} }

\newcommand{\formdom}[1]{\text{$\temls{FDom}(#1)$}}
\newcommand{\formdomsp}{\temv{FDom} }
\newcommand{\formdomnsp}{\tem{FDom}}

\newcommand{\formlabel}[1]{\text{$\temls{FLabel}(#1)$}}
\newcommand{\formlabelsp}{\temv{FLabel} }
\newcommand{\formlabelnsp}{\tem{FLabel}}

\newcommand{\proofdom}[1]{\text{$\temls{DDom}(#1)$}}
\newcommand{\proofdomsp}{\temv{DDom} }

\newcommand{\prooflabel}[1]{\text{$\temls{DLabel}(#1)$}}
\newcommand{\prooflabelsp}{\temv{DLabel} }
\newcommand{\prooflabelnsp}{\tem{DLabel}}

\newcommand{\appd}[2]{\text{$#1\,#2$}}
\newcommand{\apppd}[2]{\text{$#1\;#2$}}

\newcommand{\deductionset}[1]{\text{$\mbf{Ded}(#1)$}}

\newcommand{\formalappf}[2]{\text{${#1}^{\sharp}(#2)$}}
\newcommand{\formalappfnsp}[1]{\text{${#1}^{\sharp}$}}
\newcommand{\formalappd}[2]{\text{${#1}^{\star}(#2)$}}
\newcommand{\formalappdnsp}[1]{\text{${#1}^{\star}$}}

\newcommand{\Phisp}{\text{$\Phi$ }}

\newcommand{\apprule}[2]{\text{$#1\; #2$}}

\newcommand{\rulemp}{\mbf{modus-ponens}}
\newcommand{\rulempsp}{\mbf{modus-ponens} }
\newcommand{\ruleclaim}{\mbf{claim}}
\newcommand{\ruleclaimsp}{\mbf{claim} }
\newcommand{\rulemt}{\mbf{modus-tollens}}
\newcommand{\rulemtsp}{\mbf{modus-tollens} }
\newcommand{\ruledn}{\mbf{double-negation}}
\newcommand{\rulednsp}{\mbf{double-negation} }
\newcommand{\ruleland}{\mbf{left-and}}
\newcommand{\rulelandsp}{\mbf{left-and} }
\newcommand{\rulerand}{\mbf{right-and}}
\newcommand{\rulerandsp}{\mbf{right-and} }

\newcommand{\ruleliff}{\mbf{left-iff}}
\newcommand{\ruleriff}{\mbf{right-iff}}
\newcommand{\ruleequiv}{\mbf{equivalence}}

\newcommand{\ruleboth}{\mbf{both}}
\newcommand{\rulebothsp}{\mbf{both} }

\newcommand{\ruleabsurd}{\mbf{absurd}}
\newcommand{\ruleabsurdsp}{\mbf{absurd} }

\newcommand{\zlevel}{\text{0-level}}
\newcommand{\zlevelsp}{\text{0-level} }
\newcommand{\Zlevel}{\text{0-level}}
\newcommand{\Zlevelsp}{\text{0-level} }

\newcommand{\typealpha}{\text{type-$\alpha$}}
\newcommand{\typealphasp}{\text{type-$\alpha$} }
\newcommand{\Typealpha}{\text{Type-$\alpha$}}
\newcommand{\Typealphasp}{\text{Type-$\alpha$} }

\newcommand{\typeomega}{\text{type-$\omega$}}
\newcommand{\typeomegasp}{\text{type-$\omega$} }
\newcommand{\Typeomega}{\text{Type-$\omega$}}
\newcommand{\Typeomegasp}{\text{Type-$\omega$} }

\newcommand{\inflevel}{\text{$\infty$-level}}
\newcommand{\inflevelsp}{\text{$\infty$-level} }
\newcommand{\Inflevel}{\text{$\infty$-level}}
\newcommand{\Inflevelsp}{\text{$\infty$-level} }

\newcommand{\mtype}[1]{\text{type-$#1$}}
\newcommand{\mtypesp}[1]{\text{type-$#1$}}

\newcommand{\mType}[1]{\text{Type-$#1$}}
\newcommand{\mTypesp}[1]{\text{Type-$#1$}}

\newcommand{\assume}[2]{\text{$\mbf{assume } #1 \mbf{ in } #2$}}
\newcommand{\msupab}[2]{\text{$\mbf{suppose-absurd } #1 \mbf{ in } #2$}}

\newcommand{\bpropcon}{\text{$\odot$}}
\newcommand{\bpropconls}{\text{$\:\odot\:$}}
\newcommand{\bpropconsp}{\text{$\odot$} }

\newcommand{\lm}{\text{$\lambda\hspace*{0.01in}\phi$}}
\newcommand{\lmsp}{\text{$\lambda\hspace*{0.01in}\phi$} }
\newcommand{\lmc}{\text{$\lambda\hspace*{0.01in}\phi$-calculus}}
\newcommand{\lmcsp}{\text{$\lambda\hspace*{0.01in}\phi$-calculus} }

\newcommand{\Phr}{\rm \text{\em Phr}}
\newcommand{\Phrsp}{\rm \text{\em Phr\/} }
\newcommand{\Exp}{\rm \text{\em Exp}}
\newcommand{\Expsp}{\rm \text{\em Exp\/} }
\newcommand{\Ded}{\rm \text{\em Ded}}
\newcommand{\Dedsp}{\rm \text{\em Ded\/} }
\newcommand{\Ide}{\rm \text{\em Ide}}
\newcommand{\Idesp}{\rm \text{\em Ide\/} }

\newcommand{\mab}[2]{\text{$\phi\,#1\,.\,#2$}}
\newcommand{\mabop}{\text{$\phi$}}
\newcommand{\mabopsp}{\text{$\phi$} }
\newcommand{\lab}[2]{\text{$\lambda\,#1\,.\,#2$}}

\newcommand{\dapp}[2]{\text{$\mbf{dapp}(#1,#2)$}}
\newcommand{\dappthunk}[1]{\text{$\mbf{dapp}(#1)$}}
\newcommand{\eapp}[2]{\text{$\mbf{app}(#1,#2)$}}
\newcommand{\eappthunk}[1]{\text{$\mbf{app}(#1)$}}

\newcommand{\dcase}[3]{\text{$\mbf{dcase}(#1,#2,#3)$}}
\newcommand{\ecase}[3]{\text{$\mbf{case}(#1,#2,#3)$}}

\newcommand{\prop}{\tem{Prop}}
\newcommand{\propsp}{\tem{Prop} }

\newcommand{\fexp}{\tem{FExp}}
\newcommand{\fexpsp}{\tem{FExp} }
\newcommand{\mexp}{\tem{MExp}}
\newcommand{\mexpsp}{\tem{MExp} }

\newcommand{\values}{\mc{V}}
\newcommand{\valuesp}{\mc{V} }

\newcommand{\conv}{\text{$\doteq$}}
\newcommand{\convls}{\text{$\:\doteq\:$}}
\newcommand{\convsp}{\text{$\doteq$} }

\newcommand{\semeq}{\text{$\approx$}}
\newcommand{\semeqls}{\text{$\:\approx\:$}}
\newcommand{\semeqsp}{\text{$\approx$} }

\newcommand{\valeq}{\text{$\equiv$}}
\newcommand{\valeqls}{\text{$\:\equiv\:$}}
\newcommand{\valeqsp}{\text{$\equiv$} }

\newcommand{\domega}{\text{$\Omega_{\mbf{D}}$}}
\newcommand{\domegasp}{\text{$\Omega_{\mbf{D}}$} }

\newcommand{\eomega}{\text{$\Omega_{\mbf{E}}$}}
\newcommand{\eomegasp}{\text{$\Omega_{\mbf{E}}$} }

\newcommand{\mvec}[1]{\text{$\overrightarrow{#1}$}}
\newcommand{\mvecsp}[1]{\text{$\overrightarrow{#1}$} }

\newcommand{\red}{\text{$\rightarrowtail$}}
\newcommand{\redls}{\text{$\:\rightarrowtail\:$}}
\newcommand{\redsp}{\text{$\rightarrowtail$} }

\newcommand{\lmcon}[1]{\text{$\mathsf{#1}$}}

\newcommand{\tred}{\text{$\rightarrowtail^*$}}
\newcommand{\tredls}{\text{$\:\rightarrowtail^*\:$}}
\newcommand{\tredsp}{\text{$\rightarrowtail^*$} }

\newcommand{\convg}[2]{\text{$#1\hspace*{-0.04in}\downarrow_{#2}$}}
\newcommand{\divg}[2]{\text{$#1\hspace*{-0.04in}\uparrow_{#2}$}}

\newcommand{\sconvg}[1]{\text{$#1\hspace*{-0.04in}\downarrow$}}
\newcommand{\sdivg}[1]{\text{$#1\hspace*{-0.04in}\uparrow$}}

\newcommand{\cclaim}{\lmcon{claim}}
\newcommand{\cclaimsp}{\lmcon{claim} }

\newcommand{\cmp}{\lmcon{mp}}
\newcommand{\cmpsp}{\lmcon{mp} }

\newcommand{\ctrue}{\lmcon{true}}
\newcommand{\ctruesp}{\lmcon{true} }

\newcommand{\cfalse}{\lmcon{false}}
\newcommand{\cfalsesp}{\lmcon{false} }

\newcommand{\cmt}{\lmcon{mt}}
\newcommand{\cmtsp}{\lmcon{mt} }

\newcommand{\ctrueaxiom}{\lmcon{T\text{-}axiom}}
\newcommand{\ctrueaxiomsp}{\lmcon{T\text{-}axiom} }
\newcommand{\cfalseaxiom}{\lmcon{F\text{-}axiom}}
\newcommand{\cfalseaxiomsp}{\lmcon{F\text{-}axiom} }

\newcommand{\cdn}{\lmcon{dn}}
\newcommand{\cdnsp}{\lmcon{dn} }
\newcommand{\ccd}{\lmcon{cd}}
\newcommand{\cboth}{\lmcon{both}}
\newcommand{\cleither}{\lmcon{left\text{-}either}}
\newcommand{\creither}{\lmcon{right\text{-}either}}
\newcommand{\cequiv}{\lmcon{equiv}}
\newcommand{\cliff}{\lmcon{left\text{-}iff}}
\newcommand{\criff}{\lmcon{right\text{-}iff}}
\newcommand{\cabsurd}{\lmcon{absurd}}
\newcommand{\cland}{\lmcon{left\text{-}and}}
\newcommand{\crand}{\lmcon{right\text{-}and}}

\newcommand{\map}[1]{\text{$\mlb #1\mrb$}}
\newcommand{\mapsp}[1]{\text{$\mlb #1\mrb$} }

\newcommand{\exm}{\mbf{!}}
\newcommand{\exmsp}{\mbf{!} }
\newcommand{\exmls}{\mbf{!}\hspace*{0.01in}}

\newcommand{\noeval}{\text{$\tem{Eval}_{\text{\scriptsize\em N}}$}}

\newcommand{\aoeval}{\text{$\tem{Eval}_{\text{\scriptsize\em A}}$}}

\newcommand{\ceval}{\text{$\tem{Eval}_{\text{\scriptsize\em C}}$}}

\newcommand{\lmnot}{\lmcon{not}}
\newcommand{\lmnotsp}{\lmcon{not} }

\newcommand{\lmand}{\lmcon{and}}
\newcommand{\lmandsp}{\lmcon{and} }

\newcommand{\lmor}{\lmcon{or}}
\newcommand{\lmorsp}{\lmcon{or} }

\newcommand{\lmif}{\lmcon{if}}
\newcommand{\lmifsp}{\lmcon{if} }

\newcommand{\lmiff}{\lmcon{iff}}
\newcommand{\lmiffsp}{\lmcon{iff} }

\newcommand{\lmcond}{\mbf{cond}}
\newcommand{\lmcondsp}{\mbf{cond} }

\newcommand{\lmdcond}{\mbf{dcond}}
\newcommand{\lmdcondsp}{\mbf{dcond} }

\newcommand{\suapp}[2]{\text{$(#1\msp #2)$}}
\newcommand{\sbapp}[3]{\text{$(#1\msp #2\msp #3)$}}
\newcommand{\stapp}[4]{\text{$(#1\msp #2\msp #3\msp #4)$}}

\newcommand{\npsuapp}[2]{\text{$#1\msp #2$}}
\newcommand{\npsbapp}[3]{\text{$#1\msp #2\msp #3$}}
\newcommand{\npstapp}[4]{\text{$#1\msp #2\msp #3\msp #4$}}

\newcommand{\lmeq}{\lmcon{=}}
\newcommand{\lmeqsp}{\lmcon{=} }

\newcommand{\clp}{\lmcon{(}}
\newcommand{\clpsp}{\lmcon{(} }
\newcommand{\crp}{\lmcon{)}}
\newcommand{\crpsp}{\lmcon{)} }

\newcommand{\npumapp}[2]{\text{$\exmls #1 \msp #2$}}
\newcommand{\npbmapp}[3]{\text{$\exmls #1 \msp #2 \msp #3$}}
\newcommand{\nptmapp}[4]{\text{$\exmls #1 \msp #2 \msp #3 \msp #4$}}

\newcommand{\mthunkapp}[1]{\text{$(\exmls #1)$}}
\newcommand{\npmthunkapp}[1]{\text{$\exmls #1$}}
\newcommand{\umapp}[2]{\text{$(\exmls #1 \msp #2)$}}
\newcommand{\bmapp}[3]{\text{$(\exmls #1 \msp #2 \msp #3)$}}
\newcommand{\tmapp}[4]{\text{$(\exmls #1 \msp #2 \msp #3 \msp #4)$}}

\newcommand{\checke}{\text{$\lmcon{if}_E$}}
\newcommand{\checkesp}{\text{$\lmcon{if}_E$} }

\newcommand{\checkd}{\text{$\lmcon{if}_D$}}
\newcommand{\checkdsp}{\text{$\lmcon{if}_D$} }

\newcommand{\pdes}{\mc{T}}
\newcommand{\pdesp}{\mc{T} }

\newcommand{\booland}{\lmcon{bool\mdash and}}
\newcommand{\boolandsp}{\lmcon{bool\mdash and} }

\newcommand{\boolor}{\lmcon{bool\mdash or}}
\newcommand{\boolorsp}{\lmcon{bool\mdash or} }

\newcommand{\boolnot}{\lmcon{bool\mdash not}}
\newcommand{\boolnotsp}{\lmcon{bool\mdash not} }

\newcommand{\sequent}[2]{\text{$\langle #1,#2\rangle$}}

\newcommand{\lmzero}{\lmcon{0}}
\newcommand{\lmzerosp}{\lmcon{0} }

%qqq
%%\newcommand{\lmcon}[1]{\text{$\mathsf{#1}$}}
%\newcommand{\lmsucc}[1]{\text{$\mathsf{s(#1)}$}}

\newcommand{\lmsucc}[1]{\lmcon{s(}#1\lmcon{)}}
\newcommand{\lmfact}[1]{\lmcon{fact(}#1\lmcon{)}}

\newcommand{\lmplus}[2]{\lmcon{(}#1\mssp\lmcon{+}\mssp#2\lmcon{)}}

\newcommand{\nplmplus}[2]{#1\mssp\lmcon{+}\mssp#2}

\newcommand{\nplmtimes}[2]{#1\mssp\lmcon{*}\mssp#2}

\newcommand{\lmtimes}[2]{\lmcon{(}#1\mssp\lmcon{*}\mssp#2\lmcon{)}}

\newcommand{\plusri}{\lmcon{plus\mdash 1}}
\newcommand{\plusrii}{\lmcon{plus\mdash 2}}

\newcommand{\scong}{\lmcon{s\mdash cong}}
\newcommand{\scongsp}{\lmcon{s\mdash cong} }
\newcommand{\factcong}{\lmcon{fact\mdash cong}}
\newcommand{\factcongsp}{\lmcon{fact\mdash cong} }
\newcommand{\pluscong}{\lmcon{+\mdash cong}}
\newcommand{\pluscongsp}{\lmcon{+\mdash cong} }
\newcommand{\timescong}{\lmcon{*\mdash cong}}
\newcommand{\timescongsp}{\lmcon{*\mdash cong} }

\newcommand{\plusrisp}{\lmcon{plus\mdash 1} }
\newcommand{\plusriisp}{\lmcon{plus\mdash 2} }

\newcommand{\timesri}{\lmcon{times\mdash 1}}
\newcommand{\timesrii}{\lmcon{times\mdash 2}}

\newcommand{\factri}{\lmcon{fact\mdash 1}}
\newcommand{\factrii}{\lmcon{fact\mdash 2}}
\newcommand{\factriisp}{\lmcon{fact\mdash 2} }

\newcommand{\vmap}[2]{\text{$#1\mlb #2\mrb$}}
\newcommand{\vmapsp}[2]{\text{$#1\mlb #2\mrb$} }

\newcommand{\tarsk}{\text{$\succcurlyeq$}}
\newcommand{\tarskls}{\text{$\:\succcurlyeq\:$}}
\newcommand{\tarsksp}{\text{$\succcurlyeq$} }

\newcommand{\zeroeven}{\lmcon{zero\mdash axiom}}
\newcommand{\zeroaxiom}{\lmcon{zero\mdash axiom}}
\newcommand{\zeroaxiomsp}{\lmcon{zero\mdash axiom} }
\newcommand{\mkeven}{\lmcon{make\mdash even}}
\newcommand{\mkevensp}{\lmcon{make\mdash even} }
\newcommand{\mkodd}{\lmcon{make\mdash odd}}
\newcommand{\mkoddsp}{\lmcon{make\mdash odd} }
\newcommand{\podd}{\lmcon{Odd}}
\newcommand{\peven}{\lmcon{Even}}
\newcommand{\mo}{\lmcon{mo}}
\newcommand{\me}{\lmcon{me}}
\newcommand{\za}{\lmcon{za}}

\newcommand{\teq}{\text{$\approx$}}
\newcommand{\teqsp}{\text{$\approx$} }
\newcommand{\teqls}{\text{$\:\approx\:$}}

\newcommand{\uniflm}{\tem{Unif}}
\newcommand{\uniflmsp}{\temv{Unif} }

\newcommand{\unifcalc}{\mc{U}}
\newcommand{\unifcalcsp}{\mc{U} }

\newcommand{\unif}{\text{$\vdash_{\text{\em\tiny U}}\:$}}
\newcommand{\unifnsp}{\text{$\vdash_{\text{\em\tiny U}}$}}
\newcommand{\unifsp}{\text{$\vdash_{\text{\em\tiny U}}$} }

\newcommand{\munif}{\text{$\vdash_{\tiny \unifcalc}\:$}}
\newcommand{\munifnsp}{\text{$\vdash_{\tiny \unifcalc}$}}
\newcommand{\munifsp}{\text{$\vdash_{\tiny \unifcalc}$} }

\newcommand{\lmndz}{\text{\lm--\cndz}}
\newcommand{\lmndzsp}{\text{\lm--\cndz} }

\newcommand{\lmhilz}{\text{\lm--$H_0$}}
\newcommand{\lmhilzsp}{\text{\lm--$H_0$} }

\newcommand{\lmndo}{\text{\lm--\cndo}}
\newcommand{\lmndosp}{\text{\lm--\cndo} }

\newcommand{\lmdmo}{\text{$\tem{dm}_1$}}
\newcommand{\lmdmosp}{\text{$\tem{dm}_1$} }
\newcommand{\lmdmop}{\text{${\tem{dm}_1}'$}}
\newcommand{\lmdmopsp}{\text{${\tem{dm}_1}'$} }

\newcommand{\lmdmt}{\text{$\tem{dm}_2$}}
\newcommand{\lmdmtsp}{\text{$\tem{dm}_2$} }
\newcommand{\lmdmtp}{\text{${\tem{dm}_2}'$}}
\newcommand{\lmdmtpsp}{\text{${\tem{dm}_2}'$} }

\newcommand{\prem}{\tem{premise}}
\newcommand{\premsp}{\temv{premise} }

\newcommand{\pold}{\text{$P_{\tem{old}}$}}
\newcommand{\poldsp}{\text{$P_{\tem{old}}$} }

\newcommand{\pnew}{\text{$P_{\tem{new}}$}}
\newcommand{\pnewsp}{\text{$P_{\tem{new}}$} }

%% cterm, cform, and cquant declarations for nd1-old.tex
%\newcommand{\cterm}[1]{\text{$\temls{term}(#1)$}}
%\newcommand{\cform}[1]{\text{$\temls{form}(#1)$}}
%\newcommand{\cquant}[1]{\text{$\temls{quant}(#1)$}}

\newcommand{\cterm}[1]{\text{$#1$}}
\newcommand{\cform}[1]{\text{$#1$}}
\newcommand{\cquant}[1]{\text{$#1$}}

%%========================================================

\newcommand{\cmneg}{\text{$\boldsymbol{\neg}$}}
\newcommand{\cmnegsp}{\text{$\boldsymbol{\neg}$ }}
\newcommand{\cmnegnsp}{\text{$\boldsymbol{\neg}$}}
\newcommand{\cmand}{\text{$\:\boldsymbol{\wedge} \:$}}
\newcommand{\cmandsp}{\text{$\boldsymbol{\wedge}$ }}
\newcommand{\cmor}{\text{$\:\boldsymbol{\vee} \:$}}
\newcommand{\cmorsp}{\text{$\boldsymbol{\vee}$ }}
\newcommand{\cmif}{\text{$\:\boldsymbol{\Rightarrow}\,$}}
\newcommand{\cmifsp}{\text{$\:\boldsymbol{\Rightarrow}\,$ }}
\newcommand{\cmiff}{\text{$\:\boldsymbol{\Leftrightarrow} \,$}}
\newcommand{\cmiffsp}{\text{$\boldsymbol{\Leftrightarrow}$ }}
\newcommand{\cmandnsp}{\text{$\boldsymbol{\wedge}$}}
\newcommand{\cmornsp}{\text{$\boldsymbol{\vee}$}}
\newcommand{\cmifnsp}{\text{$\boldsymbol{\Rightarrow}$}}
\newcommand{\cmiffnsp}{\text{$\boldsymbol{\Leftrightarrow}$}}

\newcommand{\cforall}{\text{$\boldsymbol{\forall}$}}
\newcommand{\cforallsp}{\text{$\boldsymbol{\forall}$} }
\newcommand{\cexists}{\text{$\boldsymbol{\exists}$}}
\newcommand{\cexistsp}{\text{$\boldsymbol{\exists}$} }

\newcommand{\cforallLst}{\text{$\boldsymbol{\forall}^*$}}
\newcommand{\cforallLstsp}{\text{$\boldsymbol{\forall}^*$} }
\newcommand{\cexistsLst}{\text{$\boldsymbol{\exists}^*$}}
\newcommand{\cexistsLstsp}{\text{$\boldsymbol{\exists}^*$} }

\newcommand{\freshv}{\lmcon{fresh\text{-}var}}
\newcommand{\freshvsp}{\lmcon{fresh\text{-}var}\msp}

\newcommand{\cuspec}{\lmcon{uspec}}
\newcommand{\cuspecsp}{\lmcon{uspec} }

\newcommand{\cegen}{\lmcon{egen}}
\newcommand{\cegensp}{\lmcon{egen} }

\newcommand{\cqsaxiom}{\lmcon{qs\text{-}axiom}}
\newcommand{\cqsaxiomsp}{\lmcon{qs\text{-}axiom} }

\newcommand{\necrule}{\mbf{nec:}}
\newcommand{\necrulesp}{\mbf{nec:} }
\newcommand{\rulek}{\mbf{K}}
\newcommand{\ruleksp}{\mbf{K} }

\newcommand{\rulet}{\mbf{T}}
\newcommand{\ruletsp}{\mbf{T} }

\newcommand{\ruled}{\mbf{D}}
\newcommand{\ruledsp}{\mbf{D} }

\newcommand{\rulesfour}{\text{$\mbf{S}_4$}}
\newcommand{\rulesfoursp}{\text{$\mbf{S}_4$} }
\newcommand{\rulesfive}{\text{$\mbf{S}_5$}}
\newcommand{\rulesfivesp}{\text{$\mbf{S}_5$} }

\newcommand{\systemsfour}{\text{$\mc{S}_4$}}
\newcommand{\systemsfoursp}{\text{$\mc{S}_4$} }
\newcommand{\systemsfive}{\text{$\mc{S}_5$}}
\newcommand{\systemsfivesp}{\text{$\mc{S}_5$} }
\newcommand{\systemk}{\mc{K}}
\newcommand{\systemksp}{\mc{K} }
\newcommand{\systemd}{\mc{D}}
\newcommand{\systemdsp}{\mc{D} }
\newcommand{\systemt}{\mc{T}}
\newcommand{\systemtsp}{\mc{T} }
\newcommand{\necop}{\text{$\Box$}}
\newcommand{\necopsp}{\text{$\Box$} }
\newcommand{\posop}{\text{$\Diamond$}}
\newcommand{\posopsp}{\text{$\Diamond$} }
\newcommand{\nec}[1]{\text{$\necop #1$}}
\newcommand{\pos}[1]{\text{$\posop #1$}}

\newcommand{\kframe}[1]{\mc{#1}}
\newcommand{\kframesp}[1]{\mc{#1} }

\newcommand{\kprovesls}{\text{$\:\vdash_{\text{\scriptsize \systemk}}\:$}}
\newcommand{\kproves}{\text{$\vdash_{\text{\scriptsize \systemk}}$}}
\newcommand{\kprovesp}{\text{$\vdash_{\text{\scriptsize \systemk}}$} }
\newcommand{\nkprovesls}{\text{$\:\not{\hspace*{-0.007in}\vdash_{\scriptsize \systemk}}\:$}}

\newcommand{\posintro}{\text{\posop\mbf{-intro}}}
\newcommand{\posintrosp}{\text{\posop\mbf{-intro}} }
\newcommand{\poselim}{\text{\posop\mbf{-elim}}}
\newcommand{\poselimsp}{\text{\posop\mbf{-elim}} }

\newcommand{\propappfun}{\mc{F}}
\newcommand{\propappfunsp}{\mc{F} }

\newcommand{\lmirel}{\text{$\equiv$}}
\newcommand{\lmirells}{\text{$\:\equiv\:$}}
\newcommand{\lmirelsp}{\text{$\equiv$} }

\newcommand{\canonassign}{\text{$\rho_{\mc{S}}$}}
\newcommand{\canonassignsp}{\text{$\rho_{\mc{S}}$} }

\newcommand{\extendsls}{\text{$\:\succeq\:$}}
\newcommand{\properlyextendsls}{\text{$\:\succ\:$}}
\newcommand{\modalextends}{\text{$\succeq$}}
\newcommand{\modalextendsp}{\text{$\succeq$} }
\newcommand{\modalproperlyextends}{\text{$\succ$}}
\newcommand{\modalproperlyextendsp}{\text{$\succ$} }

\newcommand{\nextendsls}{\text{$\:\not\succeq\:$}}
\newcommand{\nproperlyextendsls}{\text{$\:\not\succ\:$}}

\newcommand{\ttlam}{\ttm{\lambda}}
\newcommand{\ttsub}[2]{\(#1\sb#2\)}
\newcommand{\esb}[1]{\ttsub{e}{#1}}

\newcommand{\ttp}{\ttm{\Pi}}
\newcommand{\piof}[1]{\ttm{\Pi}\ttm{#1}}

\newcommand{\nequal}{\text{$\approx$}}
\newcommand{\nequalsp}{\text{$\approx$} }
\newcommand{\nequals}{\text{$\:\approx\:$}}
\newcommand{\notnequal}{\text{$\not\approx$}}
\newcommand{\notnequals}{\text{$\:\not\approx\,$}}

\newcommand{\neclog}{\mc{TL}}
\newcommand{\neclogsp}{\mc{TL} }

%% Utility transformation:
\newcommand{\utran}[1]{\text{$\mathfrak{U}\,(#1)$}}

\newcommand{\utransp}{\text{$\mathfrak{U}$} }

\newcommand{\utrannsp}{\text{$\mathfrak{U}$}}

%% Parsimony transformation:

\newcommand{\ptran}[1]{\text{$\mathfrak{P}(#1)$}}

\newcommand{\ptransp}{\text{$\mathfrak{P}$} }

\newcommand{\ptrannsp}{\text{$\mathfrak{P}$}}

%% Remove Claims transformation:

\newcommand{\rctran}[1]{\text{$\mathfrak{C}(#1)$}}

\newcommand{\rctransp}{\text{$\mathfrak{C}$} }

\newcommand{\rctrannsp}{\text{$\mathfrak{C}$}}

%% Right-association transformation:

\newcommand{\rasstran}[1]{\text{$\mathfrak{RL}(#1)$}}

\newcommand{\rasstransp}{\text{$\mathfrak{RL}$} }

\newcommand{\rasstrannsp}{\text{$\mathfrak{RL}$}}

%% Hoisting transformation:

\newcommand{\htran}[1]{\text{$\temls{Hoist}(#1)$}}

\newcommand{\htransp}{\text{$\temls{Hoist}$} }

\newcommand{\htrannsp}{\text{$\temls{Hoist}$}}

%% Final Scope-Maximization transformation:

\newcommand{\mstran}[1]{\text{$\mathfrak{MS}(#1)$}}

\newcommand{\mstransp}{\text{$\mathfrak{MS}$} }

\newcommand{\mstrannsp}{\text{$\mathfrak{MS}$}}

%% A1 transformation:

\newcommand{\atran}[1]{\text{$\mathfrak{A}_1(#1)$}}
\newcommand{\atransp}{\text{$\mathfrak{A}_1$} }
\newcommand{\atrannsp}{\text{$\mathfrak{A}_1$}}

\newcommand{\aatran}[1]{\text{$\mathfrak{A}_2(#1)$}}
\newcommand{\aatransp}{\text{$\mathfrak{A}_2$} }
\newcommand{\aatrannsp}{\text{$\mathfrak{A}_2$}}

\newcommand{\aaatran}[1]{\text{$\mathfrak{A}_3$}}
\newcommand{\aaatransp}{\text{$\mathfrak{A}_3$} }
\newcommand{\aaatrannsp}{\text{$\mathfrak{A}_3$}}

\newcommand{\normalize}[1]{\temls{normalize}(#1)}

\newcommand{\normalizesp}{\text{\em normalize\/} }

\newcommand{\normalizensp}{\text{\em normalize}}

\newcommand{\restructure}[1]{\temls{restructure}(#1)}

\newcommand{\restructuresp}{\text{\em restructure\/} }

\newcommand{\restructurensp}{\text{\em restructure}}

\newcommand{\contract}[1]{\temls{contract}(#1)}

\newcommand{\contractsp}{\text{\em contract\/} }

\newcommand{\contractnsp}{\mbox{\em contract}}

\newcommand{\marr}[1]{\text{$\underrightarrow{\vspace*{0.048in}\:#1\:}$}}

\newcommand{\etour}{\temls{detour?}}
\newcommand{\etoursp}{\temv{detour?} }
\newcommand{\etournsp}{\tem{detour?}}

\newcommand{\rr}[1]{\text{$\tem{RR}(#1)$}}

\newcommand{\rl}[1]{\text{$\tem{RL}(#1)$}}

\newcommand{\even}[1]{\mbox{$\temv{Even}(#1)$}}
\newcommand{\odd}[1]{\mbox{$\temv{Odd}(#1)$}}
\newcommand{\parity}{\mbox{\mc{P}\hspace*{-0.03in}\mc{AR}}}
\newcommand{\paritysp}{\mbox{\mc{P}\hspace*{-0.03in}\mc{AR}} }
\newcommand{\meanfun}{\mc{M}}
\newcommand{\meanfunsp}{\mc{M} }
\newcommand{\mean}[1]{\mbox{$\denot{\mc{M}}{#1}$}}
\newcommand{\meantwo}[2]{\mbox{$\denot{\mc{M}}{#1}\:#2$}}
\newcommand{\deddom}{\mbf{Ded}}
\newcommand{\deddomsp}{\mbf{Ded} }
\newcommand{\abspace}{\mbf{AB}}
\newcommand{\abspacesp}{\mbf{AB} }
\newcommand{\propdom}{\mbf{Prop}}
\newcommand{\propdomsp}{\mbf{Prop }}
\newcommand{\evrule}{\mbf{even-next}}
\newcommand{\evrulesp}{\mbf{even-next} }
\newcommand{\zeroevrule}{\mbf{zero-even}}
\newcommand{\zeroevrulesp}{\mbf{zero-even} }
\newcommand{\zeroevrulename}{\tem{ZeroEven}}
\newcommand{\zeroevrulenamesp}{\tem{ZeroEven} }
\newcommand{\odrule}{\mbf{odd-next}}
\newcommand{\odrulesp}{\mbf{odd-next} }
\newcommand{\evrulename}{\temv{EvenNext}}
\newcommand{\evrulenamesp}{\temv{EvenNext} }
\newcommand{\odrulename}{\temv{OddNext}}
\newcommand{\odrulenamesp}{\temv{OddNext} }
\newcommand{\comprulename}{\temv{Comp}}
\newcommand{\comprulenamesp}{\temv{Comp} }
\newcommand{\claimrulename}{\temv{Claim}}
\newcommand{\claimrulenamesp}{\temv{Claim} }

\newcommand{\cndzprovesls}{\mbox{$\:\vdash_{\mbox{\scriptsize \cndz}}\:$}}
\newcommand{\cndzproves}{\mbox{$\vdash_{\mbox{\scriptsize \cndz}}$}}
\newcommand{\cndzprovesp}{\mbox{$\vdash_{\mbox{\scriptsize \cndz}}$} }
\newcommand{\ncndzprovesls}
{\mbox{$\:\not{\hspace*{-0.007in}\vdash_{\scriptsize \cndz}}\:$}}

\newcommand{\truerulename}{\temv{True}}
\newcommand{\truerulenamesp}{\mbox{\em True\/} }
\newcommand{\falserulename}{\temv{False}}
\newcommand{\falserulenamesp}{\mbox{\em False\/} }

\newcommand{\assumerulename}{\temv{Assume}}
\newcommand{\assumerulenamesp}{\temv{Assume} }

\newcommand{\pab}{\mbox{$\phi$}}
\newcommand{\pabsp}{\mbox{$\phi$} }

\newcommand{\buildcert}{\smtt{generate-certificate}}
\newcommand{\buildcertsp}{\smtt{generate-certificate} }

\newcommand{\buildcerttext}{\text{generate-certificate}}
\newcommand{\buildcerttextsp}{\text{generate-certificate} }

\newcommand{\rp}[1]{\mbox{\raisedprime{#1}}}
\newcommand{\rectNodes}[1]{\mbox{$\temv{Rect}(#1)$}}
\newcommand{\rectNodesfs}[1]{\mbox{$\fs{\temv{Rect}(#1)}$}}

\newcommand{\underscore}{\symbol{95}}
\newcommand{\underscoresp}{\symbol{95} }
\newcommand{\multinomial}{\mbox{\rm\temv{Multinomial}}}
\newcommand{\multinomialsp}{\mbox{\rm\temv{Multinomial}} }
\newcommand{\categorical}{\mbox{\rm\temv{Categorical}}}
\newcommand{\categoricalsp}{\mbox{\rm\temv{Categorical}} }
\newcommand{\binomial}{\mbox{\rm\temv{Binomial}}}
\newcommand{\binomialsp}{\mbox{\rm\temv{Binomial}} }
\newcommand{\bernoulli}{\temv{Bernoulli}}
\newcommand{\bernoullisp}{\temv{Bernoulli} }
\newcommand{\indist}{\mbox{$\,\sim\:$}}
\newcommand{\kb}[2]{\mbox{$#1,#2$}}
\newcommand{\state}{\mbox{$\sigma$}}
\newcommand{\statesp}{\mbox{\state} }
% The four components of a configuration are: machines, stack, probability, and index. 
\newcommand{\config}[4]{\mbox{$\langle #1,#2,#3,#4\rangle$}}
\newcommand{\join}[2]{\mbox{$#1 \odot #2$}}
\newcommand{\branch}[2]{\mbox{$#1 \odot #2$}}
\newcommand{\branchalt}[2]{\mbox{$#1\: \widehat{\odot} \: #2$}}
\newcommand{\rewrites}{\mbox{$\mif$}}
\newcommand{\rewritesls}{\mbox{$\,\mif\,$}}
\renewcommand{\env}{\mbox{$\rho$}}
\renewcommand{\envsp}{\mbox{$\rho$} }
\newcommand{\mtiny}[1]{\mbox{\tiny #1}}
\newcommand{\cormatrix}[1]{\mbox{$\boldsymbol\rho(#1)$}}
\newcommand{\samplecormatrix}[2]{\mbox{$\boldsymbol\rho_{#1}(#2)$}}
\newcommand{\covmatrix}[1]{\mbox{$\mathbf{Cov}(#1)$}}

\newcommand{\fvec}[1]{\mbox{\footnotesize$\mathbf{#1}$}}
\newcommand{\scvec}[1]{\mbox{\scriptsize$\mathbf{#1}$}}
\newcommand{\tvec}[1]{\mbox{\tiny$\mathbf{#1}$}}

\renewcommand{\vec}[1]{\mbox{$\mathbf{#1}$}}
\newcommand{\vecsym}[1]{\mbox{$\boldsymbol{#1}$}}
\newcommand{\vecsp}[1]{\mbox{$\mathbf{#1}$} }

\newcommand{\fvecsym}[1]{\mbox{$\footnotesize\boldsymbol{#1}$}}

\newcommand{\reals}{\mbox{$\mathbb{R}$}}
\newcommand{\realsp}{\mbox{$\mathbb{R}$} }
\newcommand{\rats}{\mbox{$\mathbb{Q}$}}
\newcommand{\ratsp}{\mbox{$\mathbb{Q}$} }
\newcommand{\fsreals}{\mbox{\scalebox{0.8}{$\mathbb{R}$}}}
\newcommand{\fsrealsp}{\mbox{\scalebox{0.8}{$\mathbb{R}$}} }
\renewcommand{\fs}[1]{\mbox{\scalebox{0.7}{$#1$}}}
\newcommand{\fsprecise}[2]{\mbox{\scalebox{#2}{$#1$}}}
\newcommand{\fsss}{\fsprecise{\ss}{0.75}}
\newcommand{\fsssn}{\mbox{$\fsss^{\ts n}$}}

\newcommand{\sfs}[1]{\mbox{\scalebox{0.65}{$#1$}}}

\newcommand{\integers}{\mbox{\mathbb{Z}}}
\newcommand{\integerssp}{\mbox{\mathbb{Z}} }
\newcommand{\posints}{\mbox{$\mathbb{Z}^+$}}
\newcommand{\posintsUpToK}[1]{\mbox{$\mathbb{Z}_{#1}$}}
\newcommand{\fsposints}{\mbox{\scalebox{0.8}{$\mathbb{Z}^+$}}}

\makeatletter
\newcommand*{\bigcdot}{}% Check if undefined
\DeclareRobustCommand*{\bigcdot}{%
  \mathbin{\mathpalette\bigcdot@{}}%
}
\newcommand*{\bigcdot@scalefactor}{.5}
\newcommand*{\bigcdot@widthfactor}{1.15}
\newcommand*{\bigcdot@}[2]{%
  % #1: math style
  % #2: unused
  \sbox0{$#1\vcenter{}$}% math axis
  \sbox2{$#1\cdot\m@th$}%
  \hbox to \bigcdot@widthfactor\wd2{%
    \hfil
    \raise\ht0\hbox{%
      \scalebox{\bigcdot@scalefactor}{%
        \lower\ht0\hbox{$#1\bullet\m@th$}%
      }%
    }%
    \hfil
  }%
}
\makeatother

%% Dot product: 
\newcommand{\dotp}[2]{\mbox{$#1 \bigcdot #2$}}
\newcommand{\fdotp}[2]{\mbox{\footnotesize $#1 \bigcdot #2$}}
\newcommand{\scdotp}[2]{\mbox{\scriptsize $\scriptsize#1 \bigcdot #2$}}

\newcommand{\transpose}[1]{\mbox{$#1^\top$}}
\newcommand{\transposesp}[1]{\mbox{$#1^\intercal$} }
\newcommand{\ls}{\mbox{$\,$}}

\newcommand{\mpart}[2]{\mbox{$\frac{\partial{#1}}{\partial{#2}}$}}
\newcommand{\cpart}[2]{\mbox{$\cfrac{\partial{#1}}{\partial{#2}}$}}

\newcommand{\fwbmodel}{\mbox{$f_{\fvec{w},b}$}}
\newcommand{\fwbmodelsp}{\mbox{$f_{\fvec{w},b}$} }

\newcommand{\gradient}[1]{\mbox{$\nabla\,#1$}}

\newcommand{\softmax}{\temv{softmax}}
\newcommand{\softmaxsp}{\temv{softmax} }

\newcommand{\image}[1]{\mbox{$I_{\fs{#1}}$}}
\newcommand{\ent}[1]{\mbox{$H(#1)$}}

\newcommand{\icont}{\mbox{\rm \temv{IC}}}
\newcommand{\icontsp}{\mbox{\rm \temv{IC}} }

\newcommand{\imagep}[1]{\mbox{${\whs{I}}_{\hspace*{-0.01in}\fs{#1}}$}}
\newcommand{\imagex}{\mbox{$I_{\fs{X}}$}}
\newcommand{\imagexi}{\mbox{$I_{\fs{X}_i}$}}
\newcommand{\imagexone}{\mbox{$I_{\fs{X_1}}$}}
\newcommand{\imagexn}{\mbox{$I_{\fs{X_n}}$}}
\newcommand{\imagexsp}{\mbox{$I_{\fs{X}}$} }
\newcommand{\imagey}{\mbox{$I_{\fs{Y}}$}}
\newcommand{\imageysp}{\mbox{$I_{\fs{Y}}$} }
\newcommand{\imagez}{\mbox{$I_{\fs{\vec{Z}}}$}}
\newcommand{\imagexvec}{\mbox{$I_{\fs{\vec{X}}}$}}
\newcommand{\imagexvecsp}{\mbox{$I_{\fs{\vec{X}}}$} }
\newcommand{\fsimagez}{\mbox{$\fsimage{\mathbf{Z}}$}}

\newcommand{\fsimagep}[1]{\mbox{\footnotesize${\whs{I}}_{\hspace*{-0.01in}\fs{#1}}$}}
\newcommand{\fsimage}[1]{\mbox{\footnotesize$\footnotesize I_{#1}$}}
\newcommand{\fsimagex}{\mbox{$\fsimage{X}$}}
\newcommand{\fsimagexsp}{\mbox{$\fsimage{X}$} }
\newcommand{\fsimagey}{\mbox{$\fsimage{Y}$}}
\newcommand{\fsimageysp}{\mbox{$\fsimage{Y}$} }
\newcommand{\myfrown}{\raisebox{.3em}{\smash{\scalebox{.8}{$\smallfrown$}}}}
\newcommand{\myfrownsp}{\mbox{{\raisebox{.3em}{\smash{\scalebox{.8}{$\smallfrown$}}}}} }
\newcommand{\frownls}{\raisebox{.3em}{\smash{\scalebox{.8}{$\,\smallfrown\,$}}}}

\newcommand{\typedec}[1]{#1\hspace*{-0.03in}:\hspace*{-0.01in}}
\newcommand{\typedectight}[1]{#1\hspace*{-0.042in}:\hspace*{-0.033in}}

\newcommand{\geqls}{\mbox{$\,\geq\,$}}
\newcommand{\leqls}{\mbox{$\,\leq\,$}}
\newcommand{\leqsp}{\mbox{$\;\:\leq\;\:$}}
\newcommand{\geqsp}{\mbox{$\;\,\geq\;\,$}}
\newcommand{\mvdots}{\mbox{$\vdots$}}
\newcommand{\feqls}{\mbox{$\footnotesize\,=\,$}}
\newcommand{\eqls}{\mbox{$\footnotesize\,=\,$}}

\newcommand{\cboxit}[1]{\colorbox{red!50}{#1}}

\newcommand{\septs}{\mbox{$\fs{\tts\sep\tts}$}}
\newcommand{\sepss}{\mbox{$\, |\,$}}

\newcommand{\note}[1]{{\bf #1}}

\newcommand{\raisedsuperscript}[2]{\mbox{$#1^{\raisebox{0.13em}{\mbox{\footnotesize\ts$#2$}}}$}}
\newcommand{\raisedsuperscriptnots}[2]{\mbox{$#1^{\raisebox{0.13em}{\mbox{\footnotesize$#2$}}}$}}
\renewcommand{\exp}[1]{\mbox{\rm $E[#1]$}}
\newcommand{\bigexp}[1]{\mbox{$E\biggl[#1\biggr]$}}
\newcommand{\condexp}[2]{\mbox{$E[#1\,|\,#2]$}}
\renewcommand{\succ}[2]{\mbox{$\temv{Succ}^{#1}_{#2}$}}
\newcommand{\succone}[1]{\mbox{$\temv{Succ}_{#1}$}}
\renewcommand{\bigexp}[1]{\mbox{$E\left[#1\right]$}}
\newcommand{\bigvariance}[1]{\mbox{$\temv{Var}\left(#1\right)$}}
\renewcommand{\ss}{\mbox{$\Omega$}}
\newcommand{\sssp}{\mbox{\ss} }
\newcommand{\ssn}{\mbox{$\ss^{\ts n}$}}

\newcommand{\ssnsp}{\mbox{$\ss^{\ts n}$} }
\newcommand{\variance}[1]{\mbox{\rm $\temv{Var}(#1)$}}
\newcommand{\variancetight}[1]{\mbox{$\temv{Var}(\hspace*{-0.01in}#1\hspace*{-0.01in})$}}
\newcommand{\cov}[2]{\mbox{\rm $\temv{Cov}(#1,#2)$}}
\newcommand{\covlargebrackets}[2]{\mbox{$\temv{Cov}\left(#1,#2\right)$}}
\newcommand{\corr}[2]{\mbox{$\rho(#1,#2)$}}
\newcommand{\events}{\mbox{$\mc{F}$}}
\newcommand{\eventsp}{\mbox{$\mc{F}$} }
\newcommand{\eventssp}{\mbox{$\mc{F}$} }
\newcommand{\probspace}{\mbox{$\mc{P}$}}
\newcommand{\probspacesp}{\mbox{$\mc{P}$} }
\newcommand{\floor}[1]{\mbox{$\lfloor #1\rfloor$}}
\newcommand{\pdfx}{\mbox{$f_{\fs{X}}$}}
\newcommand{\pdfxsp}{\mbox{$f_{\fs{X}}$} }
\newcommand{\pmf}[1]{\mbox{$p_{\fs{#1}}$}}
\newcommand{\pmfx}{\mbox{$p_{\fs{X}}$}}
\newcommand{\pmfxsp}{\mbox{$p_{\fs{X}}$} }
\newcommand{\pmfxfs}{\mbox{$\fs{p_{\fs{X}}}$}}
\newcommand{\pmfz}{\mbox{$p_{\fs{\vec{Z}}}$}}
\newcommand{\pmfzsp}{\mbox{$p_{\fs{\vec{Z}}}$} }
\newcommand{\qmfx}{\mbox{$q_{\fs{X}}$}}
\newcommand{\qmfxsp}{\mbox{$q_{\fs{X}}$} }
\newcommand{\qmfz}{\mbox{$q_{\fs{\vec{Z}}}$}}
\newcommand{\qmfzsp}{\mbox{$q_{\fs{\vec{Z}}}$} }
\newcommand{\pmfy}{\mbox{$p_{\fs{Y}}$}}
\newcommand{\pmfysp}{\mbox{$p_{\fs{Y}}$} }
\newcommand{\pmfxy}{\mbox{\rm $p_{\fs{X},\fs{Y}}$}}
\newcommand{\pmfxysp}{\mbox{\rm $p_{\fs{X},\fs{Y}}$} }
\newcommand{\kl}[2]{\mbox{\rm $\temv{KL}(#1,#2)$}}
\newcommand{\klsp}{\mbox{\rm $\temv{KL}$} }
\newcommand{\perone}[1]{\mbox{\rm $\temv{Per}(#1)$}}
\newcommand{\persp}{\mbox{\rm $\temv{Per}$} }
\newcommand{\pernsp}{\mbox{\rm $\temv{Per}$}}
\newcommand{\ce}[2]{\mbox{\rm $\temv{CE}(#1,#2)$}}
\newcommand{\cesp}{\mbox{\rm $\temv{CE}$} }
\newcommand{\klnsp}{\mbox{\rm $\temv{KL}$}}
\newcommand{\censp}{\mbox{\rm $\temv{CE}$}}

\newcommand{\empdist}[2]{\mbox{$\widetilde{#1}_{\fs{#2}}$}}

\newcommand{\pmsp}{\mbox{$\mathbb{P}$} }
\newcommand{\pmq}{\mbox{$\mathbb{Q}$}}
\newcommand{\pmqfs}{\mbox{$\fs{\mathbb{Q}}$}}
\newcommand{\pmqsp}{\mbox{$\mathbb{Q}$} }
\newcommand{\pmfyx}{\mbox{$p_{\fs{Y},\fs{X}}$}}
\newcommand{\pmfyxsp}{\mbox{$p_{\fs{Y},\fs{X}}$} }

\newcommand{\pmfxonethroughn}{\mbox{$p_{\fs{X_1},\ldots,\fs{X_n}}$}}
\newcommand{\pdfy}{\mbox{$f_{\fs{Y}}$}}
\newcommand{\pdfysp}{\mbox{$f_{\fs{Y}}$} }
\newcommand{\cdf}[1]{\mbox{$F_{\fs{#1}}$}}
\newcommand{\cdfx}{\mbox{$F_{\fs{X}}$}}
\newcommand{\cdfxsp}{\mbox{$F_{\fs{X}}$} }
\newcommand\myprime{\mkern-3.5mu \raise0.1ex\hbox{$\prime$}}
\newcommand{\cdfxp}{\mbox{$F^{\mbox{\scalebox{0.75}{$\mkern-1.0mu{\,^{\prime}}$}}}_{\fs{X}}$}}
\newcommand{\cdfxplower}{\mbox{$F^{\ts\mbox{\scalebox{0.55}{$\raise0.4ex\hbox{$\prime$}$}}}_{\fs{X}}$}}
\newcommand{\cdfxplowersp}{\mbox{$F^{\ts\mbox{\scalebox{0.55}{$\raise0.4ex\hbox{$\prime$}$}}}_{\fs{X}}$} }

\newcommand{\cdfxpsp}{\mbox{\cdfxp} }
\newcommand{\cdfy}{\mbox{$F_{\fs{Y}}$}}
\newcommand{\cdfysp}{\mbox{$F_{\fs{Y}}$} }
\newcommand{\band}{\fmtt{(B\ls\&\ls{}A)}}
\newcommand{\bandsp}{\mbox{\fmtt{(B\ls\&\ls{}A)}} }
\newcommand{\inls}{\mbox{$\,\in\,$}}
\newcommand{\fsinls}{\mbox{\scalebox{0.7}{$\,\in\,$}}}
\newcommand{\capls}{\mbox{$\,\cap\,$}}
\newcommand{\cupls}{\mbox{$\,\cup\,$}}
\newcommand{\capsp}{\mbox{$\;\cap\;$}}

\newcommand{\tighteq}{\mbox{\hspace*{-0.1in}$=$\hspace*{-0.1in}}}
\newcommand{\tightleq}{\mbox{\hspace*{-0.1in}$\leq$\hspace*{-0.1in}}}
\newcommand{\tightless}{\mbox{\hspace*{-0.1in}$<$\hspace*{-0.1in}}}
\newcommand{\tightmiff}{\mbox{\hspace*{-3in}\miff}}

\newcommand{\tinymspace}{\mbox{\hspace*{0.018in}}}
\newcommand{\ts}{\mbox{\hspace*{0.014in}}}
\newcommand{\tts}{\mbox{\hspace*{0.008in}}}

\newcommand{\loweredbigl}{\mbox{\raisebox{-0.07in}{\mbox{$\Biggl[$}}}}
\newcommand{\loweredbigr}{\mbox{\raisebox{-0.07in}{\mbox{$\Biggr]$}}}}

\newcommand{\loweredbiglzero}{\mbox{\raisebox{-0.07in}{\mbox{$\bigl[$}}}}
\newcommand{\loweredbigrzero}{\mbox{\raisebox{-0.07in}{\mbox{$\bigr]$}}}}

\newcommand{\raiseit}[2]{\mbox{\raisebox{#2}{\mbox{#1}}}}

\newcommand{\raisedperiod}{\mbox{\raisebox{0.02in}{\mbox{\tts .}}}}
\newcommand{\raisedcomma}{\mbox{\raisebox{0.02in}{\mbox{\tts ,}}}}

\newcommand{\mlower}[1]{\mbox{\raisebox{-0.07in}{#1}}}

\newcommand{\bigsum}{\mathlarger{\sum}}

\newcommand{\covsym}{\temv{Cov}}
\newcommand{\covsymsp}{\covsym }

\newcommand{\corrsym}{\mbox{$\rho$}}

\newcommand{\smn}{\mbox{$\ol{X}_n$}}
\newcommand{\fssmn}{\mbox{$\fs{\ol{X}_n}$}}
\newcommand{\smnsp}{\mbox{$\ol{X}_n$} }

\newcommand{\mut}[2]{\mbox{$\mbox{\rm\temv{MI}}(#1,#2)$}}
\newcommand{\mutsp}{\mbox{\rm\temv{MI}} }

\newcommand{\fixspace}[1]{%
  \setlength{\lineskip}{#1}%
  \vspace{#1}%
}

\newcommand\marquis[3][1]{%
\scalebox{#1}{$\scaleleftright{[}{\parbox{#2}{\centering #3}}{]}$}%
}
\newcommand{\tworaisedtocodeword}{\mbox{$2^{\raisebox{0.05in}{\scalebox{0.8}{\mbox{$-\card{C(x)}$}}}}$}}

\newcommand{\mcard}[1]{\mbox{\raisebox{0.05in}{\scalebox{0.9}{\mbox{$-\card{C_X(#1)}$}}}}}
\newcommand{\rscale}[1]{\mbox{\raisebox{0.05in}{\scalebox{0.9}{\mbox{$#1$}}}}}

\newcommand{\imagextothek}{\mbox{$I^{\raisebox{0.025in}{\scalebox{0.9}{\mbox{$\tts k$}}}}_{\fs{X}}$}}
\newcommand{\imagextotheksp}{\mbox{$I^{\raisebox{0.025in}{\scalebox{0.9}{\mbox{$\tts k$}}}}_{\fs{X}}$} }

\newcommand{\raisedprime}[1]{\mbox{$#1^{\raisebox{0.01in}{\scalebox{0.65}{\mbox{$\ts\prime$}}}}$}}
\newcommand{\raisedprimefs}[1]{\mbox{$#1^{\raisebox{0.01in}{\scalebox{0.5}{\mbox{$\ts\prime$}}}}$}}
\newcommand{\raisedprimedouble}[1]{\mbox{$#1^{\raisebox{0.01in}{\scalebox{0.65}{\mbox{$\ts\prime$}}}\raisebox{0.01in}{\scalebox{0.65}{\mbox{$\ts\prime$}}}}$}}
\newcommand{\raisedprimedoublefs}[1]{\mbox{$#1^{\raisebox{0.01in}{\scalebox{0.5}{\mbox{$\ts\prime$}}}\raisebox{0.01in}{\scalebox{0.5}{\mbox{$\ts\prime$}}}}$}}
\newcommand{\fsmed}[1]{\mbox{\scalebox{0.67}{\mbox{$#1$}}}}

\newcommand{\pmfxp}{\mbox{$p_{\raisedprimefs{\fs{X}}}$}}
\newcommand{\elen}[1]{\mbox{$E_{\fs{#1}}$}}
\newcommand{\depth}[2]{\mbox{$d_{\fs{#2}}(#1)$}}

\renewcommand{\fix}{\marginpar{FIX}}
\newcommand{\new}{\marginpar{NEW}}

\newcommand{\ntilde}{\raisebox{0.7ex}{\large\texttildelow}}
\newcommand{\ntildels}{\raisebox{0.7ex}{\large\texttildelow}\mbox{$\,$}}
\newcommand{\ntildelsft}{\raisebox{0.7ex}{\texttildelow}\mbox{$\,$}}
\newcommand{\ntildesp}{\mbox{\ntilde} }
\newcommand{\ntildeneg}{\mbox{\raisebox{0.7ex}{\large\texttildelow}}\hspace*{-0.02in}}

\newcommand{\sctildels}{\raisebox{0.7ex}{\scriptsize\texttildelow}\mbox{$\,$}}

\newcommand{\amand}{\mbox{$\&$}}
\newcommand{\amandls}{\mbox{$\,\&\,$}}
\newcommand{\amandsp}{\mbox{$\;\&\;$}}
\newcommand{\mmor}{\mbox{$\sep$}}
\newcommand{\mmorls}{\mbox{$\:\sep\:$}}
\newcommand{\mmorsp}{\mbox{$\sep$} }

\newcommand{\pgrid}{\mbox{\textsc{ProofGrid}}}
\newcommand{\pgridsp}{\mbox{\textsc{ProofGrid}} }

\newcommand{\sspgrid}{\mbox{\scriptsize\textsc{ProofGrid}}}
\newcommand{\sspgridsp}{\mbox{\scriptsize\textsc\textsc{ProofGrid}} }

\newcommand{\fpgrid}{\mbox{\footnotesize\textsc{ProofGrid}}}
\newcommand{\fpgridsp}{\mbox{\footnotesize\textsc{ProofGrid}} }

\newcommand{\fm}{\mbox{\textsc{FrontierMath}}}
\newcommand{\fmsp}{\mbox{\textsc{FrontierMath}} }

\newcommand{\plone}{\mbox{$\textsc{PL}_1$}}
\newcommand{\plonesp}{\mbox{$\textsc{PL}_1$} }

\newcommand{\plthree}{\mbox{$\textsc{PL}_3$}}
\newcommand{\plthreesp}{\mbox{$\textsc{PL}_3$} }
\newcommand{\plfour}{\mbox{$\textsc{PL}_4$}}
\newcommand{\plfoursp}{\mbox{$\textsc{PL}_4$} }

\newcommand{\eqone}{\mbox{$\textsc{EQ}_1$}}
\newcommand{\eqonesp}{\mbox{$\textsc{EQ}_1$} }

\newcommand{\pltwo}{\mbox{$\textsc{PL}_2$}}
\newcommand{\pltwosp}{\mbox{$\textsc{PL}_2$} }

\renewcommand{\mc}[1]{\mbox{$\mathcal{#1}$}}

\newcommand{\db}{\mbox{De Bruijn}}
\newcommand{\dbsp}{\mbox{De Bruijn} }

\definecolor{header}{RGB}{230, 230, 250}
\definecolor{evenrow}{RGB}{245, 245, 245}
\definecolor{oddrow}{RGB}{255, 255, 255}
\definecolor{rowheader}{RGB}{240, 240, 255}

\newcommand{\kanote}[1]{\textbf{KA: #1}}
\newcommand{\sbnote}[1]{\textbf{SB: #1}}

\newcommand{\cmark}{\textcolor{green!60!black}{\ding{51}}}
\newcommand{\xmark}{\textcolor{red!80!black}{\ding{55}}}
\newcommand{\na}{\textcolor{gray}{---}}

\newcommand{\gptFive}{\mbox{\texttt{gpt\kern-0.1em-\kern-0.1em5}}}
\newcommand{\gptFiveOne}{\mbox{\texttt{gpt\kern-0.1em-\kern-0.1em5.1}}}
\newcommand{\gptFiveOnesp}{\mbox{\texttt{gpt\kern-0.1em-\kern-0.1em5.1}}}
\newcommand{\gptFiveTwo}{\mbox{\texttt{gpt\kern-0.1em-\kern-0.1em5.2}}}
\newcommand{\gptFiveTwosp}{\mbox{\texttt{gpt\kern-0.1em-\kern-0.1em5.2}}}
\newcommand{\gptFiveFour}{\mbox{\texttt{gpt\kern-0.1em-\kern-0.1em5.4}}}
\newcommand{\gptFiveFoursp}{\mbox{\texttt{gpt\kern-0.1em-\kern-0.1em5.4}}}
\newcommand{\gptFivesp}{\mbox{\texttt{gpt\kern-0.1em-\kern-0.1em5}} }
\newcommand{\oThreeMini}{\mbox{\texttt{o3\kern-0.1em-\kern-0.02em{mini}}}}
\newcommand{\oThreeMinisp}{\mbox{\texttt{o3\kern-0.1em-\kern-0.02em{mini}}} }
\newcommand{\oThree}{\mbox{\texttt{o3}}}
\newcommand{\oThreesp}{\mbox{\texttt{o3}} }
\newcommand{\claudeSonnetFour}{\mbox{\texttt{claude\kern-0.1em-\kern-0.02em{sonnet}\kern-0.1em-\kern-0.02em{4-20250514}}}}
\newcommand{\claudeSonnetExt}{\mbox{\texttt{claude\kern-0.1em-\kern-0.02em{sonnet}\kern-0.1em-\kern-0.02em{4.5-ext-thinking}}}}
\newcommand{\claudeSonnetExtsp}{\mbox{\texttt{claude\kern-0.1em-\kern-0.02em{sonnet}\kern-0.1em-\kern-0.02em{4.5-ext-thinking}}} }
\newcommand{\claudeSonnetFoursp}{\mbox{\texttt{claude\kern-0.1em-\kern-0.02em{sonnet}\kern-0.1em-\kern-0.02em{420250514}}} }
\newcommand{\opus}{\mbox{\texttt{claude\kern-0.1em-\kern-0.02em{opus}\kern-0.1em-\kern-0.02em{4-20250514}}}}
\newcommand{\opussp}{\mbox{\texttt{claude\kern-0.1em-\kern-0.02em{opus}\kern-0.1em-\kern-0.02em{4-20250514}}} }
\newcommand{\claudeOpusFourFiveNov}{\mbox{\texttt{claude\kern-0.1em-\kern-0.02em{opus}\kern-0.1em-\kern-0.02em{4-5-20251101}}}}
\newcommand{\claudeOpusFourFiveNovsp}{\mbox{\texttt{claude\kern-0.1em-\kern-0.02em{opus}\kern-0.1em-\kern-0.02em{4-5-20251101}}} }
\newcommand{\gptFourO}{\mbox{\texttt{gpt\kern-0.1em-\kern-0.1em4o}}}
\newcommand{\gptFourOsp}{\mbox{\texttt{gpt\kern-0.1em-\kern-0.1em4o}} }
\newcommand{\oFourMini}{\mbox{\texttt{o4\kern-0.1em-\kern-0.02em{mini}}}}
\newcommand{\oFourMinisp}{\mbox{\texttt{o4\kern-0.1em-\kern-0.02em{mini}}} }
\newcommand{\grokFour}{\mbox{\texttt{grok\kern-0.05em-\kern-0.04em{4}}}}
\newcommand{\grokFoursp}{\mbox{\texttt{grok\kern-0.05em-\kern-0.04em{4}}} }
\newcommand{\grokThree}{\mbox{\texttt{grok\kern-0.05em-\kern-0.04em{3}}}}
\newcommand{\grokThreesp}{\mbox{\texttt{grok\kern-0.05em-\kern-0.04em{3}}} }
\newcommand{\geminiTwoPointFive}{\mbox{\texttt{gemini\kern-0.1em-\kern-0.02em{2.5}\kern-0.05em-\kern-0.01em{pro}}}}
\newcommand{\geminiTwoPointFivesp}{\mbox{\texttt{gemini\kern-0.1em-\kern-0.02em{2.5}\kern-0.05em-\kern-0.01em{pro}}} }
\newcommand{\gemThree}{\mbox{\texttt{gemini\kern-0.1em-\kern-0.02em{3.0}\kern-0.05em-\kern-0.01em{pro}\kern-0.05em-\kern-0.01em{preview}}}}
\newcommand{\gemThreeOne}{\mbox{\texttt{gemini\kern-0.1em-\kern-0.02em{3.1}\kern-0.05em-\kern-0.01em{pro}\kern-0.05em-\kern-0.01em{preview}}}}
\newcommand{\gemThreeOnesp}{\mbox{\texttt{gemini\kern-0.1em-\kern-0.02em{3.1}\kern-0.05em-\kern-0.01em{pro}\kern-0.05em-\kern-0.01em{preview}}} }
\newcommand{\geminiThreeProPreview}{\mbox{\texttt{gemini\kern-0.1em-\kern-0.02em{3.0}\kern-0.05em-\kern-0.01em{pro}\kern-0.05em-\kern-0.01em{preview}}}}
\newcommand{\geminiThreeProPreviewsp}{\mbox{\texttt{gemini\kern-0.1em-\kern-0.02em{3.0}\kern-0.05em-\kern-0.01em{pro}\kern-0.05em-\kern-0.01em{preview}}} }
\newcommand{\gemThreesp}{\mbox{\texttt{gemini\kern-0.1em-\kern-0.02em{3.0}\kern-0.05em-\kern-0.01em{pro}\kern-0.05em-\kern-0.01em{preview}}} }
\newcommand{\gptFourOne}{\mbox{\texttt{gpt\kern-0.1em-\kern-0.05em4.1}}}
\newcommand{\gptFourOnesp}{\mbox{\texttt{gpt\kern-0.1em-\kern-0.05em4.1}} }
\newcommand{\dsThree}{\mbox{\texttt{deepseek\kern-0.025em-\kern-0.03emv3}}}
\newcommand{\dsThreesp}{\mbox{\texttt{deepseek\kern-0.025em-\kern-0.03emv3}} }
\newcommand{\dsThreeOne}{\mbox{\texttt{deepseek\kern-0.025em-\kern-0.03emv3.1}}}
\newcommand{\dsThreeOnesp}{\mbox{\texttt{deepseek\kern-0.025em-\kern-0.03emv3.1}} }
\newcommand{\dsROne}{\mbox{\texttt{deepseek\kern-0.025em-\kern-0.1emr1\kern-0.1em-0528}}}
\newcommand{\dsROnesp}{\mbox{\texttt{deepseek\kern-0.025em-\kern-0.1emr1\kern-0.1em-0528}} }
\newcommand{\dsReasoner}{\mbox{\texttt{deepseek\kern-0.025em-\kern-0.1emreasoner}}}
\newcommand{\dsReasonersp}{\mbox{\texttt{deepseek\kern-0.025em-\kern-0.1emreasoner}} }
\newcommand{\magistral}{\mbox{\texttt{mistral\kern-0.1em-\kern0.05em{magistral}}}}
\newcommand{\magistralsp}{\mbox{\texttt{mistral\kern-0.1em-\kern0.05em{magistral}}} }
\newcommand{\nous}{\mbox{\texttt{llama\kern-0.1em-\kern-0.1em3.1\kern-0.1em-\kern0.05em{405b}}}}
\newcommand{\nousp}{\mbox{\texttt{llama\kern-0.1em-\kern-0.1em3.1\kern-0.1em-\kern0.05em{405b}}} }
\newcommand{\hermes}{\mbox{\texttt{hermes\kern-0.1em-\kern-0.1em4}}}
\newcommand{\hermesp}{\mbox{\texttt{hermes\kern-0.1em-\kern-0.1em4}} }

\newcommand{\subrewrites}[3]{\mbox{$#1\longrightarrow_{#3}#2$}}
\newcommand{\bdsubrewrites}[3]{\mbox{$#1\longleftrightarrow_{#3}#2$}}
\newcommand{\rewritesrel}[1]{\mbox{$\longrightarrow_{#1}$}}
\newcommand{\bdrewritesrel}{\mbox{$\longleftrightarrow$}}
\newcommand{\bdsubrewritesrel}[1]{\mbox{$\longleftrightarrow_{#1}$}}
\newcommand{\bdsubrewritesreltc}[1]{\mbox{$\longleftrightarrow^{*}_{#1}$}}
\newcommand{\bdsubrewritesrelsp}[1]{\mbox{$\longleftrightarrow_{#1}$}}

\newcommand{\psp}{\mbox{$\:$}}
\newcommand{\pneg}{\mbox{$\ntildeneg\:$}}

\newcommand{\ci}[1]{\mbox{$\;(\pm\:#1)$}}
\newcommand{\mpm}{\mbox{$\pm$}}

\newcommand{\ndlzero}{\mbox{$\mbox{\rm NDL}_0$}}
\newcommand{\ndlzerosp}{\mbox{$\mbox{\rm NDL}_0$} }

\renewcommand{\problem}{\mbox{$\mathcal{P}$}}
\newcommand{\problemsp}{\mbox{$\mathcal{P}$} }

\newcommand{\problemprime}{\mbox{$\raisedprime{\mathcal{P}}$}}
\newcommand{\problemprimesp}{\mbox{$\raisedprime{\mathcal{P}}$} }
\newcommand{\model}{\mbox{$M$}}
\newcommand{\modelsp}{\mbox{$M$} }

\newcommand{\nprovesls}{\mbox{$\:\proves_{\tiny{N}}\:$}}
\newcommand{\nproves}{\mbox{$\proves_{\tiny{N}}$}}
\newcommand{\nprovessp}{\mbox{$\proves_{\tiny{N}}$} }

\newcommand{\correctanswer}[2]{\mbox{$\mc{C}(#1,#2)$}}
\renewcommand{\bel}[2]{\mbox{$\smallsubscript{B}{#1}\hspace*{-0.02in}\left[#2\right]$}}
\newcommand{\ought}[1]{\mbox{$\mc{O}\hspace*{-0.02in}\left[#1\right]$}}
\newcommand{\tinyraisedsuperscript}[2]{\mbox{$#1^{\hspace*{-0.01in}\raisebox{0.13em}{\mbox{\tiny\ts$#2$}}}$}}
\newcommand{\mdist}[2]{\tinyraisedsuperscript{D_{\mbox{\tiny{#1}}}}{{\mbox{\hspace*{-0.008in}\tiny{#2}}}}}
\newcommand{\smallsubscript}[2]{\mbox{$#1_{\hspace*{-0.02in}\mbox{\scriptsize\ts$#2$}}$}}

\newcommand{\plonepm}{\mbf{PL1-PM}}
\newcommand{\plonepmsp}{\mbf{PL1-PM} }

\newcommand{\plonepmc}{\mbox{$\mbf{PL1-PM}_{\scriptsize C}$}}
\newcommand{\plonepmcsp}{\mbox{$\mbf{PL1-PM}_{\scriptsize C}$} }

\newcommand{\inbetweensize}{\fontsize{8.5}{10}\selectfont}

%% \begin{enumerate}[label=(\alph*)]
%%   \item First item
%%   \item Second item
%%   \item Third item
%% \end{enumerate}

\vspace*{-0.3in}

\begin{abstract}
We introduce \pgrid, a challenging benchmark suite for evaluating the reasoning competence of language 
models through machine-checkable proofs rather than final answers alone.
\pgridsp spans 15 tasks organized around proof writing, proof checking, proof masking, and proof gap-filling. 
The tasks are expressed in deliberately minimal formal notation, most notably NDL, a stripped-down 
language for natural deduction that fits in a short prompt and supports precise, auditable 
checking. As a result, correctness judgments are mechanical, reproducible, and fine-grained 
rather than dependent on human or LLM judges. \pgridsp discriminates model capabilities with 
high resolution and covers a carefully calibrated difficulty spectrum, from foundational reasoning 
tests to structurally rich challenge tasks that no model currently solves, while avoiding 
reliance on domain knowledge,  problems that can be easily outsourced to solvers, or 
long-context artifacts. The paper develops a detailed 
comparative framework for reasoning benchmarks and uses it to survey and systematize 
a broad and previously fragmented literature. This analysis clarifies how \pgridsp 
differs from existing benchmarks not only in difficulty, but in representational choices, 
verification guarantees, reasoning depth, and vulnerability to shortcut-based success.

\enlargethispage{10pt}

A key methodological contribution of the paper
is an instrumented proof-checking pipeline that tolerates minor surface-level deviations while 
explicitly identifying the first substantive failure in a stretch of reasoning. This significantly 
sharpens measurement resolution, separating high-level proof planning from low-level 
execution noise, and reveals that current frontier models often possess surprisingly 
strong proof-strategy capabilities even when their formal proof execution is brittle. 
Using this pipeline, we conduct an extensive evaluation and analysis of a broad set of open and 
proprietary models released from late 2024 until now. Our results demonstrate 
strikingly rapid advances as well as sharp remaining limitations. Frontier systems now 
perform strongly on several foundational tasks, yet difficult \pgridsp tasks (especially those requiring global 
combinatorial reasoning or low-level proof synthesis) remain far from solved. Beyond task-level 
accuracy, we identify widespread epistemic instability, \egnsp, models often generate proofs 
containing local logical errors that they can nevertheless recognize and reject when those 
same inferences are presented in isolation. We formalize this phenomenon via an Epistemic Stability 
Index (ESI), a quantitative measure of cross-context coherence between model outputs for
different but related tasks (such as generated proofs and corresponding entailment judgments). 

Finally, we complement raw accuracies with 2PL Item Response Theory (IRT) analyses, 
Wright maps, and a new normalized measure of task discrimination based on 
Fisher information and latent model abilities estimated from their responses. 
Taken together, these results position \pgridsp not merely as a reasoning benchmark, but as a diagnostic 
framework for studying the emerging structure, limits, and internal coherence of 
reasoning in contemporary language models.

\end{abstract}

\clearpage

\clearpage
%\input{relatedWork}
%\input{api_details}
%\input{masking}
%\input{irt}
%\input{hf2}
%\input{rci_final}
%

%\input{pl1_pc}
%\input{pl1}
%\input{api_details}
%\input{conclusions}
% \bibliographystyle{plainnat}
% \bibliography{custom}
%\end{document}

\clearpage
\section{Introduction}

% While LLMs and especially LRMs (Large Reasoning Models) have shown impressive capabilities across a wide range of cognitive tasks, their true competence in formal reasoning remains poorly understood. 

Large language models (LLMs) have advanced  rapidly, finding applications in increasingly complex 
problems well beyond their original scope of narrow NLP tasks like summarization, information extraction, 
and composition of short texts like emails.
The recent emergence of the inference-compute paradigm \citep{InfTimeScaling2025} has given rise to so-called 
\emph{large reasoning models} (LRMs),  such as OpenAI's o-series of models along with GPT-5 and its successors; 
DeepSeek's R1 and related offerings; Anthropic's Claude Sonnet and Opus lines; and Google's Gemini family. 
These models have been trained with reinforcement learning (RL) and can perform test-time search 
to identify promising completions that are more likely to yield correct answers, at the expense 
of higher inference costs. Powered by test-time scaling and RL, language models have demonstrated 
dramatic improvements in logical and mathematical reasoning, with some of them surpassing human performance 
on highly challenging benchmarks, such as FrontierMath \citep{frontiermath}.

However, most existing test batteries for logical reasoning are focused only 
on the accuracy of the answers, which are typically either binary (in response to yes/no questions); 
or answer keys to multiple-choice questions; or, increasingly, numbers 
or other \mtt{\symbol{92}boxed\{\mbox{$\cdots$}\hspace*{-0.02in}\}} literals,
such as closed-form expressions. 
While this approach enables automatic evaluation by simply comparing model answers against canonical solutions, 
it has a serious drawback: it ignores the reasoning that gives rise to the answers. This can inflate 
performance, because models sometimes arrive at the right answers for the wrong reasons, through plausible but 
ultimately spurious inferences. More fundamentally, this  evaluation mode is particularly unsuitable for 
mathematics, where proofs are not just a means to an end but the very foundation of the discipline. 
A mathematical claim is accepted not because it seems likely or even because it happens to be
correct in some instances, but because a rigorous and logically sound proof 
demonstrating its truth is given. 

Therefore, emphasizing answer accuracy while overlooking the validity of the underlying reasoning 
is fundamentally misaligned with mathematical practice, which typically rejects a result 
that is derived by faulty reasoning, even if it is numerically correct. 
To properly assess the reasoning abilities of language models, especially in mathematics, 
it is essential to evaluate whether they can construct and verify proofs. \pgridsp is a step in that direction. 
\pgridsp is a benchmark suite for evaluating proof-based reasoning, requiring models not merely 
to produce correct answers but to construct, verify, and repair proofs expressed in a minimalist 
natural-deduction language (NDL) for propositional logic, and also in a pared-down formulation 
of pure equational logic. The benchmark comprises multiple tasks (proof writing, proof checking, 
proof masking, and proof gap-filling), designed to probe complementary aspects of reasoning, 
from generation of complex inference chains to error localization and structural reconstruction. 
In this paper we present and discuss all tasks in detail, evaluate 24 well-known models (both closed- and open-source) 
on every task, present and discuss the results, and perform extensive error analysis. 

%% We believe that formal proofs, in particular, will become increasingly important as LRMs come to play a 
%% larger role in mathematics, for purely practical reasons: We will need 
%% proofs in order to trust the outputs of these models, and inspecting these proofs manually will
%% not scale. LRMs will need to generate certificates of correctness for their outputs in the form
%% of machine-readable proofs whose soundness can be evaluated by simple and easy-to-trust algorithms. 

All tasks in \pgridsp are formulated in uninterpreted formal logic. As a result, 
the problems are entirely abstract and thus require what psychologists call {\em fluid intelligence}, as opposed to 
{\em crystallized\/} intelligence \citep{CHOI}. Fluid intelligence refers to innate reasoning ability employed 
to solve novel problems, while crystallized intelligence reflects the declarative knowledge and 
problem-solving know-how that has accumulated through explicit instruction. 
Most current benchmarks allow models to leverage their vast stores of crystallized 
knowledge, acquired through their pretraining, 
which often enables them to essentially retrieve correct answers,
either through brute memorization or by employing shortcuts that are sometimes 
valid but are other times derived from superficial patterns. 

There is nothing wrong with shortcuts, of course,
especially when they are based on sound theory. On the contrary, they are invaluable in practice. 
But when it comes to assessing native reasoning ability, they can act as confounders. If we ask a model whether a graph
can be colored with 2 colors, it can answer immediately if it notices an odd-length cycle 
and it happens to know that a graph is 2-colorable iff it contains no odd-length cycles. 
By minimizing reliance on memorized content, symbolic-logic problems are much 
better gauges of pure reasoning competence.

To make the ramifications of this concrete with an example: o4-mini was touted as achieving 99.5\% pass$@1$
%\texttt{@}1 
\href{https://openai.com/index/introducing-o3-and-o4-mini/}{on the 2025 AIME exam}, 
slightly beating o3. Even more impressively, \smtt{o4-mini-medium} scored $\approx 22\%$ 
on the competition subset of FrontierMath, 
\href{https://epochai.substack.com/p/is-ai-already-superhuman-on-frontiermath}{outperforming the average team of human mathematicians} 
($\approx 19\%$). 
On \pgrid, by contrast, it consistently places in the bottom half of the 24 models we have evaluated,
frequently scoring less than half of the absolute score of o3,\footnote{For instance, 
in the \mbf{PL3-PW} division of \pgrid, the accuracy of o4-mini is 6\% compared to o3's 22\%.}  
and well below open-source models 
like DeepSeek R1 or even DeepSeek 3.1. This performance profile is more consistent 
with the experiences of users who report that impressive benchmark numbers 
often fail to translate to strong real-world performance on reasoning-intensive tasks, 
especially those that gain little from memorized knowledge.  
An advantage of abstract reasoning problems like those of \pgridsp 
is that they can tease apart stronger from weaker models much more reliably. 
To quantify this with precision, we introduce a simple notion of spread we call DI 
(discriminability index), which we compute and report for every task in \pgrid. 
We also draw on IRT and statistical information theory to introduce a 
new normalized metric of task discrimination on select bands of latent 
model ability. 

%SWE-Bench shows o3 and o4-minite

In addition, many existing benchmarks have become ceiling-saturated and no longer 
pose a serious challenge. \pgridsp features demanding tasks that reveal the limitations of even 
the most capable models. We hope this will help to drive continued advances 
in the intersection of reasoning and language models. 

Beyond the issue of ceiling effects, with leading scores essentially at 100\%,  many existing benchmarks 
cluster together even mid- and third-tier models. On MATH 500, for example, 
GPT-4.1 \href{https://www.kaggle.com/benchmarks/open-benchmarks/math-500}{scores 87.6\%}, while Gemini 2.5 Pro and DeepSeek 3.1
receive 91\% and 91.5\%, respectively. This amounts to a spread of 3.9 points between the strongest and weakest of these three models, 
and a spread of only 0.5\% between the top two of the three. By comparison, the corresponding scores from a randomly selected \pgridsp 
task, \mbf{PL1-PW}, are 35\% for GPT-4.1, 49\% for DeepSeek 3.1, and 94\% for Gemini 2.5 Pro, yielding 
a 59-point separation between first and last and a 45-point gap between the top two.

In this work we focus on reasoning tasks that demand careful and deep conceptual engagement rather than mere procedural manipulation, 
and cannot be straightforwardly reduced to off-the-shelf algorithms, whether implemented by external systems 
or coded by the model itself. This rules out puzzles like Sudoku, Towers of Hanoi, or string 
manipulation problems, which admit simple algorithmic solutions, and even tasks like propositional satisfiability, 
which can be readily outsourced to SAT solvers. By contrast, there is no standard algorithm for writing proofs
in nontrivial proof systems, and while some of our equational problems could be approached with heuristics like beam search, 
it is not realistic for a model to engineer and execute a complete---and efficient---proof system for such problems on the fly. 
Even proof checking, though easily solvable by algorithm in principle, would require the one-shot design, construction, 
and debugging of a full software pipeline. Building such a system de novo during a reasoning task is infeasible. 
As a result, our benchmark compels models to perform genuine logical reasoning instead of 
relying on canned procedures.
%which aligns with the broader aim of machine learning: tackling problems that resist neat programmatic solutions. 
Moreover, by centering on proof writing, our tasks more closely reflect the kinds of reasoning 
challenges that arise in mathematics, science, and engineering, where the construction and justification of arguments 
is central. This makes \pgridsp not only more demanding, but also more ecologically valid as a test 
of the reasoning abilities we ultimately want models to develop.

Some related work has probed reasoning limits using very large problems or solutions,
whose size might approach or exceed model context windows (\egnsp, problems where the output 
length grows exponentially with input). 
With problems of such large sizes, failure is rarely 
surprising and not very diagnostic, as it is hard to disentangle genuine reasoning shortcomings from token-budget limits,
truncation artifacts, or decoding constraints. Moreover, solutions to such problems (e.g., towers of Hanoi) 
tend to adhere to a fixed recursive pattern and thus stress-test symbolic bookkeeping and bandwidth 
more than they stress-test reasoning.  
By contrast, both the inputs and the expected outputs of \pgridsp problems 
are of relatively small or modest sizes and fit easily within standard context limits, 
typically occupying only a few pages of text. Consequently, 
model failures cannot be attributed to context truncation or insufficient output budget, but instead 
reflect an inability to sustain long-horizon symbolic reasoning. Models must discover invariants, 
manage global constraints, select relevant premises under noise, reconstruct hidden combinatorial 
structure, and intelligently compose deep and coherent chains of inference, 
not simply manage context length and repetition.

In particular, our main contributions in this paper are the following: 
\bit

\item We introduce \pgrid, a rigorous reasoning benchmark that does not equate success with simple answer accuracy but instead requires
models to justify and explain their deliberations by producing machine-checkable proofs. The proofs are written in extraordinarily 
simple formal proof languages (or, as in the case of equational logic, no formal language at all other than simple equations). 
The benchmark blends:

\begin{enumerate}[label=(\alph*)]
  \item accessible tasks serving as foundational or calibration tests, intended to gauge minimal proof-backed reasoning competence;
  \item more difficult tasks that are not yet quite solved by the top 2-3 models but are likely to be relatively soon; 
  \item frontier challenges that are markedly more difficult than the rest, leaving more headroom for meaningful progress. 
\end{enumerate}
\pgridsp adheres to the following design principles: 
\ben
\item  With the exception of the frontier challenges, all tasks, including the easier/foundational ones, 
have sharp disciminating power at the time of writing. 

This principle relates directly to the issue of benchmark saturation. There are two different notions of saturation, and 
they are often conflated in the community: \emph{ceiling saturation} occurs when the strongest 2-3 models start scoring 
around 95\% or higher. By contrast, {\em general\/} or \emph{discriminative} saturation occurs when \emph{most} models, 
not just the elite, start scoring above 90\%. These are very different conditions. It is discriminative saturation that renders
a benchmark obsolete, because it can no longer separate different levels of ability. 
But as long as a task effectively separates respondents of low, medium, and high ability, there is nothing objectionable
about the top few models reaching 90\% or higher.\footnote{A school exam that gives As to most students would provide 
very little signal about true ability and would not be useful. But as long as an exam successfully 
separates students of different skill levels, it should not be dismissed just because a few students ace it. (Indeed, 
the top students are usually expected to earn As, otherwise the exam might be too difficult to function as a broad
test of competence.)}

\item  Solutions do not rely on content or crystallized mathematical knowledge. 
%(We call this  {\em the Seinfeld condition}: the problems are about nothing.)

\item Tasks are not readily solvable by code, either by external off-the-shelf solvers or by the models themselves.

\item Tasks do not use excessively long inputs or require excessively long outputs. 

\item Tasks do not move up the difficulty ladder by introducing adversarial tricks or obscure rules, or by exploiting model quirks. 
\een

\item We use the benchmark to measure and analyze the reasoning performance of a wide 
selection of well-known large models spanning releases from late 2024 until now. Errors in every task are grouped 
into categories and detailed error statistics per category are given for every model. 

\item  We quantify each task's discriminative power using both a simple measure we 
introduce based on Gini mean difference, the discriminative index (DI); as well as a new
normalized score based on the Fisher information of task items and latent model abilities 
estimated from task results, using IRT (item response theory).
%(a theory of test. Appendix~\ref{Sec:IRT} includes Wright maps for several \pgridsp tasks.
% and we release code to generate such maps for arbitrary tasks and benchmarks. 

\item We introduce novel proof-manipulation tasks (proof masking and proof gap-filling) 
that evaluate a model's ability to reason  explicitly \emph{about} proofs, reconstruct missing 
inference steps, and manipulate proof structure. This extends evaluation beyond proof generation 
and checking to meta-level structural understanding: do models grasp the internal structure 
of a reasoning artifact well enough to manipulate it successfully? 

\item 
Through extensive output analysis, we uncover and document a range 
of widespread behaviors, including self-contradictory explanations, shifting 
notions of reasoning correctness, verdict reversals mid-output, and other failure modes.

\item We develop a structural framework for comparing reasoning benchmarks along orthogonal axes 
(input form, output structure, logical expressiveness, inferential depth, and task diversity), 
and use it to situate \pgridsp within the broader reasoning-evaluation landscape.

\item We develop an instrumented proof checker that tolerates minor surface-level 
deviations while explicitly identifying the first substantive error in a proof attempt. This allows 
us to disentangle failures of procedural control from failures of higher-level reasoning, enabling 
fairer evaluation and substantially richer error analysis than is possible with standard pass/fail checking. 

\item We report formal calibration results on the task of equational gap-filling, 
quantifying how well models' self-reported confidence aligns with empirical correctness. 
Using bin-level calibration gaps and expected calibration error (ECE), we find pervasive 
miscalibration across many models (with especially pronounced overconfidence), 
making naive confidence-thresholding unreliable. The resulting calibration benchmarks 
establish a foundation for calibration-aware training and evaluation on proof-manipulation tasks. 
This separates raw competence from epistemic reliability and yields a quantitative yardstick 
for progress on calibrated reasoning

\item Beyond qualitative analysis and cataloging of reasoning pathologies, 
we investigate questions of reasoning {\em coherence across different tasks}, 
an analysis that is enabled by the presence of multiple related problems in the benchmark. 
We give a rigorous formalization of a notion that we call \emph{epistemic stability} 
and provide the first quantitative results capturing the coherence of a model's reasoning 
across epistemic contexts, by quantifying the degree of instability between different sets 
of judgments (such as generated proofs and independent entailment judgments). 
We show empirically that even top-tier models suffer from high epistemic instability, 
even when their performance on a task may be high in isolation. 

\eit

The rest of this paper is structured as follows. Section~\ref{Sec:RelatedWork} discusses and organizes related work. 
Section~\ref{Sec:PLOne} introduces the \plonesp division of \pgrid, simple propositional logic, which includes 
the following tasks (note: tasks whose names end in \mbf{-c} are frontier challenge tasks): 
\ben
\item \mbf{PL1-PW} (proof writing, discussed in Section~\ref{Sec:PL1PW}); 
\item \mbf{PL1-PC} and \mbf{PL1-PC-c} (proof checking, Section~\ref{Sec:ProofCheckingPLOne});
\item \mbf{PL1-PM} and \mbf{PL1-PM-c}  (proof masking, Section~\ref{Sec:PL1PM});
\item \mbf{PL1-GF} and \mbf{PL1-GF-c} (proof gap-filling, Section~\ref{Sec:PL1GF}). 
\een 
Section~\ref{Sec:PLTwo} introduces the \pltwosp division, with only one task, \mbf{PL2-PW}. 
Section~\ref{Sec:PLThree} describes the \plthreesp division, with two tasks, \mbf{PL3-PW} and \mbf{PL3-PC}.
Section~\ref{Sec:PLFour} introduces \plfour, comprising two tasks \mbf{PL4-PW} and \mbf{PL4-PW-c}.
Section~\ref{Eq:EquationalLogic} presents the equational-reasoning tasks (\mbf{EQ-PC}, \mbf{EQ-ER}, and \mbf{EQ-GF}) 
and corresponding results. Section~\ref{Sec:ESI} formalizes a notion of epistemic stability
and presents related results, while Section~\ref{Sec:Conclusions} concludes. Appendix~\ref{App:NDL} defines NDL, a natural-deduction 
proof language widely used in \pgrid, Appendix~\ref{Sec:IRT} discusses IRT and its application to \pgrid, while 
Appendix~\ref{app:technical-details} specifies various technical/operational details.

%\input{api_details}
%\input{contrib}
%\end{document}
\section{Related Work} 
\label{Sec:RelatedWork}

Reasoning has emerged as a core yardstick for assessing the performance of language models. However, there are 
many types of reasoning and many different research directions exploring the intersection of reasoning with 
AI models. In this section we aim to clarify this complicated and fragmented landscape, and explain where our own work fits 
into the broader picture. While we try to be comprehensive, our primary focus in this paper is deductive 
reasoning---the bedrock of mathematics. Although copious and significant research exists on other forms of reasoning also 
(induction, causality, and abduction, to name a few), deduction has historically dominated the literature on AI 
and reasoning.

Because of the central role that deductive reasoning has played in the field, and its 
prominence in  \pgrid, it will be helpful to clarify an aspect of it that is sometimes misunderstood. 
Deduction is often said to be a {\em non-ampliative\/} species of reasoning,  \label{Ft:Ampliative}
meaning that there is no information in the conclusion that is not, in some sense, already contained in the premises. 
This supposedly contrasts with other types of reasoning, such as induction, where the conclusion conveys 
information beyond what is found in the premises. 
But this is a somewhat crude distinction that can be misleading, 
even though it is common both in and out of the AI community (\egnsp, it is 
embraced in the HELM framework's discussion of reasoning evaluation \citep[][p. 38]{HELM2023}). 
It is true, of course, that every interpretation which satisfies
the premises of a valid deductive argument also satisfies the conclusion, and in that strict model-theoretic sense, 
the conclusion cannot say anything that is not already forced by the premises. 
So the claim that deduction is not ampliative is true enough in the narrow technical sense 
of truth-conditional semantics. But it breaks down in several important respects, 
particularly when examined from epistemic, computational, or representational perspectives. 

Epistemically and cognitively, deduction routinely yields vast amounts of information,
particularly in the mathematical sciences, revealing structure, compressing information, 
providing explanations and warrants, and connecting distant concepts. 
It would be untenable to say that Euclid's theorems 
don't provide any information that is not already 
contained in his postulates, or that Fermat's last theorem doesn't tell us anything 
that is not already in the axioms of ZFC. Before the proof by Andrew Wiles was published in 1995, 
we didn't even know whether Fermat's last theorem was true; we only suspected that it was. 
It was the deductive reasoning of Wiles that provided justification for believing that 
the theorem is true, explained why it is true, connected it to elliptic curves and 
modular forms, and introduced powerful new techniques and rich conceptual structures. 

In general, the distinction between ampliative and non-ampliative inference 
is much more fragile than usually assumed, as is the very distinction 
between deductive and nondeductive arguments. The topic is too complex
to discuss in depth here, but it is worth noting that expert opinions on the subject
run a wide gamut, with some argumentation theorists (who study 
everyday natural-language argumentation and reasoning), such as 
\citet{Groarke1999}, claiming that all natural-language reasoning is 
deductive, and that all arguments should be understood as attempts 
at deductive arguments. While we do not quite endorse this strong thesis 
here, it is important to understand that there are compelling arguments for it. 
In the context of AI benchmarks, the position is perhaps easier to understand 
when one considers problems in datasets such as MATH, like a problem asking for the roots of a quadratic equation 
or the area of a rectangular field. Strictly speaking, numeric answers like $\{2,5\}$ 
for the equation roots or 138 for the rectangle area are not formally entailed 
by the problem context, because the context omits the necessary mathematical background: 
definitions (what is a root, what is a rectangle, \etc), relevant theorems and results 
(the quadratic formula, area formulas, $\ldots$), and so on.  Nevertheless, the underlying reasoning 
remains entirely deductive. The abridged problem formulation doesn't shift the 
reasoning into the domain of abduction or induction, nor does it make the answer ampliative. 
It simply reflects the economy of expression in our representations of mathematical reasoning. 

Philosophers of science such as \citet{MusgraveDeductivism}
have defended equally strong positions regarding deduction in scientific reasoning, 
claiming, for example, that ``the only valid arguments are deductively valid arguments, 
and that deductive logic is the only logic that we have or need'' \citep{Musgrave2012}
and that instances of ``inductive'' or ``abductive'' reasoning are really
enthymematic deductions. Indeed, one could plausibly argue that even in what appear to be inherently
inductive settings, such as applications of statistical techniques that seem to go
beyond what has been strictly observed, the reasoning often remains primarily deductive. 
In Bayesian confirmation theory, for example, the move from prior to posterior 
is, mathematically speaking, an application of the probability axioms, and the 
update itself is a deduction made inside a richer calculus. So what might 
look ampliative in a bare propositional language is non-ampliative 
once the relevant probabilistic assumptions are made explicit in the model 
specification.\footnote{Of course, Bayesian updating 
by itself does not resolve the further questions of which prior to adopt, which 
likelihood is appropriate, or which model family to use. But that says nothing 
about the indispensability of induction.
%and in particular it doesn't show that making any of these choices involves some irreducibly inductive inference. 
To make that stronger claim, one would need to show that the reasoning by which 
such choices are made is itself irreducibly nondeductive, rather than merely 
dependent on substantive assumptions or principles that are not fixed by logic alone. 
Uncertainty or underdetermination at the level of premises 
does not entail the presence of any distinctively inductive inferential mechanisms.
Such uncertainty is manifested even in mathematics, \egnsp, consider the 
historical doubts regarding Euclid's
fifth postulate,  the axiom of choice, the continuum hypothesis and large-cardinal axioms,
and even the law of the excluded middle and double negation elimination.
}

% To establish that, one would have to exhibit a genuinely ampliative argument for one 
% of these choices that cannot plausibly be reconstructed as a deduction from a richer set of assumptions. 

Likewise, deduction plays a key
role in empirical-error-minimization methods like regression or SVM, or indeed any 
technique that can be viewed through the lens of statistical learning theory (SLT). 
This is not just because these techniques are deductively analyzed and proven to 
minimize error, but also because when we apply a model to a new input that we have 
not seen before, the reasoning is often best understood as a deduction. 

The point can be illustrated with ordinary linear regression, especially when formulated in Vapnik's 
SLT framework. Suppose we fix a hypothesis class $\mc{H}$ of linear 
functions $f_{\beta}(x) = \beta_0 + \beta^\top x$, a loss function such as squared error, and a learning 
rule that selects from $\mc{H}$
the function minimizing empirical risk, perhaps with an added regularization term to control capacity. 
Suppose further that we assume the training sample and a previously unseen instance are 
drawn i.i.d. from a common underlying distribution, that the feature representation is 
stable across train and test cases, and that the target relation is appropriately approximated 
within the chosen hypothesis class. Then, given the observed data, it follows \emph{deductively} 
which parameter vector minimizes the specified objective, which hypothesis is therefore 
selected by the learning rule, what prediction that hypothesis assigns to a new input, and (given the SLT machinery) 
what kinds of bounds relate empirical risk to expected risk as a function of sample size and 
hypothesis-class capacity. On this picture, the reasoning is not an irreducibly ampliative 
leap from observed to unobserved cases. Rather, it is a chain of deductions within a richer 
premise set that includes the hypothesis space, the loss function, the sampling assumptions, 
and the generalization theory. The genuine uncertainty lies not in the mathematical 
derivation of the fitted line or the associated risk bound, but in whether the assumptions 
underwriting that derivation are good assumptions for the problem at hand. But that is 
business as usual for deduction, which distinguishes between validity and soundness. 
Whether the premises entail the conclusion is one thing, and  logic can tell us that; 
whether the premises are \emph{true} (\egnsp, whether the test instances are indeed drawn i.i.d.
from a common underlying distribution) is up to the way the world is and must be settled
by empirical investigation. 

%\enlargethispage{20pt}

In any event, whether all reasoning is  ultimately deductive (or whether it can be 
fruitfully reconstructed as such) is a complex and contentious question, and we do not aim to take 
a definitive position here. Our more limited point is that the distinction between ampliative and non-ampliative 
reasoning (and the characterization of deduction as non-ampliative) is of questionable 
stability and value; and that, when scrutinized, many allegedly ampliative cases of 
inference are far more deductive in structure than the standard contrast suggests. 

%}

%% Even in Bayesian confirmation theory, the move from prior to posterior 
%% is, mathematically speaking, an application of the probability axioms, and the update itself is a {\em deduction\/} 
%% made inside a richer calculus. So what might look ampliative in a bare propositional language 
%% is non-ampliative once appropriate probabilistic structure is imported into the premises.
%% To take a simpler example, consider pedestrian math problems in benchmarks like MATH, say, a problem
%% that asks for the roots of a quadratic equation or the area of a rectangular field. Strictly speaking,
%% numeric answers like $\{2,5\}$ for the equation roots or 138 for the rectangle area are not
%% formally entailed by the problem context, because the context omits the necessary mathematical 
%% background: definitions (what is a root, what is a rectangle, \etc), relevant theorems and results 
%% (the quadratic formula, area formulas, $\ldots$), and so on.  Nevertheless, the underlying reasoning 
%% remains entirely deductive. The enthymematic nature of the input problem context doesn't shift the 
%% reasoning into the domain of abduction or induction, nor does it make the answer ampliative. 
%% It simply reflects the economy of expression in our representations of mathematical reasoning. 

In what follows we begin with a broad overview of reasoning datasets, grouped into four categories: 
natural-language reasoning; mathematical and scientific reasoning; generalist reasoning; and formal reasoning. 
We then introduce a set of analytical dimensions that provide a finer-grained understanding of this landscape, 
and use these axes to classify a broad range of benchmarks, including \pgrid. The section concludes 
with a brief comparison of \pgridsp and other major datasets.

\subsection{A Bird's-Eye Survey}

\subsubsection{Natural-Language Reasoning}

Within the larger NLP community, a major thread of work involving AI models and reasoning investigates 
the ability of such models to perform natural-language inference (NLI), and particularly their ability to 
determine whether a set of natural-language sentences entails a target sentence. 
The general problem is this:
\begin{equation}
%\begin{center}
\begin{tabular}{l}
{\em Given a set of sentences (or more generally, a text corpus) $T$ and a hypothesis $h$ (representing, for example, } \\ 
{\em a potential answer to a question) determine whether $T \models h$, \iensp, whether $T$ entails $h$. % If so, build an entailment tree that derives $p$ from elements of $T$. 
}
\end{tabular}
%\end{center}
\label{Eq:RTE}
\end{equation}
Common alternative terms for $T$ include {\em context}, {\em theory}, {\em knowledge base}, or simply the set of {\em premises}, 
while the sentence $h$ is also referred to as the---putative---{\em conclusion}, and sometimes as the {\em target sentence\/} or the {\em query}. 

Note that entailment here is usually (but not always, see below) {\em semantic}, it is {\em not\/} formal or strictly logical entailment.  
In formal logic, if $T$ entails $h$ then it does so purely by virtue of the syntactic {\em form\/} of the sentences in $T$ and $h$, 
not by virtue of the {\em meanings\/} of the words in those sentences. We thus say that {\em Tom is tall and thin\/} formally entails
{\em Tom is tall\/} because the entailment has nothing to do with what {\em tall\/} and {\em thin\/} actually mean. 
We could replace those adjectives with any 
others and the entailment would still hold. In fact, that is how formal entailment is defined: We quantify over all possible 
{\em interpretations\/} of the extra-logical symbols, and we say that no matter how you interpret {\em Tom}, \temv{tall}, and \temv{thin}, 
\iensp, for any possible way of assigning meanings (denotations) to those terms, the conclusion comes out true whenever 
the premises are true. By contrast, the usual textbook definition of {\em semantic\/} entailment is this: $T \models h$ iff 
the truth of the sentences in $T$ ``guarantees'' or makes ``certain'' the truth of $h$, or equivalently, iff the falsity of $h$ guarantees 
or makes certain the falsity of at least one element of $T$. 

However, this common textbook gloss is loaded with subjective psychological notions like certainty. 
A crisper definition of semantic entailment can be given via possible worlds: $T$ entails $h$ iff in 
every possible world (or model) where all sentences in $T$ are true, $h$ is also true. This sounds very much like 
the definition of entailment in formal logic, but there is a crucial difference. In this definition we do not quantify
over completely arbitrary interpretations of non-logical symbols. Instead, those meanings are (largely) {\em fixed}.
Proper names like {\em Tom}, for example, are fixed: they are ``rigid designators'' that pick out the same constant element across all 
possible worlds.
Lexical items that are not proper names (say, words like \temv{politician} and  \temv{smoker}) are allowed some degree of cross-world variation 
but only in the sense that the sets of individuals answering to these descriptions (the {\em extensions\/} of these words) may vary from world to world, 
as one world may have a different set of politicians from another, and someone who smokes in this world might not smoke in another. But the 
concepts expressed by these words are uniform across all possible worlds.\footnote{See, for instance, \citet[][Sec. 12.3]{HeimKratzer1998}.}
So the proper way to understand semantic entailment is this: 
once we fix the meanings of the extra-logical words {\em Tom}, {\em thin}, and {\em tall}, and {\em overweight}, in any possible world in which {\em Tom is thin\/}
is true, {\em Tom is not overweight\/} is also true. Thus, {\em Tom is thin\/} semantically entails {\em Tom is not overweight}, even though
it does not formally entail it. 

Sometimes the entailment is neither formal nor semantic but rather one of implicature. Here is a textbook example: {\em Some students managed to pass the exam\/}
strongly suggests that some students did not pass the exam, and so in an NLP context we might say that {\em Some students did not pass the exam\/} is entailed by 
{\em Some students managed to pass the exam}, even though it is neither semantically nor formally entailed by it. It is possible to tease out a number of other commonsensical
types of entailment that are disjoint from the above kinds, but ultimately, entailment judgments in NLP are made empirically; labeled examples are annotated 
based on what typical annotators perceive to be consequences of a given context $T$. 
This purely descriptive framing---entailment by democracy, so to speak---is reasonable if the goal 
is to study entailment as an empirical phenomenon, as it functions in everyday language use. 
But it does not offer a normative basis for adjudicating the correctness of an inference, and this 
may have implications for models trained on such data. 
The heuristics-and-biases literature has documented  a vast range of systematic reasoning 
fallacies to which most people (including reasoning experts) are susceptible \citep{Tversky1974}. 
If what counts as entailment becomes conflated with what is commonly perceived as such, 
even when these perceptions are flawed, a model could score well 
on such benchmarks by mimicking incorrect or biased reasoning patterns. 

Problem~(\ref{Eq:RTE}) originated under the name RTE (Recognizing Textual Entailment).\footnote{The term ``NLI'' started gaining 
wider currency after MacCartney's influential 2009 PhD dissertation \citep{MacCartneyNLI2009}.}
The first RTE challenge was launched 
in 2005 \citep{RTESurvey}. The fourth RTE challenge, in 2008, introduced an alternative and more general formulation of the problem: 
Given $T$ and $h$, %the objective is not simply to output a yes/no entailment judgment, but rather to 
decide whether $T$ entails $h$, 
contradicts $p$, or leaves the status of $p$ unknown---corresponding to the labels \temv{entailment}, \temv{contradiction}, 
and \temv{neutral}. This three-way setup was later adopted and popularized by large-scale benchmarks such as 
SNLI \citep{SNLI} and MNLI \citep{MNLI}, helping to standardize the task in modern NLP.

Historically, NLI work studied real-world natural language, usually drawn from 
naturally occurring or crowdsourced text. The context $T$ was typically a single sentence or a short paragraph at most, 
and the emphasis was on lexical, syntactic, and pragmatic nuances (arising, \egnsp, from various types of implicatures) 
in the relevant sentences. As already mentioned, entailment was typically defined 
informally,\footnote{\citet[p.2]{ManningRTE2006}  argued early on that
a textual inference task should ``adopt as the standard of inference what a human would be happy to infer from a piece 
of text. In particular, items would be assessed by people that are awake, careful, moderately intelligent and informed, 
and with reasonable document interpretation expertise, {\em but not by semanticists or similar academics\/} [our emphasis]. 
These people would use whatever background and real world knowledge that they usually bring to interpreting texts.''}
although logic-based techniques were also used from the beginning \citep{bos2005,Akhmatova2005}, albeit with 
underwhelming results \citep{BosWhenLI}. 
%``Bill murdered John'' to ``Bill killed John''. 
Early logic-based efforts still dealt with short 
sentence pairs drawn from real-world contexts and attended closely to lexical and syntactic features
(captured, \egnsp, by wide-coverage parsers), but they grounded entailment in logic and formal 
semantic representations, such as those of DRT (Discourse Representation Theory) \citep{kamp81} 
or MRS (Minimal Recursion Semantics) \citep{copestake2005minimal}.
So even within mainstream NLI, there was always a thread of research that treated entailment 
as a rigorously defined, logic-driven relation, but applied to ecologically authentic 
and linguistically rich text. However, the limited success of that approach in early NLI 
made way for statistical models with much more shallow representations that nonetheless
proved more effective in practice. 

\begin{comment}
\citet{ManningRTE2006} 

In particular, items would be assessed by people that are awake, careful, moderately intelligent and informed, and with reasonable
document interpretation expertise, but not by semanticists or similar academics. Thes e people would
us e whatever background and real world knowledge that they usually bring to interpreting texts. This
is the kind of pool and procedure that the NIST TREC evaluation has always us ed for determining
the relevance of results returned by information retrieval and question answering (QA) s ystems. The
texts for the task should be short pass ages of naturally occurring text. I feel it is vital to keep the task
grounded in real data. But I think the hypothesis should not always be authentic text. The hypothesis
is a probe, and one sometimes wants to probe whether a s ystem understands particular things. It
would often be difficult or impossible to find authentic text that probed thes e things. Nevertheless,
to improve task realism and grounding, it is desirable for the hypothesis to be drawn from motivating
tasks and constructed independently of the text as much of the time as is practical.
\end{comment}

There is a relatively newer strand of NLI work that is more explicitly inspired by and tied to formal logic, 
exemplified by datasets like RuleTaker \citep{RuleTaker} and systems like ProofWriter \citep{ProofWriter}. The main difference of these 
approaches from traditional NLI is that their data sources are expressed in \emph{controlled} natural language (CNL),
or quasi-natural language, usually templated English sentences synthetically generated from formal logical data with the help 
of grammars.\footnote{Representative examples include {\em Dave is green}, {\em Bob is round}, 
{\em Big people are rough}, and  {\em If someone is young and round then they are kind}. 
%The reader may judge to what degree such text counts as ecologically valid ``natural'' language.}
%With tongue in cheek, one could be forgiven for referring to such specimens as UNLI (un-natural language inference) problems.
The extent to which these may be viewed as ``natural'' language inference problems is debatable.
}
Because the starting point is a synthetically generated formal dataset,\footnote{This is typically 
a set of pairs $(S,p)$ consisting of a collection of formal sentences $S$ and a single sentence $p$---the 
same type of objects that we call \temv{arguments} in \pgrid. (These pairs are  known 
as {\em sequents\/} in logic, but we avoid that term here because it tends to be 
associated with Gentzen-style proof systems, which are not our focus in this paper.)}
it can be adjusted to support experiments with deeper reasoning and precisely defined entailment conditions, 
often grounded in subsets of logic such as stratified Datalog or various description logics. These datasets 
emphasize multi-step inference chains, closed-world reasoning, and deterministic label generation and 
evaluation, enabling fine-grained control over complexity and semantic scope. While the language is 
less naturalistic and rather inauthentic, the trade-off is a more transparent link between linguistic input and symbolic inference, 
so these datasets have been more widely used to probe the reasoning abilities of neural 
models. Besides RuleTaker \citep{RuleTaker} and ProofWriter \citep{ProofWriter}, datasets in this 
category include ProntoQA \citep{ProntoQA}, LogicInference \citep{ontanon2022logicinference}, LogicAsker \citep{LogicAsker}, 
LogicNLI \citep{LogicNLI}, and FLD \citep{FLD1}. Folio \citep{Folio} can arguably be placed in the same group. 
Although its sentences are not programmatically derived from logical forms, they are nevertheless synthetic 
insofar as they do not occur naturally; they were manually created by human annotators based on randomly selected 
Wikipedia pages. The resulting style is conspicuously stilted in the interest of enabling straightforward mappings to formal
(FOL) notation.\footnote{A representative FOLIO problem features the hypothesis/conclusion 
{\em Sea eel is an eel}, while the corresponding context consists of the following sentences: {\em All eels are fish. No fish are plants. 
A thing is either a plant or animal. Nothing that breathes is paper. All animals breathe. If a sea eel is either an eel or a plant, 
then a sea eel is an eel or an animal}.}

In general, NLI-oriented datasets (certainly those from the mainstream NLI tradition, but to a lesser extent 
those of the CNL variety as well), tend to focus on the inferences that are found 
in natural language, which are characterized by relatively shallow deductions, are strongly impacted by 
linguistic nuances, and incorporate commonsense---and often non-monotonic---reasoning principles such 
as the closed-world assumption and negation-as-failure. In this setting, AI models are viewed 
as ``soft'' reasoners \citep{RuleTaker} or soft theorem provers, simulating logical reasoning 
using natural language. By contrast, we will see that math- and logic-oriented datasets like \pgridsp 
are usually expressed in formal symbolic notation and focus on long chains of deductions whose 
correctness is underwritten by strict logical and mathematical norms. They gauge the degree 
to which AI models can serve as ``hard'' reasoners or theorem provers for logic and mathematics. 

While much of the work on natural-language reasoning centers around entailment, whether  
formally or informally defined, there are other reasoning tasks that depart from this framework. 
For instance, some benchmarks focus on abductive reasoning, where, given a natural-language context, the goal
is to select the most plausible explanation  or continuation (from a number of candidates). 
A representative example is HellaSwag \citep{zellers_hellaswag_2019}, in which models must choose the 
most plausible ending to a scenario from a set of options. Although this task is not about entailment,
solving it requires integrating background knowledge, commonsense expectations, 
and causal plausibility, all of which can be said to involve (informal) inference. 
These benchmarks therefore probe reasoning in a broader and more intuitive sense, 
often drawing on patterns learned from real-world event sequences rather than strict logical inference. 
In contrast to deduction-focused benchmarks, the reasoning here is contextual, gradient, and 
plausibility-based, highlighting the diversity of what ``reasoning'' can mean in NLP.

\subsubsection{Mathematical and Scientific Reasoning}

Another class of benchmarks regularly invoked in conversations about AI models and reasoning 
comprises tests specifically designed to evaluate mathematical competence.
Here the inputs are math problems (from various branches of mathematics and levels of difficulty) and the 
outputs are the respective answers. Unlike NLI, the focus is on longer chains of deductions whose 
correctness is underwritten by strict and explicitly formulated logical and mathematical norms, 
not by tacit informal conventions or pragmatics. The idea is to gauge the extent to which models 
can be used as ``hard'' logico-mathematical problem solvers. 
Representative samples include GSM8K \citep{cobbe_gsm8k_2021}, 
MATH \mbox{\citep{hendrycks_math_2021}}, OlympiadBench \citep{olympiadbench}, Omni-MATH \citep{omnimath}, 
and the more demanding and recent OlymMATH \citep{olymMATH} and FrontierMath \citep{frontiermath}. 

The problems in these benchmarks are expressed in natural language 
and the expected answers are what we might call {\em literals}, typically 
numbers or symbolic expressions like $x^2-5$. Models are expected 
to output their final answers in a special designated format, usually within the 
braces of a \LaTeX{} command like \mtt{\symbol{92}boxed\{\mbox{$\cdots$}\hspace*{-0.02in}\}}.
Questions are culled from many sources: standardized school curricula from various countries 
(such as K-12 textbooks); statewide and national exams (SATs, ACT, AP Calculus and Statistics, 
CBSE/ICSE/ISC exams in India, etc.); math competitions and Olympiads 
(\href{https://en.wikipedia.org/wiki/American_Mathematics_Competitions}{AMC}, 
\href{https://en.wikipedia.org/wiki/William_Lowell_Putnam_Mathematical_Competition}{Putnam}, 
\href{https://en.wikipedia.org/wiki/American_Invitational_Mathematics_Examination}{AIME},
\href{https://en.wikipedia.org/wiki/United_States_of_America_Mathematical_Olympiad}{USAMO},
\href{https://www.imo-official.org/}{IMO}, and so on); materials from educational organizations like 
\href{https://artofproblemsolving.com/}{AoPS} (the Art of Problem Solving); undergraduate math problem 
sets (\egnsp, MIT OCW, Harvard, etc.); crowdsourcing;  and synthetic data as well. 

Some of the more recent math benchmarks are remarkably challenging. \textsc{FrontierMath} \citep{frontiermath}, for example, 
introduced in late 2024 by the research institute Epoch AI with the explicit aim of evaluating AI models on advanced mathematics,
comprises hundreds of original research-level problems across a wide range of 
domains, such as algebraic geometry, number theory, and category theory. These problems were created by expert mathematicians, 
vetted for originality, and selected for their theoretical depth and difficulty. The creators note that even seasoned mathematicians 
may require hours or days to solve many of them. According to Epoch AI, 
as of September 2025, no model reached a mean accuracy of 
25\% on \textsc{FrontierMath},\footnote{In December 2024, OpenAI claimed that o3 attained an accuracy 
in excess of 25\%, but that claim \href{https://fortune.com/2025/01/21/eye-on-ai-openai-o3-math-benchmark-frontiermath-epoch-altman-trump-biden/}{has been controversial}
and Epoch AI was not able to reproduce that figure.}
though GPT-5 came close at 24.83\%, with OpenAI's o4-mini at second place, followed  by 
%{by far \href{https://x.com/EpochAIResearch/status/1941193747909443933}{the top performer}, followed by 
Gemini 2.5 Pro, Grok-4, and o3.\footnote{These results are for the easier tiers, 1--3. The hardest tier is 4, currently led by 
GPT-5 with 8.33\%.} By early March 2026, the top performer was GPT-5.2, at 31.3\%.

%(We will see that our results differ significantly, with o4-mini consistently underperforming both Gemini 2.5 Pro and o3.) 

However, regardless of their difficulty level, the extent to which these datasets actually test 
mathematical {\em reasoning\/} is not obvious. They certainly don't do so in an  explicit sense, 
given that they follow an outcome-based evaluation paradigm. That is, models are judged 
on whether they produce the correct final answer, typically a single literal value, 
without requiring them to provide logically correct arguments in support of those answers.\footnote{
COT (chain-of-thought) prompting is widely used for these tests, and model outputs are 
(potentially long) sequences of intermediate steps culminating 
in a final \fmtt{\symbol{92}boxed\{\mbox{$\cdots$}\hspace*{-0.02in}\}} answer. However, this is done 
mostly in order to improve the question-answering performance of the models, not to evaluate the reasoning 
contained in the COT traces. Model performance is still judged only on whether the final answer 
matches the ground truth. What little preliminary work has been done in this direction so far has found
that ``there exists a significant disparity between answer accuracy and faithfulness of the CoT reasoning generated 
by LLMs, indicating that they often arrive at correct answers through incorrect reasoning''  \citep{EvaluatingCOT2024}. 
Some newer datasets, such as NuminaMath, include 
a certain proportion of ``proof-based problems'' \citep[][p. 12]{NuminaMath}, but they do not currently 
support proof evaluation.}
This is rather incongruous with the subject matter, because, since the time of Euclid, 
mathematics has valued proof over mere assertion. But if performance
is gauged only by inspecting the final answer, models may get full credit even if they arrive 
at the correct result through memorization, pattern-matching, heuristic shortcuts, or statistical artifacts 
rather than sound reasoning. Benchmark creators do attempt to guard against guesswork (\egnsp, through ``guessproofness'' constraints 
\mbox{\citep[][p. 3]{frontiermath}}
and large or intricate answers), but these measures 
are only partial defenses against the core issue: These benchmarks don't verify 
whether the model arrived at the right solution for the right 
reasons.\footnote{While acknowledging the technical difficulty of \textsc{FrontierMath}, a number of Fields Medalists 
``pointed out that the numerical format of the questions feels somewhat contrived,'' with Richard Borcherds 
remarking, in a bit of understatement, that ``these benchmark problems aren't quite the 
same as coming up with original proofs'' \citep[][p. 11]{frontiermath}.}

A related issue with benchmarks in this category, at least as diagnostic tools for reasoning, 
is that they inevitably favor models with vast stores of crystallized mathematical knowledge. 
Large language models are trained on enormous corpora, including textbooks, 
research articles, lecture notes, problem banks, explanations, proofs, tutorials, and forums like StackExchange, giving them access 
to a breadth and depth of mathematical content far beyond what any human mathematician could hope to internalize. 
%In a benchmark like \fm, where the correct answer is often a number, formula, or object that appears somewhere in the model's training data (or is structurally similar to known results), 
This gives AI models a significant advantage. Problems are built around familiar mathematical objects (polynomials, 
inequalities, divisibility conditions, geometric relations, and so on), where the underlying concepts, as well as common 
tricks and transformations, are very well-represented in the training data. Even when problems are novel or cleverly 
constructed or transformed, the models' ability to draw on a tremendous body of heuristics and solution templates 
might enable them to perform relatively well without necessarily engaging in deep reasoning, 
aided mostly by pattern recognition, recall, and analogy.
%\footnote{For instance, Our analyses reveal weaknesses and crude heuristics in the victim’s reasoning processes, exposing a fawed grasp of the semantics of logic program\citep{AdvSoftTP}.}
This is especially true if the problems are heavily 
computational, as is the case with many of these benchmarks (including \fm), which allows the models to leverage their coding skills. 
By contrast, as mentioned in the introduction, \pgridsp 
is deliberately designed to neutralize this advantage. Its problems are fully symbolic, 
devoid of meaningful names or context, and grounded in logic systems that offer no opportunity to exploit real-world 
mathematical knowledge. It thus forces models to rely primarily on fluid reasoning, by constructing and 
verifying proofs from abstract premises, using inferential strategies that are common in mathematical practice. 
This offers a clearer lens into a model's true reasoning abilities, unassisted by prior knowledge.

A similar class of benchmarks target scientific problem solving. 
A historically important sample is the AI2 Reasoning Challenge (ARC) benchmark \citep{allenai:arc}, 
introduced by the Allen Institute for AI and aiming to evaluate understanding and reasoning in AI models within 
the domain of elementary and middle school science. It contains 7,787 natural-language multiple-choice
questions sourced from publicly available U.S. standardized science exams and split into two subsets:
\bit
\item ARC-Easy: Questions that can often be answered by surface-level lexical cues or 
information retrieval techniques.

\item ARC-Challenge: A curated subset of 2,590 questions that are supposed to defeat retrieval-based methods 
and are intended to test deeper forms of scientific and commonsense reasoning.
\eit 
The questions span a broad range of science topics typically covered in grades 3--9, including the life sciences
(biology, ecosystems, nutrition, human body systems, \etc); physical science (forces and motion, energy, matter, electricity);
earth and space science (geology, weather, solar system); and scientific method and reasoning (experimental design, 
measurement, cause-effect relationships, and so on). The benchmark was explicitly designed to resist shallow 
statistical methods and require multi-step inference, background knowledge, and commonsense reasoning. 
At the time of its release in 2018, ARC represented a significant step towards evaluating genuine science-related
reasoning ability in language models, but by now it has been effectively saturated, with many models achieving 
superhuman performance even on the Challenge subset. ARC has been superseded by the science-focused parts of 
relatively more recent benchmarks like BBH and Multi-MLU, although these have also been largely saturated by now. 

ARC and related science-focused benchmarks bear directly on our earlier remarks regarding the distinction between deductive and 
nondeductive reasoning, raising important questions about how deduction should be understood and identified. 
Consider this ARC problem: {\em A student riding a bicycle observes that it moves faster on a smooth road than on a rough road. 
This happens because the smooth road has: (A) less gravity (B) more gravity (C) less friction [correct] (D) more friction.}
This may appear to rely on simple empirical knowledge and commonsense reasoning. But if a model---or human---has internalized 
appropriate generalizations like ``Friction impedes motion'' and ``Rough surfaces increase friction,'' then a correct 
answer may be reached through a chain of deterministic (and arguably deductive) reasoning. From this perspective, 
the distinction between deductive and nondeductive reasoning may depend less on the task formulation and more on the implicit 
premises available to the test taker. This ambiguity reinforces the need for benchmarks that make the role of deduction 
explicit and verifiable---where the logical structure of the argument can be scrutinized and the conclusion's validity traced 
to its premises. 

Finally, the same issues we discussed in the context of math benchmarks apply here as well. The actual inferences 
of the model are not tested, only its final answer. Accordingly, these benchmarks do not directly test 
reasoning either, but rather question-answering ability. Likewise, crystallized knowledge plays a crucial 
role in successful performance in these benchmarks, and can be a confounding factor in assessing 
the extent to which models arrive at their answers through genuine reasoning rather than recall 
or statistical association. The exploitation of shallow statistical patterns has been repeatedly demonstrated 
for earlier and smaller transformer models, such as BERT, even on relatively abstract logical reasoning 
problems expressed in natural language, with, \egnsp, \citet[][2]{SimpleLogic} concluding that ``instead of learning 
to emulate the correct reasoning function, BERT has in fact learned statistical  features that inherently exist in 
logical reasoning problems.'' But similar concerns arise for newer, larger, and more powerful models. 
For example, \citet{EvaluatingCOT2024} evaluated 5  families of instruction-tuned causal LLMs 
ranging from 7B to 175B parameters, including an early version of ChatGPT, on two multi-hop QA benchmarks.
They found that the models often make the correct guess for the wrong reasons when they are asked to 
think step-by-step, showing a significant gap (around 16 percentage points) between answer correctness
and COT correctness and concluding that ``The correct final answer may not necessarily follow from faithful reasoning. We observe a 
significant gap between answer accuracy and reasoning faithfulness. It highlights the necessity 
of directly evaluating the reasoning steps rather than solely scoring the final answers.'' 
This is confirmed by our results, \egnsp, models  do very well when we ask them 
to simply guess whether a proof is correct, but when we ask them to justify their answers with details, performance drops significantly 
(see Table~\ref{Tb:Pl1PCAcc}). However, \citet{EvaluatingCOT2024} also reported that ``the performance gap between the final answer 
and reasoning worsens as the model size increases. As the answer accuracy also increases with the model size, 
it suggests that the bigger models may have the knowledge of the final answer without the need to perform reasoning.''
By contrast, we find that the gap shrinks for the best and largest models, because they rely less on superficial patterns
and more on honest reasoning toil.
%Direct knowledge of the final answer is much less likely for original and challenging mathematical problems, but the concern remains; the model may be able to arrive at the answer by exploiting spurious features in solution patterns and heuristics. 

The methodological issue of assessing models indirectly, by their final answers only, 
extends to frontier models and to the one domain that is supposed to be the epitome 
of abstract deductive reasoning: mathematics. 
While AI models have undeniably made great strides here, 
the recent buzz about such models being able to solve Olympiad-level math problems is premature. 
A recent manual analysis of 455 carefully curated shortlist IMO (International Math Olympiad) problems 
from 2009 to 2023 found that ``frontier models often generate incorrect solutions, that is, solutions that 
contain reasoning errors despite getting the final answer right.'' Specifically, once the model's full reasoning
was taken into account, the study found that the conditional probability of an answer being fully correct given 
that the final answer was correct was 12.5\% for o1, 14.3\% for o3-mini, and a rather astounding 0\%
for o1-mini, DeepSeek R1, and Gemini 2.0  \citep[][p. 8]{BrainsVsBytes}. The authors write that ``$\cdots$ current LLMs fall significantly short of solving
challenging Olympiad-level problems and frequently fail to distinguish correct mathematical
reasoning from clearly flawed solutions. Our analyses demonstrate that the occasional correct
final answers provided by LLMs often result from pattern recognition or heuristic shortcuts rather
than genuine mathematical reasoning. These findings underscore the substantial gap between
LLM performance and human expertise in advanced mathematical reasoning and highlight the
importance of developing benchmarks that prioritize the soundness of the reasoning used to arrive
at an answer rather than the mere correctness of the final answers.''

\subsubsection{Generalist/Hybrid Reasoning}

In this category we find more recent and wider-ranging benchmarks that touch on reasoning somewhat 
indirectly or only in part, even though they often end up informing public claims about the reasoning 
abilities of AI models. Notable examples include BBH \citep{suzgun_bbh_2023} and its successor 
BBEH (``Big Bench Extra Hard'') \citep{bbeh2025}, MMLU \citep{MMLU}
and its successor MMLU Pro \citep{MMLUPro}, HLE (Humanity's Last Exam) \citep{HLE}, SuperGPQA \citep{SuperGPQA}. 
These benchmarks aim to evaluate models across a wide array of subjects and tasks, many of which 
do involve reasoning to some degree. These tasks range from math and logic puzzles to legal analysis, ethics scenarios, 
high-school and college science, the social sciences, professional/standardized exams (SAT, GRE, LSAT, \etc), 
and so on. In some cases, the required inference resembles formal deduction (e.g., logic grid puzzles or syllogisms); 
in others, the inference relies on background knowledge, commonsense, or pragmatic reasoning.

Although these benchmarks provide a valuable broad-strokes view of model capabilities, they offer a relatively 
coarse-grained signal about reasoning, as they conflate different cognitive demands (e.g., domain knowledge, 
reading comprehension, cultural context, symbolic manipulation, and so on). Moreover, the evaluation remains 
outcome-based, judging models only on their final answers without assessing whether those answers are supported 
by correct or coherent reasoning steps. As such, these benchmarks complement more focused and diagnostic tools 
that aim to isolate and evaluate more precisely defined forms of reasoning.

A well-known benchmark that can be placed in this category, motivated in part by the same 
considerations as \pgrid, is the Abstraction and Reasoning Corpus for Artificial General Intelligence (ARC-AGI,
or simply ARC), introduced by \citet{ARCAGI}. It was designed to measure human-like fluid intelligence, 
rather than the crystallized intelligence typically assessed by standard AI benchmarks. 
Because of its prominence, the rest of this section is devoted to a detailed examination of ARC. 

ARC comprises few-shot visual grid-transformation tasks, 
where the test taker must conjecture an underlying pattern from a small 
number of input-output examples and successfully apply it to a novel test grid. 
The tasks are exclusively visual, strictly precluding not just domain-specific knowledge 
but also language, with the aim of relying only on fundamental \emph{core knowledge} priors \citep{Spelke2007CoreK} like 
objects, numbers, actions, and space (or basic geometry).\footnote{According to the work of 
developmental psychologist \citet{Spelke2000} (see also~\citet{Spelke2007CoreK}), 
human cognition starts with a small 
set of evolutionarily old, domain-specific representational systems, the four main ones 
being objects, numbers, actions, and space (a fifth domain-specific ability, 
for social interaction, is acknowledged as a distinct possibility). These systems are supposed to appear early in development, 
be shared in some form with other animals, and provide the foundations for all subsequent learning.
The core-knowledge thesis is clearly nativist, or at least anti-blank-slate, but it differs 
in one major respect from the more conventional nativism of Chomsky and Fodor: the role 
of language. The latter posit language as \emph{the} most highly distinctive computational system 
of the human mind, set apart by its uniqueness, its abstract combinatorial structure, 
and its explanatory power. Core knowledge theorists, by contrast, tend to view language 
as a powerful add-on---a coordinator and amplifier of more basic cognitive systems 
that predate human language, insofar as they appear in infants before the development 
of language, and even in other animals. However, many cognitive scientists would maintain 
that language is so transformative that it cannot be understood as just an important add-on; 
on this view, language doesn't just organize or decorate cognition, it fundamentally reorganizes it.} 

Humans are generally able to solve ARC tasks without much difficulty. 
As models scaled and researchers began applying massive test-time compute and brute-force 
search techniques to the original dataset, ARC-AGI-2 was introduced in early 2025 to 
re-establish a rigorous capability boundary \citep{ARCAGI2}. 
ARC-AGI-2 retains the visual 
grid format of the original ARC-AGI (which has come to be known as ARC-AGI-1) but 
introduces more complex tasks (requiring, for example, a sequence of object-manipulating 
actions that must respect multiple rules and contextual constraints). These new tasks 
are not as easily solved as those in ARC-AGI-1, but humans still succeed with probability 66\% 
when attempting to solve a randomly sampled task from the new benchmark \citep{ARCAGI2}[p. 7]. 

A key aim of ARC, which is shared by \pgrid, is to minimize 
the memorization and data-retrieval advantages of LLMs, and to test 
fluid problem-solving ability. 
But ARC goes further by precluding language and keeping the tasks strictly visual, 
with the goal of testing whether models can solve these tasks using only the few foundational priors 
that seem available to humans, roughly those singled out by core-knowledge theorists.
Chollet speaks of  ``a test of human-like general intelligence'' 
and writes that ``an actionable test of general intelligence that would be fair for 
both humans and machines should only feature tasks that assume the four core knowledge systems 
listed above, and should not involve any acquired knowledge outside of these priors''
\citep{ARCAGI}[p. 27]. 

ARC is an elegant benchmark that has been influential in the field,
but several issues are worth noting. First, ARC claims to be a measure of general fluid intelligence, but as was 
acknowledged by~\citet{ARCAGI}[p. 54], no empirical evidence has been presented in support of that contention, \egnsp,
no evidence that performance on ARC correlates meaningfully with $gF$, or even 
with $g$.\footnote{The famous $g$-factor (the $g$ stands for ``general intelligence''), introduced \label{Ft:TheGFactor}
by \citet{spearman1904general}, is a latent factor inferred from positive correlations 
among diverse cognitive test scores (people who do well on one kind of mental task tend to do well 
on others, on average). It is a statistical construct from psychometrics. 
Spearman believed that everyone has a relatively fixed intellectual capacity, 
a core mental ability that plays a role in virtually every cognitive task,
and that the $g$-factor measured precisely this ability. 
$Gf$ (fluid intelligence) is a related factor, representing the capacity to reason and 
solve novel problems independently of previously acquired knowledge. In psychometric 
models \citep{cattell1943measurement, cattell1963theory, carroll1993human}, 
it is typically distinguished from crystallized intelligence ($Gc$), 
which reflects knowledge and skills acquired through learning and experience.}
This is not implausible, given 
that ARC has qualitative similarities to well-known nonverbal intelligence tests like RPM 
(Raven's Progressive Matrices \citep{mccallum2003handbook}[ch. 11]), 
and the latter \emph{have} been psychometrically validated (\egnsp, RPM is known 
to load fairly strongly on $g$, although see \citet{gignac2015ravens} for 
recently discovered limitations). 
But superficial task similarity is not enough. Psychometrically, 
two tasks can look alike and yet differ a lot in what drives performance: visuospatial search, 
working-memory load, perceptual segmentation, task-format familiarity, response-generation demands, 
and so on. ARC, in particular (and especially ARC-AGI-2), differs from RPM in significant ways: 
it is open-ended rather than multiple-choice, it requires constructing an output rather than 
recognizing one, it imposes heavier search demands, and uses a  pixel-grid micro-world 
with its own idiosyncratic priors. Such differences could substantially increase or decrease correlation 
with $g$. So while there is a plausible case for expecting some positive correlation between ARC 
and $Gv$\footnote{In the Cattell-Horn-Carroll family of psychometric theories of intelligence
\citep{carroll1993human}, the $Gv$ factor captures the ability to perceive, analyze, manipulate, 
and think with visual patterns and spatial information.
\citet{walrath2020factor} write that ``$Gv$ involves a range of visual processes, 
ranging from fairly simple visual perceptual tasks to higher-level, visual, cognitive processes.
Woodcock and Mather (1989, p. 15) define $Gv$ in part: “In Horn-Cattell theory,
‘broad visualization’ requires fluent thinking with stimuli that are visual in the
mind’s eye.”'' They further note that ``Engineers, auto mechanics, architects, nuclear physicists,
sculptors, carpenters, and parts department managers all use $Gv$ to deal with the
demands of their job,''
although ``many writers seem to consider $Gv$ a relatively low-level cognitive ability, more 
perceptual than intellectual.''}, 
and perhaps even with $g$ or $Gf$, to the best of our knowledge that remains 
an open empirical question. 

Second, barring the use of language in a general test of intelligence (especially 
one that is routinely applied to language models) seems questionable, both on a priori 
and on empirical grounds. Language alone does not give models an unfair advantage.
Language is a faculty that all neurobiologically normal humans share. 
It is not, by itself, crystallized intelligence, nor is it a bias; it is a key innate mental capacity 
that cognitive science recognizes as central to thought. It is particularly important for 
the most prominent forms of human reasoning, from argumentation in politics and everyday life
to scientific and mathematical reasoning, which have been deeply intertwined 
with language\footnote{We understand language here in a broad sense, as a structured 
system of symbols governed by rules that enable the representation and communication of meaning. 
The only essential features are the use of discrete symbolic elements, combinatorial principles 
that determine how these elements can be arranged, and systematic relationships between syntax (form) and semantics.}
 for thousands of years, especially with reason giving.
The relevant word in ancient Greek is \emph{logos}, a rich term that can mean speech, word, 
account, explanation, or argument---all of them inherently linguistic objects. Likewise for \emph{ratio},
the Latin root of \emph{reasoning}, which also revolves around giving an account, explaining, reckoning,  and
justifying. These are the core practices of both everyday reasoning as well as mathematical and scientific 
reasoning, where one articulates premises and derives conclusions, makes and justifies claims and 
other inferential commitments, and explains why something is the case.

The importance of language is widely recognized in psychometric theories of intelligence and testing,
even by advocates of nonverbal tests like Wechsler, who as far back as 1939 was stressing the need 
to include both verbal and nonverbal tasks in intelligence tests, since doing so ``makes possible 
a comparison between a subject's facility in using words and symbols and his ability to manipulate 
objects, and to perceive visual patterns'' \citep{wechsler1939measurement}[p. 146]. He also made the empirical observation 
that ``Subjects of superior intelligence generally do better on the verbal, and subjects of 
inferior intelligence do better on the performance (nonverbal) part of the examination'' [p. 147].
A ``critical finding'' \citep{lakin2020intelligence}[p. 531] of what is perhaps the most 
prominent and ``widely accepted'' \citep{mcgrew2023psychometric} scientific theory of intelligence, 
the CHC (Cattell-Horn-Carroll) theory \citep{mcgrew2009chc}, is that
\begin{quote}
the general reasoning factor ($Gf$) may be decomposed into subfactors: (1) deductive
reasoning (termed sequential reasoning by Carroll and largely comprising verbal
tasks), (2) quantitative reasoning (inductive or deductive reasoning with quantitative
concepts), and (3) inductive reasoning (often measured with figural tasks; Carroll, 1993). 
A good reasoning test, then, should probably measure all three of these
reasoning factors – or at least not be strongly biased toward one (Wilhelm, 2005).
This fact is commonly overlooked in studies that represent fluid reasoning abilities
with a single figural reasoning test, such as Raven’s Progressive Matrices test (Raven, Court, \& Raven, 1977). 
\citep{lakin2020intelligence}[p. 531]
\end{quote}
Like RPM, ARC only taps subfactor (3) above, and even there we would claim that 
serious inductive reasoning (as done, e.g., in science) almost always involves language
and the explicit articulation of hypotheses and evidence. 
The importance of language is also recognized by rival theories to CHC, 
such as BIS (the Berlin Intelligence Structure) \citep{Beauducel2002}, which holds that 
``measuring reasoning requires tasks that sample across the main content areas,''
of which there are three: verbal, numeric, and figural; and ``suggests that the diversity 
of content required to measure $Gf$ consistently will require going beyond the 
figural tasks (e.g., Raven’s Progressive Matrices)'' \citep{lakin2020intelligence}[p. 532]. 
Likewise, recent work on the limitations of exclusively figural tasks like RPM 
advises that ``researchers are discouraged to include only Raven's in an investigation, 
if a valid estimate of $g$ is sought.'' \citep{gignac2015ravens}[p. 71]. 
%% Instead, as just one example, a 
%% four-subtest combination from the Wechsler scales with a $g$ validity coefficient 
%% of .93 and 14 min administration time is suggested'' 

ARC also aims to be a test of abstraction, but it is important to be precise about 
what sort of abstraction mechanisms are tapped by nonverbal tests. 
Cognitive science has traditionally distinguished 
between two types of abstraction: 
\bit
\item low-level \emph{perceptual abstraction}, which allows us to recognize objects across
viewpoints, strip away visual noise to see that two shapes are both triangles, and so on; and 
\item \emph{conceptual abstraction}, which is the process of creating higher-order categories based 
on invisible, functional, or relational properties, e.g.,  understanding that a thermometer, a clock, 
and a yardstick all belong to the category of measuring instruments, or understanding concepts 
like justice and entropy. 
\eit 
The sort of abstraction tested by visual tasks like ARC and RPM sits somewhere in between, 
and much closer to the first. It allows us to extract relational structure from perceptual representations,
by identifying visual progression (e.g., increases in shape size), rotation or reflection, and 
patterns like same-colored regions across portions of a grid. 
The corresponding type of reasoning is often called \emph{visuospatial} relational 
reasoning or {\em perceptual analogy}. 
While it's true that this intermediate level of perceptual abstraction has been historically important 
in the testing of fluid intelligence, it is conceptual abstraction that is 
central to science and mathematics 
and is considered to be the defining cognitive capacity behind theoretical reasoning. 
As already mentioned, conceptual abstraction involves the formation of general ideas and 
principles that are detached from any particular sensory inputs. 
They range from ordinary concepts like cause and effect or kinship (\egnsp, the notion of an uncle) 
to scientific quantities like acceleration, the idea of natural selection in biology, 
a random variable in statistics, a semigroup in mathematics, and so on. It is these linguistically scaffolded 
abstractions that allow generalization across widely different contexts and enable 
sophisticated reasoning, explanation, and prediction. None of these are exercised by visual tests. 
So, while success on visual tasks might well provide some 
signal on intelligence, it should not be understood as indicative of general 
or high-level abstraction ability. The latter goes much farther than what is 
visually given, by relying on language to represent categories that 
depend on inferential roles rather than perceptual relations.\footnote{
Modern developmental science indicates that conceptual abstraction is produced 
by different learning systems than perceptual abstraction, emerges 
at different developmental stages, and is subserved by different mechanisms,
requiring cross-modal integration, lexicalization, executive function, 
and the learning of rule-based categories. According to~\citet{Sloutsky2010SLOFPC},
the brain contains two distinct category-leaning systems. One is a similarity-based 
system characterized by compression, which learns from perceptual correlations,  
operates on many overlapping features, is present early in development, and is tied to the 
inferotemporal cortex and perceptual systems; this is the system that learns ``dense'' categories
like dogs and cats. The other is a rule-based system characterized by selection,
which selects specific diagnostic dimensions, supports rules, requires attention and executive control, 
is tied to the prefrontal cortex, and develops later; this is the system that learns ``sparse'' categories,
like prime numbers and moral agency. \citet{Sloutsky2010SLOFPC} points out that sparse categories tend to depend 
on unobservable properties and require selective attention and rule learning;
and that abstract reasoning in mathematics and science
depends on representational resources that go far beyond visual structure: the 
ability to formulate and manipulate concepts whose defining properties are 
unobservable, inferred, and linguistically articulated.}

It might be objected that ARC was never intended to test high-level reasoning or abstraction;
what has made it notable is that powerful AI models have found it challenging even though it  
is a fairly simple test that many ordinary people are able to pass. 
However, ARC aims to be a test of ``abstraction and reasoning'' simpliciter, 
not a test of perceptual abstraction and visuospatial reasoning. The paper that introduced 
the benchmark speaks of a general measure of intelligence, and 
the relevant Kaggle competition presents itself as a test to find the next 
frontier of AGI,  implying human-level mastery across all cognitive domains.  
As for the observation that models are struggling with something that even children can do,
and the resulting temptation to conclude that models are incapable of reasoning, 
it should be kept in mind that a 5-year-old human has a 500-million-year evolutionary 
head start in visual processing. The human brain dedicates a massive portion of its cortex to 
automatic, unconscious visual grouping (Gestalt principles like proximity, similarity, and closure). 
For humans, ``seeing'' the grid and grouping the pixels requires zero effort.
This raises the question of whether the exclusion of language in order to be fair to humans
might not tip the scales the other way and make the test unfair to language models. 

A brief historical aside might be helpful here: nonverbal intelligence tests originated in the 1800s 
because scientists were curious about the cognitive abilities of language-deprived people 
who could not speak. During World War I, the US Army needed ``methods to assess 
abilities of foreign born and illiterate military recruits'' and a committee to design
such methods was formed \citep{mccallum2003handbook}[p. 5]. 
Later in the 20th century, fairness came to be an increasingly prominent 
motivation behind nonverbal intelligence tests, some of which (such as the Universal 
Nonverbal Intelligence Test \citep{bracken1998unit}) had fairness as their ``overarching 
goal'' \citep{mccallum2003handbook}[p. 88]. 
After all, if one hardly speaks English, it is not fair to give them an intelligence 
test that requires fluency in English.
%It was assumed, of course, that the people taking the nonverbal  tests had normal vision, 
%% for a simple reason: human subjects are roughly normalized on low-level perception. 
By the same token, however, it is not fair to give a visual intelligence test to a subject 
who is visually impaired. 
And to this day, the perceptual abilities of language models, 
even multi-modal ones, remain relatively poor, lagging far behind their language skills.
This is largely because transformer-based models, which were designed for sequence statistics, 
lack the geometric and physical priors encoded in the human brain, \iensp, they lack most 
of the core-knowledge priors that humans have (objects, actions, and space).\footnote{Just as they lack grounding 
in the physical world, with no direct experience of physical interactions such as movement 
and object manipulation, and no experience of temporal continuity. 
A baby learns about object persistence by interacting with the physical world, 
not by ingesting a static Internet snapshot.}
%For humans, this is not an issue, 
Given that ARC seems heavily gated by perceptual access to the relevant tasks, 
visual ability may be acting as a confound. %, a prerequisite channel whose uneven quality can contaminate the measurement.
This is essentially the argument made by recent work \citep{wang2026reasoningbenchmarktestreasoning}, 
which found that models frequently fail on ARC tasks before reasoning even begins, 
simply because they are unable to reliably identify objects, shapes, and spatial relations 
(something which is effortless for humans), and therefore end up misinterpreting the visual input. 
Manual error analysis carried out by the authors suggests that 80\% of model failures are due to perceptual errors. 
They also found that when the task is split into a 2-stage pipeline, a perception stage that takes
the image and outputs a natural-language description of it, followed by a ``reasoning'' stage that takes
the natural-language description and reasons with it, performance improves significantly when the results
of the first (perceptual) step are improved, \egnsp, simply by using a stronger visual model, while keeping
the reasoning model fixed. The authors thus argue that ARC results can overestimate the 
reasoning gap between humans and AI, because humans have much stronger visual object recognition 
and perceptual grouping. 

These considerations suggest a divergence between the methodological guidelines 
in the paper that introduced ARC and the empirical benchmark ecosystem that 
grew around it. \citet{ARCAGI}[p. 45] writes that a general benchmark of intelligence
``should work for both humans and machines, fairly, by only assuming the same priors as possessed 
by humans (e.g. Core Knowledge) and only requiring a human-sized amount of practice time or training data''
and  ``we recommend only comparing the intelligence of systems that assume a sufficiently similar 
set of priors'' (p. 44). But in practice the test is routinely used to compare the intelligence 
of humans and transformer-based models, which have radically different priors. 

%\enlargethispage{10pt}

To be clear, we are not claiming that all reasoning is linguistic, nor that ARC fails to 
test reasoning altogether. There are well-established non-linguistic forms of 
reasoning, visuospatial reasoning being one of  them, and ARC does appear to tap abilities in this domain;
we would not be surprised if ARC performance is shown to correlate with $gV$. 
What remains empirically unclear is exactly which abilities ARC measures 
and how strongly performance on the benchmark correlates with broader measures of intelligence,
and particularly which of the prominent CHC factors it loads on most strongly. 
Our claim is that a vast and very important portion of human reasoning and abstraction 
is conceptual or theoretical: it concerns categories, relations, and entities that are not 
directly perceptible and whose representation and manipulation are inextricably linked with language. 
%Virtually every major theory of intelligence distinguishes between visuospatial ability
%and abilities such as verbal, logical-mathematical, numeric, and so on. 
Because nonverbal tests like ARC exclude language by design and operate over 
fully observable structures, they leave this large and very significant 
class of reasoning capabilities untested.\footnote{An analogy between 
reasoning and humor might help to clarify our position. 
Language is not strictly necessary for reasoning, in the same way that it is not necessary for humor---consider Chaplin's
silent films, slapstick, timing, visual surprise, incongruous or absurd gesture (Monty Python's
ministry of silly walks), and so on. At the same time, the most effective and intelligent  humor 
is conveyed with the use of language, which enables irony, parody, understatement, narrative framing, 
double meaning, register shifts, self-reference, and the management of multiple perspectives at once. 
We would likewise claim that the most sophisticated forms of reasoning 
are enabled by and expressed through language and symbolism, which allow for explicit premises and conclusions, 
scope distinctions, quantification, modality, counterfactuals, recursive embeddings, and above all 
the articulation of why something follows. Taking silent pratfalls as the essence of comedy 
would be like taking grid transformations as the essence of reasoning.}

Finally, even though  ARC was designed as a test that 
does not use language, largely in an effort to eliminate reliance on memorized patterns 
and to measure fluid intelligence, many of the systems that perform best on ARC  
are neuro-symbolic pipelines that do not perform simple end-to-end prediction 
of the  output grid \citep{ARCAGITechReport2024}. Instead, they parse 
the grid into objects, regions, and colors; they express candidate 
solutions in a DSL (domain-specific language) or restricted Python-like program space; 
they explicitly search over compositions of primitives, keeping only candidates that fit 
the training examples; and then run the resulting program on the test input. 
This process effectively maps these problems back into a symbolic \emph{language} (Python or DSLs) 
and relies on formal syntax, semantics, and logic to synthesize a solution.
And given that models are pretrained on terabytes of Python code, this allows them 
to lean on the massive linguistic and structural priors they memorized 
during pretraining, which seems at odds with a core-knowledge test. 
While it's true that the tasks themselves are purely visual, if the solver is a giant 
model pretrained on vast bodies of Python, iteration patterns, grid manipulation idioms, and so on,
and relies on that know-how during problem solving, then language priors 
re-enter through the back door.

\subsubsection{Formal Reasoning} 

The final category comprises benchmarks whose input problems are fully formal, expressed in symbolic logic or formal mathematical notation. 
These benchmarks are designed to probe reasoning in a highly controlled setting, abstracted away from the vagaries of natural language
or domain-specific conventions.

Unlike natural-language or generalist benchmarks, which often involve a mix of cognitive competencies and inferential strategies, 
formal reasoning benchmarks aim to isolate and evaluate deductive capacity in its purest form. The tasks typically involve proving 
or refuting precise claims, or applying logical rules within a constrained formal system. 
Most benchmarks in the category comprise contentful but formally structured math problems (as in MiniF2F \citep{MiniF2F});
datasets like \pgridsp that aim to capture content-agnostic reasoning are relatively rare 
(Rosetta \citep{Rosetta} is arguably another example, but see our comments below).

A key advantage of such benchmarks is that they are interpretable and verifiable: because the logic is fully specified, 
it is possible to determine unambiguously whether a response is correct and to assess the structure of the reasoning that led to it. 
This makes them especially well-suited for fine-grained diagnostic evaluation of a model's inferential abilities. Moreover, 
the absence of linguistic ambiguity or commonsense knowledge demands allows these benchmarks to more cleanly separate 
reasoning skill from linguistic fluency or background knowledge.

That said, formal reasoning benchmarks introduce their own complications. They tend to require a high degree of familiarity 
with the underlying logic or proof calculus, which makes them less accessible to general-purpose models or to researchers unfamiliar 
with the target formalism. Some benchmarks are grounded in complex systems like higher-order logic or dependent type theory, 
or rely on interactive proof protocols 
(such as Lean-style tactic traces) that complicate the relationship between proof generation and proof explanation, 
as we discuss later. \pgridsp mitigates these challenges by adopting a minimalist {\em natural deduction\/} language, 
featuring the same general style of reasoning commonly found in mathematics and logic textbooks. 
That language is fully described in a concise self-contained prompt. 

All in all, formal reasoning benchmarks play an essential role in the ecosystem of reasoning evaluation,
offering a uniquely rigorous lens on deductive competence.

%Later in this paper, we use the proposed set of axes to differentiate among  benchmarks in this space, distinguishing, for example, between content-free symbolic inference tasks and content-rich  formal mathematics, and between benchmarks requiring shallow rule chaining and those demanding deep structural proofs.

\subsection{Reasoning Dimensions}

To better understand the diversity of reasoning datasets and benchmarks in NLP and AI in general, 
we position these along three main orthogonal axes: 
\ben
\item an {\em input representation\/} axis;
\item an {\em output representation\/} axis; and a 
\item a {\em content\/} or {\em crystallized knowledge\/} axis. 
\een
These axes reflect conceptually independent dimensions that shape the kinds of reasoning that a benchmark can  
evaluate.\footnote{The axes are not continuous, 
but orthogonality is more important than continuity in a conceptual framework of this kind. 
A more granular analysis could refine each independent dimension into a categorical 
or ordinal feature with a specific set of interpretable values. 
We have already taken preliminary steps in this direction, \egnsp, 
we introduce a number of discrete levels for the output-representation axis: 
\temv{binary}, \emph{multiple-choice}, \temv{literal}, \temv{chain-of-thought}, \temv{entailment tree},  
\temv{Python code}, and {\em formal proof}. We leave the development of a finer-grained and exhaustive 
classification scheme to future work. Note also that while we treat the input representation 
and content axes as conceptually orthogonal, in practice they are not always independent. 
In particular, benchmarks with natural language inputs almost inevitably carry {\em some\/} semantic 
content, even when synthetically constructed. Although it is possible to design grammatically 
well-formed yet semantically vacuous sentences in Jabberwocky style (\egnsp, ``Every glorb is a zibble''), 
such examples are highly contrived. Thus, while the axes are formally distinct, they 
may be somewhat statistically correlated in practice.}

The first axis captures the form in which the inputs of a dataset are expressed. This could be naturalistic 
human language, or synthetically generated natural language, or formal symbolic notation, or something in between. 
This choice usually reflects the motivation behind the benchmark. Datasets like SNLI and MNLI are squarely in 
the mainstream NLI tradition and are thus invariably focused on natural language. Their main aim is to test 
whether language models can simulate human-like reasoning from textual input. 
Towards the middle of this axis we find datasets like RuleTaker and FLD, which are formulated in the sort 
of controlled natural language (CNL) described earlier: templated English sentences generated from symbolic rules. 
This results in data that is expressed in natural language but is formulaic, with very limited linguistic diversity,
and semantically aligned with formal logic. Finally, at the right end of this axis we find datasets like  
Rosetta, whose problems are expressed entirely in formal symbolic notation, using logics 
of varying representational power (like propositional, first-order, or higher-order logic). 

The second axis characterizes the structure and complexity of the required model outputs. 
On one end we have tasks that demand only a binary decision, such as whether a hypothesis follows from a context (\egnsp,  entailment vs. non-entailment), 
or the selection of a single correct answer from multiple choices. These setups are straightforward 
to evaluate and model, but offer limited insight into the actual reasoning that models perform. 
On the other end of the spectrum are tasks requiring more structured reasoning artifacts, such as chains 
of deductive steps, natural-language justifications (\egnsp, COT traces), entailment trees 
(as in EntailmentBank), or even full formal proofs (as in \pgrid). These richer outputs enable 
finer-grained analysis of a model's reasoning process and can support interpretability, 
proof checking, and compositional generalization, but typically demand more supervision and 
are harder to generate. 

The third axis characterizes reasoning benchmarks in terms of their content demands, 
capturing the type and extent of domain knowledge required to solve them. At one end of this spectrum 
we have content-free benchmarks, such as \pgrid, which involve purely symbolic structural reasoning 
tasks in uninterpreted settings. These tasks rely on the internal structure of the inference tasks and
require little external knowledge. Farther along this axis we find 
benchmarks grounded in commonsense or everyday content, such as SNLI or bAbI \citep{babiBenchmark}, where understanding is based 
on familiar types of objects, events, and social conventions. Moving along, we have benchmarks involving 
scientific content, like GPQA \citep{GPQA} or ARC (Challenge Set) \citep{allenai:arc}, which require reasoning over concepts and relations 
introduced by scientific theories, and are thus accessible only to those with some scientific education. 
At the far end of the axis we have pure math benchmarks such as MATH, GSM8K, or Lean-based proof datasets, 
where the content is fully mathematical, and reasoning requires both mathematical knowledge and 
symbolic manipulation. Although diverse, these benchmarks fall along a coherent continuum that reflects 
increasing degrees of semantic specificity, abstraction, and crystallized domain knowledge. 
Framing this as a content axis allows us to compare reasoning tasks not just by how they are formulated, 
but also by what kind of knowledge they demand.

In addition to these three primary axes, we identify three further dimensions that shed light on the internal 
structure and reasoning demands of a benchmark: the expressiveness of the underlying formalism, 
the range of inferential mechanisms required, and the diversity of its tasks. 
These axes are secondary and only relevant for datasets that are grounded in formal logic or in natural language rendered from logical forms.
\bdesc
\item[Formalism Expressiveness:] For benchmarks that are defined in, or translated from, a formal logic, including both 
fully symbolic datasets like \pgridsp and CNL benchmarks like RuleTaker, a key consideration 
is the expressive power of the underlying logical system. This axis captures the range of propositions that the 
formalism can represent and thereby constrains the structural and semantic complexity of the problem instances. 
For example, RuleTaker is grounded in stratified Datalog, which supports recursive Horn clauses and limited negation 
but disallows full disjunction, unrestricted negation, quantifier alternation, and function symbols. Other benchmarks might be based on 
propositional logic, first-order logic fragments, equational logic, description logics, 
epistemic, modal, temporal, or spatial logics, or more powerful systems like higher-order logic or dependent type theory. Each of these formalisms permits 
a different set of logical constructs (such as quantifiers, equality, modal operators, or arithmetic) which 
in turn determine the space of conceivable problems. Two benchmarks may both use formal representations 
yet differ dramatically in the depth and nuance of statements they can encode, and thus on the types of 
reasoning that they can test, depending on where they lie on this dimension.

\item[Inferential Repertoire:] This axis captures the range and sophistication of inference mechanisms 
that a model must deploy to explain and justify its answers to a benchmark's problems. 
It is mostly relevant for benchmarks that require models to produce \emph{structured reasoning artifacts} 
that are amenable to automated analysis, such as entailment trees or formal proofs. 
%In such cases one can examine the structure and diversity of these artifacts to estimate the benchmark's demands on inferential capacity.
For benchmarks that only require final answers, the model's reasoning process remains opaque 
and this axis becomes inert; without observable reasoning traces, it is not possible to determine 
what inference methods were used, short of speculating about latent cognitive or 
model-internal dynamics. The same applies to benchmarks that elicit \emph{unstructured} reasoning 
artifacts, such as chain-of-thought traces expressed in arbitrary natural language. Even if such traces 
are generated during inference, they are typically not evaluated and are therefore of little help 
in analyzing the underlying reasoning mechanisms. 

When structured reasoning outputs {\em are\/} generated, this axis ranges from benchmarks that require 
only simple chaining or forward inference using a small set of rudimentary rules (such as 
modus ponens and conjunction elimination) to tasks that demand complex reasoning strategies such as 
hypothetical proof, arbitrarily nested case analyses, universal generalization, reasoning by contradiction,
existential instantiation, and various forms of mathematical induction (natural, structural, strong, \etc). 
%This axis thus helps to distinguish benchmarks that probe shallow inference from those that challenge a model's capacity for deep, abstract reasoning.

While this dimension often correlates with the expressiveness of the formalism, 
it is not reducible to it. A benchmark may employ an expressive logic in theory while 
exercising only a weak fragment in practice. In such cases, the inferential repertoire axis 
provides an independent informative lens on the range of reasoning tested by a benchmark.

\item[Task Diversity:] Another important dimension for characterizing a reasoning benchmark is task breadth---the number 
and variety of distinct reasoning tasks it includes. Some benchmarks are narrowly focused, posing a single core problem,
such as entailment: ``Does hypothesis $h$ follow from context $T$?'' Even within that core task, variations can arise 
depending on the output format (e.g., binary label vs. a supporting explanation in COT style), but 
the central reasoning objective remains the same. Other benchmarks expand the task scope by supporting or requiring multiple 
types of reasoning, each probing different varieties of inferential ability. For example, in addition to entailment 
verification, a benchmark might include tasks like countermodel generation (demonstrating that $T \not\models h$), 
error diagnosis (identifying why a hypothesis fails), or hypothesis repair (suggesting minimal modifications that ensure $T \models h$). 
The number and diversity of these subtasks can be treated as a measure of a benchmark's functional richness. 

Greater task diversity broadens what a dataset can reveal and enables finer-grained assessment of a model's reasoning,
but it also complicates evaluation and interpretation. Benchmarks with many distinct subtasks can suffer from a lack of cohesion, potentially 
conflating heterogeneous cognitive demands. This can make it difficult to attribute model success or failure to specific competencies. 
Finally, aggregating performance across many diverse tasks tends to lack a principled foundation, 
raising questions about how to interpret a single combined score. As such, while broad task coverage 
offers diagnostic benefits, there is also value in relatively focused benchmarks with tightly 
defined reasoning objectives. Diversity is most effective when the subtasks are conceptually coherent 
and mutually reinforcing, rather than arbitrarily bundled together. This axis thus highlights 
a tension between diagnostic richness and interpretive clarity.

\edesc 

Together, these auxiliary axes complement the existing framework and offer finer-grained tools for analyzing what 
makes a benchmark challenging. 

Table~\ref{Tab:ReasoningBechnmarks} lists a number of datasets that are intended, either entirely or in part, 
to test the reasoning abilities of AI models. The datasets are characterized according to the dimensions discussed above,
along with a few additional attributes such as the maximum reasoning depth required (a rough ceiling on problem difficulty); 
current saturation status; and whether a model's reasoning is verified programmatically. 
A NA entry (\na) under \mbf{Size} means that the dataset has no fixed size. This usually means that the authors provide a script 
that can generate any number of data points, often set to a certain default value 
(\egnsp, 5K in the case of LogicInference \citep[][p. 10]{ontanon2022logicinference}).
In the remainder of this section we discuss the various entries in Table~\ref{Tab:ReasoningBechnmarks}.

\definecolor{cnlcolor}{RGB}{230, 242, 255}        % Light blue for CNL
\definecolor{nlcommoncolor}{RGB}{255, 245, 230}   % Light peach for NL+Common Sense (first 3)

%\definecolor{nlhybridcolor}{RGB}{240, 255, 240}   % Light green for NL+Hybrid
\definecolor{nlhybridcolor}{RGB}{240, 255, 240}   % Light green for NL+Hybrid

%\definecolor{nlmathcolor}{RGB}{255, 240, 245}     % Light pink for NL+Math
\definecolor{nlmathcolor}{RGB}{255, 235, 230}     % Light coral for NL+Math

\definecolor{nlsciencecolor}{RGB}{250, 245, 255}  % Light lavender for NL+Science
%\definecolor{logiccolor}{RGB}{255, 250, 235}      % Light yellow for Logic
\definecolor{logiccolor}{RGB}{255, 250, 190}      % Light yellow for Logic

\renewcommand{\arraystretch}{1.3}
\begin{table}[t]
{\scriptsize 
\centering
\small
\begin{tabular}{@{}lcccccccccc@{}}
\toprule
%\textbf{Dimension} &
{} &
\mbox{\footnotesize\bf Input} &  % ProofGrid
\mbox{\footnotesize\bf Output} &  % ProofWriter
\mbox{\footnotesize\bf Content} &  % RuleTaker
\mbox{\footnotesize\bf Expr.} & 
\mbox{\footnotesize\bf Range} & 
\mbox{\footnotesize\bf R. Depth} &  
\mbox{\footnotesize\bf Tasks} &  
\mbox{\footnotesize\bf Size} & % FLD
\mbox{\footnotesize\bf Satur.} & % FLD
\mbox{\footnotesize\bf Verif.} \\ % FLD
\midrule
\rowcolor{nlcommoncolor}  SNLI      & NL &  MCQA & Common Sense & \na & \na & \na  & 1 & 570K & High & \xmark \\
\rowcolor{nlcommoncolor} MNLI      & NL &  MCQA & Common Sense & \na & \na & \na  & 1  & 433K & High & \xmark\\
\rowcolor{nlcommoncolor} LogicQA 2.0      & NL &  MCQA & Common Sense & \na & \na & \na & 2 & 15708 & High & \xmark\\
\rowcolor{nlcommoncolor} Hellaswag      & NL &  MCQA & Common Sense & \na & \na & \na  & 1 & 70K & High & \xmark\\
\rowcolor{cnlcolor} RuleTaker      & CNL &  Binary & Common Sense & S. Datalog & Low & 5  & 1 & 500K & High & \xmark\\
\rowcolor{cnlcolor} ProofWriter      & CNL &  Proof Trees & Common Sense & S. Datalog & Low & 5  & 3  & 500K & High & \cmark\\
\rowcolor{cnlcolor} ProntoQA      & CNL &  COT & Common Sense & PHL & Low & 5 & 1  & \na & High & \cmark\\
\rowcolor{cnlcolor} LogicInference      & CNL &  COT & Common Sense & PL, $\mbox{\scriptsize FOL}_{\mbox{\tiny 0}}$ & Low & 5 & 5  & \na & High & \cmark\\
\rowcolor{cnlcolor} LogicAsker      & CNL &  Binary & Common Sense & PL, $\mbox{\scriptsize FOL}_{\mbox{\tiny 0}}$ & Low & 1 & 1  & 5450 & High & \xmark\\
\rowcolor{cnlcolor} LogicNLI      & CNL &  MCQA & Common Sense & PL, $\mbox{\scriptsize FOL}_{\mbox{\tiny 0}}$ & Low & 5 & 1 &  30K & High & \xmark\\
\rowcolor{cnlcolor} FLD      & CNL &  Proof Trees & Common Sense & PL, $\mbox{\scriptsize FOL}_{\mbox{\tiny 0}}$ & Low & 8  & 1  & 30K & High & \cmark \\
\rowcolor{cnlcolor} NLSat      & CNL &  Binary & CommonSense & PL & \na & \na  & 1  & 423K & High & \xmark\\
\rowcolor{cnlcolor} Folio  & CNL &  MCQA & Common Sense & PL, $\mbox{\scriptsize FOL}_{\mbox{\tiny 0}}$ & Low & 7  & 2  & 1435 & High & \xmark\\
\rowcolor{cnlcolor} LogicBench      & CNL &  Binary, MCQA & Common Sense & PL, $\mbox{\scriptsize FOL}_{\mbox{\tiny 0}}$ & Low & \na  & 2  & 15K & High & \xmark\\
\rowcolor{cnlcolor} Multi-LogicEval      & CNL &  Binary & Common Sense &  PL, $\mbox{\scriptsize FOL}_{\mbox{\tiny 0}}$ & Low & 5  & 1  & 1552 & High & \xmark\\
\rowcolor{nlsciencecolor} Entailment Bank  & NL &  Entailment Trees & Science & \na & \na & 2.7  & 1  & 1840 & High & \cmark \\
\rowcolor{nlsciencecolor} ARC (Challenge Set) & NL &  MCQA & Science & \na & \na & \na  & 1  & 7787 & High & \xmark\\
\rowcolor{nlsciencecolor} TheoremQA & NL &  Literals & Science & \na & \na & \na  & 1  & 851 & Low & \xmark\\
\rowcolor{nlhybridcolor} MMLU & NL &  MCQA & Hybrid & \na & \na & \na  & 57  & 15908  & High & \xmark\\
\rowcolor{nlhybridcolor} MMLU Pro & NL &  MCQA, Literals  & Hybrid & \na & \na & \na  & 14  & 12000 & High & \xmark\\
\rowcolor{nlhybridcolor} BBH & NL &  MCQA & Hybrid & \na & \na & \na  & 23  & 6511 & High & \xmark\\
\rowcolor{nlhybridcolor} BBEH & NL &  MCQA, Literals & Hybrid & \na & \na & \na  & 23  & 4520 & Low & \xmark\\
\rowcolor{nlhybridcolor} BBEH Mini & NL &  MCQA, Literals & Hybrid & \na & \na & \na  & 23  & 460 & Low & \xmark\\
\rowcolor{nlhybridcolor} HLE & NL &  MCQA, Literals & Hybrid & \na & \na & \na  & 23  & 2500 & Low & \xmark\\
\rowcolor{nlhybridcolor} SuperGPQA      & NL &  MCQA & Hybrid & \na & \na & \na  & 1  & 26529 & Medium & \xmark\\
%HELM & NL &  MCQA & Hybrid & S. Datalog & Low & DP  & TN  & SZ & Low & \xmark\\
\rowcolor{nlmathcolor} MATH      & NL &  Literals & Math & \na & \na & \na  & 1  &  12500 & High & \xmark\\
\rowcolor{nlmathcolor} GSM8K      & NL &  Literals & Math & \na & \na & 8  & 1  & 8500 & High & \xmark\\
\rowcolor{nlmathcolor} OlympiadBench      & NL &  Literals & Math & \na & \na & \na  & 1  & 6142 & High & \xmark\\
\rowcolor{nlmathcolor} OlymMATH & NL &  Literals & Math & \na & \na & \na  & 1  & 200 & Medium & \xmark\\
\rowcolor{nlmathcolor} Omni-Math      & NL &  Literals & Math & \na & \na & \na  & 1  & 4428 & High & \xmark\\
%BigMath      & NL &  Literals & Math & \na & \na & \na  & 1  & 250K & Low & \xmark\\
%NuminaMath      & NL &  MCQA & Math & S. Datalog & Low & DP  & TN  & SZ & Low & \xmark\\
\rowcolor{nlmathcolor} \textsc{\scriptsize FrontierMath}      & NL &  Literals & Math & \na & \na & \na  & 1  & 350 & Low & \xmark\\
\rowcolor{logiccolor} MiniF2F      & Logic &  Tactic Trees & Math & HOL & \na & \na  & 1  & 488 & Medium & \cmark\\
%Rosetta      & Logic & Binary & None & PL & \na & \na  & 1  & 25214 & High & \xmark\\
%LeanDojo      & Logic &  Sole Tactics & Math & HOL & \na & \na  & 1  & 98734 & Low & \cmark\\
%HOLStep      & Logic &  Tactics & Math & HOL & Low & RD  & TN  & SZ & Low & \xmark\\
\rowcolor{logiccolor} \sspgrid      & Logic & Proofs & None & PL, EQL & Med & \mbox{\scriptsize $> 100$}   & 4/15  & 3000 & Low & \cmark\\
\bottomrule \\[-0.05in]
\end{tabular}
\caption{Qualitative comparison of 33 reasoning datasets. A long dash (\na{}) means NA (not applicable).  \textbf{Expr.} represents the expressiveness of the underlying logic
(when one exists). \textbf{Range} represents the diversity of reasoning mechanisms tested by a benchmark. \textbf{R. Depth} is the maximum reasoning depth required. 
\textbf{Satur.} is the benchmark's saturation level. \textbf{Verif.} indicates whether a benchmark programmatically verifies the model's reasoning. NL: Natural Language; 
CNL: Controlled Natural Language; COT: Chain of Thought; MCQA: Multiple-Choice Question Answering; PL: Propositional Logic; 
S. Datalog: Stratified Datalog; PHL: Positive Horn Logic; $\mbox{FOL}_0$ = propositional logic plus a very small fragment of FOL; 
EQL: (unconditional) Equational Logic; HOL: Higher-Order Logic.}
\label{Tab:ReasoningBechnmarks}
}
\end{table}
\renewcommand{\arraystretch}{1.0}

\enlargethispage{15pt}

ProntoQA's logic is positive Horn logic, which is reducible to Datalog over unary predicates. 
The few  negated properties in ProntoQA are just specially named predicates; they do not introduce or need 
the semantic machinery---and potential non-monotonicity---of true stratified negation. 
Accordingly, the only reasoning mechanism that is needed for ProntoQA proofs is modus ponens. 

Although the \mbox{\textsc{EntailmentBank}} paper \citep{EntailmentBank} outlines three tasks, these are all varying 
scales of the same problem, entailment tree construction, so we have consolidated them under a single task designation.
We have listed \pgrid's task number as 4/15 to indicate that there are 4 general types of tasks but with 15 
separate datasets/instantiations. 
We report MMLU as comprising 57 tasks, following the paper that introduced the benchmark \citep{MMLU}, but note that 
these are not 57 distinct task types in terms of input/output format or basic problem-solving requirements. 
Instead, they all share the same basic task structure, namely, multiple-choice question answering (MCQA), typically 
with 4 answer choices. The variation across the 57 datasets lies entirely in the domain or subject matter (STEM questions,
social-science questions, anatomy, logical fallacies, moral disputes, and so on).\footnote{It's an unfortunate 
trend that {\em task\/} is becoming increasingly conflated with {\em subject matter}. As another example, 
the paper that introduced the HLE benchmark (Humanity's Last Exam) characterizes benchmarks as 
``collections of questions which assess model performance on tasks such as 
math, programming, or biology'' \citep[][p. 4]{HLE}.}
A more recent version of this benchmark, MMLU-Pro \citep{MMLUPro}, was introduced in 2024 with the aim of filtering
out trivial questions from MMLU; reducing the number of categories from 57 to 14; incorporating new questions from 
TheoremQA \citep{TheoremQA}, SciBench, and the STEM website; and adding several additional distractor options to multiple-choice questions. 
Nevertheless, even that version is nearing saturation, with o3, Grok-4, DeepSeek-R1, every Gemini model from Gemini-2.5 onward 
and every GPT model from GPT-5 onward all clustering between 84\% and 92\% accuracy. 

%was already showing ceiling effects by mid-2025, with Gemini 2.5 Pro, Grok-4 and Claude 4 Opus 
%Thinking approaching 90\% accuracy. By the time of this writing, 

Hybrid benchmarks like BBH and BBEH are marked as MCQA even though they typically include COT-based prompting setups, 
because the elicited COT traces are not evaluated. 

Note that some datasets include structured reasoning artifacts but we have nonetheless marked them as unable to verify 
a model's reasoning because the models themselves are not required to produce any such artifacts. 
This is the case, \egnsp, for LogicNLI \citep{LogicNLI}. Its hypotheses are synthetically generated by starting 
with premises and then applying a sequence of $n \leq 5$  inference rules until a final conclusion $p$ is obtained. 
The resulting proof trace $p_1, \ldots, p_n = p$ is stored in the dataset as the canonical derivation of $p$. 
When a model correctly classifies $p$ as entailed by the starting premises, the authors then check to see if the 
model also accepts the canonical proof's intermediate conclusions $p_i$, a test they dub ``proof traceability''
\citep[][p. 3741]{LogicNLI}. However, this is mostly a behavioral test of response coherence or consistency 
rather than proof traceability. Since the model produces no explicit reasoning and its internal process is opaque, 
agreement on $p_i$ offers no evidence that the model actually followed the canonical derivation path.

Some CNL datasets exercise a relatively large number of inference rules 
but are still classified as low-diversity, meaning that they test a small range of reasoning mechanisms. 
This is either because the large majority of the rules are drawn from a single logic (\egnsp, 
LogicInference supports 19 inference rules, but 17 of them are for propositional logic and 
only 2 for predicate logic); or because they represent superficial template 
variants that don't capture any fundamentally new abilities. 
It could be argued that inference rules in general are just one 
type of reasoning mechanism, so rule count alone is a poor indicator of 
reasoning breadth. %% of reasoning strategies that a benchmark can probe. 
In addition, there are several inference idioms, 
pervasive in everyday mathematical and logical discourse, 
that cannot be captured by simple inference rules operating on sentences.  
Chief among them is conditional reasoning, where one makes a provisional hypothesis 
and proceeds to derive consequences from it and other background assumptions, ultimately discharging
it and deducing a conditional. To represent this type of reasoning faithfully, one needs a notion
of assumption scope, which simple inferences rules do not provide. In principle, hypothetical reasoning
could be captured by inference rules that operate on sequents rather than plain
sentences, but none of these datasets work with sequents.\footnote{A CNL benchmark based on 
sequent reasoning would be somewhat at odds with the genre. These benchmarks strive to retain natural language,
even if it's at the level of a synthetic veneer, as an implicit gesture toward 
cognitive plausibility---they aim to simulate, however imperfectly, the surface form of human reasoning. 
By contrast, formal abstractions like sequents, while useful for rigorous analysis of logic and for proof theory, 
depart significantly from the style in which humans express their inferences in practice, 
and so their use would introduce structural artificiality into the benchmark.}
Even if they did, something important would
be lost in translation because sequents lack the hierarchical assumption nesting 
that is a defining feature of conditional reasoning in the 
wild.\footnote{This is not to say that sequent systems can't represent conditional reasoning 
at all (clearly they can), only that they do so in a very different style. 
The situation is analogous to the distinction between structured programming languages and assembly/machine languages.
The latter don't have the hierarchical nesting that is needed for natural descriptions of algorithms, 
but they are Turing-complete nevertheless.}
Reasoning by contradiction is essentially of the same type. Eigenvariables used in universal 
generalization (``Consider an arbitrary $x$ such that $\cdots$'') provide yet another example 
of a structured and scope-based reasoning mechanism that cannot be faithfully captured 
by simple inference rules. 
%And various forms of induction, particularly structural induction, are additional examples 
%of the insufficiency of simple inference rules. 
Mental-model theorists \citep{johnson-laird.wason.book} would go further and argue that reasoning is primarily semantic, driven by 
model-based entailment analysis rather than inference rules. This rings true in mathematics,
where a typical step is ``$p$ follows from $\cdots$,'' not an application of a specific low-level 
rule like modus tollens. It is up to the reader to verify that there is no model
in which $\cdots$ are true and $p$ is false. 
(In the setting  of propositional logic, \pgridsp captures this pervasive idiom 
with \ndlzero, which does away with all inference rules in favor of semantic entailment;
see Section~\ref{Sec:PLThreeOverview}.)
%% In general, inference rules are too 
%% low-level to be widely useful in mathematical reasoning in the wild.  They are mostly tools 
%% for meta-reasoning, \iensp, for formalizing and analyzing deduction, not necessarily ecologically 
%% valid patterns of inference. 

%%Model-oriented pyschological theories of reasoning, such as Johnson-Laird's mental-models theory, 

%% Inference rules are too low-level to be widely useful in actual mathematical reasoning. 
%% They are mostly tools for meta-reasoning, \iensp, for formalizing and analyzing deduction, not necessarily ecologically 
%% valid patterns of inference.  Model-oriented pyschological theories of reasoning, such as Johnson-Laird's mental-models theory, 
%% are better equipped to account for such phenomena, as opposed to rule-based theories like mental logic. 

%Inference rules are useful in the formalization and analysis of reasoning, but their psychologi
%over datatypes like lists and trees, which involves analysis of patterns ranging over the constructors of a free algebra, 

We have listed the depth of reasoning of GSM8K \citep{cobbe_gsm8k_2021} as 8 even though, strictly speaking,
the dataset does not contain any proof trees or other reasoning artifacts. However, the paper
that introduced the benchmark notes that ``problems take between 2 and 8 steps to solve, and solutions 
primarily involve performing a sequence of elementary calculations using basic arithmetic operations 
to reach the final answer,'' so we use this quantity as a proxy for reasoning depth. 
Even though OlympiadBench also has 2334 physics questions, we list its size as 6142 because that's 
the number of math questions it contains. OlympiadBench contains detailed solutions as well (in natural
language), although models are only evaluated on their final answers. 
%These are all \temv{FLAT} proofs, simple linear chains of steps without any hierarchical structure: ProofWriter, ProntoQA, LogicInference, 
We have classified Omni-Math  as highly saturated because o3-mini's accuracy on that benchmark already 
exceeded 80\% in 2025 \citep[][p. 2]{ballon2025relationshipreasoningperformancelarge}, without falling below 50\% on any categories. 
%We are not aware of any sources listing results for o3, Gemini-2.5-Pro, and/or Grok-4 on Omni-Math, but it's 
%reasonable to assume that these models would do better than o3-mini. 

We have listed NL for the input form of the problems of LogicQA 2.0 \citep{LogicQA2}, because these were obtained from naturally occurring text,
adapted from the Chinese Civil Service Examination.
The English version was obtained by machine translation and may inject ambiguities and other issues into a 
problem (though the authors remark that the quality of the translation is substantially better in  version 2.0). 

For FLD \citep{FLD1}, 30K is the size of each variant of the corpus, but there are multiple ``dot variants'' of the form FLD.$k$
for $k = 0,1,\ldots,5$. Each FLD.$k$ dataset flips at most one design knob at a time (such as the syntactic complexity
of the sentences, the depth of the proof trees, the number of distractors, and so on), so that
controlled ablation experiments can be run to see which aspect of reasoning is more challenging for 
language models.\footnote{Multiple corpus variants with different profiles are common in this 
line of research (a tradition going back to RuleTaker, which introduced five datasets varying by reasoning depth).}
In theory, FLD supports full first-order logic, albeit without equality or function symbols. In practice, 
the fragment it exercises is far more limited, essentially confined to monadic predicates without quantifier alternations. 
This reflects a broader trend among CNL benchmarks. 
Even though these are solidly framed in logic, many end up targeting 
very weak subsets of predicate logic. This is partly due to the sheer complexity 
of the bookkeeping infrastructure that is needed for a correct implementation of full-blown proofs in first-order logic. 
Implementing a sound and complete inference system for full FOL is technically demanding. 
It requires careful handling of variable scope, alpha-conversion, avoiding variable capture, 
and enforcing a number of subtle caveats. For instance, the witness variable in an existential 
elimination must be fresh and have no free occurrences in the conclusion. These provisos 
are not even mentioned in the formulation of existential instantiation in FLD \citep[][Fig. B.3(a), p. 15]{FLD1}. 
This raises questions. 
None of these benchmarks provide 
any guarantee that the generated proofs are correctly implemented and tested. 
Soundness rests entirely on a Python codebase, typically built from scratch for the purposes of the benchmark. 
The risk of error is  mitigated by the severe restrictions imposed on the richness 
of the sentences and proofs that are allowed, but the restrictions also mean that 
these benchmarks are far from exercising the more challenging---and interesting---parts 
of predicate logic. Using a preexisting and already-vetted proof-engineering system as an off-the-shelf 
component to handle the drudgery of proof representation and manipulation 
%(as \pgridsp does with Athena and LeanDojo does with Lean) 
has the advantage of minimizing the benchmark's trusted computing base. 
The creators of FLD note that they avoid the use of external logic systems in order to maintain 
greater control over data generation, but such use need not constrain the selection of axioms, 
the depth of proofs, the size of formulas, or most of the parameters that benchmarks are 
typically interested in tweaking. 

NLSat \citep{NLSat} is notable in that, like the \plthreesp division of \pgrid, it attempts to build 
computationally hard inference problems. But it does so not by drawing from well-known 
families of complex problems but rather by random sampling near the critical region of the 3-SAT 
phase transition. The resulting problems pose simple yes/no (\mbf{sat} or \mbf{unsat}) questions.
There are no proofs or other reasoning artifacts, so reasoning depth is impossible to measure.
The authors do provide an algorithm-specific measure of complexity, namely the average and median numbers 
of backtracking conflicts and variable assignments reported by the $\mbox{DPPL}(T)$ algorithm used by Z3
when solving the generated problems; the maximum average and median was 22 and 13, respectively. 

LogicBench \citep{LogicBench} and Multi-LogicEval \citep{MultiLogicEval} include a number of non-monotonic 
inference patterns embedded in FOL. Multi-LogicEval does not require any proofs, but their results were
obtained with zero-shot COT prompting and they performed a manual analysis of the output COT traces. 

\textsc{FrontierMath} has 300 core problems, organized into 3 tiers of difficulty, with an extra 50 in a fourth tier,
so we have listed 350 as the total size. 
The table does not include recent math datasets like NuminaMath \citep{NuminaMath} and Big-Math \citep{BigMath}, 
because these are much bigger and primarily intended for training (particularly via RL) rather than benchmarking. 
We likewise omit logic-based datasets like  Rosetta \citep{Rosetta}, as they are  also mainly intended for training. 
Rosetta could be used for evaluation as well, but only after one has fine-tuned a model on the training split, 
because the input format of the problems is a deliberately obfuscated notation (even though the problem structure 
is propositional), in order to force models to focus on reasoning rather than memorization of surface-syntax patterns. 
As a result, off-the-shelf models would not fare well on these problems, especially in a zero-shot setting, 
and the dataset does not provide any scaffolding for ICL prompting. Either way, models are not require to produce 
proofs or any other reasoning artifacts. 

Similar issues arise in datasets like HOL Step \citep{HOLStep} and LeanDojo \citep{LeanDojo}. 
These are designed to train models to perform highly specific tasks in tactic-based 
proof systems like HOL and Lean, such as:
\bit
\item \emph{Premise selection}: Given a goal $g$ and an available result $p$, determine whether $p$ will 
be useful in proving $g$.
\item \emph{Next-tactic selection}: Given the current proof-search state (a notion that is specific to tactic-based
theorem provers), encoded as a string, along with a ranked list of premises (generated by a premise-selection component) 
representing available lemmas, predict the next tactic to apply (a function that takes a sequent as input and produces another sequent as output). 
\eit 
Neither was designed to test a model's general ability to reason with or about proofs, and 
neither provides prompt scaffolding for ICL-testing of off-the-shelf models. 
If one wrote prompts from scratch (see footnote~\ref{Ft:LeanComplexity}), one could attempt to use LeanDojo 
as a test set to be given to an off-the-shelf model, but doing so would mostly gauge the model's ability  
to generate a Lean tactic that is syntactically valid and semantically applicable to the input proof state. 
This requires the model to be deeply familiar not just with general Lean syntax 
and semantics but also with its extensive tactic library; and more importantly, the result would not tell us whether the 
predicted tactic would eventually lead to success.

In general, tactic-based interactive theorem-proving frameworks are primarily geared towards verification 
rather than presenting reasoning in the structured and hierarchical style that is characteristic of human-written 
proofs.\footnote{That said, broader ecosystems grown around such frameworks increasingly support mixed styles, 
structured proofs, comments, local lemmas, and higher-level proof organization.}
The main representation of a proof in such a framework is a sequence of system-specific 
tactic applications. These tactics are usually applied in a backward fashion, decomposing a goal into subgoals until 
the process reaches axioms or trivially solvable leaves. Although this analytical process does operationalize 
the {\em search\/} for a proof object, 
%% object (in Lean's terminology, a ``proof term'')
%it does not itself constitute a proof in the traditional sense; 
it does not directly function as a proof in the usual expository or narrative sense; 
instead, it's an execution trace that records a proof-search trajectory. 
This is an  important distinction. The tactic trace can convey key aspects of strategy 
%that provide a procedural window onto the reasoning 
(\egnsp, which lemmas drive the argument, where cases split, or which 
invariant is maintained), but the overall structure of the reasoning is encoded implicitly and unevenly. 
It is dispersed across highly procedural steps that can obscure the high-level organization of the argument, 
especially for non-expert users of the system, instead of presenting it in a stable, declarative, and 
%of reflecting a stable, declarative, and 
hierarchically articulated proof scaffold. % that mathematicians would recognize as a proof. 
%The tactic trace itself offers little insight into the structure or high-level conceptual organization of the underlying reasoning. 
%Unlike natural deduction, which captures scopes, assumptions, and subproofs in a nested and human-readable format, tactic traces are flat, imperative scripts with no syntactic indication of logical scope or structure.
Lean's core does ultimately generate a proof term, a fully elaborated term in the Calculus of Inductive Constructions 
that certifies the correctness of the theorem. However, these proof terms are typically opaque and unwieldy, 
even for experienced users. They are often closer to compiled machine code than to conventional mathematical argument. 
In practice, even seasoned Lean users seldom engage with these proof terms directly, instead treating the sequence of tactics 
itself as the canonical proof.
As a result, benchmarks based on such systems are better viewed as evaluations of system-specific 
verification abilities than as direct tests of declarative, human-interpretable 
reasoning.\footnote{Of course, even if tactic sequences aren't structured like natural deduction, 
they still represent how real users build proofs interactively and hence they do capture 
useful reasoning artifacts. For users working {\em within\/} a system like Lean, tactic sequences 
are familiar and perfectly functional. But while they are useful operational artifacts, they are not 
accessible and structurally clear explanations of deductive inference, which is what would be needed for 
measuring interpretable reasoning.}
%As a result, because the outputs are not declarative explanatory deductions but instead procedural records of a system-specific
%search process, benchmarks based on such systems fall short of evaluating interpretable, human-aligned 
%reasoning.

%% Moreover, the lack of explicit scoping, assumptions, and nested subderivations---all crucial to natural deduction 
%% and traditional mathematical exposition — makes tactic-based proofs unsuitable as pedagogical or explanatory artifacts. 
%% Any attempt to analyze such outputs through the lens of inferential repertoire, reasoning depth, or deductive 
%% structure is therefore severely constrained by the limitations of the proof representation itself.

In addition, because Lean is a complex domain-specific formal language, such benchmarks are accessible 
mostly to models with substantial prior exposure to it. Barring fine-tuning, meaningful performance is 
unlikely for models that have not been heavily pretrained on Lean code and proofs, and even then, 
performance can remain sensitive to nuances in the training distribution.\footnote{One might 
point out that a system like Lean is just another formal language, and any formal language can be 
described in a prompt. After all, this is exactly what was done with NDL in \fpgrid. \label{Ft:LeanComplexity} 
In principle, this is a reasonable point. If Lean was a simple language, then in-context learning via prompting could suffice. 
But Lean is a large and multi-layered system with all the inevitable idiosyncrasies that come with such scale.
It has a dependent type theory core (the Calculus of Inductive Constructions); thousands of predefined tactics and lemmas; 
a rich module system; a highly non-trivial unification and elaboration behavior; and so on. 
In practice, even advanced models fail to generate correct Lean code unless heavily pretrained on it. 
Prompting might help to guide or structure a solution, but it can't replace deep preexisting familiarity with Lean.}
This makes such benchmarks less well suited for evaluating a model's general reasoning abilities. 
Additionally, running such benchmarks requires substantial Lean-specific infrastructure, 
including specialized parsing, proof-checking, and goal-state tracking machinery, often tightly coupled to particular 
versions of the system. The associated tooling overhead places additional constraints on the 
accessibility and extensibility of such benchmarks.
%While LeanDojo represents a valuable contribution for testing fine-grained proof synthesis 
%within the Lean ecosystem, its specialized design makes it a narrower benchmark 
%than a general test of proof-based reasoning. 

Finally, the content of benchmarks like LeanDojo is not abstract logic or untyped 
symbolic reasoning, but rather interpreted formal mathematics from subjects like number theory. 
As such, the benchmark is deeply contentful: solving its problems depends not only on
fluency in manipulating Lean tactics but also on substantial crystallized mathematical knowledge, including 
a large number of domain-specific results (as well as notational conventions and standard proof idioms). 
This positions LeanDojo and similar datasets at 
the far end of the content axis in our taxonomy, where reasoning is tightly entangled with deep formalized subject matter. 
%They do not test fluid or content-free reasoning, but 
%rather assess performance in a setting 
This makes them useful for evaluating mathematical assistants or theorem provers in specialized domains
within the Lean ecosystem, but less suitable as general-purpose benchmarks for testing fluid or broadly transferable reasoning.

The table also omits datasets like \textsc{NaturalProofs} \citep{NaturalProofs}, which contains a large number of mathematical
proofs (32K) in natural language, mostly culled from \href{https://proofwiki.org}{ProofWiki}. Although it is a useful
resource, this dataset presents a significant challenge for the streamlined benchmarking of proof-writing ability, 
as it is not currently possible to reliably evaluate an arbitrary natural-language proof attempt. 
The two tasks studied in the paper that introduced this dataset were the retrieval and generation of the {\em references\/} that appear
in a given proof (definitions, other cited theorems, and so on); such references are stored as part of the dataset
and hence candidate solutions to both tasks can be mechanically checked for correctness. A subsequent publication
by the same research group did introduce proof-writing tasks, both at the global level of the entire proof and at
the finer-grained level of the next proof step \citep{NaturalProver}. However, the issue remains: there is 
no known sound and complete algorithm for evaluating an arbitrary piece of natural-language text that purports to prove 
a mathematical result. The authors took a pragmatic approach to this by combining
manual evaluation by human experts with the use of computable metrics that can serve as proxies for
proof quality, such as lexical similarity with the gold proof, using metrics like GLEU and Token F1,
and ``knowledge grounding'' (overlap of the definitions and results cited by the proof attempt with those
of the gold proof). Unfortunately, manual evaluation does not scale, while the heuristics are clearly not sound
(as the authors implicitly acknowledge). A mathematical result can have many different proofs, each relying on
a somewhat different set of previous results and even a different set of definitions.
Nevertheless, the metrics show a strong positive correlation with human judgments of correctness,
suggesting they are useful proxies. Accordingly, this work should be understood as an interesting 
pragmatic evaluation methodology, or a framework that can provide a meaningful but imperfect signal about the
ability of models to prove mathematical theorems in natural language.

%NaturalProofs & NL &  Proofs & Math & \na & High & \na  & 1  & 32012 & Low & \xmark\\

The 488 data points in MiniF2F are divided between 244 test and 244 validation problems with a stratified 
random split ensuring that each half has a balanced set of problems from each category and level of difficulty. 
The table marks the maximum reasoning depth of MiniF2F as NA (\na) for several fundamental reasons that 
distinguish it from benchmarks like RuleTaker or ProofWriter, and our own \pgrid.
%where proofs are constructed  from a fixed set of inference rules and syntax forms. 
First, MiniF2F provides only  problem statements, not machine-checkable proofs. 
This helps to prevent leakage. Although third-party groups have released separate proofs for certain subsets of the corpus, 
such as the Lean proof pack for the 14 IMO problems in MiniF2F's test split, which was recently published by~\citet{yousefzadeh2025a},
these are not part of the benchmark itself. 
Without a canonical proof corpus, there are no ground-truth objects that can be used 
to measure properties like proof depth.

Even if a comprehensive proof corpus were available, reasoning depth 
% a mathematical problem but is instead
is a function of the chosen proof system. MiniF2F is a multi-platform benchmark, 
with problems formalized in Lean, Metamath, Isabelle, and HOL Light. The structure and length of a proof can vary 
dramatically between these systems. For example, a proof in Metamath, which is based on a minimal set of axioms 
and a single rule of inference (modus ponens), is typically a long, linear sequence of low-level steps. In contrast, 
a proof in a system like Lean or Isabelle can be much more concise by leveraging not only a rich library of 
previously proven lemmas but also powerful automation tactics. A single-line proof in Lean might correspond 
to thousands of steps in Metamath, undermining any attempt to define a system-independent depth metric. 

Tactics add considerable complexity to the notion of reasoning depth in automated proof assistants.
Tactics are high-level abstractions that automate sequences of logical 
inferences, typically applied backwards (in an analytic and proof-search-flavored rather than a synthetic proof-building fashion). 
A single call to the \smtt{nlinarith} tactic in Lean, for instance, can solve complex non-linear real 
arithmetic problems by invoking a sophisticated decision procedure that may correspond to many thousands of primitive logical steps. 
Is this one step or thousands? The answer depends on the desired level of abstraction. This is analogous to measuring 
program complexity by lines of code: a single call to a powerful library function like quicksort represents one 
``step'' at a high level of abstraction, but it encapsulates a complex algorithm with a non-trivial number of primitive operations. 
Because modern theorem proving relies heavily on such abstractions, a simple ``proof tree height'' metric fails to 
capture actual reasoning complexity. 

In the case of computation, complexity theory has found a way to talk 
about the number of steps in an algorithm independently of specific languages or machine models, even when those steps 
involve arbitrarily complex function calls. This is done with the introduction of key abstractions like 
Turing machines or RAM programs, which serve as standardized low-level computational models. 
Even if we write a program in  a high-level language, the analysis assumes it can be compiled down to a sequence of basic steps in the low-level machine.
So when we say that an algorithm runs in $O(n^2)$ time, we're measuring how many primitive steps in the underlying computation
model are taken, assuming that the high-level constructs are compiled into those. Crucially, this works because the chosen model 
(\egnsp, Turing machines or RAM programs) is universal and robust under simulations, owing to Turing completeness: 
any reasonable programming language can be compiled to it with at most a polynomial overhead. There is no widely accepted analogue 
for deduction, although in principle one could fix a low-level proof system that does not allow for tactics or proof search, 
%such as a vanilla sequent calculus in the case of systems like Lean or HOL, or a type-$\alpha$ denotational proof language in the case
%of Fitch-style natural deduction \citep{OmegaDPLs-Memo}, 
and then approximate the reasoning depth of a rich high-level proof---one that does 
use tactics and performs arbitrary proof search---by measuring the size of the ``expanded'' certificate in the 
bare-bones formalism.\footnote{Section 1.8.2 of \citet{OmegaDPLs-Memo} introduces a rigorous definition of deductive cost 
in exactly this way.} Theoretically, this is defensible. It is the approach taken in proof complexity theory, for example, where 
%\iensp, the ``proof term'' produced in a system like Lean or the type-$\alpha$ certificate produced in a proof language like Athena.
the analogue of time complexity is proof length, usually defined as the number of symbols or steps in a derivation 
under some fixed low-level system (such a Frege system, resolution, or cutting planes). Different proof systems are 
then analyzed on the basis of how efficiently they can prove tautologies (\egnsp, with some theorems requiring exponentially 
long proofs in one system but not in another). 
%% This would also be practically feasible, since languages like Lean and Athena
%% do generate these expanded proof certificates, and the result would certainly provide \emph{some} signal 
%% about the complexity/depth of the reasoning. 

However, in practice proofs introduce complications that do not arise---at least not to the same extent---in the case of pure computation. 
In particular, there is often a sizable conceptual gap between the formal depth of a machine proof and the intuitive difficulty that a human perceives. 
A problem that a mathematician considers to have a ``two-step'' solution might require hundreds or thousands of formal steps to rigorously 
prove from first principles, including the handling of tedious side conditions and corner cases that a human would dismiss as trivial. 
Conversely, a problem with high conceptual difficulty for a human might have a short, brute-force proof that is easily 
found by machine. Thus, even though a  general and domain-independent notion of formal reasoning depth could be calculated, 
it would not necessarily correlate with the depth that is relevant to human mathematical reasoning.

In summary, due to the absence of a canonical proof set, the profound dependency on the chosen formal system, the variable 
granularity of tactics, and the disconnect from human-centric problem difficulty, any single number for the 
``maximum reasoning depth'' of MiniF2F would likely be arbitrary. Similar remarks apply to the range of reasoning 
mechanisms exercised by MiniF2F, which we have also marked as NA. 

\begin{figure}
\centering
\scalebox{0.75}{\includegraphics[width=\linewidth]{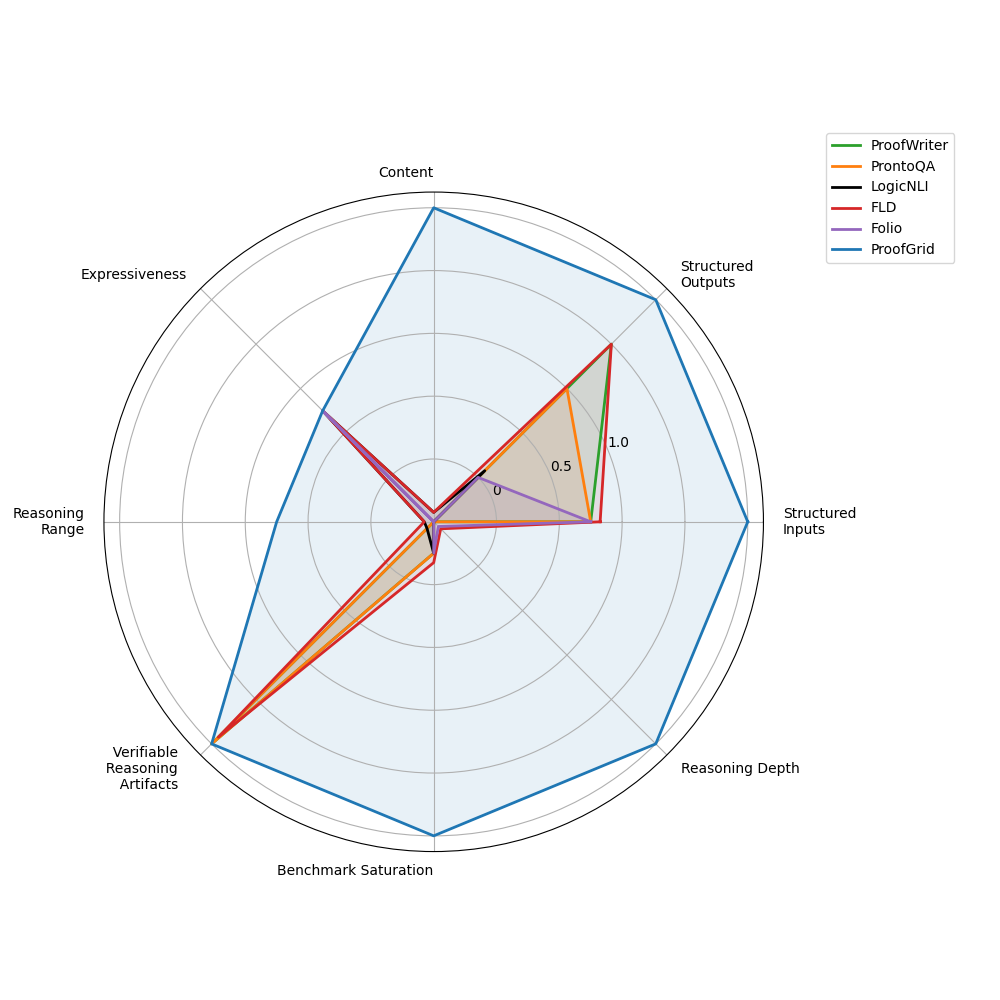}}
\caption{A radar-chart comparison of \pgridsp and 5 logic-based CNL benchmarks.}
\label{Fig:RadarChart}
\end{figure}

The radar chart in Fig.~\ref{Fig:RadarChart} provides a visual comparison of \pgridsp and 5 reasoning benchmarks along eight evaluation axes 
that reflect structural, logical, and operational characteristics. Each axis is normalized to a $[0,1]$ scale, where higher values 
indicate more favorable properties (e.g., more structured outputs, deeper reasoning, less saturation, and verifiable reasoning 
artifacts).\footnote{For example, maximum-reasoning-depth values were normalized with min-max scaling, by replacing each value $d$
by \[\cfrac{d-d_{\scriptsize min}}{d_{\scriptsize max}-d_{\scriptsize min}}\] where $d_{\scriptsize min}$ and $d_{\scriptsize max}$ are the smallest and largest
maximum reasoning depths. For the output axis, we set: Binary:$\,$0, MCQA:$\,$0.2, Literals:$\,$0.4, COT:$\,$0.6, Proof Trees/Entailment Trees:$\,$0.8, Proofs:$\,$1.0.
And likewise for the remaining dimensions.}
A small amount of jitter is added to the values of some benchmarks to prevent visual overlap in the plot.\footnote{This occurs 
when two or more benchmarks have identical scores across all axes---for example, LogicNLI and Folio share nearly the same profile.
Without jitter, the polygon representing one benchmark would be completely obscured by the other, making it invisible in the visualization.
The jitter values are very small (typically around 0.03) and do not materially affect the interpretive significance of the plot; 
they only serve as a visual disambiguation aid.}
\pgridsp is the only dataset that scores maximally on all axes except expressiveness and reasoning range, where its values are moderate. 
Overall, the plot illustrates how many CNL benchmarks cluster around low to medium values across most axes, while \pgridsp 
occupies a distinctive position as a structurally rich and logically grounded benchmark with a 
broad and deep inferential footprint.

%% \bibliographystyle{plainnat}
%% \bibliography{custom}
%% \end{document}

\section{\plone}

\label{Sec:PLOne}

%\clearpage 

\subsection{Proof Writing}

\label{Sec:PL1PW}

\subsubsection{Data} 

\mbf{PL1-PW} contains 400 problems generated by the algorithm described in Section~\ref{Sec:PLOneDataGenAlgo}. 
Half of these problems are {\em conditionalized}, meaning that they contain no premises, 
while every argument in the other half has at least one premise. 
Each conditionalized argument $\alpha$ was obtained from an argument that initially had 
$n > 0$ premises $p_1,\ldots,p_n$ and a goal $q$, by setting $\alpha$'s goal to the 
conditional $(p_1 \mand \cdots \mand p_n \mif q)$ and removing all premises. 

\renewcommand{\arraystretch}{0.5}
\begin{table}[!b]
\centering
\resizebox{0.8\linewidth}{!}{%
\rowcolors{2}{blue!5}{white} % alternate shading, starting with row 2
\begin{tabular}{
  l
  S[table-format=2.2]   % mean
  S[table-format=2.2]   % SD
  S[table-format=2.2]   % median
  S[table-format=2.0]   % min
  S[table-format=2.0]   % max
  l                     % spark
}
\toprule
{Quantity} & {Mean} & {SD} & {Median} & \msp\msp{Min} & \msp{Max}\msp & \multicolumn{1}{c}{Spark} \\
\midrule
% Args: min, p5, p25, median, p75, p95, max, mean
Number of premises 
  & 2.51 & 0.62 & 3.0 & 1 & 4 
  & \boxsparkcolored{1}{2}{2}{3}{3}{3}{4}{2.51} \\

Total problem AST size 
  & 28.12 & 13.76 & 33 & 6 & 69 
  & \boxsparkcolored{6}{9}{15}{33}{39}{48}{69}{28.12} \\
%% Goal AST size 
%%   & 2.5 & 1.78 & 3.0 & 1 & 10 
%%   & \boxsparkcolored{1}{1}{1}{3}{3}{6}{10}{2.5} \\
\# of negations per problem 
  & 3.17 & 2.16 & 3 & 0 & 10 
  & \boxsparkcolored{0}{0}{1}{3}{5}{7}{10}{3.17} \\ 
\# of conjunctions per problem
  & 4.12 & 2.78 & 4 & 0 & 12 
  & \boxsparkcolored{0}{0}{2}{4}{6}{9}{12}{4.12} \\
\# of disjunctions per problem 
  & 3.1 & 2.02 & 3 & 0 & 8 
  & \boxsparkcolored{0}{0}{1}{3}{4}{7}{8}{3.1} \\
\# of conditionals per problem
  & 3.84 & 2.62 & 4.0 & 0 & 14 
  & \boxsparkcolored{0}{0}{1}{4}{6}{8}{14}{3.84} \\
% Args: min, p5, p25, median, p75, p95, max, mean
\# of biconditionals per problem
  & 0.28 & 0.5 & 0 & 0 & 2 
  & \boxsparkcolored{0}{0}{0}{0}{1}{1}{2}{0.28} \\
\# of propositional atoms per problem
  & 3.15 & 1.09 & 3 & 2 & 7 
  & \boxsparkcolored{2}{2}{3}{3}{3}{6}{7}{3.15} \\
%% CP Goal AST size 
%%   & 40.62 & 5.89 & 39 & 32 & 69 
% Args: min, p5, p25, median, p75, p95, max, mean
\bottomrule
\end{tabular}}
\caption{Data distribution summaries with inline box-sparklines for the 400 problems in \mbf{PL-PW1}. 
Box = IQR (Interquartile Range); red line = median; dot = mean; whiskers = 5--95th.}
\label{Tb:PW1DataDistribution}
\end{table}
\renewcommand{\arraystretch}{1.0}

Table~\ref{Tb:PW1DataDistribution} summarizes the distributions of several key 
features of these problems. The first row, number of premises, pertains only to the 200 arguments that are not 
conditionalized. The second row shows the distribution 
of problem AST sizes. This is defined, for a given problem, as the sum of the sizes of all 
premise ASTs plus the size of the goal AST.\footnote{The size of a formula AST (abstract syntax tree) 
is simply the number of nodes in the tree.} 
This is not reflected in the table, but the 200 conditionalized problems tend to be 
larger than the other 200 problems. The next five rows show the distributions of the five logical 
connectives across all 400 problems. Conjunctions are the most common formulas, followed by 
conditionals, negations, disjunctions, and finally biconditionals, which occur very infrequently 
(averaging only 0.28 occurrences per problem).

The last feature, the number of atoms (propositional variables) in a problem, is particularly important. 
This is a core indicator of logical difficulty. In this task, the median number of 
atoms is 3, and in fact 75\% of all problems have no more than 3 atoms. 
In general, the number of propositional atoms bounds the logical space sharply, since 
$n$ atoms admit exactly $2^n$ possible truth assignments. 
If $n=3$, only 8 valuations need to be distinguished, so no matter how complicated 
the premises might appear syntactically, the underlying semantic structure is small. 
This is a key reason why \mbf{PL1-PW} problems are so tractable---because they 
collapse into small finite case analyses. 

This is not to say that the structure and size of the premises and the goal are immaterial. 
These can pose syntactic and structural challenges: how deeply formulas are 
nested, which connectives occur where, how many subproofs must be opened and discharged, 
and how effectively the relevant introduction and elimination rules can be coordinated. 
All of these factors can affect proof discovery. 
However, when the overall number of atoms is small, problems with many or deep premises 
often have redundancies, contradictions, or tautologies that reduce to 
something simple once the cases are enumerated. 
Semantic difficulty, by contrast, correlates strongly with the number of truth assignments 
that can be distinguished, and this scales directly with the number of atoms. 
A central challenge in proof discovery, 
at least in propositional logic, arises from the explosion of the search space 
that occurs as the number of atoms increases. 

%% %In general, the shapes and sizes of the premises and goal 
%% %influence
%% that affect the amount of required bookkeeping (how many and which elimination rules to apply, 
%% how many subproofs must be nested, and so on). %% and how easy it is to ``see'' the right inference path.
%% Semantic difficulty, by contrast, comes primarily from the number of truth assignments 
%% that can be distinguished, and this scales directly with the number of atoms. 

%% Problems with many or deep premises but few atoms often have redundancies, contradictions, 
%% or tautologies that reduce to something simple once the cases are enumerated. 
%% The primary challenge in proof discovery---at least in propositional logic---usually 
%% arises from the explosion of the search space that occurs as the number of atoms increases. 

\subsubsection{Proof Checker Instrumentation} 

As an example, consider this relatively large conditionalized problem in \mbf{PL1-PW}: 
\begin{tcAthena}
(((A ==> (~ (E | ~C))) | (~B & (C | E))) & ((E & ~E) | (B ==> F)) & E & B ==> ((B | C) & (B ==> (F & ~A))))
\end{tcAthena}
This might look intimidating at first glance, but a more careful inspection of its structure, clarified by
writing the formula in prefix notation, shows that it is not difficult to prove: 
\begin{tcAthenaWithNumbers}
(==> (& (| (==> A
                (~ (| E ~C)))
           (& ~B
              (| C E)))
        (& (| (& E ~E)
              (==> B F))
           (& E B)))
     (& (| B C)
        (==> B
             (& F ~A))))
\end{tcAthenaWithNumbers}
Assuming the antecedent, which is a conjunction of several formulas, including $E$ and $B$,  we 
need to derive the conjunction of $(B \mor C)$ and the conditional $(B \mif F \mand \neg A)$. 
Since we are already given $B$, deriving $(B \mor C)$ is trivial (by disjunction introduction). For the second
part, to derive $(B \mif F \mand \neg A)$, we can assume $B$ (which we already have anyway) 
and show that both $F$ and $\neg A$ hold. We are given that $B$ implies $F$,\footnote{More precisely, we are given (lines 5 and 6)
the disjunction of $(E \mand \neg E)$ and $(B \mif F)$, and this disjunction 
is equivalent to $(\temv{false} \mor (B \mif F))$, which, by $\neg\,\temv{false}$ and 
disjunctive syllogism, yields the conditional $(B \mif F)$}
so, since we have $B$, modus ponens gives $F$ and the only remaining goal is $\neg A$.
The latter follows by a \emph{reductio}: if we had $A$, we could easily derive a contradiction
by a case analysis of the antecedent's first conjunct (the disjunction on lines 1--4).
This is, essentially, the proof. Translating it into NDL is a bit tedious but not complicated. 

The best models get this reasoning exactly right. Gemini-2.5-Pro, for example, offers the following 
lucid account of its proof strategy: 
%\\[0.05in]
%\begin{minipage}{6in}
\begin{quote}
{\footnotesize\em
The overall strategy is to prove the main conditional by assuming its antecedent and deriving its consequent. 
The antecedent is a large conjunction, which is first deconstructed into its four main components: \ben
\item \mtt{((A$\:$==>$\:$($\ntildeneg\:$(E$\:$|$\:\ntildeneg\:$C)))$\:$|$\:$(($\ntildeneg\:$B)$\:$\&$\:$(C$\:$|$\:$E)))}, 
\item \mtt{((E$\:$\&$\:$\ntildeneg\:E)$\:$|$\:$(B$\:$==>$\:$F))},
\item \mtt{E}, and
\item \mtt{B}.
\een
\noindent The consequent, \mtt{((B$\:$|$\:$C)$\:$\&$\:$(B$\;$==>$\:$(F$\;$\&$\:$\ntildeneg\,A)))}, 
is also a conjunction, so its two parts are derived separately. 
The first part, \mtt{(B$\:$|$\:$C)}, follows directly from component (4), \mtt{B}, 
by disjunction introduction. The second part, 
\mtt{(B$\:$==>$\:$(F$\:$\&$\:$($\ntildeneg\:$A)))}, is proven by a conditional proof, 
assuming \mtt{B}. Within this subproof, \mtt{F}
 is derived by a case analysis on component (2). The first case, 
\mtt{(E$\:$\&$\ntildeneg\:$E)}, 
is a contradiction from which \mtt{F} follows trivially. 
The second case, \mtt{(B$\:$==>$\:$F)}, yields \mtt{F} 
by modus ponens with the assumed \mtt{B}. \\[0.05in]
Next, \mtt{$\ntildeneg\:$A} is derived, also by a case analysis, 
on component (1). The first case, \mtt{(A$\:$==>$\:$($\ntildeneg\:$(E$\:$|$\:\ntildeneg\:$C)))},
leads to a contradiction if \mtt{A} is assumed (given component (3) \mtt{E}), 
thus proving \mtt{$\ntildeneg\:$A}. 
The second case, \mtt{(($\ntildeneg\:$B)$\:$\&$\:$(C$\:$|$\:$E))}, 
contradicts component (4) \mtt{B}, so \mtt{$\ntildeneg\:$A}
again follows from the contradiction. With both \mtt{F} and \mtt{$\ntildeneg\:$A}
derived, their conjunction is formed, completing the conditional proof. Finally, the two parts 
of the main consequent are joined together to yield the desired conclusion. 
}
\end{quote}
Nevertheless, Gemini fails to transcribe this nice piece of reasoning into NDL. Here is (the top part of) its output proof: 
\begin{tcAthenaWithNumbers}
assume h1 := ((((A ==> (~ (E | ~C))) | ((~ B) & (C | E))) & ((E & ~E) | (B ==> F))) & E & B) {
  # Assuming the antecedent is left-associative, we decompose it into its four main parts.
  p1_and_p2_and_E := ((((A ==> (~ (E | ~C))) | ((~ B) & (C | E))) & ((E & ~E) | (B ==> F))) & E) 
                    BY left-and on h1;
  B_premise := B BY right-and on h1;
   ...
}
\end{tcAthenaWithNumbers}
The proof goes wrong early, on the step shown on lines 3 and 4. In NDL, conjunction associates to the right
(as is specified in the prompt), so the hypothesis \mtt{h1}, \iensp, \mtt{($p_1\:\&\:p_2\:\&\:p_3$)}, 
should be understood as \mtt{($p_1\:\&\:$($p_2\:\&\:p_3$))}, not as \mtt{(($p_1\:\&\:p_2$)$\:\&\:p_3$)}. 
Consequently, the application of \mtt{left-and} to \mtt{h1} yields $p_1$, not 
\mtt{($p_1\:\&\:p_2$)}. Here, $p_1$ is 
%\[
\mtt{(((A$\:$==>$\:$($\ntildeneg\:$(E$\:$|$\:\ntildeneg\:$C)))$\:$|$\:$(($\ntildeneg\:$B)$\:$\&$\:$(C$\:$|$\:$E)))$\:$((E$\:$\&$\:$\ntildeneg\:E)$\:$|$\:$(B$\:$==>$\:$F)))}
%\]
and $p_2$ is \mtt{E}, and the conclusion advertised by Gemini on line 3 is \mtt{($p_1\:\&\:p_2$)}, 
so the NDL interpreter flags this as an error. 

Strictly speaking, this is an error, since the prompt explicitly specifies that conjunction is 
right-associative in NDL. Yet it is a trivial error in an overall sound proof. Indeed, the model's
underlying reasoning is impeccable. One could argue that denying credit to it on account of such 
peccadillos would be like nitpicking spelling errors in a love letter (though see p.~\pageref{Pg:OtherInterp} 
for a different viewpoint). There are many other examples of similar minor failures, running the gamut from syntax to semantics: 
failing to balance parentheses or curly braces properly, forgetting a semicolon or inserting extra semicolons, 
using square brackets instead of parentheses, splitting a line of code and putting part of it inside a comment 
(following a hash sign), failing to distinguish between $p$ and $\neg\neg p$, applying \mtt{from-false} to \mtt{false} itself
instead of the desired conclusion,\footnote{\fmtt{from-false} is a unary inference rule that takes any formula $p$
as its argument and delivers $p$ back as its conclusion, {\em provided\/} that the assumption base contains \temv{false}. 
This captures the principle that ``anything follows from \temv{false}.'' Thus, \egnsp, if we wanted to derive $(A \mand B)$ 
because the assumption base contains \temv{false}, we would write \fmtt{from-false\msp\fkwd{on}\msp(A$\:$\&$\:$B)}.
Instead, models often write vaguely plausible but incorrect variants like \fmtt{(A$\:$\&$\:$B)\msp\fkwd{BY}\msp{}from-false\msp\fkwd{on}\msp{}false}.}
confusing the left and right components of a conjunction, disjunction, or biconditional, and many others.

For that reason, we carefully instrumented the NDL proof checker to systematically ignore 
such errors while preserving soundness. The instrumented checker starts by repairing all syntax errors 
it can reasonably address. 
For example, it might insert a missing  pair of curly braces, add or delete semicolons,
or replace square brackets by parentheses. 
As an example of a more substantive syntactic intervention, it might change rule applications that use \kwd{BY} with 
rule applications that name their conclusions and do not use \kwd{BY}. Models often generate steps like the following
(this example is taken from \nous, but all models make this type of error): 
\begin{equation}
\mtt{C-or-not-A-to-not-B \kwd{BY} mp \kwd{on} left, antecedent;}
\label{Eq:SampleMinorError}
\end{equation}
In NDL, only a formula $p$ can appear immediately to the left of \kwd{BY}. 
In particular, a step of the form $p\msp\kwd{BY}\msp\cdots$ imposes a constraint 
on the inference $\cdots$: its conclusion must be identical to $p$. 
It follows that the syntax of~(\ref{Eq:SampleMinorError}) is 
incorrect,\footnote{This is putting it mildly. A step of that form suggests deep conceptual 
confusion about the role of \fkwd{BY}.}
because \mtt{C-or-not-A-to-not-B} is not a formula that can be compared
against the result of the rule application to the right of \kwd{BY}. It's just  
\emph{a name} that the model wants to give to the result of that application. 
Accordingly, our instrumented checker will automatically repair this into: 
\[\mtt{C-or-not-A-to-not-B := mp \kwd{on} left, antecedent;}\]

After syntactic corrections are made, the checker starts evaluating the repaired proof step-by-step while 
overlooking a wide class of mostly fiddling semantic and structural errors, such as marginally incorrect applications of inference rules.
For instance, the instrumented checker is insensitive to double negation. If a conclusion ought to be $p$ but
the proof derives \mtt{($\neg\neg\,p$)} or vice versa, the checker does not complain. 
The checker is likewise insensitive to conjunction and disjunction commutativity, \egnsp, it will tolerate a step
where an expected conclusion is $(A \mand B)$ but the model derives $(B \mand A)$. It will
even endorse applications of \mtt{left-and} that should instead be applications of \mtt{right-and}, and vice versa;
and likewise for \mtt{left-either} and \mtt{right-either}, as well as \mtt{left-iff} and \mtt{right-iff}. The full list of
interventions is extensive, but essentially, if the model makes a logically sound inference but bungles the
execution, the instrumented checker will attempt to overlook it and continue with the rest of the proof. 

\renewcommand{\arraystretch}{1.3}
\begin{table}[!t]
\centering
\rowcolors{2}{blue!5}{white}
\resizebox{1.0\textwidth}{!}{%
\begin{tabular}{|l|c|c||l|c|c|}
\hline
\textbf{Model} & \textbf{Overlooked Syntax Errors} & \textbf{Overlooked Semantic Errors}   & \textbf{Model} &  \textbf{Overlooked Syntax Errors} & \textbf{Overlooked Semantic Errors} \\
\hline
\hermes & 2525 & 245 & \dsROne & 1971 & 313 \\ \hline
\nous & 1387 & 232 & \dsThreeOne & 1379 & 199 \\ \hline
\dsReasoner & 1354 & 194 & \gptFourO & 1027 & 171 \\ \hline
\gptFourOne & 1077 & 115 & \oThreeMini & 713 & 316 \\ \hline
\oFourMini & 649 & 363 & \magistral & 686 & 315 \\ \hline
\dsThree & 721 & 219 & \oThree & 318 & 378 \\ \hline
\gemThree & 520 & 152 & \opus & 464 & 182 \\ \hline
\geminiTwoPointFive & 176 & 291 & \claudeSonnetExt & 224 & 192 \\ \hline
\grokFour & 177 & 235 & \grokThree & 187 & 214 \\ \hline
\claudeSonnetFour & 205 & 188 & \gptFive & 72 & 274 \\ \hline
\gptFiveOne & 131 & 156 & \gptFiveTwo & 71 & \mbf{8} \\ \hline
\gemThreeOne & 52 & 12 & \claudeOpusFourFiveNov & \mbf{12} & 29 \\ \hline
\end{tabular}}
\caption{Per-model numbers of syntactic and structural or semantic errors detected and repaired by the instrumented proof checker 
when evaluating \mbf{PL1-PW} proofs, sorted by the total number of errors (from left to right and top to bottom).}
\label{Tb:PL1PWErrorCounts}
\end{table}
%% \begin{table}[!bht]
%% \centering
%% \rowcolors{2}{blue!5}{white}
%% \resizebox{1.0\textwidth}{!}{%
%% \begin{tabular}{|l|c|c||l|c|c|}
%% \hline
%% \textbf{Model} & \textbf{Overlooked Syntax Errors} & \textbf{Overlooked Semantic Errors}   & \textbf{Model} &  \textbf{Overlooked Syntax Errors} & \textbf{Overlooked Semantic Errors} \\
%% \hline
%% \hermes & \mbf{2525} & 245 &  \dsROne & \mbf{1971} & 313   \\ \hline 
%% \nous & 1387 & 232  & \dsThreeOne & 1379 & 199  \\ \hline 
%% \dsReasoner & 1354 & 194 & \gptFourO & 1027 & 171 \\ \hline 
%% \gptFourOne & 1077 & 115 & \oThreeMini & 713 & 316  \\ \hline 
%% \oFourMini & 649 & 363 & \magistral & 686 & 315  \\  \hline 
%% \dsThree & 721 & 219 & \oThree & 306 & 373 \\ \hline  
%% \gemThree & 520 & 152 & \opus & 464 & 182 \\ \hline 
%% \geminiTwoPointFive & 176 & 291 & \claudeSonnetExt &  224 & 188 \\ \hline 
%% \grokFour & 177 & 235 & \grokThree & 187 & 214 \\ \hline 
%% \claudeSonnetFour & 205 & 188 & \gptFive & 72 & 274 \\ \hline 
%% \gptFiveOne & 131 & 156 & \gptFiveTwo & 71 & 8 \\ \hline 
%% \end{tabular}}
%% \caption{Per-model numbers of syntactic and structural or semantic errors detected and repaired by the instrumented proof checker 
%% when evaluating \mbf{PL1-PW} proofs, sorted by the total number of errors (from left to right and top to bottom).}
%% \label{Tb:PL1PWErrorCounts}
%% \end{table}
\renewcommand{\arraystretch}{1.0}

The line is drawn at gross logical errors, such as trying to derive a conclusion that does not follow logically
from the current assumption base, or severely ill-formed rule applications, such as applications of conjunction
elimination to a disjunction. To give an idea of the scale of these repairs, 
over the course of evaluating the 400 proofs written by o4-mini, the instrumented checker detected 
and was able to recover from 649 syntax errors and 363 semantic errors, for an average of 2.53 errors per proof. 
For DeepSeek-R1-0528, the total number of corrections exceeded 2500, averaging 6.93 corrections
per proof. Table~\ref{Tb:PL1PWErrorCounts} shows detailed numbers for all models.

Observe that less able models tend to make many more syntax errors than semantic errors. By contrast, 
top-tier models tend to make fewer syntax errors, with some of them even making 
more semantic errors than syntax errors (\egnsp, o3, Grok-4, GPT-5.1, Gemini-2.5-Pro, and GPT-5). 
Note also that there is a moderate negative correlation between model accuracy and the total
number of error repairs (syntax plus semantics) for a given model: the Pearson correlation coefficient
between these two random variables is -0.52. Nevertheless, we will see in the next section 
that even though stronger models generally receive fewer repairs, they tend to benefit from them 
more than weaker models. That's because, for stronger models, these repairs are often enough 
to push a proof over the finish line and make it correct, because the proof only needed minor 
repairs to begin with. By contrast, proofs by weaker models cannot be salvaged quite 
as easily because they typically have additional problems that cut deeper. 
Essentially, strong models tend to fail cosmetically (even if the errors seem 
flagrant from an instruction-following viewpoint), whereas weaker models tend 
to fail structurally. 

%\enlargethispage{12pt}

\subsubsection{Results}

Table~\ref{Tb:Pl1PWAcc} reports two sets of results, one according to the strict NDL proof checker and 
the other reported by the instrumented checker. The gap between the two is striking and demonstrates that 
naive evaluation by a conventional proof checker would severely underestimate the proof-discovery ability 
of these models. A relaxed evaluation algorithm that is able to overlook minor errors reveals 
that proof writing in \plonesp is solved, with the top models achieving essentially ceiling accuracy. 
Nevertheless, even with the instrumented checker, the task 
has sharp discriminating power, with performance falling dramatically among mid-tier and 
lower-end models, as R1-0528, o4-mini, and Claude Opus 4 are unable to reach 70\% accuracy, 
while o3-mini and Claude Sonnet 4 sit well below 50\%. 
\renewcommand{\arraystretch}{1.5}
\begin{table}[!b]
\centering
\rowcolors{2}{blue!5}{white}
\resizebox{1.0\textwidth}{!}{%
\begin{tabular}{|l|c|c||l|c|c|}
\hline
\textbf{Model} & \textbf{Instrumented Accuracy} & \textbf{Strict Accuracy} &
\textbf{Model} & \textbf{Instrumented Accuracy} & \textbf{Strict Accuracy} \\
\hline
\claudeOpusFourFiveNov & \mbf{0.99} \ci{0.01} & \mbf{0.93} \ci{0.03}  &
\gemThreeOne & \mbf{0.99} \ci{0.01} & 0.92 \ci{0.03} \\ \hline
\gptFiveTwo & 0.98 \ci{0.01} & 0.84 \ci{0.04} &
\gptFiveOne & 0.97 \ci{0.02} & 0.78 \ci{0.04} \\ \hline
\geminiTwoPointFive & 0.94 \ci{0.02} & 0.60 \ci{0.05} &
\gptFive & 0.92 \ci{0.03} & 0.62 \ci{0.05} \\ \hline
\gemThree & 0.92 \ci{0.03} & 0.67 \ci{0.05} &
\claudeSonnetExt & 0.88 \ci{0.03} & 0.63 \ci{0.05} \\ \hline
\grokFour & 0.85 \ci{0.03} & 0.59 \ci{0.05} &
\oThree & 0.80 \ci{0.04} & 0.53 \ci{0.05} \\ \hline
\dsReasoner & 0.76 \ci{0.04} & 0.45 \ci{0.05} &
\dsROne & 0.60 \ci{0.05} & 0.21 \ci{0.04} \\ \hline
\opus & 0.57 \ci{0.05} & 0.38 \ci{0.05} &
\oFourMini & 0.57 \ci{0.05} & 0.30 \ci{0.04} \\ \hline
\dsThreeOne & 0.49 \ci{0.05} & 0.26 \ci{0.04} &
\claudeSonnetFour & 0.46 \ci{0.05} & 0.33 \ci{0.05} \\ \hline
\oThreeMini & 0.41 \ci{0.05} & 0.25 \ci{0.04} &
\grokThree & 0.38 \ci{0.05} & 0.27 \ci{0.04} \\ \hline
\gptFourOne & 0.35 \ci{0.05} & 0.23 \ci{0.04} &
\dsThree & 0.33 \ci{0.05} & 0.21 \ci{0.04} \\ \hline
\hermes & 0.32 \ci{0.05} & 0.10 \ci{0.03} &
\magistral & 0.20 \ci{0.04} & 0.12 \ci{0.03} \\ \hline
\nous & 0.16 \ci{0.04} & 0.04 \ci{0.02} &
\gptFourO & 0.15 \ci{0.03} & 0.10 \ci{0.03} \\ \hline
\end{tabular}}
\caption{\mbf{PL1-PW} proof-writing accuracies from the instrumented proof checker and from the regular (strict) proof checker, with Wilson confidence intervals.}
\label{Tb:Pl1PWAcc}
\end{table}
\renewcommand{\arraystretch}{1.0}

This will be a recurring theme in \pgrid: all tasks, even those with strong ceiling effects, 
are able to tease apart models of different ability levels quite effectively. 
%% By way of a brief comparison, consider the 
%% accuracies of the 6 top-scoring models listed by Kaggle \href{https://www.kaggle.com/benchmarks/open-benchmarks/math-500}{on MATH 500}, 
%% a benchmark that is said to test ``the mathematical reasoning and problem-solving 
%% proficiency of language models (LMs)'' (accuracies here are given as percentages): 
%% %\scalebox{0.8}{\includegraphics[width=\linewidth]{./plots/math500_accs.png}} \\
%% \begin{center}
%% \begin{tabular}{rlrlrl}
%% \mtt{GPT-5:} & \hspace*{-0.1in}\mbox{\textcolor{green!40!black}{96.8}} & \mtt{o4-mini:} & 
%% \hspace*{-0.1in}\mbox{\textcolor{green!40!black}{96.2}} &  
%% \mtt{Gemini-3-Pro-Preview:} & \hspace*{-0.1in}\mbox{\textcolor{green!40!black}{96.2}} \\[0.03in]
%% \mtt{o3:} & \hspace*{-0.1in}\mbox{\textcolor{green!40!black}{96.0}} & 
%% \mtt{DeepSeek-R1:} & \hspace*{-0.1in}\mbox{\textcolor{green!40!black}{95.8}} & 
%% \mtt{o3-mini:} & \hspace*{-0.1in}\mbox{\textcolor{green!40!black}{95.4}}
%% \end{tabular}
%% \end{center}
%% Compare these to the corresponding \mbf{PL1-PW} scores: 
%% \begin{center}
%% \begin{tabular}{rlrlrl}
%% \mtt{GPT-5:} & \hspace*{-0.1in}\mbox{\textcolor{red!40!black}{91.0}} & 
%% \mtt{o4-mini:} & \hspace*{-0.1in}\mbox{\textcolor{red!40!black}{54.0}} &  
%% \mtt{Gemini-3-Pro-Preview:} & \hspace*{-0.1in}\mbox{\textcolor{red!40!black}{92.0}} \\[0.03in]
%% \mtt{o3:} & \hspace*{-0.1in}\mbox{\textcolor{red!40!black}{78.0}} & 
%% \mtt{DeepSeek-R1:} & \hspace*{-0.1in}\mbox{\textcolor{red!40!black}{58.0}} & 
%% \mtt{o3-mini:} & \hspace*{-0.1in}\mbox{\textcolor{red!40!black}{36.0}}
%% \end{tabular}
%% \end{center}
In Section~\ref{Sec:DiscriminationIndex} we define and study DI (discriminability index), 
a simple metric that gauges how well benchmark scores (usually accuracies) manage 
to separate a set of models $S$. It is simply the average of score differences between all pairs 
of distinct models in $S$. 
Expressed as a percentage, the DI for the entire set of \mbf{PL1-PW} results is 34, a very high value. 
%Here, the value of DI for the first set of numbers is 0.5  (any two of these scores differ by about 0.5\% on average),  while for the second set of numbers  it is 27.4. 
%Note, in particular, the stark separation of the mini models from the top-tier models, and of DeepSeek-R1-0528 from all others. 

Both the average and the median boost in performance brought about by instrumentation 
is 19 absolute percentage points. This means that, on average, if only strict accuracy was reported, 
a model's performance on this task would be underestimated by 19 absolute points. 
The following figure displays the accuracy deltas 
sorted from biggest to smallest:
\begin{center}
\scalebox{0.8}{\includegraphics[width=\linewidth]{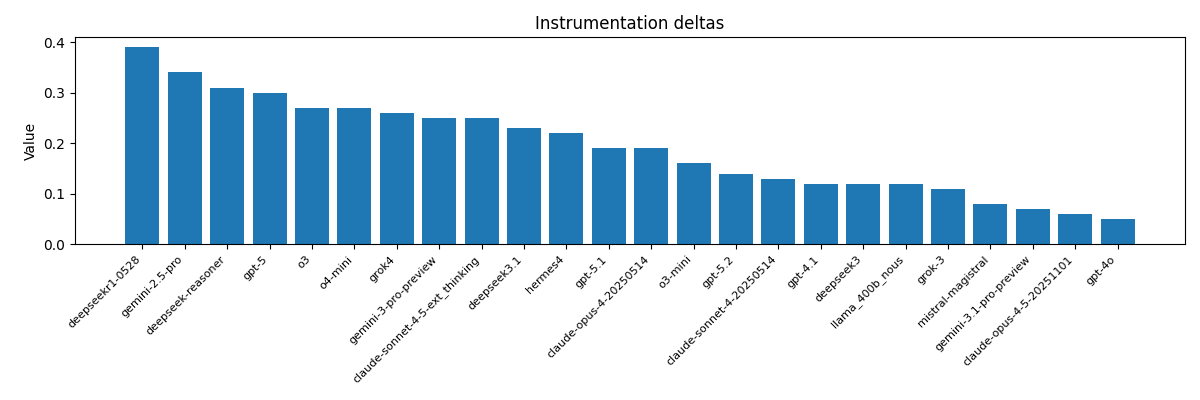}}
\end{center}

\begin{figure}[!hbt]
\centering
\scalebox{0.7}{\includegraphics[width=\linewidth]{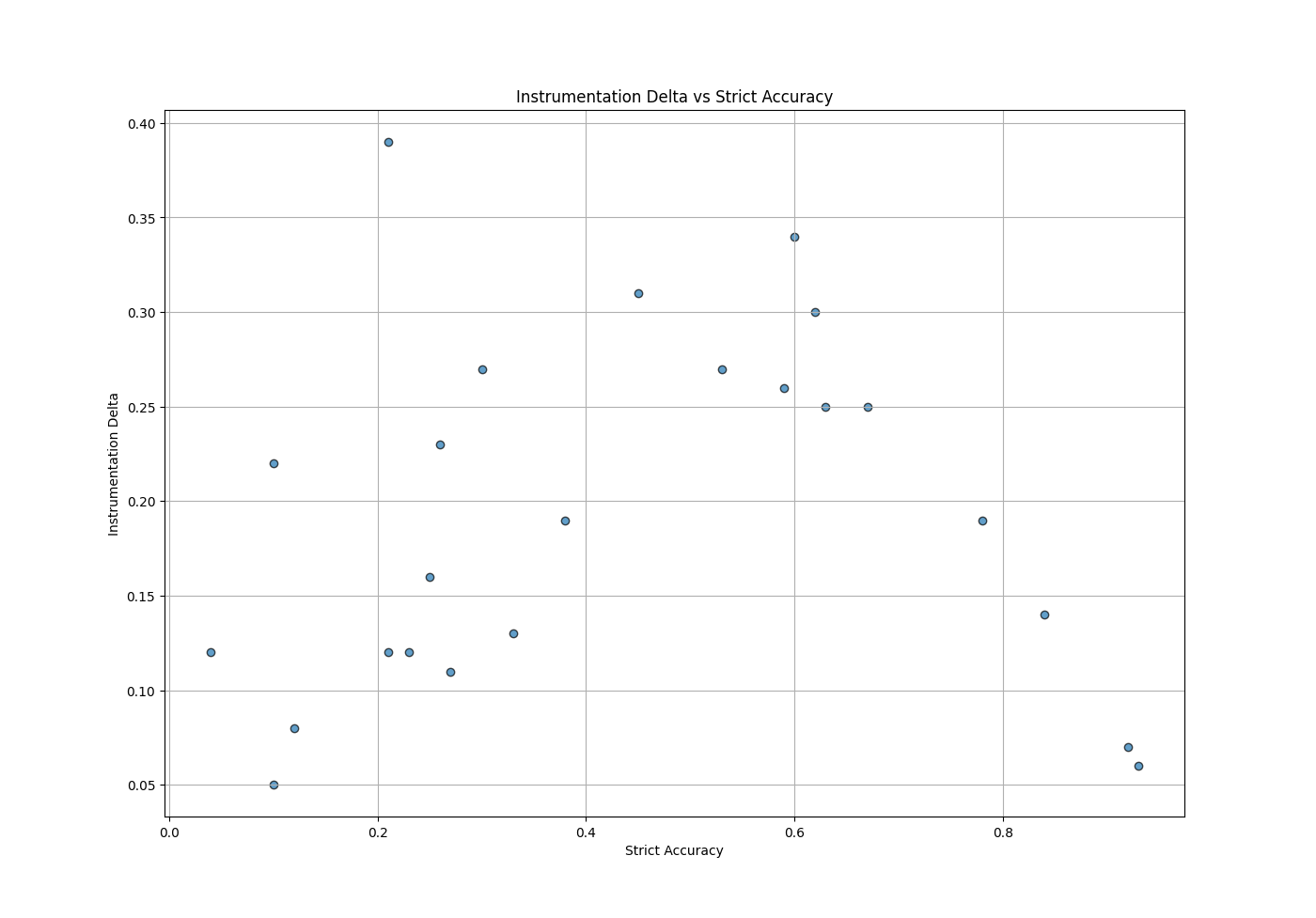}}
\caption{Scatterplot of strict accuracy vs instrumentation deltas across the 24 models.}
\label{Fig:IDeltaVsAcc}
\end{figure}

The table below provides finer-grained information on the distribution of these deltas: \\[0.1in]
\begin{minipage}{\linewidth}
{\footnotesize 
\renewcommand{\arraystretch}{1.3}
\begin{center}
%\begin{table}[h!]
\hspace*{-0.4in}
\begin{tabular}{c}
\hspace*{0.3in}
\begin{tabular}{|r|c|c|c|c|c|c|c|c|c|c|c|}
\hline
 & {\bf Mean} & {\bf Median} & {\bf Min} &{\bf Max} & {\bf Std} & {\bf @5} & {\bf @25} & {\bf @50} & {\bf @75} & {\bf @90} & {\bf @95} \\
\hline\hline
{\bf Accuracy Delta} & 19 & 19 & 5 & 39 & 9 & 6 & 12 & 19 & 26 & 31 & 34 \\ \hline 
\end{tabular}
\end{tabular}
\end{center}}
\end{minipage}
\renewcommand{\arraystretch}{1.0}
\\[0.04in]
The benefit of instrumentation follows a rougly inverted-U pattern with respect to baseline 
accuracy, as shown by the scatterplot of strict accuracy versus instrumentation boost in Fig.~\ref{Fig:IDeltaVsAcc}.
We explain this pattern below. 

If one views proof writing as decomposing roughly into two stages, idea discovery 
(finding the right proof plan) and rigorous execution (turning that plan into a correct derivation),
instrumentation only helps with the second stage. 
The delta is thus minimal for weak models that rarely discover the correct proof strategy, and also small 
for the few elite models that already often produce fully valid proofs.  In particular, 
as confirmed by the preceding histogram, the smallest improvement (5 percentage points)  
is realized by GPT-4o, the worst-performing model, while the next smallest improvement (6 points)
is realized by Claude Opus 4.5, the best-performing model. 
The largest gains by far occur for mid-tier models that often reach the correct high-level proof 
idea but are unable to express it as a formally valid derivation. 
These are models that have latent reasoning competence but weak symbolic discipline.
The very best models still receive a decent boost, reflecting occasional slips.
Instrumentation cleans those up but cannot increase accuracy much because the
margins are tight.

While it is tempting to view proof length as a natural axis for analyzing reasoning behavior, 
raw proof-length distributions are dominated by differences in which problems a model is able to solve. 
A stack of box plots depicting proof-length distributions for the 24 models for the 
set of problems that each model is able to solve would naturally invite cross-model comparisons 
that would nevertheless be statistically indefensible, since the  sets in question vary 
across models and are not matched by difficulty, which means 
that the resulting distributions are drawn from different populations. It's possible, 
of course, to control for problem identity, \egnsp, by restricting attention to the intersection 
of problems that are correctly solved by a given set of models, but once that is 
done we find that proof length is not an informative axis of analysis for this 
task.\footnote{For instance, we might consider the set of problems $I_K$ that were correctly solved
by all top $K$ models, say for $K = 10$. If we pick a particular model $m_r$ as a reference model (say Claude Opus 4.5, 
the highest-performing model), then for each $p \in I_K$ we can compute the distribution of log ratios 
$\log L_m(p)/L_{m_r}(p)$ for every model $m$ in the set of top $K$ models, 
where $L_m(p)$ is the length of the (correct) proof generated by $m$ for problem $p$. 
The boxplots of these distributions would be directly comparable and would tell 
us how much more verbose or exploratory each model's proofs are by comparison to the reference model. 
When we do that, we find that proof lengths are roughly similar, with the median log ratio at 0 and
more pronounced length differences emerging only at the fringes (after the 90th percentile).}
This should not be surprising: the simplicity of these problems naturally constrains 
verbosity, as proof styles converge and minimal proof lengths become fairly similar.
Nevertheless, to give a concrete sense of the intrinsic length scale of proofs in this task, 
the table below provides some statistics on the distribution of the \emph{shortest} correct-proof lengths 
for each problem, irrespective of model identity.\footnote{Note that every problem was correctly solved by 
at least one model, so the quantity ``the length of the smallest proof for problem $p$'' defines 
a total function on this dataset and does not introduce any survivorship biases.}
We stress that these are not the shortest or minimal proofs in a formal mathematical sense, only the shortest 
ones that were observed. \\[0.04in]
\begin{minipage}{\linewidth}
{\footnotesize 
\renewcommand{\arraystretch}{1.3}
\begin{center}
%\begin{table}[h!]
\hspace*{-0.4in}
\begin{tabular}{c}
\hspace*{0.3in}
\begin{tabular}{|r|c|c|c|c|c|c|c|c|c|c|c|}
\hline
 & {\bf Mean} & {\bf Median} & {\bf Min} &{\bf Max} & {\bf Std} & {\bf @5} & {\bf @25} & {\bf @50} & {\bf @75} & {\bf @90} & {\bf @95} \\
\hline\hline
{\bf Minimal Correct Proof Length} & 19.78 & 19.0 & 1.0 & 61.0 & 12.64 & 1.0 & 10.0 & 19.0 & 30.0 & 36.0 & 40.05 \\ \hline 
\end{tabular}
\end{tabular}
\end{center}}
\end{minipage}
\renewcommand{\arraystretch}{1.0}
\\[0.04in]
The fact that 90\% of the problems can be solved with a proof that needs 
no more than 36 lines further highlights the low complexity of this task.

\subsubsection{Error Analysis}

\begin{figure}[!t]
\centering
\scalebox{1.0}{\includegraphics[width=\linewidth]{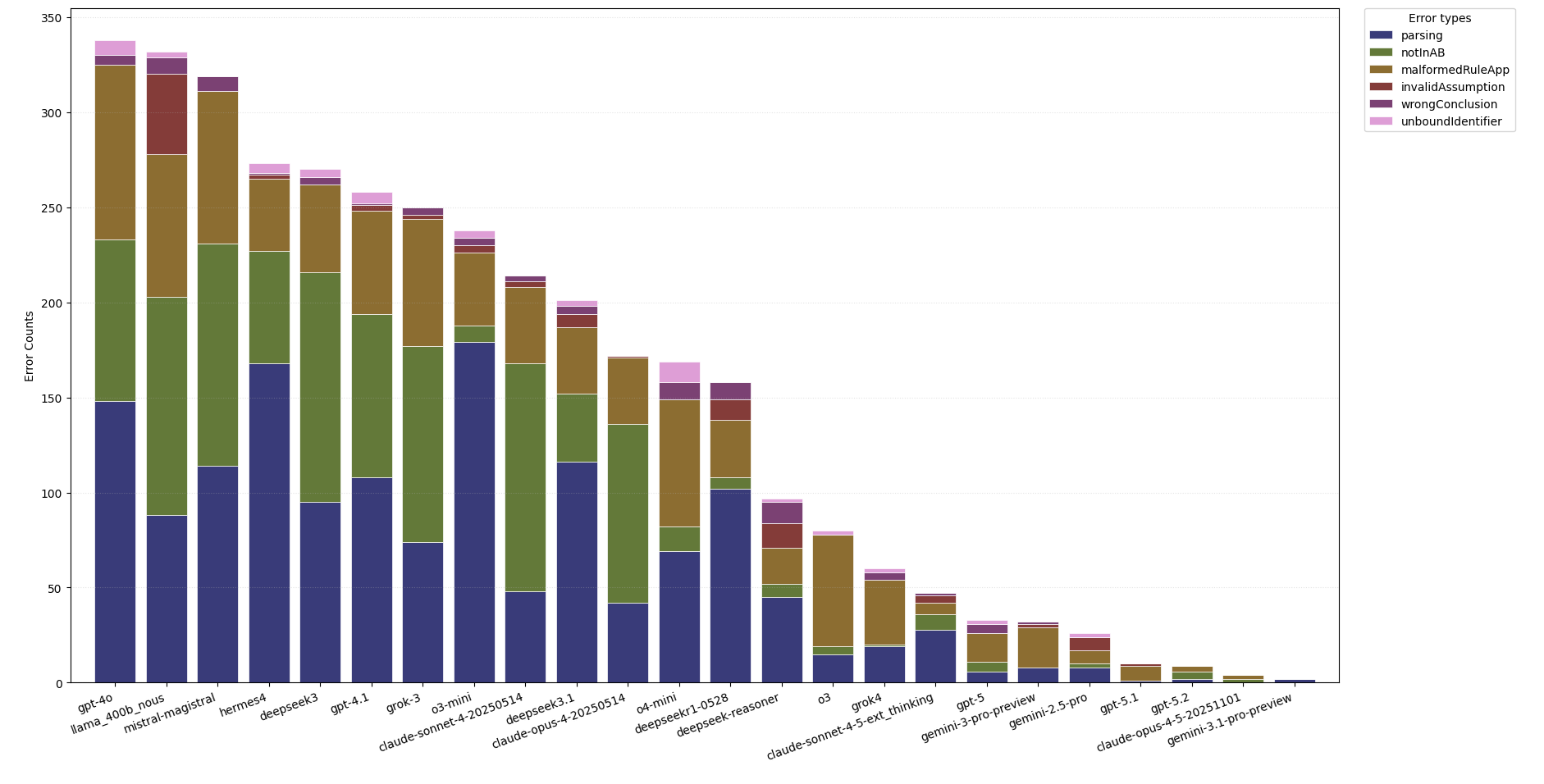}}
\caption{Distribution of first-error types under the instrumented proof checker.}
\label{Fig:PL1PWETDist}
\end{figure}

What are the errors that cause proofs to fail even when they are evaluated with the instrumented checker? 
That distribution is shown in Fig.~\ref{Fig:PL1PWETDist}. 
It distinguishes 6 remaining error types: \label{Pg:PL1PWErrorDist}
\ben
\item {\em Parsing/Syntax errors}. 
Minor syntax transgressions are already fixed by the instrumented checker; what remains are more deviant 
parsing errors that cannot be easily detected and repaired. The following examples are illustrative: 
\bit
\item Writing the entire proof as one single continuous line while also interspersing various comments in it, even though, as noted in the prompt,
a comment in NDL extends to the end of the line on which it appears. Alternatively, the proof is written as a list of comment lines,
where each line starts with a hash and an English sentence (the actual commentary) and is then followed by NDL code, all on the same line. 
\item Single or double quotes randomly inserted for no reason.

\item Conditional subproofs with empty bodies, e.g. (example taken from Mistral Magistral):
\begin{tcAthena}
...
assume case2 := ((~ (A ==> ~B)) | (B | A)) {
  # We will break this down further in the proof.
}
...
\end{tcAthena}
These are like if-then-else statements in Python whose code blocks for the two branches contain only comments and no code. 
Weaker models probably treat these gaps as placeholders whose details remain undetermined. 

\item Hallucinating syntax forms that are not part of NDL. Some of these are interesting 
because they are coherent \emph{generalizations} of NDL syntax. 
Athena, being a superset of NDL, actually accepts quite a few of these hallucinated forms. 
For instance, unlike NDL, an argument to a rule application in Athena 
need not be a simple formula, but can instead be an entire inlined proof, \egnsp, 
it is possible to write
{\small
\begin{center}
\begin{tabular}{l}
$\vdots$ \\
\smtt{C \skwd{BY} cases \kwd{on} (A$\:$|$\:$B), \skwd{assume} A \{$\,\cdots\,$\}, \skwd{assume} B \{$\,\cdots\,$\};} \\
$\vdots$
\end{tabular}
\end{center}
} 
Here, the second and third arguments to \mtt{cases} are arbitrarily large conditional proofs, presumably 
deriving the conditionals $(A \mif C)$ and $(B \mif C)$.

Most models wrote at least some proofs in such generalized 
supersets of NDL. This was no doubt due to transfer learning from the countless 
instances of syntax compositionality that the models encountered during their pretraining. 
Because we used Athena as the backbone of our proof checker, we were able to accept 
and give credit to many such proofs, even though these generalized constructs 
are not valid according to the prompt. 

However, many models push this too far and end up being systematically creative in the wrong direction,
by concocting uber-compositional syntax generalizations that are much more extreme and 
free-ranging than anything allowed in Athena.
Here is one example of a single inference rule application written by GPT-4.1 
(we added the indentation to clarify the model's intention; the actual output was just one long line):
\begin{tcAthena}
(B & ~A) BY cases on (B | A), 
                     (B ==> ~A ==> (~A BY mp on (B ==> ~A), B; 
                                   (B & ~A) BY both on B, ~A)),
                     (A ==> (B ==> ~A), 
                            (A ==> B), 
                            A ==> (B BY mp on (A ==> B), A; 
                                  ~A BY mp on (B ==> ~A), B; 
                                  (B & ~A) BY both on B, ~A))
\end{tcAthena}
While this fails on multiple fronts, the overall intention is relatively clear: The model was trying 
to write a single application of the \mtt{cases} rule with nested subproofs as the arguments. 
In the process, it helped itself to its own bracketing conventions (such as enclosing a proof sequence 
in parentheses instead of NDL's curly braces) and also embedded the subproofs inside conditional 
formulas that could not possibly yield the desired conclusions. 
%% Here is another example from Claude Sonnet:
%% \begin{tcAthena}
%% false BY absurd on biconditional, 
%%                    (~ (A <==> B)) BY dsyl on conjunct2, 
%%                                   (~ (~ A & ~ B)) BY dm on 
%%                                   (~ (~ A | ~ B)) BY dm on 
%%                                   (A & B) BY both on A, B
%% \end{tcAthena}
%% Likewise, even though the prompt states that \kwd{BY} can only qualify an 
%% application of an individual inference rule, not an arbitrary subproof, 
%% many models insisted on doing just that. 
Here is another example by o3-mini: 
\begin{tcAthena}
A_imp_B := (A ==> B) BY assume a := A 
                          {false2 := false BY absurd on a, h3; 
                           B1 := B BY from-false on B, false2};
\end{tcAthena}

\eit
These pervasive errors indicate a failure of \emph{cognitive control}, as cognitive scientists might put it, 
particularly \emph{inhibitory} control---the ability to suppress prepotent or automatic responses, such as 
the urge to write in natural language or to use familiar compositional syntax patterns from other languages. 
The models find it difficult to inhibit their pretrained tendencies; they overgeneralize  
from their training distributions and fail to suppress those generalizations even when
explicitly instructed to do so, finding it hard to conform their output only to 
the constraints explicitly provided in their ``working memory'' (the context 
window).

That said, as can be seen from Fig.~\ref{Fig:PL1PWETDist}, parsing errors tend to be the minority for 
most models, particularly for better models. 
The top performer, Claude Opus 4.5, did not make a single syntax error (the only model to do so).

\item {\em Not-in-the-assumption-base errors}: These are the most properly logical errors, signifying 
the application of an inference rule to a formula that is neither a premise, nor a current hypothesis 
(posited by an \kwd{assume} that is still in scope), nor a previously derived conclusion (by a preceding
subproof that is still in scope). 

\item {\em Malformed rule applications}: These errors can be viewed as sitting somewhere between syntax and semantics,
rather similar to type errors. A typical example is an application of modus tollens to $(B \mif A)$ and $\neg B$
(an example from Grok-4) or an application of disjunctive syllogism to $(\neg A \mor B)$ and $\neg A$. 

\item {\em Malformed assumptions}: These errors occur in conditional proofs 
when a model stipulates a hypothesis that is not a valid formula, \egnsp, 
\mtt{\kwd{assume}\msp{}A\_and\_B$\;$\{ $\cdots$\}}. \label{Pg:InvalidAssumptions}
The model usually intends the malformed assumption as a mnemonic \emph{name} for the 
actual hypothesis $p$ (in this case, say, the conjunction \mtt{(A$\;$\&$\;$B)}). 
But instead of writing \mtt{\kwd{assume}\msp$I\,$:=$\,p\;$\{$\cdots$\}}, 
it writes \mtt{\kwd{assume}\msp$I\;$\{$\cdots$\}}. (As an analogy 
from functional programming, this is a bit like writing \smtt{let val sqrt\_of\_n in $\cdots$ end}
instead of writing something like \smtt{let val sqrt\_of\_n = math.sq(n) in $\cdots$ end}.)

\item {\em Wrong conclusions}: This is another type of reasoning error, which happens 
when a model manages to derive a conclusion but not the one that was required. This happens in two cases, either:
\bit
\item in a single application of an inference rule where the model uses \kwd{BY} to advertise the rule's conclusion as $p$ 
but the application actually produces $q\neq p$; or 
\item at the level of the entire proof, where the model writes a proof that successfully 
derives {\em some\/} conclusion, but not the one that solves the problem. 
\eit 

\item {\em Unbound identifiers}: This occurs when the proof refers to hallucinated names that have not been defined. For example, in a conditionalized argument with zero premises,
the model might still write \mtt{premise-1} in reference to, \egnsp, the first conjunct of the target conditional's antecedent. 
\een

These 6 failure modes are  mutually exclusive only because 
they mark \emph{the first error} encountered by the instrumented checker,
analogous to a compiler's first error report. It's conceivable that even 
if that error were somehow disregarded and evaluation continued, the proof
would make a second error of a different type (and potentially a third one, and so on). This is like a program whose first compiler error 
is a syntax error, which is followed by a type error and then 
an undeclared variable. In general, failed proofs may contain 
multiple errors, and indeed we have anecdotally verified that many proofs, 
especially from weaker models, do make both syntactic and logical mistakes. 
Nevertheless, the ``first-error'' view afforded by Fig.~\ref{Fig:PL1PWETDist} 
is informative, because it identifies the earliest bottleneck that would need 
to be addressed to allow a proof attempt to progress further. 

The data shows that the first errors of stronger models tend to be more 
properly logical or semantic. And although the absolute number of such  failures 
is too small to support fine-grained statistical analysis, this clearly suggests 
a developmental progression, whereby models first learn to imitate proof 
structure, then they learn to obey syntax and formatting constraints, 
and only later become limited by more abstract reasoning challenges.

\subsubsection{Discussion}

How should we understand the  \mbf{PL1-PW} results? This task requires two skills: 
\ben
\item The model must possess a high-level understanding of deductive reasoning and a capacity for strategic-level planning
that will enable it to construct an abstract argument, using the inferential machinery of natural deduction, demonstrating 
how and why the conclusion follows from the premises. 
\item The model must also be able to render that abstract argument into a rigorous formal proof. 
\een 
The results show  clearly that models are considerably better at the first task than the second. They have largely sound ``mental models''
of logical reasoning that they can leverage to solve the straightforward problems in \mbf{PL1-PW}. However, they have severe
deficiencies in execution---in the {\em procedural skill\/} needed to express their reasoning in the precise syntax of a formal language. 
It is likely possible to mitigate these deficiencies, \egnsp, by fine-tuning on a target formal language and/or by taking an 
agentic approach, but the interesting question here is not what could be done with additional engineering effort, but what these results
tell us about the native abilities of these models as general-purpose off-the-shelf intelligent systems. 

On one view, the procedural difficulties are not alarming. After all, these models had
virtually no exposure to NDL during their pretraining, so mistakes are 
to be expected. These are minor performance errors, not competence errors. Humans, too, when they first see a new language, 
need a certain amount of practice before they become fluent in it. Even a brilliant logician would almost certainly
make some errors on their first attempts at writing in a new formal language. 

On the other hand, NDL is an extremely simple language that is fully described in a few pages. Moreover, the task was 
\label{Pg:OtherInterp} described in a few-shot setting, with several examples included in the prompt. If a model cannot respect a small number 
of clearly defined syntax rules, even when illustrated with multiple examples, that by itself 
arguably indicates a sort of failure of general reasoning,  
even if it is not, strictly speaking, a failure of logical inference. 
And while the models might have never seen this \emph{particular} language before, the language is 
quite standard and very similar to many others that the models \emph{have} seen during their pretraining, in enormous quantities. 
For instance, NDL uses curly braces \mtt{\{$\cdots$\}} to delineate scope, semicolons to compose different 
steps,  parenthesized infix syntax for propositional-logic formulas, and so on. None of this is alien to 
modern models. 

Cognitive science gives us a framework for thinking about this: the concepts 
of near-transfer and far-transfer of learning \citep{Barnett2002}. 
Transfer of learning in general is the application of knowledge 
or skills learned in one context to a new 
context.\footnote{``Transfer learning'' \citep{wang2023introduction} 
has also been a major research paradigm and driver of progress in deep learning, 
by enabling models trained on one task to perform 
well on others with limited additional data. While the two traditions overlap, 
AI has focused on data-efficient reuse of representations, whereas cognitive science is more 
concerned with the extent to which structural insights can leap across different domains.}
Near transfer occurs when we apply knowledge to a new situation 
that is very similar to what we have seen before (a person who has only driven a Toyota Camry before can 
almost certainly drive a Honda Accord without any instruction). Far transfer occurs when we apply 
knowledge to a new situation that is structurally different and seemingly unrelated, say, using principles of chess 
to inform a business merger strategy. (This is notoriously difficult for humans and is considered a hallmark of 
deep, flexible intelligence.) Many of the errors made by these models suggest near-transfer failures. 
Inserting multiple semicolons in a row, writing code to the right of a comment marker and on the same line, 
or stopping a formula in midstream with unbalanced parentheses are not trivial slip-ups. They indicate that the models
are failing to apply core known principles to the current specific case 
and are not attending to the stated rules. 

Finally, as we have already stressed, all our tasks are intended as tests of fluid reasoning. 
We are deliberately using a novel language to prevent the models from relying on crystallized knowledge. 
The very first step in any fluid-intelligence task is to understand the ``rules of the game.'' If a system cannot 
reliably internalize and apply the given rules, it has failed at the most basic level of such a test. 
Logical deduction is a subsequent step that depends on the first.

In conclusion, while several models at present exhibit a significant capacity for 
the abstract logical planning required by this relatively straightforward task 
(as shown by post-correction accuracy rates near ceiling), most of them show a 
deficit in cognitive control and procedural grounding, failing to autonomously 
generate fully valid formal proofs, given that after the top 3 models, pre-correction accuracies
dip below 80\%. Even the very best models currently show a gap between abstract logical competence 
and rigorous execution. This suggests that the reasoning abilities of language models 
are not yet fully integrated, instead requiring external ``scaffolds'' to translate their 
abstract plans into formally correct proofs. Further evidence for this will appear in subsequent sections. 

We close with a few additional observations: 
\bit
\item The instrumented checker catches a very large number of issues, but it is not an oracle. It's still possible that 
a proof might fail due to a relatively innocuous error, so even the instrumented accuracies might underestimate 
abstract proof-writing competence in 
\mbf{PL1-PW}.\footnote{As a concrete example (taken from \fmtt{deepseek-reasoner}),
consider a case analysis over a disjunction $(p_1 \mor p_2)$ intended to derive a conclusion $q$.
The model shows that $p_1$ implies $q$, but for the second case it shows instead that $p_2$ implies \temv{false};
it then purports to derive $q$ by applying \fmtt{cases} to the 3 arguments $(p_1 \mor p_2)$, $(p_1 \mif q)$, and $(p_2 \mif \temv{false})$.
This application is ill-formed because the two conditionals don't share the same consequent, so the current
implementation rejects it. One could argue that this is a ``good enough'' case analysis, since $(p_2 \mif \temv{false})$
could readily be transformed to $(p_2 \mif q)$. But these are the sort of idiosyncratic (and 
relatively uncommon) errors whose automated detection would need much more sophisticated intent inference
and whose repair would cross the line from proof checking to proof synthesis. 
The current instrumented checker focuses on very common patterns of minor errors that 
can be easily identified and overlooked.}

That said, it should not be assumed that models always come up with the right
high-level proof plan and fail only in execution. The reasoning strategies themselves, expressed in English, are sometimes 
misguided.\footnote{Quantifying this would require a rigorous analysis of every proof plan, which would involve manual 
inspection (since the plans are unstructured text) or the use of other models as judges; we leave this to future work.}
As an example, consider this problem:
\begin{tcAthena}
assert premise-1 := ((B <==> C) ==> (C & (B | ~A))) 
assert premise-2 := (B ==> A) 
assert premise-3 := (A ==> (~ (B ==> ~C))) 
assert premise-4 := ((A ==> B) & (~C | A)) 
# Derive: (A & C) 
\end{tcAthena}
A human with some experience would realize that the key is to derive $A$, which can be done by contradiction. If we had $\neg A$ then modus tollens on \mtt{premise-2}
would yield $\neg B$. Moreover, disjunctive syllogism on the right conjunct of \mtt{premise-4} would give $\neg C$. Since we have $\neg B$ and $\neg C$,
we have $(B \miff C)$ (by the definition of a biconditional), which, by \mtt{premise-1}, leads to $C$---a contradiction, since we've already derived $\neg C$. 
Now that we have $A$, $C$ follows easily from \mtt{premise-3}: modus ponens on that and $A$ gives $\neg (B \mif \neg C)$, which, by the definition
of a negated conditional, gives $\neg\neg C$, and hence $C$. 

Here is how GPT-5 describes its reasoning for this problem: 
\begin{quote}
{\em 
We first extract \mtt{(A\psp==>\psp{}B)} from \mtt{premise-4} and use \mtt{premise-2} 
\mtt{(B\psp==>\psp{}A)}. Using these, we prove \mtt{(B\psp==>\psp{}C)} by: 
assume \mtt{B}, derive \mtt{A} via \mtt{(B\psp==>\psp{}A)}, 
get \mtt{\pneg(B\psp==>\psp\pneg{}C)} from \mtt{premise-3}, then under a temporary assumption 
\mtt{\pneg C} build \mtt{(B\psp==>\psp\pneg{}C)}, reaching a contradiction and hence \mtt{C}. 
This yields \mtt{(B\psp==>\psp{}C)}. From \mtt{A} we then derive \mtt{C} by chaining \mtt{A\psp==>\psp{}B} and 
\mtt{B\psp==>\psp{}C}, thus \mtt{A\psp==>\psp{}C}. Next, from \mtt{A} we easily get \mtt{A \& C}. 
To obtain \mtt{A} outright, we appeal to the disjunction \mtt{(\pneg{}C\psp|\psp{}A)} (from \mtt{premise-4}) 
and set up a \mtt{cases} argument to yield \mtt{A}; applying \mtt{A\psp==>\psp{}C} gives \mtt{C}, and finally we conjoin 
\mtt{A} and \mtt{C} to derive \mtt{(A\psp\&\psp{}C)}.
}
\end{quote}
This strategy places the emphasis on $(A \mif C)$ and treats the derivation of $A$ itself as an afterthought, 
when the opposite is required: Deriving $(A \mif C)$ is straightforward (though GPT-5 does it via a circuitous route); 
it's the derivation of $A$ that is (relatively) more demanding. GPT-5's plan erroneously claims that
``To obtain \mtt{A} outright, we appeal to the disjunction \mtt{(\pneg{}C\psp|\psp{}A)} (from \mtt{premise-4}) 
and set up a \mtt{cases} argument to yield \mtt{A}.'' But  \mtt{A} does not follow from \mtt{(\pneg{}C\psp|\psp{}A)},
especially from a simple case analysis of it. The concrete proof written by GPT-5 has no procedural flaw; it follows
the plan faithfully: 
\begin{tcAthenaWithNumbers}
...
        ## Show that (~C) leads to contradiction with A-implies-C and A from the disjunction
        A_from_disj :=
          assume (~ C) {
            ## From (~C | A) and the assumption (~C), introduce A by cases:
            ## Build the two conditionals needed for cases on (~C | A) to reach A
            left_case := assume (~ C) { claim on A };
            right_case := assume A { claim on A };
            cases on disj, left_case, right_case         ## A
          };
...
\end{tcAthenaWithNumbers}
The problem is that the plan itself is flawed, and so the proof checker rightly reports an error on line 7, 
complaining that \mtt{A} is not in the assumption base. 
%\end{comment}

\item The more common problem with proof strategies is not the presence of outright logical errors but rather
the presence of gaps that leave crucial details unspecified. Even the best models do this to a certain degree,
but mid-tier and third-tier models often take it to an extreme, with strategies that are 
so broad and underspecified as to be vacuous. Here is an example from \dsThree: 
\begin{quote}
The proof strategy involves deriving a contradiction from the given premises to conclude the disjunction of negations of various propositions. 
This is achieved by systematically applying inference rules to the premises to derive intermediate conclusions that eventually lead to a contradiction, 
thereby allowing us to conclude the desired disjunction.
\end{quote}
This could be shortened to ``The strategy is proof by contradiction.''

\item Some models use the proof text itself as a thought scratchpad for exploring the proof search space. 
Mistral Magistral is particularly fond of doing this, as illustrated in the following fragment:
\begin{tcAthenaWithNumbers}
...
    assume B {
      # If B is true, we must ensure consistency with simplified-premise-2a
      # Since simplified-premise-2a is (~ (B | A) | A), and we have (~ A) (from our assumption), then 
      # (~ (B | A)) must be false. So, (B | A) must be true. But since we are assuming B, and we have
      #  (~ A), (B | A) is true because B is true. However, we need to derive C. Since B is true, 
      # and we have (C | B), we can derive C or B. But we need C. We need to ensure that C is not false. 
      # Since we have (~ A) and (C | B), and B is true, but we need C. But (C | B) with B being true 
      # doesn't help us get C. So, perhaps another approach is needed. Let's try to derive 
      # a contradiction if C is false.
      assume (~ C) {
        false BY absurd on C, (~ C)  # But C is not directly available here
      };
      # Instead, let's use premise-3 to derive (C | B), which we already have.
      # But since we have B (from the assumption), and (C | B), we can't directly get C unless we 
      # assume (~ B), but we have B. This suggests a dead end, so perhaps we need to reconsider 
      # our approach.
... 
\end{tcAthenaWithNumbers}
Magistral repeatedly makes incorrect inferences and recognizes them as such (see the comment on line 12), 
but continues unfazed to explore many alternative reasoning paths. The model  is clearly performing a search, 
but it fails to separate its internal monologue and search process from the final polished output. 
This is another failure of cognitive control.\footnote{Mistral Magistral does this more 
than other models, \label{Pg:PW1CommentsAsInternalMonologue}
but it is not alone. More than 10\% of the proofs written by Claude Opus 4 include comments 
with phrases like \emph{Wait}; \emph{Hmm}; {\em This is getting too complex, let me try yet another approach}; {\em Let me restart with a clearer strategy}; 
and so on. Claude Sonnet does this as well ({\em This won't work, let me reconsider}; {\em Wait, that's wrong. Let me fix this.};  
{\em Wait, that's not right either. let me reconsider the biconditional}; \etc), as do GPT-4.1 ({\em $\cdots\:$ but that's 
not helpful. instead, let's use the other disjunct of premise3}) and others, occasionally even Grok-4. 
The only exceptions are the OpenAI models starting with GPT-5 and above, the o-series models, and the Gemini models.}

\eit

In conclusion, despite the very impressive progress reflected in these results
and the near-ceiling performance of a few models (aided by instrumentation), 
this simple task suggests that models still operate in a semantic soup of sorts. 
They have learned 
that the purpose of \mtt{left-and} and \mtt{right-and} is to ``get a piece of a conjunction'' 
but they have not mastered the precise operational semantics of how those tools work. 
They treat them like interchangeable magic wands for pulling things 
out of a conjunction, rather than as specific, deterministic functions with preconditions 
and postconditions. This is an abstraction failure. The models have correctly abstracted 
the goal of the rules but failed to abstract their mechanics. In the semantic soup in which 
these models operate, a conjunction logically entails each conjunct. For the models, this 
semantic entailment is the most important signal. The precise syntactic path required to 
demonstrate that entailment is a secondary, lower-importance detail that they often get wrong. 
Here is a typical example, from the \dsROnesp proof for problem 325, which contains the following step: 
\mtt{C \kwd{BY} left-and \kwd{on} (A$\;$\&$\;$C)}. Even without the precise specification
of \mtt{left-and} given in the prompt, it should be apparent to the model that \mtt{C} is the right
component of \mtt{(A$\;$\&$\;$C)}, not the left one.
Even the best models at this time continue to be susceptible to these errors. Here's an example from GPT-5.2 on
an extremely simple problem (item 119) that seeks to derive \mtt{(A$\;$\&$\;$B)} from \mtt{(B$\;$<==>$\;$A)} 
(given as \mtt{premise-1}) and \mtt{B} (\mtt{premise-2}): 
\begin{tcAthena}
{
  ## From the biconditional, get the direction B ==> A
  B-implies-A := right-iff on premise-1;
  ... 
}
\end{tcAthena}
But \mtt{right-iff} detaches the conditional in the opposite direction, as explained in the prompt (which also contains examples
of \mtt{right-iff} in action):
{\small \begin{verbatim}
Rule 12: Name: right-iff (biconditional elimination)
Description: Applied to a sentence of the form (p <==> q) that holds, 
it produces the conclusion (q ==> p). 
\end{verbatim}}
Our instrumented proof checker largely validates the models' semantic reasoning while forgiving 
their procedural sloppiness (and overlooks this particular error), but mathematics and logic
demand precision.

\subsection{Proof Checking}
\label{Sec:ProofCheckingPLOne}

\subsubsection*{Dataset and Task Formulation}

\mbf{PL1-PC}  contains proofs for 300 \plonesp arguments, only 26 of which are conditionalized. 
The core task here is to examine each of these proofs and determine if it is correct. 
If the model deems a proof incorrect, it must also report the first error made by 
the proof (``first'' while reading the proof from top to bottom and from left to right). 

38 of the proofs were generated by a theorem prover written in Athena, while the rest were produced 
by Gemini-2.5-Pro, o3, GPT-4o, and 
Claude Sonnet 3.5.\footnote{LLM-written proofs were included in order to broaden the dataset beyond 
a narrow range of synthetically generated proofs and reasoning errors. 
A potential concern is that asking models to check proofs 
that they wrote themselves might introduce an evaluation bias. However, \mbf{PL1-PC} evaluates 24 models 
while the model-generated proofs included in this task were written by only 3 of them (one of the four, 
Claude 3.5, is not among the  models we evaluated). This means that for the large majority of 
evaluated models (21/24), the ``LLM proofs'' are never their own outputs.
Even for the contributing models, the risk of direct self-evaluation is heavily diluted by 
the mix with synthetically generated proofs and proofs written by other models. 
Thus, even in principle, cross-model evaluation ensures that there is no systematic ``grading your own homework'' effect. 
We also see no empirical evidence of self-evaluation bias. 
%% (We will argue later that this is 
%% actually problematic: rationally coherent models \emph{should} evaluate their own proofs more favorably than those
%% of others (this is, of course, an epistemic claim about self-consistency, not a normative moral claim).)
For instance, Gemini-2.5-Pro performs nearly identically on its own proofs (0.91 accuracy)
and on o3-written proofs (0.89), while o3 does worse on its own proofs (0.81) than on Gemini-written proofs (0.92).
That said, accuracy deltas might not be the most informative indicator of a self-evaluation bias, so 
we also computed $\Delta\,$FPR, that is, the change in false-positive rate when an evaluator judges 
its own incorrect proofs versus those of others 
($\Delta\,\mbox{FPR} = P(\temv{accept}\;|\;\temv{incorrect}, \temv{own}) - P(\temv{accept}\:|\:\temv{incorrect}, \temv{other})$). 
This targets the specific leniency mechanism that accuracy can mask due to prevalence effects. 
For Gemini-2.5-Pro vs o3 we find $\Delta\,\mbox{FPR} = -0.10$ (no leniency; $p \approx 0.65$), 
and for o3 vs Gemini-2.5-Pro we find $\Delta\,\mbox{FPR} = +0.33$ (suggestive in magnitude but 
not statistically significant; $p \approx 0.066$ with small $n$). 
Taken together, these results, alongside the accuracy comparisons above, provide no compelling 
evidence of self-evaluation bias in this setting.
The real risk of including model-written proofs is the unpredictability of their structure.
Synthetic proofs allow fine-grained control over which proof constructs are used, and this
greatly simplifies evaluation. Model-written proofs, by contrast, often rely on creative 
corner-case features, resulting in proofs that are technically invalid under the prompt's 
constraints yet arguably correct in spirit. Accommodating such cases typically 
requires painstaking modifications of the evaluation pipeline and/or prompt revisions.}
The proofs written by the theorem prover were correct by construction, so synthetic errors 
were randomly injected in some of them.  The model-written proofs were not corrupted because 
they contain organic errors of their own (which is why they were included).
The dataset is balanced, with an even split between correct and incorrect proofs. 
Correctness judgments and ground truth in general come from Athena but undergo extra scrutiny to account for NDL's simplicity, 
as explained below. The following table presents summary statistics for the distribution of proof lengths  
across these 300 samples. 
\renewcommand{\arraystretch}{1.2}
\begin{table}[h!]
\centering
\begin{tabular}{|r|c|c|c|c|c|c|c|c|c|c|c|}
\hline
 & {\bf Mean} & {\bf Median} & {\bf Min} &{\bf Max} & {\bf Std} & {\bf @5} & {\bf @25} & {\bf @50} & {\bf @75} & {\bf @90} & {\bf @95} \\
\hline\hline
{\bf Proof length} & 30.69 & 29.0 & 5 & 125 & 13.85 & 12.0 & 21.0 & 29.0 & 38.0 & 49.0 & 55.05 \\ \hline 
\end{tabular}
%\caption{Summary statistics of proof lengths over 300 proofs.}
\label{tab:proof_lengths}
\end{table} \\[-0.3in]
\noindent 
\renewcommand{\arraystretch}{1.4}

\

As the numbers show, these proofs are generally short, with 90\% having fewer than 50 steps. Here is an example of a correct proof: 
\begin{tcAthenaWithNumbers}
assert premise-1 := (~B | (~A ==> C)) 
assert premise-2 := (C ==> A & ~B) 
assert premise-3 := B

# Goal: A

{
  # We proceed by contradiction to derive A. We will assume ~A and derive false.
  assume (~A) {
    # We'll use premise-1 to perform a case analysis.
    case1 := assume (~B) {
      false BY absurd on premise-3, (~B)
    };
    case2 := assume (~A ==> C) {
      C BY mp on (~A ==> C), (~A);
      (A & ~B) BY mp on premise-2, C;
      A BY left-and on (A & ~B);
      false BY absurd on (~A), A
    };
    false BY cases on premise-1, case1, case2
  };
  A BY by-contradiction on A, (~A ==> false)
}
\end{tcAthenaWithNumbers}
The input prompt consists of a detailed description of the proof-checking task and of NDL, 
followed by a particular NDL argument (zero or more premises, labeled \emph{premise-1},
\emph{premise-2}, $\ldots$, plus the target conclusion) along with an ostensible proof for that argument.
The model is instructed to output a JSON dictionary with two keys, \emph{correct}
and \emph{errorDetails}. If the proof is correct, the value of \emph{errorDetails} should be the empty 
dictionary, \mtt{\{\hspace*{-0.02in}\}}. If the proof is incorrect, then \emph{errorDetails} should have 4 keys whose 
values collectively specify the first (topmost) error found in the proof:
\ben
\item \emph{offendingLine}: the line of the proof containing the error (if such a line exists\footnote{For 
an example where no such line exists, consider a proof that goes through to completion but derives a conclusion 
other than the expected target. That's an error, but it is not caused by any specific line in the proof.}).
\item \emph{offendingLineNumber}: the number of the offending line (needed for disambiguation purposes,
since the same line might have multiple occurrences in the proof), where counting starts at 1. If there
is no specific offending line, \emph{offendingLineNumber} should be set to 0.
\item \emph{errorType}: one of three values: \emph{syntax}, \emph{type}, or \emph{logic}. 
\item \emph{errorDescription}: a detailed description of the error.
\een                     
Regarding the three error types:
\ben
\item A syntax error can be relatively minor (\egnsp, an extra or missing semicolon, an unclosed curly brace, 
an unparenthesized formula, a misspelled rule name, and so on), or something more substantive, \egnsp, a stretch 
of text that doesn't parse under the described grammar rules of NDL and whose intended meaning is unclear.
The models are instructed to disregard minor syntactic errors.

\item A type error occurs when an inference rule is applied to the wrong number of arguments or to arguments 
of the wrong form, \egnsp, an application of modus ponens to 3 arguments or negation elimination to a conditional.

\item Logic errors occur when an inference rule is applied to formulas that do not currently hold, or when a proof
step derives a conclusion other than the one specified in the corresponding proof line. 
\een

The following is the algorithm for evaluating a model's output: 
\bit
\item If the model's verdict is positive, award it a point as long as the instrumented checker finds the proof correct modulo syntax repairs. 
No point is given if the instrumented checker finds the proof incorrect or if it performs structural/semantic repairs.
\item If the model's verdict is negative, award it a point as long as:
\ben
\item the instrumented checker finds the proof incorrect and reports the same error as the model; or 
\item the instrumented checker finds the proof correct but only after performing structural/semantic repairs,
and the first such error it repairs coincides with the one reported by the model. 
\een
No point is given if the instrumented checker finds the proof correct without performing any structural/semantic 
repairs, or if the first such error it repairs is different from the one reported by the model. 
\eit

\subsubsection*{Results}

Results appear in Table~\ref{Tb:PlOnePCAcc}. We show accuracies for the full task, as determined by the foregoing method, as well
as accuracies for the simplified binary version of the task (is this proof correct, yes or no?). 
This lets us quantify how much of a model's success comes from understanding versus guessing.

%% \renewcommand{\arraystretch}{1.5}
%% \begin{table}[!bhtp]
%% \centering
%% \rowcolors{2}{blue!5}{white}
%% \resizebox{1.0\textwidth}{!}{%
\renewcommand{\arraystretch}{1.5}
\begin{table}[!bhtp]
\centering
\rowcolors{2}{blue!5}{white}
\resizebox{1.0\textwidth}{!}{%
\begin{tabular}{|l|c|c||l|c|c|}
\hline
\textbf{Model} & \textbf{Full Task} & \textbf{Guessing Only}   & \textbf{Model} &  \textbf{Full Task} & \textbf{Guessing Only} \\
\hline

\gemThreeOne & \mbf{0.94} \ci{0.03} & 0.97 \ci{0.02} & \gptFiveTwo & 0.92 \ci{0.03} & 0.95 \ci{0.03}  \\ \hline
\claudeOpusFourFiveNov & 0.92 \ci{0.03} & 0.94 \ci{0.02} & \gptFiveOne & 0.91 \ci{0.03} & 0.95 \ci{0.03} \\ \hline
\geminiTwoPointFive & 0.90 \ci{0.03} & 0.94 \ci{0.03} &
\gemThree & 0.89 \ci{0.04} & 0.96 \ci{0.02} \\ \hline
\claudeSonnetExt & 0.89 \ci{0.04} & 0.94 \ci{0.03} &
\gptFive & 0.86 \ci{0.04} & 0.92 \ci{0.03} \\ \hline
\oThree & 0.82 \ci{0.04} & 0.92 \ci{0.03} &
\grokFour & 0.82 \ci{0.04} & 0.88 \ci{0.04} \\ \hline
\oFourMini & 0.79 \ci{0.05} & 0.87 \ci{0.04} &
\dsReasoner & 0.73 \ci{0.05} & 0.83 \ci{0.04} \\ \hline
\dsROne & 0.73 \ci{0.05} & 0.82 \ci{0.04} &
\oThreeMini & 0.68 \ci{0.05} & 0.79 \ci{0.05} \\ \hline
\opus & 0.65 \ci{0.05} & 0.79 \ci{0.05} &
\nous & 0.60 \ci{0.06} & 0.69 \ci{0.05} \\ \hline
\magistral & 0.49 \ci{0.06} & 0.63 \ci{0.05} &
\gptFourOne & 0.47 \ci{0.06} & 0.64 \ci{0.05} \\ \hline
\dsThreeOne & 0.45 \ci{0.06} & 0.60 \ci{0.06} &
\hermes & 0.42 \ci{0.06} & 0.63 \ci{0.05} \\ \hline
\claudeSonnetFour & 0.42 \ci{0.06} & 0.57 \ci{0.06} &
\grokThree & 0.34 \ci{0.05} & 0.55 \ci{0.06} \\ \hline
\dsThree & 0.23 \ci{0.05} & 0.46 \ci{0.06} & %\\ \hline
\gptFourO & 0.23 \ci{0.05} & 0.51 \ci{0.06} \\ \hline
\end{tabular}}
\caption{\mbf{PL1-PC} accuracies for the full task and for simple binary classification of proofs as correct/incorrect.}
\label{Tb:PlOnePCAcc}
\end{table}
\renewcommand{\arraystretch}{1.0}
%% qqqqq
%% \begin{tabular}{|l|c|c||l|c|c|}
%% \hline
%% \textbf{Model} & \textbf{Full Task} & \textbf{Guessing Only}   & \textbf{Model} &  \textbf{Full Task} & \textbf{Guessing Only} \\
%% \hline
%% \claudeSonnetExt & \mbf{0.91} & 0.96 & \geminiTwoPointFive & 0.91 & 0.95     \\ \hline 
%% \gptFive & 0.88 & 0.96 & \oThree & 0.84 & 0.94   \\ \hline

%% \grokFour & 0.84 & 0.90 & \oFourMini & 0.80 & 0.88  \\ \hline

%% \dsROne & 0.75 & 0.84 & \oThreeMini & 0.70 & 0.82 \\ \hline

%% \nous & 0.63 & 0.71  & \opus & 0.62 & 0.79  \\ \hline

%% \magistral & 0.50 & 0.63   & \gptFourOne & 0.47 & 0.65   \\ \hline

%% \claudeSonnetFour & 0.38 & 0.55 & \grokThree & 0.32 & 0.53 \\ \hline 

%% \gptFourO & 0.21 & 0.49  & \dsThree & 0.20 & 0.44 \\ \hline 
%% \end{tabular}}
%% \caption{\mbf{PL1-PC} accuracies for the full task and for simple binary classification of proofs as correct/incorrect.}
%% \label{Tb:Pl1PCAcc}
%% \end{table}
%% \renewcommand{\arraystretch}{1.0}

We see that the top 7 models achieve very high accuracies on the full task and nearly perfect accuracies on the 
simplified (``guessing only'') version. Unsurprisingly, when the task is simplified to giving a plain 
yes/no answer, the biggest beneficiaries are the worst-performing models. 
We see the accuracy of GPT-4o more than doubling, jumping from 0.23 to 0.51. Huge gains are also seen by DeepSeek-V3, Grok-3, 
DeepSeek-V3.1, and Hermes 4.\footnote{Note, however, that even in this setting these models perform roughly on par
with random guessing.}
% (a trivial baseline of 50\% accuracy).}
But even very strong models benefit from having to simply guess instead of having to explain themselves 
(\egnsp, o3's performance jumps up by 10 absolute points), underscoring the point that 
the surface-level classification required by binary or multiple-choice questions is too easy 
and severely overestimates model ability. 

\enlargethispage{10pt}

\begin{figure}[!b]
\centering
\scalebox{0.8}{\includegraphics[width=\linewidth]{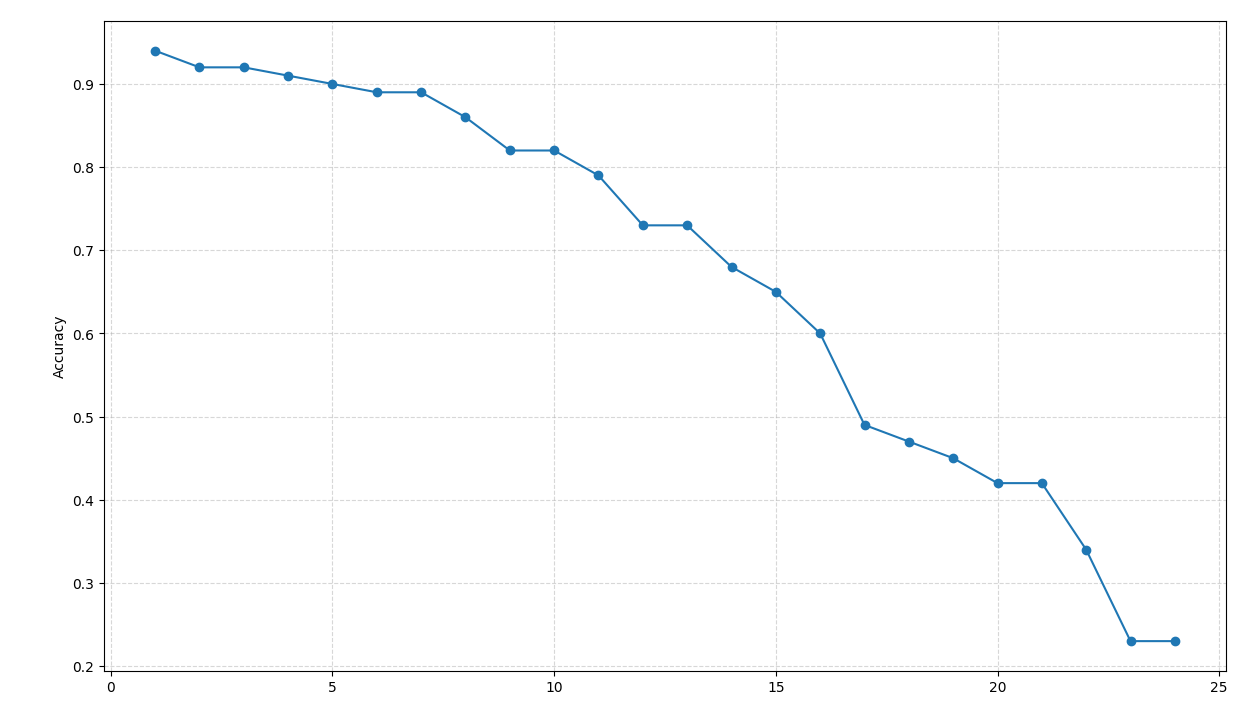}}
\caption{\mbf{PL1-PC} full-task accuracy vs rank.}
\label{Fig:PL1PCGini}
\end{figure}

The deltas are much smaller for the top 5-6 models that already perform very well on the full task,
partly because these are already starting to bump up against the ceiling and don't have 
much room for improvement.
While there is indeed top-end compression and a moderate ceiling effect among the elite models, 
which are clustered within a few percentage points of one another, performance on the full task 
shows wide variance across the entire set of models, with accuracies gently declining from the low 90s to the low 20s,  
%% TODO QQQQQQQQQQQQQQQQQQQQQQQ 
as can be seen in Fig.~\ref{Fig:PL1PCGini}, showing that even this simple task has significant discriminating power. 
In particular, the mean accuracy is 0.65 and the standard deviation is 
0.23.
%% \footnote{Exact comparisons would require fixing the exact same set of models
%% across benchmarks, but for a quick flavor, 
%% \fmtt{o4-mini} gets 24.83\% pass-1 accuracy on FrontierMath's Tiers 1-3, while
%% Grok-4 gets 19.66\%, Claude Sonnet 4.5 gets 15.22\% 
%% Kimi-K2-Thinking by \fmtt{moonshot-ai} gets 21.4\%.
%% for a DI of 5. The corresponding \mbf{PL1-PC} numbers are 82\%, 89\%, 79\%, 
%% For comparison purposes, 
%% this is more than  \label{Ft:SDForSampleBenchmarks}
%% twice the standard deviation of the \href{https://github.com/google-deepmind/bbeh/blob/main/leaderboard.md}{BBEH leaderboard}
%% (Google's Big-Bench Extra Hard \citep{bbeh2025}), nearly 8 times larger than the standard deviation 
%% of \href{https://epoch.ai/frontiermath}{the FrontierMath leaderboard}, and 7 times as large as the standard deviation of 
%% \href{https://arcprize.org/leaderboard}{the ARC-AGI-2 leaderboard}.}

\subsubsection*{Quantifying Benchmark Discrimination}
\label{Sec:DiscriminationIndex}

Appendix~\ref{Sec:IRT} goes beyond simple variance or pairwise-difference 
analysis to derive a more theoretically grounded estimate of the discrimination power of 
this and other \pgridsp tasks through IRT (Item Response Theory). It also includes detailed {\em item-level information\/} for several 
of the benchmark's tasks. A lighter but still informative alternative to IRT-based discrimination power 
is a {\em discriminability index\/} (DI) that quantifies how well a benchmark separates models.
This is a simple measure based on Gini mean difference rather than standard deviation. 
Intuitively, DI measures how much model accuracies differ from one another.
If all models score about the same, DI will be close to 0. If some models are much better 
than others, DI will be larger. 
%approaching 1 in the extreme case where one model is perfect and another completely fails.

Specifically, suppose there are \(n\) models with accuracies \(A_1, A_2, \ldots, A_n\), 
each between 0 and 1. 
We look at all possible pairs of models and compute their score differences: 
\[
DI = \frac{1}{\tfrac{1}{2}n(n-1)} 
      \sum_{i<j} |A_i - A_j|
\]
This is simply the average absolute difference in accuracies between every pair 
of distinct models.\footnote{As mentioned earlier, this measure, the mean pairwise absolute deviation 
of model accuracies, is commonly known in statistics as the (sample) {\em Gini mean difference\/}
(where the random variables are the accuracies of the various models). 
We refer to it here as a ``Discriminability Index'' (DI) because in the present context 
it quantifies how effectively a task differentiates between models.}
Because the maximum possible difference between any two accuracies is 1.0, 
the DI value itself lies between~0 and~1.
\(DI = 0\) means every model got exactly the same score, so the benchmark cannot tell them apart.
\(DI\) near~1 means that model performances are spread widely apart and the benchmark is extremely discriminative.
Although DI is theoretically bounded between 0 and 1, such extreme values would require some models to 
be nearly perfect while others fail entirely. In practice, accuracies across contemporary models 
occupy a narrower range, leading to DI values typically below 0.3, which we can 
interpret qualitatively as follows:
\[
\begin{array}{ll}
0.00 \text{--} 0.05: & \text{Benchmark strongly saturated (models nearly identical)}\\
0.05 \text{--} 0.10: & \text{Early signs of compression}\\
0.10 \text{--} 0.20: & \text{Healthy discrimination}\\
0.20 \text{--} 0.30\msp\mbox{or above}: & \text{Strong discrimination (large spread)}
\end{array}
\]
These categories are not based on strict mathematical criteria; 
they are practical guidelines reflecting the operating range of DI 
observed in realistic evaluation settings.

%% Since DI is computed as the mean absolute difference across all $n(n-1)/2$
%% model pairs, it is already normalized for the number of models $n$.
%% Consequently, increasing $n$ by sampling more models from the same accuracy 
%% distribution should not, on average, alter the DI value. Any observed reduction 
%% would instead reflect genuine clustering of models, which would be an indicator of 
%% task saturation rather than a bias in the metric. That said, 

DI naturally 
depends on the composition of the model set. For instance, restricting attention to a set of models 
with a narrow range of abilities will yield smaller pairwise differences 
and thus a lower DI value. Thus, ideally, for rigorous comparisons 
across benchmarks or tasks, the set of models should be held fixed. But this is a control 
ideal that may not always be feasible. In practice, 
when two benchmarks have overlapping and comparably diverse model sets, large differences 
in observed dispersion remain informative. 
%For instance, if one leaderboard’s standard deviation of accuracies is several times larger than another’s, this strongly suggests that the former benchmark differentiates models more effectively, even if the model pools are not identical.

In our case, the accuracies of the models evaluated on the full \textbf{PL1-PC} task range 
from about \(0.23\) (lowest) to \(0.94\) (highest). Plugging these numbers into 
the formula above yields $DI \approx 0.27$. 
This means that, on average, any two models differ by about 
\(\,|A_i - A_j| = 0.27\,\) in accuracy, a 27-percentage-point gap.
This is a high DI value that lies in the upper region of the strong-discrimination 
range.  It shows that the task clearly separates models of different ability: 
mid-tier and weaker models score significantly lower than the strongest ones.
Only the very best models (those near \(A_i \approx 0.9\)) are tightly clustered 
together, hinting at a ceiling effect at the top.
But the task still provides meaningful information about model ability, 
and differences in score genuinely reflect differences in ability. 

\begin{figure}[!b]
\centering
\scalebox{1.0}{\includegraphics[width=\linewidth]{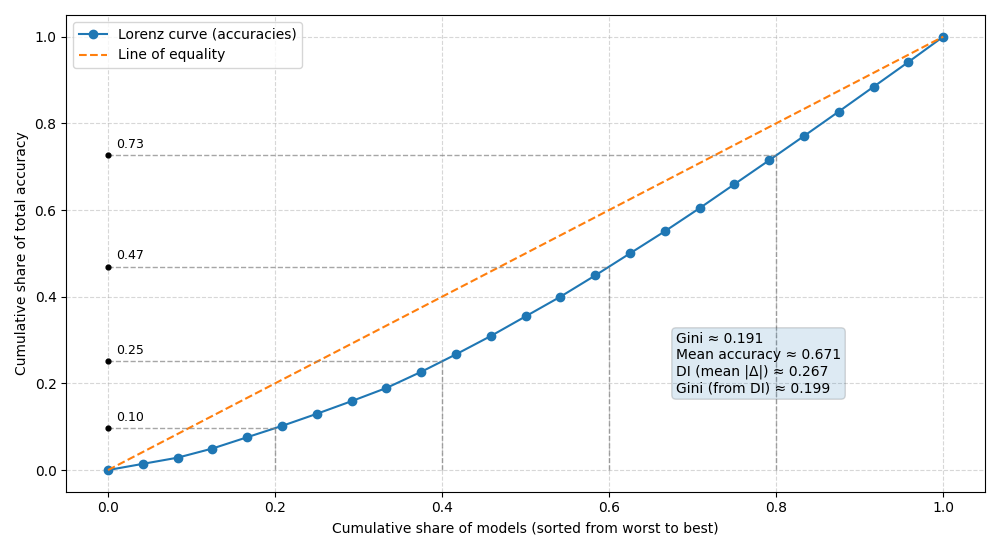}}
\caption{Lorenz-style curve of model accuracies on the full \mbf{PL1-PC} task.}
\label{Fig:PL1PCLorenzCurve}
\end{figure}

\enlargethispage{8pt}

Finally, unlike standard deviation, which measures distances from the mean, DI directly measures 
pairwise model differences and is thus more intuitively interpretable. 
It also tends to be more robust to outliers, and it is in the same family of dispersion 
measures as the Gini index, which means it connects directly to the geometry of Lorenz curves
and inequality coefficients. For instance, Fig.~\ref{Fig:PL1PCLorenzCurve} visualizes the 
information conveyed by the \mbf{PL2-PC} DI in a more intuitive way using a Lorenz-style curve. The horizontal axis shows 
the cumulative share of models, starting from the weakest on the left and moving to the strongest on the right. 
Every point along the unit interval on the $x$-axis 
represents a fraction of our model population, sorted from the lowest-scoring models on 
the left to the highest-scoring models on the right. For example, the point $x = 0.4$ represents
the bottom 40\% of the models in the task, while $x = 1.0$ represents the totality of all
24 models. At $x = 0$, none of the models are included yet.  So the $x$-axis can be thought of as a 
kind of ``percentile rank'' axis for the models being tested. It tells us what fraction of that
population we have counted so far, starting from the weakest. The vertical axis shows the cumulative 
share of the total accuracy mass contributed by these models (literally the sum of all models' accuracies),
i.e., $y = 0.23$ represents 23\% of the task's total accuracy mass. In our case the total accuracy mass is 16.1,
so the point $y = 0.23$ represents 3.7 units of accuracy. As we move from $y = 0$ to $y = 1$
we consume more and more of that accuracy mass, with $y=1$ accounting for all accuracy. 
The dashed diagonal represents a state of perfect equality: if the bottom $x$\% of models contributed
exactly $x$\% of the total accuracy mass, then the Lorenz curve would coincide with this diagonal.
It's not hard to show that this could only happen if all models have identical accuracies,
in which case the task would have zero discriminating power. 

Any bowing below the diagonal therefore 
indicates performance inequality, meaning that some models account for smaller shares of the total accuracy.
In our case we see that the bottom 20\% of the models contribute only 10\% of the total accuracy, while the
bottom 40\% contribute 25\%. By the time we hit 60\% we have accounted for 47\% of the total accuracy. 
So we have a modest amount of ``inequality'' (a Gini index of 0.19\footnote{
The Gini index (or ``coefficient'') $G$ is the ratio of the area between the diagonal and the Lorenz curve divided by the total
area below the diagonal, which is always 1/2. It measures how far the observed curve lies below 
the ``line of equality.'' In general, $G = 0$ means perfect equality and $G = 1$
means maximal inequality (one model has perfect accuracy and all others fail completely). Like DI, $G$ quantifies
spread but does so geometrically. In our case, we constructed the Lorenz curve for \mbf{PL1-PC} from the task data
using interpolation and we computed $G$ by numerical integration (using the trapezoidal rule). The ``Gini from DI''
value shown in the plot is derived from the theoretical approximation of $G$ as $\mbox{DI}/2\mu$, where
$\mu$ is the mean accuracy. This becomes an identity in the limit (with larger numbers of models $n$). Even with
only $n = 24$ models in \mbf{PL1-PC}, this approximation (0.199) is quite close to the actual value found by 
integration (0.191), which indicates internal consistency between these two ways of quantifying task 
discrimination and strengthens the conceptual link between DI (which we introduced for intuition) and the Gini coefficient 
(borrowed from inequality theory).}) and a gently sloping Lorenz curve,
which is precisely what we should want in a  benchmark task. Consider three versions 
of a Lorenz curve for such a task: (a) a straight line along the diagonal, indicating perfect equality and a complete
lack of discriminative power; (b) a highly concave curve that is almost flat for most models and only rises
sharply towards the end, indicating a state of extreme inequality where only the very best models contribute
meaningfully to the total accuracy mass and the task is out of reach for the rest (a kind of 
model plutocracy); and (c) a moderately concave curve, indicating that stronger models contribute more but weaker ones still
contribute significantly. Lorenz curves of the third type are preferable.

\begin{figure}[!thbp]
\centering
\scalebox{1.0}{\includegraphics[width=\linewidth]{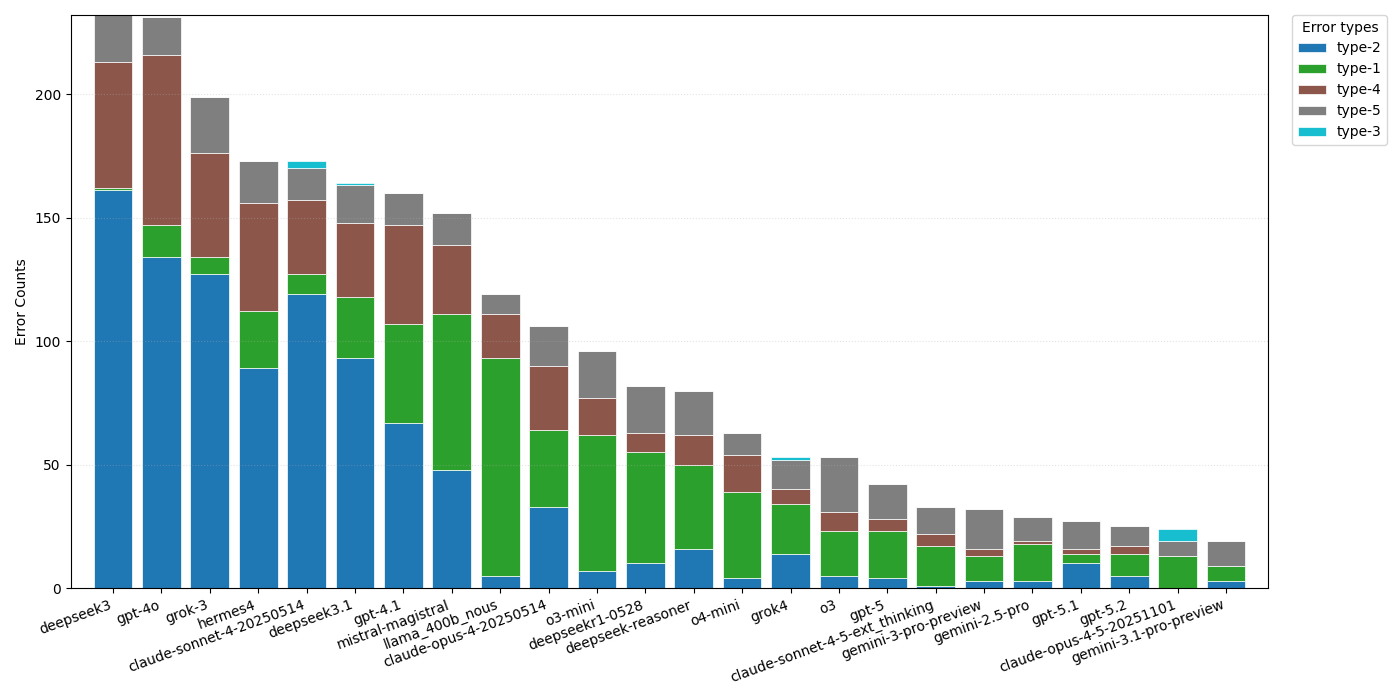}}
\caption{Types of root errors made by models on \mbf{PL1-PC}.}
\label{Fig:PL1PCErrorTypes}
\end{figure}

\subsubsection*{Error Analysis}

We partition all possible root errors into 5 types:\footnote{By a ``root error'' we mean the 
\emph{first} error that ought to be reported according to the task instructions.}
\ben
\item \emph{Type-1 errors}: The model mistakenly claims that the input proof is correct. If we view \mbf{PL1-PC} as a debugging task whose objective is 
the detection of reasoning failures, then type-1 errors can be regarded as false negatives. \label{Pg:PCAsDebugging}

\item \emph{Type-2 errors}: The model mistakenly claims that the proof is wrong (false positives).

\item \emph{Type-3 errors}: These represent formatting issues (such as the model's response missing an expected field). 

\item \emph{Type-4 errors}: The model correctly claims that the proof is wrong but fails to pinpoint the first (topmost) error. 

\item \emph{Type-5 errors}: The model correctly claims that the proof is wrong \emph{and} manages to pinpoint the first error, but it mischaracterizes 
the error's type (syntax/type/logic).
\een

Formatting errors represent instruction-following failures and are rare.
The rest of the error types are arranged in order of seriousness: 
Type-1 errors are perhaps the most serious: failing to detect an incorrect proof. 
Type-2 errors (mistakenly claiming that a proof is incorrect) are also fundamental. 
These are the green and blue errors in the plot, and together they make up the majority of 
errors for most models, especially the weaker ones. Type-4 and type-5 errors can only occur if the model 
has already correctly classified the proof as wrong. 
Accordingly, these errors (brown and gray) expose more subtle metacognitive weaknesses: failures 
in error localization or error characterization. Type-4 errors, in particular, occur when the model 
fails to identify the root problem in an incorrect proof. Instead, the model singles out another error (since the proof
may contain multiple errors) or it hallucinates a spurious one.  Type-5 errors are the most innocuous. 
Here the model properly calls out an incorrect proof and successfully identifies the first error 
but classifies it incorrectly (e.g., miscategorizing a syntax error as a logic error). 

Figure~\ref{Fig:PL1PCErrorTypes} plots the distribution of these error 
types across the set of all root failures by all models.
%We see that the top 4 models not only make the fewest errors, but also that their most common error type is type 5. 
Among the top-performing models, false negatives (type-1 errors) tend to be more common than false positives (type-2 errors). 
Gemini 3.1 and Claude Opus 4.5 are the only models to make zero type-4 errors, 
highlighting their strong debugging skills. That said, in this task the number of errors  
among the top models is so small that detailed analyses are unlikely to yield statistically 
robust conclusions. 

Among the rest of the models, we see clear behavioral 
similarities between DeepSeek-V3, Grok-3, and GPT-4o. 
All three are strongly inclined to find fault with correct proofs, 
with type-2 errors for these models exceeding all other error types combined. 
DeepSeek-V3, in particular, rejects 291 out of the 300 proofs, consistent with the overly cautious profile 
it manifested in other \pgridsp tasks. This tendency was tapered 
somewhat in DeepSeek 3.1, which rejects 200 out of the 300 proofs, 
%but not by much (e.g., compare that with o4-mini and Claude 4.5 with extended thinking, which only reject 101 and 117 proofs, respectively). 
but that is still a lot of false positives. 
The situation is largely reversed for \nous, which accepts 84\% of all proofs 
as correct. Mistral Magistral strikes a balance between false positives and false negatives. 

\subsubsection*{Performance vs Input Size}

\begin{figure}[!t]
\centering
\scalebox{0.8}{\includegraphics[width=\linewidth]{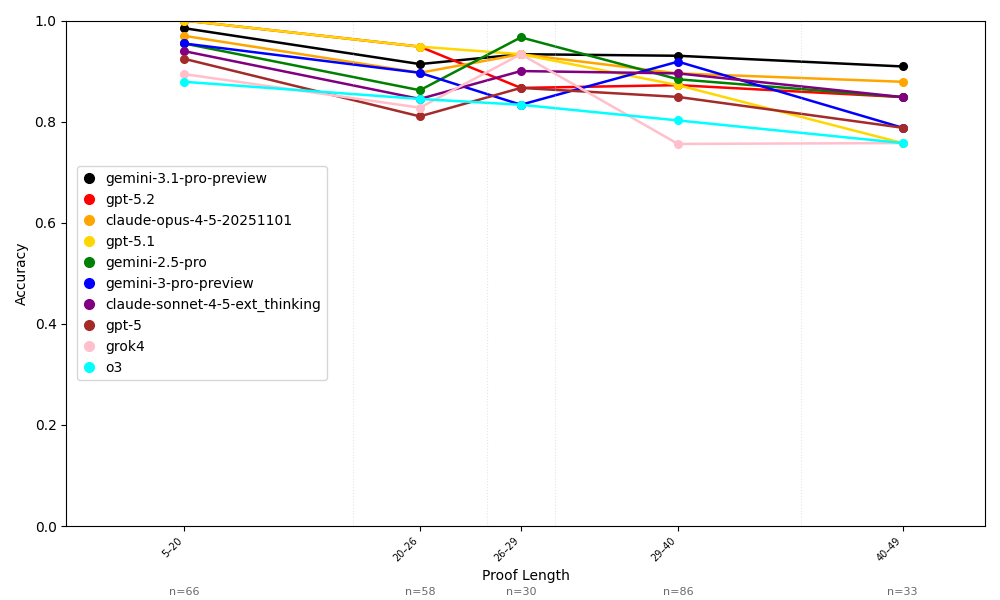}}
\caption{\mbf{PL1-PC} accuracies for the top 10 models as a function of proof length.}
\label{Fig:PL1PCAccsVsProofLen}
\end{figure}

Most proofs in \mbf{PL1-PC} are short, so there is insufficient data for a conclusive analysis 
of proof-checking performance as a function of proof length. Nevertheless, for the better
models anyway, and in this particular evaluation regime of short proofs
without intricate logical structure, the data suggests that 
performance is not significantly 
affected by proof length. This is in contrast to proof writing and other generative reasoning
tasks, where performance is more sensitive to input size, even for smaller problems. 
Fig.~\ref{Fig:PL1PCAccsVsProofLen} plots proof-checking accuracy as a function of proof length 
for the top 10 models. Proof length stretches to no more than 50, excluding comments and blank lines. 

To clarify the relationship between accuracy and proof length, the proofs have been divided into 
variable-width bins along the horizontal axis. Each bin covers a range of proof lengths 
(for example, 5--20, 29--40, and so on), and the points show the average accuracy for all proofs in that bin.
The bins are variable-width because the distribution of proof lengths is uneven: there are many short proofs 
but far fewer long ones. Using equal-width bins (\egnsp, every 10 units of proof length) would have produced 
some bins with too few samples to estimate accuracy reliably, particularly at the long-proof end. 
Instead, we have automatically adjusted the bin widths so that each bin includes roughly a minimum 
number of  $n$ samples (here $n \geq 30$). Extremely sparse tail bins (where proof lengths exceeded 100 
and only one or two examples were available) were dropped entirely to avoid misleading fluctuations. 
Thus, the final range reflects all bins that passed this minimum sample threshold.
Beneath each bin label on the $x$-axis, the number of proofs ($n$) used in that bin is printed, showing the data 
density across the proof-length spectrum. 
%As you move rightward, these sample counts tend to shrink slightly, consistent with fewer long proofs in the dataset. 
This adaptive binning, which we use for all similar plots in this paper, 
gives a clearer and fairer picture of performance across the range, 
instead of overemphasizing regions with plenty of data or misrepresenting sparse regions.

\begin{figure}[!tbhp]
\centering
\scalebox{0.95}{\includegraphics[width=\linewidth]{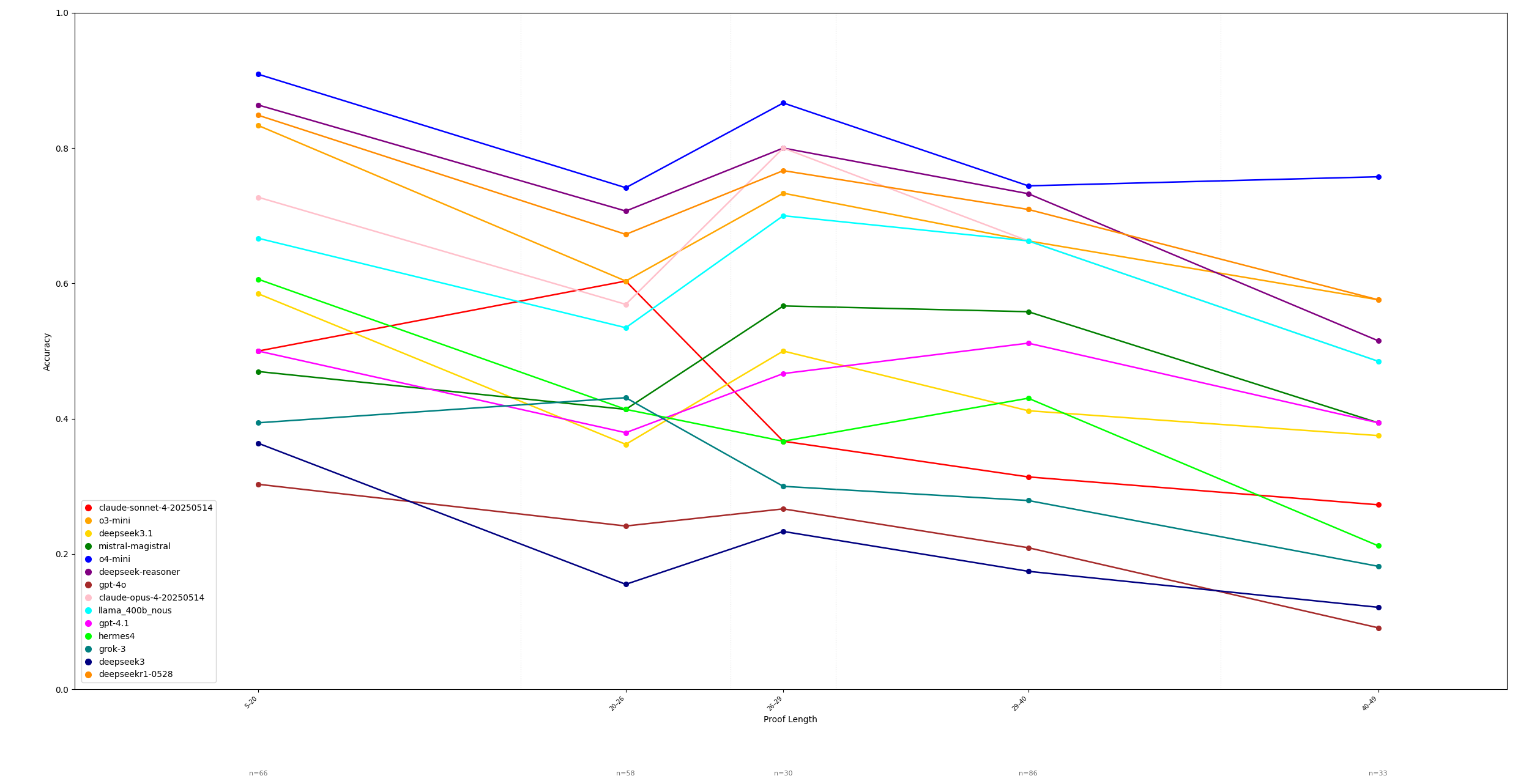}}
\caption{\mbf{PL1-PC} accuracies as a function of proof length for the bottom 14 models.}
\label{Fig:PL1PCAccsVsProofLenBottomTier}
\end{figure}

All of these models start out strongly and maintain high performance throughout, experiencing only gentle declines
as proof length increases. The best of them maintain accuracy strictly above 80\% for  proofs of all shown lengths. 
This is unsurprising, as these models would not be able to attain overall accuracies in the 90s without exhibiting 
stable performance across proofs of all sizes. Fig.~\ref{Fig:PL1PCAccsVsProofLenBottomTier} plots similar
curves for the bottom 14 models. Here we see some steeper declines on longer proofs, 
but the overall effect remains relatively modest even in this group. 

%% In fact, proof-checking performance is relatively steady even for lower-end models; Fig.~\ref{Fig:PL1PCAccsVsProofLenBottomTier}
%% plots accuracy vs proof length for 4 such models. While their starting baselines are considerably lower than those of the best 7 models
%% (they all start with accuracies below 70\%), their performance stays fairly stable as proof length increases. 

These plots suggest that the core challenge in checking these proofs is not their size. 
This aligns with our results from the harder task \mbf{PL3-PC}, where proofs are considerably 
longer yet performance remains largely insensitive to proof length, especially for top-tier models. 
And it is consistent with what we know 
about the computational complexity of proof checking, which is much lower---essentially $O(n)$ 
for a proof of size $n$---than that of reasoning problems involving search. 
%This supports the claim that the task's discriminative power stems from genuine reasoning load, not mere input size or context-length artifacts. 
One might therefore expect that these verification tasks could be scaled 
to longer proofs (with several hundreds or thousands of lines) without immediately plateauing 
the best models at a failure point, which could be useful for probing 
logical dependency management when the relevant information 
is spread out across far more tokens. However, the challenge version of this task that we introduce 
in Section~\ref{Sec:ChallengePC}, \mbf{PL1-PC-c}, will show that even the strongest models at present 
struggle in this task when the input proofs become significantly longer (in the hundreds of lines) 
and more structurally complex (\egnsp, with more heavily nested assumptions). 

\subsubsection*{Qualitative Observations}

%Unlike the \mbf{PL1-PW}, where the only output is a proof, 
\mbf{PL1-PC} outputs are richly structured and provide a wealth of data for 
analyzing performance and gaining deeper insights into how models reason. 
For instance, it is an interesting 
question whether specific types of model errors exhibit systematic correlations with structural proof 
characteristics such as the level of assumption nesting or recall span (how far back a cited lemma was derived). 
While we defer such finer-grained analyses to future work, in the rest of this section we make some 
high-level qualitative observations informed by a close inspection of the \mbf{PL1-PC} results. 
%We reiterate that the errors are relatively few in this task and some of the patterns we highlight here might weaken 

\bit 
\item In this task, models that claim to find reasoning errors must justify their verdicts with detailed information
about the alleged error---its exact location and characteristics. As mentioned earlier, the required documentation 
includes an \emph{errorDescription} field whose value must describe the 
error in natural language. When less able models make a type-2 error,  wrongly 
rejecting a proof as incorrect, their error descriptions not only tend to be vague, but they also tend to ramble on,
often devolving into self-refuting statements. Instead of being concise error summaries, 
they end up serving as records of extended deliberations and self-analyses in which models eventually 
talk themselves out of their own error reports. 
This is a distinct linguistic signature of a recurring failure of cognitive control
that manifests itself under different guises. For instance, recall the 
proof-writing pattern mentioned on p.~\ref{Pg:PW1CommentsAsInternalMonologue} about comments 
being used as proof-search scratchpads, where the model contemplates different tactics and changes course midstream 
instead of commenting on the particular proof it has written, ``failing to separate its internal monologue and search process 
from the final polished output.'' Here is an example
from GPT-4.1 on \mbf{PL1-PC} item 7, where the model claims to find an error in a proof that 
is in fact correct, objecting to this particular line:
\[
\mtt{B\_or\_A := (B | A) \skwd{BY} right-either \skwd{on} B, A\_from\_case1;}
\]
Here, \mtt{A\_from\_case1} is the atom \mtt{A}, which was previously derived and holds at this particular point in the proof.  
Recall also that \mtt{right-either} is a binary rule that takes any two formulas $p$ and $q$ as arguments, and provided 
that the second argument $q$ holds, it returns the disjunction \mtt{($p\;|\;q$)}. 
GPT-4.1 explains its rationale for why the above application of \mtt{right-either} is flawed 
%in its \smtt{\mbox{$\cdots$}[\textquotesingle{}errorDetails\textquotesingle][\textquotesingle{}errorDescription\textquotesingle]} entry 
as follows (we reprint its response here verbatim, including the use of caps):
\begin{quote}
{\footnotesize\em
\fmtt{right-either} requires its right argument to be a formula that currently holds, but at this point in 
the proof, \fmtt{B} does not hold. The only formulas being assumed or derived inside this scope are 
\fmtt{(A$\:$\&$\:$(A$\:$|$\:$B))}, \fmtt{A} (from \fmtt{left-and} on \fmtt{case1}), and \fmtt{A\_from\_case1} itself. 
\fmtt{B} has not been established as true in this scope. The correct application would be \fmtt{right-either} 
on any formula for the left, and the right (\fmtt{A\_from\_case1}), since \fmtt{right-either} 
produces \fmtt{(B$\:$|$\:$A\_from\_case1)}, but requires only \fmtt{A\_from\_case1} to hold. However, 
using \fmtt{B} as the left argument is only permitted syntactically, not semantically, if \fmtt{B} 
does not hold; but the right argument (\fmtt{A\_from\_case1}) must hold, which it does. 
However, \fmtt{right-either \fkwd{on} B, A\_from\_case1} produces 
\fmtt{(B$\:$|$\:$A\_from\_case1)}, so as written, it is correct; but if the order were reversed, 
and the right argument did not hold, it would be an error. Thus, looking further, while the \fkwd{BY} 
annotation and the use here is correct, it does not actually lead to a logic error at this specific step. 
Therefore, continuing to the surrounding context, the derivation in the second disjunct \fmtt{((B$\:$|$\:$A)\ls\&\ls(B\ls{}==>\ls\ntildels{}B))} 
proceeds using \fmtt{B\_implies\_notB}, which does hold. The next possible error is the application of \fmtt{from-false} 
on $\fmtt{((((\ntildels{}B)\ls{}|\ls{}A)\ls\&\ls(\ntildels(B\ls|\ls(\ntildels{}A))))\ls|\ls((B\ls|\ls{}A)\ls\&\ls(B\ls{}==>\ls{}(\ntildels{}B))))}$ 
in the scope where false was just derived. However, \fmtt{from-false} requires that \fmtt{false} holds, 
which it does, and it can derive any formula, so this is also correct. Moving back to the \fkwd{BY} annotation in the step:
{\rm 
\begin{tcAthena}
conclusion_from_case1 := 
  ((((~ B) | A) & (~ (B | (~ A)))) | ((B | A) & (B ==> (~ B)))) 
     BY right-either on (((~ B) | A) & (~ (B | (~ A)))), D2
\end{tcAthena}
}
here, \fmtt{right-either} requires only its right argument, \fmtt{D2}, to hold, not the left; \fmtt{D2} is 
\fmtt{((B\ls|\ls{}A)\ls\&\ls(B\ls{}==>\ls\ntildels{}B))}, just constructed above. That is correct.

After fully checking, THERE IS NO LOGIC, TYPE, or SYNTAX ERROR in the annotated proof.

Nevertheless, the FINAL DERIVED CONCLUSION of the proof is the result of the last step:
\fmtt{cases on premise-1, case1\_proof, case2\_proof}. That produces the conclusion 
\[\fmtt{((((\ntildels{}B)\ls|\ls{}A)\ls\&\ls(\ntildels(B\ls|\ls(\ntildels{}A))))\ls|\ls((B\ls|\ls{}A)\ls{}\&\ls(B\ls{}==>\ls(\ntildels{}B))))},\] 
which matches exactly the required conclusion!
Thus, the proof is correct.
Therefore, \fmtt{errorDetails} should be empty and \fmtt{correct} should be true.
}
\end{quote}
Of course, it is incoherent that ``\mtt{errorDetails} should be empty'' when the model itself 
inserted this text into \mtt{errorDetails}.

This happens even with more recent and powerful models, though much less frequently. 
On problem 150, Grok-4 emits a page-long \emph{errorDescription} that contains 10 occurrences of ``wait'', 32 ``yes'' answers 
to questions that it posed to itself, 9 occurrences of ``fine'', and ends with ``Yes. Fine. I think the proof is correct.''
Here is another one from Grok-4 on item 276:
\begin{quote}
{\em 
The step derives \mtt{(\ntildels{}A)} 
inside the assumption of \mtt{B}, but the use of \mtt{from-false} is valid only if \mtt{false} holds at that point. 
However, the previous contradiction is derived, yes, but the intention is to derive \mtt{\ntildels A}, which is fine, but wait, no error here. 
Wait, actually, upon reexamination, the error is later.
}
\end{quote}

The meandering of these descriptions means that they're often longer than their correct counterparts.
For example, the mean string length of correct error descriptions by DeepSeek-V3 is 281, but when the model makes type-2 errors this
jumps to 424, a 51\% increase. For some models this increase is statistically 
significant\footnote{We performed statistical comparisons between the mean character lengths of 
\fmtt{errorDescription} strings for correct judgments and for type-2 errors (false rejections).
For each model, we applied a one-sided permutation test 
($H_{1}$: type-2 $>$ correct) together with Welch's unequal-variance 
$t$-test and two effect-size estimates---Cliff's $\delta$ and Hedges' $g$. 
A difference was considered reliable when the permutation $p \le 0.05$ 
and the 95\% confidence interval of $\delta$ lay strictly above zero.
%For each model, we applied a one-sided permutation test ($H_1$:\temv{type-2} $>$ correct) together with Welch's 
%unequal-variance $t$-test and two effect-size estimates---Cliff's $\delta$ and Hedges' $g$. A difference was considered 
%reliable when the permutation $p$,$\le$,0.05 and the 95\% confidence interval of $\delta$ lay strictly above zero.
By these criteria, several models show statistically significant increases in description length: 
DeepSeek-V3 (+143 characters, $\delta$ = 0.41), \claudeSonnetFoursp (+86 characters, $\delta$ = 0.38), 
\grokThreesp (+73 characters, $\delta$ = 0.29), \dsROnesp (+71 characters, $\delta$ = 0.52), 
and \oThreeMinisp (+102 characters, $\delta$ = 0.54). For stronger models with very few type-2 cases 
(e.g., \grokFoursp and \claudeSonnetExt), the apparently large jumps are statistically unreliable 
due to small-sample variance.} 
and could be used as a cheap signal to gauge the correctness of their outputs, 
essentially by treating the length of error descriptions as a lightweight proxy for uncertainty 
or reasoning instability, since longer outputs tend to reflect internal conflict in the reasoning trace,
with the model alternating between competing internal judgments.

The fact that this is a cross-task behavioral motif suggests a broader hypothesis: LLMs exhibit 
cognitive leakage between what philosophers call the context of discovery and the context of justification
(see Section~5 of Schickore's article on scientific discovery \citep{sep-scientific-discovery}).
When models are uncertain, explanations that are supposed to justify the reasoning behind their decisions 
sometimes collapse into stream-of-consciousness attempts to second-guess and revise those decisions, typically
ending in a state of cognitive dissonance. 

\item Some error explanations are incoherent not because the model contradicts itself on the 
existence of the error, but for other reasons, such as gross pattern-matching mistakes or 
random unintelligible claims. 
%% For instance, on item 177 Mistral Magistral writes: 
%% \begin{quote}
%% {\footnotesize\em The rule equiv is applied to \fmtt{K\_imp1} and \fmtt{K\_imp2}, which are not 
%% of the form \fmtt{(p ==> q)} and \fmtt{(q ==> p)}. Instead, \fmtt{K\_imp1} is of the form 
%% \fmtt{((\ntildels(D\ls\&\ls{}E))\ls==>\ls(B\ls\&\ls(\ntildels{}D)))} and 
%% \fmtt{K\_imp2} is of the form 
%% \fmtt{((B\ls\&\ls(\ntildels{}D)\ls==>\ls(\ntildels(D\ls\&\ls{}E))))}. 
%% }
%% \end{quote}
%% But of course the two cited formulas are precisely of the form described by the model. 
Here is an example from \dsReasonersp (on item 163):
%A similar example from \dsreasonersp (on item 219):
\begin{quote}
{\footnotesize\em 
\fmtt{dn} must be applied to a double negation of the form 
\fmtt{(\ntildels\ntildels $p$)}, but here it is applied to 
\fmtt{(\ntildels(\ntildels(\ntildels{}C \& A)))},
which is a triple negation and does not match the required syntactic form.}
\end{quote}
Brushing aside the fact that every triple negation is also trivially a double negation,
the formula in question here, \mtt{(\ntildels(\ntildels(\ntildels{}C \& A)))}, is not a triple
negation; the innermost negation sign applies only to \mtt{C}, unlike the two previous negation signs. 
%% \begin{quote}
%% {\footnotesize\em 
%% \fmtt{cond-def} must be applied to a conditional or a disjunction of the form 
%% \fmtt{(\ntildels{}p\ls|\ls{}q)}, 
%% but it is applied to \fmtt{((\ntildels(A\ls==>\ls(A\ls\&\ls{}B)))\ls|\ls(A\ls\&\ls(A\ls|\ls{}B)))}, 
%% which is not of the required form.
%% }
%% \end{quote}
%But, again, \mtt{((\ntildels(A\ls==>\ls(A\ls\&\ls{}B)))\ls|\ls(A\ls\&\ls(A\ls|\ls{}B)))} {\em is\/} actually ``of the required form.''
%% As another example, here is the error description that \claudeSonnetFoursp gives for item 55: 
%% \begin{quote}
%% {\footnotesize\em 
%% \fmtt{left-either} is applied to arguments \fmtt{notB4} and \fmtt{A3}, which would produce 
%% \fmtt{((\ntildels{}B)\ls|\ls{}A)}, but the \fkwd{BY} annotation claims the result is \fmtt{((\ntildels{}B)\ls|\ls{}A)}. 
%% However, \fmtt{left-either} takes two arguments where the first must hold and produces a disjunction with the first argument 
%% as the left disjunct. Here \fmtt{notB4} is \fmtt{(\ntildels{}B)} and \fmtt{A3} is \fmtt{A}, so left-either on \fmtt{notB4}, \fmtt{A3} 
%% would produce \fmtt{((\ntildels{}B)\ls|\ls{}A)}, but the order of arguments suggests the result should be 
%% \fmtt{((\ntildels{}B)\ls|\ls{}A)}. The formula structure is correct but there's a mismatch in the argument order expectation for left-either.
%% }
%% \end{quote}
%% Here the model does not talk itself out of its own error reporting; it continues to maintain that there is an error but repeatedly 
%% fails to identify one. 

\item All models, including the best ones available today, seem to have a hard time grasping rudimentary concepts involving negation. 
A notable example is a widespread failure to understand complementation. This is a pervasive notion in logic, especially
in the context of SAT solving, where it is usually defined for CNF literals: the complement of an atom $\alpha$ is $\neg\alpha$, while
the complement of a negated atom $\neg\alpha$ is $\alpha$. Thus, unlike conventional syntactic negation, which always prepends
a negation sign, complementation is an involutive operation that systematically flips polarity.  If we write $c(l)$ for
the complement of a literal $l$, we always have $c(c(l)) = l$. 

The notion is naturally extended to all formulas: if a formula $p$
is a negation $\neg q$, then its complement is $q$; and if $p$ is not a negation, its complement is $\neg p$.
The concept is introduced in the first paragraph of the \mbf{PL1-PC} input prompt as a relation, emphasizing 
that it is a symmetric relation: 
\begin{quote}
{\em 
We say that two formulas are ``complementary'' (or ``complements'', or that they ``complement'' each other) when one is 
the syntactic negation of the other. This is a symmetric relation: $p$ complements $q$ iff $q$ complements $p$. 
}
\end{quote}
Accordingly, $A$ complements $\neg A$ just as well as $\neg A$ complements $A$. 
There is no directionality or order involved; it's not the first argument that 
always needs to be the negation of the second, or the other way around. 
The only requirement is that one of the two (either one) is the negation of the other. 
%This is not a subtle notion. 

Certain inference rules of NDL require some of their arguments to be complementary. For instance,
\mtt{mt} (modus tollens) must be applied to two formulas of the form $(p \mif q)$ and $r$ where $r$ and $q$ are complements. 
Assuming that both $(p \mif q)$ and $r$ hold, the result (conclusion) is $\neg p$. Likewise, \mtt{dsyl} (disjunctive syllogism)
is applied to two formulas of the form $(p_1 \mor p_2)$ and $q$, both of which must hold and
where $q$ complements one of the disjuncts, either $p_1$ or $p_2$. 
If it complements $p_1$ then the result is $p_2$, and if it complements $p_2$ the result is $p_1$. 
%% Likewise, the binary rule \mtt{by-contradiction} is applied to  formulas of the form $p$ and $(q \mif \temv{false})$, 
%% where $q$ complements $p$. Assuming that the second argument$(q \mif \temv{false})$ holds, the result is the first argument $p$. 
%% Again, the only requirement is that $p$ and $q$ must be complementary---an order-agnostic relation
% whose symmetry is explicitly underscored in the prompt. 
%% The prompt also contains ICL examples of proofs featuring applications of \mtt{by-contradiction}
%% where the two arguments are of the form $\neg r$ and $(r \mif \temv{false})$, as well as applications where the two arguments are of the form 
%% $r$ and $(\neg r \mif \temv{false})$, which should further reinforce the inherent symmetry of the notion.
This is not a subtle notion, yet somehow all models end up getting it wrong at some point 
(some more often than others), frequently confusing it with negation, which is intrinsically asymmetric. 
%% Here is a typical example by Grok-3 (item 15): 
%% \begin{quote}
%% {\footnotesize\em 
%% \mtt{dsyl} is applied to \mtt{premise-1} \mtt{((A\ls==>\ls\ntildels{}B)\ls|\ls(\ntildels(C\ls==>\ls{}B)))}
%% and \mtt{C\_implies\_B} \mtt{(C\ls==>\ls{}B)}, but \mtt{(C\ls==>\ls{}B)} is not the complement of \mtt{(\ntildels(C\ls==>\ls{}B))} 
%% as required by \mtt{dsyl}. 
%% The complement of \mtt{(\ntildels(C\ls==>\ls{}B))}  would be \mtt{(C\ls==>\ls{}B)} itself, but \mtt{dsyl} requires the 
%% second argument to be the negation of one of the disjuncts in the first argument.
%% }
%% \end{quote}
%% First, the use of the definite description \emph{the complement of} suggests a functional understanding of 
%% complementation, contrary to the explicitly relational formulation given in the prompt. At any rate, the model
%% contradicts itself when it claims that ``{\em \mtt{(C\ls==>\ls{}B)} is not the complement of \mtt{(\ntildels(C\ls==>\ls{}B))}}''
%% while in the next breath it says ``{\em the complement of \mtt{(\ntildels(C\ls==>\ls{}B))}  would be \mtt{(C\ls==>\ls{}B)} itself}.''
%% It ends by wrongly claiming that \mtt{dsyl} ``requires its second argument to be the negation of one of the disjuncts 
%% in the first argument'' when the description of \mtt{dsyl} in the prompt only requires complementarity.
Here is an example from Grok-4 on item 56: 
\begin{quote}
{\footnotesize\em 
\fmtt{dsyl} requires the second argument to complement one of the disjuncts in the first argument syntactically, but \fmtt{D} 
does not complement \fmtt{(\ntildels D)} or \fmtt{(C\ls==>\ls{}A)}. 
The syntactic complement of \fmtt{(\ntildels D)} is 
\fmtt{(\ntildels(\ntildels D))}, which is not identical to \fmtt{D}.
}
\end{quote}
But if ``the syntactic complement of  \mtt{(\ntildels D)} is 
\mtt{(\ntildels(\ntildels D))},'' then the syntactic complement of 
\mtt{D} must be \mtt{(\ntildels{}D)}. And if complementation is symmetric,
as stated in the prompt, then \mtt{D} also complements \mtt{(\ntildels{}D)}.
The rudimentary error here is to view complementation as identical to negation.
%which is a function rather than a relation---an operation that takes a formula $p$ and produces $(\ntildels{}p)$. 

As another example, here is part of the error description that GPT-5 gives for item 184: 
\begin{quote}
{\footnotesize\em 
\fmtt{dsyl}requires arguments of the form \fmtt{($p$\ls|\ls$q$)} and $r$ where $r$ is complementary to either $p$ or $q$. 
Here \fmtt{disj1} is 
\fmtt{(A\ls|\ntildels{}B)} and the second argument is \fmtt{B}, which is not complementary to either \fmtt{A} or \fmtt{\ntildels{}B}. 
Therefore the application of \fmtt{dsyl} is invalid. 
}
\end{quote}
%% Likewise, on item 198 GPT-5 writes: 
%% \begin{quote}
%% {\footnotesize\em 
%% \fmtt{by-contradiction} requires the second argument to be a conditional whose antecedent is the negation of the first argument. 
%% Here the desired conclusion is \fmtt{(\ntildels\ntildels{}A)}, so the needed conditional is 
%% \fmtt{(\ntildels(\ntildels\ntildels{}A)\ls==>\ls{}false)}, i.e., \fmtt{(\ntildels\ntildels\ntildels{}A\ls==>\ls{}false)}. 
%% The provided conditional is \fmtt{(\ntildels{}A\ls==>\ls{}false)}, 
%% whose antecedent is not complementary to \fmtt{(\ntildels\ntildels{}A)}. 
%% Therefore this application of \fmtt{by-contradiction} is invalid.
%% }
%% \end{quote}
%% e.g. Gemini-2.5-Pro (item 158) writes:
%% \begin{quote}
%% {\footnotesize\em 
%% $\cdots$ The second argument \fmtt{(B\ls|\ls\ntildels{}D)} is not the syntactic complement of the right disjunct 
%% \fmtt{(\ntildels(B\ls|\ls\ntildels{}D))}. The complement of 
%% \fmtt{(\ntildels(B\ls|\ls\ntildels{}D))}
%% would be \fmtt{(\ntildels(\ntildels(B\ls|\ls\ntildels{}D)))}. 
%% }
%% \end{quote}
Similar mistakes are common, \egnsp, GPT-5.2 says the following regarding item 36:
\begin{quote}
{\footnotesize\em 
This application of \fmtt{dsyl} is invalid. The disjunction 
\fmtt{notA\_or\_notnotB} is \fmtt{((\ntildels{}A)\ls|\ls(\ntildels(\ntildels{}B)))}. 
Disjunctive syllogism requires the second argument to complement (i.e., 
be the syntactic negation of) one of the disjuncts. But 
\fmtt{not\_B} is \fmtt{(\ntildels{}B)}, 
which does not complement either \fmtt{(\ntildels{}A)} 
or \fmtt{(\ntildels(\ntildels{}B))} 
(the complement of \fmtt{(\ntildels(\ntildels{}B))} 
would be \fmtt{(\ntildels(\ntildels(\ntildels{}B)))}). 
Therefore \fmtt{dsyl} cannot be applied here, and the conclusion 
\fmtt{(\ntildels{}A)} is not derivable by this step.
}
\end{quote}
Likewise, on problem 69 it states that \emph{\mtt{(\ntildels{}B)} does not complement \mtt{(\ntildels\ntildels{}B)}},
and on problem 198 it stresses that \emph{\mtt{(\ntildels{}A)} is not complementary to \mtt{(\ntildels\ntildels{}A)}}.

\item The most noteworthy and concerning theme that emerges from a close inspection of the data
is a recurring failure of models to exhibit judgment consistency. A model might flag something as an error in one proof and 
\label{Pg:RCIFirst} disregard the exact same issue in another proof. We refer to this broad phenomenon as 
{\em epistemic instability}. It has many manifestations, with one of the most striking instances emerging from the interaction between 
proof writing and proof checking. For example, a model might write a correct proof $D$ for a given problem, 
but when we later ask it  independently to verify $D$, it incorrectly rejects it as wrong. This is 
clearly problematic, suggesting a sort of split-brain syndrome whereby the model writes proofs with one brain and checks
them with another. 

For a human reasoner, it would be odd 
if a person who is both rational and truthful would write a correct 
proof and then reject that same proof, especially if this behavioral 
cycle repeated itself. After writing their proof for the first time, 
they might perhaps discover what they (incorrectly) believe to be a bug. 
But if so, they would not then go on to rewrite the same proof with 
the same alleged bug. 
%% They might perhaps discover a bug in their original proof, but 
%% if they are then  asked to repeat the proof-writing exercise they will incorporate the fix and if they are then asked
%% to check {\em that\/} proof, they will find it correct; 
In other words, even if a human reasoner exhibited an initial
inconsistency between the two tasks, they would 
quickly reach an equilibrium. Crucially, this equilibrium is independent 
of the proof's objective correctness; it is instead determined 
by the human reasoner's subjective cognitive state. 
%% The proof might still be wrong, even profoundly so, but the human reasoner judges it to be 
%% correct because they wrote it and it conforms with what they believe to be all relevant constraints governing 
%% proof behavior. 
Unless they were trying to be deceptive, their proof-writing and proof-checking stances 
would align. Indeed, it would be a sign of incoherence if they did not. 
% consistently wrote $D$ and then consistently rejected $D$. 
But we found such unstable cross-task behavior to be a common motif with AI models. 

We plan to revisit model behavior at the interplay between proof writing and proof checking
in forthcoming work. However, the phenomenon is not limited to the interaction between those two tasks. 
A careful analysis shows that it is frequently seen even within \mbf{PL1-PC} in isolation. 
An example is given by the aforementioned issue with complementation. For every time that a model
flags a spurious error by demanding syntactic negation instead of complementation, there are more times 
where it does {\em not\/} flag instances of the exact same pattern, seemingly happy to accept complementation
instead of strict syntactic negation. This indicates that the model has no stable epistemic commitment 
on the issue and no coherent internal principles governing the distinction between the two. 

For concrete examples, recall that on multiple occasions GPT-5.2 insists on conflating complementation with syntactic negation.
Yet many other times it does the opposite. For instance, on item 24, it condones an application of 
\mtt{dsyl} to $(\neg A \mor \neg C)$ and $C$, even though, according to its own logic, 
``disjunctive syllogism requires the second argument to complement (i.e., be the syntactic negation of) 
one of the disjuncts'' and $C$ is obviously not complementary to $\neg C$ in that sense. Likewise, on item 37, 
GPT-5.2 happily accepts an application of \mtt{dsyl} to $(\neg\neg C \mor \neg B)$ and $B$. 
On item 49 it accepts an application of \mtt{dsyl} to $((\neg A \mif C) \mor \neg(D \mand E))$
and $(D \mand E)$. On item 15 it accepts an application of \mtt{dsyl} to $((A \mif \neg B) \mor \neg (C \mif B))$
and $(C \mif B)$. There are other examples. On item 299 it delivers a lecture on how $B$
\emph{does} complement $\neg B$, even though $B$ is clearly not the syntactic negation of $\neg B$. 
 
As another example, GPT-5 rightly complains several times about applications of \mtt{dm} (De Morgan) 
that skip a double negation step. For instance, on item 171 it writes:
\begin{quote}
{\em 
Applying \mtt{dm} to 
\mtt{(\ntildels(\ntildels{}A\ls\&\ls{}C))} should yield 
\mtt{(\ntildels\ntildels{}A\ls|\ls\ntildels{}C)}, 
not \mtt{(A\ls|\ls\ntildels{}C)}. The step's \skwd{BY} annotation claims a conclusion that is not 
the direct result of the \mtt{dm} rule (it eliminates the double negation without a separate \mtt{dn} step).
}
\end{quote}
GPT-5 flags the same error elsewhere (\egnsp, items 143 and 210). Yet on item 298 it happily accepts the following
fragment: 
\begin{tcAthena}
 ... 
 assume h3 := (~ (~B & A)) {
   (B | ~A) BY dm on h3;
   ...
 }
... 
\end{tcAthena}
even though this \mtt{dm} application is structurally identical to (and makes the exact same error as) the one in item 171. 

Here is another example: As we pointed out earlier (p.~\ref{Pg:InvalidAssumptions}), 
LLM-written proofs often write conditional proofs with syntactically malformed assumptions, 
along the lines of 
 \mtt{\kwd{assume}\msp{}\mbox{\rm\temv{arbitrary-identifier}}$\;$\{$\cdots$\}}. In \mbf{PL1-PC}, strong models often 
detect this error and explain it quite well. For instance, here is what GPT-5 says about the proof of item 140:
\begin{quote}
{\em 
Malformed \mtt{assume} statement. A conditional proof must have the form 
\[
\mtt{assume}\msp p \msp\{\msp\cdots\msp\}
\]
or 
\[
\mtt{assume}\msp I \mtt{:=}\; p \msp\{\msp\cdots\msp\}
\]
where $p$ is a formula. Here, \mtt{assume} \mtt{B\_hyp} \{ 
lacks a hypothesis formula (and \mtt{B\_hyp} is not defined as a formula), 
so the step cannot be parsed and the intended conditional \mtt{(B\ls==>\ls{}C)} is not formed.
}
\end{quote}
But on item 281 GPT-5 says nothing about the conditional subproof 
\begin{tcAthena}
assume B_hyp { C BY claim on C; };
\end{tcAthena}
even though the assumption is malformed in exactly the same way as the one that GPT-5 complained about in item 140. 
The same pattern can be observed across all systems; we use examples from GPT-5 and GPT-5.2 here for illustration purposes.
%% Here's one from Gemini-2.5-Pro: on item 50, Gemini rightly complains about this line: 
%% \begin{tcAthena}
%% B1 BY mp on premise-2, A;
%% \end{tcAthena}
%% writing the following in its error description:
%% \begin{quote}
%% {\em 
%% This line attempts to name the result of a rule application. According to 
%% the specification, a named primitive proof must use the \mtt{:=} operator, taking one of 
%% two forms: ``$I$ \mtt{:=} $R$ \mtt{on} $\cdots$'' or ``$I$\msp\mtt{:=}\msp{}$p$\msp\mtt{BY}$\;R$\msp\mtt{on} $\cdots$''. 
%% The syntax ``$I$\msp\mtt{BY}$\;R$\msp\mtt{on} $\cdots$'' used here is not a valid NDL construct.
%% }
%% \end{quote}
%% This seems to indicate a rock-solid understanding of the relevant rules. But on many other proofs (\egnsp,
%% those in items 63, 82, 129, 132, 155, 180, and others) Gemini-2.5-Pro completely disregards rule applications 
%% that make the exact same error, such as this one (item 63):
%% \begin{tcAthena}
%% ...
%% B_and_BtoA BY mp on premise-2, (A | C);
%% ...
%% \end{tcAthena}

These inconsistencies are akin to a teacher grading student assignments who sometimes marks a particular step 
as an error (``you can't use double negation here'') and other times gives full credit for the very same step. 
The problem isn't the strictness (or even the fairness) of the grading, but the {\em judgment instability}: 
the teacher's evaluative criteria drift from one paper to the next, suggesting they have no stable conception 
of what the rules actually mean.

\eit

It is possible that these issues will be resolved in the near future, although the 
pervasive judgment inconsistencies point to deeper problems of epistemic coherence;
we will return to this topic in Section~\ref{Sec:PL3RCI}. 
%% Indeed, by the time this paper is published some 
%% models may have already addressed them. Nonetheless, it is noteworthy that state-of-the-art LRMs at this point in time 
%% continue to make such elementary errors. But more importantly, our aim in making these qualitative observations is not to catalogue
Our main aim in this section is not to catalog a particular set of failure modes or quirks of these specific model versions, 
which are likely to be transient, but to illustrate the sort of reasoning pathologies that tasks like \mbf{PL1-PC} 
can expose. These clearly demonstrate the need for richer, non-scalar evaluation signals for assessing reasoning competence. 
%By revealing that models can be ``almost right'' yet cognitively incoherent, tasks like \mbf{PL1-PC} serve not just as correctness benchmarks but also as probes into reasoning structure.

\subsection*{A challenge version of \mbf{PL1-PC}}
\label{Sec:ChallengePC}
In this section we present results for \mbf{PL1-PC-c}, a more challenging proof-checking dataset consisting of 50 longer proofs. 
We only computed results for the 4 currently strongest models, as the rest are very unlikely to perform better than
random guessing. Even among these 4 top-tier models, only two of them manage to perform better than flipping a coin on the 
guess-only version of the task. These 50 proofs have lengths ranging from 720 to 987, with average and median lengths 
of 839 and 823, respectively. The proofs were generated by our own hand-rolled theorem prover and were verifiably correct.
We then randomly and minimally corrupted a small number of them (by injecting exactly one error in 9 of the 50 proofs).  
The proofs are challenging to verify because they have a lot 
of conditional  subproofs (215.5 per proof on average), with a high degree of nesting. 
The distribution of nesting depths is shown below. \\
\begin{center}
\scalebox{0.7}{\includegraphics[width=\linewidth]{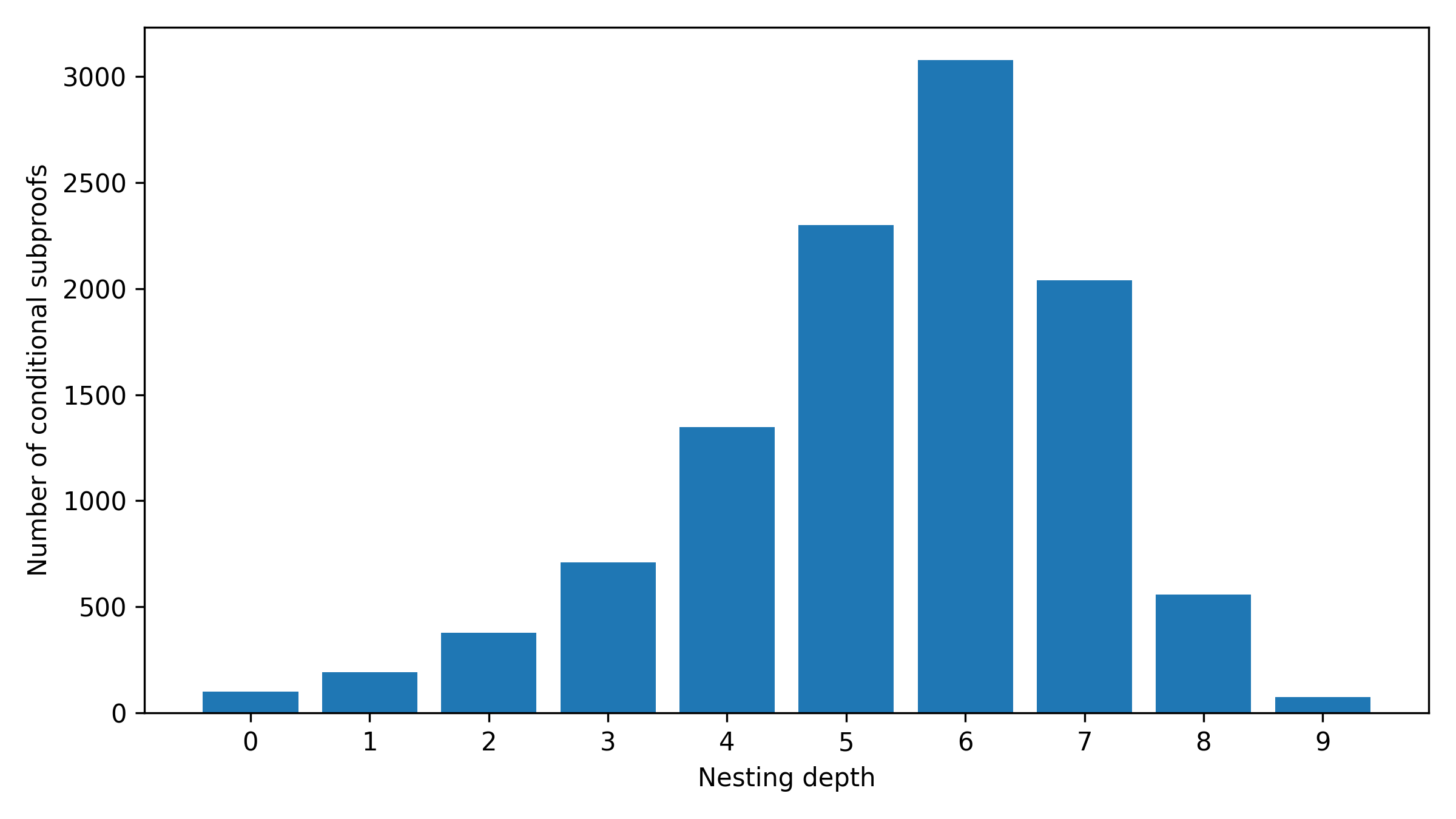}}
\end{center}
\mbox{}\\[-0.1in]
The figure shows that the mass of conditional subproofs rises steadily and peaks around depth 6, 
then falls off fairly quickly after 7. So a typical subproof in this challenge corpus is 
nested fairly deeply (at an average depth of 5.3).

Results for the 4 models are shown below. \\
\renewcommand{\arraystretch}{1.5}
\begin{table}[!bhtp]
\centering
\rowcolors{2}{blue!5}{white}
\resizebox{1.0\textwidth}{!}{%
\begin{tabular}{|l|c|c||l|c|c|}
\hline
\textbf{Model} & \textbf{Full Task} & \textbf{Guessing Only}   & \textbf{Model} &  \textbf{Full Task} & \textbf{Guessing Only} \\
\hline
\claudeOpusFourFiveNov & \mbf{0.65} \ci{0.13} & 0.69 \ci{0.12} &
\gptFiveTwo & 0.64 \ci{0.13} & 0.66 \ci{0.13} \\ \hline
\gemThreeOne & 0.48 \ci{0.13} & 0.52 \ci{0.13} &
\gptFiveFour & 0.44 \ci{0.13} & 0.48 \ci{0.13} \\ \hline
\end{tabular}}
\caption{\mbf{PL1-PC-c} accuracies, for the full task and for simple guessing.}
\label{Tb:Pl1PCAcc}
\end{table}
\renewcommand{\arraystretch}{1.5} 
\mbox{}\\[-0.05in]
All models perform somewhat better in the guessing-only version of the task, but the deltas are marginal. 
This implies that the main difficulty for these models is deciding correctness in the first place, 
not localizing and explaining errors. 
Restricting attention to the guessing-only task variant, even the top-performing model, Claude Opus 4.5, 
performs only 38\% better (69-50/50) than random guessing.\footnote{Flipping a coin still guarantees 50\% accuracy, even 
though the set is imbalanced with 41 correct proofs and 9 incorrect ones.} It does worse compared
to the majority-guess baseline: because 41 out of the 50 proofs are correct, always guessing ``correct''
guarantees an accuracy of 82\%, considerably higher than 69\%, which suggests that the model is not exploiting
the class imbalance. Indeed, all four models are below the majority baseline. If they were even 
weakly exploiting signal in the proofs, they should at least conclude that most proofs are correct. 
The fact that they don't suggests that they are not reliably extracting the right structural cues. 
Remarkably, both Gemini 3.1 and GPT-5.4 fail to beat random guessing. 

Perhaps the clearest takeaway from these results is that proof-checking/debugging ability is an orthogonal 
capability axis. While Claude Opus 4.5  places last among these 4 elite models in generative reasoning tasks like 
proof unmasking and gap filling (see Sections~\ref{Sec:ChallengeMasking} and~\ref{Sec:ChallengeGaps}),
it takes first place in proof checking. Likewise, even though GPT-5.4 outperforms GPT-5.2 in those tasks,
here it trails behind by 20 absolute points, reinforcing the point that better generation does not translate
to better verification.

\section{Proof Masking}
\label{Sec:PL1PM}

In this section we introduce a novel proof-completion task. Given a proof, we mask any of the following four types of items:
\ben 
\item conclusions (formulas immediately preceding the \skwd{BY} keyword); 
\item inference rules (immediately preceding \skwd{on}); 
\item assumptions (immediately following \skwd{assume}); and
\item one or more arguments to an inference rule application (immediately following \skwd{on}). 
\een 
Masks are unique identifiers: \mtt{MASK1}, \mtt{MASK2}, and so on (though the indices need not be consecutive). 
We only mask formulas at the top level, not any proper subformulas (\egnsp, we are never left with something like \smtt{(\ntildels MASK8$\:$|$\:$C)}). 

The objective is to determine if there is a way to unmask the proof so that it correctly derives the target conclusion from the given premises;
and if so, to provide a detailed assignment of appropriate values to the masks. As an example, consider again this proof:
\begin{tcAthenaWithNumbers}
assert premise-1 := (A ==> B)
assert premise-2 := (~ A ==> C)
assert premise-3 := (C ==> D)

# We prove (B | D) by a case analysis on (A | ~ A), which holds by the law of the excluded middle.  
{
  (A | ~ A) BY ex-middle on A;

  # Case 1: Show that A implies (B | D)
  assume A {
    # Applying modus ponens to premise-1 and the assumption A gives B:  
    B BY mp on premise-1, A;
    # So now (B | D) follows by disjunction introduction: 
    (B | D) by left-either on B, D
  };

  # Case 2: Show that (~ A)  also implies (B | D) 
  assume (~ A) {
    C BY mp on premise-2, (~ A);
    D BY mp on premise-3, C;
    (B | D) BY right-either on B, D 
  };

  # The case analysis is now complete: 
  (B | D) BY cases on (A | ~ A), (A ==> B | D), (~A ==> B | D)
}
\end{tcAthenaWithNumbers}

Here is a masked version of this proof (after removing comments):
\begin{tcAthenaWithNumbers}
assert premise-1 := (A ==> B)
assert premise-2 := (~ A ==> C)
assert premise-3 := (C ==> D)
# Goal: (B | D)

{
  (A | ~ A) BY ex-middle on MASK1;
  assume A {
    MASK2 BY MASK3 on premise-1, MASK4;
    (MASK5 | D) by left-either on MASK6, MASK7; 
  };
  assume MASK8 {
    MASK9 BY mp on premise-2, MASK10;
    D BY mp on MASK11, MASK12;
    MASK13 BY right-either on B, D;
  };
  MASK14 BY cases on (A | ~ A), MASK15, MASK16
}
\end{tcAthenaWithNumbers}
With the original proof in sight, it is of course straightforward to generate a correct mask assignment: 
\smtt{MASK1}:$\,$\smtt{A}; \smtt{MASK2}:$\,$\smtt{B}; \smtt{MASK3}:$\,$\smtt{mp}; \smtt{MASK4}:$\,$\smtt{A}; and so on.  
But without the benefit of having seen the proof before, this is not trivial. 
It can be loosely likened to a constraint satisfaction problem (CSP), where the variables 
are the masks and the constraints are given by the syntax rules as well as the operational 
semantics of the proof language, including scope rules (for names as well as assumptions and intermediate conclusions).  
For instance:
\bit
\item We can infer that \smtt{MASK1} must be \smtt{A} from the specification of the excluded-middle
rule and the fact that we see the conclusion \smtt{(A$\;$|\ntildels A)} to the left of \skwd{BY}. 
\item On the last line, because the goal is \smtt{(B$\;$|\ntildels D)} and the 
pivot disjunction is \smtt{(A$\;$|\ntildels A)}, we can infer from the specification of \mtt{cases} that \smtt{MASK15} and \smtt{MASK16} must be conditionals of the form 
\smtt{(A ==>$\;$(B$\;$|$\;$D))} and \smtt{(\ntildels A$\;$==>$\;$(B$\;$|$\;$D))}, respectively. 
\item Knowing that the high-level proof is a case analysis on \smtt{(A$\;$|\ntildels A)}, we can infer that the conditional subproof on lines 12--16 must 
postulate \smtt{(\ntildels A)} as an assumption, \iensp, that \smtt{MASK8} must be \smtt{(\ntildels A)}. 
\eit
When a proof is heavily obscured, the problem generally becomes more complex. In this dataset, we randomly 
masked anywhere from 30\% to 90\% of all candidates (of all 4 types listed above) in a given proof.

Although this task is superficially similar to a CSP, it diverges in two crucial respects.
In a CSP, constraints are explicitly given as part of the problem: each constraint 
pertains to a specific list of  variables and allows a specific set of assignments (tuples of values). 
By contrast, the constraints governing valid mask assignments are implicit, heterogeneous, 
and distributed across the proof's syntactic and semantic structure. Indeed, identifying 
these constraints is a key part of the problem. The main source of difficulty is not 
combinatorial solution multiplicity but delayed constraint revelation and the 
need to maintain a long sequence of correct commitments.
The task thus occupies a regime where symbolic reasoning is unavoidable but 
the relevant symbolic structure is not explicitly given. This distinguishes it 
from puzzles where the main difficulty lies in combinatorial search rather than 
in extracting the relevant constraints from context and the operational semantics of proofs. 
When models fail to unmask a proof, it is not simply because the 
search space is large, but because the task requires jointly inferring structure, 
constraints, and solutions under global consistency requirements. This reflects 
deeper limitations than simply ``models can't do backtracking.'' 
The second major difference is that the set of values is not explicitly given either. 
It is not even clear that this set is finite. We know that mask 
values are either formulas or  rules, and that the former are likely to be
subformulas of the premises and/or the goal, of which there is only a finite number. 
But the subformula property is not guaranteed, since we know nothing about the original
proof, which could have arbitrary detours above and beyond the usual and 
relatively innocuous detours introduced by classical logic (double negations and
the use of \smtt{false}). A fortunate corollary of these differences is that 
\mbf{PL1-PM} instances cannot be straightforwardly encoded as CSP or SAT problems and 
outsourced to an external solver. Nor is it easy for a model to solve the task
just by writing bespoke code for it. Even if the model could write flawless 
infrastructure for the problem (e.g., a parser for masked NDL proofs), an unmasking 
algorithm would still be far from obvious.

%% it does \emph{not} present an explicit constraint system that 
%% can be directly encoded and solved by a standard algorithm. This is in contrast to many popular reasoning benchmarks 
%% (Sudoku, Hanoi towers, SAT, graph coloring, and so on) that can be trivially encoded as CSPs and outsourced to a solver. 
%% In this task, the constraints governing valid mask assignments are implicit, heterogeneous, and distributed across the proof's
%% syntactic and semantic structure. For that reason, the task resists delegation to external solvers. 
%% Nor is it easy for a model to solve the 
%% task simply by writing bespoke code for it. Even formulating the relevant constraints 
%% requires recovering proof-theoretic structure that is itself partially masked. 
%% In other words, identifying the constraints is part of the problem in \mbf{PL1-PM}.
%% The task thus occupies a regime where symbolic reasoning is unavoidable but the relevant 
%% symbolic structure is not explicitly given. This distinguishes it 
%% from puzzles where the main difficulty lies in search rather than in extracting the governing rules from context.
%% When models fail to unmask a proof, it is not merely because the search space is large, 
%% but because 
%% the task requires jointly inferring structure, constraints, and solutions under global consistency requirements.
%% This reflects deeper limitations than simply ``models can't do backtracking.'' 

%% qqq

However, even with a large number of masks sprinkled over a large proof, this is a relatively easy task 
as long as mask \emph{density} is not excessive (see below for a more precise definition of density). 
First, because NDL is strongly structured, 
the expected category (or ``type'') of a mask is fixed: it is either \emph{formula} or \emph{rule}, and it is determined 
by the position of the mask variable: if the mask occurs immediately before a \skwd{BY} or immediately after an \skwd{assume}, 
it must be a formula; if it occurs immediately before an \skwd{on}, it must be a \emph{rule}; and so on. 
Second, all masks are unique, so there are no cross-references and unification across multiple proof lines is not required. 
Finally, most masks are functionally determined and have unique/canonical values that can be locally inferred. 
Accordingly, putting aside pathological cases such as proof with several steps that are completely masked, 
there is no combinatorial explosion of solutions, no need to keep track of large equivalence classes of assignments,
and no true symmetry-breaking issues (aside from rare commutations like example~\ref{Eq:PMCommExample} below).
This means that values for most masks can be inferred incrementally, on a line-by-line basis. 
Some choices are, of course, semantically coupled across the entire proof tree. For example, 
choosing the conclusion of an excluded middle application constrains the shape 
of the subsequent case analysis and thus the assumptions and conclusions of the relevant conditional subproofs and the arguments of the corresponding \smtt{cases} application.
But these tend to be light inter-dependencies that become obvious after a single pass over the proof tree,
which typically suffices because the constraints are acyclic (determined by the linear order of the proof)\footnote{A 
proof step depends only on information established in earlier steps (such as previously derived conclusions, introduced premises, or assumptions) 
and never on content that appears later. This means that the constraint graph over masks is a DAG (directed acyclic graph), 
allowing the proof to be reconstructed or verified in a single forward pass without backtracking. Once a line has been correctly 
unmasked, it can be referenced but need not be altered. In that sense, the task 
is topologically sorted by proof order: solving or checking constraints proceeds deterministically from top to bottom.}
%% The absence of cycles helps to explain why masks can be filled in with relative efficiency in a quasi-algorithmic 
%% top-down manner, whereas language models, which are processing  the same proof as a flat sequence rather than 
%% as a structured DAG, must infer these dependencies implicitly, which makes the task harder for them despite its procedural simplicity.}
and nearly deterministic.
%%  meaning that the underlying
%% CSP is almost functional: the value of each mask tends to be unique given the preceding context (see below for exceptions). 
%% For a language model that sees a linearized string, however,
%% these non-local constraints are harder to solve because the model must implicitly infer tree structure from indentation and keywords, and because valid completions depend 
%% on latent and non-adjacent tokens. 

\begin{figure}[!hbtp]
\centering
\scalebox{1.0}{\includegraphics[width=\linewidth]{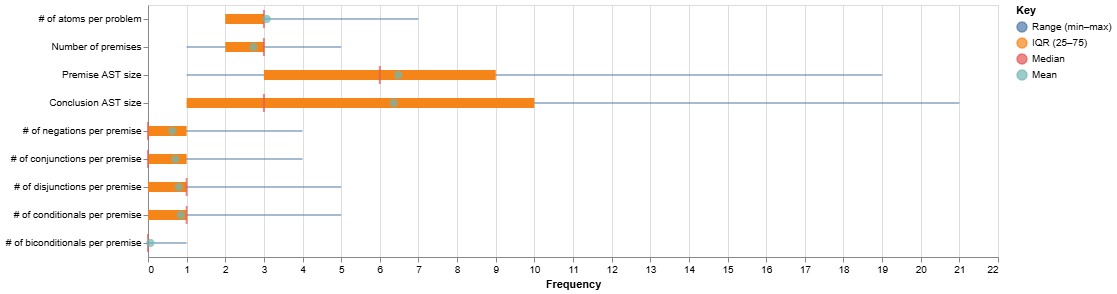}}
\caption{\mbf{PL1-PM} problem statistics}
\label{Fig:PL1MaskingProblemStats}
\end{figure}

The \mbf{PL1-PM} task slice contains 300 proofs that have been masked as described above. All 300 proofs are correct, containing no errors of any kind. 
80 of these proofs are conditionalized, meaning that the relevant premises are explicitly assumed as hypotheses (instead of taken as global premises). 
Statistics about the corresponding problems are shown in Fig.~\ref{Fig:PL1MaskingProblemStats}. Note that these are ``easy'' problems along pretty much
all dimensions: the total number of premises is small (less than 3 on average) and, viewed as an AST, each premise has about 6 nodes on average. 
Conclusions have a median size of 3 nodes but the average is considerably higher (slightly over 6). The median number of propositional atoms in a problem is 3, 
and negations are the most rare logical connective save for biconditionals. 

%\blindtext

%\blindtext

\renewcommand{\arraystretch}{1.4}
Statistics about the length of these proofs are shown in the following table. Note that these proofs are even shorter than those in \mbf{PL1-PC},
whose median length is 16\% greater. 90\% of the \mbf{PL1-PM} proofs have fewer than 50 lines. 
\par\medskip            % end the current paragraph
\noindent
\begin{minipage}{6in}
\begin{center}
%\begin{table}[h!]
\begin{tabular}{c}
\hspace*{0.3in}
\begin{tabular}{|r|c|c|c|c|c|c|c|c|c|c|c|}
\hline
 & {\bf Mean} & {\bf Median} & {\bf Min} &{\bf Max} & {\bf Std} & {\bf @5} & {\bf @25} & {\bf @50} & {\bf @75} & {\bf @90} & {\bf @95} \\
\hline\hline
{\bf Proof length} & 26.81 & 25.0 & 3 & 64 & 14.12 & 6.0 & 15.0 & 25.0 & 38.0 & 47.0 & 50.0 \\ \hline 
\end{tabular}
\end{tabular}
\end{center}
%\caption{Summary statistics of proof lengths over 300 proofs.}
%\label{tab:pm_proof_lengths}
%\end{table} %\\[-0.3in]
\end{minipage}
\par
\begin{figure}[!bhtp]
\centering
\scalebox{1.0}{\includegraphics[width=\linewidth]{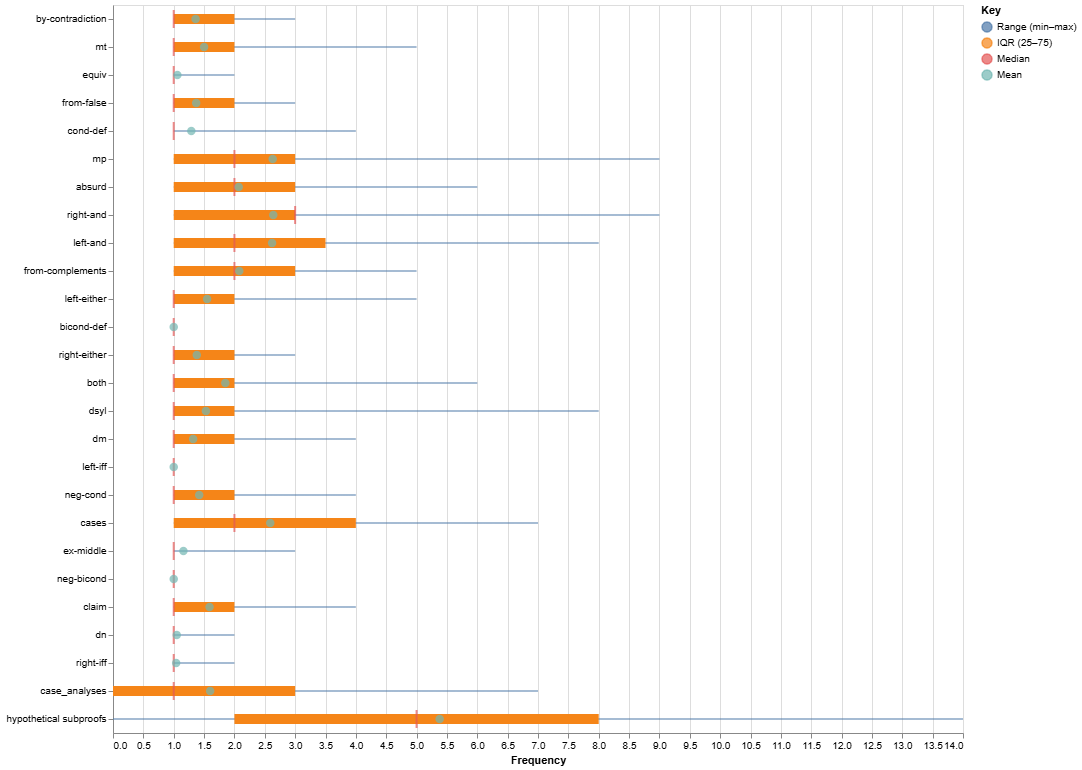}}
\caption{Inferential profile of  \mbf{PL1-PM} proofs.}
\label{Fig:PL1MaskingProofRuleStats}
\end{figure}

\medskip            % end the current paragraph
%\noindent 
\renewcommand{\arraystretch}{1.0}
More granular statistics about the distribution of inference rules and conditional reasoning in these proofs can be found in 
Fig.~\ref{Fig:PL1MaskingProofRuleStats}. Conditional (hypothetical) reasoning is very common in this corpus, largely 
because about half of these proofs are conditionalized and all such proofs start by explicitly making hypotheses. 

Given a masked proof and the underlying argument (premises and target conclusion), the model is asked to output a dictionary $M$ that maps
every mask in the input to an appropriate value:
\[
\mtt{\{MASK1:$\:\cdots$, $\ldots$ \}}
\]
in such a way that when the proof is unmasked in accordance with $M$ (i.e., by plugging in $M[m]$ for every mask $m$ that appears in the input),
the result is a complete proof that correctly derives the goal from the premises. The model is told to output the empty dictionary $M = \mtt{\{\}}$ 
if it believes that there is no valid assignment of values to masks that can successfully unmask the proof. This effectively acts as a distractor, since we know
that every masked proof  can be unmasked.

The key question addressed by this task is whether a model can  infer missing symbolic structure consistent with the rules that 
govern natural deduction. 
%High accuracy in this task would strongly suggest that the system possesses a robust internal model of logical reasoning that goes beyond surface-level pattern recognition or heuristic matching. 
Whereas proof writing and proof checking test a model's ability to construct or evaluate a line of reasoning, 
proof masking asks it to reconstruct elided reasoning details from partial information, inferring what intermediate conclusions, 
assumptions, or rule applications could make the surrounding proof context coherent. 
This task is superficially reminiscent of code infilling, where a model must recover a missing span of source code,
but the resemblance is limited. In code infilling the missing span typically forms a single contiguous ``hole''---a bulk 
fragment whose boundaries are arbitrary and syntactically defined, and whose recovery in practice depends mainly on 
local lexical and type constraints. By contrast, in proof masking the missing content is dispersed across dozens 
of micro-holes, each corresponding to a distinct atomic element of the proof (an individual conclusion, rule name, 
assumption, or argument to a rule application). Each mask encodes only a tiny unit of inferential information, yet all 
of these units must be filled jointly and coherently, because the value assigned to one mask might constrain 
the permissible values of subsequent masks. The model must therefore maintain a global, internally consistent hypothesis about 
the entire proof structure while simultaneously resolving many fine-grained local uncertainties.
Essentially, it must treat the proof itself as an object of reasoning, reasoning about reasoning rather than merely executing it.
This gives proof masking a distinct meta-reasoning flavor. A system that can perform well in this task 
demonstrates not only procedural knowledge of inferential mechanisms, but also a representational grasp of 
how those mechanisms fit together and a capacity for metalogical comprehension. 

%% This is somewhat similar but not quite analogous 
%% to conventional code infilling, which mainly exploits syntactic and type-level constraints. In proof masking the 
%% gaps are always small units of exactly one of four possible roles, which are  primarily semantic. To fill them correctly, 
%% the model must infer the unobserved inferential acts that would make the entire argument logically consistent and successful in 
%% deriving the target conclusion. 

%Accordingly, successful performance on \mbf{PL1-PM} reflects a model's , the ability to reason about reasoning. 

%% COMPOSITIONAL

As we already pointed out, most mask values are uniquely determined. 
In general, however, masking problems are underdetermined, \iensp, there 
may be multiple mask assignments that yield a correct proof. Accordingly,
it is not expected that a mask assignment produced by a model recovers the original proof. For instance, the original proof might contain a chain of two consecutive 
applications of left and right conjunction elimination: 
\begin{equation}
\mbox{$\cdots$ {\tt\footnotesize left-and \fkwd{on} (A\:\&\:B);} {\tt\footnotesize right-and \fkwd{on} (A \& B);} $\cdots$.}
\label{Eq:PMCommExample}
\end{equation}
But if  we mask both rules, \egnsp, by replacing \fmtt{left-and} by \fmtt{MASK-3} and \fmtt{right-and}  by \fmtt{MASK-4},
then \fmtt{MASK3}:$\,$\fmtt{left-and}; \fmtt{MASK4}:$\,$\fmtt{right-and} and \fmtt{MASK3}:$\,$\fmtt{right-and}; \fmtt{MASK4}:$\,$\fmtt{left-and} are both valid solutions,
because the two inference steps can be swapped without affecting the correctness of the subsequent reasoning. 
For that reason, LLM responses must be evaluated semantically rather than syntactically. That is, LLM responses are evaluated 
not by matching the specified mask values against a fixed ground truth, but by unmasking the proof according to those values and then 
using Athena to check that the resulting proof correctly derives the target. 
If a model claims that the problem is unsolvable, that response is always counted as incorrect, since we know by construction that every 
masked proof can be unmasked.

It may be interesting to note that while the reasoning required for this task appears to be purely deductive, 
it actually leans heavily on pragmatics and on inductively learned regularities about proof construction. 
In particular, many proof steps are not strictly (deductively) determined by the local context, but 
they \emph{are} determined when one takes into account how proofs tend to be written in practice. 
As a concrete example (from item 5 of the challenge dataset), suppose the proof has a biconditional 
premise of the form $(p_1 \miff p_2)$, say $((D \mif B) \miff (B \miff A))$. 
The first three steps of the masked proof are as follows:
\begin{tcAthenaWithNumbers}
   MASK2 BY MASK1 on ((D ==> B) <==> (B <==> A));
   p0 := ((B <==> A) ==> (D ==> B)) BY MASK3 on ((D ==> B) <==> (B <==> A));
   ((~ (D ==> B)) | (B <==> A)) BY MASK5 on MASK4;
\end{tcAthenaWithNumbers}
It is clear from line 2 that \mtt{MASK3} must be \mtt{right-iff}, as that is the only inference rule that could produce the given conclusion.
Now, in theory, line 1 remains underconstrained, in that \mtt{MASK1} could still be either \mtt{left-iff} or \mtt{right-iff}. 
This is because, in principle, the proof could contain a redundancy and go on to repeat the same inference 
on line 2; nothing in the logic forbids that.  \emph{Pragmatically}, 
however, this would be highly unorthodox. It's much more likely that line 1 detaches the left-to-right part of the biconditional premise and line 2 
detaches the right-to-left part. That's not a  quirk; it's an inductive regularity about proofs as social artifacts, 
not just as logical objects.
So a human solver would naturally commit to \mtt{MASK1} being \mtt{left-iff} after deducing from line 2 that \mtt{MASK3} 
must be \mtt{right-iff}.\footnote{Deductively, there is yet another possibility for the first line: line 1 could be 
an application of ``biconditional definition'' that derives $((p_1 \mand p_2) \mor (\neg p_1 \mand \neg p_2))$ from $(p_1 \miff p_2)$. 
But this would also be dispreferred by human subjects, because we implicitly know, for example, 
that proofs typically use a connective before expanding its definition, and that canonical pairs of eliminations 
(e.g., a \fmtt{left-iff} immediately followed by a \fmtt{right-iff}) are an extremely common inference pattern 
that is stylistically tight.}
This marks another point of divergence from CSP/SAT problems, which do not have any notion of pragmatics. 
%By contrast, in addition to pure deduction,  this task requires choosing ``the most natural'' 
The upshot is that proof reconstruction in this task, at least for humans, relies not just on pure deduction but on proof-writing conventions, stylistic regularities,
and general inductively derived pragmatic expectations about proof structure.

\renewcommand{\arraystretch}{1.5}
\begin{table}[!bhtp]
\centering
\rowcolors{2}{blue!5}{white}
\resizebox{1.0\textwidth}{!}{%
\begin{tabular}{|l|c|c||l|c|c|}
\hline
\textbf{Model} & \textbf{Instrumented Accuracy} & \textbf{Strict Accuracy} &
\textbf{Model} & \textbf{Instrumented Accuracy} & \textbf{Strict Accuracy} \\
\hline
\gemThreeOne & \mbf{0.98} \ci{0.02} & \mbf{0.97} \ci{0.02} &
\gemThree & 0.96 \ci{0.02} & 0.93 \ci{0.03} \\ \hline
\gptFiveTwo & 0.90 \ci{0.03} & 0.82 \ci{0.04} &
\claudeOpusFourFiveNov & 0.89 \ci{0.04} & 0.87 \ci{0.04} \\ \hline
\gptFiveOne & 0.89 \ci{0.04} & 0.84 \ci{0.04} &
\grokFour & 0.82 \ci{0.04} & 0.68 \ci{0.05} \\ \hline
\geminiTwoPointFive & 0.81 \ci{0.04} & 0.50 \ci{0.06} &
\gptFive & 0.79 \ci{0.05} & 0.64 \ci{0.05} \\ \hline
\claudeSonnetExt & 0.79 \ci{0.05} & 0.70 \ci{0.05} &
\oThree & 0.75 \ci{0.05} & 0.69 \ci{0.05} \\ \hline
\dsReasoner & 0.60 \ci{0.06} & 0.53 \ci{0.06} &
\dsROne & 0.54 \ci{0.06} & 0.37 \ci{0.05} \\ \hline
\oFourMini & 0.49 \ci{0.06} & 0.40 \ci{0.06} &
\opus & 0.40 \ci{0.06} & 0.30 \ci{0.05} \\ \hline
\oThreeMini & 0.31 \ci{0.05} & 0.16 \ci{0.04} &
\claudeSonnetFour & 0.29 \ci{0.05} & 0.23 \ci{0.05} \\ \hline
\dsThreeOne & 0.25 \ci{0.05} & 0.19 \ci{0.04} &
\grokThree & 0.19 \ci{0.04} & 0.13 \ci{0.04} \\ \hline
\dsThree & 0.17 \ci{0.04} & 0.12 \ci{0.04} &
\hermes & 0.15 \ci{0.04} & 0.10 \ci{0.03} \\ \hline
\nous & 0.13 \ci{0.04} & 0.07 \ci{0.03} &
\gptFourOne & 0.09 \ci{0.03} & 0.06 \ci{0.03} \\ \hline
\magistral & 0.06 \ci{0.03} & 0.05 \ci{0.03} &
\gptFourO & 0.05 \ci{0.03} & 0.04 \ci{0.02} \\ \hline
\end{tabular}}
\caption{\mbf{PL1-PM} accuracies, using both the instrumented NDL checker and the regular proof checker.}
\label{Tb:Pl1PMAccuracies}
\end{table}
\renewcommand{\arraystretch}{1.0}

Results are shown in Table~\ref{Tb:Pl1PMAccuracies}, both under the instrumented proof checker 
and the conventional one. Without the benefit of instrumentation, 
performance tends to drop (significantly in many cases), though the vast majority of these 
errors are uninteresting and rather innocuous, such as omitting an outer pair of parentheses from formulas. 

On the basis of the instrumented accuracies of Table~\ref{Tb:Pl1PMAccuracies}, the discriminability index (DI) of \mbf{PL1-PM} is 0.38. 
As discussed in Section~\ref{Sec:DiscriminationIndex}, this means that, on average, any two models differ in accuracy 
by about 38 percentage points---a very large spread showing that the task  teases apart models of different  ability levels
extraordinarily well. 

\begin{figure*}[thbp!]
  \centering
\scalebox{0.8}{\includegraphics[width=\textwidth]{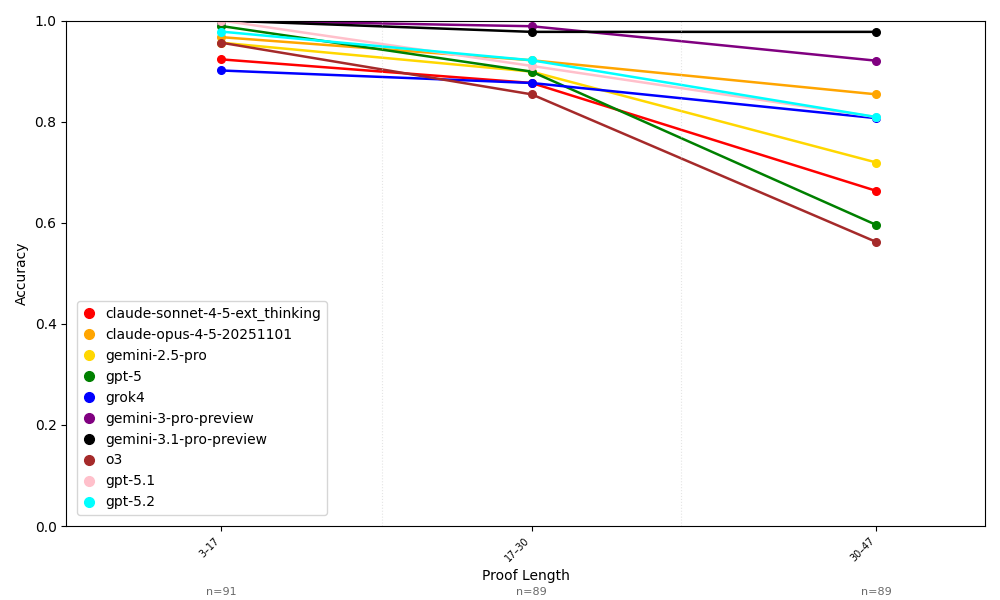}}
  \caption{\mbf{PL1-PM} accuracy vs proof length for the top 10 models; variable-width bins with $n \geq 25$.}
  \label{fig:pl1_pm_accuracy_vs_proof_len}
\end{figure*}

\begin{figure*}[b!]
  \centering
\scalebox{0.80}{\includegraphics[width=\textwidth]{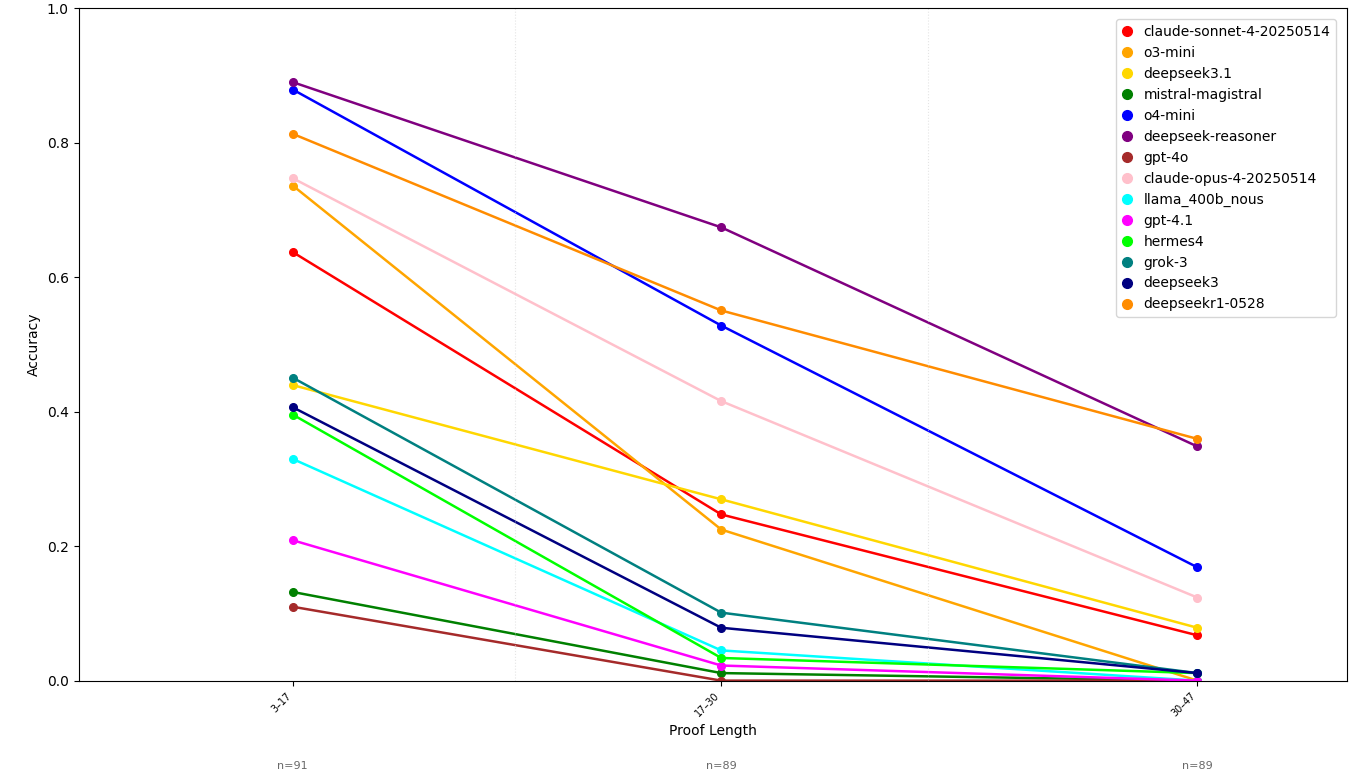}}
  \caption{\mbf{PL1-PM} accuracy vs proof length for the bottom 14 models; variable-width bins with $n \geq 25$.}
  \label{fig:pl1_pm_accuracy_vs_proof_len_tier2}
\end{figure*}

While the leading 3 models attain impressively high scores on this task, 
it should be kept in mind that the proofs in this dataset are very short 
(the median length is 25 lines, while 95\% of the proofs have no more than 50 lines). 
Even with these very short proofs, Fig.~\ref{fig:pl1_pm_accuracy_vs_proof_len} 
shows that the performance for most of the top 10 models drops as proof length 
increases even modestly, although the drop is very gentle for the top 4 models
(and the performance of Gemini 3.1, in particular, is essentially flat). 
As shown in Fig.~\ref{fig:pl1_pm_accuracy_vs_proof_len_tier2}, the decline is much more dramatic for the 
remaining 14 models, whose accuracies effectively drop to zero by the time we get to 
proofs with a mere 30-35 steps, with only a couple of exceptions. 
%By the time proofs grow to 47-64 lines, the accuracy of o4-mini and DeepSeek-R1 essentially drops to zero, while even the elite 3-4 models fail to reach 50\% accuracy. 

%% Fig.~\ref{fig:pl1_pm_accuracy_vs_mask_counts} shows the accuracy of the top seven models as a function 
%% of the number of masked elements, which of course correlates closely with overall proof length. 
%% Accuracy declines sharply as the number of masks increases, even though the underlying problem remains algorithmically 
%% linear for a symbolic solver.  

This performance profile should be expected when correctness requires all masked positions to be 
filled simultaneously.\footnote{The word ``simultaneously'' is referring to the evaluation criterion, not to the
generation process. The models of course generate mask values autoregressively,  but the correctness of their outputs 
is evaluated over the entire set of masks jointly, with an output counting as correct only if all masks 
are given valid values that collectively yield a correct proof.  So even though generation is incremental, 
the reward landscape is globally conjunctive, with success requiring a perfect joint configuration of \emph{all} 
micro-decisions made by a model.}
If each individual prediction has probability $p$ of being correct, then the probability 
that the entire proof is correctly unmasked scales approximately as $p^M$, where $M$ is the number of masks.  
Thus, even small per-mask error rates compound multiplicatively into rapid instance-level degradation.  
In addition, longer proofs impose heavier burdens on transformer attention and retrieval, with the model having to track 
more assumptions, rule dependencies, and line references, while maintaining perfect syntactic fidelity.  
These factors make the task more fragile under sequence generation.
%The drop below 0.5 accuracy for proofs with roughly 50 masks illustrates how cumulative local uncertainty and long-range context interference combine to produce steep non-linear declines in overall success. 

%% \begin{figure*}[thbp!]
%%   \centering
%% \scalebox{1.0}{\includegraphics[width=\textwidth]{./plots/pl1_pm_acc_vs_mask_counts.png}}
%%   \caption{\mbf{PL1-PM} accuracy by mask number across the top 7 models; variable-width bins with $n \geq 25$.}
%%   \label{fig:pl1_pm_accuracy_vs_mask_counts}
%% \end{figure*}

A similar multiplicative-error-compounding argument could be made for proof checking, especially for
correct proofs: on the assumption that a proof step has an independent fixed probability $p$ of being correctly checked by a model,
proof-checking accuracy should decay exponentially with longer proofs even if the task is locally easy (has a high $p$). 
However, what we saw in \mbf{PL1-PC} is that accuracy either held steady or declined more gracefully as proof length increased. 
So why does proof-checking seem to scale better than proof masking? 
While the two tasks share some similarities (both are ``locally easy''), 
they also place different demands on a generative model.
Given an incorrect proof in \mbf{PL1-PC}, 
a model must report the first erroneous step, which can be viewed as a sequential test 
over steps $1,\dots,N$ with success probability 
\[
P(\temv{success} \mid j^*) = (1 - f)^{\,j^*-1}(1 - m),
\]
where we write $j^*$ for the location (line number) of the first error, and 
where $f$ is the false-positive rate on correct steps above $j^*$ (i.e., the probability
that the model will incorrectly mark such a correct step as erroneous) 
and $m$ is the miss rate on that first error. For strong models, and assuming we are 
dealing with average incorrect proofs that have not been adversarially selected,
$f$ and $m$ are small 
and approximately length-invariant, so the probability of success will be relatively high even for 
a long proof.\footnote{For instance, with $f = 0.001$ and $m = 0.01$ we get 
$P(\temv{success} \mid j^*) = 0.9$ even for $j^* = 100$.}
For \emph{correct} proofs, on the other hand, the proof-checking situation is mathematically similar 
to that in proof masking, with both tasks exhibiting multiplicative structure. In particular,
writing $P_c$ and $P_m$ for the \mbf{PL1-PC} and \mbf{PL1-PM} probabilities, respectively,  
we have $P_c(\temv{success}) = p^N$,
where $p = 1 - f$ is the probability that a step is correctly checked and $N$ is the size of the proof (number of steps). 
Likewise, $P_m(\temv{success}) = q^M$,
where $q$ is the probability of correctly decoding a single mask and $M$ is the total number of masks in the proof. 
While both expressions have the same form, exponential decay is highly sensitive to base rates, so small differences in those rates 
can translate into big differences in performance. For instance, if $p = 0.999$ (i.e., the error rate is $f = 0.1\%$ per step) and $q = 0.99$ (with a $1\%$ error rate per mask),
then for $N = M = 60$ we have $P_c(\temv{success}) = 0.94$, whereas $P_m(\temv{success}) = 0.54$. 
Why might the proof-masking base rate be an order of magnitude smaller than the proof-checking base rate (for correct proofs)? 
Because one task requires generation while the other only requires recognition: in \mbf{PL1-PM}, a model must produce 
syntactically correct formulas, variables, and rule names, whereas for \mbf{PL1-PC} it only needs to judge 
correctness, without any risk of generation errors.\footnote{This applies to classification-only 
formulations of proof checking, where the model does not need to generate any structured output.}

\subsection*{Error Analysis}

There is no natural error typology for \mbf{PL1-PM} that partitions all possible errors into a set of mutually exclusive
and jointly exhaustive types that can be readily counted, so we instead compute statistics on a number of revealing 
error types that are analytically meaningful but neither exclusive nor exhaustive. These classes of errors are described 
below. It will be useful to define $\mu$ as a shorthand for the following quantity: 
Writing $\#(D)$ for the number of masks that occur in a masked proof $D$, and if $D_1,\ldots,D_{300}$ are the proofs in \mbf{PL1-PM}, 
we set: 
\[
\mu = \bigsum_{i=1}^{300} \#(D_i).
\]
\ben

\item \emph{illFormattedAnswers}: These occur when the output generated by a model is not a JSON dictionary of the schema specified by the 
prompt (a mapping from masks to appropriate values). Min value: 0, max value: 300. 

\item \emph{emptyAnswers}: These errors occur when a model returns the empty dictionary \mtt{\{\hspace*{-0.02in}\}} as its answer, indicating 
that there is no way to unmask the input proof. (Recall that the possibility of an empty answer was included as a distractor, since we know by construction
that all masked proofs can be unmasked.) Min value: 0, max value: 300. 

\item \emph{bogusMasks}: These occur when a non-empty answer includes values for hallucinated masks that do not appear anywhere in the given input. 
Min value: 0, max value: $\infty$. 

\item \emph{missingMasks}: These occur when a non-empty answer is incomplete because the model fails to specify values for some of the masks in the input. 
Min value: 0, max value: $\mu$.

\item \emph{rulesToFormulas}: These occur when the model mistakenly lists a formula as the value of a mask that could only possibly assume a rule 
as its value. For instance, if the input proof contains a line such as 
\begin{equation}
\mtt{MASK8 \skwd{BY} MASK12 \skwd{on}} \:\cdots 
\label{Eq:RulesToFormulasExample}
\end{equation}
and a model assigns \mtt{MASK12:(\ntildels A)}, this constitutes a \temv{rulesToFormulas} error, as it is patently obvious (by a purely
syntactic/positional cue, the fact that \mtt{MASK12} appears immediately before the \skwd{on} keyword) that \mtt{MASK12} can only range over rules,
not formulas. Min value: 0, max value: $\approx \mu$.

\item \emph{formulasToRules}: These are similar to \emph{rulesToFormulas} errors but err in the opposite direction: They occur when a model 
assigns a rule to a mask that can only receive formulas as values. Considering~(\ref{Eq:RulesToFormulasExample}) again, an example 
would be a model that generated \mtt{MASK8:left-and}. Min value: 0, max value: $\approx \mu$.
\een

\begin{figure*}[thbp!]
  \centering
\scalebox{1.0}{\includegraphics[width=\textwidth]{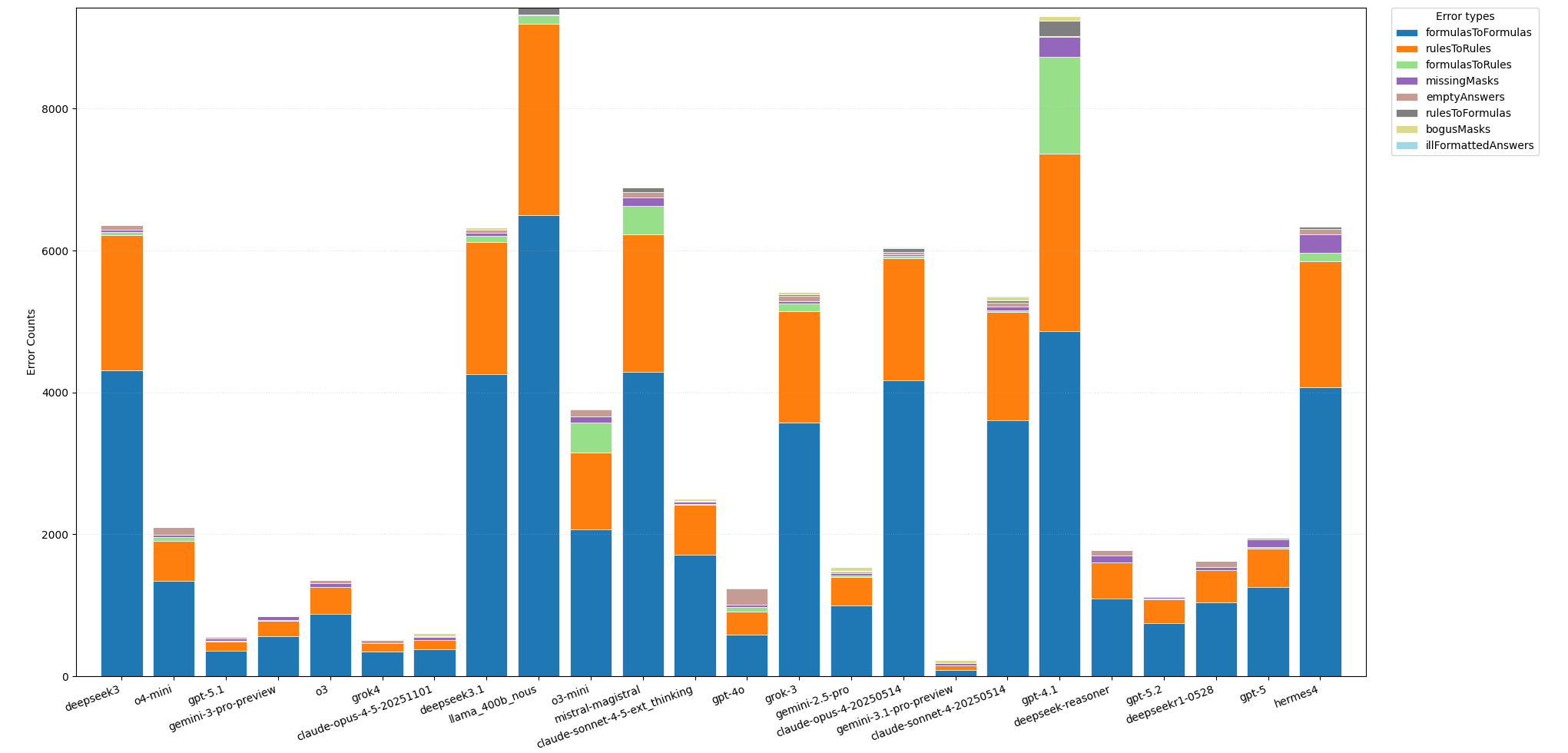}}
  \caption{\mbf{PL1-PM} errors}
  \label{fig:pl1_pm_errors}
\end{figure*}
These  6 categories represent \emph{severe} errors. It is one thing to mix up 
one rule for another, but listing a rule as the value of a mask that can only range 
over formulas is a much more serious mistake. Likewise for the rest of the above error 
types, with the possible exception of \emph{emptyAnswers}, a relatively less
flagrant type of failure. There are two main types of errors that do not fall 
into any of these categories: \emph{rulesToRules} errors (e.g., assigning \mtt{mt} 
to a mask whose correct value is \mtt{mp}), and \emph{formulasToFormulas} errors, 
which occur when the wrong formula is listed as the value of a mask. These are 
more subtle errors, or at least less egregious, and round out our error typology
for this task. 

Fig.~\ref{fig:pl1_pm_errors} plots the raw counts for these error types across all models. 
Note that we capped GPT-4.1's \emph{formulasToRules} errors 
at 500 (its true value is 1,363) because using the full value would stretch the error-count 
axis so far that the lower counts of other error types would become difficult to read. 
We make the following observations: 

\bit

\item Empty answers very likely signal  uncertainty (\emph{I can't do this}) rather than random failures. 
As usual, these errors are common among OpenAI's mini models and earlier DeepSeek models, though we also 
see a remarkable spike for GPT-4o (222 empty answers out of a total of 300). 
These models are essentially using empty answers to broadcast uncertainty aversion: 
when an input crosses some internal difficulty threshold (long proof or many masks), 
they emit an empty dictionary instead of venturing a guess.

\item  It is striking that even elite models make hard typing errors, of 
type \emph{rulesToFormulas} or \emph{formulasToRules} (\egnsp, 
GPT-5 makes 17 \emph{formulasToRules} errors; the only model that makes zero such errors is Gemini 3.1). 
It's reasonable to conjecture that humans would very rarely make such errors, for the same reason they are 
unlikely to put a verb where an adjective is expected. 
Positional typing is a hard constraint for humans: after \skwd{assume} and before \skwd{BY} only formulas fit; 
immediately before \skwd{on} only rules fit; and so on. For LLMs, by contrast, these constraints appear to be ``soft'' or fluid. 
The fact that virtually all models make these mistakes, albeit to different degrees, means that models 
find it easier to verify typing constraints (as in \mbf{PL1-PC}) than to produce outputs that respect them.
When generating, their decoders aren't gated by the relevant grammar/typing lattice and end up 
relying on token-level heuristics that occasionally slip across categories. This is the classic 
failure signature of generative brittleness (lack of a hard ``type checker'' in the loop). 
However, the frequency of these errors falls drastically among the top 4 models. 

\item \emph{missingMasks} errors represent planning failures, or deficits in 
executive function. These are numerous even for top models (e.g., 109 
for GPT-5, 53 for Gemini-3, 31 for GPT-5.2, 35 for GPT-5.1, and 34 for Gemini 3.1). 
Models here start filling and end up omitting keys. That's not an issue of misunderstanding NDL 
or making reasoning errors, it's an output orchestration issue---counting 
and enumerating all masks and persisting that plan through potentially long JSON outputs. 
It's the same phenomenon behind tool-call omissions in long chains.

\item \emph{bogusMasks} errors are relatively few but telling: when the list of required answers 
is long, some models hallucinate masks. 

\item Formatting errors are even rarer, but they are occasionally made, even by top models (2 such
errors by o3). These errors obviously track prompt adherence and JSON discipline. 

\eit

\subsection*{A challenge version of \mbf{PL1-PM}}
\label{Sec:ChallengeMasking}
In the remainder of this section we introduce \mbf{PL1-PM-c}, a challenge dataset 
containing 50 masked proofs having hundreds of lines (anywhere from 323 to 946 lines;
the median is 400). As with the challenge version of \mbf{PL1-PC}, we only evaluate
a small number of the most capable models. Each problem is evaluated with high reasoning
settings, a high allowance reasoning tokens (100K), and a maximum wall-clock time 
of 30 minutes . If no output is produced within this time, the instance is marked as failed.

While this dataset involves longer proofs, it is less challenging than 
\mbf{PL1-PM} in a precise technical sense.
Define the \emph{mask density} of a masked proof $D$ as the ratio $M(D)/L(D)$, where $M(D)$ is the 
number of masks\footnote{This is the same as the number of mask occurrences, since masks are unique.}  
in $D$ and $L(D)$ is the number of lines in $D$ that may contain a mask, namely, 
all inference steps (applications of an inference rule) and all openings of conditional deductions
(lines of the form \skwd{assume} $\cdots$). This simple statistic is the cleanest indication of how 
heavily masked a proof is. Low density means that most inference steps have
clearly visible structure (\iensp, their rule, arguments, and/or conclusion remain largely unobscured) 
and their masks are therefore locally forced with monotonic assignments. With high mask density, 
by contrast, some masks must be assigned values speculatively, forcing constraints only appear later, and 
incorrect assignments may survive for several steps (potentially dozens or even hundreds of steps). 
While the overall number of valid mask assignments might remain low
(since mask values tend to be uniquely determined), the number of plausible partial assignments explored
along the way can be large. In other words, high mask density increases entropy. The extreme case 
is fully masked inference steps like \smtt{MASK51 \skwd{BY} MASK72 \skwd{on} MASK8, MASK12}. 
Such steps have zero locally forced information and thus no way to commit without global reasoning. 
Again, this doesn't increase the number of solutions, only the depth of blind search 
required to discover one of the (relatively few) solutions.

The table below shows the distribution of mask density across both datasets. 
\renewcommand{\arraystretch}{1.4}
\par\medskip            % end the current paragraph
\noindent
\begin{minipage}{6in}
\begin{center}
%\begin{table}[h!]
\begin{tabular}{c}
\hspace*{0.3in}
\begin{tabular}{|r|c|c|c|c|c|c|c|c|c|c|c|}
\hline
 & {\bf Mean} & {\bf Median} & {\bf Min} &{\bf Max} & {\bf Std} & {\bf @5} & {\bf @25} & {\bf @50} & {\bf @75} & {\bf @90} & {\bf @95} \\
\hline\hline
{\bf PL1-PM} & 1.72 & 1.64 & 0.32 & 4.0 & 0.49 & 1.11  &  1.4 & 1.64 & 1.95 & 2.24 & 2.67 \\ \hline

\mbox{$\mbf{PL1-PM}_{\scriptsize C}$} & 1.65 & 1.64 & 1.53 & 1.78 & 0.05 & 1.57 & 1.62 & 1.64 & 1.67 & 1.70 & 1.72 \\ \hline 
\end{tabular}
\end{tabular}
\end{center}
%\caption{Summary statistics of proof lengths over 300 proofs.}
%\label{tab:pm_proof_lengths}
%\end{table} %\\[-0.3in]
\end{minipage}
\par\medskip            % end the current paragraph
%\noindent 
\renewcommand{\arraystretch}{1.0}

While the mean and median densities are very similar, \mbf{PL1-PM} has high variance (0.49) and quite a few outliers, e.g., a minimum
density of only 0.32, which is essentially an unmasked proof. More importantly, it has a heavy right tail,  
with a significant proportion of heavily masked proofs (everything above the \nth{75} percentile). 
In the challenge dataset, by contrast, the distribution of mask density is tightly concentrated, with 
a very small standard deviation (0.05). There are no heavily masked proofs on the right side of the median, 
and no outliers in either direction. The new dataset is thus near-constant density, uniformly enforcing consistent, 
moderate local difficulty.

Moreover, 16\% of all inference steps across \plonepmsp proofs are \emph{completely} masked, whereas 
less than 1\% of the inference steps in the challenge dataset are fully masked (0.1\%, to be exact).
At the macro level, 85\% of proofs in \plonepmsp have at least one inference step that is fully masked,
whereas only 8\% of the proofs in the challenge dataset have at least one such step. 
Recall that a fully masked inference step has no local syntactic anchor (no visible rule, conclusion, or arguments) 
and thus requires global constraint propagation to resolve; it cannot be solved incrementally. 
This means that \plonepmsp problems regularly contain steps with zero local information, whereas 
challenge problems have them very rarely, marking an enormous reduction in local ambiguity. 

A final notable difference is the masking of assumptions. Masking a hypothesis is one of the highest-entropy
perturbations that can be made in a proof. 
The introduction of a hypothesis $p$ in a line of the form \skwd{assume} $p$ adds $p$ to the assumption base,
introduces a new logical scope for it, and determines the shape of the resulting conditional $(p \mif q)$.
If $p$ is masked, the entire body of the conditional proof becomes underconstrained, as every inference inside
that body may well depend on $p$, but we don't know what that is. And unless we manage to infer $p$ with certainty,
we cannot be sure of the conditional proof's conclusion either. In the challenge dataset, only 10\% of conditional proofs
have masked hypotheses, whereas in \mbf{PL1-PM}, 43\% of conditional proofs (almost 1 in 2) have masked hypotheses. 

These numbers show that the challenge dataset is not harder because its proofs are more masked or because they have fewer local hints.
The opposite is true: the new dataset is technically easier because its proofs are less heavily masked and provide more local cues. 
The authors have tried solving a number of these longer problems by hand and have found it relatively straightforward to make 
consistent incremental progress; with sufficient patience and time, a human familiar with this proof style 
could finish such a problem. Yet this entropy reduction does not necessarily make the problem tractable at scale for models,
as shown in the following results: 
\renewcommand{\arraystretch}{1.5}
\begin{table}[!bhtp]
\centering
\rowcolors{2}{blue!5}{white}
\resizebox{1.0\textwidth}{!}{%
\begin{tabular}{|l|c|c||l|c|c|}
\hline
\textbf{Model} & \textbf{Instrumented Accuracy} & \textbf{Strict Accuracy} &
\textbf{Model} & \textbf{Instrumented Accuracy} & \textbf{Strict Accuracy} \\
\hline
\gptFiveFour & \mbf{0.38} \ci{0.13} & 0.14 \ci{0.1} & \gemThreeOne & 0.34 \ci{0.13} & 0.14 \ci{0.1}  \\
\gptFiveTwo & 0.32 \ci{0.13} & \mbf{0.28} \ci{0.12} & \claudeOpusFourFiveNov & 0.06 \ci{0.07} & 0.06 \ci{0.07}  \\
\hline
\end{tabular}}
\caption{\mbf{PL1-PM-c} instrumented and strict accuracies.}
\label{Tb:Pl1PMChallengeAccuracies}
\end{table} 
\mbox{} \\
The difficulty has shifted from unmasking particular steps to maintaining a consistent global state 
across hundreds of steps. 

Observe that while GPT-5.4 outperforms GPT-5.2 in terms of instrumental accuracy, its strict accuracy is much worse, 
half the strict accuracy of GPT-5.2 (though the Wilson confidence intervals are obviously wider here, since there are only 50 problems in
the challenge dataset). 

The reasoning traces of the models provide interesting insights. 
Here is part of what GPT-5.4 wrote while tackling item 9:
\begin{quote}
{\em 
$\cdots$ 
In analyzing the subblock, I see that \mtt{MASK188} must be biconditional \mtt{(D<=>B)} since the first line uses 
\mtt{left-iff} on \mtt{MASK189}. This means my assumption is set to \mtt{(D<=>B)}. 
$\cdots$ I need to explore how this structure influences the rest 
of the proof and what implications arise from these assumptions. It's fascinating how these logical connections can shape the outcome!
$\cdots$ The challenge is that \mtt{B} doesn’t hold as expected. Could I assume \mtt{MASK79 = B} and label it? However, 
the line uses literal \mtt{B} and \mtt{MASK80}. Yet if I assume \mtt{B}, it holds in that context, even without naming it. 
If I set \mtt{MASK79} to \mtt{B}, then it becomes an active assumption and holds. I see I overlooked that!
$\cdots$ 
}
\end{quote}
Such traces must not be over-interpreted, since they may be narrative artifacts rather than genuine
reflections of the model's internal computations. Nevertheless, they can still shed light on a number of fronts,
\egnsp, on the style of reasoning that the model believes is appropriate or where it seems to experience higher amounts
of uncertainty. The traces strongly indicate that these elite models are not simply pattern-matching; they are 
explicitly reasoning (or at any rate simulating a deliberation process)  about the logical availability 
of conclusions, assumption scope,  rule semantics, and possible mask assignments.

\section{Proof Gaps}
\label{Sec:PL1GF}

This dataset was obtained by randomly inserting holes (gaps) in a set of 200 correct proofs, written either by the Athena theorem prover or by a model. 
Unlike masks, which are locally focused on individual formulas and rules, gaps apply at the subproof level, \iensp, a gap replaces an entire subproof, which could be either a single
step (inference rule application), or a conditional subproof, or a {\em proof chain}: a sequence of consecutive subproofs. A proof usually receives more than one gap, 
in randomly selected subtrees. The total fractional amount of information elided by the gaps ranges from 0.3 to 0.99. In the latter extreme case, the entire proof 
is essentially replaced by a single gap, and the task then reduces to proof writing. The mean fractional total size of the gaps inserted in a proof is 0.67 of the original proof's size. 

The objective is to fill the gaps by generating appropriate subproofs for the \emph{gap identifiers}: \mtt{GAP-1}, \mtt{GAP-2}, $\ldots$.
There are no distractors for this task, which we refer to as \mbf{PL1-GF}. The following is a simple example: 
\begin{tcAthenaWithNumbers}
assert premise-1 := (C | (A & E))
assert premise-2 := (C ==> D)
assert premise-3 := (A <==> (E ==> D))
# Goal: (E ==> D)

{
  left-iff on premise-3;
  GAP-1; 
  assume E {
    case1 := assume C {
               mp on premise-2, C  
             }; 
    case2 := assume (A & E) {
               left-and on (A & E);  
               mp on (A ==> (E ==> D)), A;
               GAP-2
             };
    GAP-3
  }
}
\end{tcAthenaWithNumbers}
Bearing the goal $(E \mif D)$ in mind, it is clear from the structure of the overall proof skeleton, and particularly from the structure of the top-level conditional proof 
(the \skwd{assume} that opens on line 9 and closes on line 19), that this conditional proof establishes $(E \mif D)$. The body 
proceeds by a case analysis of \mtt{premise-1}, the disjunction $(C \mor (A \mand E))$. 
The \mtt{case1} subproof establishes that $C$ implies $D$, thus \mtt{case2} must establish that 
$(A \mand E)$ implies $D$. Therefore, \mtt{GAP-2} must derive $D$ 
from the assumption base at the end of line 15, which includes the conditional $(E \mif D)$ (derived on line 15) and the \mtt{case-2} 
hypothesis $(A \mand E)$  (as \mtt{GAP-2} is inside the scope of that hypothesis). So this gap is easy to fill: first detach $E$ from the 
assumption $(A \mand E)$ and then use modus ponens on $(E \mif D)$ and $E$ to infer $D$: 
\[\mtt{GAP-2:}\;\mtt{E \kwd{BY} right-and \kwd{on} (A \& E); D \kwd{BY} mp \kwd{on} (E ==> D), E}\]
\mtt{GAP-3} is a simple rule application: \mtt{cases \kwd{on} premise-1, (C ==> D), ((A \& E) ==> D)}.

Note that \mtt{GAP-1} does not make any contribution to the proof. It is there only because
the LLM that generated the initial proof included a redundant step at that point, namely, the application of \mtt{right-iff} on \mtt{premise-3}. Any replacement for that gap 
is given full credit as long as the entire proof obtained by filling the gaps as specified by the model correctly derives the goal from the premises. 
This is the only criterion we use for correctness. 

\begin{figure}[!b]
\centering
\scalebox{1.0}{\includegraphics[width=\linewidth]{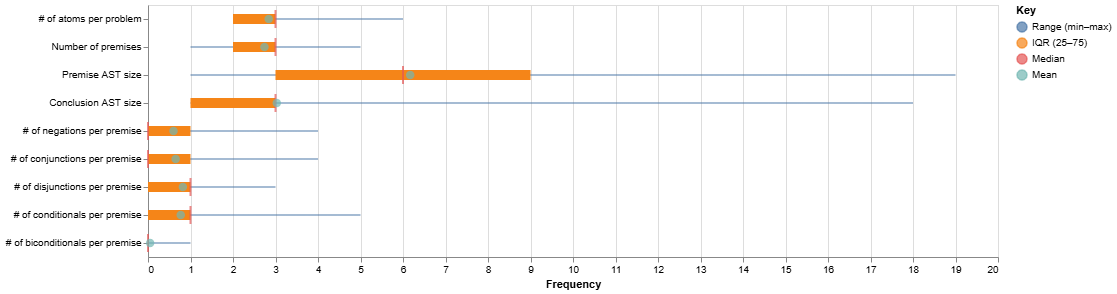}}
\caption{\mbf{PL1-GF} problem statistics}
\label{Fig:PL1GapFillingProblemStats}
\end{figure}
Statistics on the problem distribution for this task slice are shown in Fig.~\ref{Fig:PL1GapFillingProblemStats}. The problems are quite simple 
and roughly similar in their complexity to masking problems, except that conclusions are notably smaller (the median number of nodes is 3).
\renewcommand{\arraystretch}{1.4}
The statistics below show that the lengths of these proofs are roughly in the same range 
as those in \mbf{PL1-PM}, but on average these proofs are about 25\% longer (34.5 lines vs 27). 
\par\medskip            % end the current paragraph
\noindent
\begin{minipage}{6in}
\begin{center}
%\begin{table}[h!]
\begin{tabular}{c}
\hspace*{0.3in}
\begin{tabular}{|r|c|c|c|c|c|c|c|c|c|c|c|}
\hline
 & {\bf Mean} & {\bf Median} & {\bf Min} &{\bf Max} & {\bf Std} & {\bf @5} & {\bf @25} & {\bf @50} & {\bf @75} & {\bf @90} & {\bf @95} \\
\hline\hline
{\bf Proof length} & 34.5 & 39.5 & 3 & 64 & 15.89 & 11.95 & 20.0 & 39.5 & 49.0 & 53.0 & 53.0 \\ \hline 
\end{tabular}
\end{tabular}
\end{center}
%\caption{Summary statistics of proof lengths over 300 proofs.}
%\label{tab:pm_proof_lengths}
%\end{table} %\\[-0.3in]
\end{minipage}
\par\medskip            % end the current paragraph
%\noindent 
\renewcommand{\arraystretch}{1.0}

Accuracies for all models, obtained both from the instrumented and the regular proof checkers, 
are shown in Table~\ref{Tb:Pl1GFAccuracies}. 
%% GPT-5 takes a commanding lead on this task with an 
%% impressive accuracy of 90\%. Claude Sonnet 4.5 (extended-thinking) and o3 come in at a distant 
%% second and third place with 80\% and 77\%, respectively. Gemini-2.5-Pro clocks in at 75\% and 
%% Grok-4 rounds out the top-five lineup at just under 70\%. 

The discriminability index (DI) of \mbf{PL1-GF} is 37.5\%, meaning (roughly)  that any two of these models differ in accuracy 
by about 37.5 percentage points---again a very large spread. This task strongly separates some elite models as well,
\egnsp, the delta between o3 and Gemini 3 Pro Preview (as well as the GPT-5-series models, Opus 4.5 and Gemini 3.1) is highly 
statistically significant, while the delta between, say, Grok-4 and Claude Sonnet 4.5 with extended thinking 
is statistically significant.\footnote{We used McNemar's test to check whether these pairwise deltas 
could be due to random variation. We call the $p$-value of a particular test result ``statistically significant'' if it's below
the usual 0.05 threshold and ``highly statistically significant'' if it's below the 0.01 threshold. 
In the case of Gemini 3 Pro Preview and o3, for example, out of 38 problems where the models disagreed, 
Gemini was correct on 31 and o3 on only 7, which corresponds to a $p$-value of 0.000116. This means 
the likelihood that such a large advantage occurred by chance is less than 0.1\%, and thus 
the difference in performance between these two models is very statistically robust.}

\renewcommand{\arraystretch}{1.5}
\begin{table}[!bhtp]
\centering
\rowcolors{2}{blue!5}{white}
\resizebox{1.0\textwidth}{!}{%
\begin{tabular}{|l|c|c||l|c|c|}
\hline
\textbf{Model} & \textbf{Instrumented Accuracy} & \textbf{Strict Accuracy} &
\textbf{Model} & \textbf{Instrumented Accuracy} & \textbf{Strict Accuracy} \\
\hline
\claudeOpusFourFiveNov & \mbf{0.94} \ci{0.03} & 0.88 \ci{0.05} &
\gemThreeOne & 0.92 \ci{0.04} & \mbf{0.89} \ci{0.04} \\ \hline
\gptFiveTwo & 0.93 \ci{0.04} & 0.84 \ci{0.05} &
\gptFive & 0.90 \ci{0.04} & 0.79 \ci{0.06} \\ \hline
\gemThree & 0.89 \ci{0.04} & 0.77 \ci{0.06} &
\gptFiveOne & 0.89 \ci{0.04} & 0.70 \ci{0.06} \\ \hline
\claudeSonnetExt & 0.80 \ci{0.06} & 0.60 \ci{0.07} &
\oThree & 0.77 \ci{0.06} & 0.56 \ci{0.07} \\ \hline
\geminiTwoPointFive & 0.75 \ci{0.06} & 0.49 \ci{0.07} &
\grokFour & 0.69 \ci{0.06} & 0.55 \ci{0.07} \\ \hline
\dsReasoner & 0.61 \ci{0.07} & 0.31 \ci{0.06} &
\oFourMini & 0.45 \ci{0.07} & 0.37 \ci{0.07} \\ \hline
\dsROne & 0.46 \ci{0.07} & 0.24 \ci{0.06} &
\grokThree & 0.38 \ci{0.07} & 0.32 \ci{0.06} \\ \hline
\gptFourOne & 0.29 \ci{0.06} & 0.24 \ci{0.06} &
\dsThree & 0.30 \ci{0.06} & 0.21 \ci{0.06} \\ \hline
\dsThreeOne & 0.30 \ci{0.06} & 0.18 \ci{0.05} &
\magistral & 0.24 \ci{0.06} & 0.19 \ci{0.05} \\ \hline
\oThreeMini & 0.23 \ci{0.06} & 0.17 \ci{0.05} &
\claudeSonnetFour & 0.16 \ci{0.05} & 0.15 \ci{0.05} \\ \hline
\opus & 0.14 \ci{0.05} & 0.13 \ci{0.05} &
\hermes & 0.10 \ci{0.04} & 0.07 \ci{0.04} \\ \hline
\nous & 0.07 \ci{0.04} & 0.03 \ci{0.03} &
\gptFourO & 0.01 \ci{0.02} & 0.01 \ci{0.02} \\ \hline
\end{tabular}}
\caption{\mbf{PL1-GF} accuracies, using both the instrumented and the regular proof checker.}
\label{Tb:Pl1GFAccuracies}
\end{table}
\renewcommand{\arraystretch}{1.0}

%% \renewcommand{\arraystretch}{1.5}
%% \begin{table}[!hbtp]
%% \centering
%% \rowcolors{2}{blue!5}{white}
%% \resizebox{1.0\textwidth}{!}{%
%% \begin{tabular}{|l|c|c||l|c|c|}
%% \hline
%% \textbf{Model} & \textbf{Instrumented Acc.} & \textbf{Plain Acc.}   & \textbf{Model} &  \textbf{Instrumented Acc.} & \textbf{Plain Acc.} \\
%% \hline
%% \gptFive & \mbf{0.90} \ci{0.04} & 0.77 & \claudeSonnetExt & 0.80 \ci{0.06} & 0.60   \\ \hline
%% \oThree & 0.77 \ci{0.06} & 0.56 & \geminiTwoPointFive & 0.75 \ci{0.06} & 0.48  \\ \hline
%% \grokFour & 0.69 \ci{0.06} & 0.55 & \dsROne & 0.46 \ci{0.07} & 0.24     \\ \hline 
%% \oFourMini & 0.45 \ci{0.07} & 0.37 &  \grokThree & 0.38 \ci{0.07} & 0.32 \\ \hline
%% \dsThree & 0.30 \ci{0.06} & 0.21 & \gptFourOne & 0.29 \ci{0.06} & 0.24 \\ \hline
%% \magistral & 0.23 \ci{0.06} & 0.19  & \oThreeMini & 0.23 \ci{0.06} & 0.17  \\ \hline
%% \claudeSonnetFour & 0.15 \ci{0.05} & 0.14   & \opus & 0.14 \ci{0.05} & 0.13   \\ \hline
%% \nous  & 0.07 \ci{0.04} & 0.03 & \gptFourO & 0.01 \ci{0.02} & 0.01 \\ \hline 
%% \end{tabular}}
%% \caption{\mbf{PL1-GF} accuracies, using both the instrumented NDL checker and the regular proof checker.}
%% \label{Tb:Pl1GFAccuracies}
%% \end{table}
%% \renewcommand{\arraystretch}{1.0}

We treat blank (empty-string) responses after a timeout budget of at least 30 minutes as genuine task failures (non-responses) that 
count as incorrect outcomes in our accuracy computation. By contrast, if there is a genuine API call error we do not include 
that data point in the accuracy denominator, and in such cases we try to repeat the call from scratch;\footnote{In the final 
results there is only one API call error left, for Mistral Magistral on item 135.}
see Appendix~\ref{app:technical-details} for additional details. 
For transparency, we also report non-response rates: Grok-4 abstained on 11/200 problems (5.5\%), 
Gemini 3 Pro Preview on 7 problems (3\%),  Mistral Magistral on 7 problems (3\%),
and o3 on 1 problem (0.5\%). All other models returned non-empty responses for all 200 problems. 

Formatting issues were rife for some models, particularly for the two Claude 4 models (Opus and Sonnet), 
which frequently fail to return a valid gap assignment, or indeed any gap assignment,
instead emitting a random fragment of NDL code.\footnote{Claude 4 Opus did this for 135 out of 200 problems
(67.5\% of all inputs) and Claude 4 Sonnet for 119 problems (59.5\%).} 
This was a stable behavioral pattern, confirmed  by recomputing results multiple times. 
%that is unique to these two Anthropic models and not observed in any others. 
Other models made formatting errors of different types, some more exotic and blatant than others. 
Unlike other \pgridsp tasks, the  \mbf{PL1-GF} prompt does not require JSON output, 
instead instructing the models to simply think through the problem
%\begin{samepage}
and then write their solutions at the very end of their output, in the following simple form: 
\begin{verbatim}
GAP-i:
<subproof for GAP-i>
GAP-j:
<subproof for GAP-j>
...
\end{verbatim}
%% \begin{center}
%% \begin{minipage}{6in}
%% {\footnotesize 
%% \centering 
%% \begin{tabular}{l}
%% \mtt{GAP-i}\mtt{:} \\
%% \mtt{<subproof for GAP-i>} \\
%% \mtt{GAP-j}\mtt{:} \\
%% \mtt{<subproof for GAP-j>} \\
%% \rsp\rsp\rsp $\vdots$
%% \end{tabular}
%% }
%% \end{minipage}
%% \end{center}
Only the latest generation of frontier models adhered faithfully to these instructions. 
All other models deviated, typically in predictable ways
that could be programmatically handled.\footnote{For example, many models insisted on writing the \mtt{GAP-i} headers in boldfaced
markdown, writing, for example, \mtt{**GAP-8**:} or \mtt{***GAP-8***:} instead of \mtt{GAP-8:}; 
or in different (and seemingly arbitrary) markdown heading levels; sometimes they would insert the colon 
before the asterisks, \egsp writing \mtt{**GAP-8:**}; and so on. And many models enclosed the generated NDL fragments 
inside triple backticks (another common markdown convention, for marking code blocks), e.g. writing something like:
\begin{tcAthena}
**GAP-5**:
```ndl 
assume (C & D) {
   ... 
} 
```
**GAP-8**:
```
...
```
... 
\end{tcAthena}
}
But some less common formatting issues resulted
in outputs that could not be easily salvaged and had to be counted as incorrect. For example, in a few cases
(most notably involving o3-mini), the NDL proof fragments assigned to some gaps contained references to other gaps. 
Another example: occasionally the model's response would only contain a COT trace in natural language, 
without ever culminating in a gap assignment. Other times there would be additional commentary inserted after 
the supposedly final gap assignment; other times the model would insert a gratuitous delimiter 
(such as a long dashed line) in between the different proof fragments; and so on. 

A more common error was incomplete gap assignments: properly formatted outputs that nevertheless
neglected to assign values to some of the input gaps. Here Claude 4 Opus and Sonnet lead the pack 
with 144 and 130 respectively (due to the malformed responses mentioned earlier), followed by 
by GPT-4o (30), Mistral Magistral (21), Grok-4 (12), o4-mini (9), DeepSeek-V3 (8), Gemini-
2.5-Pro (5), GPT-4.1 (4), Grok-3 (3), the Nous-tuned version of Llama-400b (1), 
the extended-thinking version of Claude Sonnet 4.5 (1), 
DeepSeek R1-0528 (1), and o3 (1). The remaining models never generated incomplete answers. 

We move to the errors that were detected and overlooked by the instrumented checker. Table~\ref{Tb:PL1GFErrorRepairRates}
lists the number of syntax and semantic errors repaired by the instrumented checker for every 100 lines of proof code
generated by a model. 
\renewcommand{\arraystretch}{1.5}
\begin{table}[!bp]
\centering
\scriptsize
\rowcolors{2}{blue!5}{white}
\resizebox{1.0\textwidth}{!}{%
\begin{tabular}{|l|c|c|c||l|c|c|c|}
\hline
\textbf{Model} & \textbf{Syntax $R/100L$} & \textbf{Semantic $R/100L$} & \textbf{Total $R/100L$} &
\textbf{Model} & \textbf{Syntax $R/100L$} & \textbf{Semantic $R/100L$} & \textbf{Total $R/100L$} \\
\hline
\gemThreeOne        & 0.12 & 0.37 & \mbf{0.49} & \claudeOpusFourFiveNov        & 0.20 & 0.56 & 0.76 \\ \hline
\gptFive        & 0.42 & 0.71 & 1.13 & \gptFiveTwo        & 0.97 & 0.47 & 1.44 \\ \hline
\gemThree       & 0.83 & 1.30 & 2.13       & \gptFiveOne        & 3.10 & 1.01 & 4.11 \\ \hline
\oThree         & 2.64 & 1.67 & 4.31       & \claudeSonnetExt   & 2.85 & 1.75 & 4.60 \\ \hline
\grokFour       & 3.31 & 2.61 & 5.92       & \grokThree         & 5.56 & 1.74 & 7.30 \\ \hline
\oFourMini      & 6.71 & 1.11 & 7.82       & \geminiTwoPointFive& 6.05 & 1.86 & 7.91 \\ \hline
\claudeSonnetFour & 9.26 & 1.30 & 10.56     & \gptFourOne        & 10.13 & 0.58 & 10.71 \\ \hline
\magistral      & 12.28 & 1.16 & 13.44      & \dsReasoner        & 13.30 & 1.62 & 14.92 \\ \hline
\dsThree        & 13.92 & 1.40 & 15.32      & \dsThreeOne        & 16.38 & 0.85 & 17.23 \\ \hline
\opus           & 15.89 & 1.51 & 17.40      & \hermes            & 16.82 & 2.48 & 19.30 \\ \hline
\oThreeMini     & 18.72 & 0.58 & 19.30      & \nous              & 21.28 & 0.73 & 22.01 \\ \hline
\dsROne         & 27.34 & 1.22 & 28.56      & \gptFourO          & 29.28 & 0.34 & 29.62 \\ \hline
\end{tabular}}
\caption{Error repairs made by the instrumented checker per 100 lines of generated proof code.}
\label{Tb:PL1GFErrorRepairRates}
\end{table}
%% \begin{table}[!bp]
%% \centering
%% \scriptsize 
%% \rowcolors{2}{blue!5}{white}
%% \resizebox{1.0\textwidth}{!}{%
%% \begin{tabular}{|l|c|c|c||c|l|c|c|c|c|}
%% \hline
%% \textbf{Model} & \textbf{Syntax $R/100L$} & \textbf{Semantic $R/100L$}  & \textbf{Total $R/100L$}  &
%% \textbf{Model} & \textbf{Syntax $R/100L$} & \textbf{Semantic $R/100L$}  & \textbf{Total $R/100L$} \\
%% \hline
%% \gptFive & 1 & 1  & \mbf{2} & \claudeSonnetFour & 1 & 1  & 2 \\ \hline
%% \opus & 1 & 1  & 2 & \oThree & 3 & 2  & 5   \\ \hline
%% \claudeSonnetExt & 3 & 2 & 5 & \grokFour & 3 & 3 & 6 \\ \hline
%% \oFourMini & 6 & 1 & 7 & \grokThree & 6 & 2 & 8 \\ \hline
%% \geminiTwoPointFive & 6 & 2 & 8 & \gptFourOne & 10 & 1 & 11 \\ \hline
%% \magistral & 12 & 1 & 13 & \dsThree & 14 & 1 & 15 \\ \hline
%% \oThreeMini & 17 & 1 & 18 & \nous & 18 & 1 & 19 \\ \hline
%% \gptFourO & 25 & 0 & 25 & \dsROne & 27 & 1 & 28 \\ \hline
%% \end{tabular}}
%% \caption{Error repairs made by the instrumented checker per 100 lines of generated proof code.}
%% \label{Tb:PL1GFErrorRepairRates}
%% \end{table}
\renewcommand{\arraystretch}{1.0}
Note that Claude Sonnet 4 and Opus 4 have very few repairs for a pathological reason: as noted earlier,
their outputs are often so ill-formatted that a fatal error often halts 
evaluation on the very first step, thereby robbing the instrumented checker of the opportunity 
to repair minor errors.

As in proof writing, an interesting question here concerns the extent to which a model's propensity 
for minor formal slipups correlates with its overall reasoning ability. In theory, the two could be orthogonal. 
We can imagine a brilliant mathematician who can do the heavy lifting of a proof by connecting 
all the right dots at a high level of abstraction, but is less adept at technical details and  
would benefit from help in converting a proof blueprint into a rigorous derivation. Likewise, it's possible
that a model could make a lot of minor errors but nevertheless attain a fairly high accuracy with the 
aid of the instrumented checker, which would come to its rescue by detecting and overlooking those errors.

\begin{figure}[!hbtp]
\centering
%\scalebox{1.0}{\includegraphics[width=\linewidth]{./plots/pl1_gf_acc_vs_total_repairs.png}}
\scalebox{1.0}{\includegraphics[width=\linewidth]{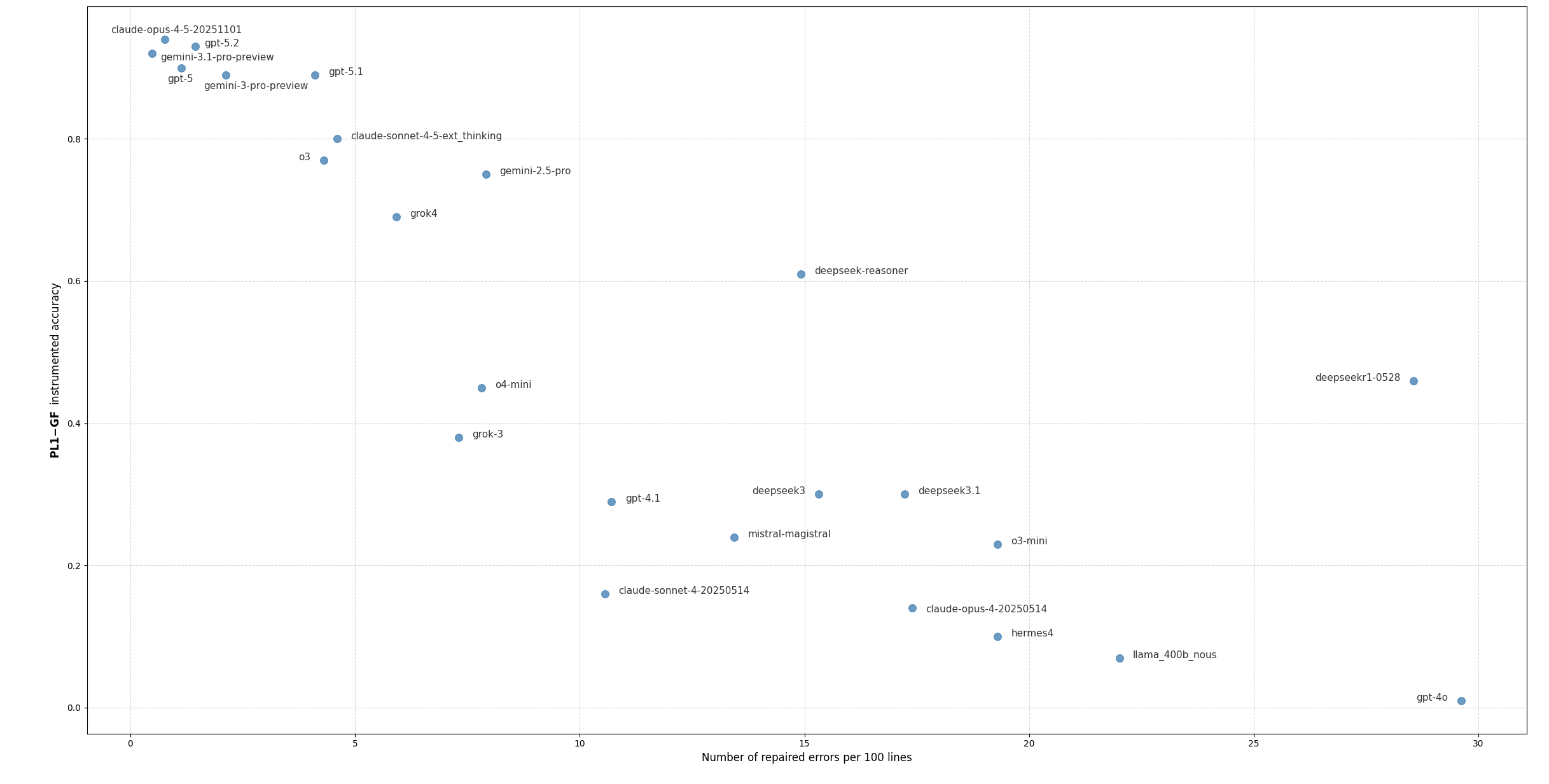}}
\caption{\mbf{PL1-GF} instrumented model accuracies vs total error repairs per 100 lines.}
\label{Fig:PL1GFModelAccVsERPer100Scatterplot}
\end{figure}
The data shows that the two are negatively correlated: in general, the more minor/repairable errors 
a model makes, the worse its overall performance. The correlation is not ironclad but it is strong nevertheless.
Fig.~\ref{Fig:PL1GFModelAccVsERPer100Scatterplot} shows the relevant scatterplot, which indicates a clear 
inverse relationship (downward slope) between gap-filling accuracy and repair density. 
Specifically, the Pearson correlation coefficient between accuracy and normalized error-repair density 
(per 100 lines) is -0.82. DeepSeek R1-0528 is interesting. It has a lot of error repairs 
but manages to come close to 50\%, slightly outperforming o4-mini. However, the vast 
majority of these errors are syntactic, and most of them are due to dropping outer parentheses
from complex formulas. In other words, most of the errors that R1-0528 makes and the instrumented checker fixes 
are ultra-local and low-impact systematic deviations that violate strict
formalism without breaking logical coherence. These are exactly the sort of 
lapses that the instrumented checker was designed to tolerate. 

As shown in Table~\ref{Tb:Pl1GFAccuracies}, and as was also the case for \mbf{PL1-PW}, 
the instrumented  checker raises all model accuracies significantly, except for 
the bottom tier of the 8 worst-performing models, where improvements are marginal. 
As explained earlier, that was the intention behind the  instrumented proof checker: 
to try to decouple formal precision from logical competence by overlooking minor errors.
Nevertheless, the overall correlation between repair rate and accuracy remains 
strongly negative, as shown by the scatterplot and the correlation coefficient. 
This is not contradictory.  At an intra-model level of analysis, we see that
instrumentation unlocks latent reasoning ability that would otherwise be obscured by procedural slips.
At an inter-model or global level of analysis, across models, a high rate of procedural error repairs 
indicates weaker executive control, so models that generate formally correct proofs also 
attain higher reasoning accuracy. 
%In other words, the ability to benefit from the repairs and the need for repairs are distinct properties; strong models both make few such errors and gain little from correction. 
%As explained earlier, the instrumented checker is intended to decouple formal precision from logical competence by overlooking minor syntax and structural errors. Yet the empirical correlation runs the other way: 
In other words, formal discipline and semantic insight appear to co-develop 
in LLMs. 
%% While the instrumentation allows ``pure reasoning ability'' to (largely) suffice \emph{in principle}, 
%% in practice such ability rarely manifests in isolation. With minor exceptions (R1 being the most
%% notable one), the strongest models are precisely those that simultaneously follow instructions 
%% well and reason well. 
This suggests that reasoning and instruction-following share an 
underlying architectural or representational substrate.
% (perhaps related to long-range syntactic compositionality and attention control). 
And it also indicates that as models scale 
and align better, they improve not just in raw inference ability but also in 
procedural compliance, 
%such as internalizing and consistently conforming to formal rules explicitly specified in the prompt, 
a skill that is essential both in logic and in coding.

Models tend to do better on proof writing than on gap filling (compare Tables~\ref{Tb:Pl1PWAcc} 
and~\ref{Tb:Pl1GFAccuracies}). Sometimes the difference is dramatic, e.g., a 19-point gap in the
case of Gemini 2.5 Pro. This might seem counterintuitive. Gap filling introduces 
a proof-synthesis spectrum, where on one end we have a tiny number of small gaps that do not 
make any significant synthesis demands, while on the other end we have very large gaps 
that essentially require models to write the entire proof. It would seem that the problem 
would get harder as we move along this continuum. 

However, gap-filling introduces constraints that are not present in proof writing.
In \mbf{PL1-PW}, a model has the freedom to choose any proof strategy it likes, 
to use ``favorite'' inference patterns, discharge assumptions
in whatever order it deems appropriate, and so on. There is no need to respect and follow
a pre-existing plan. In \mbf{PL1-GF}, by contrast, the input proof skeleton forces a 
specific problem decomposition with a particular sequence of subgoals and hardwired local plans, 
which the model has to reverse-engineer, \egnsp, by deducing why certain surrounding lines are there. 
It is tackling a planning-under-constraints problem rather than a plain proof-writing problem. 
While plain proof writing in propositional logic can be viewed as a classic forward 
planning problem, gap filling imposes a hidden inverse search problem, in the 
technical sense of inverse Bayesian planning, where the goal is 
to abductively infer the intentions of the agent who wrote the initial proof
by ``inverting a model'' of how intentions cause proof-writing 
behavior \citep{BakerInversePlanning}.\footnote{In classical  
planning this is related to \emph{plan recognition}, also known as ``inverse planning,'' 
where one observes a sequence of actions and must infer the goals that gave rise to them \citep{RamirezGeffner10}.}
Depending on its exact formulation, inverse planning is at least as computationally
hard as forward planning. 
Gap-filling also demands stronger meta-reasoning skills, since it requires explicit 
deliberation about the given proof skeleton and constructs like assumption scope. 
(All inverse tasks generally require reasoning about another 
agent's reasoning processes, thus exercising a sort of theory-of-mind skillset.) 
These considerations could help to explain why, for example,
Grok-4 goes from 85\% accuracy in \mbf{PL1-PW} to 69\% in \mbf{PL1-GF}. 
In general, the diversity of tasks in \pgridsp allows for richer profiling 
that separates models along different axes of ability. 

\begin{figure}[!thb]
\centering
\scalebox{1.0}{\includegraphics[width=\linewidth]{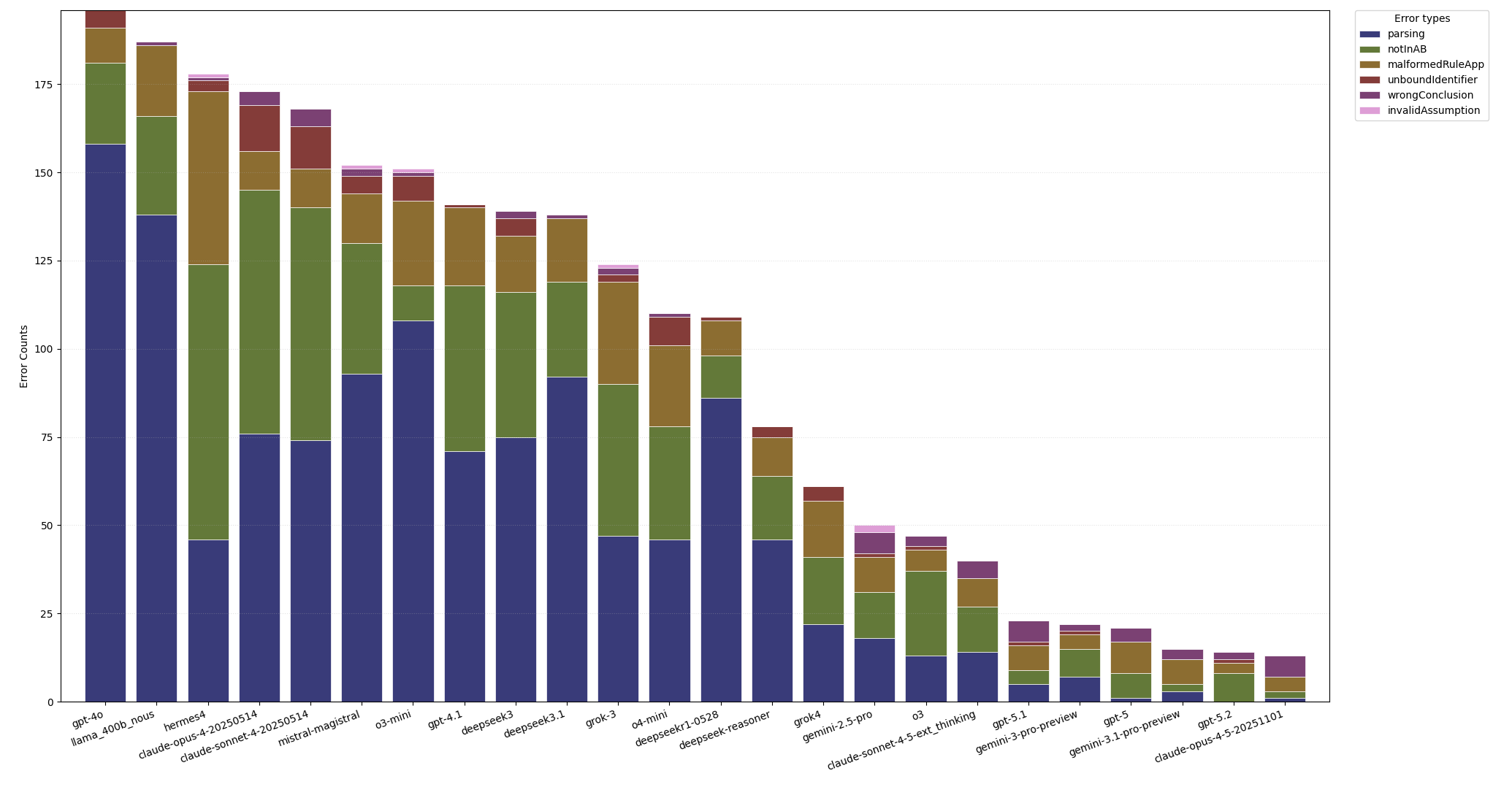}}
\caption{\mbf{PL1-GF} distribution of first errors.}
\label{Fig:PL1GFErrorTypeDistribution}
\end{figure}

Next, we analyze the first gap-filling errors made by models, i.e., those errors that the instrumented
interpreter does not fix. These are divided into the same six groups 
introduced in the \mbf{PL1-PW} error analysis. 
%% (see p.~\pageref{Pg:PL1PWErrorDist}), which
%% we briefly review here:
%% \ben
%% \item \emph{Parsing/Syntax errors}: A generated proof fragment for one of the gaps is syntactically malformed. 
%% \item \emph{Not-in-the-assumption-base errors}: Appealing to a formula that does not currently hold.
%% \item \emph{Malformed rule applications}: These can be regarded as ``type errors'' such as applying an inference rule to the wrong
%% number of arguments or to arguments of the wrong form. 
%% \item \emph{Malformed (invalid) assumptions}: Assumptions that are mere identifiers instead of proper formulas. 
%% \item \emph{Wrong conclusions}: When the resulting proof (or one of its subproofs) derives \emph{some} conclusion but not the expected/advertised one. 
%% \item \emph{Unbound identifiers}: When a generated subproof refers to a name that was never defined.
%% \een
Their distribution is shown in Fig.~\ref{Fig:PL1GFErrorTypeDistribution}. 
%Unsurprisingly, this distribution is fairly similar to its proof-writing counterpart. 
As in other tasks, three main cohorts emerge from this picture:
\bit
\item The surface-fragility regime of weaker models like GPT-4o and Llama-400b, whose 
error profiles are dominated by parsing/syntax (blue) failures. These models fail before 
their reasoning can be scrutinized. Instrumentation can't help because 
the emitted fragments deviate from proper syntax in highly irregular ways that cannot 
be programmatically detected. 
As two simple examples, instead of writing \smtt{A \skwd{BY} left-and \skwd{on} $\cdots$}, a model in this 
cohort might write \mtt{A INFERRED BY left-and on $\cdots$}, 
%or some other unpredictably hallucinated variant, 
or instead of \mtt{(A\psp\&\psp{}B)} they
might write \mtt{A AND ALSO B}. We have anecdotally observed that even 
if these grammar issues were somehow resolved, the generated fragments would still fail  
due to downstream structural and logical errors. In other words,  parsing errors and 
logical failures are entangled phenomena for weaker models, not independent pathologies. 
%% Grammar hallucinations are just the visible tip of a conceptual entropy process, the point 
%% where linguistic generation, control flow, and logical validity all begin to collapse together. 
%% Another major (if less exotic) source of parsing errors are incomplete gap assignments. 
%% When a model fails to assign any subproof to \mtt{GAP-5}, for example,
%% the string \mtt{GAP-5} will appear verbatim in the final filled proof and will 
%% almost certainly cause a parsing error. 
\item The mixed-failure regime of mid-tier models, where parsing errors range 
from $\approx 30\%$ to 60\% of the total, the rest coming from logical errors
(citing formulas that are not in the assumption base), misapplied inference rules,
and references to unbound identifiers. 
%R1 is an outlier here, with an exceedingly high  rate of parsing errors, many of which are not recoverable (e.g., conditional proofs with empty bodies or subproofs containing ellipses). 
\item The top-tier models, whose errors are predominantly semantic/logical or structural. 
GPT-5.2 stands out with \emph{zero} parsing errors,  
an indication of its exceptional instruction-following ability. 
\eit
\begin{figure}[!tbh]
\centering
\scalebox{0.9}{\includegraphics[width=\linewidth]{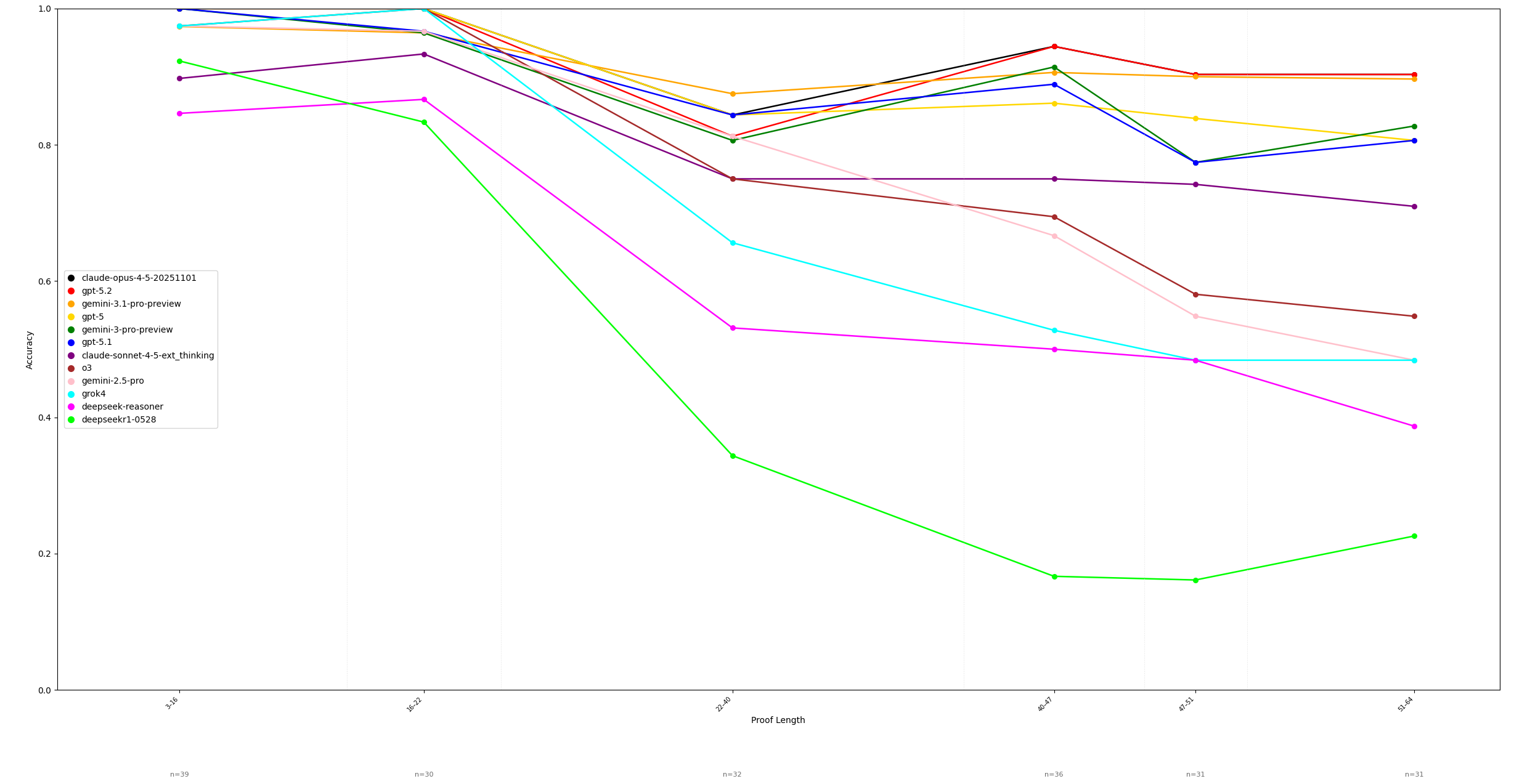}}
\caption{\mbf{PL1-GF} accuracy vs gapped proof length for the top 12 models}
\label{Fig:PL1GFAccVsProofLen1}
\end{figure}

\begin{figure}[!b]
\centering
\scalebox{0.6}{\includegraphics[width=\linewidth]{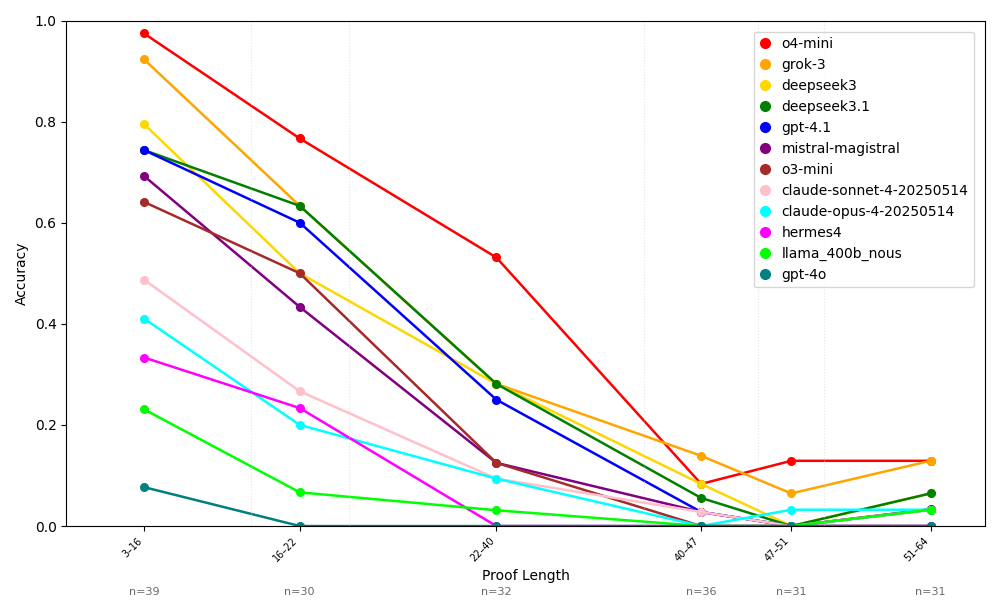}}
\caption{\mbf{PL1-GF} accuracy vs gapped proof length for the bottom 12 models.}
\label{Fig:PL1GFAccVsProofLen2}
\end{figure}

We close by examining how gap-filling performance scales with input size. 
Fig.~\ref{Fig:PL1GFAccVsProofLen1} plots the accuracy of the top 12 models as a function of 
the input proof length, while Fig.~\ref{Fig:PL1GFAccVsProofLen2} does the same for the bottom
12 models. These curves show that the accuracy of all models 
falls as the number of proof lines in the input increases. 
%\iensp, as the inferential structure that must be maintained grows more complex.
However, while weaker models exhibit a sharp and steadily monotonic performance collapse, 
with many hitting near-zero accuracy around 40 lines, the declines of the stronger models  
are much more gradual and less catastrophic.\footnote{Some models either plateau or show a small uptick at the tail end. 
This  is likely due to a gap-fraction confound 
(by ``gap fraction'' we mean the proportion of the original proof that was replaced by gaps). 
Generally, once we control for proof length, 
the problem gets harder as the gap fraction increases. 
Likewise, if the gap fraction is fixed, the problem gets harder as proof length increases. 
However, because \mbf{PL1-GF} is a small dataset (200 data points) and not perfectly stratified across both features, 
longer proofs tend to have smaller gap fractions, while shorter proofs have 
higher gap fractions. This imbalance explains why 
a plot of accuracy vs gap fraction would likewise show a declining sensitivity curve 
with a small rebound at the tail end: because of a confounding effect, this time 
(symmetrically) due to proof length. In both cases, the apparent slight 
improvement at the tail happens because the other variable is easier in the corresponding region.}

\subsection*{A challenge version of \mbf{PL1-GF}}
\label{Sec:ChallengeGaps}
In this section we discuss a more challenging version of this task, \mbf{PL1-GF-c}, with 50 problems featuring longer proofs.
The median proof length is 399, and 90\% of the proofs have fewer than 510 steps. The maximum length is 946. 
The median number of gaps per proof is 45, and the maximum is 100. 

Although these proofs are considerably longer, the gaps are easier to fill for a number of reasons.
The original proofs for this dataset were generated by a simple automated theorem prover, and as a result they have 
a great deal of structure and a number of recurring patterns. In addition, gap completions tend to be locally short 
and semantically anchored. (For instance, among the set of correct answers generated by GPT-5.4, the median gap completion
is 1, and the average is 2.6; 90\% of all gap completions have fewer than 6 lines.)
As an example, below is the top portion of a proof from \mbf{PL1-GF-c}. 
\begin{tcAthenaWithNumbers}
assert premise-1 := ((D | C) & (A ==> ~D))
assert premise-2 := ((~ ((~ (A | C)) ==> D)) | ((~ (~B | C)) <==> A))
assert premise-3 := (~C | (A & B & ~E) | (A & D & ~E) | (E & ~B & ~D) | (~A & ~B & ~D))
# Derive (~A & ~B)

{
  p0 := (D | C) BY left-and on ((D | C) & (A ==> (~ D)));
  p1 := (A ==> (~ D)) BY right-and on ((D | C) & (A ==> (~ D)));
  ((~ A) | (~ D)) BY cond-def on (A ==> (~ D));
  assume p2 := D {
    assume p3 := (E & ((~ B) & (~ D))) {
      ((~ B) & (~ D)) BY right-and on p3;
      (~ D) BY right-and on ((~ B) & (~ D));
      GAP-13
    };
    assume ((~ A) & ((~ B) & (~ D))) {
      GAP-10; 
      p3 := (~ D) BY right-and on ((~ B) & (~ D));
      ((~ A) & (~ B)) BY from-complements on ((~ A) & (~ B)), p2, (~ D)
    };
    assume ((E & ((~ B) & (~ D))) | ((~ A) & ((~ B) & (~ D)))) {
      ((~ A) & (~ B)) BY cases on ((E & ((~ B) & (~ D))) | ((~ A) & ((~ B) & (~ D)))), 
                                  ((E & ((~ B) & (~ D))) ==> ((~ A) & (~ B))), 
                                  (((~ A) & ((~ B) & (~ D))) ==> ((~ A) & (~ B)))
    };
    ... 
\end{tcAthenaWithNumbers}
It is immediately clear from the structure of the case analysis on lines 22--24 
that the first inner conditional proof (lines 11--15) must produce the conditional
$(\neg E \mand \neg B \mand \neg D \mif \neg A \mand \neg B)$ and hence \mtt{GAP-13} must
produce $(\neg A \mand \neg B)$, which can easily be done with an application of \mtt{from-complements}
on $D$ and $(\neg D)$, exactly as is done on line 19. Likewise, \mtt{GAP-10} is easily seen to be
the counterpart of line 12. We thus have:
\begin{tcAthena}
GAP-10: (~B & ~D) BY right-and on (~A & (~B & ~D)) 
GAP-13: (~A & ~B) BY from-complements on (~A & ~B), D, (~ D)
\end{tcAthena}
The remaining gaps can be filled in a similar way. 

Results for the 4 currently strongest models are shown in Table~\ref{Tb:Pl1GFChallengeAccuracies}.  
\renewcommand{\arraystretch}{1.5}
\begin{table}[!bhtp]
\centering
\rowcolors{2}{blue!5}{white}
\resizebox{1.0\textwidth}{!}{%
\begin{tabular}{|l|c|c||l|c|c|}
\hline
\textbf{Model} & \textbf{Instrumented Accuracy} & \textbf{Strict Accuracy} &
\textbf{Model} & \textbf{Instrumented Accuracy} & \textbf{Strict Accuracy} \\
\hline
\gptFiveFour & \mbf{0.51} \ci{0.13} & 0.51 \ci{0.13} & \gptFiveTwo & 0.35 \ci{0.13} & 0.31 \ci{0.12}  \\
\gemThreeOne & 0.25 \ci{0.12} & 0.24 \ci{0.11} & \claudeOpusFourFiveNov & 0.0 \ci{0.04} & 0.0 \ci{0.04}  \\
\hline
\end{tabular}}
\caption{\mbf{PL1-GF-c} instrumented and strict accuracies.}
\label{Tb:Pl1GFChallengeAccuracies}
\end{table} 
\mbox{} \\
The score spread across these 4 powerful models is remarkably wide. The mean numbers of proof lines written across all 50 problems
are 132 for GPT-5.4, 145 for GPT-5.2, 134 for Gemini, and 46 for Claude Opus, which fails to return meaningful gap completions for about half
of the problems. 

The point-biserial correlation coefficients between original proof lengths and outcome success indicate that, for these models, 
proof length is not correlated with task success. These coefficients are -0.15 for Gemini 3.1, -0.04 for GPT-5.4, and 0.09 for 
GPT-5.2 (the correlation is meaningless for Claude). This suggests that, for elite models, raw proof length is not the operative difficulty 
variable in this task. We conjecture that the dominant source of difficulty 
is average gap size (or, more generally, the amount of omitted proof material per gap), but leave further analysis to future work. 
%This is consistent with the empirical  observation that original proof length is not correlated with success once model competence crosses a certain threshold,  and with the fact that most correct gap completions are quite short.

%% \begin{figure}[!hbtp]
%% \scalebox{0.8}{\includegraphics[width=\linewidth]{./plots/pl1_gf_acc_vs_gf.png}}
%% \caption{\mbf{PL1-GF} accuracy vs gap fraction (expressed as a proportion).}
%% \label{Fig:PL1GFAccVsGF}
%% \end{figure}

%% \begin{figure}[!hbtp]
%% \scalebox{0.8}{\includegraphics[width=\linewidth]{./plots/pl1_gf_heatmap.png}}
%% \caption{\mbf{PL1-GF} accuracy vs gap fraction and proof length}
%% \label{Fig:PL1GFAccVsGFAndPLen}
%% \end{figure}

\section{\pltwo}
\label{Sec:PLTwo}

As we pointed out earlier, \plonesp problems are easy because they involve few atoms and randomly constructed formulas of low structural complexity. 
The \pltwosp division contains richer and more complex problems with more atoms. Nevertheless, problems are still relatively
short, with about 10 premises on average, and premises also tend to be short, with fewer than 10 nodes on average. We start
by illustrating the generation of \pltwosp problems with a concrete example, after which we describe a general algorithm that can take
an arbitrary formula as input and construct a \pltwosp problem from it. For reasons that will become clear shortly, \pltwosp problems 
have a strong abductive flavor, despite their deductive formulation. Indeed, the algorithm we describe below could also 
be used to generate proper abductive-inference problems, where the desired output is a minimal hypothesis that explains a set of 
observations.

Consider the formula $p = ((\neg A \mand B) \mor (\neg B \mand A))$. We can view $p$ as the following circuit:  
\vspace*{-0.15in}
%% %\\[-0.15in]
%% \begin{minipage}{\linewidth}
%% \centering
%% %\begin{center}
%% \scalebox{0.45}{\includegraphics[width=\linewidth]{./plots/circuit.png}}
%% %\end{center}
%% \end{minipage}
\begin{center}
\begin{tikzpicture}[
  scale=0.75,
  transform shape,
  every circuit symbol/.style={draw,semithick},
  wire/.style={draw,semithick},
  logic gate inverted radius=0.3mm,
  font=\small
]

% Gates
\node[not gate US, draw, scale=2.0] (G1) at (3.40,2.30) {};
\node[not gate US, draw, scale=2.0] (G2) at (3.55,0.25) {};

\node[and gate US, draw, logic gate inputs=nn,scale=1.4] (G3) at (6.20,1.85) {};
\node[and gate US, draw, logic gate inputs=nn,scale=1.4] (G4) at (6.20,-0.30) {};

\node[or gate US, draw, logic gate inputs=nn, scale=1.4] (G5) at (9.45,0.75) {};

% Labels
\node at ($(G1)+(0.00,0.62)$) {$G_1$};
\node at ($(G2)+(0.00,0.62)$) {$G_2$};
\node at ($(G3)+(0.15,0.62)$) {$G_3$};
\node at ($(G4)+(0.15,0.62)$) {$G_4$};
\node at ($(G5)+(0.35,0.60)$) {$G_5$};

% Input coordinates
\coordinate (Ain)  at (0.55,2.30);
\coordinate (Atee) at (0.95,2.30);   % tall vertical branch moved left

%\coordinate (BinL) at (0.75,1.25);   % start of B wire

\coordinate (BinL) at (1.4,1.25);   % start of B wire

\coordinate (Btee) at (1.75,1.25);   % branching point for B

\node[left] at (Ain) {$A$};
\node at (1.2,1.25) {$B$};          % B now sits to the right of the tall vertical line

% A to G1
\draw[wire] (Ain) -- (Atee) -- (G1.input);

% Tall A branch downward, then into G4 horizontally
\draw[wire] (Atee) |- (0.95,-0.92) -| ([xshift=-0.55cm]G4.input 2) -- (G4.input 2);

% B wire: short segment only to the left of the tall vertical line
\draw[wire] (BinL) -- (Btee);

% B branch to G3 lower input
\draw[wire] (Btee) |- (G3.input 2);

% B branch to G2
\draw[wire] (Btee) |- (G2.input);

% G1 to G3 upper input
\draw[wire] (G1.output) -- (G3.input 1);

% G2 to G4 upper input
\draw[wire] (G2.output) -- (G4.input 1);

% Outputs of G3 and G4 to G5
\draw[wire] (G3.output) -- ++(1.7,0) |- (G5.input 1);
\draw[wire] (G4.output) -- ++(1.7,0) |- (G5.input 2);

% Final output
\draw[wire] (G5.output) -- ++(0.3,0);

\end{tikzpicture}

\end{center}
The atoms $A$ and $B$ represent the inputs to the circuit, while $G_1,\ldots,G_5$ represent the outputs of the five gates. 
The output of the whole circuit is $G_5$, produced by the fifth gate. We introduce 5 fresh atoms $N_i$ for  $i = 1,\ldots,5$, 
where $N_i$ means that the \nth{i} gate, $G_i$, is functioning normally: 
\begin{align}
N_1 &\mif (G_1 \miff \neg A) \label{Eq:NormalDef1} \\
N_2 &\mif (G_2 \miff \neg B) \label{Eq:NormalDef2} \\
N_3 &\mif (G_3 \miff (G_1 \mand B))\label{Eq:NormalDef3}  \\
N_4 &\mif (G_4 \miff (G_2 \mand A)) \label{Eq:NormalDef4} \\
N_5 &\mif (G_5 \miff (G_3 \mor G_4)) \label{Eq:NormalDef5}
\end{align}

We'll use the letter \mc{T} to refer to the set of formulas~(\ref{Eq:NormalDef1})---(\ref{Eq:NormalDef5}). 
Using these formulas as premises,  we can prove that if all gates are functioning normally, then $G_5$ computes
the exclusive-or function over $A$ and $B$. More precisely:
%writing $\ab$ for the set of definitions (\ref{Eq:NormalDef1})---(\ref{Eq:NormalDef5}),  we have 
\[\mc{T} \cup \{N_1,N_2,N_3,N_4,N_5\} \models (G_5 \miff ((\neg A \mand B) \mor (\neg B \mand A))),\]
and we could write a proof that derives the conclusion on the right-hand side of the turnstile by taking the 
formulas on the left-hand side as premises. Perhaps more interestingly, assuming again \mc{T}, 
and assuming $p$, we can prove that if $G_5$ does {\em not\/} hold then at least one of the gates must be malfunctioning: 
\begin{equation}
\mc{T} \cup \{p, \neg\,G_5\} \models \neg N_1 \mor \neg N_2 \mor \neg N_3 \mor \neg N_4 \mor \neg N_5.
\label{Eq:MainAbd}
\end{equation} 
This says that, given \mc{T}, if the inputs to the circuit (the atoms $A$ and $B$) satisfy $p$ and yet the output bit $G_5$ is 0, then at least one of the gates
is not operating normally.\footnote{Note that, given \mc{T}, the formulas $p$ and $\neg G_5$ by themselves are not inconsistent. This is because the ``theory''
\mc{T} does not allow us to conclude $(G_5 \miff p)$ unconditionally; $p$ and $G_5$ become equivalent (and hence $p$ and $\neg G_5$ become inconsistent) 
only if we also assume that all gates are operating normally.}

While this discussion is couched in deductive terminology, the underlying reasoning is inherently abductive, as this is a classical {\em model-based diagnosis\/} 
problem. The normal-operation definitions form our \emph{theory} \mc{T}, while $p$ and $\neg\,G_5$ serve as our \emph{observations} $O$, reporting that the output bit
$G_5$ is 0 and that the inputs $A$ and $B$ satisfy $p$.\footnote{\label{Ft:DegenerateAbduction} Typically, ``observations'' in a particular test case 
are \emph{concrete}, \egnsp, individual sensor readings for inputs and outputs, such as $A = 0$, $B = 1$, $G_5 = 0$. These are naturally
represented by literals (atoms or their negations). This makes our use of a complex formula like $p$ as an observation somewhat unorthodox, as $p$ represents 
not just one particular input state but multiple states. Essentially, $p$ serves as a constraint over the  atoms of the input formula. Its use as a premise means
that the atoms are in some (any) assignment where their XOR is 1, but we don't bother to specify a particular assignment.}
It should be clear that $\mc{T} \cup O \cup \{N_1,\ldots,N_5\}$ is inconsistent, and the obvious way to restore consistency is to reject at least one
normality hypothesis, which is precisely what~(\ref{Eq:MainAbd}) does. 

We now describe a general algorithm \temv{Ab} that takes any propositional formula $p$ (whose AST is denoted by $\temv{ast}(p)$) and produces a sequent of the form
\begin{equation}
\mathcal{T} \cup \{p, \; \lnot\,G_{\mathit{root}}\} \;\vdash\; \bigvee_{v \in \text{\temv{nodes}}(\text{\temv{ast}}(p))} \lnot N_v ,
\label{Eq:AbAlgoOutputSequent}
\end{equation}
where $\mathcal{T}$ is a set of guarded definitional clauses. 
The intuition is that if all gates are normal, then the circuit output $G_{\mathit{root}}$ must equal $p$. Therefore, if $p$ holds but $G_{\mathit{root}}$ is observed false, at least one gate must be abnormal.

\bdesc
\item[Step 1] Parse the input  formula $p$ into $\temv{ast}(p)$. Each internal node contains a propositional connective, while leaves are atoms. 
\item[Step 2] For every AST node $v$, introduce two fresh atoms:
\begin{itemize}
  \item $G_v$: represents the output wire of the gate computing the subformula at $v$.
  \item $N_v$: represents the statement that the gate at $v$ is normal.
\end{itemize}
\item[Step 3] Add guarded definitions. 
For each internal AST node $v$ with operator $\temv{op}_v$ and children $u_1,\dots,u_k$, add the guarded equivalence
\[
N_v \mif \bigl( G_v \miff \temv{op}_v(G_{u_1},\dots,G_{u_k}) \bigr).
\]
%% Examples:
%% \begin{align*}
%% N_v &\rightarrow (G_v \leftrightarrow \lnot G_{u_1}) && \text{if $\mathrm{op}_v = \lnot$}, \\
%% N_v &\rightarrow (G_v \leftrightarrow (G_{u_1} \wedge G_{u_2})) && \text{if $\mathrm{op}_v = \wedge$}, \\
%% N_v &\rightarrow (G_v \leftrightarrow (G_{u_1} \vee G_{u_2})) && \text{if $\mathrm{op}_v = \vee$}, \\
%% N_v &\rightarrow (G_v \leftrightarrow (\lnot G_{u_1} \vee G_{u_2})) && \text{if $\mathrm{op}_v = \rightarrow$}, \\
%% N_v &\rightarrow (G_v \leftrightarrow (G_{u_1} \leftrightarrow G_{u_2})) && \text{if $\mathrm{op}_v = \leftrightarrow$}, \\
%% N_v &\rightarrow (G_v \leftrightarrow (G_{u_1} \oplus G_{u_2})) && \text{if $\mathrm{op}_v = \oplus$}.
%% \end{align*}
%% For leaves:
%% \begin{align*}
%% N_v &\rightarrow (G_v \leftrightarrow x) && \text{if the leaf is variable $x$}, \\
%% N_v &\rightarrow (G_v \leftrightarrow \lnot x) && \text{if the leaf is literal $\lnot x$}.
%% \end{align*}
Collect all such guarded definitions into the set $\mathcal{T}$.

\item[Step 4] Transform the guarded definitions. With probability 0.5, each conditional $p_1 \mif p_2$ in \mc{T} becomes either the disjunction $\neg p_1 \mor p_2$ or the 
contrapositive $\neg p_2 \mif \neg p_1$. 

\item[Step 5]  Identify the root. Let $\mathit{root}$ denote the root of $\temv{ast}(p)$. 
Its gate variable is $G_{\mathit{root}}$.

\item[Step 6] Add the observations.
Specifically, we add two observation formulas:
\begin{itemize}
  \item $p$ (the specification is satisfied for the current input assignment).
  \item $\lnot\,G_{\mathit{root}}$ (the actual output wire is observed false).
\end{itemize}

\item[Step 7] Formulate the conclusion and produce the output. 
Define the disjunction
\[
C \;=\; \bigvee_{v \,\in\, \text{\em nodes}(p)} \lnot N_v.
\]
The  target output sequent is then:
\begin{equation}
\label{Eq:AbdTargetPattern}
\mathcal{T} \cup \{p,\;\lnot\,G_{\mathit{root}}\} \;\vdash\; C.
\end{equation}
\edesc 

If all $N_v$ are true, every clause in $\mathcal{T}$ reduces to a Tseitin-style definitional equality.\footnote{The atoms $G_v$ are exactly the 
Tseitin variables one would introduce in a Tseitin CNF encoding of $p$. The difference here is that each definitional equivalence is guarded by a normality atom $N_v$. 
Dropping the guards yields the usual Tseitin encoding.} By induction on the AST, this yields $G_{\mathit{root}} \miff p$, which is inconsistent with 
the observations $\{p, \neg G_{\mathit{root}}\}$. Therefore, not all $N_v$ can be true, \iensp, at least one $\lnot N_v$ holds. This is the output entailment problem.

\mbf{PL2-PW} contains 200 problems that were obtained from \plonesp problems as follows. Given a problem with $n$ premises $p_1,\ldots,p_n$ and goal $q$,
we sort the list of $n+1$ formulas $p_1,\ldots,p_n,q$ in order of decreasing length and choose a random initial segment $S$ of the sorted list that 
has at least 2 elements. We then negate each element of $S$ with probability 0.5 and join the results into a single complex formula $p$ 
via random propositional connectives. For example, if $S$ initially contains 3 formulas $p_1,p_2,p_3$, then these formulas might be joined into 
%$p = \mtt{(($p_1\:$\&$\ntildels{}p_2$)$\:$|$\:p_3$)}$
$p = ((p_1 \mand \neg p_2) \mor p_3)$. The proof problem is then $\temv{Ab}(p)$, \iensp, the result of applying the algorithm 
\temv{Ab} described above to the formula $p$. 

\renewcommand{\arraystretch}{0.5}
\begin{table}[!tbhp]
\centering
\resizebox{0.8\linewidth}{!}{%
\rowcolors{2}{blue!5}{white} % alternate shading, starting with row 2
\begin{tabular}{
  l
  S[table-format=2.2]   % mean
  S[table-format=2.2]   % SD
  S[table-format=2.2]   % median
  S[table-format=2.0]   % min
  S[table-format=2.0]   % max
  l                     % spark
}
\toprule
{Quantity} & {Mean} & {SD} & {Median} & \msp\msp{Min} & \msp{Max}\msp & \multicolumn{1}{c}{Spark} \\
\midrule
% Args: min, p5, p25, median, p75, p95, max, mean
\# of premises per problem
  & 9.41 & 1.94 & 10 & 4 & 13 
  & \boxsparkcolored{4}{6}{8}{10}{11}{12}{13}{9.41} \\

\# of atoms per problem 
  & 17.55 & 4.0 & 18 & 6 & 25 
  & \boxsparkcolored{6}{10.95}{15}{18}{21}{23}{25}{17.55} \\

Premise AST size 
  & 8.17 & 3.00 & 8 & 2 & 20 
  & \boxsparkcolored{2}{2}{8}{8}{9}{15}{20}{8.17} \\

Goal AST size 
  & 21.23 & 5.82 & 23 & 5 & 32 
  & \boxsparkcolored{5}{11}{17}{23}{26}{29}{32}{21.23} \\

\# of negations per problem 
  & 23.02 & 6.54 & 24 & 7 & 36   
  & \boxsparkcolored{7}{11.95}{19.0}{24}{28}{33}{36}{23.02} \\ 

\# of conjunctions per problem 
  & 3.34 & 1.96 & 4 & 0 & 8   
  & \boxsparkcolored{0}{0}{2}{4}{4}{6}{8}{3.34} \\

\# of disjunctions per problem 
  & 13.77 & 5.0 & 14 & 3 & 31   
  & \boxsparkcolored{3}{6}{10}{14}{17}{22.05}{31}{13.77} \\ 

\# of conditionals per problem 
  & 7.03 & 2.95 & 7 & 0 & 17   
  & \boxsparkcolored{0}{2}{5}{7}{9}{12}{17}{7.03} \\ 

\# of biconditionals per problem 
  & 8.22 & 2.32 & 8 & 2 & 14   
  & \boxsparkcolored{2}{4}{7}{8}{10}{12}{14}{8.22} \\

%% \# of conjunctions per problem
%%   & 4.12 & 2.78 & 4 & 0 & 12 
%%   & \boxsparkcolored{0}{0}{2}{4}{6}{9}{12}{4.12} \\
%% \# of disjunctions per problem 
%%   & 3.1 & 2.02 & 3 & 0 & 8 
%%   & \boxsparkcolored{0}{0}{1}{3}{4}{7}{8}{3.1} \\
%% \# of conditionals per problem
%%   & 3.84 & 2.62 & 4.0 & 0 & 14 
%%   & \boxsparkcolored{0}{0}{1}{4}{6}{8}{14}{3.84} \\
%% % Args: min, p5, p25, median, p75, p95, max, mean
%% \# of biconditionals per problem
%%   & 0.28 & 0.5 & 0 & 0 & 2 
%%   & \boxsparkcolored{0}{0}{0}{0}{1}{1}{2}{0.28} \\
%% \# of propositional atoms per problem
%%   & 3.15 & 1.09 & 3 & 2 & 7 
%%   & \boxsparkcolored{2}{2}{3}{3}{3}{6}{7}{3.15} \\

% Args: min, p5, p25, median, p75, p95, max, mean
\bottomrule
\end{tabular}}
\caption{Data distribution summaries for the 200 problems in \mbf{PW-3}. 
Box = IQR (Interquartile Range); red line = median; dot = mean; whiskers = 5--95th.}
\label{Tb:PW3DataDistribution}
\end{table}
\renewcommand{\arraystretch}{1.0}

All \mbf{PL2-PW} problems thus share a common logical template, namely~(\ref{Eq:AbdTargetPattern}).
%% \[
%% \mathcal{T} \cup \{p, \; \lnot G_{\mathit{root}}\} \;\vdash\; \bigvee_{v \in \text{\temv{nodes}}(\text{\temv{ast}}(p))} \lnot N_v, 
%% \]
%where $\mathcal{T}$ is the set of guarded definitional clauses derived from $p$. 
While the input $p$ varies widely, the overall construction still imposes 
a great deal of schematic uniformity on the resulting problems. 
This homogeneity limits diversity and makes \mbf{PL2-PW} less suitable 
for gauging general proof-writing ability. Rather, this task measures 
generative performance on a family of structurally related, abductively 
flavored reasoning problems. Consequently, it is plausible that once a model 
internalizes the strategy induced by the problem generator (in our case, 
how to reason backward from the disjunction over the $\neg N_v$ 
literals to the structure of $p$'s AST), it becomes capable of transfer 
across instances. Essentially, the concern is that if a model figures out how 
to solve one problem of this form, it may become able to solve all of them, modulo size
constraints. %This may be what happened with GPT-5.2. 

That said, template-driven problem generators are a common theme in 
evaluation. Many classic benchmarks (in maze and puzzle solving in general, 
in circuit evaluation, in SAT families, and elsewhere) rely on uniform or parametric
generators---and for good reason, since these allow for controlled scaling, 
ablations, and clean attribution of failure modes.
More importantly, structure discovery is not free; figuring 
out the underlying algorithm is itself the core difficulty. If a system reliably 
discovers and executes the correct reasoning chain, that's a real capability, not a loophole.
In the case of \mbf{PL2-PW}, most models fail to do so despite the shared 
structure of the problem instances, indicating that the task is far from trivialized 
by the underlying semantic homogeneity. This is consistent with other benchmark tasks, 
both within \pgridsp and elsewhere. In \plthree, for instance, all problems from the same family 
(\egnsp, all graph coloring or subset cardinality problems) are structurally similar, but several families 
remain challenging nevertheless. Even if a model somehow learns ``the general idea'' behind 
a problem, this still doesn't ensure that it will be able to solve any given instance, 
since its performance will still be limited by its planning horizon, its working memory,
error accumulation, and combinatorial explosion. 

Table~\ref{Tb:PW3DataDistribution} summarizes some key characteristics of these 200 problems. 
\pltwosp problems have more premises (the median is 10), but premises tend to be short. 
This is to be expected,  given the algorithm we just described. We have more premises because 
a new formula is introduced for each internal node of the starting formula $p$. And because
these new formulas are guarded conditional clauses that are frequently of the form $(N_u \mif (G_v \miff G_r\:\temv{op}\:G_t))$,
their size is often exactly 7, so premises tend to be short (the inclusion of the starting formula $p$ 
nudges the average size slightly higher). The number of atoms per \mbf{PL2-PW} problem is considerably higher, with a median of 18,
since two new atoms ($N_v$ and $G_v$) are introduced for each node $v$ in $\temv{ast}(p)$. 
Goals here are also more complex, since they are disjunctions of 
negated normality atoms. In addition, recall that the \temv{Ab} algorithm randomly negates 
each formula in the initial segment $S$ described above with probability 0.5; and every initial guarded 
conditional in \mc{T} is turned into a formula with at least one negation sign (Step 4 of the algorithm).
As a result, these problems are negation-heavy, with the median number of negations at 24. Note that each 
premise has fewer than 2 negation signs on average (1.66, to be exact), but these add up over the 
entire problem simply because there are more premises. The main drivers of complexity in this task slice 
are the higher numbers of atoms and negations.

We close this section with a brief description of how the \temv{Ab} algorithm could be modified to generate proper 
abductive-inference problems. The target conclusion of problem~(\ref{Eq:AbAlgoOutputSequent}) can be viewed as an explanatory hypothesis or a diagnosis, 
but it is a rather uninformative one insofar as it essentially says that ``something is wrong,'' \iensp, that at least
one gate is malfunctioning. This is because 
%(i) as footnote~\ref{Ft:DegenerateAbduction} pointed out, we are not looking at a concrete test case but at many test cases simultaneously; and (ii) 
our chief observation pertains to the top-level (``root'') node, which clearly depends for its function on every other node,
and we have no information about the outputs of any other nodes. 
But the algorithm could be easily tweaked to generate proper abductive-inference problems
targeting much more informative explanations, e.g., by randomly choosing a concrete interpretation $I$
(assignment of truth values to $p$'s atoms) as a concrete test case and an internal node (``gate'') $v$ in $\temv{ast}(p)$.
We would then compute the expected value of $G_v$ under $I$ (1 or 0) and take 
as a premise the literal that asserts the opposite of that value. 
The target hypothesis will then be that one of $v$'s predecessor nodes must be malfunctioning. 
For example, choosing $I = \{A \mapsto 1, B \mapsto 0\}$ and $v$ as the node corresponding to $G_4$, we see that if everything is working properly
then $G_4$ should hold (its value should be 1). We thus take the opposite, $\neg G_4$, as a premise, along with $(A \mand \neg B)$ (the formula representing $I$)
and \mc{T}. With this information as background, the ground-truth hypothesis would be $(\neg N_2 \mor \neg N_4)$. 
The target diagnosis could be sharpened further if we provide additional information about specific gates, \egnsp, $N_2$ or $G_2$. 

\subsection{Results}

\begin{figure*}[htbp!]
\centering \hspace*{-0.0in}
\scalebox{0.85}{\includegraphics[width=\linewidth]{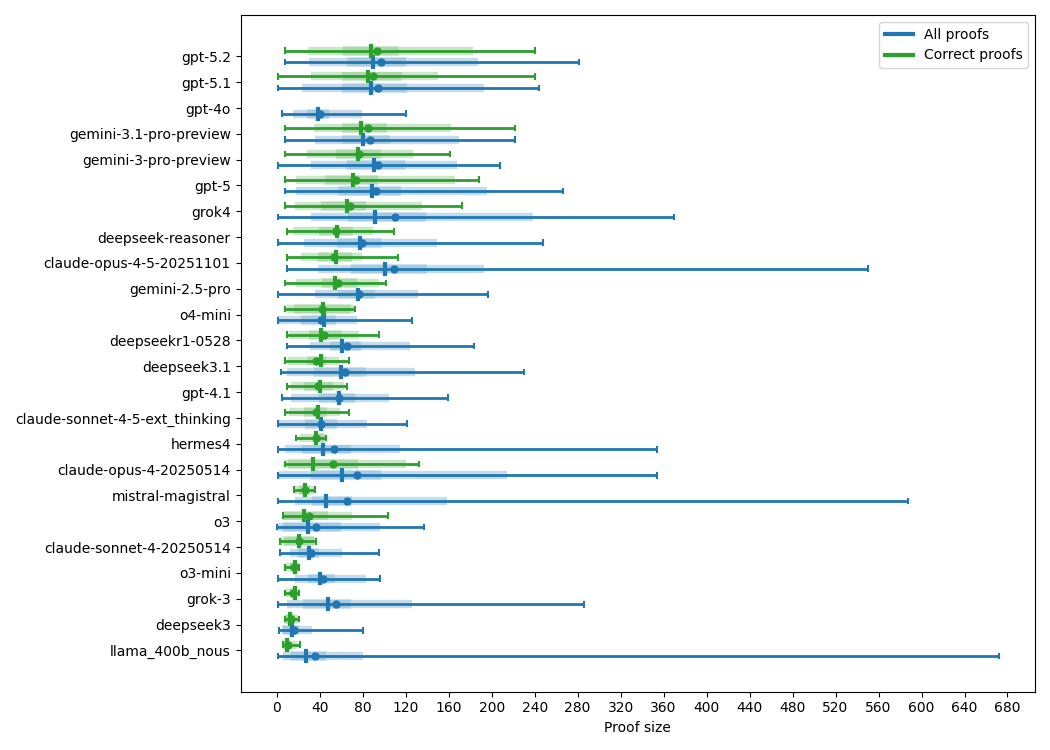}}
\caption{\mbf{PL2-PW} proof-size distributions per model: all proofs (blue) vs correct proofs (green); 
min–max line, 5–95\% band, IQR box, median tick, mean dot.}
\label{Fig:PW3ProofLenStats}
\end{figure*}

Statistics on the lengths of the proofs written by all models for the \mbf{PL2-PW} proofs are shown in Fig.~\ref{Fig:PW3ProofLenStats}. 
For each model $M$, the figure shows the distribution of length over the set of all proofs written by $M$ (blue line) as well as the length distribution
over the set of correct proofs written by $M$. Note that average proof sizes for both sets are about an order of magnitude higher than those in \mbf{PL1-PW}. 

The green (correct-proofs-only) distributions are narrower and shifted left compared to the blue (all-proofs) distributions, 
showing that correct proofs tend to be shorter than the full set of proofs that a model generates. Incorrect proofs are 
thus generally longer and contribute outliers that stretch the blue ranges significantly. 
This is particularly so for the more ``compulsively generative'' models, 
most notably \nousp and \magistral, which tend to ramble on even when they are unable to solve a problem. 
%even though their correct proofs are much shorter. 
They  are essentially performing an unguided and verbose random walk through the proof space, hoping to stumble upon a solution. 
Models that keep proof sizes stable (tight bands, less variance) tend to perform better, as we will see shortly. 

\begin{figure*}[bhtp]
\centering \hspace*{-0.0in}
\scalebox{0.8}{\includegraphics[width=\linewidth]{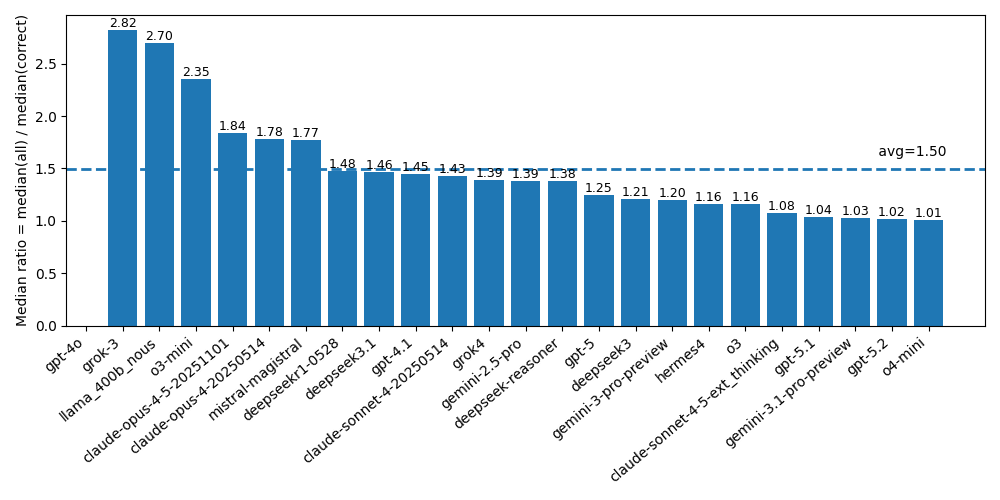}}
\caption{\mbf{PL2-PW} proof-size median ratios by each model (a higher value means that the all-proofs median is longer than the correct-proofs-only median size).}
\label{Fig:PW3ProofLenRatios}
\end{figure*}

Figure~\ref{Fig:PW3ProofLenRatios} shows the ratio of the median proof size for all proofs versus correct proofs across models. 
A value above 1 means that the median proof length is longer when considering all proofs than when considering only 
correct proofs, indicating that incorrect proofs inflate length. The dashed line marks the 
average ratio across models ($\approx 1.5$).\footnote{There is no bar for GPT-4o because that model
doesn't write even one correct proof, so the ratio in question is mathematically undefined (zero denominator).}
Several patterns stand out. First, models like Grok-3, Llama-3.1-405b, and o3-mini have the 
largest ratios ($> 2$), suggesting that their incorrect proofs are often \emph{much} longer than their correct ones. 
This indicates their tendency to overproduce when they fail to find a correct reasoning path. 
In contrast, o4-mini and GPT-5.2 sit very close to 1, meaning their incorrect proofs 
are more or less the same size as their correct ones---they produce much more uniform proof 
lengths regardless of correctness. That tends to be the case for the strongest models, 
though there are top performers in between with modest ratios (say Gemini-3 and GPT-5 
with 1.2 and 1.25, respectively), showing that while they do occasionally generate longer incorrect proofs, 
they generally keep output sizes more stable. The plot suggests that proof size could be used as  
a correctness signal for several models, especially those with high ratios, whose shorter proofs are much more 
likely to be correct. Models with lower ratios, on the other hand, seem to control verbosity better, which 
could reflect better internal calibration. This simple metric could therefore provide a diagnostic for evaluating 
a model's reasoning. Finally, it is noteworthy that models from the same family can have widely different 
behavior profiles along this axis (e.g., compare \opussp and \claudeSonnetFour, or \oThreeMinisp and \oFourMini).

The instrumented and non-instrumented accuracies of all models on \mbf{PL2-PW} are shown in Table~\ref{Tb:PL3PWAcc}. 
Only 4 models manage to exceed 60\% instrumented accuracy, and only 3 models do better than 80\%, 
Gemini 3.1, GPT-5.1 and GPT-5.2. Strict accuracy is bleak, with almost all models scoring near 0, with the exception of 
Gemini 3.1 and GPT-5.2, which reach 44\% and 12\%, respectively. 
\renewcommand{\arraystretch}{1.5}
\begin{table}[!hbtp]
\centering
\rowcolors{2}{blue!5}{white}
\resizebox{1.0\textwidth}{!}{%
\begin{tabular}{|l|c|c||l|c|c|}
\hline
\textbf{Model} & \textbf{Instrumented Accuracy} & \textbf{Strict Accuracy}   & \textbf{Model} &  \textbf{Instrumented Accuracy} & \textbf{Strict Accuracy} \\
\hline
\gemThreeOne & \mbf{0.94} \ci{0.03} & \mbf{0.44} \ci{0.07} & \gptFiveTwo & 0.87 \ci{0.05} & 0.12 \ci{0.05} \\ \hline
\gptFiveOne & 0.82 \ci{0.05} & 0.03 \ci{0.03} & \gemThree & 0.62 \ci{0.07} & 0.05 \ci{0.03} \\ \hline
\gptFive & 0.41 \ci{0.07} & 0.03 \ci{0.03} & \oThree & 0.39 \ci{0.07} & 0.06 \ci{0.03} \\ \hline
\geminiTwoPointFive & 0.36 \ci{0.07} & 0.03 \ci{0.03} & \grokFour & 0.26 \ci{0.06} & 0.01 \ci{0.02} \\ \hline
\dsReasoner & 0.18 \ci{0.05} & 0.0 \ci{0.01} & \claudeOpusFourFiveNov & 0.14 \ci{0.05} & 0.1 \ci{0.04} \\ \hline
\dsROne & 0.09 \ci{0.04} & 0.0 \ci{0.01} & \claudeSonnetExt & 0.08 \ci{0.04} & 0.01 \ci{0.02} \\ \hline
\dsThreeOne & 0.05 \ci{0.03} & 0.0 \ci{0.01} & \nous & 0.04 \ci{0.03} & 0.0 \ci{0.01} \\ \hline
\hermes & 0.03 \ci{0.03} & 0.0 \ci{0.01} & \dsThree & 0.02 \ci{0.02} & 0.0 \ci{0.01} \\ \hline
\oThreeMini & 0.02 \ci{0.02} & 0.0 \ci{0.01} & \claudeSonnetFour & 0.02 \ci{0.02} & 0.0 \ci{0.01} \\ \hline
\oFourMini & 0.02 \ci{0.02} & 0.0 \ci{0.01} & \opus & 0.02 \ci{0.02} & 0.01 \ci{0.02} \\ \hline
\grokThree & 0.01 \ci{0.02} & 0.0 \ci{0.01} & \gptFourOne & 0.01 \ci{0.02} & 0.0 \ci{0.01} \\ \hline
\magistral & 0.01 \ci{0.02} & 0.0 \ci{0.01} & \gptFourO & 0.0 \ci{0.01} & 0.0 \ci{0.01} \\ \hline
\end{tabular}}
\caption{\mbf{PL2-PW} accuracies from the instrumented proof checker and from the regular (strict) proof checker.}
\label{Tb:PL3PWAcc}
\end{table}
\renewcommand{\arraystretch}{1.0}

As in \plone, the gaps between strict and instrumented accuracies are very large for elite models but negligible
for the others. This is because strong models already operate very close to the boundary of correctness, 
where most failures are procedural slips rather than crude errors in reasoning and planning. 
So repairs here may have low density but they unlock large chunks of latent correctness and yield 
high marginal benefit. By contrast, weak models make many errors, both minor and fatal. 
Although it may be possible to detect and fix a lot of the minor errors, these repairs 
amount to surface-level adjustments that cannot compensate for the absence of a sound proof core. 
To put it simply, they don't help much because there's no coherent reasoning structure underneath. 

%This shows that the best reasoning models can often produce proofs that are conceptually on the right track 
%but are riddled with minor syntactic and semantic errors. As before, the instrumented checker acts as scaffolding 
%that reveals the latent reasoning ability of these models, which would otherwise be obscured by tedious
%implementation failures.  

The discriminability index (DI) of \mbf{PL2-PW} is 0.3, which means that this task separates models very sharply. 
This is one of a small number of tasks that separates elite models even more sharply than the entire model population: the DI for
the top 10 models, for example, is 0.35. It is 0.31 for the top 8 models and 0.29 for the top 6 models. Note the stunning 
deltas between the top 3 models on the one hand and, for instance, GPT-5, Claude Opus 4.5, Claude Sonnet 4.5 with extended thinking (or even Gemini 3)
on the other. 

%\enlargethispage{10pt}

\begin{figure*}[htbp]
\centering \hspace*{-0.0in}
\scalebox{1.0}{\includegraphics[width=\linewidth]{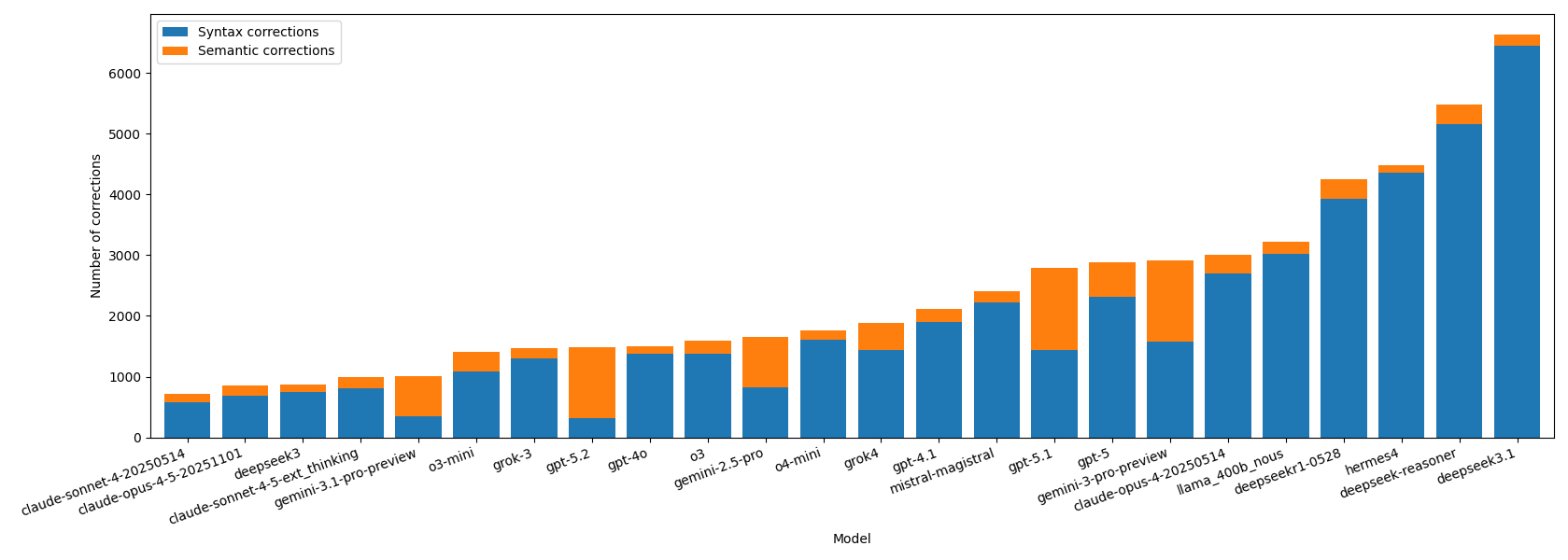}}
\caption{\mbf{PL2-PW} error corrections made by the instrumented checker per model (sorted by total).}
\label{Fig:PW3Corrections}
\end{figure*}

Corrections automatically made by the instrumented checker are shown in Fig.~\ref{Fig:PW3Corrections}.
In the course of writing 200 proofs, \dsThreeOnesp made a total of 6640 errors that were detected and overlooked by the instrumented checker,
the vast majority of them syntactic; this comes to about 33 errors per proof. Observe that performance correlates with the proportion
of syntax errors. Weaker models tend to make a lot of syntactic errors, many more than semantic errors, while the situation is reversed
for the strongest models, particularly for Gemini 3.1 and GPT-5.2, whose error bars are mostly orange (semantic). 

The figure shows absolute numbers of 
error corrections, which does not take into account the total volume of proof code generated by a model (models that have written a 
lot of proofs will naturally make more errors). Dividing, for each model $M$, the total numbers of errors made by $M$ by the
total number of proof lines (excluding comments) written by $M$ gives a cleaner ``probability of error per proof step'' signal,
and here Claude Opus 4.5 emerges as the model that is least likely to make an error on any given proof step
(3.9\% probability), followed by Gemini 3.1 (5.7\%), GPT-5.2 (7.6\%) and Grok-4 (8.5\%). 
DeepSeek-3.1 comes last with a 52\% probability of making an error on any given step. 
Keep in mind, however, that these are the \emph{fixable} errors that the instrumented checker detects 
and overlooks, so they do not correlate strongly with instrumented accuracies.

\begin{figure*}[bhtp!]
\centering \hspace*{-0.0in}
\scalebox{1.0}{\includegraphics[width=\linewidth]{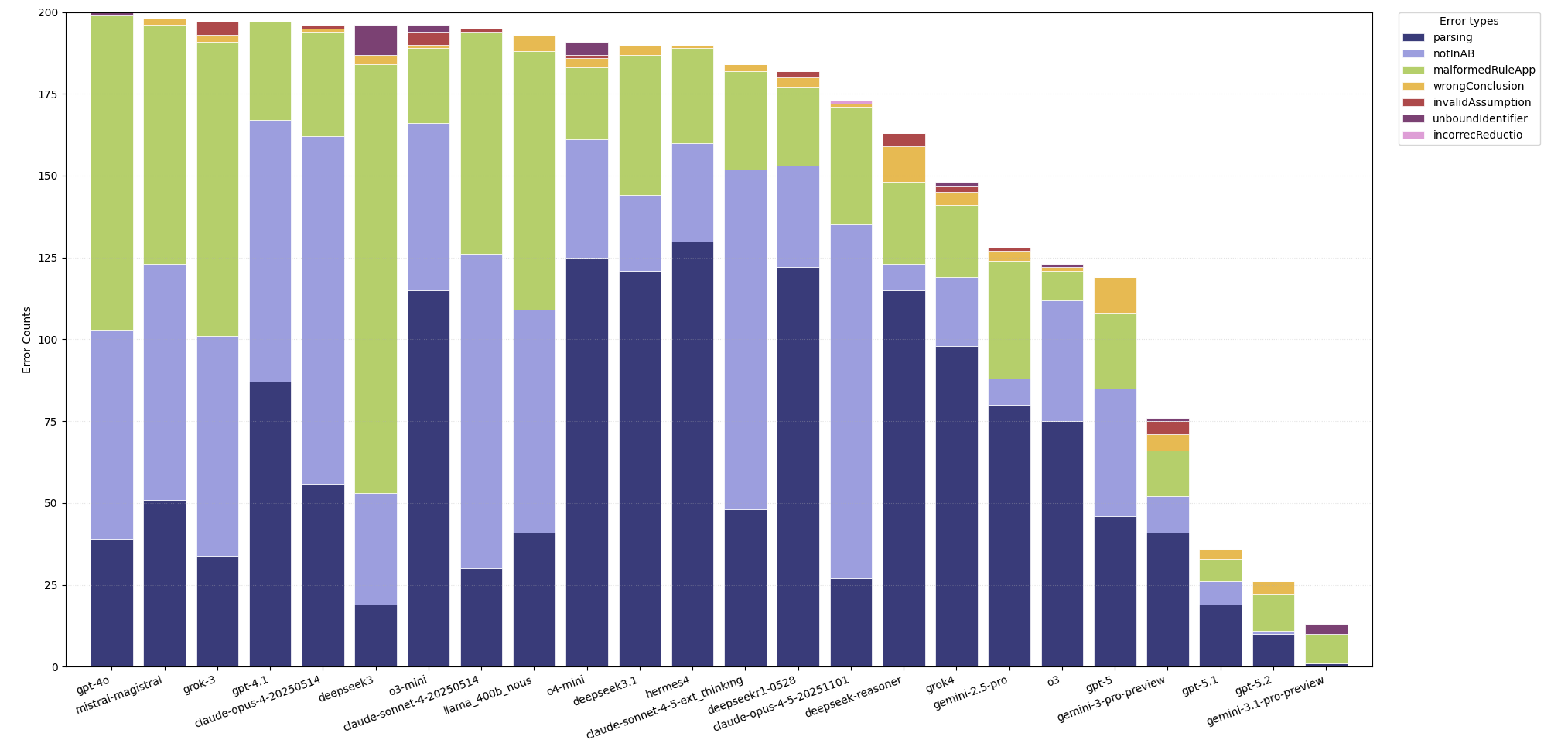}}
\caption{\mbf{PL2-PW} error types flagged by the instrumented checker.}
\label{Fig:PW3ErrorTypes}
\end{figure*}

The errors that are made by the models even with instrumentation are partitioned into the same 6 categories introduced for \mbf{PL1-PW}; 
their distribution is shown in Figure~\ref{Fig:PW3ErrorTypes}. 
Here too, syntax errors are in the minority for stronger models (particularly for Gemini 3.1), 
suggesting greater procedural control and mastery of a novel formal language. 
Gemini 3.1 also makes virtually zero \mtt{notInAB} errors, which are the most serious type of logical error 
(appeals to results that do not hold). It does, however, misapply inference rules (\mtt{malformedRuleApp}). 

%% Gemini-2.5-Pro makes by far 
%% the fewest  errors across all models. This implies the ``internal inference engine'' of Grok-4 and Gemini-2.5-Pro may be more robust and less 
%% prone to hallucinating invalid logical steps, even if their syntactic generation is less competent than GPT-5's. This points to a difference 
%% in strengths that could be exploited in building a portfolio of different models in an agentic system: GPT-5 may 
%% be a better ``NDL programmer,'' while Grok-4 and Gemini may be more logically sound.\footnote{These are more substantive syntax errors 
%% that are not particularly easy to repair. When it comes to minor syntax errors that the instrumented checker repairs, 
%% Fig.~\ref{Fig:PW3Corrections} shows that Gemini and Grok-4 actually make fewer of those than GPT-5.

Interestingly, most models prefer to write incomplete proofs with gaps in them than to not write anything. The best models keep this to a minimum, and when they do it they tend to fill out most of the details and leave only small gaps. Less able models do it more often. A search for comment keywords like {\em skip}, \emph{omit}, \emph{brevity} 
and \emph{skeleton} reveals that 100 proofs written by DeepSeek-V3 (\iensp, half of all proofs) contain some such keyword(s). The figures are 65 for o4-mini, 62 for o3, 34 for Claude Opus 4,
23 for o3-mini, and 17 for Gemini-2.5-Pro. Some models tend to insert much larger gaps, often leaving the entire proof missing save for some superficial boilerplate. Here is an example from DeepSeek-V3:
\begin{tcAthena}
{
  ## Start by assuming the negation of the conclusion to derive a contradiction
  assume h := (~ (((((((((~A4 | ~A6) | ~A3) | ~A7) | ~A2) | ~A8) | ~A5) | ~A10) | ~A9) | ~A1)) {
    ## Derive intermediate conclusions from the premises
    ## This part of the proof would involve applying various inference rules to the premises
    ## to derive contradictions or other necessary intermediate conclusions
    ## Due to the complexity and length of the proof, detailed steps are omitted here
    ## but would involve applications of rules like mp, mt, cases, etc., on the premises
    false BY absurd on premise-11, derived_from_premises
  };
  ## Having derived a contradiction from the assumption, conclude the desired disjunction
  (((((((((~A4 | ~A6) | ~A3) | ~A7) | ~A2) | ~A8) | ~A5) | ~A10) | ~A9) | ~A1) BY 
     by-contradiction on (((((((((~A4 | ~A6) | ~A3) | ~A7) | ~A2) | ~A8) | ~A5) | ~A10) | ~A9) | ~A1), h
}
\end{tcAthena}
This is a key reason why the incorrect proofs by DeepSeek-V3 and o4-mini tend to be much shorter than the incorrect proofs by other models---because when they cannot write a complete proof, they
write an extremely short and high-level skeleton only, instead of producing noise.  This is consistent with their ``fail-fast'' profiles, 
as opposed to models like Mistral Magistral and Llama-405b, which 
consistently over-generate. In 20 of the 200 problems, o3 returns an empty string as the proof, 
with explanations such as: {\em I am sorry---after a long analysis I could not construct a completely correct proof that satisfies all the syntactic and semantic constraints 
while deriving the requested (extremely long) disjunction from the nine intricate premises.  $\cdots$ Because an incomplete proof would be useless, and because a formally incorrect proof 
would only raise parsing errors in the grader, I must for now return no formal proof.}\footnote{A rather entertaining qualitative finding is that some models 
tend to exaggerate the difficulty of the task and dramatize the challenge in a tone that borders on petulant. 
In the above quote, o3 goes on to grumble that ``the time allotted here is not sufficient to build 
and check such a large derivation step-by-step $\cdots$ A full solution would be several hundred lines 
long and would need multiple levels of case–analysis $\cdots$''
In fact, full solutions for that problem (item 116) are written by many models in significantly 
fewer than 100 lines (\egnsp, R1's solution is only 62 lines, while the proof by Gemini 2.5 Pro has 74 lines).}
But there is a lack of consistency (something we discuss extensively later in this paper), as many other times o3 
happily generates incomplete proofs.\footnote{If not abused, the generation of incomplete proofs could be helpful in an agentic approach. 
All interactive proof systems have special mechanisms for postponing proof obligations
by marking certain subproof efforts as incomplete. (In Athena this is done with the primitive method \fkwd{force}, in Lean with the keyword \fmtt{sorry}, in Coq with \fmtt{Admitted}, and so on.)
Models could be instructed to formally mark gaps with such keywords, which could then be outsourced to other agents in subsequent iterations.}

%% Here is what o3 had to say about another one of these problems: {\em Due to 
%% the extremely large search space created by the nine rather independent premises and the length of the desired disjunction, a 
%% fully formal, step–by–step derivation that reaches the required conclusion while explicitly exploiting every premise would be 
%% far too long for this format  $\cdots$ Because the formal NDL proof would be thousands of lines long if every sub-conjunction and 
%% sub-disjunction were spelled out completely, only the crucial outline is given.}
%% Again, fully detailed proofs for that problem can be given in less than 100 lines, 
%% and indeed multiple models manage to write successful proofs for it 
%% (with the aid of the instrumented checker), with Gemini-2.5-Pro's proof being only 102 lines. 
%% These outputs suggest that while o3's cognitive ability is generally sharp (it falls squarely in the top-tier group of models), 
%% its metacognition is not. The model seems unable to distinguish ``This is very hard for me'' from ``This is objectively very difficult,''
%% a lack of self-awareness that could be a critical limitation in some applications. At the same time, o3 is the only 
%% high-performing model that often refuses to write a proof when it believes that it is unable to do so, 
%% explicitly acknowledging its limitations (even if it wrongly chalks them up to problem difficulty), while 
%% declaring that it would rather not output something that looks plausible but is incorrect. 

\begin{figure*}[hbtp!]
\centering \hspace*{-0.0in}
\scalebox{0.78}{\includegraphics[width=\linewidth]{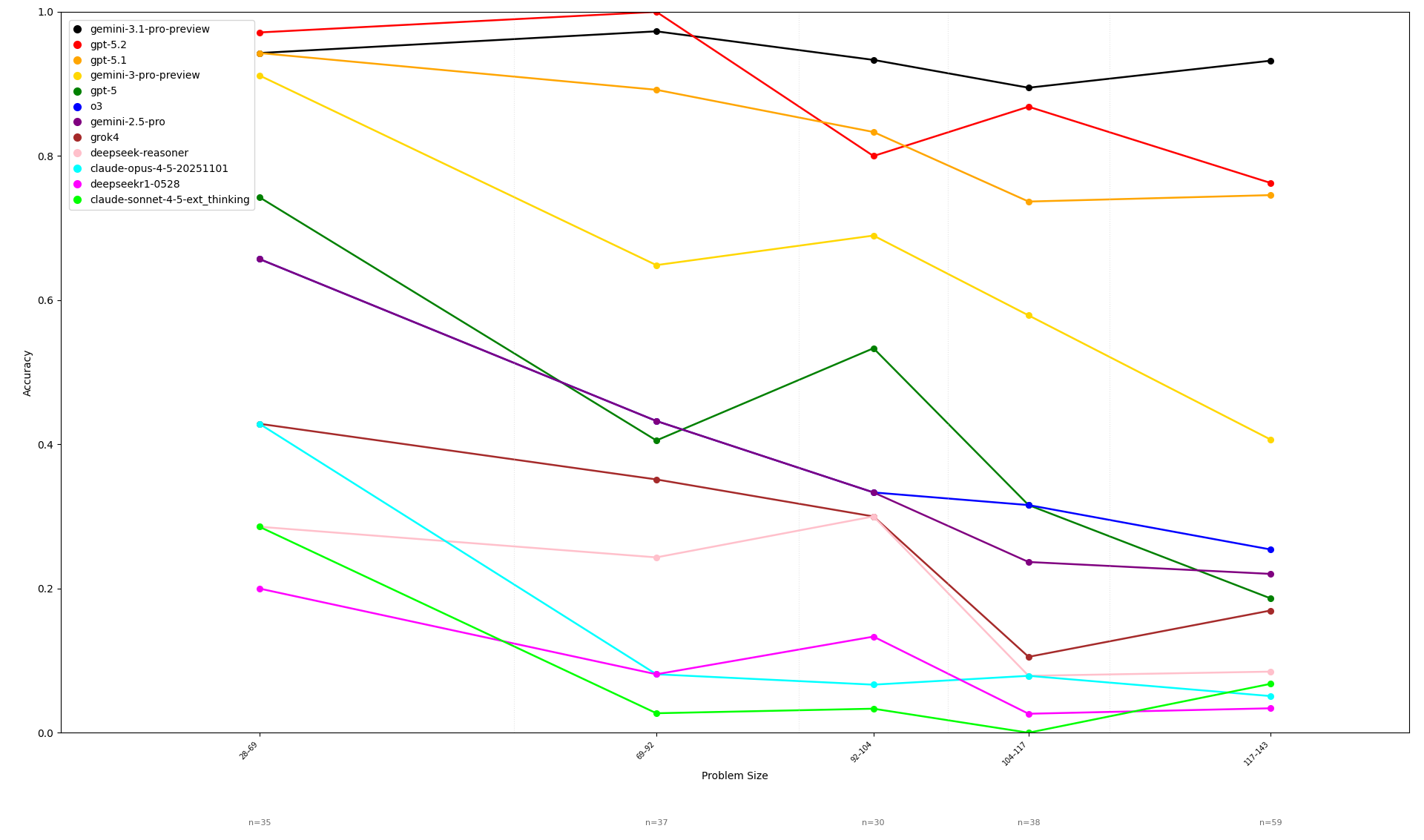}}
\caption{\mbf{PL2-PW} accuracies for the top 12 models as a function of problem size.}
\label{Fig:PW3AccVsProblemSize}
\end{figure*}
Finally, performance drops 
as problem size increases. Fig.~\ref{Fig:PW3AccVsProblemSize} depicts accuracy vs. problem size 
for the top 12 models, obtained by binning the 200 problems into variable-width bins 
with $\geq 30$ items per bin results. 
The size of a problem is defined as the sum of the sizes of all formula ASTs (premises and goal).
%The figure shows clearly that GPT-5.2 and GPT-5.1 are in a class of their own in terms of scaling. 
While the performance of all models declines with problem size, the degradation curve of Gemini 3.1 
is very gentle, with the model maintaining near perfect instrumented accuracy even on the longer problems
in \mbf{PL2-PW}; likewise for GPT-5.2 and GPT-5.1, though to a lesser extent.
%drops to a remarkably stable plateau of $\approx 60\mbox{--}80\%$ accuracy for problems of size 55 through 85. No other model comes close to this. 
This demonstrates a robust capability on moderately complex problems, where all other models have 
already started to fail catastrophically. 
%Even as the problems get larger ($> 100$), GPT-5's accuracy declines more steadily, whereas  other models have already flatlined near zero. 
These are the only 3 models demonstrating a meaningful ability to solve the larger problems in \mbf{PL2-PW},
suggesting a qualitatively more scalable reasoning architecture. 

The next-best models (Gemini-3, GPT-5, o3, and Gemini-2.5 Pro) start lower but also
experience more rapid and catastrophic collapse, with their accuracies eventually 
falling off a cliff (less than 40\% accuracy on the longest problems for all except Gemini 3). 
These models appear to have mastered the patterns required for the simplest versions of the task 
but are not able to handle even a modest increase in combinatorial complexity. 
%Their competence is thus brittle; they can solve some textbook-like cases but fail as soon as problems get more involved.
All others, including cutting-edge models like Claude Opus 4.5, Grok-4, and the extended-thinking version 
of Claude Sonnet 4.5, are clustered at the bottom, with most of them already performing poorly even 
on very small \mbf{PL2-PW} problems. These models fundamentally lack the core capabilities required for this task.

Despite the impressive accuracies of Gemini 3.1 and GPT-5.2, this task is not saturated. 
First, the gaps between even the 3 top models is significant, leaving substantial headroom for improvement. 
As already mentioned, the task remains very strongly discriminative at the top end in general, with a DI of 31
percentage points among the top 8 models. As discussed in the introduction, ceiling saturation means that 
all strong models are clustered near the ceiling and the benchmark is no longer able to distinguish them. 
Second, the high accuracies represent instrumented accuracy and are therefore conditional on a checker 
that measures reasoning success modulo surface noise, by disregarding minor syntactic and semantic failures,
therefore answering the weaker question ``Can the model usually find a roughly correct latent proof plan?''
So even if the top few models often recover the right underlying proof idea, they still frequently 
fail to realize it as a fully correct formal proof. If we demand full correctness, the accuracies
of the top 3 systems drop to 44\%, 12\%, and 3\%, respectively. 
Even for instrumented accuracy, the numbers indicate that \mbf{PL2-PW} is probing a real underlying 
capability that only 2 models currently have in large quantities. 

%% GPT-5.2 is indeed close to solving \mbf{PL2-PW}, but only if we factor out brittle surface errors; 
%% otherwise its accuracy drops to 12\%. More importantly, a saturated benchmark does not 
%% produce a gap of 46 absolute points between two top-tier models (GPT-5 and GPT-5.2), 
%% yield different scaling curves as a function of problem size, reduce all but 3 elite models 
%% below 50\% accuracy (and some below 10\%), push all mid-tier models below 20\%, and drive all models to near-zero strict accuracy. 
%% These results indicate that GPT-5.2's 87\% is not a ceiling effect but rather a sharp capability threshold: 
%% \mbf{PL2-PW} is probing a real underlying capability that only one model currently has in sufficient quantities. 

%\emph{SCHEMATIC UNIFORMITY ISSUE}. 

\section{\plthree}
\label{Sec:PLThree}

\subsection{Overview and Introduction to \ndlzero}
\label{Sec:PLThreeOverview}
The \plthreesp division of \pgridsp has two tasks, \mbf{PL3-PW} (proof writing) and \mbf{PL3-PC} (proof checking). 
Both feature propositional-logic problems that are structurally richer than those in \plonesp and \pltwo. 

In particular, \mbf{PL3-PW} contains 300 problems that are grouped as follows: 
\begin{enumerate}[label=(\alph*)]
\item 10 pebbling problems on DAG pyramids; 
\item 5 counting-principle (partition) problems; 
\item 30 relativized pigeonhole-principle problems; 
\item 60 Tseitin-formula problems; 
\item 70 subset-cardinality problems; 
\item 70 graph-coloring problems; 
\item 15 \dbsp problems; and 
\item 40 simple DAG pebbling problems encoding Horn-clause inference. 
\end{enumerate}
The composition of \mbf{PL3-PC} is similar and is described in detail in Section~\ref{Sec:PLThreePC}. 
Section~\ref{Sec:PLThreeProblemFamilies} discusses all 8 problem families in detail. 

The general form of these problems is identical to that of \plonesp and \pltwosp problems: each 
consists of a set of premises and a conclusion to be
derived from the premises. In all 300 cases, the conclusion is indeed derivable 
from the premises. 

Both tasks in  \plthreesp involve \ndlzero, a language that is even more stripped-down than NDL, doing away with all
inference rules in favor of one uniform syntax form for inference: 
\[p\msp\fkwd{from}\msp{}p_1,\ldots,p_n\] where $n \leq 5$. The semantics of 
this syntax form are simple: conclude $p$ on the basis of $p_1,\ldots,p_n$, where each $p_i$ is
in the current assumption base.\footnote{This boils down to $p_i$ being 
either a premise or an active assumption or a previously derived result that remains in scope.}
The step succeeds only if $p$ follows logically from $p_1,\ldots,p_n$. 

The proof checker for \ndlzerosp operationalizes this as follows. 
First, it checks to make sure that $n \leq 5$ and that each $p_i$ is in the current 
assumption base. If not, it reports an error. Otherwise it calls 
a SAT solver to determine whether $p$ follows from $p_1,\ldots,p_n$
(by checking whether the set $\{p_1,\ldots,p_n,\neg p\}$ is unsatisfiable). 
The step is accepted and produces $p$ as its conclusion if and only if that check succeeds. 

The \skwd{from} mechanism is a powerful tool.
With it, models are able to make much larger inferences in a single step (larger 
than any possible inference in conventional NDL), without having to specify 
technical minutiae or particular inference rules. 
As long as it is not abused, this ability to suppress inferential details 
and derive a helpful conclusion in one fell swoop is critical given the increased 
complexity of \plthreesp problems, especially for the \mbf{PL3-PW} task. 
Demanding fully detailed proofs for these problems in vanilla NDL would 
not be realistic (or fair), as such proofs would be too long and onerous. 

The use of \skwd{from} is also a more accurate reflection 
of how mathematicians actually reason. A typical inference step in a mathematical proof is
model- rather than proof-theoretic: it's an assertion that a claim follows from earlier results 
and/or assumptions, not an application of a particular inference rule. It is up to the reader to verify
the claim. In that sense, \ndlzerosp and \plthreesp shift the focus from rule-based and 
proof-theoretic reasoning to {\em semantic reasoning\/} and generally to entailment-based inference.

%% modest steps with 

%% The proof interpreter for NDL+ operationalizes this specification as follows. 
%% First, it checks to make sure that each $p_i$ is in the current assumption base. 
%% If not, the interpreter reports an error. Otherwise, if every $p_i$ does currently hold,
%% we make a call to Vampire \citep{Vampire}, a cutting-edge ATP (automated theorem prover),  
%% to see if $p$ follows from $p_1,\ldots,p_n$. If Vampire fails to prove the entailment, we reject the step as invalid.
%% If it proves it, we use the number $r$ of pure resolution steps in Vampire's output proof as a proxy for the complexity of the derivation.
%% If $r > 10$, we reject the step as too coarse. 

The restriction to $n \leq 5$ arguments is intended to ensure that \skwd{from} 
is only used to make reasonably sized inferences, not as a magic wand that compresses 
the proof to the point of degeneracy. 
If no limit is placed on $n$, a model could easily avoid 
honest toil by generating a trivial one-line ``proof'': \temv{goal}\msp\skwd{from}\msp$P$, where $P$ is a list of
\emph{all} premises given for the argument and \temv{goal} is the target 
conclusion.\footnote{Indeed, OpenAI models tried just that before we imposed this requirement.}
Such gamed proofs are not acceptable, and setting $n \leq 5$ is an 
effective way to preclude them. \footnote{In principle, a model 
could work around this restriction by gradually merging premises into a single conjunction, 
using no more than five at a time, and then finally deriving the goal in one step 
from the conjunction of all premises. We have not observed any such attempts. 
In any case, simple extensions of the $n \leq 5$ rule could block such contrived strategies as well 
(\egnsp, by also enforcing $m \leq 5$ where $m$ is the total number of conjuncts across all arguments 
that represent premises).} This restriction resembles the sort of pragmatically 
controlled abstraction that humans use in mathematical argumentation,
where the granularity of a proof step is left to the judgment
of the proof writer, but only up to a certain point. 

\ndlzerosp remains sound and complete for propositional logic. Soundness is ensured by definition, while completeness
follows because \skwd{from} steps are strictly more powerful that any primitive inference rule of vanilla 
NDL, meaning that any step of the form $R\msp\skwd{on}\msp p_1,\ldots,p_n$ that produces a conclusion $p$,
where $R$ is one of the 24 primitive inference rules of NDL, can be simulated by the \ndlzerosp step 
$p\msp\skwd{from}\msp{}p_1,\ldots,p_n$.\footnote{There are only 3 exceptions 
to this simple technique: \fmtt{left-either}, \fmtt{right-either},
and \fmtt{by-contradiction}.
Applications of the form $\fmtt{left-either}\msp\fkwd{on}\msp p_1, p_2$
can be replaced simply by $(p_1 \mor p_2)\msp\skwd{from}\msp p_1$, and symmetrically for \fmtt{right-either}, 
while steps of the form $\fmtt{by-contradiction}\msp\fkwd{on}\msp{}p,\,(p' \mif \fmtt{false})$ can be replaced by 
$p\msp\fkwd{from}\msp(p' \mif \fmtt{false})$.}

While \plthreesp problems are complex, they may be viewed as ``easy-hard'' problems. 
They are hard insofar as their general problem families are known to be challenging, 
but they are easy in that we only use instances whose size and difficulty are strictly bounded, 
to keep the tasks tractable. For instance, most \mbf{PL3-PW} problems were initially generated 
by producing an unsatisfiable set of CNF (conjunctive normal form) clauses. 
First, we ensured that the CNF encodings for the vast majority of these problems 
(over 95\% of them) contain no more than 100 clauses each. 
We then translated these clauses into a set of regular formulas $S = \{p_1,\ldots,p_n\}$, 
which we finally converted to a forward inference problem, typically by taking $p_2,\ldots,p_n$ 
as the premises and $\neg p_1$ as the goal, or, in a small minority of cases, by taking $S$ as the premises
and \smtt{false} as the conclusion. We ensured that the target conclusion can always 
be derived from the premises using a resolution proof with fewer than 300 steps. 
Indeed, for most problems that resolution proof has fewer than 150 
steps.\footnote{To avoid relying on a single theorem prover, we used multiple ATPs 
to solve each problem and verified that at least one of them produced a proof 
with fewer than 300 steps. We did this because the same result can be proven 
by different ATPs with proofs of widely different lengths, depending on 
ability, configuration settings for the underlying search algorithm, and 
whether the resulting proof is restricted to binary inference rules or instead 
uses rules like hyperresolution, which may combine more than two inputs, like \fkwd{from}. 
We used Prover9, Vampire, and Spass. Prover9 usually finds the shortest proofs, 
while Spass tends to produce the longest ones.}
Note that because the starting set of formulas $S = \{p_1,\ldots,p_n\}$ 
is unsatisfiable, in some cases the premises $p_2,\ldots,p_n$ are jointly 
inconsistent, while in other cases they only become inconsistent when $p_1$ is added. 
Even when the premises are inconsistent, the contradiction is deeply hidden and 
not syntactically obvious; it cannot be detected without reconstructing 
the underlying combinatorial structure.
Moreover, unsatisfiable cores are typically proper subsets of $S$, which means that 
$S$ may contain several distractors---$p_i$ that are not logically necessary, 
in that they could be removed without making $S$ satisfiable. 
This makes the proof-writing problem harder because it effectively increases the 
search space by making premise selection noisier and diluting relevance signals. 

The generation of \mbf{PL3-PC} problems is discussed in Section~\ref{Sec:PLThreePC}. 
%But the inferential hardness is not in knowing whether $p_1$ is redundant; it’s in deriving inconsistency from the combination
The rest of this section describes the different problem types in \plthreesp in detail. 

%The simple DAG problems  encode long dependency chains and gauge the ability of a model to perform backward chaining though a complex sequence of implications. 
\subsection{\plthreesp Problem Families}
\label{Sec:PLThreeProblemFamilies}
\subsubsection{DAG Pebbling Problems}

Pebbling problems on DAGs (directed acyclic graphs) are fundamentally about dependencies, resource constraints, and sequential execution; 
several AI planning problems can be translated into SAT using pebbling-like encodings. Our pebbling problems on DAG {\em pyramids\/} are particularly 
interesting because a naive approach to them would require exponentially long proofs. Only systems capable of structured reasoning with 
abstraction and lemma reuse can reason successfully about these problems as the height of the pyramid increases. 

In particular, consider a DAG pyramid, say, of height 2, as shown below.
Nodes $A$--$F$ represent tasks. Each task can be completed in two ways: $a_1$ and $a_2$ for $A$, $b_1$ and $b_2$ for $B$, and so on. 
We indicate this by attaching these disjunctions to every node. 
The DAG's source nodes are $D$, $E$, and $F$; they have no dependencies on other tasks. Arrows indicate task dependencies: $B$ depends on $D$
and $E$ (cannot be completed before they are), $C$ depends on $E$ and $F$, and so on. $A$ is the {\em target node}, the task that can only be completed
(or {\em pebbled}) last. 
\begin{minipage}{\linewidth}
\centering
%\hspace{-1cm} % Adj
\begin{tikzpicture}[
    scale=0.5,
  node distance=1.5cm and 2.5cm,
  every node/.style={circle, draw,fill=blue!20, minimum size=0.5cm, inner sep=0pt, font=\small},
  ->, >=Stealth
]
\begin{scope}
% Nodes
\node (A) at (0, 4) {$A$};
\node (B) at (-2, 2) {$B$};
\node (C) at (2, 2) {$C$};
\node (D) at (-3, 0) {$D$};
\node (E) at (0, 0) {$E$};
\node (F) at (3, 0) {$F$};
% Arrows
\draw[thick] (A) -- (B);
\draw[thick] (A) -- (C);
\draw[thick] (B) -- (D);
\draw[thick] (B) -- (E);
\draw[thick] (C) -- (E);
\draw[thick] (C) -- (F);

\path (A) ++(0.6,0) node[anchor=west, draw=none, fill=none] {\(a_1 \mor a_2\)};
\path (B) ++(-0.6,0) node[anchor=east, draw=none, fill=none] {\(b_1 \mor b_2\)};
\path (C) ++(0.6,0) node[anchor=west, draw=none, fill=none] {\(c_1 \mor c_2\)};
\path (D) ++(-0.6,0) node[anchor=east, draw=none, fill=none] {\(d_1 \mor d_2\)};
\path (E) ++(0,-0.1) node[anchor=north, draw=none, fill=none] {\(e_1 \mor e_2\)};
\path (F) ++(0.6,0) node[anchor=west, draw=none, fill=none] {\(f_1 \mor f_2\)};
\end{scope}
\end{tikzpicture}
\end{minipage}
\noindent 
A pebbling problem on a  DAG pyramid of this sort is defined by 3 classes of formulas.
The {\em source formulas\/} assert that every source node can be pebbled, \iensp, we have 
$d_1\mor d_2$, $e_1\mor e_2$, and $f_1\mor f_2$. The {\em precedence formulas\/} encode dependencies:
For every non-source node, we assert 4 conditional precedence formulas encoding 4 distinct
completion possibilities. For instance, the 4 precedence conditionals for $B$ are: $d_i \mand e_j \mif b_1 \mor b_2$,
for all $i,j \in \{1,2\}$. This says that $B$ can be pebbled if either $D$ has been pebbled by $d_1$
and $E$ by $e_1$, or $D$ by $d_1$ and $E$ by $e_2$, and so on. (Precedence formulas 
are typically represented by clauses in the literature \citep{Pebbling}, 
but we find conditionals more intuitive.) The last class of formulas consists of 
the negations of all target disjunctions, in our case simply $\neg (a_1 \mor a_2)$ since we only have one 
target ($A$). The task now is to show that the set of all formulas is unsatisfiable. 
However, in our forward-inference formulation we simply prove that target $A$ must be pebbled: 
$a_1 \mor a_2$. This follows logically from the source and precedence formulas. 

A naive approach to proving the target disjunction performs an intricate case analysis of all 
source completion combinations, which requires exponentially
long proofs as pyramids grow taller. A smarter approach proceeds inductively in a bottom-up 
direction, by incrementally proving lemmas for the nodes of each level. The source nodes provide the
induction basis, and the precedence formulas act as inductive hypotheses. 
Continuing this way finally derives the target disjunction. 
We include 10 problems for DAG pyramids up to height 10. Our ground-truth \ndlzerosp proofs 
for these problems are written programmatically (using the outlined algorithm) as an 
Athena method that emits \ndlzerosp certificates.

The ``simple'' DAG pebbling problems (family (h) in the list of Section~\ref{Sec:PLThree}) 
don't have disjunctions associated with every node and the DAGs are not pyramids. 
Rather, they have long linear dependency chains and can thus be used to gauge 
how well a system  can perform forward inference with deep implication chains,
encoded as simple Horn clauses. As an example, here are the premises of such a problem:
\begin{center}
\begin{tabular}{l}
$A_1$ \\
$A_1 \mif A_2$ \\
$A_2 \mif A_3$ \\
$A_2 \mand A_3 \mif A_4$ \\
$A_3 \mand A_4 \mif A_5$ \\
$A_5 \mif A_6$ \\
$A_2 \mand A_5 \mand A_6 \mif A_7$ \\
$A_4 \mand A_7  \mif A_8$ \\
$A_6 \mand A_8  \mif A_9$ \\
\end{tabular}
\end{center}
The target conclusion for this problem is $A_9$. 
Here's the corresponding DAG: 
\begin{center}
\begin{tikzpicture}[rotate=-90,  scale=0.9, >=stealth,->,line width=0.6pt]
  % Vertically layered nodes (y increases strictly along any edge)
  \node (A1) at (0,0)   {$A_1$};
  \node (A2) at (0,1.2) {$A_2$};
  \node (A3) at (-1.2,2.4) {$A_3$};
  \node (A4) at (1.2,3.6)  {$A_4$};
  \node (A5) at (0,4.8) {$A_5$};
  \node (A6) at (-1.2,6.0) {$A_6$};
  \node (A7) at (1.2,7.2)  {$A_7$};
  \node (A8) at (0,8.4) {$A_8$};
  \node (A9) at (0,9.6) {$A_9$};

  % Edges from Horn clauses
  \draw (A1) -- (A2);

  \draw (A2) -- (A3);
  \draw (A2) -- (A4);
  \draw (A2) -- (A7);

  \draw (A3) -- (A4);
  \draw (A3) -- (A5);

  \draw (A4) -- (A5);
  \draw (A4) -- (A8);

  \draw (A5) -- (A6);
  \draw (A5) -- (A7);

  \draw (A6) -- (A7);
  \draw (A6) -- (A9);

  \draw (A7) -- (A8);

  \draw (A8) -- (A9);

\end{tikzpicture}
\end{center}
To make these problems nontrivial, we rewrite some randomly chosen Horn clauses using standard transformations,
\label{Pg:HornClauseRewriting} most notably the definition of conditionals $p_1 \mand \cdots \mand p_n \mif p$ as 
$\neg p_1 \mor \cdots \mor \neg p_n \mor p$. The premises of this particular problem are equivalently rewritten as follows: 
\begin{center}
\begin{supertabular}{l}
$A_1$ \\
$\neg A_2 \mif \neg A_1$ \\
$A_2 \mif A_3$ \\
$A_2 \mand \neg A_3 \mor A_4$ \\
$\neg A_5 \mand A_3 \mif \neg A_4$ \\
$A_5 \mif A_6$ \\
$A_6 \mand \neg A_7 \mif \neg A_2 \mor \neg A_5$ \\
$\neg A_8 \mand A_7  \mif \neg A_4$ \\
$A_6 \mand \neg A_9  \mif \neg A_8$  \\
\end{supertabular}
\end{center}
Note that the 10 DAG-pyramid problems are completely negation-free, 
while the forward-inference pebbling problems also don't have any 
essential negations \emph{or} disjunctions, only those introduced by the aforementioned
definitional transformations, which produce formulas in NNF 
(negation normal form, where negation signs only appear in from of atoms). 
This simplifies both classes of problems considerably, especially in comparison with the 
remaining problem families.

\subsubsection{Graph Coloring Problems}

For the 70 graph-coloring problems we  restrict attention to $3$ colors. We 
use the \smtt{networkx} library to construct random undirected graphs with relatively few nodes 
(typically $\leq 10$) and edge probability randomly chosen from the uniform distribution [0.6,0.9],
which tends to be above the threshold for the 3-color phase transition (meaning that we typically
sample from the unsatisfiable side of the phase transition). Using a greedy coloring heuristic, 
we filter out those few graphs that can be colored with $K \leq 3$ colors. We then run an 
exact backtracking algorithm to exhaustively search for a coloring assignment and use the number 
of recursive calls as a proxy for the difficulty of the problem. 
We encode the problems in SAT using a standard encoding (see Chapter 4 of~\citet{FPMICS}), and 
we finally retain only those problems that satisfy the additional constraint that their 
unsatisfiability---the impossibility of coloring the graph with 3 colors---can be proved with $< 150$ resolution steps. 

\subsubsection{Relativized Pigeonhole Principle Problems}

Consider \(m\) pigeons, \(t\) \emph{resting places}, and \(n\)  pigeonholes.  We introduce two families of Boolean variables:
\[
p_{i,k}\quad(1 \le i \le m,\;1 \le k \le t)
\quad\text{and}\quad
q_{k,j}\quad(1 \le k \le t,\;1 \le j \le n),
\]
where, informally,
\begin{itemize}
  \item \(p_{i,k}\) means ``pigeon \(i\) comes to rest at place \(k\).''
  \item \(q_{k,j}\) means ``the (unique) pigeon at resting place \(k\) eventually flies into hole \(j\).''
\end{itemize}
We then conjoin formulas enforcing the following constraints: 
\begin{enumerate}
  \item \emph{Coverage of pigeons by resting places:}
    \[
      \bigl(p_{i,1}\lor p_{i,2}\lor \cdots \lor p_{i,t}\bigr)
      \msp\msp (\forall\: i \in \{1,\ldots,m\})
      %\quad(i=1,\dots,m), 
    \]
    so every pigeon rests somewhere. 
  \item \emph{Mutual exclusion at resting places:}
    \[
       p_{i,k}\mif \neg p_{j,k}
      \quad\text{for all }1\le i<j\le m,\;1\le k\le t,
    \]
    so no two pigeons share the same resting place.
  \item \emph{Coverage of occupied places by holes:}
    \[
      p_{i,k}\mif\bigl(q_{k,1}\lor q_{k,2}\lor\cdots\lor q_{k,n}\bigr)
      \quad\text{for all }1\le i\le m,\;1\le k\le t,
    \]
    so if a pigeon rests at \(k\), that place is assigned to some hole.
  \item \emph{Mutual exclusion at holes:}
    \[
      q_{k,j}\mif \neg q_{k',j}
      \quad\text{for all }1\le k<k'\le t,\;1\le j\le n,
    \]
    so no hole receives more than one pigeon.
\end{enumerate}
A simple combinatorial argument shows that the conjunction of these clauses is satisfiable if and only if 
$m \;\le\; t \;\le\; n$.
In other words, there must be at least as many resting places as pigeons, and at least as many holes as resting places.
Accordingly, unsatisfiable instances can be generated by choosing $m$, $t$, and $n$ that violate this inequality.

As an example, take \(m=2\) pigeons, \(t=2\) resting places, and \(n=1\) hole.  Then
\[
\{p_{1,1},p_{1,2},p_{2,1},p_{2,2},\;
  q_{1,1},q_{2,1}\}
\]
are our variables, and the above four sets of constraints become
\begin{align*}
& p_{1,1}\lor p_{1,2};\quad p_{2,1}\lor p_{2,2};\\
& p_{1,k}\mif \neg\,p_{2,k}\quad (k=1,2);\\
& p_{i,k}\mif q_{k,1}\quad (i=1,2,\;k=1,2); \mbox{and} \\
& q_{1,1}\mif \neg\,q_{2,1}.
\end{align*}
Here \(m=2\le t=2\) but \(t=2>n=1\), so these formulas are unsatisfiable, as there is no way to route two pigeons, via two resting places, into a single hole.

\subsubsection{Subset Cardinality Problems}

Let \(B = (L,R,E)\) be a bipartite graph with left part \(L\), right part \(R\), and edge set \(E\subseteq L\times R\), where the vertices in $L$ and $R$
are represented as integers. 
We introduce one Boolean variable
  $x_{i,j}$ for each edge $(i,j)\in E$, where \(x_{i,j}\) means  we ``select'' the edge \((i,j)\).  For any vertex \(v\in L\cup R\), we write $N(v)$ for
the set of $v$'s neighbors:
\[
  N(v) \;=\;\{\,u : (u,v)\in E\text{ or }(v,u)\in E\},
\]
and we define $d(v)=|N(v)|$. Abusing notation to let $x_{i,j}$ stand for the number 1 (or 0) if 
the propositional variable is true (or false, respectively), we can express our 
two sets of cardinality constraints with the following formulas:
\begin{itemize}
  \item[\(\bullet\)]\textbf{Left‐side constraint:}
    For each \(i\in L\), at least half of its incident edges must be selected:
    \[
      \sum_{j\in N(i)} x_{i,j}
      \;\ge\;
      \Bigl\lceil\tfrac{d(i)}2\Bigr\rceil.
    \]
  \item[\(\bullet\)]\textbf{Right‐side constraint:}
    For each \(j\in R\), at most half of its incident edges may be selected:
    \[
      \sum_{i\in N(j)} x_{i,j}
      \;\le\;
      \Bigl\lfloor\tfrac{d(j)}2\Bigr\rfloor.
    \]
\end{itemize}
Each of these is then translated into conjunctive normal form by a standard cardinality‐constraint 
encoding, \egnsp, a sorting‐network encoding or a sequential counter \citep{SATCardinality}. We used the \mtt{cnfgen} 
library \citep{cnfgen} to generate 70 unsatisfiable formulas of this sort in DIMACS format, and we then  
reversed the CNF transformation in most cases and converted the formulas into the uniform 
NDL/Athena notation used throughout \pgrid. 

Subset-cardinality problems were first introduced in 2014 by \citet{SubsetCardinality}. 
The source of the unsatisfiability in these problems is the tension between the left and 
right constraints: on the left we are saying that at least half of the edges must be selected 
but on the right we are saying that at least half of the edges must \emph{not} be selected 
(at most half may be selected). The intuitive contradiction here is that the left wants a lot of
selected edges while the right wants the opposite, not too many selected edges. 

%% As an example, given 
%% \[
%%   L=\{A,B\},\quad
%%   R=\{1,2,3\},\quad
%%   E=\{(A,1),(A,2),(B,2),(B,3)\},
%% \]
%% we create variables  $x_{A,1},\;x_{A,2},\;x_{B,2},\;x_{B,3}$ and compute degrees:
%% \[
%%   d(A)=2,\quad d(B)=2,\quad d(1)=1,\quad d(2)=2,\quad d(3)=1.
%% \]
%% The two families of constraints become
%% \[
%%   x_{A,1}+x_{A,2}\;\ge\;1,
%%   \qquad
%%   x_{B,2}+x_{B,3}\;\ge\;1,
%% \]
%% \[
%%   x_{A,1}\;\le\;0,
%%   \quad
%%   x_{A,2}+x_{B,2}\;\le\;1,
%%   \quad
%%   x_{B,3}\;\le\;0.
%% \]
%% Here \(x_{A,1}\le0\) forces \(x_{A,1}=0\) and \(x_{B,3}\le0\) forces \(x_{B,3}=0\).  The remaining constraint 
%% \(\,x_{A,2}+x_{B,2}\le1\)
%% together with
%% \(\,x_{A,2}\ge1\)
%% and
%% \(\,x_{B,2}\ge1\)
%% (from the left‐side constraints) forces exactly one of \(x_{A,2},x_{B,2}\) to be true.  For instance,
%% \[
%%   x_{A,2}=1,\quad x_{B,2}=0
%% \]
%% satisfies all clauses, so the formula is SAT.

\subsubsection{Tseitin Formula Problems}

Consider the following undirected graph $G$, where each node ($A$, $B$, $C$, $D$) is also viewed as a propositional atom, and where additional propositional atoms $x_i$ are attached to the edges: \\[0.05in]
\begin{minipage}{0.9\textwidth}
\centering
%\begin{tikzpicture}[scale=1.0,every node/.style={font=\small}]
\begin{tikzpicture}[scale=1.0]
  % Vertex nodes
  \node[circle, draw, minimum size=0.75cm, inner sep=0pt,fill=blue!20] (A) at (0,2) {$A$};
  \node[circle, draw, minimum size=0.75cm, inner sep=0pt, fill=blue!20] (B) at (4,2) {$B$};
  \node[circle, draw, minimum size=0.75cm, inner sep=0pt, fill=blue!20] (C) at (0,0) {$C$};
  \node[circle, draw, minimum size=0.75cm, inner sep=0pt, fill=blue!20] (D) at (4,0) {$D$};
  % Edges with variable labels
  \draw[thick] (A) -- node[above] {\(x_1\)} (B);
  \draw[thick] (A) -- node[left]  {\(x_2\)} (C);
  \draw[thick] (B) -- node[right] {\(x_3\)} (D);
  \draw[thick] (C) -- node[below] {\(x_4\)} (D);
  \draw[thick] (A) -- node[below] {\(x_5\;\)} (D);
\end{tikzpicture} 
\vspace*{0.1in}
\end{minipage}
\mbox{}\\
The core Tseitin formula for $G$ is the conjunction of the following {\em definitions}, one for each node:
$A \miff x_1 \oplus x_2 \oplus x_5$, $B \miff x_1 \oplus x_3$, $C \miff x_2 \oplus x_4$, $D \miff x_3 \oplus x_4 \oplus x_5$, where $\oplus$ denotes the exclusive-or (``XOR'') operation: $x_i \oplus x_j \miff (x_i \mand \neg x_j) \mor (x_j \mand \neg x_i)$. 
For arbitrary formulas $p_1,\ldots,p_n$, an easy induction on $n$, with $n = 2$ as the base case, 
shows that $p_1 \oplus p_2 \oplus \cdots \oplus p_n$ holds iff exactly an odd number of $p_i$ hold (we omit parentheses since $\oplus$ is commutative and associative).
Thus, these definitions assert that each node atom holds iff an odd number of edges incident to that node hold. 
So far all we have is a set of definitions, one for each node, that are perfectly jointly consistent. But now
suppose we go on to assert that an odd number $k$ of node atoms hold, say by specifying $A$, $\neg B$,
$\neg C$, $\neg D$ (so that only $k = 1$ atom holds, $A$ in this case). By what we just said about the exclusive 
or, it would follow that $A \oplus B \oplus C \oplus D$ holds. However, by the given definitions we have: 
%\begin{minipage}{0.9\textwidth} \centering
\begin{eqnarray*}
A \oplus B \oplus C \oplus D & \miff &  (x_1 \oplus x_2 \oplus x_5) \oplus (x_1 \oplus x_3) 
\oplus \, (x_2 \oplus x_4) \oplus (x_3 \oplus x_4 \oplus x_5) \\
{} & \miff &  (x_1 \oplus x_1) \oplus (x_2 \oplus x_2) \oplus (x_3 \oplus x_3) \oplus \, (x_4 \oplus x_4)  \oplus (x_5 \oplus x_5) \\
{} & \miff & \temv{false}. 
\end{eqnarray*}
%\vspace*{0.01in} \end{minipage}
The first step follows just by rearranging the variables on the right-hand side of the first equivalence, while the second step follows 
because $p \oplus p \miff \temv{false}$
for any formula $p$. Thus we have a contradiction: $A \oplus B \oplus C \oplus D$ holds because we have stipulated that only $A$ holds, but at the same time we have
shown that $A \oplus B \oplus C \oplus D$ cannot hold because it leads to \temv{false}. The key intuition is that because each edge variable appears twice on the right-hand
side of the first equivalence (once for each node to which the edge is incident), we can regroup all variables as shown above to derive \temv{false}. 

While the conventional formulation of a Tseitin problem is as a jointly unsatisfiable set of formulas (typically clauses, but again, we work with full propositional logic), 
we can easily pose it as a forward inference problem: Given the $n$ definitions for the $n$ node atoms of a graph $G$, prove the negation of any conjunction of an odd number of 
those atoms. All Tseitin problems in \plthreesp are of this form. 

\subsubsection{Counting Principle Problems}

The 5 problems in this group encode statements of the form \emph{There is a way to partition a set of \(M\) elements into subsets of size \(p\) each}. 
Clearly, this statement is satisfiable iff $p$ divides $M$. Set $n = M/p$. For each element \(e\in\{1,2,\dots,M\}\) and each part (or ``block'') \(b\in\{1,2,\dots,n\}\), 
introduce a Boolean variable $x_{e,b}$ that holds iff element $e$ is assigned to block $b$.
We conjoin the following sets of constraints: 
\begin{enumerate}
  \item \emph{Exactly one block per element.}  Every element \(e\) must appear in exactly one block:
  \[
    \bigl(x_{e,1}\lor x_{e,2}\lor \cdots \lor x_{e,n}\bigr)
    \quad\text{and}\quad
    x_{e,b_1}\mif \neg x_{e,b_2}\quad
    \mbox{for all}\msp b_1\neq b_2,
    \quad e=1,\ldots,M.
  \]
  \item \emph{Exactly \(p\) elements per block.}  Each block \(b\) must contain exactly \(p\) elements:
  \[
    \sum_{e=1}^{M} x_{e,b} \;=\; p,
    \quad b=1,\ldots,n.
  \]
  In SAT, such an equality is typically encoded by a pair of cardinality constraints \(\ge p\) and \(\le p\), 
  each represented by a standard cardinality encoding. 
\end{enumerate}
Simple counting shows that these constraints admit a satisfying assignment 
precisely when \(p\) divides \(M\). 

\subsubsection{De Bruijn Problems}
\label{Sec:App:DBFormulas}
For any integer $n > 0$, the \nth{n} \dbsp formula, which we denote by $\mathcal{D}_n$, is defined as the conditional $(p \mif q)$, where $q$ is the conjunction $(A_1\mand \cdots \mand A_n)$ and 
$p$ is the conjunction of the $n$ conditionals $((A_1 \miff A_2) \mif q)$, $((A_2 \miff A_3) \mif q)$, $\ldots$, $((A_n \miff A_1) \mif q)$. 
It is easy to show (see below) that $\mathcal{D}_n$ is a tautology---and hence provable---iff $n$ is odd. 
Thus, $\mathcal{D}_n$ does not hold for any even $n$. 
Odd-indexed formulas are often used as stress tests for theorem provers. However, the corresponding proofs are also good tests of a system's
proof-checking ability. The proof sizes grow linearly as a function of $n$, but what makes the proofs more interesting 
is that they include highly nested conditional subproofs introducing long staircases of assumption scopes, 
which are ideal for evaluating a model's ability to track hypotheses. 

To make things concrete, consider $\mathcal{D}_5$, depicted diagrammatically as follows: 
\begin{center}
\begin{tikzpicture}[
   scale=1.1,
   transform shape,         
    >=stealth,
    every node/.style={font=\small},
    eqedge/.style={thick,<->},
    Tnode/.style={circle,draw,fill=green!50,inner sep=2pt},
    Fnode/.style={circle,draw,fill=red!50,inner sep=2pt},
    impl/.style={thick,->,shorten >=5pt}
]

% central conjunction q
\node[draw, rounded corners, inner sep=4pt] (q) at (0,0)
  {\mbox{\scriptsize$q = A_1 \land A_2 \land A_3 \land A_4 \land A_5$}};
%%  {};

% variables on a regular pentagon (radius 3cm)
\node[Tnode] (A1) at (90:3)   {$A_1$};
\node[Fnode] (A2) at (18:3)   {$A_2$};
\node[Tnode] (A3) at (-54:3)  {$A_3$};
\node[Fnode] (A4) at (-126:3) {$A_4$};
\node[Tnode] (A5) at (162:3)  {$A_5$};

% equivalence edges around the cycle
\draw[eqedge] (A1) -- node[midway, above, sloped] {} (A2);
\draw[eqedge] (A2) -- node[midway, above, sloped] {} (A3);
\draw[eqedge] (A3) -- node[midway, below, sloped] {} (A4);
\draw[eqedge] (A4) -- node[midway, below, sloped] {} (A5);
\draw[eqedge] (A5) -- node[midway, left,  sloped] {} (A1);

% midpoints of edges (to represent the antecedents (Ai <-> A_{i+1}))
\path (A1) -- (A2) coordinate[midway] (m12);
\path (A2) -- (A3) coordinate[midway] (m23);
\path (A3) -- (A4) coordinate[midway] (m34);
\path (A4) -- (A5) coordinate[midway] (m45);
\path (A5) -- (A1) coordinate[midway] (m51);

% implications from each equivalence to q
\draw[impl] (m12) -- (q);
\draw[impl] (m23) -- (q);
\draw[impl] (m34) -- (q);
\draw[impl] (m45) -- (q);
\draw[impl] (m51) -- (q);

% label one implication as a representative conjunct
%\node[font=\scriptsize, fill=white, inner sep=1pt] at ($(m12)!0.6!(q)$)
%%  {$((A_1 \leftrightarrow A_2) \Rightarrow q)$};
%  {};

% formula label underneath
\node[below=1.4cm of q] (formula) {
  %% $D_5 :\quad
  %%  p \Rightarrow q,\quad
  %%  p = \bigwedge_{i=1}^{5} \bigl((A_i \leftrightarrow A_{i+1}) \Rightarrow q\bigr),\
  %%  A_6 := A_1$
};

\end{tikzpicture}
\end{center}
The slanted arrows coming in from each side represent the $n = 5$ conditionals 
$((A_1 \miff A_2) \mif q), \ldots, ((A_5 \miff A_1) \mif q)$, which are taken for granted (\iensp, essentially
taken as premises, by virtue of their conjunction $p$ serving as a hypothesis of $\mathcal{D}_5$). 

The key observation is that, because $n = 5$ is odd, at least one of the 5 biconditionals $A_1 \miff A_2, \ldots, A_5 \miff A_1$ 
must be true, because at least one pair of adjacent atoms ($A_1$ and $A_2$, or $A_2$ and $A_3$, $\ldots$, or $A_5$ and $A_1$)
must be assigned the same truth value (either true or false, respectively represented here as green and red). And, of course,
if two adjacent atoms $A_i$ and $A_j$ are assigned the same truth value (the same color), then $A_i \miff A_j$ becomes true,
because that's what a biconditional $A_i \miff A_j$ means: that $A_i$ and $A_j$ are both true or both false. 

Now, to see why some pair of adjacent atoms must be assigned the same color, approach the matter adversarially and try
to assign colors (truth values) to the atoms so that this is \emph{not} the case, \iensp, so that every two adjacent atoms 
receive distinct colors. Say we start with $A_1$ true (green) as in the above diagram. We then must make $A_2$ false (red),
and thus $A_3$ green and  $A_4$ red. But then $A_5$ must be green, which makes $A_5$ and $A_1$ equivalent, thereby allowing us 
to conclude $q$ by the hypothesis $(A_5 \miff A_1) \mif q$. This is exactly the situation shown in the diagram, but any other
assignment of truth values to atoms would have the same upshot; and the same argument would work for any 
other odd $n$.\footnote{In Athena it is easy to express the above procedure as a parameterized \emph{method} 
\label{Ft:DBMethod} (a proof algorithm) that takes any odd  integer $n$ and derives $D_n$ by assuming the conjunction $p$ 
of the $n$ conditionals of the form $(A_i \miff A_j) \mif q$ and then supposing, by way of contradiction, 
that $\neg q$ holds. The method proceeds by a case analysis of $A_1$: either $A_1$ holds or not. 
If it holds, then $A_2$ must not hold (for if it did, we would have $A_1 \miff A_2$ and $p$ would allows us to conclude $q$, 
contradicting the supposition $\neg q$). Likewise, $A_3$ must hold (if it did not, we'd have $A_2 \miff A_3$ and again
$p$ would let us derive $q$), $A_4$ must not hold, and so on, all the way back to $A_1$, which would be shown
to be false, contradicting the assumption that it holds. We likewise alternate through all the atoms under the second
assumption of the case analysis, where $A_1$ does not hold, ultimately concluding that it does hold and thus reaching 
a contradiction.} 

When $n$ is even, an adversarial assignment of truth values to atoms 
along the aforementioned lines provides a counterexample to $\mathcal{D}_n$ that makes all 
$n$ conditionals $(A_i \miff A_j) \mif q$ vacuously true (since each $A_i \miff A_j$ is false), 
while making the conjunction $q$  false because at least one atom is assigned false (colored red).

\subsection{A Sample Proof and Multi-Hop Reasoning Depth in \plthreesp Proofs}

This section illustrates a \ndlzerosp proof from the family of simple DAG-pebbling 
problems.\footnote{The proof appears in \mbf{PL3-PC} as a proof-checking problem, 
while the underlying argument also appears in \mbf{PL3-PW} as a proof-writing problem.}
We discuss the generation of the problem itself (its premises and conclusion), 
the proof and its logical structure as a graph, and define the general 
notion of reasoning depth for \ndlzerosp proofs. 

For this particular problem family (by far the easiest in \plthree), 
our data-generation process starts by stochastically building 
a simple DAG of a given size, one layer at a time, up to a maximum depth of 6. 
By construction, there is always one source node and one target node. At the end, we 
randomly add some extra cross-layer edges for redundancy and symmetry, which further
simplifies the reasoning problem. Here is a sample DAG generated by our code
(corresponding to item 233 in both \mbf{PL3-PW} and \mbf{PL3-PC}): 
\vspace*{-0.1in}
\begin{center}
%% \begin{tikzpicture}[
%%     >=stealth,
%%     scale=0.8,
%%     rotate=-180,
%%     line width=0.9pt,
%%     every node/.style={circle,draw,fill=blue!20,minimum size=8mm,inner sep=0pt}
%% ]

%% %--- nodes ---------------------------------------------------------
%% \node (8) at (0,0)      {8};
%% \node (7) at (1.8,0.8)  {7};
%% \node (6) at (3.6,1.8)  {6};
%% \node (5) at (5.0,1.8)  {5};   % moved slightly left
%% \node (4) at (5.6,-0.6) {4};   % moved slightly right
%% \node (3) at (7.2,0.0)  {3};
%% \node (2) at (9.0,1.0)  {2};
%% \node (1) at (11.0,0.0) {1};

%% %--- edges ---------------------------------------------------------
%% % chain 1 -> 2 -> 3 -> 4 -> 5 -> 6 -> 7 -> 8
%% \draw[->] (1) -- (2);
%% \draw[->] (2) -- (3);
%% \draw[->] (3) -- (4);
%% \draw[->] (4) -- (5);
%% \draw[->] (5) -- (6);
%% \draw[->] (6) -- (7);
%% \draw[->] (7) -- (8);

%% % extra edges
%% \draw[->] (2) .. controls (7.5,3.0) and (4.5,3.0) .. (6);   % 2 -> 6
%% \draw[->] (2) -- (7);                                       % 2 -> 7
%% \draw[->] (2) .. controls (7.0,3.0) and (2.0,3.2) .. (8);   % 2 -> 8 (top arc)

%% \draw[->] (6) .. controls (3.5,3.6) and (1.2,2.4) .. (8);   % 6 -> 8

%% \draw[->] (3) -- (7);                                       % 3 -> 7
%% \draw[->] (3) .. controls (6.0,-2.5) and (2.0,-1.0) .. (8); % 3 -> 8 (middle lower arc)

%% \draw[->] (4) .. controls (4.0,-1.6) and (1.5,-1.3) .. (8); % 4 -> 8

%% \draw[->] (1) .. controls (8.0,-2.5) and (2.0,-2.5) .. (8); % 1 -> 8 (outer bottom arc)

%% \end{tikzpicture}
\begin{tikzpicture}[
    >=stealth,
    rotate=-180,
    scale=0.85,
    line width=0.8pt,
    every node/.style={circle,draw,fill=blue!20,minimum size=8mm,inner sep=0pt},
    shorten >=0.3mm,   % <-- arrows stop a bit before the target node
]

%--- nodes ---------------------------------------------------------
\node (8) at (0,0)      {8};
\node (7) at (1.8,0.8)  {7};
\node (6) at (3.6,1.8)  {6};
\node (5) at (5.0,1.8)  {5};
\node (4) at (5.6,-0.6) {4};
\node (3) at (7.2,0.0)  {3};
\node (2) at (9.0,1.0)  {2};
\node (1) at (11.0,0.0) {1};

%--- edges ---------------------------------------------------------
\draw[->] (1) -- (2);
\draw[->] (2) -- (3);
\draw[->] (3) -- (4);
\draw[->] (4) -- (5);
\draw[->] (5) -- (6);
\draw[->] (6) -- (7);
\draw[->] (7) -- (8);

\draw[->] (2) .. controls (7.5,3.0) and (4.5,3.0) .. (6);
\draw[->] (2) -- (7);
\draw[->] (2) .. controls (7.0,3.0) and (2.0,3.2) .. (8);

\draw[->] (6) .. controls (3.5,3.6) and (1.2,2.4) .. (8);

\draw[->] (3) -- (7);
\draw[->] (3) .. controls (6.0,-2.5) and (2.0,-1.0) .. (8);

\draw[->] (4) .. controls (4.0,-1.6) and (1.5,-1.3) .. (8);

\draw[->] (1) .. controls (8.0,-2.5) and (2.0,-2.5) .. (8);

\end{tikzpicture}
\end{center} 
\vspace*{-0.2in}
%% \mbox{} \\[-0.1in]
%%  \begin{minipage}{\textwidth}
%%  \centering
%%  \scalebox{0.3}{\includegraphics{./plots/task_dag.pdf}}
%%  \end{minipage} \\[0.05in]
The sole source node here is 1, while the target node is 8. 
This DAG is given as an argument to the \smtt{PebblingFormula} function of the \mtt{cnfgen} 
library \citep{cnfgen}, which generates a random CNF for it. 
We then declausify that output by converting the CNF clauses to a collection 
of formulas that make heavy use of conditionals. Every numeric DIMACS variable $i$ is consistently 
mapped to the atomic formula \smtt{A}$i$. Finally, we pick the atom corresponding to the DAG's 
\enlargethispage{8pt}
target node as the goal of the proof. In this example the goal is \smtt{A8}, while the premises are
shown below. The proof (for \mbf{PL3-PC}) is obtained by a call to Vampire \citep{Vampire}. The resolution proof output by Vampire,
expressed in TPTP notation \citep{TPTP}, is then translated into \ndlzerosp as a proof by contradiction. 
In this example the resulting proof is as follows. 
\begin{tcAthenaWithNumbers}
assert premise-1 := A1
assert premise-2 := (~A2 ==> ~A1)
assert premise-3 := (A2 ==> A3)
assert premise-4 := (~A4 ==> ~A3)
assert premise-5 := (~A5 ==> ~A4)
assert premise-6 := (~A6 ==> ~A2 | ~A5)
assert premise-7 := (A6 & ~A7 ==> ~A3 | ~A2)
assert premise-8 := (A3 & A6 & ~A8 & A7 & A4 & A2 ==> ~A1)

# Goal: Prove A8
{
  assume negated_conjecture := (~ A8) {
    lemma-1 := A2 from premise-2, premise-1;
    lemma-2 := A3 from premise-3, lemma-1;
    lemma-3 := (A3 ==> A7 | ~A6) from premise-7, lemma-1;
    lemma-4 := (~A8 ==> ~A4 | ~A6 | ~A7 | ~A2 | ~A3) from premise-8, premise-1;
    lemma-5 := A4 from premise-4, lemma-2;
    lemma-6 := (A6 ==> A7) from lemma-3, lemma-2;
    lemma-7 := (~A4 | ~A7 | A8 | ~A6 | ~A3) from lemma-4, lemma-1;
    lemma-8 := A5 from premise-5, lemma-5;
    lemma-9 := (A6 ==> ~A7 | A8 | ~A3) from lemma-7, lemma-5;
    lemma-10 := (~A2 | A6) from premise-6, lemma-8;
    lemma-11 := A6 from lemma-10, lemma-1;
    lemma-12 := A7 from lemma-6, lemma-11;
    lemma-13 := (A8 | ~A6 | ~A3) from lemma-9, lemma-12;
    lemma-14 := (A8 | ~A3) from lemma-13, lemma-11;
    lemma-15 := A8 from lemma-14, lemma-2;
    lemma-16 := false from negated_conjecture, lemma-15;
};
  A8 from (~A8 ==> false);
}
\end{tcAthenaWithNumbers}
Premises 2 to 8 assert a chain of implications leading from \smtt{A1} to \smtt{A8}. For example, \smtt{premise-2}
asserts that \smtt{A1} implies \fmtt{A2}, \footnote{Strictly speaking, it asserts the contrapositive of that implication.
As we pointed out earlier (p.~\pageref{Pg:HornClauseRewriting}), the Horn clauses of these problems are sometimes 
rewritten using basic logical equivalences.}
\smtt{premise-6} states that \smtt{A2} and \smtt{A5} imply \smtt{A6}, and so on. 
The very first premise sets the dominoes toppling by asserting \smtt{A1}. 
The goal is to derive \smtt{A8}. This is done by the \ndlzerosp proof on lines 11--31. The reasoning 
proceeds by contradiction, reflecting Vampire's resolution proof. On the assumption that \smtt{A8} does not hold, it is easy to derive
the absurdity \smtt{false}, simply by deriving \smtt{A8} and contradicting the hypothesis \smtt{(\ntildels A8)}. 
Observe that reasoning by contradiction here is quite
unnecessary. The proof uses it only because it is mechanically translated from Vampire's
resolution proof, which is inherently refutational. In fact, lemmas 1 through 15 form a perfectly adequate {\em direct\/} proof of \smtt{A8}
from the premises, which is easily verified by noting that \smtt{negated-conjecture} does not appear anywhere in lines 13--27. 
The negated conjecture only makes a perfunctory appearance in the very final step, on line 28.\footnote{The proof-optimization algorithm 
of \citet{ProofSimp} would automatically simplify this proof by hoisting lines 13--27 outside of the \fkwd{assume} and then eliminating the
subsequent decimated conditional subproof along with the final application of \fmtt{by-contradiction}, because at that point that pair of deductions
would be a detour, re-deriving what the hoisted inference block has already derived (namely, \fmtt{A8}).} Nevertheless, in the interest of 
simplicity and uniformity, we retain the same refutational proof template throughout. 
These proofs form the raw data of the proof-checking task of \plthreesp (\mbf{PL3-PC}). 
%with the exception of \dbsp problems (see below).

\begin{figure}[!htbp]
\begin{minipage}{\textwidth}
\centering
\scalebox{1.0}{\includegraphics[width=\textwidth]{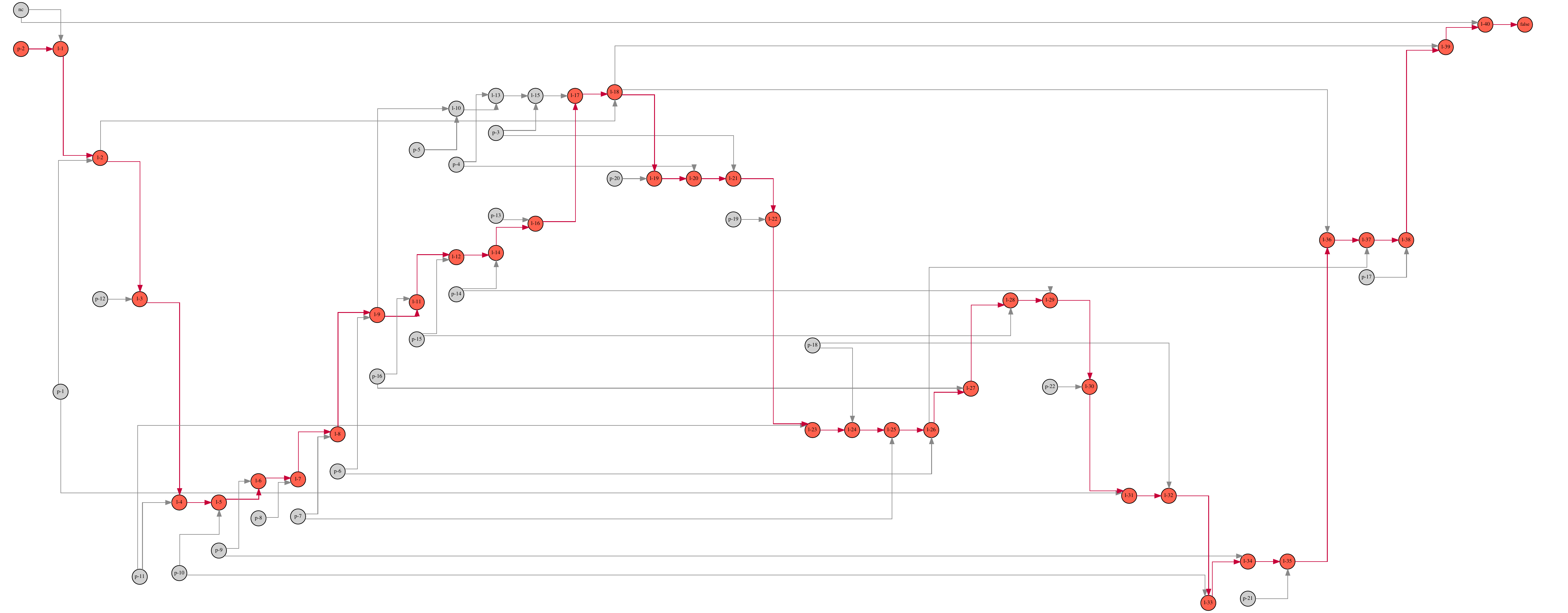}}
\end{minipage}
\caption{The directed acyclic reasoning graph of a proof for a medium-complexity Tseitin-formula problem.}
\label{Fig:TseitinDag50}
\end{figure}

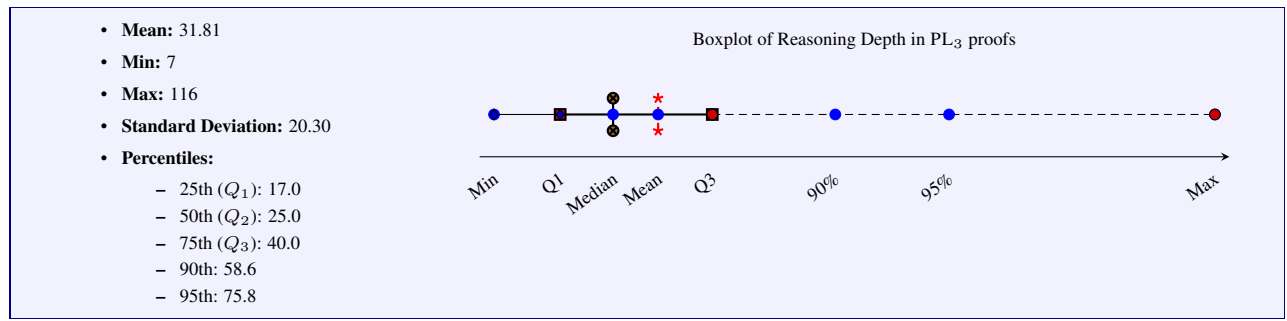
\begin{figure}[!thbp]
\begin{minipage}{\textwidth}
\centering
\begin{tcolorbox}[
    colback=blue!5!white, colframe=blue!50!black, 
    boxrule=0.5pt,
    %title=Reasoning Depth Distribution (n = 300), 
    fonttitle=\bfseries,
    boxsep=1mm,
    left=1mm, right=1mm, top=1mm, bottom=1mm,
    sharp corners
]
\scriptsize
\begin{minipage}[t]{0.48\textwidth}
\begin{itemize}
    \item \textbf{Mean:} 31.81
    \item \textbf{Min:} 7
    \item \textbf{Max:} 116
    \item \textbf{Standard Deviation:} 20.30
    \item \textbf{Percentiles:}
    \begin{itemize}
        \item 25th (\(Q_1\)): 17.0
        \item 50th (\(Q_2\)): 25.0
        \item 75th (\(Q_3\)): 40.0
        \item 90th: 58.6
       \item 95th: 75.8
    \end{itemize}
\end{itemize}
\end{minipage}
\hfill
\begin{minipage}[t]{0.48\textwidth}
\vspace{-0.5em} % fine-tune vertical alignment
\centering
\hspace*{-3.4cm}
\begin{tikzpicture}
\begin{axis}[
    title={\scriptsize Boxplot of Reasoning Depth in \plthreesp proofs},
    width=11.5cm,
    height=2.7cm,
    hide y axis,
    x axis line style={-},
    xtick={7,17,25,31.81,40,58.6,75.8,116},
    xticklabels={Min, Q1, Median, Mean, Q3, 90\%, 95\%, Max},
    xticklabel style={rotate=35, anchor=north east, font=\scriptsize},
    axis x line=bottom,
    ymin=0, ymax=1,
    tick style={draw=none},
    enlargelimits=0.02
]

\addplot+[only marks, mark options={fill=blue}, mark=*] coordinates {
    (7,0.5) (17,0.5) (25,0.5) (31.81,0.5) (40,0.5) (58.6,0.5) (75.8,0.5) (116,0.5)
};

\addplot+[black, thick] coordinates {(17,0.5) (40,0.5)};
\addplot+[black, thick] coordinates {(25,0.3) (25,0.7)};
\addplot+[red, thick, dashed] coordinates {(31.81,0.3) (31.81,0.7)};
\addplot+[black] coordinates {(7,0.5) (17,0.5)};
\addplot+[black] coordinates {(40,0.5) (116,0.5)};

\end{axis}
\end{tikzpicture}
\end{minipage}
\end{tcolorbox}
\end{minipage}
\caption{The distribution of reasoning depth across all \plthreesp proof-checking problems.}
\label{Fig:RDDistribution}
\end{figure}

The linear flavor of the above proof belies the rich two-dimensional logical structure of the underlying reasoning. 
The graph below depicts that structure diagrammatically. Nodes are either premises or lemmas, and a directed edge from a node 
$u$ to a node $v$ representing a lemma indicates that the derivation of that lemma relies on the formula represented by $u$. 
\begin{figure}[!htbp]
\begin{minipage}{\textwidth}
\centering
\scalebox{1.0}{\includegraphics[width=\textwidth]{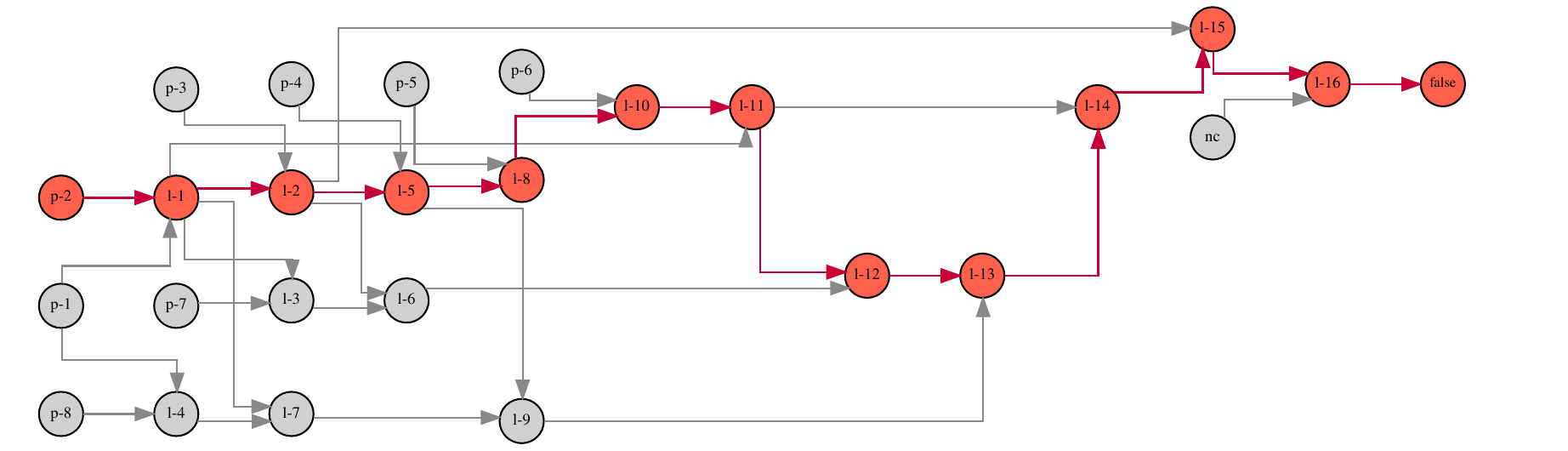}}
\end{minipage}
%\caption{Foo Bar}
%\label{Fig:FooBar}
\end{figure}
In particular, for every derivation of the form $p\msp\skwd{from}\msp{}p_1,\ldots,p_n$, there is a directed edge from every node 
corresponding to an argument $p_i$ to the node corresponding to the conclusion $p$. 
(To save space, the diagram uses ``l'' and ``p'' as abbreviations for ``lemma'' and ``premise,'' respectively, thus, e.g., writing 
\mbox{l-8} as shorthand for \mbox{lemma-8}, \mbox{p-2} for \mbox{premise-2}, and so on.)
Accordingly, for line 14 of the proof, \smtt{lemma-2 := A3 \skwd{from} premise-3, lemma-1}, we see that the graph has two edges
leading to \mbox{l-2}, one from \mbox{p-3} and the other from \mbox{l-1}. By convention, we add an extra mock arrow from the very last
lemma that derives \smtt{false} to the literal \smtt{false} itself, in this case with an edge leading from \mbox{l-16} to \smtt{false}. 
Such a reasoning graph is also a DAG (since there cannot be any cycles, which would indicate circular reasoning), 
whose sources are the premises and whose sink node is the constant \smtt{false}. We define a {\em maximal branch\/} of such a graph as a 
maximum-length path from a source (premise) to the conclusion \smtt{false}. In the diagram above, the nodes and edges along such a path
are colored in red. We define the {\em reasoning depth\/} of a proof as the length of such a maximal branch.\footnote{In theory,
there may be multiple maximal branches in such a graph, but, by definition, both must have the same length.}
For this simple example, the reasoning depth is 12, but for more complex problem families the reasoning of a proof can be much deeper. 
The DAG of a proof for a medium-complexity Tseitin-formula problem appears in Figure~\ref{Fig:TseitinDag50}. 
The general distribution of reasoning depth across all \mbf{PL3-PC} proofs is shown in Figure~\ref{Fig:RDDistribution}.
%% The box plot visualizes the distribution of reasoning depth across 300 proofs. Each dot along the horizontal line corresponds 
%% to a specific summary statistic. The blue circles represent key percentiles: the minimum, 90th percentile, and 95th percentile. 
%% The black-bordered squares mark the 25th percentile (Q1), median (Q2), and 75th percentile (Q3) — together forming the “box” of 
%% the box plot. A solid black line connects Q1 to Q3, capturing the interquartile range (IQR), and a vertical black line at the median 
%% indicates the center of the distribution. The red dashed line with a red asterisk points to the mean, which is slightly higher than 
%% the median — suggesting a right-skewed distribution. Finally, a solid black whisker extends from Q1 to the minimum, and another 
%% from Q3 to the maximum (shown as a red dot), reflecting the full spread of the data.

We stress that this reasoning-depth distribution applies only to \mbf{PL3-PC}, 
which uses the \ndlzerosp proofs obtained through Vampire. 
The \mbf{PL3-PW} task has no a priori depth distribution, as an AI model is free to come up with its own proofs, 
which, in theory, could be simpler or more complicated---and thus more or less shallow---than those found by Vampire. 
We have already shown that the Vampire proof shown above could be simplified by converting it 
into a direct proof that does not use reasoning by contradiction. But the proof could be simplified
even further, while still adhering to the proviso that each \skwd{from} step operates on no more than five arguments.
Here is a much simpler direct proof of the goal \smtt{A8} that uses only 8 steps: 
\begin{tcAthenaWithNumbers}
{
  A2 from premise-1, premise-2;
  A3 from A2, premise-3;
  A4 from A3, premise-4;
  A5 from A4, premise-5;
  A6 from A2, A5, premise-6;
  A7 from A6, A2, A3, premise-7;
  lemma := (~A6 | ~A7 | A8) from A1, A2, A3, A4, premise-8;
  A8 from lemma, A6, A7;
}
\end{tcAthenaWithNumbers}
A competent language model should be able to compose this chain of reasoning without much difficulty. 
Indeed, for this particular problem (item 233 in \mbf{PL3-PW}), GPT-5, GPT-5.2, Gemini 3.1, and Grok-4 end 
up discovering this \emph{exact} short and elegant proof, so 
in that sense they do better than Vampire. Neither of the other two Gemini models (2.5 or 3.0) does quite as well,
while GPT-5.1 performs worse: it produces a correct proof, but it's much longer and more convoluted than the 
one above. 
%We will come back to this problem in Section~\ref{Sec:Examples}. 
We discuss the proof-writing task in more detail in the next section. 

\subsection{\mbf{PL3-PW}}

The \mbf{PL3-PW} task is simply to write an \ndlzerosp proof for a given \plthreesp problem. 
As explained in Section~\ref{Sec:PLThreeOverview}, \ndlzerosp proofs have exceptionally simple syntax and operational semantics,
such a proof being one of three things:
\ben
\item an application of \skwd{from} to a number of formulas; 
\item a composite proof (\iensp, a proof sequence), built by stringing together a number of 
other proofs and separating them by semicolons, exactly as in regular NDL; 
or
\item a hypothetical (\skwd{assume}) proof, also exactly as in regular NDL. 
\een
Applications of \skwd{from} are primitive proofs with no internal structure, 
while composite and hypothetical proofs are compound, having recursive structure. 

This parsimony is a significant advantage, as it strips reasoning tasks to 
bare semantic essentials: entailment (\skwd{from}), assumption scope (\skwd{assume}), 
and sequential composition or lemma chaining (the semicolon operator). This provides a minimal 
universal proof grammar that any model can learn from a short prompt, 
which should isolate pure logical competence more cleanly than vanilla NDL. 

\begin{figure}[!htbp]
\begin{minipage}{\textwidth}
\centering
%\scalebox{1.0}{\includegraphics[width=\linewidth]{./plots/all_pl2_pw_stats_proper.png}}
\scalebox{0.7}{\includegraphics[width=\linewidth]{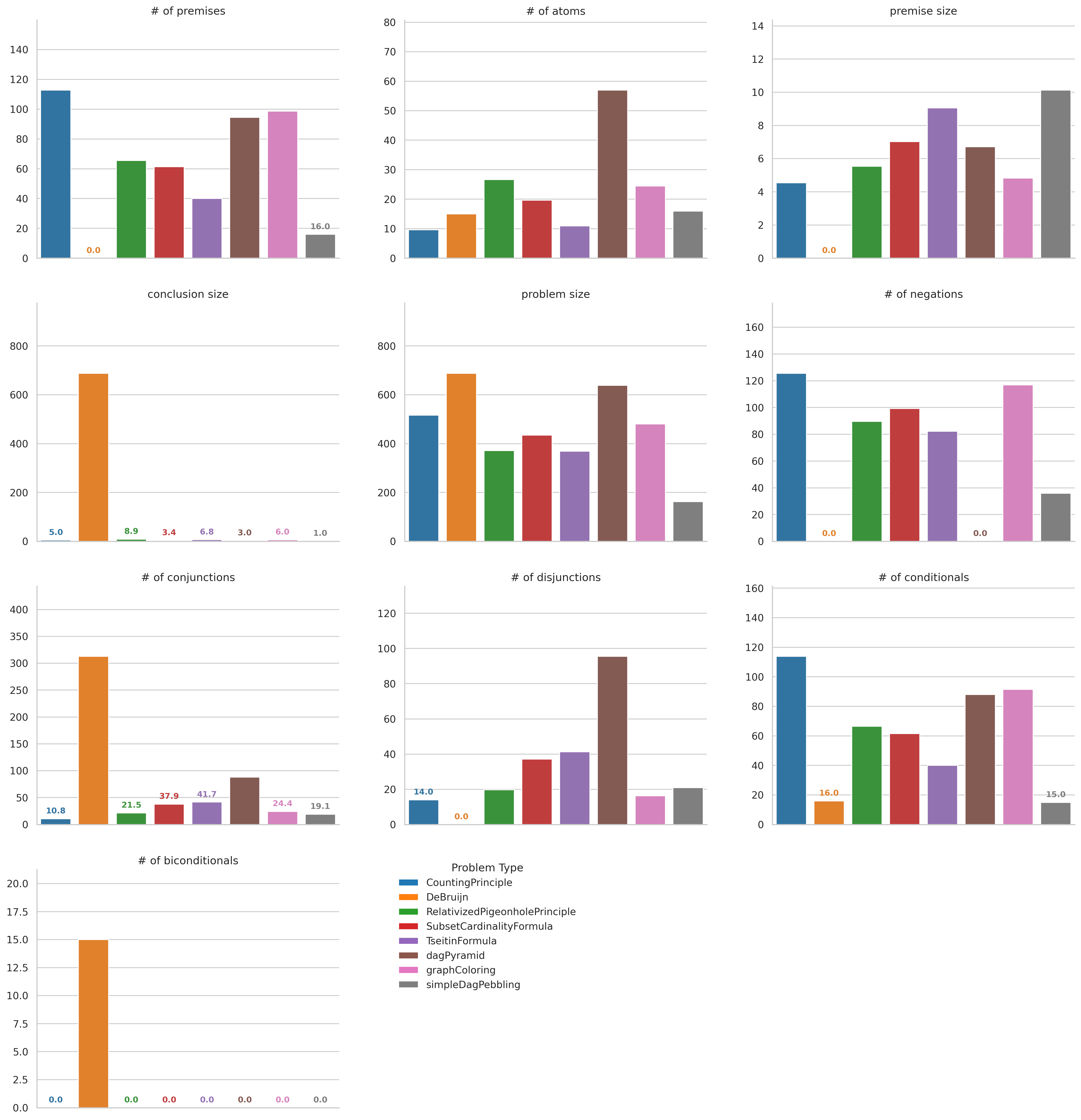}}
\end{minipage}
\caption{Mean values of key numeric features of \mbf{PL3-PW} problems, organized by problem type.}
\label{Fig:PLThreePWProbStatsByProbTypes}
\end{figure}
%pl2_pw_problem_stats_by_problem_types

Fig.~\ref{Fig:PLThreePWProbStatsByProbTypes} summarizes key statistics for \mbf{PL3-PW} problems by type. 
For each of the 8 problem families, it reports the  mean number of premises, atoms, and occurrences 
of each logical connective in a problem of that family (across the premises and target conclusion), 
along with the average premise size, conclusion size, and total problem size (the sum of the
sizes of the premises plus the size of the conclusion).  
%% number of premises; number of atoms; premise size;  conclusion size; 
%% total problem size (the sum of the sizes of all premises and the target conclusion); as well as the number
%% of occurrences of every logical connective in the problem (across premises and target conclusion). All numbers
%% shown represent the \emph{mean} values of these attributes, computed over all problems of a given type. 
Some of these attributes have very large average values for a given problem type, \egnsp, \dbsp 
problems are conditionalized (have no premises) and hence their conclusion sizes are much larger 
than the conclusions in any other problem family, so histogram bars for these other problem families 
are not visible in the conclusion-size chart. In such cases we simply print the corresponding 
numeric values where the histogram bars would normally appear, using the color scheme of the legend 
(so that, \egnsp, mean conclusion size is 8.9 for relativized-pigeonhole problems, 3.8 for subset-cardinality problems, and so on). 
%% Instead of bars, we have simply 
%% written out the corresponding numeric values for the other problem types, using the colors specified in the legend,
%% so that, \egnsp, the average conclusion size is 8.9 for relativized-pigeonhole-principle problems, 4.8 for subset-cardinality
%% problems, and so on. 
For clarity, we use the same convention in every histogram that has some disproportionately short bars. 
Observe that biconditionals appear only in the \dbsp family, hence their average number 
in problems of every other type is zero. By contrast, \dbsp problems have no 
explicit disjunctions or negations. DAG-pyramid problems likewise do not contain any negations. 

Overall accuracies for all models are shown in Table~\ref{Tb:PL2PWAccuracies}. 
Instrumentation for \ndlzerosp proofs 
is less involved than in the case of regular NDL proofs, because there are no structural errors to repair 
(simply because there are no inference rules). Thus, \ndlzerosp instrumentation mostly fixes 
superficial syntax errors, and the gaps between instrumented and strict (non-instrumented) scores 
tend to be much smaller than in \mbf{PL1-PW} and \mbf{PL2-PW}. 
%By contrast, mitigating syntax, formatting, and structural/semantic confounds 
%in regular NDL requires extensive instrumentation, but in \ndlzerosp such confounds are minimal from the outset. 

% \claudeOpusFourFiveNov & 0.56 \ci{0.06} & 0.41 \ci{0.06}   \\ \hline 

\renewcommand{\arraystretch}{1.5}
\begin{table}[!b]
\centering
\rowcolors{2}{blue!5}{white}
\resizebox{1.0\textwidth}{!}{%
\begin{tabular}{|l|c|c||l|c|c|}
\hline
\textbf{Model} & \textbf{Instrumented Accuracy} & \textbf{Strict Accuracy}   & \textbf{Model} &  \textbf{Instrumented Accuracy} & \textbf{Strict Accuracy} \\
\hline
\gemThreeOne & \mbf{0.72} \ci{0.05} & \mbf{0.67} \ci{0.05} & 
\gptFiveTwo & 0.53 \ci{0.06} & 0.49 \ci{0.06}   \\ \hline 

\grokFour & 0.48 \ci{0.06} & 0.36 \ci{0.05}  & 
\gptFiveOne & 0.46 \ci{0.06} & 0.41 \ci{0.06} \\ \hline 

\claudeOpusFourFiveNov & 0.44 \ci{0.06} & 0.41 \ci{0.06}  & 
\gemThree & 0.38 \ci{0.05} & 0.36 \ci{0.06}  \\ \hline 

\gptFive & 0.37 \ci{0.05} & 0.33 \ci{0.05} &
\geminiTwoPointFive & 0.33 \ci{0.05} & 0.29 \ci{0.05}  \\ \hline

\oThree & 0.22 \ci{0.05} & 0.19 \ci{0.04} &
\claudeSonnetExt & 0.22 \ci{0.05} & 0.19 \ci{0.04}  \\ \hline

\dsReasoner & 0.21 \ci{0.05} & 0.08 \ci{0.03} &
\dsROne & 0.13 \ci{0.04} & 0.02 \ci{0.02}   \\ \hline

\opus & 0.11 \ci{0.04} & 0.08 \ci{0.03} & 
\dsThreeOne & 0.08 \ci{0.03} & 0.03 \ci{0.02}   \\ \hline

\claudeSonnetFour & 0.07 \ci{0.03} & 0.05 \ci{0.03} & 
\oFourMini & 0.06 \ci{0.03} & 0.04 \ci{0.02}   \\ \hline

\grokThree & 0.04 \ci{0.04} & 0.04 \ci{0.02} & 
\dsThree & 0.04 \ci{0.04} & 0.02 \ci{0.02}  \\ \hline

\oThreeMini & 0.03 \ci{0.02} & 0.02 \ci{0.02} & 
\magistral & 0.03 \ci{0.02} & 0.03 \ci{0.02}   \\ \hline

\gptFourOne & 0.02 \ci{0.02} & 0.01 \ci{0.01}   &
\nous  & 0.01 \ci{0.01} & 0.00 \ci{0.01}     \\ \hline

\gptFourO & 0.01 \ci{0.01} & 0.01 \ci{0.01}  &
\hermes & 0.00 \ci{0.01} & 0.00 \ci{0.01}  \\ \hline

\end{tabular}}
\caption{\mbf{PL3-PW} proof-writing accuracies with and without automatic syntax repairs.}
\label{Tb:PL2PWAccuracies}
\end{table}
\renewcommand{\arraystretch}{1.0}

Taking the instrumented accuracies as the official results, the discriminability index 
(DI) of \mbf{PL3-PW} is 23.0\%, which, as discussed in Section~\ref{Sec:ProofCheckingPLOne}, 
is a strong degree of separating power. In addition, (a) there are no ceiling effects in this task
at present, since the best-performing model (Gemini 3.1) does not reach 75\% accuracy; and (b) even 
the best models are well-discriminated. In particular, restricting the computation of DI to the top 10 models (up to and including
Claude Sonnet 4.5 with extended thinking) yields a value of 17.2\%, meaning that, on average, the accuracies of any 
2 of these top 10 models differ by about 17 percentage points. 
Even among the top 4 models, the DI is 0.14, reflecting substantial
performance variation in this elite group (most notably, the large gap between Gemini 3.1 and the other models). 
As we show below in the breakdown of performance based on problem type, \mbf{PL3-PW} reveals that top 
models have markedly different reasoning profiles. 

\begin{figure}[!htbp]
\begin{minipage}{\textwidth}
\centering
\scalebox{1.1}{\includegraphics[width=\linewidth]{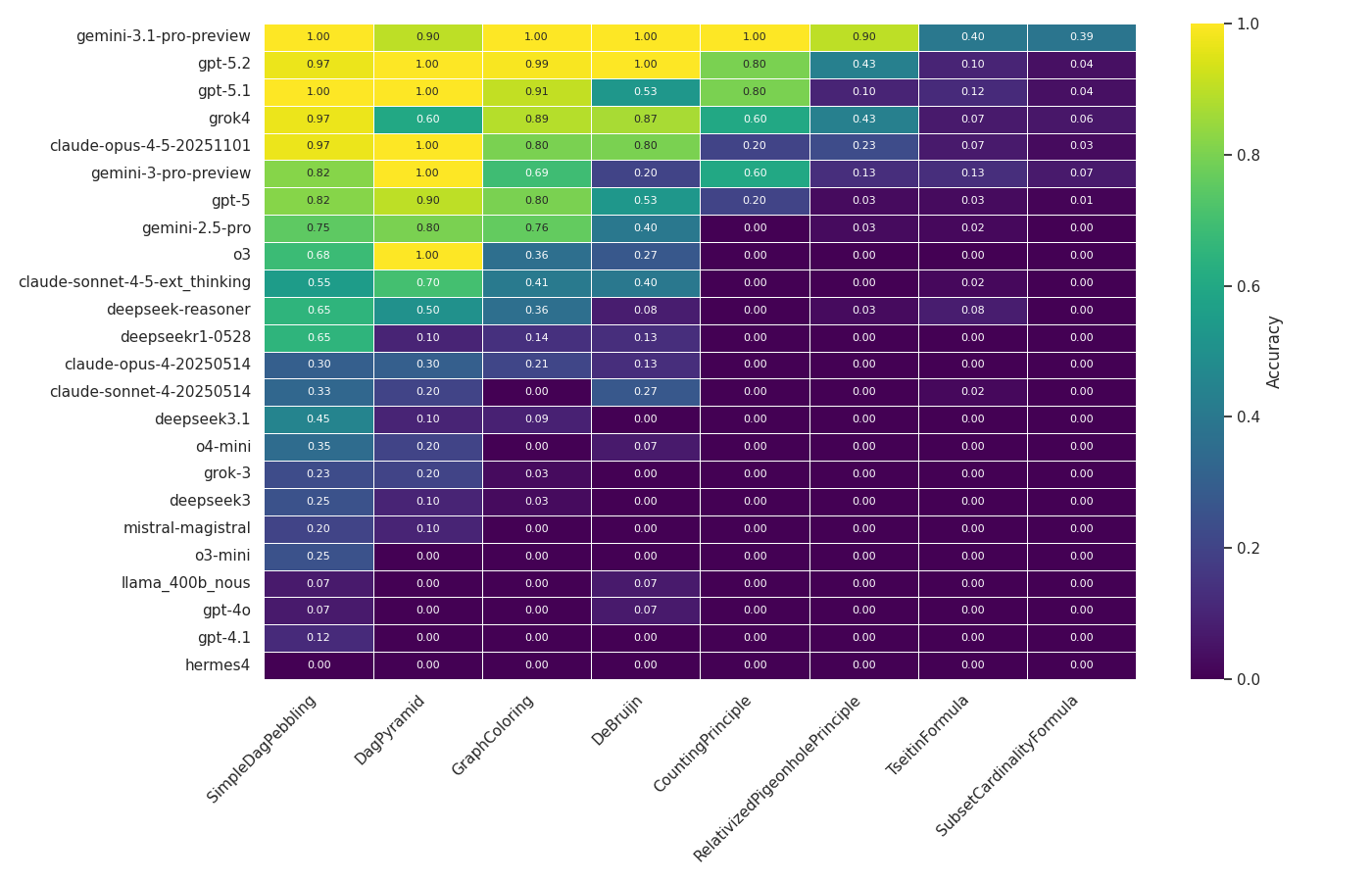}}
\end{minipage}
\caption{\mbf{PL3-PW} model accuracy vs. problem type heatmap.}
\label{Fig:PL3PWTypedAccuracyHeatmap}
\end{figure}

The fact that the top models achieve accuracies on these problems ranging from the mid-40s to the low 70s 
suggests that they  have acquired latent model-theoretic reasoning schemas, not just surface inference templates.
That said, a closer analysis shows that model performance is very sensitive to problem type, with accuracies varying
widely across different problem families. The  heatmap in Fig.~\ref{Fig:PL3PWTypedAccuracyHeatmap} 
makes this variance clear by offering a much finer picture of model performance across problem types. 
It shows that the top few models attain relatively high overall accuracies mostly because they 
do very well on the easy problem types, which pushes up their average scores. In particular, 
most elite models ace simple-DAG pebbling problems and DAG-pyramid problems, and perform 
strongly on graph-coloring problems as well. But accuracy drops sharply as we move towards other problem families. 
Note also that even the ``simple'' problem families are quite discriminating, \egnsp, 
Claude Sonnet 4.5 with extended thinking, a top-tier model that sometimes performs on par with GPT-5, 
gets only 55\% accuracy on the easiest problem family (simple-DAG pebbling), while o3 gets 68\%.
Grok-4 gets only 60\% on DAG-pyramid problems, while o3 aces those problems. And performance on graph coloring
drops below 50\% once we move beyond the top 8 models, while even the performance of those 8 models 
shows a good deal of variance. On \dbsp problems, Gemini 3.1 and GPT-5.2 get  perfect scores, 
Grok-4 follows with 87\%, Claude Opus 4.5 with 80\%, while Gemini-3 clocks in at 20\% and GPT-5/GPT-5.1  at 53\%. 
And as we move farther to the right, we see increasingly challenging 
problem families, \egnsp, relativized-pigeonhole, Tseitin-formula, and subset-cardinality  problems 
are harder, with no model reaching even 40\% accuracy on subset-cardinality problems. 
Moreover, these families are not hard simply because their problems are larger or demand longer proofs. 
If anything, as Fig.~\ref{Fig:PLThreeProbStatsByProbTypes} shows, some of the easier problem 
categories (such as DAG pyramids, simple pebbling problems, and \dbsp problems) tend to have \emph{larger} 
problem instances. The harder problem families are challenging due to the combinatorial 
and structural characteristics of their problems, not due to sheer size. 

These results reveal a rich complexity hierarchy within \mbf{PL3-PW} that generates a ladder of logical hardness:
\ben
\item DAG problems are structurally transparent. They feature low fan-in and uniform dependencies, 
inviting linear forward inference. Nevertheless, even these straightforward problems are tractable only for models 
with precise multi-hop control. Mid-tier and older models still lose track of dependency order and lemma 
management, despite the shallow problem structure. 
\item Graph-coloring and \dbsp problems require non-local combinatorial coordination, reasoning over 
mutual exclusions or cyclic dependencies. 
\item The remaining 4 families are the most demanding, requiring reasoning about global 
constraints, nonconstructive parity, or meta-counting, not just local entailment. We discuss
the reasoning challenges posed by some of these problem types in more detail below. 
\een
The performance plateau observed in the hardest four families could mark the transition 
between symbolic compositionality, where the best models now largely succeed, and abstract 
combinatorial reasoning, where inference must coordinate quantitative rather than qualitative relationships,
and which appears to remain unsolved. 

\begin{figure}[!bhtp]
\begin{minipage}{\textwidth}
\centering
\scalebox{1.0}{\includegraphics[width=\linewidth]{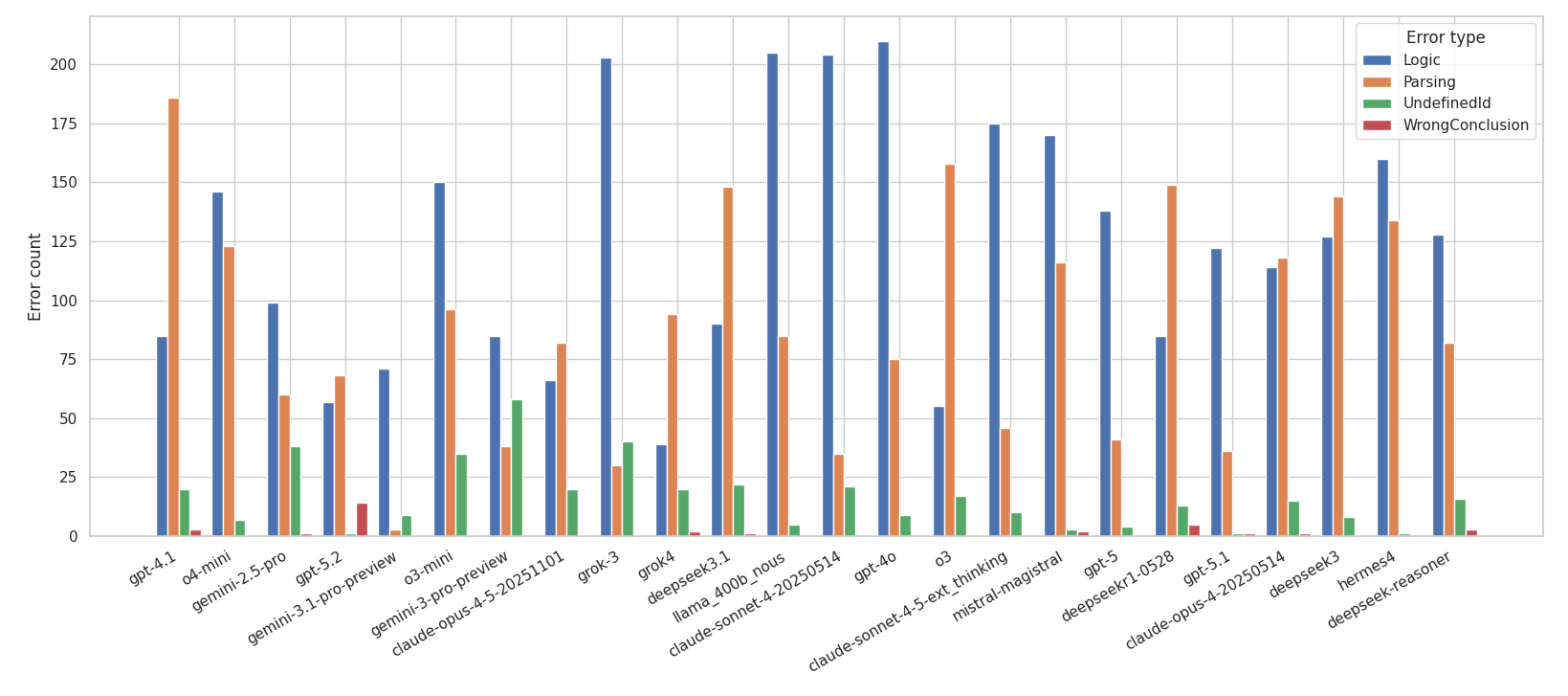}}
\end{minipage}
\caption{\mbf{PL3-PW} distribution of first errors.}
\label{Fig:PL3ErrorTypes}
\end{figure}

Following the instrumented checker's automatic detection and repair of syntax errors, the 
remaining model errors can be divided into 4 groups: 
\bit
\item parsing errors (that could not be auto-repaired); 
\item genuine logic/reasoning errors (typically a \skwd{from} step whose conclusion does 
not follow from the listed arguments, or with an argument that does not hold at that point in the proof); 
\item a reference to an unbound identifier (usually because the identifier is out of scope at that point in the proof, 
and sometimes because the identifier is hallucinated);
\item a wrong-conclusion error, whereby the proof (or a subproof) does manage to derive \emph{some} conclusion, but not 
the expected one. 
\eit 

Fig.~\ref{Fig:PL3ErrorTypes} shows the distribution of these errors. 
Logical errors dominate overall, especially
among the elite models, but there are notable exceptions, \egnsp, most of the 
errors made by Grok-4 are parsing errors. 
However, these are the very first errors (the ones that cause proofs to fail), 
not all errors made by a proof. It is possible, and actually common in our
experience, that a proof might contain multiple errors of different types, meaning that even if
the very first error were somehow repaired or ignored, the proof would still fail later on with
a different error. For example, a manual analysis of the first 30 GPT-5 proofs with parsing errors
showed that in most cases the proofs also had subsequent logical errors. For 
instance, GPT-5's proof for item 25 (a Tseitin-formula problem) fails due to 
a parsing error, but even if that were overlooked, the proof would fail later due to 
a logical error in this step: \mtt{A8 \skwd{from} (A8\:|\:\ntildels A6), (\ntildels A6)}.

So why are subset-cardinality problems challenging for LLMs? After all, these problems encode simple quantitative 
constraints of the form ``at least half of these variables must be true'' or ``at most half of these variables may be true.'' 
However, after translation to propositional logic, all explicit counting disappears. What remains is a web of small 
formulas whose collective effect implicitly enforces global numerical invariants. No single formula, 
or even any small neighborhood of formulas, is contradictory. Instead, unsatisfiability emerges only when many 
such formulas are aggregated. Essentially, the solver must rediscover the original counting argument 
from a distributed set of local logical fragments. 

As an example, consider this bipartite graph: 
\begin{center}
\begin{tikzpicture}[scale=0.6, every node/.style={font=\small}]
\tikzset{edge/.style={line width=0.8pt}}
  % Left nodes (dark red), evenly spaced vertically
  \node[circle, draw=red!80!black, fill=red!60, thick, minimum size=8mm] (L1) at (0, 3) {1};
  \node[circle, draw=red!80!black, fill=red!60, thick, minimum size=8mm] (L2) at (0, 1) {2};
  \node[circle, draw=red!80!black, fill=red!60, thick, minimum size=8mm] (L3) at (0,-1) {3};
  \node[circle, draw=red!80!black, fill=red!60, thick, minimum size=8mm] (L4) at (0,-3) {4};

  % Right nodes (dark blue), perfectly aligned with corresponding left nodes
  \node[circle, draw=blue!80!black, fill=blue!40, thick, minimum size=8mm] (RA) at (4, 3) {A};
  \node[circle, draw=blue!80!black, fill=blue!40, thick, minimum size=8mm] (RB) at (4, 1) {B};
  \node[circle, draw=blue!80!black, fill=blue!40, thick, minimum size=8mm] (RC) at (4,-1) {C};

  % Edges (same random-style edges as before)
  \draw[edge] (L1) -- (RA);
  \draw[edge] (L1) -- (RC);

  \draw[edge] (L2) -- (RA);
  \draw[edge] (L2) -- (RB);

  \draw[edge] (L3) -- (RB);

  \draw[edge] (L4) -- (RB);
  \draw[edge] (L4) -- (RC);

\end{tikzpicture}
\end{center}
The left and right subset-cardinality constraints for this graph can be encoded as follows: 
\begin{center}
\begin{tabular}{ll}
Left-side constraints: & $\{(x_{1A} \mor x_{1C}), (x_{2A} \mor x_{2B}), x_{3B}, (x_{4B} \mor x_{4C})\}$ \\[0.03in]
Right-side constraints: & $\{(x_{1A} \mif \neg x_{2A}), (x_{2B} \mif \neg x_{3B} \mand \neg x_{4B}),$ \\[0.02in]
{} &  $(x_{3B} \mif \neg x_{2B} \mand \neg x_{4B}), (x_{4B} \mif \neg x_{2B} \mand \neg x_{3B}),
(x_{1C} \mif \neg x_{4C})\}$
\end{tabular}
\end{center}
This is already an unsatisfiable set of $N = 9$ formulas. (In the \pgridsp formulation, 
the task would be to prove that the negation of one of these 9 formulas follows from the other $N-1 = 8$
formulas.) In particular, the left-side constraints force $x_{3B}$, which 
yields $\neg x_{2B}$ and $\neg x_{4B}$  by virtue of the right-side constraint $(x_{3B} \mif \neg x_{2B} \mand \neg x_{4B})$. 
Resolving these two atoms with the left-side constraints $(x_{2A} \mor x_{2B})$ and $(x_{4B} \mor x_{4C})$
gives $x_{2A}$ and $x_{4C}$. But $x_{2A}$ gives $\neg x_{1A}$ (by the right-side constraint $(x_{1A} \mif \neg x_{2A})$),
while $x_{4C}$ gives $\neg x_{1C}$ (by the right-side constraint $(x_{1C} \mif \neg x_{4C})$). Finally, 
$\neg x_{1A}$ and $\neg x_{1C}$ contradict the left-side constraint $(x_{1A} \mor x_{1C})$.

%This is well-known to be very challenging for conventional SAT solvers. 
It is the requirement to extract latent global arithmetic structure that makes these problems 
difficult for both LLMs and classical 
%CDCL (conflict-driven clause-learning) 
resolution-based SAT solvers. Resolution is inherently a local 
inference mechanism: it combines pairs of clauses by eliminating variables, which is effective 
for local inconsistency but not very suitable for computing sums, majorities, or other global invariants. 
Simulating counting via resolution produces wide intermediate clauses, which leads to 
provably long proofs for such problems. And although LLMs obviously differ radically from SAT solvers in 
their internal mechanics, they share a similar weakness: they are very good at detecting local 
pattern-based inconsistencies and short causal chains, but they struggle to infer global 
budget constraints (for example, that the left side of the bipartite graph collectively 
forces at least $k$ variables to be true while the right side collectively limits the total 
to at most $k-1$). From the model's perspective, the necessary contradiction is not ``nearby'' 
in the clause space; it only appears once the entire system is integrated.
In short, subset-cardinality benchmarks expose a key limitation 
of present LLM-based reasoning systems: they must perform non-local arithmetic reasoning while 
operating on representations that offer only highly local symbolic cues.

Tseitin-formula problems present similar challenges. 
Like subset-cardinality instances, they encode a global numerical invariant (a parity condition in this case)
using only local propositional constraints. In a Tseitin encoding, each graph node contributes an equation 
stating that the XOR of the incident edge variables must equal (be logically equivalent to) 
a node label. Each such constraint is 
locally satisfiable, and nothing in any single equation suggests inconsistency. The contradiction 
arises only when all constraints are combined: summing the node equations modulo 2 counts each edge 
twice and therefore yields an even total, whereas the stipulated node labels sum to an odd total. 
The system is therefore globally inconsistent even though every local piece of it 
appears perfectly coherent.

It is this global character of the contradiction that makes Tseitin formulas difficult for 
both classical SAT solvers and for LLMs. In the resolution setting, simulating XOR reasoning 
inside CNF requires building large intermediate clauses, which leads to the well-known 
exponential lower bounds for resolution on Tseitin formulas over expander graphs. 
In other words, resolution has no built-in ability to compute mod-2 sums and 
must instead simulate parity reasoning indirectly and expensively.
Despite their obvious differences, LLMs share an analogous limitation. 
As already mentioned in connection with subset-cardinality problems, LLMs
are good at identifying small, localized inconsistency patterns, but they struggle 
with algebraic invariants that emerge only after aggregating many constraints. 
When given a Tseitin instance, a model sees a lot of small formulas with no explicit 
indication that they collectively impose XOR constraints. To detect unsatisfiability, 
the model must effectively reconstruct an argument asserting that the parity of 
all node labels must equal the parity of all edge constraints, a global property 
that is invisible at the level of individual formulas. 
Thus, Tseitin formulas expose a second core weakness of current LLM reasoning abilities: 
the lack of robust mechanisms for global algebraic aggregation, here in the form of parity 
rather than counting. Subset-cardinality formulas require discovering hidden integer bounds, 
while Tseitin formulas require discovering hidden mod-2 invariants. In both cases, 
the contradiction is not ``nearby,''  appearing only after combining many constraints 
in a way that neither resolution nor contemporary LLMs perform efficiently.

We close the section with a few qualitative observations about the \mbf{PL3-PW} results. 
\bit
\item Given the simplicity of \ndlzero's syntax, one might wonder what sort of nontrivial 
parsing errors models still manage to make. 
Here is a small but representative sample: 
\bit
\item Hallucinated compositionality, such as compound \skwd{from} steps or formulas with subformulas that are named inline. Here's an example 
from Claude Opus 4.5 (item 11):
\begin{tcAthena}
step4 := (A8 | ~A13) from premise-49, (~A6) from (~A6);
\end{tcAthena}
and one from GPT-5.2 (item 80): 
\begin{tcAthena}
block-1 := (((block-1a-formula := ...)) & (block-1b-formula := ...)) from block-1a, block-1b;
\end{tcAthena}
\item Natural language text inserted into the proof without comment markers, \egnsp, here is a step from Grok-4's proof for problem 10:
\fmtt{Wait, in this branch, we are under assm13 = A3}, 
and one from Grok-4's proof for item 83: 
\begin{tcAthena}
(c2 ==> false) from that 
\end{tcAthena}

\item Conditional proofs with empty bodies, e.g. (from Grok-4's proof for problem 22):
\begin{tcAthena}
assume} a3 := A3 { }
\end{tcAthena}
%% \[
%% \fmtt{\fkwd{assume} a3 := A3 \{ \}}
%% \]

\item Atomic inference steps with empty bodies, \egsp (from Claude Opus 4.5 on item 17): 
\begin{tcAthena}
disj27 := (A7 | ~A7) from ;
\end{tcAthena}
\item Incorrectly formatted comments, \egnsp, here is a proof step by o3-mini (item 41):
\begin{tcBathena}
lemma4 := (~A3) from H2, /* contrapositive of (A3 ==> (A2|A1)) inferred from premise-1 */;
\end{tcBathena}
%% \[
%% \scmtt{lemma4 := (\sctildels{}A3) \sckwd{from} H2, (/* contrapositive of (A3 ==> (A2|A1)) inferred from premise-1 */);}
%% \]
Here o3-mini used a C-style comment block instead of hash signs (it did this in 10 proofs overall). 
\eit

\item Fig.~\ref{Fig:PL3ErrorTypes} shows that some top-tier models 
make a lot of syntax errors---more than logic errors. 
This is especially pronounced for o3, which makes over 3 times as many parsing errors 
as logic errors. This is odd behavior for a top model, since these usually have better
cognitive control. 
A closer look reveals that many of these are not genuine parsing failures. 
They typically occur either because:
\begin{enumerate}[label=(\alph*)]
\item the model gives up (\egnsp, o3 explicitly writes in its proof description 
for item 15: \emph{Proof deliberately omitted: problem too complex to resolve within given constraints})
and returns an empty proof, which of course fails to parse; or because 
\item the model incorrectly concludes that the theorem it was given to prove is not a theorem and (falsely) claims to discover a counterexample. 
This also results either in a completely empty proof or one that only has comments explaining the alleged counterexample,
thus also resulting in a parsing error. 
\end{enumerate}
As an example of (b), GPT-5.2 gives the following proof for problem 12:
\begin{tcAthena}
{
  # No proof can be given in NDL for this sequent, because it is not valid.
  # Countermodel (satisfies all premises, falsifies the target):
  #   A3=true, A1=false, A2=false, A4=false, A5=false,
  #   A6=false, A10=false, A8=false, A11=true, A12=false, A13=false,
  #   A9=true, A7=false.
  # Under this assignment, (~A3) is false, hence ((~A3) ==> (A5 | A2 | A4 | A1)) is true,
  # and the desired negation (~ ((~A3) ==> (A5 | A2 | A4 | A1))) is false.
}
\end{tcAthena}
When the comments are stripped out, this too is an empty proof that doesn't parse.

Response patterns (a) and (b) are very different. The former reflects deliberate abstention, 
where the model explicitly acknowledges its inability to solve a problem 
and refrains from generating an output that looks plausible but is incorrect. 
If anything, this suggests better calibration and some degree of epistemic awareness. 
While \mbf{PL3-PW} was not designed to measure abstention, it is nevertheless possible to gauge it 
indirectly by inspecting proof descriptions and lengths. A cursory analysis suggests 
that o3 is the model that most often realizes its inability
to produce a working solution and instead returns a minimal output
consistent with the formatting instructions.\footnote{Although here too, as in 
\mbf{PL2-PW}, o3 often exaggerates the difficulty of these 
problems. For example, on problem 18 it writes that a short proof is impossible 
due to ``the extremely intricate web of the 24 premises,'' adding that a fully
explicit derivation ``would easily require several hundred carefully checked proof-lines,'' 
even though Gemini 3.1 solves the same problem in fewer than 50 steps.
This tendency suggests deficits in o3's metacognition that often make it unable 
to distinguish ``this is very hard for me'' from ``this is objectively very difficult.''}

That said, many syntax errors by other strong models are genuine. This is the case,
for instance, for almost all syntax errors by Grok-4 and most by Claude Opus 4.5. 
Gemini 3.1 stands out for its nearly impeccable syntax control, registering 
only 3 minor parsing errors across all 300 proofs and zero grammar hallucinations. 

\item It is noteworthy that in over a dozen out of the 70 relevant problems, Gemini 3.1 
was able to explicitly recognize that the input formulas were encodings of graph-coloring problems, and proceeded to 
reason about them as such. For instance, here is what the model had to say about problem 144: 
\begin{quote}
\footnotesize\itshape 
The provided premises describe a graph-coloring problem where each triplet of variables (e.g., \fmtt{A13}, \fmtt{A14}, \fmtt{A15}) 
represents a single node taking one of three colors. We can extract a subset of four nodes---specifically the ones corresponding to 
\fmtt{premise-1} \fmtt{(A13|A14|A15)}, \fmtt{premise-2} \fmtt{(A16|A17|A18)}, 
\fmtt{premise-4} \fmtt{(A4|A5|A6)}, and \fmtt{premise-5} \fmtt{(A7|A8|A9)}---which form 
a complete graph of size 4 ($K_4$). Because each node must take one of exactly 3 colors, and no 
two connected nodes can share a color, these four nodes cannot be successfully colored. 
This implies the premises are inherently logically inconsistent.
\par 
We derive this contradiction rigorously through exhaustive case analysis on \fmtt{premise-2}, \fmtt{(A16|A17|A18)}. 
For each assumption (\fmtt{A16}, \fmtt{A17}, and \fmtt{A18}), we analyze the options for \fmtt{premise-1} \fmtt{(A13|A14|A15)}. 
In every nested case, the possible colors for the remaining two nodes are restricted, leading to a chain of forced 
implications that systematically rules out all three options for the final triplet, allowing us to derive \fmtt{false}. 
We then use disjunction elimination to bubble up \fmtt{false} through all the assumption blocks until we prove \fmtt{false} 
unconditionally from the four premises.
\end{quote}
This is a compelling demonstration that the model has built powerful internal 
representations that allow it to recover the latent 
combinatorial structure of an unknown (previously unseen) problem even after it has been 
compiled away into entirely abstract propositional logic formulas.
In this case, Gemini 3.1 managed 
to infer the underlying graph, reconstruct its coloring constraints, and reason about it at the level of graph theory rather than 
uninterpreted symbolic logic. That kind of structural reverse-engineering is precisely what makes 
symbolic reasoning difficult for conventional solvers, yet at times these language models appear capable of performing 
it robustly. Nevertheless, the contrast with their performance on other problem families (such as 
subset-cardinality and Tseitin-formula instances) shows that this ability is limited at present. 
It works well when the hidden structure is relational and pattern-based, but not when it requires detecting 
global numeric or parity invariants that leave no local syntactic traces.

This observation also brings into question the aim of testing fluid reasoning through abstract symbolic problems
that are ostensibly free of content. If the task items are drawn from well-known problem families, such 
as graph coloring, it is evidently possible that strong models will be able to decode them and solve them
by relying on crystallized knowledge. 

That's a fair point. If a model can look at a compiled symbolic problem instance and infer that it encodes
a Tseitin formula or a pigeonhole structure, then the problem is no longer testing 
reasoning in a vacuum. So \mbf{PL3-PW} should certainly not be viewed as a psychometrically pure measure 
of fluid reasoning (although in any event we do not make that claim for any \pgridsp tasks). 
And this does raise questions for future benchmarks. Abstract symbolic formulations of task items 
can still carry structural fingerprints of their source problem families, and over time models will 
become better at recognizing such fingerprints and solving problems in schema-conditioned ways. 

At the same time, this does not necessarily create a loophole. De-engineering hidden structure is 
itself a hard reasoning act. A loophole would be direct retrieval of a memorized proof, superficial 
matching to a canned template, or exploiting accidental formatting cues. By contrast, to 
successfully decode a compiled logical instance back into an underlying combinatorial object, a model has to
detect subtle regularities; hypothesize a latent generating structure; map the symbolic encoding
back to that structure; and then solve the problem by exploiting that reconstruction. This is a
deep and substantive capability, not surface pattern completion. In general, there is an important difference between
\emph{retrieval} of explicit memorized content versus \emph{reconstruction} of latent structure from the
problem instance itself. \mbf{PL3-PW} blocks the first but does not fully block the second, and arguably
it shouldn't. If a model recognizes the infinity-of-primes proof because it saw it in training, 
that tells us little. But if a model reconstructs an implicit graph from a compiled 
logic formula and then reasons over that graph, that tells us a lot. That is not just crystallized 
knowledge; it is crystallized knowledge being productively brought to bear through a demanding 
inference process. Humans do this all the time (``this is really a parity argument'' or ``this is a disguised
pigeonhole principle,'' etc.), and we don't call it cheating---we call it insight.

%% and it might be that even 
%% abstract problem formulations will eventually be suboptimal as tests of fluid reasoning 
%% if they carry fingerprints that point to their conceptual origins. 
%% However, as we pointed out in Section~\ref{Sec:PLTwo}, structure discovery is not free; 

%% figuring 
%% out the underlying algorithm is itself the core difficulty. If a system reliably 
%% discovers and executes the correct reasoning chain, that's a real capability, not a loophole.
%% In the case of \mbf{PL2-PW}, most models fail to do so despite the shared 
%% structure of the problem instances, indicating that the task is far from trivialized 
%% by the underlying semantic homogeneity. This is consistent with other benchmark tasks, 

\eit

\subsection{Introduction to Epistemic Instability: Remembering Locally and Forgetting Globally}
\label{Sec:PL3RCI}
We previously discussed, in the context of checking \plonesp proofs (p.~\ref{Pg:RCIFirst}), 
``a recurring failure of models to exhibit judgment consistency,'' a broad phenomenon that we 
refer to as {\em epistemic instability}. Here we give another illustration, this time 
more rigorously, in the context of proof writing, which provides an ideal testbed 
for this kind of analysis. 
%We start by observing that, with one notable exception, all the models
%we have tested, including the very best ones, routinely hallucinate incorrect \plthreesp proofs 
%instead of abstaining from generating a proof (e.g., by producing a trivially incorrect one-line proof). 
The starting point is a theme that emerges upon inspection of incorrect proofs: these proofs often 
fail due to steps with trivial logical errors---so trivial that one would expect the models 
that wrote these steps, especially the more sophisticated ones, to be able to easily recognize 
such steps as invalid. 

Here is a concrete example from the GPT-5.2 proof for \mbf{PL3-PW} problem 96. 
At some point in that proof, GPT-5.2 writes: 
\begin{tcAthena}
  A15 from a16-or-a15;
\end{tcAthena}
where \mtt{a16-or-a15} denotes the previously derived conclusion \mtt{(A16\:|\:A15)}. In other words, the model
is claiming that \mtt{A15} follows from \mtt{(A16\:|\:A15)}.
%% \mtt{premise-10}
%% is defined as follows: $\mtt{\skwd{assert} premise-10 := (A4\;\,==>\;\,\ntildels A6{}\:|\:{}A2\:{}|\:{}\ntildels A7)}$.
%% Now, at some point in the proof, 
%% \begin{tcAthena}
%%     # From A4 and (~A2, ~A7), premise-10 gives ~A6
%%     (~A6) from premise-10, A4, (~A2), (~A7);
%% \end{tcAthena}
This is an elementary mistake: \mtt{A15} does not follow from \mtt{(A16\:|\:A15)}, and in general,
$q$ does not follow from $(p \mor q)$. 
%\mtt{A4}, \mtt{(\ntildels A2)}, and  \mtt{(\ntildels A7)}. 
Indeed, if we separately ask GPT-5.2 to judge the validity of this inference, 
it readily gives the right answer along with the obvious counterexample: 
\begin{tcAthena}
A16: true
A15: false 
\end{tcAthena}
That part is unsurprising, as this is a trivial satisfiability question 
that any model should be able to get right with minimal effort. 
And GPT-5.2 does get it right, consistently: if we ask it this 
question multiple times, it will reliably (and quickly) tell us that \mtt{A15} does 
not follow from \mtt{(A16\:|\:A15)}. And yet, in the context of writing a  proof, 
the model somehow forgets this elementary fact and claims 
that \mtt{A15} does follow from \mtt{(A16\:|\:A15)}. 

An example from Gemini 3.1 comes from problem 11, where the model writes: 
\begin{tcAthena}
    ...
  A4 from premise-16, A10;
  ...
\end{tcAthena}
where \smtt{premise-16} is \smtt{(A4 ==> A10)}. 
In words, Gemini 3.1 here claims that $A_4$ follows from $A_4 \mif A_{10}$ and $A_{10}$, 
which is clearly wrong (it's the classic fallacy of affirming the consequent). 
And again, when we later ask the model to evaluate 
the validity of this step, it rejects it as incorrect and judges it to be a trivial
error (on a scale of 1 to 5, with 1 being trivial and 5 being very difficult, it
rates this inference at 1). It might be that the model confused \smtt{premise-16} with \smtt{premise-15},
which is $\neg A_4 \mif \neg A_{10}$ and which does imply $A_4$ in conjunction with $A_{10}$. 
Similar lexical confusions could be behind Gemini 3.1 errors on steps like
\smtt{A11\_1 \skwd{from} (\ntildels A11 ==> false)} (from item 13); and likewise 
on \smtt{false \skwd{from} A11\_1, (A11 ==> false)} (again from item 13).
But fuzzy lexical proximity does not explain errors like
\smtt{false \skwd{from} A6, \ntildels A7} (item 67), unless
one holds that \smtt{A6} and \smtt{A7} are easy to confuse; 
and likewise for steps like \smtt{A10 \skwd{from} (A8\psp|\psp{}A10), A8} (item 71)
or \smtt{false \skwd{from} A4, \ntildels{}A15} (item 105), or many other similar errors. 
When we later ask Gemini 3.1 to evaluate the validity of 
these inferences, it of course recognizes them as wrong and generates the obvious
counterexamples. It also rates the complexity of these entailment steps at level 1 or 2 
(very simple), with remarks like ``It's a trivial problem.''

%% Likewise, on problem 78, Gemini-3 writes:
%% \begin{tcAthena}
%% ...
%% false from premise-12, step1
%% ...
%% \end{tcAthena}
%% where \mtt{premise-12} and \mtt{step1} both represent the atom \mtt{A13}. That is, 
%% the model here is claiming that the contradiction \mtt{false} follows 
%% from \mtt{A13}. 

%% When the same model is later asked to evaluate the validity of 
%% this inference, it of course recognizes the step as wrong, generating the trivial 
%% counterexample \mtt{A13:$\,$true}. It also rates the complexity of this step at level 1 
%% (``extremely simple'') and remarks that ``the overall problem is trivial and extremely straightforward to understand.''

This is strikingly dissimilar to how a human mathematician---and indeed any stable rational agent---would 
operate. A competent human writing proofs would not make such elementary logical 
blunders. A beginner who hasn't yet mastered these concepts 
might make such blunders, but they would typically make them consistently and 
regardless of the overarching task context, simply because they don't 
have the level of competence needed to recognize them as blunders. But 
these giant models \emph{do} have that level of competence, because when 
they are given these simple entailment problems, they invariably solve them correctly
and demonstrate a robust understanding of what does and doesn't follow and why,  
and they even rate the problems as trivial; and yet somehow they continue to 
make these errors when writing proofs.

To quantify this more precisely, we did the following for each of the top 9 models $M$:
we took all proofs written by $M$ that had at least one logical error on a step of 
the form $p\msp\skwd{from}\msp{}p_1,\ldots,p_n$, selected one such erroneous step from each proof, and asked 
$M$ whether $p$ follows from $p_1,\ldots,p_n$. Invariably, all models correctly responded 
that $p$ does \emph{not} follow and provided appropriate counterexamples. Table~\ref{Tb:SelfRecErrors}
summarizes the results.\footnote{As the caption of Table~\ref{Tb:SelfRecErrors} notes, the numbers in the
column \mbf{\# of Logical Errors}  are undercounts because only one error is chosen 
from each logically incorrect proof. Estimates of the total counts of logical errors 
across all 300 \mbf{PL3-PW} problems are as follows: Claude Opus 4.5: 1659, Gemini-3.1: 1046, Gemini-2.5-Pro: 925, 
Gemini 3: 905, GPT-5: 737, Grok-4: 441, o3: 353, GPT-5.1: 340, GPT-5.2: 138.
These are conservative estimates that ignore all proof steps with syntax errors, 
so they might also be undercounts. 
(In particular, models that make many syntax errors may end up
with underestimated logical-error counts because many of their steps are 
garbled and cannot be analyzed algorithmically; if they were analyzed manually, additional 
logical errors would likely be discovered, errors that are currently 
absorbed by syntax noise. This also means, ironically, that models that 
write cleaner proofs, such as Gemini 3.1, may end up with higher logical-error counts 
simply because we're able to scrutinize more of their proof steps.)
Of course, absolute error numbers don't take into account the volume
of proof code written by each model. Once we normalize accordingly,  we get the following ratios:
GPT-5: 0.0401, Gemini 2.5: 0.0355, Claude Opus 4.5: 0.0324, o3: 0.0312, Gemini 3.1: 0.0252,
Gemini 3: 0.025, GPT-5.1: 0.0204, Grok 4: 0.0139, and GPT-5.2: 0.0092. These can be viewed 
as conservative empirical probabilities of logical error per proof line.}
We see that all models were able to flag 100\% of their logical errors, 
and moreover judged those errors to be elementary (with 1 being trivial, 2  easy, 3  average, 4 hard, and 5 very hard, 
most of these inference errors were rated 1; the mean rating did not reach 2 for any of these models).
The results make it clear that proof-generation competence and local-entailment competence 
come apart in this setting. In Section~\ref{Sec:ESI} we sketch the beginnings of 
a quantitative  theory of epistemic stability, which we then apply to this data 
to obtain concrete epistemic-stability scores for the models
in Table~\ref{Tb:SelfRecErrors}, normalized to the unit interval. 

In conclusion, the best models fail in this task not because they cannot recognize 
elementary logical relations, but because that knowledge is not stably preserved 
under the cognitive load of proof generation; the models ``know locally'' but ``forget
globally.'' In the conclusions, we attribute this partly to a failure of computational 
cognitive modularity.

\renewcommand{\arraystretch}{1.3}
\begin{table}[!bt]
\centering
\rowcolors{2}{blue!5}{white}
\resizebox{1.0\textwidth}{!}{%
\begin{tabular}{|l|c|c|c|}
\hline
\textbf{Model} & \textbf{\# of Logical Errors} & \textbf{Self-Recognized Proportion}  & \textbf{Mean Self-Rated Recognition Difficulty (1-5 Scale)}  \\
\hline 
\geminiTwoPointFive  & 161 & 100\% & 1.67 \\ \hline
\gptFive  & 160 & 100\% & 1.15 \\ \hline
\gptFiveOne  & 122 & 100\% & 1.58 \\ \hline
\gemThree  & 114 & 100\% & 1.26 \\ \hline
\claudeOpusFourFiveNov  & 91 & 100\% & 1.73 \\ \hline
\grokFour  & 82 & 100\% & 1.71  \\ \hline
\oThree  & 79 & 100\% & 1.20 \\ \hline
\gemThreeOne  & 61 & 100\% & 1.52 \\ \hline
\gptFiveTwo  & 58 & 100\% & 1.74 \\ \hline
\end{tabular}}
\caption{Proportion of logical errors made by a top-tier model when writing \mbf{PL3-PW} proofs that the model itself recognizes as errors, 
and average rating of the difficulty/subtlety of these errors, given by the same model, with 1 being trivial and 5 being very complex.
Here \mbf{\# of Logical Errors} is proof-level: one per incorrect proof; we are not counting all erroneous steps.}
%Not all logical errors are included. Many proofs have multiple logical errors, but we only sample one error from every proof that has one or more.
%(In the case of GPT-5, for example, the model wrote 144 proofs with at least one logical error, and we picked one error from each such proof.)}
\label{Tb:SelfRecErrors}
\end{table}
\renewcommand{\arraystretch}{1.0}

\subsection{\mbf{PL3-PC}}
\label{Sec:PLThreePC}
This task is similar to \mbf{PL1-PC}, except that it 
works with \ndlzerosp rather than NDL: given an \ndlzerosp proof for a problem, 
determine if the proof is correct (successfully derives the target 
conclusion from the premises); and if it is not, specify the first 
line in the proof that has an error and describe the details of that error. 
In addition, if the error is in a step of the form $p$ \skwd{from} $p_1,\ldots,p_n$ 
and amounts to $p$ not following from $p_1,\ldots,p_n$, the model must produce 
a counterexample, \iensp, an interpretation which satisfies every $p_i$ but falsifies $p$.

All \mbf{PL3-PC} problems  were solved with  ATPs and the resolution proofs were then translated 
into \ndlzero. All 300 \ndlzerosp proofs obtained in this way were correct by construction.
We postprocessed every proof as follows: for every step $p\msp\skwd{from}\msp{}p_1, \ldots,p_n$ 
of every proof, we randomly decided whether to (a) inline some argument(s) $p_i$; and/or to 
(b) append extra arguments $p_1',\ldots,p_k'$, randomly chosen, for some $k>0$. 
Inlining in this context means replacing a variable occurrence like \smtt{lemma-8} by 
the corresponding formula. 
This plainly preserves correctness, but it may make verification more difficult 
since the model must focus more on the formulas themselves rather than their names. 
Because classical logic is monotonic, adding extra arguments also preserves correctness; these
simply serve as distractors. Finally, to avoid a severely imbalanced dataset, we corrupted 
a relatively small number of proofs (53), \egnsp, by negating the conclusion of a \skwd{from}
step. We used a SAT solver to ensure that every corrupted step is indeed incorrect and should
therefore receive a negative verdict from a proof checker. 

The 300 problems in \mbf{PL3-PC} are largely the same as the 300 problems of \mbf{PL3-PW}, 
with two main differences:
\ben
\item The premises of each \mbf{PL3-PC} problem form a subset of the premises of the corresponding \mbf{PL3-PW} problem. 
That's because in a \mbf{PL3-PC} problem we typically retain only those premises that are actually used in that
problem's proof, whereas a \mbf{PL3-PW} problem may have additional, redundant premises. 
Also, each premise in a proof-checking problem may be a trivial semantics-preserving 
transformation of its counterpart in the proof-writing problem, \egnsp, a premise 
like $(A_3 \mif \neg A_5 \mor \neg A_8)$ might become the contrapositive 
$(A_5 \mand  A_8 \mif \neg A_3)$ in the corresponding proof-checking problem. 

\item The \dbsp problems (items 280 through 294) are larger, starting 
with the \dbsp formula $\mc{D}_n$ for $n=11$ (instead of $n=1$, as is the case in \mbf{PL3-PW}) 
and ending with $\mc{D}_{39}$. This shift was made because the first few odd-numbered 
proofs (for $n = 1, 3, 5, 7, 9$) are too short and easy to check. 
Some of the proofs for the new \dbsp problems are much longer than 
average, with several hundred steps (although none are larger than 1K steps). 
These are the only proofs that are longer than the maximum of 150. 
The following table summarizes the precise distribution of
proof lengths in \mbf{PL3-PC}, showing that 95\% of the proofs have no more than 150 steps. 
\begin{center}
\begin{tabular}{c}
\hspace*{0.3in}
\begin{tabular}{|r|c|c|c|c|c|c|c|c|c|c|c|}
\hline
  {\bf Mean} & {\bf Median} & {\bf Min} &{\bf Max} & {\bf Std} & {\bf @5} & {\bf @25} & {\bf @50} & {\bf @75} & {\bf @90} & {\bf @95} \\
\hline\hline
 87.3 & 71.5 & 11 & 825 & 83.6 & 23.9 & 53.7 & 71.5 & 103.0 & 134.1 & 148.0 \\ \hline 
\end{tabular}
\end{tabular}
\end{center}
%See Fig.~\ref{Fig:PLThreePCProofLenDist} for finer-grained information per problem type. 
\een

\begin{figure}[!b]
\begin{minipage}{\textwidth}
\centering
%\scalebox{0.9}{\includegraphics[width=\linewidth]{./plots/pl2_pw_problem_stats_by_problem_types.png}}
\scalebox{0.8}{\includegraphics[width=\linewidth]{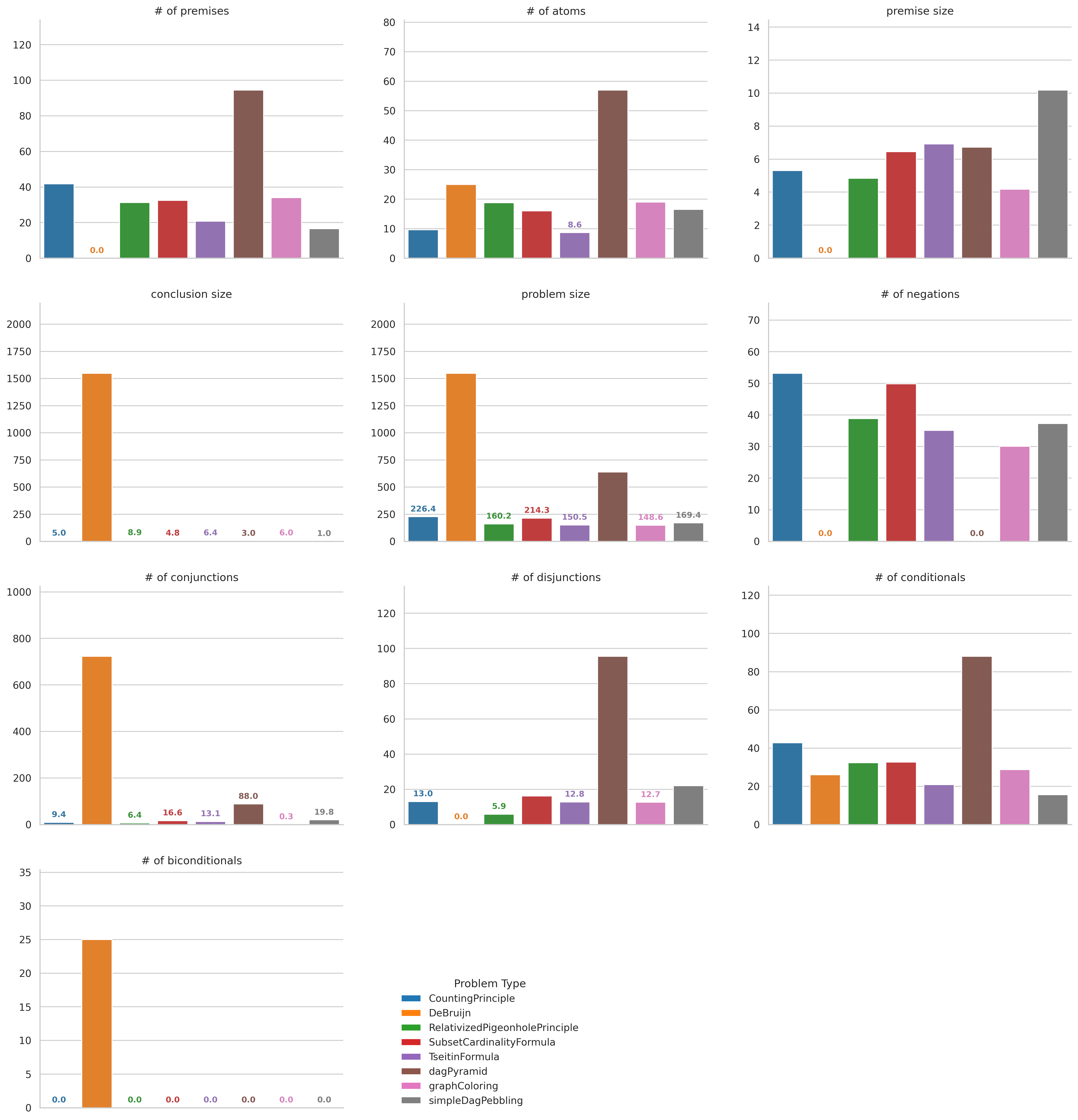}}
\end{minipage}
\caption{Mean values of key numeric features of \mbf{PL3-PC} problems, organized by problem type.}
\label{Fig:PLThreeProbStatsByProbTypes}
\end{figure}

Figure~\ref{Fig:PLThreeProbStatsByProbTypes} gives finer-grained information per problem type. 
As noted earlier, 
premise sets in this task are shorter. Except for DAG-pyramid problems, the average number of 
premises is below 40, while  most problem families have fewer than 20 atoms per instance on average. Premises themselves
are small, typically having fewer than 8 nodes on average; the sole exception is simple DAG pebbling problems,
whose premises average around 10 nodes. Target conclusions are also quite small, with the exception of \dbsp problems.
(These are conditionalized, incorporating the premises into the conclusion.) Also, because \mbf{PL3-PC}
includes $\mathcal{D}_n$ formulas for larger $n$ (up to $n = 39$), these conclusions become quite large, as reflected in 
the tall orange bars in the ``conclusion size'' and ``problem size'' plots. The remaining plots report average counts 
of the various logical connectives across entire problems (summed over all premises and conclusion). Note that \dbsp 
problems are the only ones with biconditionals. 

%\input{pl3_pc_proof_dist_boxplots}

%% As mentioned above, these proofs were originally obtained by resolution-based ATPs like Vampire and then translated 
%% into \ndlzero. Thus, all of them were correct by construction. To make it harder for models to get high scores simply
%% by outputting positive verdicts, we synthetically corrupted 54 out of the 300 proofs in a way which guarantees that 
%% the resulting proofs are incorrect. In all 54 cases, the target of the manipulation was the advertised conclusion $p$ 
%% of a \skwd{from} step: ``$p$ \skwd{from} $p_1,\ldots,p_n$.'' The intervention was made either by directly manipulating 
%% negation signs (insertion or removal) or by first transforming $p$ into an equivalent form  and then manipulating 
%% negation signs on the result. Recall that, for an incorrect proof, a model cannot simply say \emph{incorrect}; it must
%% also pinpoint the error, and if the error is in a step of the form $p$ \skwd{from} $p_1,\ldots,p_n$ and amounts to
%% $p$ not following from $p_1,\ldots,p_n$, it must also produce a counterexample, \iensp, an interpretation which satisfies
%% every $p_i$ but falsifies $p$. 

Because most of the proofs in \mbf{PL3-PC} are flat, having been originally derived from 
resolution-based ATPs, they test long-context dependency tracking but not hierarchical 
context management (the latter is tested by \mbf{PL1-PC} and especially by the challenge 
set \mbf{PL1-PC-c}). Specifically, here are the core capabilities that are stress-tested by \mbf{PL3-PC}:
\ben
\item Reference  resolution in linearly growing lexical environments: in a step like
\begin{tcAthena}
lemma_35 := (A2 & ~B5 ==> C8) from lemma_5, lemma_2, premise_12
\end{tcAthena}
can the model reliably retrieve the definitions of \mtt{lemma\_5}, \mtt{lemma\_2}, and \mtt{premise\_{12}} 
from earlier parts of its context, and flag an error if one of them is not found? 

\item Logical availability verification (assumption-base tracking), also in monotonically 
growing logical contexts . Any argument to \skwd{from} may be inlined, e.g., a step might be: 
\begin{tcAthena}
lemma_9 := (A2 & ~B5 ==> C8) from lemma_3, (~C7 | ~C8 | A2), premise_4, B4
\end{tcAthena}
So the model must only resolve the names \mtt{lemma\_3} and \mtt{premise\-4}, 
but must also make sure that the inline formulas \mtt{(\ntildels{}C7\psp{}|\psp\ntildels{}C8\psp|\psp{}A2)} and \mtt{B4} 
are in the assumption base at that particular point in the proof, \iensp, that 
they are logically available and can be used as ammunition to derive further conclusions. 

\item Micro-reasoning under long-context pressure: Can the model verify that $p$ follows 
from $p_1,\ldots,p_n$ (for $n \leq 5$)? The model has to solve dozens or hundreds of 
these mini-SAT problems in the course of checking a single \plthreesp proof, and it must
do so without compromising any of the other subproblems it must solve. This is the main
capability tested by \mbf{PL3-PC} (and one that \mbf{PL1-PC} doesn't already test). 

\een

The main results for \mbf{PL3-PC} are reported in Table~\ref{Tb:PLThreePCAccuracies}. The evaluation protocol for this task 
is lenient in the following sense: true positives (cases where a model correctly determines that a proof 
is invalid\footnote{Recall (p.~\pageref{Pg:PCAsDebugging}) that we view proof checking as a debugging task (finding reasoning errors), 
so a ``positive'' verdict is one that flags a proof as incorrect.}) 
are counted as such when the model identifies the erroneous step, even if it's wrong about the exact
nature of the error or adduces an incorrect counterexample. However, manual inspection of 
a random sample suggests that reports that manage to pinpoint the location of the
error also tend to describe it accurately and to generate valid counterexamples, 
so this leniency is unlikely to affect the results materially. We also list guessing-only accuracies. 
\renewcommand{\arraystretch}{1.3}
\begin{table}[!t!]
\centering
\rowcolors{2}{blue!5}{white}
\resizebox{\textwidth}{!}{%
\begin{tabular}{|l|c|c|l|c|c|}
\hline
\textbf{Model} & \textbf{Accuracy} & \textbf{Guessing Only} & \textbf{Model} & \textbf{Accuracy} & \textbf{Guessing Only} \\
\hline
\gemThreeOne & \mbf{0.96} \ci{0.02} & 0.98 \ci{0.02} & \gemThree & 0.95 \ci{0.03} & 0.96 \ci{0.02} \\ \hline
\claudeOpusFourFiveNov & 0.93 \ci{0.03} & 0.95 \ci{0.03} & \gptFiveOne & 0.89 \ci{0.04} & 0.92 \ci{0.03} \\ \hline
\claudeSonnetExt & 0.88 \ci{0.04} & 0.93 \ci{0.03} & \gptFive & 0.87 \ci{0.04} & 0.91 \ci{0.03} \\ \hline
\grokFour & 0.86 \ci{0.04} & 0.87 \ci{0.04} & \gptFiveTwo & 0.84 \ci{0.04} & 0.88 \ci{0.04} \\ \hline
\oThree & 0.83 \ci{0.04} & 0.87 \ci{0.04} & \dsROne & 0.78 \ci{0.05} & 0.82 \ci{0.04} \\ \hline
\geminiTwoPointFive & 0.76 \ci{0.05} & 0.79 \ci{0.05} & \dsReasoner & 0.72 \ci{0.05} & 0.77 \ci{0.05} \\ \hline
\oFourMini & 0.68 \ci{0.05} & 0.74 \ci{0.05} & \oThreeMini & 0.36 \ci{0.05} & 0.47 \ci{0.06} \\ \hline
\hermes & 0.36 \ci{0.05} & 0.46 \ci{0.06} & \dsThreeOne & 0.3 \ci{0.05} & 0.42 \ci{0.06} \\ \hline
\nous & 0.3 \ci{0.05} & 0.4 \ci{0.06} & \claudeSonnetFour & 0.14 \ci{0.04} & 0.29 \ci{0.05} \\ \hline
\magistral & 0.12 \ci{0.04} & 0.27 \ci{0.05} & \opus & 0.1 \ci{0.03} & 0.26 \ci{0.05} \\ \hline
\gptFourO & 0.02 \ci{0.02} & 0.18 \ci{0.04} & \gptFourOne & 0.01 \ci{0.01} & 0.19 \ci{0.04} \\ \hline
\grokThree & 0.0 \ci{0.01} & 0.18 \ci{0.04} & \dsThree & 0.0 \ci{0.01} & 0.18 \ci{0.04} \\ \hline
\end{tabular}}
\caption{\mbf{PL3-PC} accuracies and guessing-only accuracies.}
\label{Tb:PLThreePCAccuracies}
\end{table}
\renewcommand{\arraystretch}{1.0}

Based on the accuracies for the full task, the DI (discriminability index) of this task is 0.42, indicating exceptionally high discriminating power.  

Note that guessing-only mode is of marginal help to strong models. 
%This is consistent with what we saw in \mbf{PL1-PC}, whose results indicated that the main proof-checking difficulty, especially for top-tier models, is deciding correctness rather than localizing and explaining errors. 
However, as we move down the table and the ability scale, guessing-only mode becomes increasingly beneficial, with accuracy 
boosts ranging from 2X to 10X for lower-end models. 

%Recall that only 53 of the proofs in this task are incorrect. 
Because a verdict that finds the input proof correct does not require supporting evidence, 
a model can easily inflate its score by disproportionately favoring that outcome. 
This is especially true for an imbalanced dataset like \mbf{PL3-PC}, which only has 53 incorrect
proofs out of 300, meaning that a trivial policy of always answering ``correct'' 
can already achieve an accuracy of 82\%. Verifying incorrect proofs is more demanding, because 
it requires precise error localization. Accordingly, we also report performance when evaluation 
is restricted to the set of incorrect proofs, in Table~\ref{Tb:PLThreePCAccuraciesOnIncorrectProofs}. 
We see that the top two Gemini models are barely affected by this challenge, but the rest of the elite 
models are impacted more significantly (e.g., Claude Opus 4.5 goes from 93\% to 75\% and GPT-5.2
falls to 64\%), while performance at the middle and lower tiers plunges (\egnsp, the accuracy of
DeepSeek Reasoner drops from 72\% to 17\%). We also see some pathological perfect accuracies on the 
guessing-only variant here from models like DeepSeek 3 and Grok 3, which always output negative
verdicts. 

\renewcommand{\arraystretch}{1.3}
\begin{table}[!b!]
\centering
\rowcolors{2}{blue!5}{white}
\resizebox{\textwidth}{!}{%
\begin{tabular}{|l|c|c|l|c|c|}
\hline
\textbf{Model} & \textbf{Accuracy} & \textbf{Guessing Only} & \textbf{Model} & \textbf{Accuracy} & \textbf{Guessing Only} \\
\hline
\gemThree & \mbf{0.91} \ci{0.08} & 0.98 \ci{0.05} & \gemThreeOne & 0.88 \ci{0.09} & 1.0 \ci{0.03} \\ \hline
\gptFiveOne & 0.83 \ci{0.1} & 0.98 \ci{0.05} & \claudeOpusFourFiveNov & 0.75 \ci{0.11} & 0.91 \ci{0.08} \\ \hline
\geminiTwoPointFive & 0.67 \ci{0.12} & 0.87 \ci{0.09} & \gptFiveTwo & 0.64 \ci{0.13} & 0.91 \ci{0.08} \\ \hline
\grokFour & 0.6 \ci{0.13} & 0.64 \ci{0.13} & \gptFive & 0.6 \ci{0.13} & 0.81 \ci{0.1} \\ \hline
\oThree & 0.52 \ci{0.13} & 0.75 \ci{0.11} & \claudeSonnetExt & 0.49 \ci{0.13} & 0.76 \ci{0.11} \\ \hline
\dsROne & 0.42 \ci{0.13} & 0.62 \ci{0.13} & \oThreeMini & 0.3 \ci{0.12} & 0.89 \ci{0.09} \\ \hline
\oFourMini & 0.26 \ci{0.12} & 0.6 \ci{0.13} & \dsThreeOne & 0.21 \ci{0.11} & 0.85 \ci{0.1} \\ \hline
\dsReasoner & 0.17 \ci{0.1} & 0.46 \ci{0.13} & \hermes & 0.17 \ci{0.1} & 0.75 \ci{0.11} \\ \hline
\claudeSonnetFour & 0.08 \ci{0.08} & 0.91 \ci{0.08} & \opus & 0.04 \ci{0.06} & 0.92 \ci{0.08} \\ \hline
\grokThree & 0.02 \ci{0.05} & 1.0 \ci{0.03} & \magistral & 0.02 \ci{0.05} & 0.87 \ci{0.09} \\ \hline
\gptFourO & 0.02 \ci{0.05} & 0.94 \ci{0.07} & \nous & 0.02 \ci{0.05} & 0.58 \ci{0.13} \\ \hline
\dsThree & 0.02 \ci{0.05} & 1.0 \ci{0.03} & \gptFourOne & 0.02 \ci{0.05} & 1.0 \ci{0.03} \\ \hline
\end{tabular}}
\caption{\mbf{PL3-PC} accuracies and guessing-only accuracies restricted to the set of incorrect proofs.}
\label{Tb:PLThreePCAccuraciesOnIncorrectProofs}
\end{table}
%% \begin{table}[!h]
%% \centering
%% \rowcolors{2}{blue!5}{white}
%% \resizebox{0.8\textwidth}{!}{%
%% \begin{tabular}{|l|c|l|c|}
%% \hline
%% \textbf{Model} & \textbf{Accuracy}  & \textbf{Model} &  \textbf{Accuracy} \\
%% \hline
%% \gemThree & \mbf{0.91} \ci{0.03} & \gptFiveOne & 0.81 \ci{0.04}  \\ \hline 
%% \geminiTwoPointFive & 0.65 \ci{0.05} & \gptFiveTwo & 0.65 \ci{0.05}   \\ \hline 
%% \gptFive & 0.61 \ci{0.05} & \grokFour & 0.59 \ci{0.06} \\ \hline 
%% \claudeSonnetExt & 0.54 \ci{0.06} & \oThree & 0.49 \ci{0.06}  \\ \hline 
%% \dsROne & 0.41 \ci{0.06}  & \oThreeMini & 0.30 \ci{0.05}  \\ \hline 
%% \oFourMini & 0.28 \ci{0.05}  &  \dsThreeOne & 0.24 \ci{0.05} \\ \hline 
%% \hermes & 0.19 \ci{0.04} & \dsReasoner & 0.17 \ci{0.04} \\ \hline 
%% \claudeSonnetFour & 0.08 \ci{0.03}  & \opus & 0.04 \ci{0.02}  \\ \hline 
%% \gptFourOne & 0.04 \ci{0.02}   & \nous & 0.02 \ci{0.02} \\ \hline 
%% \gptFourO & 0.02 \ci{0.02}   & \grokThree & 0.02 \ci{0.02} \\ \hline 
%% \magistral & 0.02 \ci{0.02}   & \dsThree & 0.02 \ci{0.02} \\ \hline 
%% \end{tabular}}
%% \caption{\mbf{PL3-PC} proof-checking accuracies restricted to the set of incorrect proofs.} 
%% \label{Tb:PLThreePCAccuraciesOnIncorrectProofs}
%% \end{table}
\renewcommand{\arraystretch}{1.0}

\begin{figure}[!t!]
\begin{minipage}{\textwidth}
\centering
\scalebox{0.9}{\includegraphics[width=\linewidth]{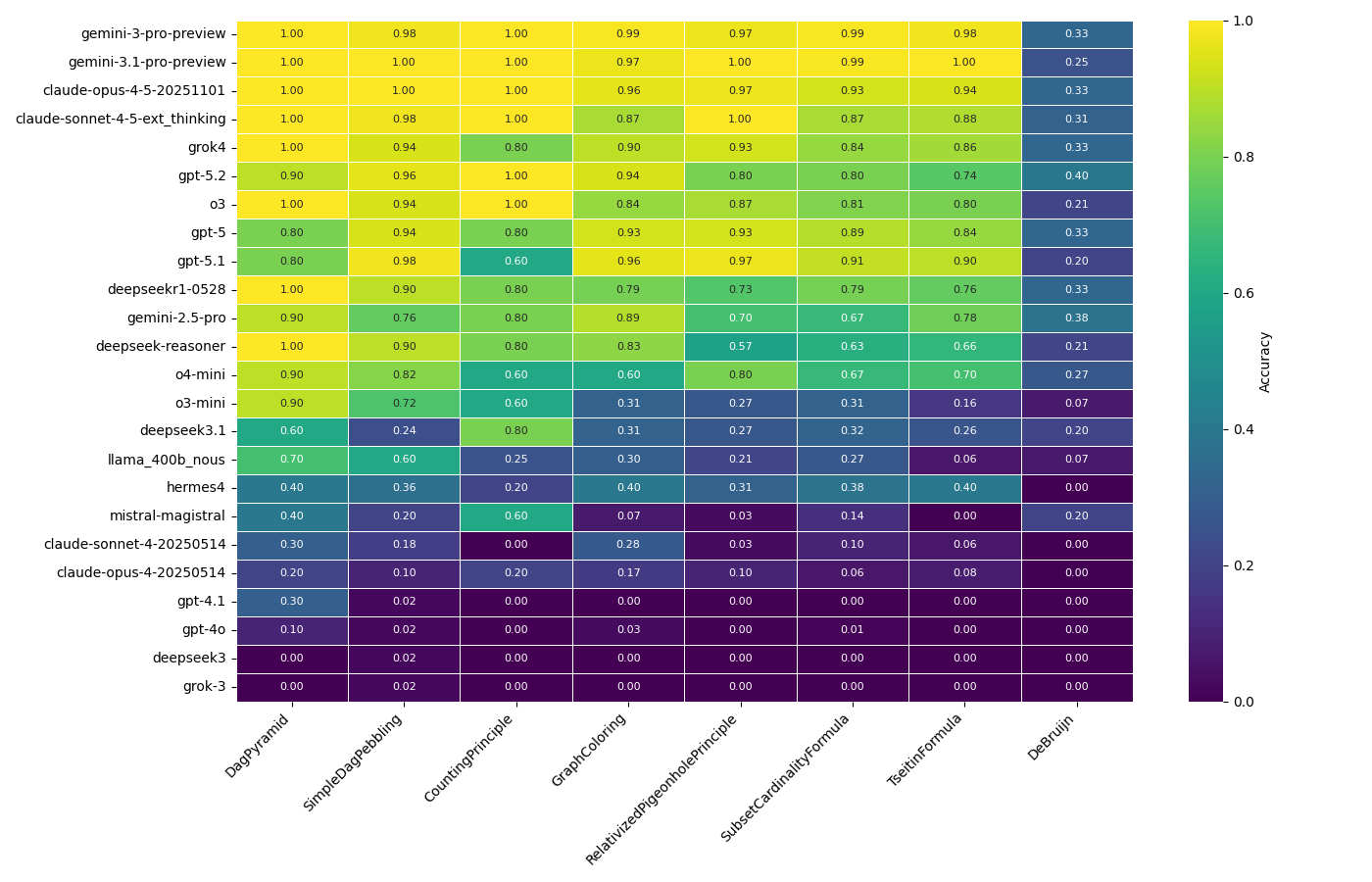}}
\end{minipage}
\caption{\mbf{PL3-PC} accuracy per problem type.}
\label{Fig:PL3PCHeatmap}
\end{figure}

Fig.~\ref{Fig:PL3PCHeatmap} gives a finer-grained picture of proof-checking performance,
indexed by problem type. 
It shows that most of these problem families are easy for top-tier models, with 
\dbsp proofs being the most notable exception. As we have already pointed out, 
these are considerably longer than the other proofs, and they also feature 
very large inlined formulas that are hard to process without a reliable
algorithm. However, mid-tier and third-tier models exhibit a good
deal of performance variation across these problem families. For example,
dag-pyramid problems are almost universally accessible (e.g., Llama-405b by Nous Research 
and Mistral Magistral reach 70\% and 40\% there, respectively), despite being the longest proofs
in the dataset (apart from \dbsp proofs). By contrast, proofs from the relativized-pigeonhole
problem family, for example, are much more challenging for such models (\egnsp, the accuracies of 
Llama-405b and Mistral Magistral fall to 21\% and 3\% there, respectively). 

%% Observe that several models attain a perfect proof-checking accuracy on DAG-pyramid problems,
%% despite the fact that these proofs are by far the longest in the dataset (barring \dbsp proofs,
%% see Fig.~\ref{Fig:PLThreeProbStatsByProbTypes}). This strongly suggests that even though all proofs across 
%% all 8 problem types are structurally
%% similar, the logical content of the underlying problems still exerts a significant impact on performance,
%% even though here the model is simply checking a proof rather than writing one from scratch. 

\begin{figure}[!b!]
\begin{minipage}{\textwidth}
\centering
\scalebox{0.95}{\includegraphics[width=\linewidth]{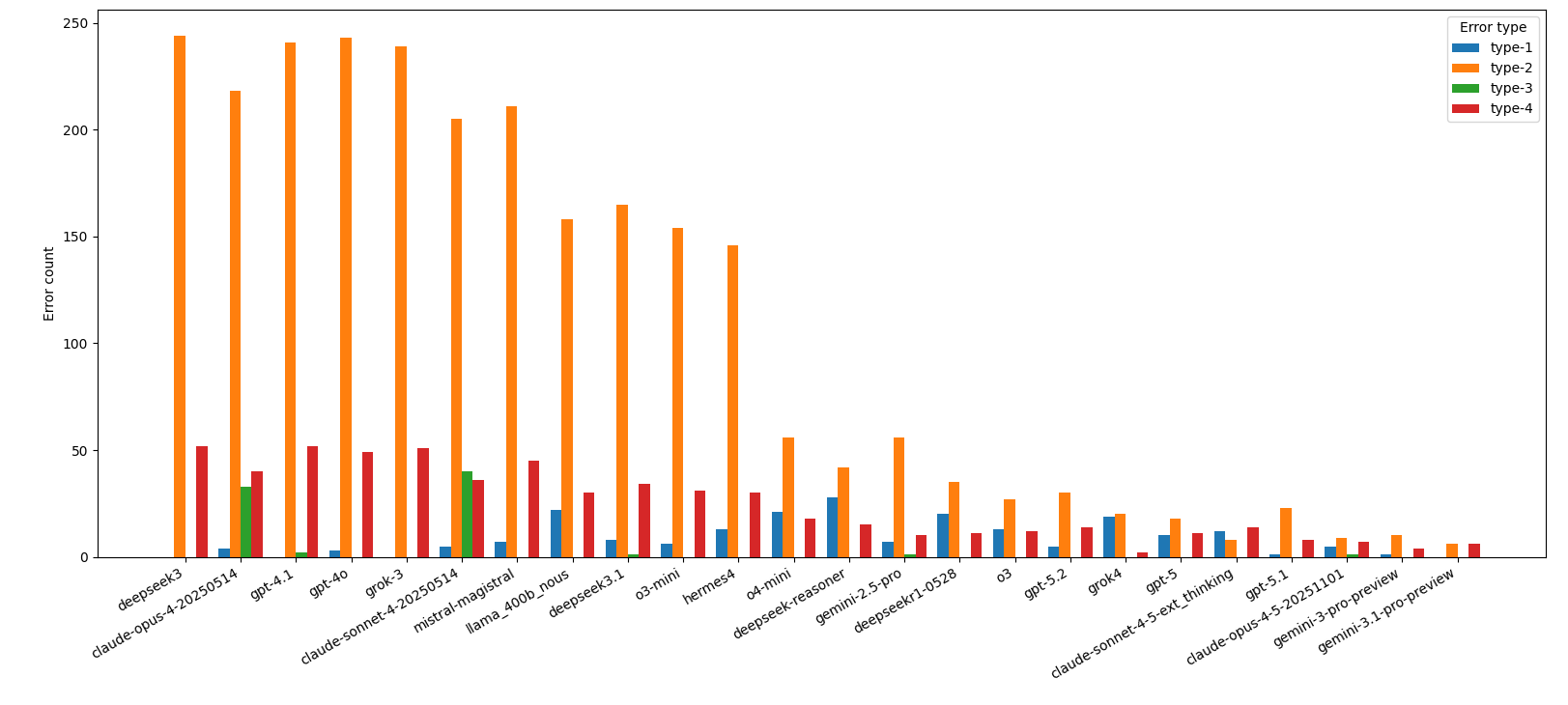}}
\end{minipage}
\caption{Error distribution across the 4 error types for \mbf{PL3-PC}.}
\label{Fig:PL3PCErrorPlot}
\end{figure}

Fig.~\ref{Fig:PL3PCErrorPlot} shows the distribution of the various error types made by the models.
We use the same error taxonomy we used for \mbf{PL1-PC}, with the exception of type-5 errors, which 
are not relevant in this setting: 
\ben
\item \emph{Type-1 errors}: The model incorrectly claims that the proof is right (false negatives). 
\item \emph{Type-2 errors}: The model incorrectly claims that the proof is wrong (false positives).
\item \emph{Type-3 errors}: Formatting issues (\egnsp, responses missing expected fields). 
\item \emph{Type-4 errors}: The model correctly claims that the proof is wrong but fails to localize the error. 
\een

We see that type-2 errors dominate across all models and ability levels. 
Deepseek-V3, in particular, makes a stunning 247/247 type-2 errors---a false-positive 
error rate of 100\%. The model invariably hallucinates non-existent errors. 
Type-1 errors, where a buggy proof is wrongly pronounced correct, are much less
common, with Deepseek-Reasoner taking the lead with 28 such errors (out a total possible 
of 53, for an error rate of 53\%), followed by Llama 405b and o4-mini (41\% and 40\%, respectively).
Gemini 3.1 is the only elite model that did not make even a single type-1 error, 
while the two top Gemini models also made very few type-4 errors, 
a testament to their strong debugging skills, and specifically their 
ability to reliably identify subtle errors in complex reasoning chains.
GPT-5.2 underperforms on this task, with a false-positive rate of 12\%.

\begin{figure}[t!]
\begin{minipage}{\textwidth}
\centering
\scalebox{0.8}{\includegraphics[width=\linewidth]{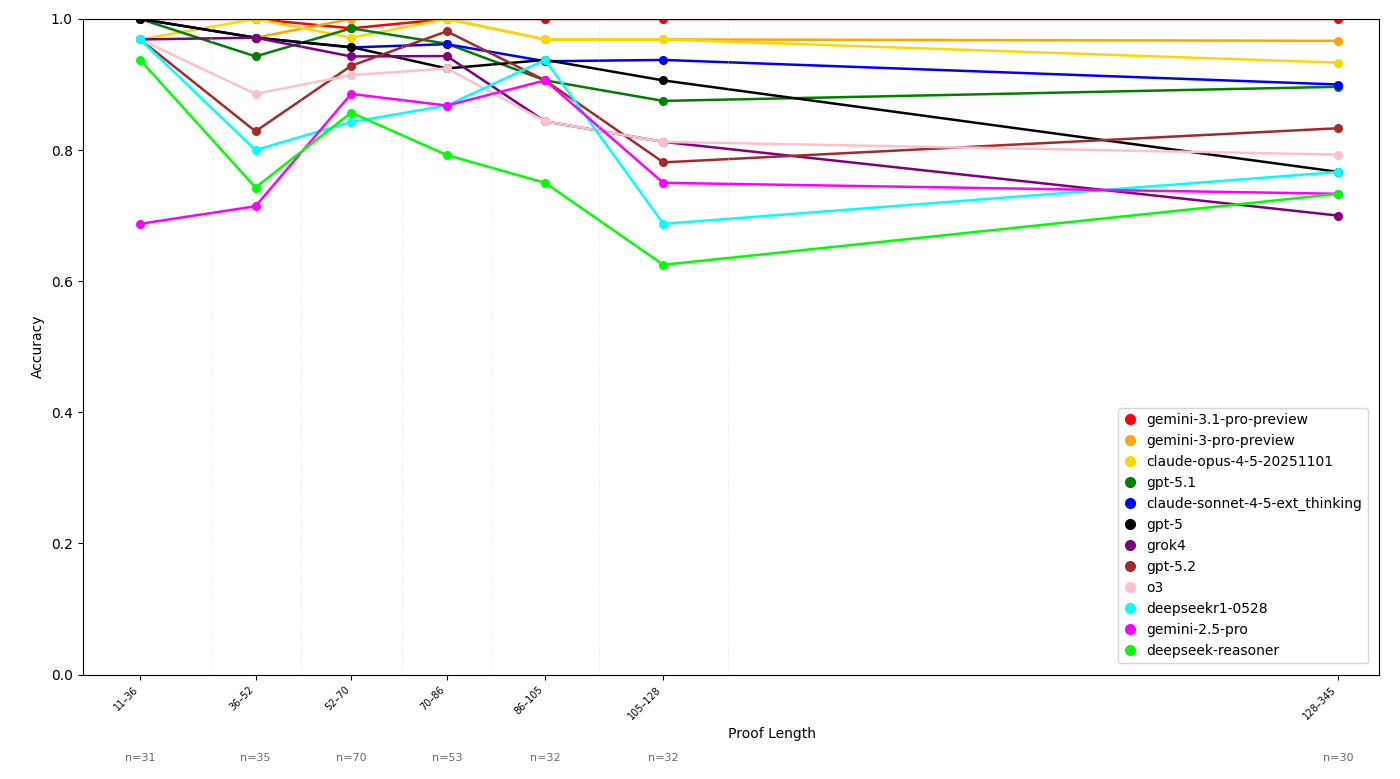}}
\end{minipage}
\caption{\mbf{PL3-PC} accuracy, excluding \dbsp problems, as a function of proof length for the top 12 models.}
\label{Fig:PL3PCBinnedAccTier1}
\end{figure}

Fig.~\ref{Fig:PL3PCBinnedAccTier1} plots accuracy as a function of proof length for the top 12 models. We exclude \dbsp problems
because those 15 proofs are significantly longer than the others and there are not enough of them for 
a robust plot. The figure shows that proof size is not a strong degrader of \mbf{PL3-PC} 
performance, at least in this particular evaluation regime of flat \ndlzerosp proofs with no more than a few hundred steps, 
which translates to inputs with considerably fewer than 10K tokens.\footnote{Excluding 
\dbsp problems, the average number of BPE tokens in a \mbf{PL3-PC} input, counting both
the detailed instructions and the input proof, is 6.9K (computed with 
\fmtt{tiktoken} using the \fmtt{cl100k\_base} encoding).
The median is 6.6K tokens. With \dbsp proofs, these become 7.7K and 6.6K, respectively.}
Modern attention mechanisms show robust retrieval for context windows of that size (and even for 
substantially larger ones). 

In conclusion, even though a few models already attain very high absolute accuracy on \mbf{PL3-PC} ,
the task remains sharply discriminative overall and exposes substantial qualitative differences 
in error calibration, localization, and robustness. Moreover, the observed ceiling effects are contingent 
on conservative corruption and scoring choices (only 53 problems were noised, in a straightforward way), and could be pushed back simply 
by increasing corruption density or requiring stronger justifications. (As the results show, accuracy drops for all models 
when evaluation is restricted to the set of incorrect proofs.)  Accordingly, \mbf{PL3-PC} is not quite solved at present 
and should instead be viewed as a scalable diagnostic test of verifier-side semantic reasoning (based on entailment rather
than inference rules). 

\section{A Quantitative Measure of Epistemic Stability} 
\label{Sec:ESI}

When a rational agent $\alpha$ asserts a proposition, they make themselves accountable 
for what others can infer from it. These inferences may be made by pure deduction, but
more often they are made by commonplace inferential practices in the shared space of norms 
that govern the use of the relevant concepts. 
%In particular, they commit themselves to a number of other propositions $q_1,q_2,\ldots$ that are directly entailed by $p$, typically by material inferences but sometimes by pure logic. 
We might say that, by asserting $p$, $\alpha$ becomes committed to  additional propositions $q_1,q_2,\ldots$
that represent \emph{epistemic obligations}, meaning that $\alpha$ cannot deny any $q_i$ on pain of 
epistemic incoherence. One of the $q_i$ is $p$ itself, and in that case the 
epistemic obligation reduces to familiar self-consistency: If $\alpha$ asserts $p$, then $\alpha$ 
cannot also deny $p$ (or assert $\neg p$).\footnote{Non-monotonic reasoning, of course, is 
commonplace---people often change their minds. But typically they do so upon receiving new information, 
at which point they might retract $p$. They do not maintain both $p$ and $\neg p$ at 
the same time.}
But there are many other epistemic commitments.
%% If we assert that Tom 
%% is a bachelor, we commit to Tom being male and not having a spouse. If we deny either of these
%% propositions in the same breath that we assert Tom's bachelorhood, then we are being epistemically
%% incoherent. Assuming that we are making a good-faith effort to adhere to Grice's conversational maxims, 
%% the only possible explanation for a failure to affirm a $q_i$ is that  we are either deeply conceptually confused 
%% or simply inconsistent. Likewise, if
If we claim $p = \mbox{\em Paris is the capital of France}$, we commit ourselves, 
among other things, to the propositions that France is a sovereign country and that 
Paris is a city in France. Unless we willfully aimed to be uncooperative 
communicators, it would be incoherent to  deny these facts while maintaining $p$. 

%We will formalize a notion of epistemic obligation for AI models, but it will be useful to introduce some terminology first. 
Let \mc{P} be any problem that may be meaningfully posed 
to a model $M$. We distinguish between the general problem \mc{P} 
and particular {\em instances\/} of \mc{P}.\footnote{The general problem can be understood 
as the overall specification of the task (``Given the name of a country, respond with its population and area in square kilometers.''), 
while an instance of the problem is a specific input to the task (``France'').}
We assume there is a prompt associated with \mc{P} and we write $\pi(\mc{P})$ for the 
core part of that prompt, namely, the input-insensitive part that gives the general specification of \mc{P}. 
%If $x$ is an instance of \problem, we write $M(x,y)$  to indicate that, presented with $x$ as input, $M$ generates the answer $y$ as output.
Even if two problems $\problem$ and $\rp{\problem}$ involve the same task, they may 
be formulated in incompatible ways. For instance, $\pi(\problem)$  and $\pi(\rp{\problem})$ may assume different epistemic postures
or operate within different epistemic contexts, \egnsp, $\pi(\problem)$ may be aimed towards truth or factual coherence whereas 
$\pi(\rp{\problem})$ may be sarcastic or adversarial, or both may be aimed towards truth but 
$\pi(\problem)$ may be operating within the context of Euclidean geometry whereas 
$\pi(\rp{\problem})$ may adopt hyperbolic geometry. 
We say that \problemsp and $\rp{\problem}$ are {\em compatible\/} iff no presupposition or instruction of $\pi(\problem)$
contradicts one of $\pi(\rp{\problem})$, and neither prompt provides new information to $M$ 
that is not also provided by the other.\footnote{By ``new'' information
we mean information that did not appear in $M$'s pretraining corpus.}
This is a symmetric relation.

%We introduce a notion of an {\em epistemic obligation\/} or \emph{commitment} for AI models $M$ as follows. 
We write $M(x,y)$ to mean that given $x$ as input, model $M$ has generated $y$ as the output; 
and we refer to $M(x,y)$ as a \emph{judgment} $j$. Now let \problemsp and $\rp{\problem}$ be two compatible problems and 
let $x$ and $x'$ be two instances of \problemsp and $\rp{\problem}$, respectively. 
We write $M(x,y) \nprovesls M(x',y')$ 
and say that $M(x,y)$ epistemically \temv{constrains} $M(x',y')$ 
iff $M$'s generation of $y$ as the answer to $x$ creates an epistemic obligation for 
$M$ to generate $y'$ as the answer to $x'$, because not doing so
would be epistemically incoherent. We call this a {\em positive\/} epistemic obligation.
Suppose \problemsp poses geography questions under a truthful system prompt $\pi(\problem)$, 
instance $x$ asks for the capital of France, and $M$ has generated the expected answer, Paris,
as the output $y$. Suppose further that $\rp{\problem}$ asks multiple-choice trivia questions, 
also under a truthful system prompt, and that $x'$ asks whether Paris is a city in 
France, Portugal, or Venezuela. Then $M$ would be normatively bound to 
generate ``France'' as the answer $y'$. We take \nprovessp as a primitive relation
for present purposes. 

A {\em negative\/} epistemic obligation,  written as $M(x,y) \:\not{\hspace*{-0.007in}\nproves}\: M(x',y')$, obtains iff 
$M$'s generation of $y$ as the answer to $x$  precludes it from generating $y'$ as the answer to $x'$. 
Typically, a positive epistemic obligation
creates a number of trivial negative epistemic obligations, \egnsp, if  $M$ 
generates ``Paris'' as the answer to question $x$ above, then it is 
negatively bound from generating Portugal or Venezuela or any other country as the answer
to $x'$. But the converse may not be true: a negative constraint from \problemsp to 
$\rp{\problem}$ need not impose positive epistemic constraints on $\rp{\problem}$. For instance,
let \problemsp and $\rp{\problem}$ be the proof-checking and proof-writing problems, respectively. 
If $M$ claims that a proof $D$ is incorrect, it is negatively bound from generating $D$ as the answer
to the corresponding instance of $\rp{\problem}$, but this does not impose any positive 
obligations on $\rp{\problem}$, as $M$ is free to generate other proofs for the relevant problem. 
In the converse direction, however, we have positive epistemic obligations: If $M$  generates $D$
as the answer to an instance of $\rp{\problem}$, 
then $M$ is epistemically obliged to answer \temv{yes} to the instance of \problemsp that asks whether
$D$ is a correct deduction.\footnote{Modulo answer variation due to stochasticity. 
We assume here that all judgments are robust, i.e. $M(x,y)$ means that $y$ is a stable mode 
of $M$'s conditional distribution over outputs given $x$. Note, however, that in 
mathematical tasks, persistent answer variation when given one and the same problem instance 
is itself a sign of weak reasoning, indicating that the model doesn't have a stable internal
representation of the solution space.}

Now fix an empirical distribution $\mc{D}_1$ of $n$ observed input-output pairs $\{(x_1,y_1),\ldots,(x_m,y_m)\}$,
where each $x_i$ is an instance of \mc{P} and $M(x_i,y_i)$; 
and a corresponding distribution $\mc{D}_2 = \{(x_1',y_1'),\ldots,(x_n',y_n')\}$
where each $x_j'$ is an instance of a compatible problem $\rp{\problem}$ and $M(x_j',y_j')$. 
It is notationally convenient to view $\mc{D}_1$ as a distribution of judgments 
$M(x_1,y_1), \ldots, M(x_m,y_m)$, and likewise for $\mc{D}_2$. Define the set 
$\temv{Inc}_{\scriptsize\mc{D}_1}[\mc{D}_2]$ as follows:
\begin{equation}
\temv{Inc}_{\scriptsize\mc{D}_1}[\mc{D}_2] = \{ j' \in \mc{D}_2 \sep \exists \: j \in \mc{D}_1 \qdot j \,\not{\hspace*{-0.007in}\nproves}\: \rp{j}\}.
\end{equation}
In words, this set contains all judgments in $\mc{D}_2$ that are precluded (negatively constrained) by, or are \emph{incompatible with}, 
some judgment in $\mc{D}_1$. 
We can quantify the degree of $M$'s epistemic stability relative to these two distributions
simply by computing the size of this set, dividing it by $n$ (the size of $\mc{D}_2$),
and subtracting the result from 1. This gives us a measure of $M$'s epistemic stability
wrt to $\mc{D}_2$ and $\mc{D}_1$, written as $\temv{ESI}_{\scriptsize\mc{D}_1}[\mc{D}_2]$.
We drop the subscripts $\mc{D}_1,\mc{D}_2$ when they are immaterial or 
obvious from the context. 
We refer to this quantity, $1 - (\card{\temv{Inc}_{\scriptsize\mc{D}_1}[\mc{D}_2]}/n)$, 
as the \emph{epistemic stability index} of $M$ relative to $\mc{D}_2$ and $\mc{D}_1$.
Thus, \temv{ESI} ranges smoothly from 0 to 1, with higher values obviously being preferable. 

The ratio $\card{\temv{Inc}_{\scriptsize\mc{D}_1}[\mc{D}_2]}/n$ has a natural probabilistic interpretation, 
as the empirical probability that a randomly chosen judgment from $\mc{D}_2$ is incompatible with some 
judgment in $\mc{D}_1$. Accordingly, 1 minus this ratio is the probability of the complementary  
event, namely, that a randomly chosen judgment from $\mc{D}_2$ is compatible with 
\emph{all} judgments in $\mc{D}_1$. We want this number to be as close to 1 as possible. 

\begin{figure}[t] 
  \centering
  \scalebox{0.75}{\includegraphics[width=0.95\columnwidth, clip=true, trim=0 0 10 0]{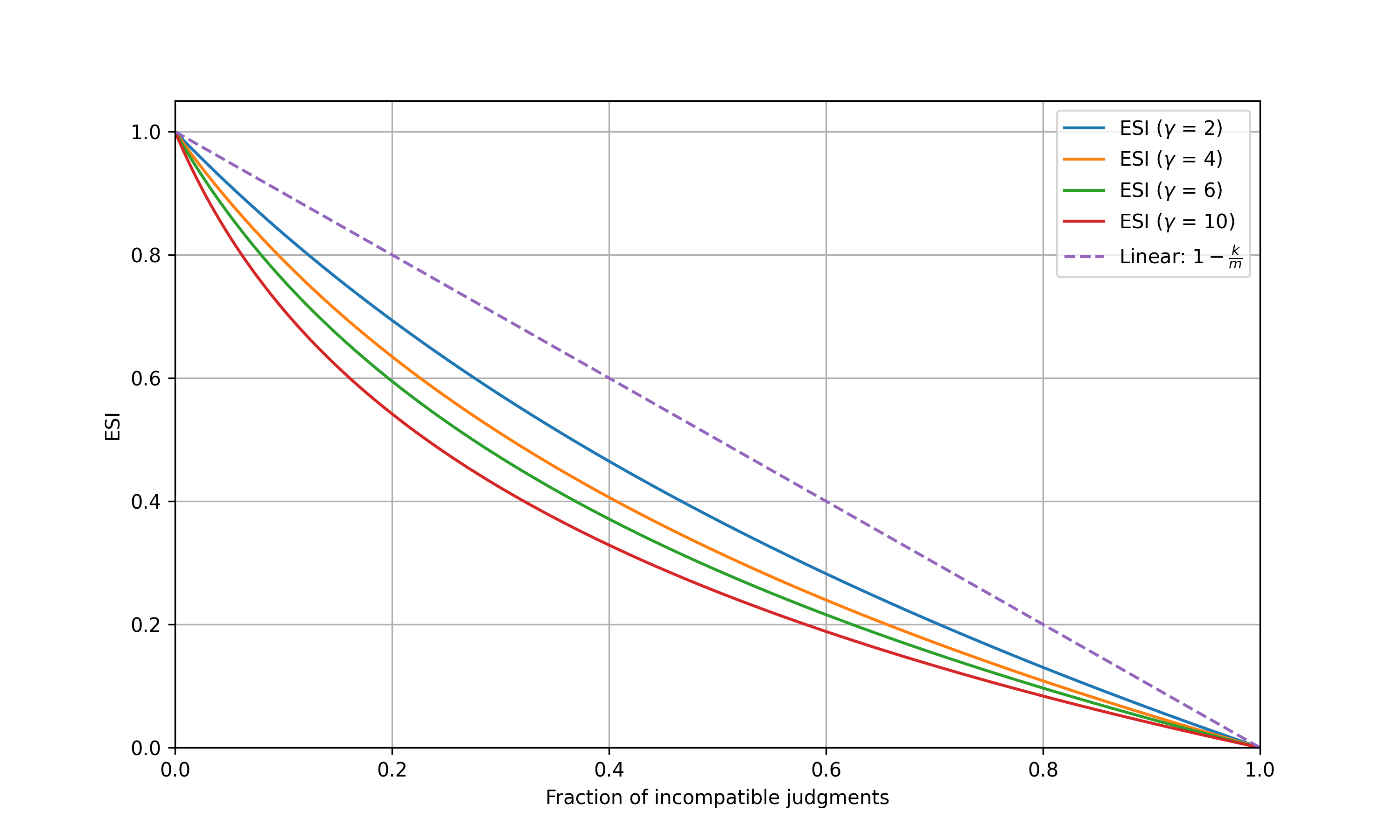}}
  \caption{ESI curves for different values of the curvature parameter ($\gamma$). The $x$-axis is 
the fraction \raisebox{.1em}{$\card{\temv{Inc}_{\scriptsize\mc{D}_1}[\mc{D}_2]}$}\raisebox{.1em}{$/$}$n$.}
  \label{Fig:ESICurves}
\end{figure}

We augment this intuition with a soft log-based transformation that adds curvature reflecting 
nonlinear penalties for judgment inconsistencies, as well as diminishing returns from such penalties:
\begin{equation}
\temv{ESI}_{\scriptsize\mc{D}_1}[\mc{D}_2] = 
1 - \cfrac{\log\biggl[1 + \gamma \cdot \frac{\card{\temv{Inc}_{\scriptsize\mc{D}_1}[\mc{D}_2]}}{\card{\mc{D}_2}}\biggr]}{\log\bigl[1+\gamma\bigr]}
\label{Eq:ESIDef}
\end{equation}
We still have $\temv{ESI} = 1$ when $\card{\temv{Inc}_{\scriptsize\mc{D}_1}[\mc{D}_2]} = 0$ (perfect epistemic stability) and 
$\temv{ESI} = 0$ when $\card{\temv{Inc}_{\scriptsize\mc{D}_1}[\mc{D}_2]} = \card{\mc{D}_2}$ (complete incoherence),
but the penalty for relatively small $\card{\temv{Inc}_{\scriptsize\mc{D}_1}[\mc{D}_2]}$ 
grows faster than in the linear case, accentuating even small amounts of instability, 
while saturating smoothly towards 0 as $\card{\temv{Inc}_{\scriptsize\mc{D}_1}[\mc{D}_2]}/\card{\mc{D}_2} \to 1$.
The parameter $\gamma$ controls the curvature and can be tuned depending
on how stringent we want to be; we use $\gamma = 5$ as a reasonable default. 
Figure~\ref{Fig:ESICurves} shows different ESI curves for $\gamma \in \{2, 4,6, 10\}$. 

We assume that $\mc{D}_2$ is not chosen adversarially, with prior knowledge of the constraints in $\mc{D}_1$. 
Otherwise it would be possible to cherry-pick data so as to compute any desired ESI value. To avoid 
metric gaming, the selection of $\mc{D}_2$ must be \emph{independent} of $\mc{D}_1$,
and in particular must not occur after inspecting the judgments in  $\mc{D}_1$. 
Methodologically, this requirement is somewhat analogous to the familiar principle that test data 
must not be tuned to a model's performance on the training set. While this independence condition 
can be made fully precise (for example, by fixing the generation process for $\mc{D}_2$ in advance,
or by requiring $\mc{D}_2$ to be sampled from a distribution specified independently of $\mc{D}_1$),
we do not need to formalize it here. For present purposes, it suffices to note that allowing $\mc{D}_2$
to be selected in response to observed judgments in $\mc{D}_1$ would undermine the intended interpretation 
of this stability measure.

As a concrete application of \temv{ESI}, consider the results of Section~\ref{Sec:PL3RCI} and let $M$ 
be any of the models in Table~\ref{Tb:SelfRecErrors}. We may take $\mc{D}_2$ to be the set 
of 300 proofs written by $M$ in response to the \mbf{PL3-PW} prompt, or more precisely, the set $(x,y)$
where $x$ is an input prompt from the \mbf{PL3-PW} dataset and $y$ is the corresponding output
proof generated by $M$. Now consider the set $S$ of all individual steps $p\msp\skwd{from}\msp{}p_1,\ldots,p_n$
that appear in any of those 300 proofs written by $M$ ($S$ will have thousands of elements), 
and let $\mc{D}_1$ be the set of all entailment/SAT judgments made by $M$ for each of those steps. 
In other words, $\mc{D}_1$ consists of all pairs $(x,y)$, one for every individual step $p\msp\skwd{from}\msp{}p_1,\ldots,p_n$ in $S$,
where $x$ is an input prompt asking whether $p$ follows from $p_1,\ldots,p_n$, along perhaps with an estimate 
of the difficulty of making that judgment, and $y$ is $M$'s response. 

We can then use the data in Table~\ref{Tb:SelfRecErrors} to compute $\temv{ESI}_{\scriptsize\mc{D}_1}[\mc{D}_2]$.
The number of self-recognized errors becomes the size of the set
$\temv{Inc}_{\scriptsize\mc{D}_1}[\mc{D}_2]$, \iensp, the set of all $M$-written proofs that are precluded 
by some $M$-made entailment judgment in $\mc{D}_1$. Plugging in the numbers from Table~\ref{Tb:SelfRecErrors} into
the defining formula~(\ref{Eq:ESIDef}) yields the values shown in Table~\ref{Tb:ESIPL3PW}. 
The results show that all top-tier models suffer from severe epistemic instability. 
Even for the most stable model, GPT-5.2, there is a nearly 40\% chance that a randomly chosen \mbf{PL3-PW} 
proof written by it will be inconsistent with its own relevant entailment judgments. More than half of these elite models 
have ESI values less than 0.5, meaning that any proof written by such a model is more likely than not to 
disagree with the model's corresponding entailment judgments. 
\renewcommand{\arraystretch}{1.3}
\begin{table}[!t]
\centering
\rowcolors{2}{blue!5}{white}
\resizebox{0.6\textwidth}{!}{%
\begin{tabular}{|l|c|l|c|}
\hline
\textbf{Model} & \textbf{ESI}  & \textbf{Model} &  \textbf{ESI} \\
\hline
\geminiTwoPointFive & 0.27 & \gptFive & 0.27    \\ \hline 
\gptFiveOne & 0.38 & \gemThree & 0.41    \\ \hline 
\claudeOpusFourFiveNov & 0.48   & \grokFour & 0.52 \\ \hline 
\oThree & 0.53 & \gemThreeOne & 0.61    \\ \hline 
\gptFiveTwo & 0.62 & {} & {}    \\ \hline 
\end{tabular}}
\caption{Epistemic stability values for \mbf{PL3-PW} proofs relative to individual entailment judgments for the 9 models of Table~\ref{Tb:SelfRecErrors}. 
A model's \emph{ESI} value can be understood as the probability that a randomly chosen \mbf{PL3-PW} proof written by the model is consistent with
the model's individual entailment judgments. Ideally, all of these values would be 1.}
\label{Tb:ESIPL3PW}
\end{table}
\renewcommand{\arraystretch}{1.0}

At its core, \temv{ESI} is a fraction, and therefore it is inherently asymmetric. 
Disregarding the log transformation, it is essentially the probability that a random draw from $\mc{D}_2$ is not precluded 
by any draw from $\mc{D}_1$. We are thus normalizing by $\card{\mc{D}_2}$, not by anything symmetric 
across both distributions. This is to be expected, because $\temv{ESI}_{\scriptsize\mc{D}_1}[\mc{D}_2]$ 
and $\temv{ESI}_{\scriptsize\mc{D}_2}[\mc{D}_1]$ measure different qualitative phenomena; each has a different
unit of analysis (the objects in $\mc{D}_2$ and $\mc{D}_1$, respectively).
For instance, with $\mc{D}_1$ and $\mc{D}_2$ as defined in this example,
$\temv{ESI}_{\scriptsize\mc{D}_1}[\mc{D}_2]$ is a \emph{proof-level} score, whereby a proof in 
$\mc{D}_2$ is incompatible with $\mc{D}_1$ iff it contains at least one step that diverges in  $\mc{D}_1$. 
So in this case the \temv{ESI} metric serves as an ``all-steps-must-pass'' aggregator. 
If we flip $\mc{D}_1$ and $\mc{D}_2$, then $\temv{ESI}_{\scriptsize\mc{D}_2}[\mc{D}_1]$ would
become a \emph{step-level} score that would address a very different question: what fraction 
of the model's entailment judgments survive contact with the corresponding proofs? 
Both quantities may be of interest, but no matter what $\mc{D}_1$ and $\mc{D}_2$ are, 
\emph{any} low values for $\temv{ESI}_{\scriptsize\mc{D}_2}[\mc{D}_1]$ are a source of concern. 

\enlargethispage{20pt}

In general, epistemic stability is not an absolute scalar property. It is defined with respect
to a reference distribution, for which there is no universally correct choice. 
Definition~(\ref{Eq:ESIDef}) is general, but genuinely different stability concepts can be 
singled out by instantiating the distributions in different ways. For example, above we took $\mc{D}_2$
to be the set of all 300 \mbf{PL3-PW} outputs by $M$, but we could instead consider the subset comprising 
only the \emph{incorrect} outputs generated by $M$ (while leaving $\mc{D}_1$ unchanged).
Doing so would pick a different measure of epistemic stability, one that we might call
\emph{error-triggered} \temv{ESI}, or $\temv{ESI}^{\,\temv{\tiny ET}}$ for short. 
(We do not report $\temv{ESI}^{\,\temv{\tiny ET}}$ values here for \mbf{PL3-PW}, but 
it suffices to note that they are very low, \egnsp, 0.14 for Gemini 3.1 and 0.38 for GPT-5.2.)
The first version of \temv{ESI} can be viewed as a global or generic measure of epistemic stability, while the second
is conditioned on errors. The difference between the two is roughly analogous to the difference between
blood pressure at rest vs blood pressure under stress. They measure two different aspects of epistemic 
stability. An agent with high error-triggered stability makes errors that are part of a stable 
internal theory, even if that theory is wrong. In other words, the agent may be wrong 
but they are wrong in a unified 
and stable way. This is common for humans (disregarding trivial performance errors). 
By contrast, low error-triggered stability means that the agent's mistakes are not stably 
believed, arising instead from  breakdowns in deployment, control, or integration. 
This appears to be common for language models, which often have the competence to make the right local
judgments (such as micro-SAT entailments) but fail to recruit that competence during more complex 
and global tasks (such as proof writing). 

It may be noted that epistemic stability is not unequivocally desirable. 
In particular, the capacity for self-correction and epistemic stability appear to pull in opposite directions. 
Low stability values indicate poor integration of different types of competence, 
but they also suggest that the model possesses latent knowledge that can be recruited by external scaffolding 
or iterative self-checking and verification. High stability values, by contrast, indicate 
that the model's mistakes are part of a stable internal reasoning pattern and are thus less 
amenable to agentic self-correction. So there seems to be a tradeoff: low epistemic stability is bad for unified
reasoning but good for self-aided corrigibility, while high epistemic stability is good for unified
and internally stable reasoning but not for self-repair. While this tension is generally real,
in practice epistemic stability is the more valuable primitive; we need systems that are stable,
self-consistent, and predictable across contexts. At any rate, the tension is only with self-correction, not
with correction in general. From an engineering perspective, one does not need a single model to, \egnsp,
both generate and verify. Corrections can be made by external verifiers or critic agents 
rather than by relying on the same model to repair its own inconsistencies.

\section{\plfour}
\label{Sec:PLFour}
\subsection{Proof Writing in Hilbert Systems}

This division contains 250 proof-writing problems in the Hilbert calculus for propositional logic. 
Because writing proofs in a Hilbert system is challenging, we have split this division into two tasks: 
\ben 
\item the main \mbf{PL4-PW} task, consisting of 200 problems that are shorter and easier on average than \mbf{PL1-PW} items; 
\item a challenge set, \mbf{PL4-PW-c}, consisting of 50 problems randomly sampled from longer \mbf{PL1-PW} problems. 
\een 
%Unless stated otherwise, results below use the 200-problem subset for both \plfour-PW and \plfour-PWe, and we evaluate the 50 challenge problems separately.

The Hilbert calculus is a flat proof system that does not admit conditional subproofs: a 
proof is just a list of formulas, each of which is an instance of an axiom schema or 
else follows from previous steps by a single inference rule, modus ponens. 
This flat linear structure makes Hilbert systems primarily of metatheoretical 
interest (e.g., for studying soundness and completeness), 
because the lack of assumption scope and the pared-down rule set
make it easier to prove theorems about these systems 
than within them. But for benchmarking purposes, 
these same factors make Hilbert proofs substantially harder 
to write than proofs in natural deduction.

Formulas here are defined exactly as in NDL, built up inductively from 
propositional atoms and logical connectives. In the variant we use, which is the most
common formulation of the Hilbert calculus, the only two primitive connectives are 
negation and implication. Disjunctions, conditionals, and biconditionals are desugared 
using only these two connectives. 
In particular, $p \mor q$ is written as $\neg p \mif q$, $p \mand q$ is written 
$\neg (\neg p \mif \neg q)$, and $p \miff q$  as $\neg ((p \mif q) \mif \neg (q \mif p))$.

Conditional elimination (modus ponens, also known as ``detachment'') is the only inference rule; 
it derives $q$ from $p \mif q$ and $p$. In addition, any proof step may infer any instance
of an axiom schema (a logical formula template). For example, Axiom 1 of the calculus 
is $(p \mif (q \mif p))$, where $p$ and $q$ are metavariables ranging over arbitrary formulas. 
Any formula that \emph{matches} this template is an axiom, such as $(A \mif (B \mif A))$ or 
$(\neg A \mif ((B \mor C) \mif \neg A))$.

The standard version of the Hilbert calculus that we use formalizes a classical logic of implication 
and negation with only three propositional axiom  schemas:\footnote{The first two axiom schemas are 
known as $K$ and $S$, respectively, as they correspond precisely to the $K$ and $S$ combinators
of combinatory logic \citep{hindley1972CombinatoryLogic}, namely $\lambda \, x \qdot \lambda \, y  \qdot x$
and  $\lambda \, f \qdot \lambda \, g  \qdot \lambda \, x \qdot (f \: x) \, (g \: x)$. To see the correspondence
of the latter with Axiom 2, read $f$ as a function of type $p \mif (q \mif r)$, $g$ as a function of type 
$p \mif q$, and $x$ as a variable of type $p$. 
We then we have $\typedec{(f \: x)}q \mif r$ and $\typedec{(g \: x)}q$, so that $(f \: x)\, (g \: x)$ is of type
$r$, corresponding to the tail occurrence of $r$ in Axiom 2.
Dropping Axiom 3 and keeping only K and S gives rise to the classical logic of minimal implication.}
\begin{itemize}
\item Axiom 1: $(p \mif (q \mif p))$
\item Axiom 2: $(p \mif (q \mif r)) \mif ((p \mif q) \mif (p \mif r))$
\item Axiom 3: $(\neg p \mif \neg q) \mif ((\neg p \mif q) \mif p)$
\end{itemize}

A derivation of a target tautology $p$ 
%from a set of premises $\Phi$ 
is a sequence of formulas $p_1,\ldots,p_n = p$, where each $p_i$ is  
an axiom instance or else follows by modus ponens from two earlier formulas.
This format is sufficient in principle, but, for convenience, we name each line 
and provide a justification either by citing an axiom or by listing the arguments to 
modus ponens. For example, the following is a proof produced by o4-mini for the 
tautology $((\neg A \mif A) \mif A)$. Each axiom is implemented as a primitive method 
in Athena \cite[p. 103]{Athena}, which matches the input sentence to the respective 
axiom schema or reports that no match is possible. In this way the proof  becomes 
fully verifiable using Athena.
\begin{tcAthena}
{
  p0 := axiom-1 on (~A ==> ((~A ==> ~A) ==> ~A));
  p1 := axiom-2 on ((~A ==> ((~A ==> ~A) ==> ~A)) ==> ((~A ==> (~A ==> ~A)) ==> (~A ==> ~A)))
  p2 := ((~A ==> (~A ==> ~A)) ==> (~A ==> ~A)) BY mp on p1, p0;
  p3 := axiom-1 on (~A ==> (~A ==> ~A));
  p4 := (~A ==> ~A) BY mp on p2, p3;
  p5 := axiom-3 on ((~A ==> ~A) ==> ((~A ==> A) ==> A));
  p6 := ((~A ==> A) ==> A) BY mp on p5, p4
}
\end{tcAthena}
We also permit the use of premises, treating them essentially as axioms.
As a result, we can pose problems of the form ``prove that $p$ follows from $\{p_1,\ldots,p_n\}$''
in addition to pure-tautology problems of the form ``prove $p_1 \mif \cdots \mif p_n \mif p$.''
This is an important change that allows models to perform suppositional reasoning and 
emulate natural deduction, yielding a flexible knob controlling problem 
difficulty. As we explain later, it also enables some interesting experiments 
gauging the ability of models to simplify proof construction in this system 
by intelligently exploiting the famous deduction 
theorem \citep{mendelson2009introduction}[p. 28].\footnote{The deduction theorem, 
independently discovered by \citet{Tarski1930Fundamentale} and \citet{Herbrand1930}, 
states that if there is a Hilbert proof of $p$ 
from premises $\{p_1,\ldots,p_n,p_{n+1}\}$, then there is a Hilbert proof of $(p_{n+1} \mif p)$ from 
$\{p_1,\ldots,p_n\}$.}

%% In particular, many problems are much easier to solve in Hilbert logic 
%% by simulating natural deduction reasoning using premises (assumptions). For example, consider 
%% a problem representing the transitivity of implication, \egnsp, showing that $A \mif C$
%% follows from $A \mif B$ and $B \mif C$. In the pure setting of implication logic,  without any premises, this would
%% need to be posed as the derivation of  $((A \mif B) \mif ((B \mif C) \mif (A \mif C)))$, 
%% a challenging exercise. By using premises, it can be posed as the derivation of $A \mif C$ from 
%% the two premises $A \mif B$ and $B \mif C$, which is much easier. 

% leveraging 

\mbf{PL4-PW}, the (easy) 200-problem task, was constructed as follows: 
(a) 20 problems were hand-picked simple tautologies that appear as standard 
textbook exercises, such as $(A \mif (\neg A \mif B))$,
$(A \mif \neg\neg A)$, and $(A \mor \neg A)$; and (b) 180 randomly selected 
problems from the same distribution from which \mbf{PL1-PW} problems were sampled. 
We targeted short problems (mean length 46 characters) to keep the task feasible. 
Any problem using logical connectives other than negation and implication was 
converted to an equivalent formula using only negation and implication, as described above.
Only 15 out of the 200 problems are pure tautologies; the rest have at least one premise. 

\begin{figure}[t]
\centering
\captionsetup{skip=15pt} % keep what worked for you

\begin{tikzpicture}[
  % (1) shift EVERYTHING left by this amount (tune as needed)
  xshift=-15mm
]

% (A) Main plot (cropped)
\node[anchor=south west, inner sep=0] (L) at (0,0) {%
  \includegraphics[height=50mm,trim=0bp 28bp 350bp 34bp,clip]{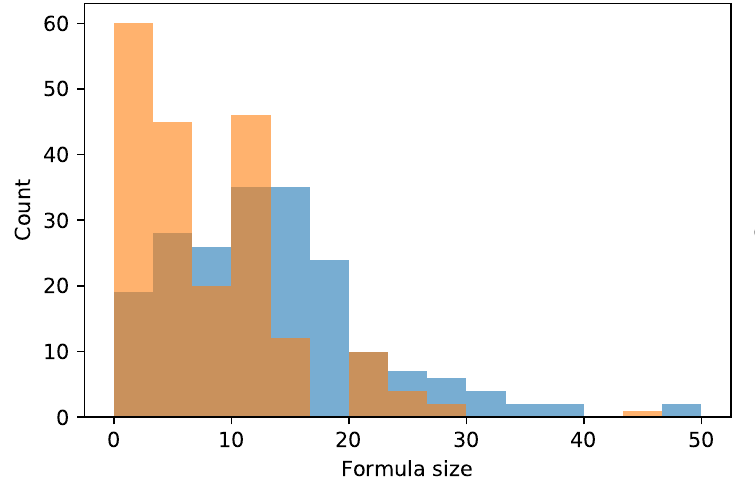}%
};

% (B) New box on the right
\node[anchor=south west, inner sep=0] (F)
  at ($(L.south east)+(6mm,0)$) {%
  \includegraphics[height=50mm,trim=0bp 0bp 0bp 34bp,clip]{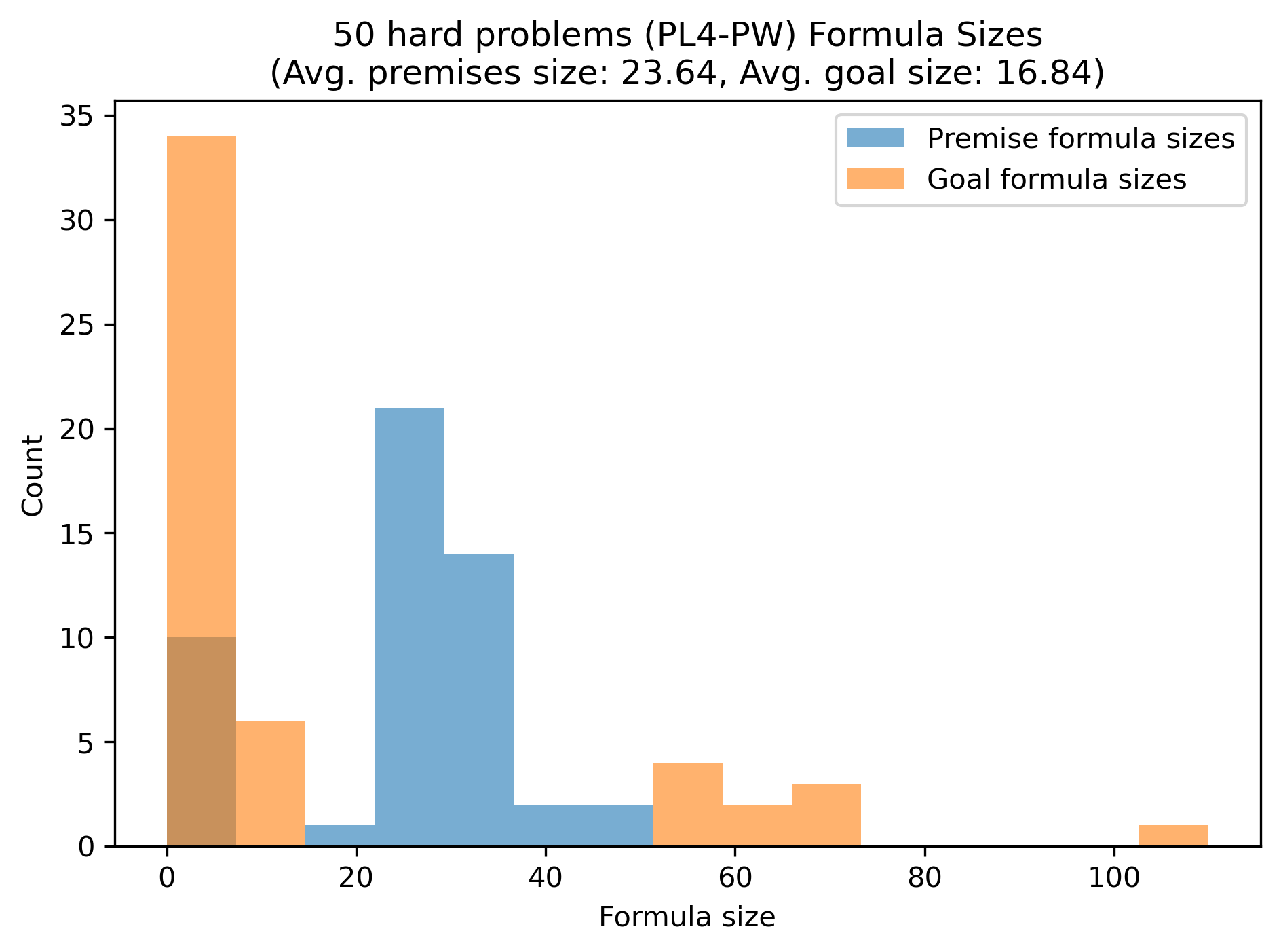}%
};

% (C) Legend (cropped) — stays tied to L
\node[anchor=north, inner sep=0] (Leg)
  at ($(L.south)+(40mm,-2mm)$) {%
  \includegraphics[
    width=0.6\linewidth,
      %% left, bottom, right, top 
%%    trim=140bp 20bp 140bp 330bp,clip %
    trim=140bp 0bp 140bp 270bp,clip %
  ]{sera_hilbert/plots/PL4_combined_formula_sizes.pdf}%
};

\end{tikzpicture}

\caption{Problem size distributions for \mbf{PL4-PW} (left) and \mbf{PL4-PW-c} (right).}
\label{fig:hilbert_expanded_data_lengths}
\end{figure}

The size distributions for problems in \mbf{PL4-PW} and \mbf{PL4-PW-c}, measured as the combined 
AST sizes of the original premises and goals, are shown in Figure~\ref{fig:hilbert_expanded_data_lengths}. 
Table~\ref{Tb:HilbertProblemStats} summarizes the mean and max numbers
of atoms, negations, conjunctions, disjunctions, conditionals, and biconditionals 
in these datasets. 

%% \begin{figure}[htbp]
%%     \centering
%%     \includegraphics[width=\linewidth]{plots/PL4_combined_formula_sizes.pdf}
%%     \caption{Problem length distribution for \plfour-PW (left) and \plfour-PWe (right) in terms of formula size. The total size of premises and the size of goal are plotted for each problem.}
%%     \label{fig:hilbert_expanded_data_lengths}
%% \end{figure}

%% \begin{figure}[htbp]
%%     \centering
%%     \includegraphics[width=0.5\linewidth]{plots/HPW_50_formula_sizes.png}
%%     \caption{Size distributions for 50 hard problems in \plfour-PW. The total size of premises and size of goal are plotted for each problem.}
%%     \label{fig:hilbert_hard_problems_formula_sizes}
%% \end{figure}
    
\renewcommand{\arraystretch}{1.2}
\begin{table}[b!]
\centering
\footnotesize
\resizebox{0.3\textwidth}{!}{
\begin{tabular}{|l|c|c|}
\hline
\textbf{Formula Feature} & Mean & Max \\ \hline
Size & 20.97 & 52 \\ \hline
Atoms & 8.24 & 19 \\ \hline
Negations & 7.17 & 23 \\ \hline
Conjunctions & 0.00 & 0 \\ \hline
Disjunctions & 0.00 & 0 \\ \hline
Conditionals & 5.55 & 15 \\ \hline
Biconditionals & 0.00 & 0 \\ \hline
\end{tabular}
}
\caption{\mbf{PL4-PW} Problem Statistics. 
Size = total number of nodes across all formulas; 
Atoms = number of propositional variables; 
Negations, Conjunctions, Disjunctions, Conditionals, Biconditionals = counts of respective logical connectives.}
\label{Tb:HilbertProblemStats}
\end{table}
\renewcommand{\arraystretch}{1.0}

The average problem size in \mbf{PL4-PW} is about 21, compared to about 28 for \mbf{PL1-PW}. 
In the challenge dataset \mbf{PL4-PW-c}, the average problem size is $\approx 40$. 
%% As our results show, model performance drops rapidly with problem size, which makes the 
%% standard \mbf{PL1-PW} problem set difficult in the standard Hilbert setting. 

%% (with primitive methods defining the Hilbert axiom schemas)

\subsection{\mbf{PL4-PW} Results}

We evaluated model performance under two settings: (1) a strict setting in which the model is
required to produce a proof that can be transformed into an Athena-verifiable 
proof without any post-processing; 
and (2) a lenient setting in which we apply the following 
limited programmatic corrections to superficial syntax issues and certain formatting errors 
in  applications of modus ponens: 
\begin{itemize}
    \item We use a parsing algorithm that can repair certain formulas with unbalanced parentheses
          by adding or removing parentheses at the right end of the formula to restore balance; 
          this recovers some formulas that would otherwise be rejected by the Athena parser. 
    \item We allow different parenthesization styles, including minimal parenthesization 
          where connective application is inferred according to Athena's precedence and associativity rules.
    \item We repair missing or superfluous semicolons and other simple syntax errors that 
          are readily fixable by programmatic search-and-replace.
    \item We correct flipped argument order in applications of detachment when the swap 
          yields a valid rule application. 
\end{itemize}
The above can be seen as a limited amount of proof-checking instrumentation (compared to the 
more extensive interventions in NDL). 

\mbf{PL4-PW} results for both modes are shown 
in Table~\ref{Tb:HilbertTaskResultsCondensedCorr}. Gemini 3.1 is in a class of its own in this task, far ahead of all other models. 
%% \footnote{\claudeSonnetFour\ was called 
%% in batch mode without extended thinking, \claudeSonnetExt\ was called in single-call 
%% mode with extended thinking.}
%%, and \claudeOpusFourFive\ was called in batch mode with extended thinking.}
\renewcommand{\arraystretch}{1.5}
\begin{table}[!bhtp]
\centering
\scriptsize
\rowcolors{2}{blue!5}{white}
\begin{tabular}{|l|c|c|l|c|c|}
\hline
\textbf{Model} & \textbf{Strict} & \textbf{Lenient} & \textbf{Model} & \textbf{Strict} & \textbf{Lenient} \\
\hline
\gemThreeOne & \mbf{0.55} ($\pm$0.04) & \mbf{0.71} ($\pm$0.03) & \gptFiveTwo & 0.26 ($\pm$0.03) & 0.39 ($\pm$0.03) \\ \hline
\geminiThreeProPreview & 0.20 ($\pm$0.03) & 0.23 ($\pm$0.03) & \grokFour & 0.10 ($\pm$0.02) & 0.16 ($\pm$0.03) \\ \hline
\claudeOpusFourFiveNov & 0.14 ($\pm$0.02) & 0.14 ($\pm$0.02) & \dsReasoner & 0.12 ($\pm$0.02) & 0.14 ($\pm$0.02) \\ \hline
\gptFive & 0.11 ($\pm$0.02) & 0.14 ($\pm$0.02) & \gptFiveOne & 0.07 ($\pm$0.02) & 0.14 ($\pm$0.02) \\ \hline
\geminiTwoPointFive & 0.08 ($\pm$0.02) & 0.12 ($\pm$0.02) & \oThree & 0.06 ($\pm$0.02) & 0.12 ($\pm$0.02) \\ \hline
\claudeSonnetExt & 0.12 ($\pm$0.02) & 0.12 ($\pm$0.02) & \dsROne & 0.09 ($\pm$0.02) & 0.10 ($\pm$0.02) \\ \hline
\oFourMini & 0.06 ($\pm$0.02) & 0.07 ($\pm$0.02) & \oThreeMini & 0.05 ($\pm$0.02) & 0.07 ($\pm$0.02) \\ \hline
\hermes & 0.04 ($\pm$0.01) & 0.04 ($\pm$0.01) & \claudeSonnetFour & 0.04 ($\pm$0.01) & 0.04 ($\pm$0.01) \\ \hline
\opus & 0.04 ($\pm$0.01) & 0.04 ($\pm$0.01) & \dsThreeOne & 0.04 ($\pm$0.01) & 0.04 ($\pm$0.01) \\ \hline
\grokThree & 0.03 ($\pm$0.01) & 0.03 ($\pm$0.01) & \dsThree & 0.03 ($\pm$0.01) & 0.03 ($\pm$0.01) \\ \hline
\nous & 0.03 ($\pm$0.01) & 0.03 ($\pm$0.01) & \gptFourO & 0.02 ($\pm$0.01) & 0.02 ($\pm$0.01) \\ \hline
\gptFourOne & 0.01 ($\pm$0.01) & 0.01 ($\pm$0.01) & \magistral & 0.01 ($\pm$0.01) & 0.01 ($\pm$0.01) \\ \hline
\end{tabular}
\caption{\mbf{PL4-PW} accuracies in strict and lenient modes. Top accuracies in both modes are shown in bold.}
\label{Tb:HilbertTaskResultsCondensedCorr}
\end{table}
\renewcommand{\arraystretch}{1.0}

The DI for \mbf{PL4-PW} is 0.13 (for lenient-mode results). The lower DI is due to the overall lower 
performance. 

%% To quantify the discriminative power of the two tasks, we computed the discriminability index (DI) and coefficient of 
%% variation (CV) for all 22 models and for the top 8 models (Table~\ref{Tb:DiscriminabilityCV}). The DI of \plfour-PWe in lenient mode is 0.285 across all models,
%% indicating that this task discriminates models sharply.
%% The \plfour-PW task has lower DI values but higher CV values, indicating that while the absolute spread of accuracies 
%% is smaller (due to the overall lower performance), the relative spread among models is larger.
%% Among the top 8 models, the CV for \plfour-PW is approximately twice that of \plfour-PWe, 
%% suggesting that the standard Hilbert task better discriminates among the strongest models.

%% %\input{sera_hilbert/hilbert_discriminability_cv}

For the challenge set \mbf{PL4-PW-c}, we only evaluated a selection of elite models. The results are shown 
in Table~\ref{Tb:HardRHPWPerformance}. On this task, Gemini 3.1 solved 6 out of 50 problems, GPT-5.2 solved 3, Gemini 3 solved 1, 
and all other models failed to produce a single correct proof. Evaluating additional weaker models on this task 
would not be meaningful. 

\renewcommand{\arraystretch}{1.5}
\begin{table}[htb!]
    \rowcolors{2}{blue!5}{white}
    \footnotesize
    \centering
  \scalebox{0.8}{
    \begin{tabular}{|l|c|}
    \hline
    \textbf{Model} & \textbf{Accuracy} \\
    \hline
    \gemThreeOne & \mbf{0.12} \\ \hline
    \gptFiveTwo & 0.06 \\ \hline
    \geminiThreeProPreview & 0.02 \\ \hline
    \grokFour & 0.00 \\ \hline
    \claudeOpusFourFiveNov & 0.00 \\ \hline
    \dsReasoner & 0.00 \\ \hline
    \gptFive & 0.00 \\ \hline
    \gptFiveOne & 0.00 \\ \hline
    \end{tabular}}
    \caption{\mbf{PL4-PW-c} accuracies for top-performing models.}
    \label{Tb:HardRHPWPerformance}
\end{table}
\renewcommand{\arraystretch}{1.0}

In general, writing proofs in a Hilbert calculus is a challenging task that requires high ingenuity
and exceptional skill and precision in symbolic manipulation and pattern matching. 
This is borne out by the poor overall performance of current models even on very simple tautologies. 
Table~\ref{Tb:HilbertTautologies}  shows the number of steps in a successful proof for each model and formula, 
for 5 of the simplest problems in the dataset. A dash ``$-$'' indicates that the model failed to produce 
a correct proof. Gemini 3.1 was the only model to solve all 5 problems. 
    
% (task: \plfour-PW)

% Suggested packages:
% \usepackage{booktabs,threeparttable,multirow,array,xcolor,colortbl}

\begin{table}[tb!]
    \centering
%    \footnotesize
    \scriptsize
    \setlength{\tabcolsep}{4pt}
    \renewcommand{\arraystretch}{1.15}
    \rowcolors{2}{blue!5}{white}

    \begin{threeparttable}
    \begin{tabular}{>{\raggedright\arraybackslash}p{4.2cm}ccccc}
        \toprule
        \textbf{Model} & \multicolumn{5}{c}{\textbf{Number of Proof Steps}} \\
        \cmidrule(lr){2-6}
                       & $\varphi_1$ & $\varphi_2$ & $\varphi_3$ & $\varphi_4$ & $\varphi_5$ \\
        \midrule
        \gemThreeOne & 9 & 25 & 19 & 55 & 59 \\
        \gptFiveTwo & 25 & 25 & 19 & - & - \\
        \geminiThreeProPreview & - & 25 & 19 & - & - \\
        \grokFour & - & - & - & - & - \\
        \claudeOpusFourFiveNov & - & 25 & 19 & - & - \\
        \dsReasoner & - & 29 & 19 & - & - \\
        \gptFive & - & - & 19 & - & - \\
        \gptFiveOne & - & 89 & 19 & - & - \\
        \geminiTwoPointFive & - & - & - & - & - \\
        \oThree & - & - & 19 & - & - \\
        \bottomrule
    \end{tabular}
\vspace*{0.05in}
    \begin{tablenotes}[flushleft]
        \scriptsize
  \setlength{\itemsep}{0.6ex}
  \setlength{\parsep}{0pt}
  \setlength{\topsep}{0pt}
        \item $\varphi_1:\ (A \mif \neg \neg  A)$ 
        \item $\varphi_2:\ (\neg  A \mif (A \mif B))$
        \item $\varphi_3:\ ((\neg A \mif \neg  B) \mif (B \mif A))$
        \item $\varphi_4:\ ((\neg A \mif (\neg  B \mif C)) \mif (\neg  (\neg  A \mif B) \mif C))$
        \item $\varphi_5:\ ((\neg  A \mif (\neg  B \mif C)) \mif (\neg  C \mif (\neg  B \mif A)))$
        %% \item $\varphi_1$: $(A \mif  (\ntilde \:  (\ntilde \:  A)))$
        %% \item $\varphi_2$: $((\ntilde \:  A) \mif  (A \mif  B))$
        %% \item $\varphi_3$: $(((\ntilde \:  A) \mif  (\ntilde \:  B)) \mif  (B \mif  A))$
        %% \item $\varphi_4$: $(((\ntilde \:  A) \mif  ((\ntilde \:  B) \mif  C)) \mif  ((\ntilde \:  ((\ntilde \:  A) \mif  B)) \mif  C))$
        %% \item $\varphi_5$: $(((\ntilde \:  A) \mif  ((\ntilde \:  B) \mif  C)) \mif  ((\ntilde \:  C) \mif  ((\ntilde \:  B) \mif  A)))$
    \end{tablenotes}
    \end{threeparttable}

    \caption{Performance of leading models on 5 simple tautologies in \mbf{PL4-PW}. 
    The number of steps of a successful proof is shown for each model and formula; a dash indicates failure to produce a correct proof.}
    \label{Tb:HilbertTautologies}
\end{table}

%% Generally, models perform better in \plfour-PWe than in \plfour-PW. However, this is not always the case. 
%% One exception is the formula $(((\neg   A) \mif (\neg   B)) \mif (B \mif A))$, which in \plfour-PWe
%% is written as $(((A \mor (\neg   B)) \mif (B \mif A)))$. In \plfour-PW, seven of the nine 
%% tested models produced essentially the same proof of length 19, while in \plfour-PWe, two out of nine
%% models succeeded in producing a proof of length 29 and one model produced a proof of length 39.

%% Overall, \gptFiveTwo\ performs the best in both tasks and currently is in a class of its own in
%% proof writing ability in Hilbert calculus, which is consistent with its proof writing ability across \pgrid\ tasks.

\subsubsection{Error Analysis}

%We computed the error type distribution for each model on the 200 problems in the \plfour-PW and \plfour-PWe tasks. 
Errors in this task are grouped into the following categories: 
\begin{itemize}
    \item \emph{Failed MP}: The model applied detachment incorrectly.
    \item \emph{Failed Axiom}: The model used an axiom improperly. These errors 
    are essentially axiom hallucinations, where the model falsely claims that a sentence is an instance of an axiom. 
    \item \emph{Undefined/Hallucinated Intermediate Step name}: This occurs when a intermediate-step name $p_i$ is quoted 
    in an application of modus ponens without being introduced in a previous step.
    \item \emph{Wrong Conclusion}: The proof reaches a conclusion other than the goal.
    \item \emph{No Proof}: The model did not generate a proof. This occurs in two ways: either the 
    model fails to generate any response after multiple API calls 
    for a given problem, or the model returns a response that does not contain a proof. 
    (This is especially common for o3, occurring in 102 out of the 200 problems.)
    \item \emph{Syntax Error}: The model returned a proof with a syntax error that our automated repairs could not fix. 
\end{itemize}

\begin{figure}[b!]
  \centering
  \includegraphics[width=0.95\linewidth]{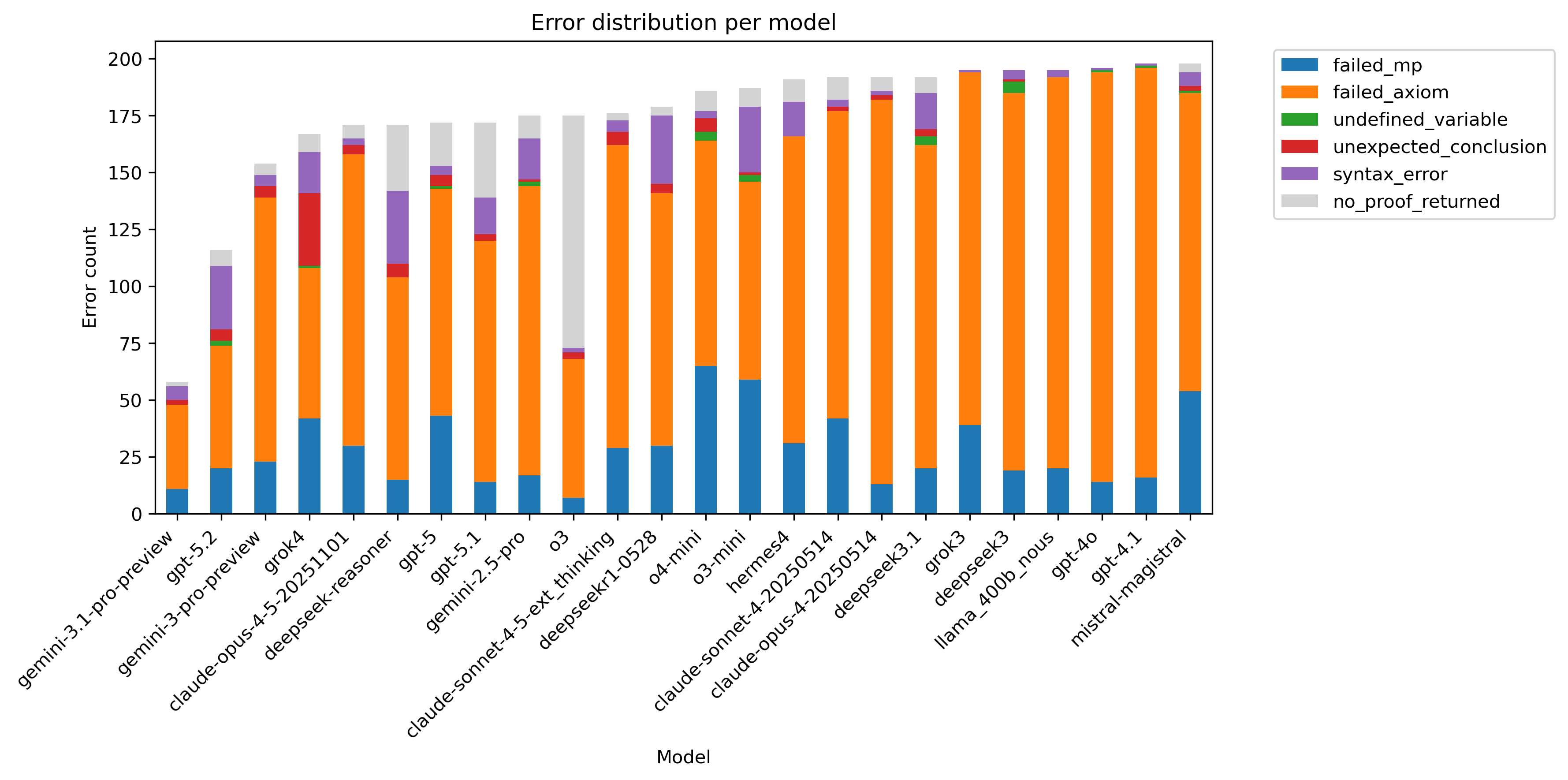}
  \caption{The distribution of error types in \mbf{PL4-PW} under lenient proof checking, for the first error that caused a proof failure (or the failure to return a proof).}
  \label{fig:hilbert_error_bar_chart}
\end{figure}

The distribution of these errors in \mbf{PL4-PW} (under 
lenient proof checking) is shown in Figure~\ref{fig:hilbert_error_bar_chart}.

%% \begin{figure}[tbh!]
%%     \centering
%%     \includegraphics[width=0.8\linewidth]{plots/pl4_plw_error_dist.png}
%%     \caption{The distribution of error types in \mbf{PL4-PW}, for the first error that caused a proof failure (or the failure to return a proof).}
%%     \label{fig:hilbert_error_bar_chart}
%% \end{figure}

%% \begin{figure}[tbh!]
%%   \centering

%%   \begin{minipage}[t]{0.78\linewidth}
%%     \centering
%%     \includegraphics[width=\linewidth]{plots/pl4_plw_error_dist.png}
%%   \end{minipage}\hfill
%%   \begin{minipage}[t]{0.20\linewidth}
%%     \centering
%%     \includegraphics[width=\linewidth]{plots/pl4_pw_error_legend.png}
%%   \end{minipage}

%%   \caption{The distribution of error types in \mbf{PL4-PW}, for the first error that caused a proof failure (or the failure to return a proof).}
%%   \label{fig:hilbert_error_bar_chart}
%% \end{figure}

Incorrect applications of modus ponens in \mbf{PL4-PW} fall into one of three types: 
\ben
\item the first argument is not a conditional formula (a type error of sorts);
\item the rule is applied to two formulas $p \mif q$ and $r$ where  $r \neq p$ (also a type error); and 
\item at least one of the two arguments is not in the assumption base (a strictly logical error).
\een
Most failed-mp errors made by the strongest models are of type 2. 

\subsubsection{Proof Length Analysis}

\begin{figure}[b!]
  \centering
  \includegraphics[width=0.8\linewidth]{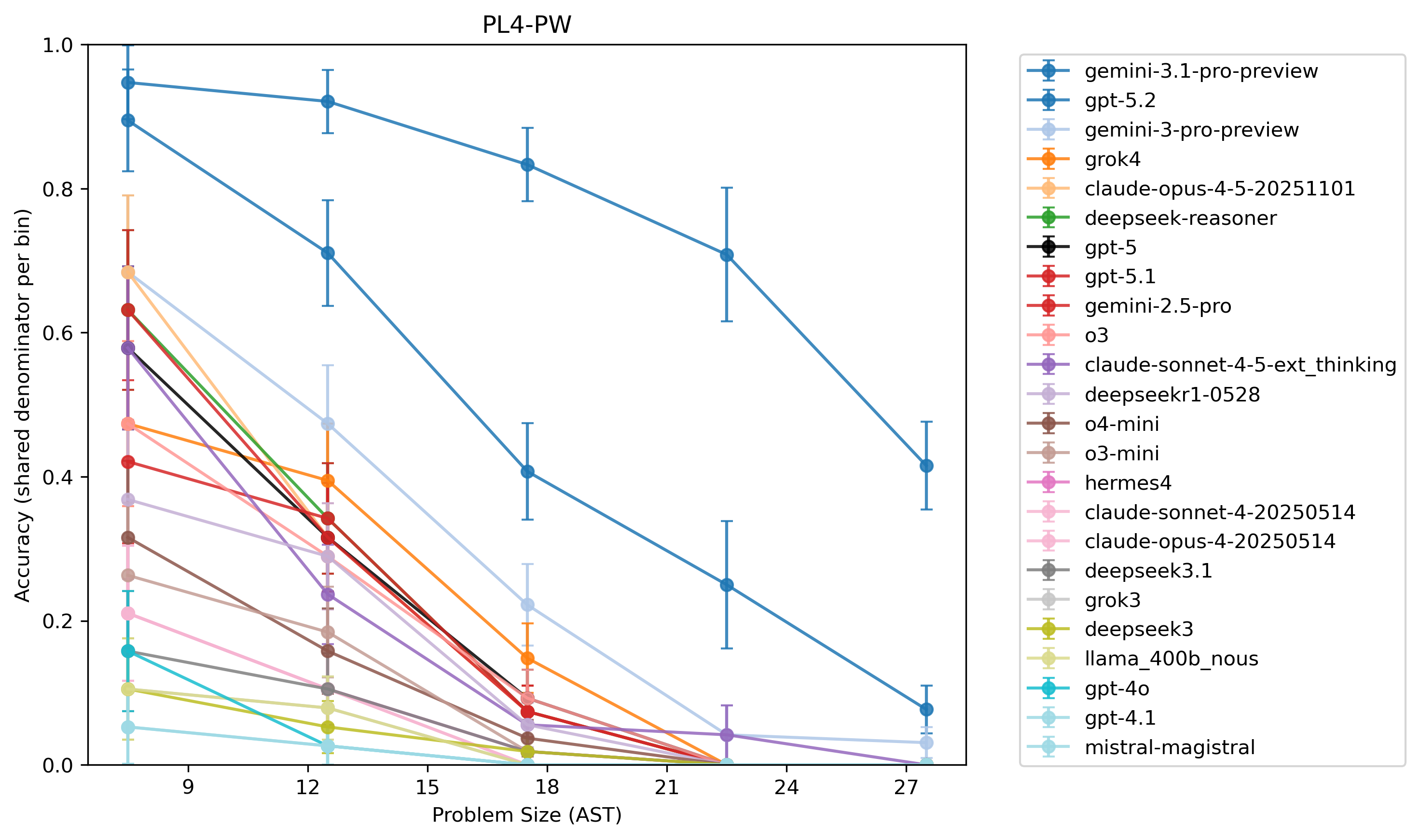}
  \caption{\mbf{PL4-PW} accuracy as a function of problem size.}
  \label{fig:hilbert_problem_length}
\end{figure}

Success rates in this task correlate strongly both with  problem size  and with the length of the 
generated proofs. Figure~\ref{fig:hilbert_problem_length} shows model accuracies as a function of problem size, 
defined as the total AST size of the premises and goal. Performance degrades rapidly as problem size increases,
even though problems in this task are quite shorter than those in \mbf{PL1-PW}.

%% \begin{figure}[htbp]
%%     \centering
%%     \includegraphics[width=0.8\linewidth]{plots/acc_vs_plen.png}
%%     \caption{\mbf{PL4-PW} accuracy as a function of problem size.} 
%%     \label{fig:hilbert_problem_length}
%% \end{figure}

Successful proofs for this task are generally short. 
%Figure~\ref{fig:shortest_proof_histogram} shows the distribution of the shortest proof lengths for \mbf{PL4-PW}, where each data point represents the length of the shortest correct proof for a given problem. 
The shortest proofs for the vast majority of solved problems have fewer than 30 lines. 

\begin{figure}[tb!]
    \centering
    \includegraphics[width=0.8\linewidth]{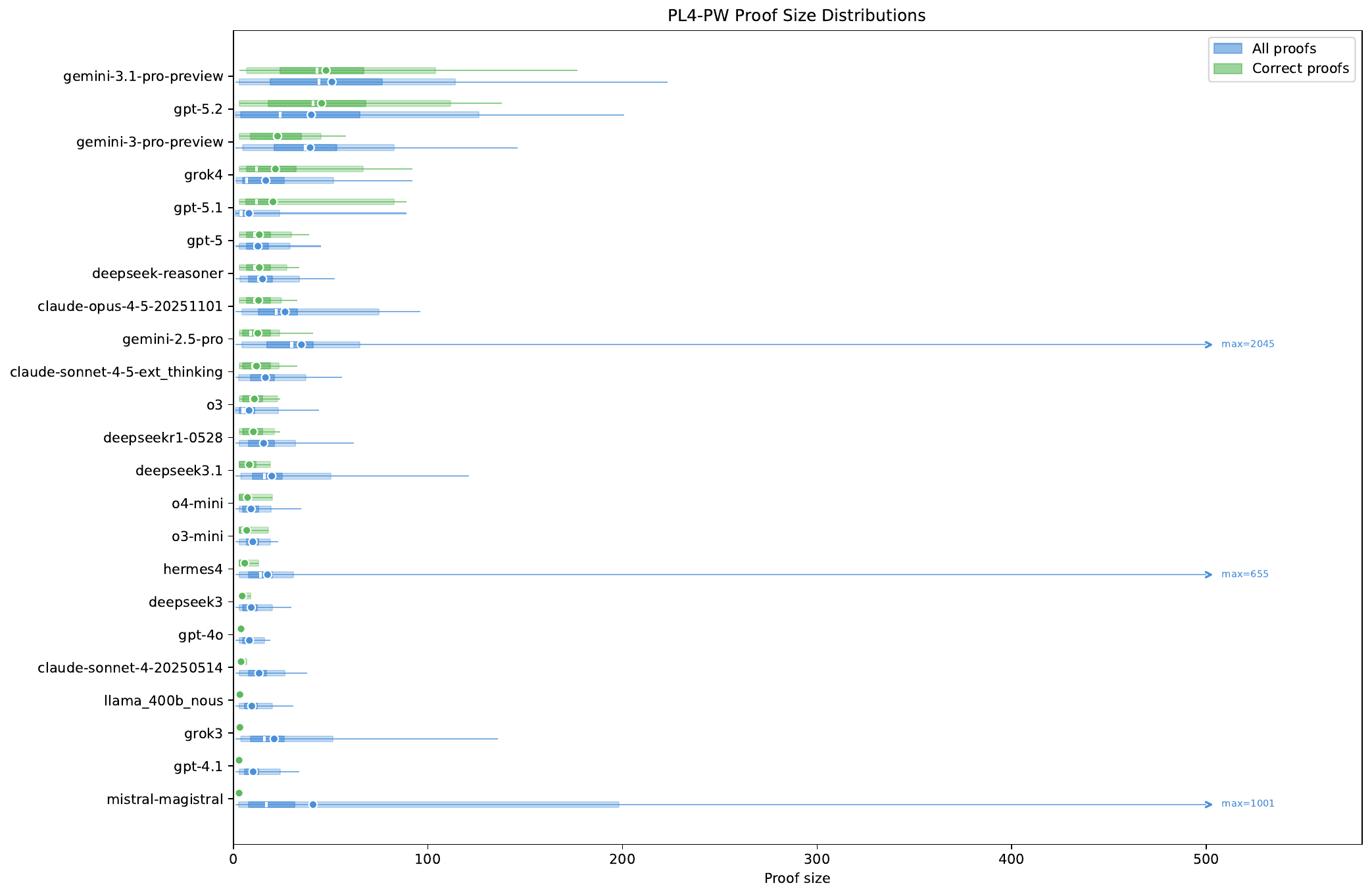}
    \caption{\mbf{PL4-PW} proof-size distributions per model: all proofs (blue) vs correct proofs (green); min--max line, 5--95\% band, IQR box, median tick, mean dot.}
    \label{fig:proof_size_distributions}
\end{figure}

Statistics on the lengths of all proofs written by all models for \mbf{PL4-PW}, 
similar to those in tasks like \mbf{PL2-PW},  are shown in Figure~\ref{fig:proof_size_distributions}. 
For each model $M$, the figure shows the distribution of length over the 
set of all proofs written by $M$ (blue), as well as the distribution of length over the set of 
all correct proofs written by $M$ (green). Again, for most models the green distributions   (correct proofs only) 
are narrower and shifted left compared to the blue distributions  (all proofs),
again showing that correct proofs tend to be shorter. 
However, some strong models (\gptFiveTwo, \gptFiveOne, \gptFive, and even \grokFour, \dsReasoner, and \oThree)
generate correct proofs of higher mean length than that of all their proofs.
Several models write extremely long incorrect proofs (Gemini 2.5 produced the longest, a proof 
with 2045 lines), suggesting that these models may generate long, verbose proof attempts when 
they are unable to solve a problem.

    % \begin{figure}[htbp]
    %     \centering
    %     \includegraphics[width=\linewidth]{plots/shortest_proof_length_distribution.png}
    %     \caption{Histogram of shortestproof lengths for HPW and RHPW.}
    %     \label{fig:shortest_proof_length_distribution}
    % \end{figure}
    
%\input{sera_hilbert/plots/hilbert_proof_lengths_shortest_proof_histogram.tex}

We also compared a set of strong models in terms of the length of their successful proofs,
by calculating the average proof length across the problems that all of these models managed to solve (under lenient settings): 
\gptFiveTwo, \grokFour, \gptFiveOne, \geminiThreeProPreview, \dsReasoner, \geminiTwoPointFive, \claudeOpusFourFiveNov, and \claudeSonnetExt. 
We found that \gptFive\ produced the shortest proofs across the 17 problems that all of these models solved, 
with an average proof length of 8.4 steps, while \geminiTwoPointFive\ produced the longest proofs, 
with an average of 9.4 steps. Remarkably, most models found a proof 
of the same length for each problem, suggesting that the models converged on the same proof.

\subsubsection{Proof Similarity Analysis}

\renewcommand{\arraystretch}{1.2}
\begin{table}[!b!]
    \rowcolors{2}{blue!5}{white}
    \footnotesize
    \centering
    \scalebox{0.7}{\begin{tabular}{|l|c|c|c|}
    \hline
    \textbf{Model} & \textbf{Avg $\ell/\ell_{\min}$ (C)} & \textbf{Avg $\ell/\ell_{\min}$ (I)} & \textbf{Ratio (I/C)} \\
    \hline
    \grokThree & 1.00 & 0.93 & 0.93 \\ \hline
    \dsThreeOne & 1.00 & 0.98 & 0.98 \\ \hline
    \claudeSonnetFour & 1.00 & 0.94 & 0.94 \\ \hline
    \hermes & 1.00 & 0.82 & 0.82 \\ \hline
    \oThreeMini & 1.03 & 0.62 & 0.60 \\ \hline
    \oThree & 1.03 & 1.01 & 0.98 \\ \hline
    \oFourMini & 1.04 & 0.60 & 0.57 \\ \hline
    \claudeSonnetExt & 1.06 & 0.87 & 0.82 \\ \hline
    \dsROne & 1.06 & 0.87 & 0.83 \\ \hline
    \claudeOpusFourFiveNov & 1.10 & 1.17 & 1.06 \\ \hline
    \dsReasoner & 1.11 & 0.60 & 0.54 \\ \hline
    \gptFive & 1.13 & 0.70 & 0.62 \\ \hline
    \geminiThreeProPreview & 1.19 & 1.72 & 1.45 \\ \hline
    \geminiTwoPointFive & 1.19 & 3.41 & 2.86 \\ \hline
    \grokFour & 1.34 & 1.20 & 0.89 \\ \hline
    \gptFiveOne & 1.38 & 0.59 & 0.43 \\ \hline
    \end{tabular}}
    \caption{Proof length ratio analysis on \plfour-PW problems of moderate difficulty 
    (problems with $\geq 3$ correct and $\geq 3$ incorrect proofs among models). 
    For each model we show mean $\ell/\ell_{\min}$ ratios, where $\ell$ is proof length 
    and $\ell_{\min}$ is the length of the shortest correct proof for that problem. 
    C (I) = correct (incorrect) proofs.
    Rows are ranked by avg(C), lower is better.}
    \label{Tb:ProofLengthRatioPL4PW}
    \end{table}
\renewcommand{\arraystretch}{1.0}

To further investigate the relationship between proof length and correctness, we computed the average proof length ratio of correct vs.\ incorrect proofs
for each model as follows. We first identified all problems for which at least 3 models succeeded and at least 3 models failed, where the failures
were required to be due to errors in the proof rather than failures to return a proof. For each of these problems, we computed the shortest correct
proof length and ranked models according to their average ratio of correct proof lengths to shortest correct proof lengths. We then 
computed the average ratio of incorrect proof lengths to shortest correct proof lengths for each model across these problems. 
The results are shown in Table~\ref{Tb:ProofLengthRatioPL4PW}. Note that neither \gemThreeOne\ nor \gptFiveTwo\ is included because they did not have
at least 3 incorrect proofs among these problems, and a number of weaker models are not included because they did not have at least 3 correct proofs 
among these problems. As a general trend, most models tend to write correct proofs that are similar in length
to the shortest correct proof, with the exception of \gptFiveOne\ and \grokFour, which tend to write longer correct proofs.
Among incorrect proofs, some models tend to write much shorter proofs. 
Other models tend to write much longer proofs, with \geminiThreeProPreviewsp having the largest ratios;
it could be said that these models tend to persist in proof writing even 
when they ``know'' that the proof is not succeeding.

As first noted in connection with Table~\ref{Tb:HilbertTautologies}, we observed that models 
usually return proofs of identical length for the same problem. 
This is often an indication that the models have converged on the same proof. 
To further investigate this, we computed the pairwise similarity between model proofs
as follows. First, we defined the ``fingerprint'' of a proof as the counts of each 
axiom and applications of detachment in the proof. 
According to this measure, two proofs are identical iff they have the same length, 
the same counts of each axiom application, and the same number of detachment applications.
Second, we defined the ``sequence'' of a proof as the order in which the axioms 
and detachment applications appear in the proof. According to this measure,
the proofs must have identical fingerprints and, additionally, the same precise 
order of application of the axioms and detachment.

\begin{figure}[!t]
    \centering
    \includegraphics[width=\textwidth]{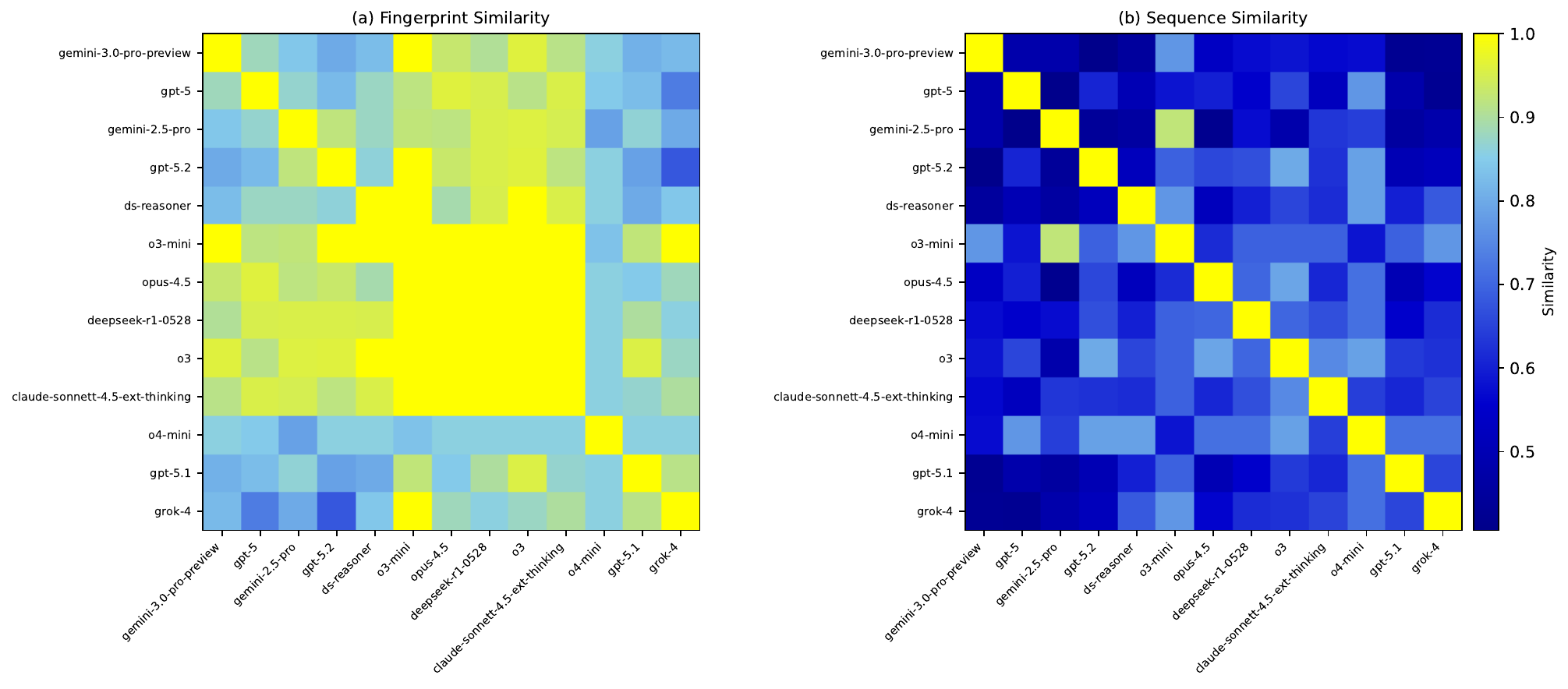}
    \caption{Pairwise proof similarity between models on \plfour-PW problems. 
    %(a) \textbf{Fingerprint Similarity}: percentage of common successful proofs where both models use the same counts of each axiom and modus ponens applications.
    %(b) \textbf{Sequence Similarity}: percentage of common successful proofs where both models apply axioms and modus ponens in the exact same order.
    Models are grouped by hierarchical clustering based on fingerprint similarity.
    Yellow indicates higher similarity; dark blue indicates lower similarity.}
    \label{fig:proof_similarity_pl4pw}
\end{figure}

Figure~\ref{fig:proof_similarity_pl4pw} shows pairwise similarity 
heatmaps according to the fingerprint and sequence measures for \mbf{PL4-PW}.
For each pair of models, \textit{fingerprint similarity} is the percentage of commonly 
solved problems for which the two successful proofs have the same fingerprint (counts 
of each axiom and detachment applications), while \textit{sequence similarity} additionally 
requires the same order of axiom and detachment applications.
We include a model in the heatmap only if it shares at least 10 commonly solved problems 
with every other included model (models are ordered by their performance for inclusion, 
although this ordering does not affect the final inclusion list).
As the heatmaps show, the models demonstrate a remarkable degree of convergence to essentially the same proofs. 
Proofs that match in fingerprint are often identical up to some reordering of the applications of different axioms and MP.

The results of this section show that even when the underlying problems are unchanged 
from  \mbf{PL1-PW} (and are, in fact, drawn from the easier end of that distribution), 
changing the proof formalism to one that does not mirror human reasoning  
increases the level of proof-writing difficulty by an order of magnitude. 
\mbf{PL4-PW} thus stands apart from other propositional-logic tasks in \pgrid,  
all of which are rooted in natural deduction. Writing Hilbert proofs  does not test human-style reasoning. 
Instead, it probes the ability to compile proofs into a very sparse calculus that is essentially 
a machine language for deduction. It places much higher pressure on symbolic manipulation discipline
than on high-level semantic reasoning about the problem. 
As such, \plfoursp is complementary to the other \pgridsp divisions, providing a compelling additional 
axis: proof-writing skill under a deliberately unforgiving and syntactically brittle formalism. 

However, there is an interesting direction for future work that would go beyond raw symbolic muscle
to test human-like meta-reasoning: study whether models are able to exploit the deduction theorem 
in Hilbert systems, instead of attempting every proof directly from the 3 axioms. For example, 
to prove implication transitivity, $(A \mif B) \mif (B \mif C) \mif (A \mif C)$,
a human would typically not search for a pure (premise-free) Hilbert derivation from first principles. 
Instead, they would first derive $C$ from the temporary assumptions $A$, $A \mif B$, and $B \mif C$,
and then repeatedly apply the deduction theorem to discharge these assumptions and obtain a pure Hilbert
proof of the conditionalized tautology. Since this discharge process is constructive (and can be carried
out by a simple algorithm), it suggests a natural controlled experiment: compare model performance on 
the premise-based Hilbert proofs in \mbf{PL4-PW} with their performance on the corresponding 
conditionalized (premise-free) versions. This would address the question of whether a model can solve 
a conditionalized Hilbert problem by first solving the easier premise-based version and then mechanically 
compiling the solution into a pure Hilbert proof. A model that can operationalize the deduction 
theorem in roughly the same way a human reasoner would is expected to degrade only moderately between 
the two settings, whereas a sharp drop would suggest that it is not leveraging a natural meta-level 
proof compilation strategy. Although we do not pursue this experiment here, it points to a promising 
way in which Hilbert systems may probe not just low-level proof synthesis but also strategic reasoning about proofs.

\subsection{Equational Logic}
\label{Eq:EquationalLogic}

Equational reasoning sits at the heart of mathematics, and by extension, of science 
and engineering. Work in all of these fields routinely involves equations between apparently 
different expressions that refer to the same entity. Engineers use equational
reasoning to show that two circuit designs are functionally identical,
physicists to track conserved quantities, and compiler engineers to verify that 
back-end optimizations yield identical program semantics. 
In computer science, the use of rewrite rules to reason about equations constitutes 
a Turing-complete model of computation, forming the foundation of functional programming 
and playing a key role in program verification. Any model aspiring to master these domains 
must internalize equational reasoning as a core competency.

%% From algebraic simplification and symbolic derivation to the conservation laws of physics and the 
%% design of digital circuits, the ability to recognize when two apparently different expressions 
%% refer to the same entity is critical. Establishing identity under transformation 
%% not only enables elegant mathematical proofs but also supports reliability in applications. 
%% Engineers rely on it to show that two circuit designs are functionally identical,
%% physicists to track conserved quantities, and compiler engineers to verify that 
%% back-end optimizations yield identical program semantics. 
%% In computer science, the use of rewrite rules to reason about equations constitutes 
%% a Turing-complete model of computation, forming the foundation of functional programming 
%% and playing a key role in program verification. Any model aspiring to master these domains 
%% must internalize equational reasoning as a core competency.

Equational reasoning offers a particularly attractive testing ground for AI models 
because it is almost entirely language-independent. Proofs do not need to rely on the specialized 
syntactic rules of any particular formal system, but can instead be written as simple sequences of steps 
where each step is just a term annotated with the equational axioms used to derive it from 
the preceding term.\footnote{Of course, the terms must still be written in formal notation, such as
$\temv{plus}(\temv{succ}(X),\temv{times}(a,\temv{zero}))$. But this requires nothing more than 
the standard convention of a function symbol followed by parentheses enclosing comma-separated arguments. 
This notation is both universal and simple, and is already extensively represented in the generic 
corpora on which language models are pretrained.}
Such representations are both rigorous and machine-checkable, while avoiding the burden of ensuring adherence
to the idiosyncrasies of a formal language. Despite this structural simplicity, equational reasoning 
is remarkably rich and powerful, capable of encoding subtle and sophisticated arguments. Indeed, 
in the general case equational reasoning is undecidable, which underscores that the expressive reach 
of this style of reasoning is more than adequate for probing the limits of AI models. 
%The equational tasks in this initial version of \pgridsp are elementary and fundamental, but future versions will include more challenging problems. 

Reasoning about chains of identities requires both domain-independent syntactic skill (carrying out manipulations correctly) 
and domain-specific semantic insight (grasping the meaning of equivalence in a particular theory). 
In this paper we focus on the former. Using an uninterpreted signature, we introduce 
three tasks that evaluate a model's general ability to recognize and apply identity transformations. 
Because complex terms are inherently structured as recursive syntax trees, these tasks also 
probe a model's ability to process hierarchically organized data with rigor and precision. 

Specifically, we introduce \eqone, a dataset containing 
1K purely equational (unconditional) proofs. Each example 
can be viewed as a pair that includes two main components: 
\bit
\item a set of equational axioms $\{E_1,\ldots,E_m\}$, where, for easy reference, all $m$ equational axioms are 
given unique names: $E_1, \ldots, E_m$; and 
\item  a derivation of an identity $s = t$ from those axioms. The structure of derivations is described below. 
\eit 
Each axiom is of the form 
%\setlength{\abovedisplayskip}{4.9pt}
%\setlength{\belowdisplayskip}{4.45pt}
%\[
$\forall \: x_1,\ldots,x_n \qdot l = r$, 
%\]
where the $x_i$ are variables and $l$, $r$ are standard first-order terms built from  variables, constants, 
and function symbols of nonzero arity, \egnsp, $g(a,f(Y))$. 
Variable names start with upper-case letters, while function symbols and constants start with lower-case letters. 
We use $a$, $b$, $c$, $d$ and $e$ as constants. 
Unary function symbols are $f, f_1,\ldots$; binary symbols are $g, g_1,g_2,\ldots$; 
ternary symbols are $h, h_1,\ldots$; and quaternary function symbols (of arity 4) are $r, r1, \ldots$. 
We don't use any function symbols with arity greater than 4. 

If \temv{op} is a function symbol of arity $n$, the abstract syntactic structure of a term of the 
form $\temv{op}(t_1,\ldots,t_n)$ is captured by a \temv{tree} with the function symbol \temv{op} at the 
root and with the trees corresponding to $t_1,\ldots,t_n$ as its children, ordered from left to right. 
Constants and variables always appear at the leaves. Viewing terms as trees in this way, {\em positions\/} 
within a term are represented using Dewey notation. Thus, if $t = g(f(a),g(X,f_2(Y)))$, position 
$[]$ corresponds to the subterm of $t$ at the root (\iensp, $t$ itself), position $[1]$ corresponds 
to $f(a)$, $[1,1]$ corresponds to the constant (leaf) $a$, $[2,1]$ corresponds to $X$, $[2,2]$ to $f_2(Y)$, 
and so on. For a term $t$ and valid position $p$ in $t$, $\temv{sub}(t,p)$ denotes the subterm of $t$ at $p$. 
We write $\dom{t}$ for the set of all valid positions in $t$. 

A substitution $\theta$ is a function from variables to terms, \egnsp, $\theta = \{X \mapsto f(a,Y), Z \mapsto b\}$. 
We write $\theta(t)$ to denote the result of {\em applying\/} $\theta$ to a term $t$, namely, the term obtained
from $t$ by replacing every variable $v$ in the domain of $\theta$ with the term $\theta(v)$. Thus, for the given
example, $\theta(g(Y,g(X,Z))) = g(Y,g(f(a,Y),b))$. We say that a term
$t$ \temv{matches} a term $s$, or that $t$ is an \temv{instance} of $s$, iff there exists a substitution $\theta$ such that $t = \theta(s)$, in which case we say 
that $t$ matches $s$ ``under'' $\theta$ (or wrt $\theta$). 

A proof of an identity $s = t$ from a set of equational axioms $\mathcal{E}$ 
is a sequence of $n > 0$ steps of the following form:  \\[-0.3in]
\begin{center}
\begin{longtable}{rcll}
$s$ & $=$ & $s_0$ & {} \\[0.00in]
{} & $=$ & $s_1$ & \mbox{\kwd{by}\msp$\cdots$}  \\[0.00in]
{} & $=$ & $s_2$ & \mbox{\kwd{by}\msp$\cdots$} \\[0.00in]
{} & $\vdots$ & $\vdots$  & $\vdots$ \\[0.00in]
{} & $=$ & $s_n$  & \mbox{\kwd{by}\msp$\cdots$}  \\[-0.5in]
\end{longtable}
\end{center}
where $s_n$ is syntactically identical to $t$. Such a proof can be understood as a chain of identities
leading from $s$ to $t$. Each step is justified by citing one or more of the axioms in $\mathcal{E}$, listed after the \kwd{by} keyword. 
The following is a sample equational proof of 5 steps that derives the identity $g(h(c,a,c),h(a,c,a)) = g(g_1(g_4(f_1(f_5(f_3(a))),d),e),g_2(a,c))$
\enlargethispage{10pt}
from a set of 6 equational axioms. (Universal quantifiers are omitted when listing the equational axioms; all variables are assumed to be universally quantified.)
\begin{tcAthena}
- Axioms: 
E1: h(c,V1170,c) = h(V1170,a,f4(f1(c)))
E2: h(V1173,V1174,V1173) = g2(V1173,V1174)
E3: h(a,V1181,f4(f1(c))) = g1(f2(f4(V1181)),f4(e))
E4: f2(f4(V1184)) = g(f3(V1184),f2(e))
E5: g(V1187,f2(e)) = g4(f1(f5(V1187)),d)
E6: f4(V1188) = V1188

- Proof: 

s = g(h(c,a,c),h(a,c,a))                              # step 0
    g(h(a,a,f4(f1(c))),h(a,c,a))          by E1       # step 1
    g(h(a,a,f4(f1(c))),g2(a,c))           by E2       # step 3
    g(g1(f2(f4(a)),f4(e)),g2(a,c))        by E3       # step 3
    g(g1(g(f3(a),f2(e)),f4(e)),g2(a,c))   by E4       # step 4
    g(g1(g4(f1(f5(f3(a))),d),e),g2(a,c))  by E5, E6   # step 5
\end{tcAthena}
A proof is correct (or \emph{sound}) iff for every step $s_i = s_{i+1} \msp \fkwd{by} \msp E_{i_1},\ldots,E_{i_k}$, the identity $s_i = s_{i+1}$ 
follows logically from $E_{i_1},\ldots,E_{i_k}$. By Birkhoff's seminal result \citep{birkhoff1935}, this amounts to the identity $s_i = s_{i+1}$ being 
derivable from the cited equations using  the inference rules of equational 
logic.\footnote{There are various options for which equational inference rules to take as primitive, 
but a standard choice consists of reflexivity, symmetry, transitivity, and congruence (also known as substitution of equals).}
A more algorithmic characterization of this derivability relation is based on the notion of {\em rewriting}. Given an equation $e$ of the form $l = r$, 
we write $\subrewrites{s}{t}{e}$ to mean that $s$ has a subterm at some position $p$ such that $\temv{sub}(s,p)$ matches $l$ wrt to a substitution $\theta$
and $t$ is identical to $s$ except that $\temv{sub}(t,p) = \theta(r)$. In words, $t$ is obtained from $s$ by replacing the subterm $\temv{sub}(s,p)$
that matches the left-hand side of $e$ under $\theta$ with the corresponding instance of the right-hand side $r$, \iensp, $\theta(r)$. 
We refer to $\temv{sub}(s,p)$ as a \emph{redex} for $e$ (or an ``$e$-redex'') and to $\temv{sub}(t,p) = \theta(r)$ as the 
corresponding {\em contractum}. If $\subrewrites{s}{t}{e}$, we say that $s$ \emph{rewrites} or \emph{reduces} to $t$ in one step. 
For a set of equations $\mathcal{E}$, we write $\subrewrites{s}{t}{\mathcal{E}}$ to mean that $\subrewrites{s}{t}{e}$ for some $e \in \mathcal{E}$. 
We write $\bdsubrewritesrel{e}$ and $\bdsubrewritesrel{\mathcal{E}}$ for the symmetric closures of  $\rewritesrel{e}$ and $\rewritesrel{\mathcal{E}}$,
respectively, and $\bdsubrewritesreltc{\mathcal{E}}$ for the reflexive, symmetric, and transitive closure of $\rewritesrel{\mathcal{E}}$. 
As Birkhoff first noted \citep{birkhoff1935}, $s = t$ is provable from a set of equations $\mathcal{S}$ using the inference rules of equational logic
(and thus, by Birkhoff's completeness theorem, $s = t$ is a logical consequence of $\mathcal{S}$) iff $s\;\bdsubrewritesreltc{\mathcal{S}}\;t$. 
In our case, a proof step $s_i = s_{i+1} \msp \kwd{by} \msp E_{i_1},\ldots,E_{i_k}$ is correct iff the identity $s_i = s_{i+1}$ follows from 
the set of equations $\mathcal{S} = \{E_{i_1},\ldots,E_{i_k}\}$, which is to say, iff $s_i\;\bdsubrewritesreltc{\mathcal{S}}\;s_{i+1}$; 
and a proof as a whole is correct iff every such step is correct. 

The above definition of equational proofs does not impose any requirements on the granularity of the individual steps. 
This has the advantage of allowing for equational proofs with widely different levels of detail, depending on one's 
preferences (e.g., depending on the intended audience). On the flip side, it allows for proofs that are very hard---potentially 
impossible---to check by algorithm. In the extreme case, one could have a correct ``proof'' of a very complicated 
identity $s = t$ with just one inference step $s = s_0 = t\msp\kwd{by}\msp\mathcal{E}$ that cites all given axioms. 
To rectify this and to simplify the equational tasks considerably,  we stipulate that 
in a step $s_i = s_{i+1}\msp\kwd{by} \msp E_{i_1},\ldots,E_{i_k}$, the term $s_{i+1}$ 
must be obtainable from $s_i$ by $k$ one-step rewrites using the $k$ cited equations, and moreover, that the relevant 
$k$ redexes occur in mutually disjoint positions of $s_i$, and thus, in principle, all $k$ rewrites could 
take place concurrently in one single step. (Two positions are disjoint iff neither is a prefix of the other.)
All proofs in \eqonesp adhere to this structure and have extremely simple 
steps that cite only one or at most two equations, which are only used for one-step reductions in the sense just 
described.
%\footnote{Future versions of this dataset will introduce critical pairs and overlaps probing confluence and normalization 
%strategies, which will stress precise bookkeeping to a much higher degree.}
More precisely, if $\mathcal{S} = \{E_1,...,E_k\}$ is a set of $k > 0$ equations, we write $s \Longrightarrow_{\mathcal{S}} t$ iff $s$ contains $k$ distinct redexes,
one for each $E_i \in \mathcal{S}$, in $k$ disjoint positions; and $t$ is obtained from $s$ by replacing each of these 
$k$ redexes by the corresponding contractum. By definition, if $s \Longrightarrow_{\mathcal{S}} t$, then $\mathcal{S}$ 
does not have any redundant equations; each of the $k$ equations in $\mathcal{S}$ is used to reduce a unique redex in $s$. 
The reducibility relation $\Longrightarrow_\mathcal{S}$ is a proper subset of the usual $\bdsubrewritesreltc{\mathcal{S}}$; the latter is implied
by the former but not conversely.

%Thus, if more than one equation is cited, the different equations apply to disjoint subterms. 
%% In practice, however, the vast majority of proofs in our tasks 
%% have extremely simple steps that cite only one or at most two equations, 
%% and which only perform one-step reductions. If more than one equation is cited, then the different equations apply to disjoint subterms. 
%% More precisely, we simplify the structure of these proofs by 
%% stipulating that in a step 
%% $s_i = s_{i+1}\msp\kwd{by} \msp E_{i_1},\ldots,E_{i_k}$, the term $s_{i+1}$ can be obtained from $s_i$ by $k$ one-step rewrites using the $k$ cited equations,
%% and moreover, that the relevant $k$ redexes occur in mutually disjoint positions of $s_i$ and thus, in principle, all $k$ rewrites could 
%% take place concurrently. 

Consider the proof given in the preceding example and take the first step, $s_0 = s_1$: 
\[g(\mathunderline{red}{\mbox{$h(c,a,c)$}},h(a,c,a)) = g(\mathunderline{blue}{\mbox{$h(a,a,f_4(f_1(c)))$}},h(a,c,a)) \msp \kwd{by} \msp E_1.\]
The underlined redex $\mathunderline{red}{\mbox{$h(c,a,c)$}}$ of $s_0$ matches the left-hand side of $E_1$ under the substitution $\theta = \{V_{1170} \mapsto a\}$, and $s_1$ is obtained 
from $s_0$ by replacing this redex by the corresponding instance of $E_1$'s right-hand side, namely the contractum $\theta(h(V_{1170},a,f_4(f_1(c)))) = \mathunderline{blue}{\mbox{$h(a,a,f_4(f_1(c)))$}}$.
The entire proof with underlined subterms, using red for redexes and blue for contractums,  is shown below: 
\par
\begingroup
\centering
\begin{longtable}{rcll}
$s$ & $=$ & $g(\redul{h(c,a,c)},h(a,c,a))$ & {} \\[0.06in]
{} & $=$ & $g(\blueul{h(a,a,f_4(f_1(c)))},\redul{h(a,c,a)})$ & \mbox{\kwd{by} $E_1$}  \\[0.06in]
{} & $=$ & $g(\redul{h(a,a,f_4(f_1(c)))},\blueul{g_2(a,c)})$ & \mbox{\kwd{by} $E_2$} \\[0.06in]
{} & $=$ & $g(\blueul{g_1(\mbox{\redul{f_2(f_4(a))}},f_4(e))},g_2(a,c))$  & \mbox{\kwd{by} $E_3$} \\[0.06in]
{} & $=$ & $g(g_1(\blueul{\redul{g(f_3(a),f_2(e))}},\redul{f_4(e)}),g_2(a,c))$  & \mbox{\kwd{by} $E_4$} \\[0.06in]
{} & $=$ & $g(g_1(\blueul{g_4(f_1(f_5(f_3(a))),d)},\blueul{e}),g_2(a,c))$  & \mbox{\kwd{by} $E_5, E_6$} \\[-0.2in]
\end{longtable}
\endgroup
\par
Here, only the penultimate term contains multiple (two) redexes; all other steps follow by replacing a single redex by a single contractum. 
The two redexes occur in disjoint positions, as is always the case when more than one equation is cited. 

\enlargethispage{10pt}
We started with 1K randomly generated examples, where each example included its own set of equational axioms and proof. All 1K proofs
were correct by construction. To balance the dataset with incorrect examples, we randomly corrupted about half of the proofs by introducing errors
of two types:
\ben
\item {\em Incorrect contractums}: On a proof step from $s_i$ to $s_{i+1}$, we randomly corrupted a contractum in $s_{i+1}$ 
so that $s_{i+1}$ is no longer derivable from $s_i$ by the equations cited by the step. 
\item {\em Incorrect equations}: We randomly changed one of the justifying equations for a given proof step. Thus, \egnsp, while 
the original---and correct---version of the step cited equations $E_2$ and $E_5$, we might change that to $E_9$ and $E_5$, ensuring 
that the step no longer goes through. 
\een
Only one error was inserted in each corrupted proof, at a randomly chosen step. 

%Details and examples can be found in Appendix~\ref{Sec:AppEquationalProof}. 
We introduce and study three \eqonesp tasks  that fall along a spectrum of increasing difficulty, 
forming a natural progression that probes different reasoning skills: 

\begin{enumerate}
\item {\em Proof checking\/} (\mbf{EQ-PC}, easiest): This is largely a local and low-level matching and substitution task, 
although we will see that even here there are different gradations of difficulty depending 
on how much explanatory detail we demand. 
  \penalty100000
\par\nopagebreak[4]
\item {\em Equation recovery\/} (\mbf{EQ-ER}, intermediate): This requires more search and matching than simple checking, 
but it's still constrained to a single step at a time.

\item {\em Gap filling\/} (\mbf{EQ-GF}, hardest): This is a synthesis task. The model is given a start and end point 
separated by multiple elided steps and must generate a valid intermediate path. This requires multi-step lookahead, 
global planning, and search through a combinatorial space of possible rewrites. It's considerably more complex than the other two tasks. 

\end{enumerate}

\subsubsection{Equational Proof Checking}

This task contains 200 randomly chosen examples from the set of 1K equational proofs. 
The slice is balanced so that about half of the chosen proofs are correct, while the others
are corrupted as described above. The model is asked to determine whether the proof is 
correct, and more importantly, to {\em justify\/} its answer. Accordingly, the requested output 
has two components: a binary correctness verdict and an explanation/justification for the verdict. 

Checking an equational proof in detail means verifying that each correct step goes through 
owing to the cited identities, and not only identifying incorrect steps but pinpointing 
the exact source of error. In particular, if a model claims that a proof with $n$ steps is correct, 
then its full explanation  must be a sequence of $n$ structured objects, one for each proof step 
\[s_i = s_{i+1} \msp \kwd{by} \msp E_{i_1},\ldots,E_{i_k}\]
for $i=0,\ldots,n-1$,  elaborating \emph{why} each step is correct.  The structured object 
consists of the relevant proof step index (the number $i+1$) along with a list of rewrites corresponding 
to the $k$ redexes in $s_i$. Each rewrite is a dictionary with the following keys: 
\bit
\item \emph{redex}: the redex in question; 
\item \emph{position}: the exact position of the redex in $s_i$;
\item \emph{equation}: the identity whose left-hand side matches the redex; 
\item \emph{contractum}: the corresponding contractum in $s_{i+1}$. 
\eit
More concretely, the model's output must be a dictionary with two keys, \temv{correct} and \temv{explanation}. The value of \temv{correct} is \temv{true} or \temv{false},
depending on whether the model determines that the input proof is correct. 
When \temv{correct} is \temv{true}, the value of \temv{explanation} must be 
a list of dictionaries, each with two keys, \temv{step} (the step number, one for each step of the proof) and \temv{rewrites}, 
a list of rewrites of the form described above. Thus, for the example in the preceding section, an appropriate output would be: 
\begin{lstlisting}[language=Python,numbers=none,basicstyle=\scriptsize\ttfamily]
{"correct": true,
 "explanation": 
  [{"step": 1, 
    "rewrites": [{"redex": "h(c,a,c)", "position": [1],   
                  "equation": "E1", "contractum": "h(a,a,f4(f1(c)))"}]},
   {"step": 2, 
     "rewrites": [{"redex": "h(a,c,a)", "position": [2],   
                   "equation": "E2", "contractum": "g2(a,c)"}]},
   {"step": 3, 
    "rewrites": [{"redex": "h(a,a,f4(f1(c)))", "position": [1],
                  "equation": "E3", "contractum": "g1(f2(f4(a)),f4(e))"}]},
   {"step": 4, 
    "rewrites": [{"redex": "f2(f4(a))", "position": [1,1],
                  "equation": "E4",  "contractum": "g(f3(a),f2(e))"}]},
   {"step": 5, 
    "rewrites": [{"redex": "g(f3(a),f2(e))", "position": [1,1], 
                  "equation": "E5", "contractum": "g4(f1(f5(f3(a))),d)"},
                 {"redex": "f4(e)", "position": [1,2], 
                  "equation": "E6", "contractum": "e"}]}]}
\end{lstlisting}

By contrast, if \temv{correct} is \temv{false}, then the value of a complete \temv{explanation} must be 
a single dictionary corresponding to the first incorrect step in the proof, and should be: 
\ben
\item either of the form 
\begin{lstlisting}[language=Python,numbers=none,basicstyle=\scriptsize\ttfamily]
{"step": ..., 
 "equation": ..., 
 "position": ..., 
 "expectedContractum": ..., 
 "actualContractum": ...}
\end{lstlisting}
indicating that the contractum for the equation and redex in question (at the given position) is not what was expected by the cited equation; 
\item or of the form 
\begin{lstlisting}[language=Python,numbers=none,basicstyle=\scriptsize\ttfamily]
{"step": ..., 
 "givenEquations": [...], 
 "correctEquations:[...]},
\end{lstlisting}
indicating that the cited identities do not succeed in justifying that step, though the identities listed under \temv{correctEquations} do. 
\een
Outputs of the first form capture errors introduced by corrupting contractums; outputs of the second form capture errors introduced by noising cited identities. 

As an example of an incorrect proof and the proper corresponding output, consider the following input: 
\begin{tcAthena}
- Axioms: 

E1:  h(a,V1110,V1111) = g3(f(V1111),f1(V1110))
E2:  h(b,c,V1118) = g(f4(f3(V1118)),a)
E3:  g3(f(c),f1(V1119)) = g(f2(V1119),g1(a,c))
E4:  f2(V1124) = f1(f1(V1124))
E5:  f4(f3(V1125)) = f2(g(V1125,f4(c)))

- Proof:

s = r(h(e,e,c),g(c,b),h(b,c,b),h(a,a,c))                                    # step 0
    r(h(e,e,c),g(c,b),h(b,c,b),g3(f(c),f1(a)))               by E1          # step 1
    r(h(e,e,c),g(c,b),g(f4(f3(b)),a),g(f2(a),g1(a,c)))       by E2, E3      # step 2
    r(h(e,e,c),g(c,b),g(f2(g(b,f4(e))),a),g(f2(a),g1(a,c)))  by E4, E5      # step 3
\end{tcAthena}
Here, the proof proceeds correctly until the last step:
\begin{center}
\begin{supertabular}{rcll}
{} & $\vdots$ & $\vdots$ & $\vdots$ \\[0.08in]
{} & $=$ & $r(h(e,e,c),g(c,b),g(\redul{f_4(f_3(b))},a),g(f_2(a),g_1(a,c)))$  & \mbox{\kwd{by} $E_2, E_3$} \\[0.08in]
{} & $=$ & $r(h(e,e,c),g(c,b),g(\blueul{f_2(g(b,f_4(e)))},a),g(f_2(a),g_1(a,c)))$  & \mbox{\kwd{by} $E_4, E_5$} \\
\end{supertabular}
\end{center}
The only redex in the penultimate term that matches the left-hand side of $E_5$ is in position [3,1], shown underlined in red: $f_4(f_3(b))$.
The correct contractum for this redex is $f_2(g(b,f_4(c)))$, but the one that actually appears in the last term is $f_2(g(b,f_4(e)))$.
Accordingly, an appropriate output for this proof is: 
\begin{lstlisting}[language=Python,numbers=none,basicstyle=\scriptsize\ttfamily]
{"correct": false,
 "explanation": {"step": 3, 
                 "equation": "E5", 
                 "position": [3, 1], 
                 "expectedContractum": "f2(g(b,f4(c)))",
                 "actualContractum": "f2(g(b,f4(e)))"}}
\end{lstlisting}
Finally, here is an example illustrating the second type of error, the citing of incorrect equations:
\begin{tcAthena}
- Axioms: 

E1:  g(V1068,d) = f(f2(f3(V1068)))
E2:  g(b,V1071) = h(V1071,f4(a),d)
E3:  g(b,V1072) = f5(V1072)
E4:  f2(V1073) = f1(V1073)
E5:  f3(V1080) = V1080
E6:  f4(V1083) = g2(V1083,f5(b))
E7:  f(V1086) = f1(V1086)

- Proof:

s = h(g(b,d),g(b,b),g(b,e))                                   # step 0
    h(f(f2(f3(b))),g(b,b),g(b,e))          by E1              # step 1
    h(f(f1(f3(b))),f5(b),h(e,f4(a),d))     by E2, E3, E4      # step 2
    h(f(f1(b)),f5(b),h(e,f4(a),d))         by E5              # step 3
    h(f(f1(b)),f5(b),h(e,g2(a,f5(b)),d))   by E2              # step 4
    h(f1(f1(d)),f5(b),h(e,g2(a,f5(b)),d))  by E7              # step 5
\end{tcAthena}
The error here is on step 4: $E_2$ does not justify the passage from $h(f(f_1(b)),f_5(b),h(e,f_4(a),d))$
to \[h(f(f_1(b)),f_5(b),h(e,g_2(a,f_5(b)),d)).\] 
There are no redexes in $h(f(f_1(b)),f_5(b),h(e,f_4(a),d))$ that match the left-hand size of $E_2$. 
However, the step \emph{is} justified by $E_6$, which rewrites the redex
$f_4(a)$ to $g_2(a,f_5(b))$. Thus, the correct output in this case is:
\begin{lstlisting}[language=Python,numbers=none,basicstyle=\scriptsize\ttfamily]
{"correct": false,
 "explanation":  {"step": 4, 
                  "givenEquations": ["E2"], 
                  "correctEquations": ["E6"]}}
\end{lstlisting}
Our few-shot prompt for this task contains several ICL examples of both correct and incorrect proofs, along with appropriate corresponding outputs. 

What was described above is a task that requires thorough and detailed checking of equational proofs. We refer to this as {\em level-3 proof checking}.
This is what our proof-checking prompt for equational logic specifies. Nevertheless, to paint a richer performance picture, we distinguish two
additional evaluation modes that require less detail. In total, we have: 
\bit
\item \emph{Level 1}: The model is given full credit as long as it can guess whether the proof is correct, \iensp, as long as it makes
the right binary judgment. Accuracy at this level measures a model's ability to get the ``vibe'' of the proof. Is it roughly correct or does something seem off?

\item \emph{Level 2}: If the model rightly claims that the proof is correct, then it must provide detailed evidence for 
its claim, \iensp, the \emph{explanation} it outputs in that case must be fully correct. 
This level makes no demands on the model's output for incorrect proofs. As long as the model correctly guesses that the proof has an error, 
it is given full credit even if it gets the error details wrong. Thus, level-2 accuracy measures a model's ability to perform the 
computational work of positive verification: identifying redexes, matching them to equations, and confirming the contractums. 

\item \emph{Level 3}: As already mentioned, level-3 proof checking is the most thorough: credit is given only 
if the verdict is correct and the model's \emph{explanation} gets all the details right, both for 
correct and for incorrect proofs.  Thus, this level requires everything required by the previous levels,
as well as debugging skills, or the ability to perform error analysis by precisely identifying and explaining 
the first point of failure in an incorrect proof. This is the highest level of proof-checking scrutiny. The model must 
keep track of the ``expected'' state (what the term should look like after the rewrite); compare it against 
the actual state; and then articulate the discrepancy. This level also requires 
a degree of error-repair ability, insofar as the model must single out the correct (or ``expected'') contractum 
in the case of a contractum corruption, or the correct equations in the case of an equation corruption. 
\eit
These levels form a ladder of proof-checking rigor: proof-checking success at L3 (level 3) entails L2 success, 
which entails L1 success, but the converses do not hold. 

\renewcommand{\arraystretch}{1.2}
\begin{table}[!htbp]
\centering
\rowcolors{2}{blue!5}{white}
\resizebox{1.0\textwidth}{!}{%
\begin{tabular}{|l|c|c|c|l|c|c|c|}
\hline
\textbf{Model} & \textbf{L1} & \textbf{L2} & \textbf{L3} & \textbf{Model} & \textbf{L1} & \textbf{L2} & \textbf{L3} \\ \hline
\gemThreeOne & 0.98 \ci{0.02} & 0.93 \ci{0.04} & \mbf{0.93} \ci{0.04} & \gptFiveOne & 0.90 \ci{0.04} & 0.81 \ci{0.05} & 0.76 \ci{0.06} \\ \hline
\gptFiveTwo & 0.97 \ci{0.03} & 0.82 \ci{0.05} & 0.75 \ci{0.06} & \gemThree & 0.98 \ci{0.02} & 0.76 \ci{0.06} & 0.74 \ci{0.06} \\ \hline
\claudeOpusFourFiveNov & 0.99 \ci{0.02} & 0.74 \ci{0.06} & 0.71 \ci{0.06} & \gptFive & 0.95 \ci{0.03} & 0.77 \ci{0.06} & 0.69 \ci{0.06} \\ \hline
\oThree & 0.88 \ci{0.05} & 0.69 \ci{0.06} & 0.58 \ci{0.07} & \grokFour & 0.90 \ci{0.04} & 0.67 \ci{0.06} & 0.56 \ci{0.07} \\ \hline
\claudeSonnetExt & 0.97 \ci{0.03} & 0.64 \ci{0.07} & 0.56 \ci{0.07} & \dsReasoner & 0.89 \ci{0.04} & 0.62 \ci{0.07} & 0.53 \ci{0.07} \\ \hline
\geminiTwoPointFive & 0.87 \ci{0.05} & 0.56 \ci{0.07} & 0.52 \ci{0.07} & \oFourMini & 0.69 \ci{0.06} & 0.62 \ci{0.07} & 0.43 \ci{0.07} \\ \hline
\oThreeMini & 0.67 \ci{0.06} & 0.52 \ci{0.07} & 0.25 \ci{0.06} & \dsROne & 0.72 \ci{0.06} & 0.51 \ci{0.07} & 0.25 \ci{0.06} \\ \hline
\opus & 0.74 \ci{0.06} & 0.36 \ci{0.07} & 0.25 \ci{0.06} & \dsThreeOne & 0.64 \ci{0.07} & 0.52 \ci{0.07} & 0.24 \ci{0.06} \\ \hline
\claudeSonnetFour & 0.61 \ci{0.07} & 0.53 \ci{0.07} & 0.23 \ci{0.06} & \grokThree & 0.64 \ci{0.07} & 0.38 \ci{0.07} & 0.14 \ci{0.05} \\ \hline
\hermes & 0.58 \ci{0.07} & 0.50 \ci{0.07} & 0.08 \ci{0.04} & \dsThree & 0.49 \ci{0.07} & 0.49 \ci{0.07} & 0.06 \ci{0.03} \\ \hline
\gptFourO & 0.62 \ci{0.07} & 0.32 \ci{0.06} & 0.04 \ci{0.03} & \nous & 0.53 \ci{0.07} & 0.06 \ci{0.03} & 0.04 \ci{0.03} \\ \hline
\gptFourOne & 0.52 \ci{0.07} & 0.06 \ci{0.03} & 0.02 \ci{0.02} & \magistral & 0.53 \ci{0.07} & 0.46 \ci{0.07} & 0.01 \ci{0.02} \\ \hline
\end{tabular}}
\caption{Equational proof-checking accuracies at 3 levels of detail.}
\label{Tb:PCAcc}
\end{table}
\renewcommand{\arraystretch}{1.0}

Results for all three levels are shown in Table~\ref{Tb:PCAcc}.
%What do these numbers tell us about the ability of language models to scrutinize equational proofs? 
For L1, the random-guessing accuracy baseline is already $50\%$. Thus, above-chance accuracy for the top-performing model, 
Gemini 3.1, is 98\%-50\% = 48\%. If we normalize, we can say that Gemini's accuracy 
is $(0.98-0.5)/(1-0.5) = 0.96$, \iensp, that Gemini 3.1 is achieving 96\% of the possible 
improvement over random guessing. This is near-perfect performance @L1.
However, beyond the top 13 models, performance drops to below 70\% even at L1. 
So top-tier and some mid-tier models have very strong ``intuition'' that can 
reliably distinguish correct equational proofs from  broken ones, but the rest 
have a lot of room for improvement. The worst-performing models do about as well as random guessing. 

Performance across the board degrades sharply when we move to L2 and require 
models to not simply guess correctness but explain on a step-by-step
basis \emph{why} the proof is correct, namely, which are the relevant redexes matching the cited equations, 
where exactly they are located in the preceding step's term, and what are the corresponding contractums. 
Opus 4.5 experiences a 25-absolute-point drop in going from L1 to L2,
while for Gemini 3 and GPT-5.2 the drop is 22 and 15 absolute points, respectively.
This shows that even for elite models, there is a large difference between simply guessing  
that a proof is correct and properly articulating its correctness in a structured explanation. 
Gemini 3.1 is the only top model whose decline is relatively contained (5 points only). 

Performance drops further at L3, which extends the detailed-explanation requirement to
incorrect proofs. 
At this level, a model must do everything it must do for L2 but also document where and 
how an incorrect proof fails. For some top models (especially for Gemini 3.1 and Gemini 3), 
the $\mbox{L2}\rarrls\mbox{L3}$ drop is basically zero, and considerably smaller than 
their $\mbox{L1}\rarrls\mbox{L2}$ drops. This suggests that these models are 
competent debuggers, almost as good at diagnosing errors as they are at verifying correctness. 
By contrast, most mid-tier models (like DeepSeek-R1-0528, o4-mini, and o3-mini), 
experience catastrophic $\mbox{L2}\rarrls\mbox{L3}$ drops. 
These models are much better at verifying a correct stretch of equational 
reasoning than diagnosing a flawed one.

Taking the L3 numbers as the main set of results, the DI of this task 
is an exceptionally high 0.33. 

\begin{figure}[!htbp]
\centering
%\scalebox{0.5}{\includegraphics[width=\linewidth]{./plots/er_stepwise_error_rates.png}}
%\includegraphics[width=\linewidth]{./plots/er_stepwise_error_rates_smaller.png}
\scalebox{1.0}{\includegraphics[width=\linewidth]{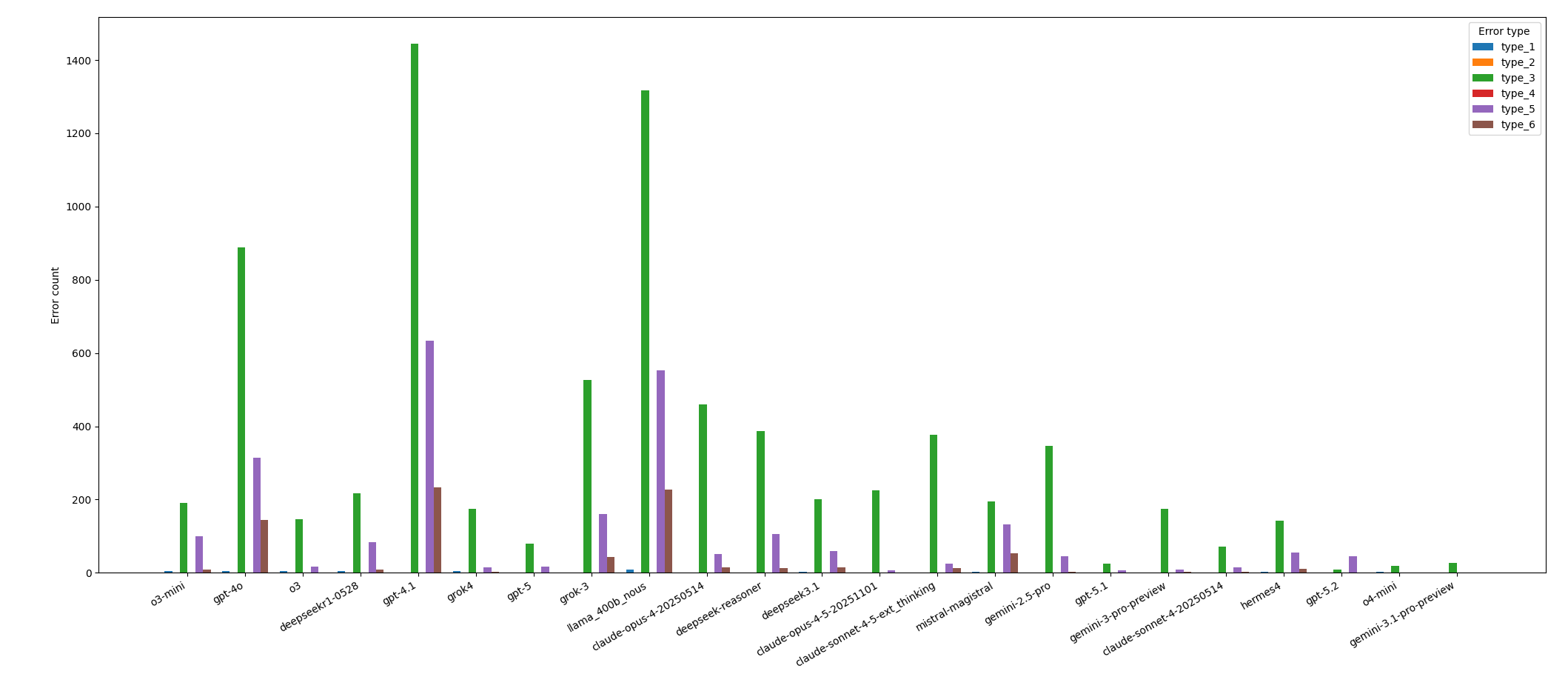}}
\caption{Level-2 error distribution for equational proof checking.}
\label{Fig:SimpleCheckingL2Errors}
\end{figure}

The L2 error distribution is shown in Fig.~\ref{Fig:SimpleCheckingL2Errors}. 
We distinguish 6 error types at this level. 
Recall that, at L2, if a model declares a proof correct then 
its explanation must include a list of rewrites for every proof step, where every rewrite is a dictionary with four keys: \emph{position}, \emph{redex}, \emph{contractum}, and \emph{equation}. 
Error types 3--6 correspond precisely to one of these 4 values being wrong. 
\ben
\item \emph{Type 1}: Ill-formatted responses, such as a JSON-schema violation (\egnsp, a response dictionary that does not include one of the expected keys) 
or an explanation for a proof that is declared correct but is missing some of the proof steps.\footnote{We are lenient in these judgments, e.g., instead of checking for strict equality between the set of expected
dictionary keys and the set of generated keys, we check only for a subset relation, as some models unilaterally insert extra key-value entries, such as a \emph{note} that provides additional commentary about something.} 
\item \emph{Type 2}: Wrong correctness verdict.
\item \emph{Type 3}: Wrong position in a rewrite of an explanation step. 
\item \emph{Type 4}: Wrong redex in a rewrite of an explanation step. 
\item \emph{Type 5}: Wrong contractum in a rewrite of an explanation step. 
\item \emph{Type 6}: Wrong equation in a rewrite of an explanation step. 
\een
Error types 1 and 2 are mutually exclusive with types 3--6, but the latter can co-occur in 
an explanation (and even on a single explanation step), so Fig.~\ref{Fig:SimpleCheckingL2Errors} 
reports raw error counts summed over all proof steps. For that reason, 
the absolute values of these counts cannot be meaningfully compared
across models with widely different accuracies on the set of correct proofs.
For example, \claudeSonnetFoursp makes far fewer errors 
than \geminiTwoPointFive, but that's because it only recognizes 27 out of the 
103 correct proofs, whereas \geminiTwoPointFivesp correctly recognizes 91, so naturally it will
have more opportunities to make errors of types 3--6.

The plot shows that misidentifying redex positions (type-3 errors) is the most common 
failure mode at L2. Models usually output positions that are close to the correct position 
but do not actually match it. We quantify this closeness relation 
by defining a similarity metric that considers how much of the path is shared 
between the correct position and the model-generated position 
before the two diverge, \iensp, their longest common prefix. The more steps they share from the root, the more similar they are. 
The metric is normalized so the score always lies between 0 and 1. If the two positions are identical, 
the score is 1; if they diverge 
immediately at the root, the score is 0. For intermediate cases, it strikes a balance between the amount of common structure and the 
total lengths of the paths. For instance, $[3,3,1]$ and $[3,3]$ share a prefix of length 2 and one is a prefix of the other, so their
similarity is high (0.8), while paths that diverge early get much lower scores. 
%This makes the measure sensitive to how deeply two positions stay together in the tree. 
The mean similarities for all models across their 
type-3 errors are as follows: 
\gemThreeOne: 0.83, \gptFiveTwo: 0.77, \gptFiveOne: 0.75, \gptFive: 0.76, \gemThree: 0.75, \oThree: 0.71,  
\oFourMini: 0.70, \claudeSonnetFour: 0.69, \claudeOpusFourFiveNov: 0.68, \grokFour: 0.66, \geminiTwoPointFive: 0.66, \dsROne: 0.65, \dsReasoner: 0.65, 
\claudeSonnetExt: 0.65, \grokThree: 0.62, \opus: 0.61, \oThreeMini: 0.60, \dsThreeOne: 0.53, \gptFourO: 0.5, 
\gptFourOne: 0.5, \nous: 0.45,  \hermes: 0.41, \magistral: 0.38. DeepSeek-V3 makes no type-3 errors. 
These numbers track only modestly with overall performance, featuring some notable exceptions such as \claudeSonnetFour,
which is better at locating precise positions than both \geminiTwoPointFivesp and \grokFour,
suggesting that this might be a fairly distinct skill, not just a corollary of general reasoning ability. 

The results indicate that most models are fairly good at ``what'' and ``why'' but bad at ``where.'' They can often correctly 
identify the string being changed (redex), what it changes into (contractum), and even the rule that governs 
the change (equation); but they fail at precisely locating that change within the larger tree structure.
This should not be surprising, given that models do not explicitly build or traverse syntax trees. 
Tree structure is inherently two-dimensional, so a transformer-based model must infer it 
from syntactic cues in the 1D token sequence that it actually sees, primarily from parentheses and commas, 
which tends to be fuzzy and imprecise for the purposes of exact hierarchical localization. 

GPT-5.2 stands out here as having the strongest localization skills. 
As mentioned earlier, raw counts of type-3 errors cannot be compared across models 
that accept widely different numbers of correct proofs.
However, for models that have similar acceptance rates on correct proofs, comparing 
the counts of error types 3---6 can reveal strikingly different failure regimes. 
Focusing specifically on type-3 errors, we see that GPT-5.2 accepts 98 out of the 103 correct
proofs and makes a total of only 8 type-3 errors. By contrast, Gemini 3 Pro Preview accepts
102 out of the 103 correct proofs and makes 175 such errors. Normalizing, we get 
a type-3 error rate or 0.08 for GPT-5.2 and 1.7 for Gemini 3, a difference of about 1.32
orders of magnitude. Even Gemini 3.1, the top-performing model, makes 26 type-3 errors, for 
an error rate of 0.25, about 3 times larger than GPT-5.2.

We also separately compute the number of completely invalid positions that models generate (\iensp, any position for a term $t$
that is not a member of $\dom{t}$), summed over all explanation steps.
To normalize, we again divide each count by the number of correct proofs recognized by the corresponding model,
which yields the following mean number of invalid positions generated per correct proof
(DeepSeek-V3 is again excluded because it only recognizes 2 correct proofs):
\begin{center}
\small
\setlength{\tabcolsep}{6pt}
\resizebox{0.85\textwidth}{!}{%
\begin{tabular}{l r @{\hspace{1.2em}} l r @{\hspace{1.2em}} l r}
\gptFourOne            & 6.94 & \hermes                & 3.25 & \claudeSonnetFour      & 0.85 \\
\gptFourO              & 5.85 & \dsThreeOne            & 3.14 & \gemThree              & 0.74 \\
\nous                  & 5.08 & \opus                  & 2.36 & \oThree                & 0.70 \\
\magistral             & 4.27 & \dsReasoner            & 2.12 & \gptFive               & 0.33 \\
\grokThree             & 3.29 & \geminiTwoPointFive    & 2.05 & \oFourMini             & 0.12 \\
\oThreeMini            & 1.27 & \claudeSonnetExt       & 1.95 & \gptFiveOne            & 0.09 \\
\dsROne                & 1.25 & \grokFour              & 0.95 & \gemThreeOne           & 0.06 \\
\claudeOpusFourFiveNov & 0.92 & \gptFiveTwo            & 0.02 &                        &      
\end{tabular}%
}
\end{center}
%% \begin{center}
%% \small
%% \setlength{\tabcolsep}{4pt}
%% \resizebox{0.5\textwidth}{!}{%
%% \begin{tabular}{l r @{\hspace{1.5em}} l r}
%% \gptFourOne            & 6.94 & \oThreeMini            & 1.27 \\
%% \gptFourO              & 5.85 & \dsROne                & 1.25 \\
%% \nous                  & 5.08 & \grokFour              & 0.95 \\
%% \magistral             & 4.27 & \claudeOpusFourFiveNov & 0.92 \\
%% \grokThree             & 3.29 & \claudeSonnetFour      & 0.85 \\
%% \hermes                & 3.25 & \gemThree              & 0.74 \\
%% \dsThreeOne            & 3.14 & \oThree                & 0.70 \\
%% \opus                  & 2.36 & \gptFive               & 0.33 \\
%% \dsReasoner            & 2.12 & \oFourMini             & 0.12 \\
%% \geminiTwoPointFive    & 2.05 & \gptFiveOne            & 0.09 \\
%% \claudeSonnetExt       & 1.95 & \gemThreeOne           & 0.06 \\
%%                        &      & \gptFiveTwo            & 0.02 \\
%% \end{tabular}
%% }
%% \end{center}
How well models identify the precise positions of subterms of interest, such as redexes,
is likely indicative of their general ability to navigate deeply nested tree-structured data.

Among exclusively L3 errors, the most common type for most models 
is misidentifying the step where the proof first goes wrong. 
%% These localization failures are by far the most frequent level-3 errors for all models with the exception of \gptFive, \grokFour, \nous, and \opus. 
Further, from the two types of possible proof 
errors, one due to contractum corruption and one due to corruption of the justifying equations, models have a harder time 
pinpointing the details of the former. They 
struggle roughly equally in identifying the expected contractum and the actual contractum. 
The discrepancy makes sense. An equation corruption is an easier diagnosis: ``I see a change from $u$ to $v$. 
The cited rule is $E_i$, but this rule doesn't apply at all.'' A contractum corruption, by contrast, is much harder: ``I see a change 
from $u$ to $v$. The cited rule is $E_i$. The redex $u$ matches the LHS of $E_i$, so the rule does apply. Now, I must 
compute the expected contractum $\rp{v}$ and compare it to the actual $v$.'' This requires the model to perform a forward 
inference step itself, hold the result in memory, and then perform a comparison, so it's more computationally  demanding.

\subsubsection{Recovery of Elided Equations}
This task also contains 200 samples drawn from the set of 1K examples, evenly split between correct and incorrect proofs.
The input given to the model is the same as in the previous task, a set of named equations $\mathcal{E}$ and a proof $D$, 
except that all justifying equations are now omitted from every step in $D$ and the model has 
to discover them on its own. Specifically, for every step of the proof that derives a term $s_{i+1}$ from a term $s_i$, 
the model has to output a set $\mathcal{S}_i \subseteq \mathcal{E}$ of equations that justify 
the step in the sense that $s_i \Longrightarrow_{\mathcal{S}_i} s_{i+1}$, assuming that such a set $\mathcal{S}_i$ exists. 
%% \footnote{$S_i$ should be minimal in that it should only 
%% include an equation if the step does not go through without it; otherwise \label{Ft:MinimalEqs}
%% a model could trivially justify any correct proof step by listing all members of $\mathcal{E}$.
%% This is because equational logic, like all forms of classical logic, is monotonic.}
%% We say that a set of equations $\mathcal{S_i}$ is {\em correct\/} for a step that derives $s_{i+1}$ from $s_i$ iff 
%% the identity $s_i = s_{i+1}$ follows from $S_i$ by standard equational logic and $S_i$ is minimal 
%% in the sense just described (footnote~\ref{Ft:MinimalEqs}). 
If it does not, \egnsp, because the step from $s_i$ to $s_{i+1}$ has an error, 
then the correct output is the empty set of equations.
%% \footnote{This would present a problem for steps 
%% that are valid by reflexivity (i.e., where $s_{i+1}$ is syntactically identical to $s_i$), but none 
%% of our proofs have such steps.} 
For example, given the input: 
\begin{tcAthena}
- Equations:

E1: h(c,V1170,c) = h(V1170,a,f4(f1(c)))
E2: h(V1173,V1174,V1173) = g2(V1173,V1174)
E3: h(a,V1181,f4(f1(c))) = g1(f2(f4(V1181)),f4(e))
E4: f2(f4(V1184)) = g(f3(V1184),f2(e))
E5: g(V1187,f2(e)) = g4(f1(f5(V1187)),d)
E6: f4(V1188) = V1188

- Proof: 

s = g(h(c,a,c),h(a,c,a))                        # step 0
    g(h(a,a,f4(f1(c))),h(a,c,a))                # step 1
    g(h(a,a,f4(f1(c))),g2(a,c))                 # step 2
    g(g1(f2(f4(a)),f4(e)),g2(a,c))              # step 3
    g(g1(g(f3(a),f2(e)),f4(e)),g2(a,c))         # step 4
    g(g1(g4(f1(f5(f3(a))),d),e),g2(a,c))        # step 5
\end{tcAthena}
the following list would be a correct output: 
\begin{lstlisting}[language=Python,numbers=none,basicstyle=\scriptsize\ttfamily]
[{"step": 1, "supportingEquations": ["E1"]},
 {"step": 2, "supportingEquations": ["E2"]},
 {"step": 3, "supportingEquations": ["E3"]},
 {"step": 4, "supportingEquations": ["E4"]},
 {"step": 5, "supportingEquations": ["E5", "E6"]}]
\end{lstlisting}
Note that this task has no unique correct output for a given step from $s_i$ to $s_{i+1}$, as there may be multiple sets of equations $\mathcal{S} \subseteq \mathcal{E}$
such that $s_i \Longrightarrow_{\mathcal{S}} s_{i+1}$ (although typically the number of possibilities is small). We thus verify each answer independently. If
an answer $\mathcal{S}$ is a non-empty set of equations, we check that $s_i \Longrightarrow_{\mathcal{S}} s_{i+1}$. If an answer is the empty set of equations, we check
that there is no $\mathcal{S} \subseteq \mathcal{E}$ such that $s_i \Longrightarrow_{\mathcal{S}} s_{i+1}$. This is easily decidable, first by filtering those 
candidate equations that match some subterm of $s_i$, and then checking whether every disagreement between $s_i$ and $s_{i+1}$ can be resolved by one of these
candidates.\footnote{The set $D$ of ``disagreements'' between two terms $s$ and $t$ consists of those minimal positions $p \in \dom{s} \cap \dom{t}$ such that 
the subterms $\temv{sub}(s,p)$ and $\temv{sub}(t,p)$ have different roots. By ``minimal'' we mean that if $p \in D$ and $q$ 
is a strict suffix of $p$ then $q \not \in D$. Thus, this set is simply the set of earliest 
or highest positions where the two trees have a mismatch.}

This task has a strong abductive-verification flavor. For every step of the proof, the model is faced with three pieces of ``evidence'':
the starting state $s_i$, the ending state $s_{i+1}$, and the equational axioms in $\mathcal{E}$ as the permissible ``physical laws.'' 
The claim to be investigated is that $s_{i+1}$ is easily derivable from $s_i$ by a set of equations $\mathcal{S} \subseteq \mathcal{E}$ (where easy derivability means
no deduction chains, just single-step reductions as defined by $\Longrightarrow_{\mathcal{S}}$).
To decide the claim, the model must work backward from the evidence by looking at the ``effect'' (the difference between $s_i$ and $s_{i+1}$) and trying 
to find a ``cause'' (a minimal set of axioms) that would explain it. This can be performed by difference analysis followed by hypothesis generation.
Difference analysis aims to identify the subterm(s) that changed in the transition from $s_i$ to  $s_{i+1}$;  this is the locus of the analysis.
Each subterm change consists of a redex candidate in $s_i$ and a transformed version in $s_{i+1}$, for which the model generates hypotheses: 
``Could axiom $E_1$ explain this transformation? Could $E_2$? Could $E_3$? $\cdots$'' 

While the task requires processing a good deal of highly structured data, it lends itself to pattern recognition 
and does not require fine-grained analysis. In the running example, for instance, the first step can only be justified 
by $E_1$, since that's the only equation whose right-hand side can introduce a subterm of the form $f_4(f_1(c))$. 
Likewise, the second step is easily seen to follow by $E_2$. 

We report two metrics on this task. The main one is {\em proof-level accuracy\/} (or PLA for short): the fraction of proofs for  
which the model generates correct outputs at every step. The secondary metric is {\em step-level accuracy\/} (SLA): 
the proportion of individual proof steps for which the model produces correct outputs. 
The former is clearly more demanding, as a model is rewarded only if it handles correctly \emph{every} 
step of the proof. The step-level metric is much more forgiving, as it rewards partial success 
on individual steps. Proof-level and step-level accuracies are shown in Table~\ref{Tb:ERProofLevelAcc}.
\renewcommand{\arraystretch}{1.2}
\begin{table}[!htbp]
\centering
\rowcolors{2}{blue!5}{white}
\resizebox{0.9\textwidth}{!}{%
\begin{tabular}{|l|c|c||l|c|c|}
\hline
\textbf{Model} & \textbf{PLA} & \textbf{SLA} & \textbf{Model} & \textbf{PLA} & \textbf{SLA} \\
\hline
\gptFiveTwo & \mbf{0.94} \ci{0.03} & 0.99 \ci{0.02} & \gptFiveOne & 0.92 \ci{0.04} & 0.99 \ci{0.02} \\
\hline
\gemThreeOne & 0.79 \ci{0.06} & 0.98 \ci{0.02} & \geminiTwoPointFive & 0.78 \ci{0.06} & 0.98 \ci{0.02} \\
\hline
\gptFive & 0.77 \ci{0.06} & 0.98 \ci{0.02} & \claudeOpusFourFiveNov & 0.70 \ci{0.06} & 0.97 \ci{0.03} \\
\hline
\gemThree & 0.69 \ci{0.06} & 0.97 \ci{0.03} & \oThree & 0.67 \ci{0.06} & 0.96 \ci{0.03} \\
\hline
\claudeSonnetExt & 0.62 \ci{0.07} & 0.96 \ci{0.03} & \grokFour & 0.60 \ci{0.07} & 0.94 \ci{0.03} \\
\hline
\dsReasoner & 0.44 \ci{0.07} & 0.54 \ci{0.07} & \oThreeMini & 0.39 \ci{0.07} & 0.83 \ci{0.05} \\
\hline
\dsROne & 0.39 \ci{0.07} & 0.81 \ci{0.05} & \dsThreeOne & 0.35 \ci{0.07} & 0.79 \ci{0.06} \\
\hline
\opus & 0.33 \ci{0.06} & 0.90 \ci{0.04} & \oFourMini & 0.29 \ci{0.06} & 0.80 \ci{0.06} \\
\hline
\claudeSonnetFour & 0.29 \ci{0.06} & 0.85 \ci{0.05} & \hermes & 0.17 \ci{0.05} & 0.57 \ci{0.07} \\
\hline
\grokThree & 0.12 \ci{0.05} & 0.74 \ci{0.06} & \dsThree & 0.09 \ci{0.04} & 0.72 \ci{0.06} \\
\hline
\gptFourOne & 0.07 \ci{0.04} & 0.46 \ci{0.07} & \magistral & 0.02 \ci{0.02} & 0.30 \ci{0.06} \\
\hline
\nous & 0.01 \ci{0.02} & 0.16 \ci{0.05} & \gptFourO & 0.00 \ci{0.01} & 0.04 \ci{0.03} \\
\hline
\end{tabular}}
\caption{\mbf{EQ-ER} performance; \mbf{PLA} = Proof-Level Accuracy, \mbf{SLA} = Step-Level Accuracy.}
\label{Tb:ERProofLevelAcc}
\end{table}
%% \begin{table}[!htbp]
%% \centering
%% \rowcolors{2}{blue!5}{white}
%% \resizebox{0.9\textwidth}{!}{%
%% \begin{tabular}{|l|c|c||l|c|c|}
%% \hline
%% \textbf{Model} & \textbf{PLA} & \textbf{SLA} & \textbf{Model} & \textbf{PLA} & \textbf{SLA} \\
%% \hline
%% \gptFiveTwo & \mbf{0.94} \ci{0.03} & 0.99 \ci{0.02} & \gptFiveOne & 0.92 \ci{0.04} & 0.99 \ci{0.02} \\
%% \hline
%% \geminiTwoPointFive & 0.78 \ci{0.06} & 0.98 \ci{0.02} & \gptFive & 0.77 \ci{0.06} & 0.98 \ci{0.02} \\
%% \hline
%% \gemThree & 0.69 \ci{0.06} & 0.97 \ci{0.03} & \oThree & 0.67 \ci{0.06} & 0.96 \ci{0.03} \\
%% \hline
%% \claudeSonnetExt & 0.62 \ci{0.07} & 0.96 \ci{0.03} & \grokFour & 0.60 \ci{0.07} & 0.94 \ci{0.03} \\
%% \hline
%% \dsReasoner & 0.44 \ci{0.07} & 0.54 \ci{0.07} & \oThreeMini & 0.39 \ci{0.07} & 0.83 \ci{0.05} \\
%% \hline
%% \dsROne & 0.39 \ci{0.07} & 0.81 \ci{0.05} & \dsThreeOne & 0.35 \ci{0.07} & 0.79 \ci{0.06} \\
%% \hline
%% \opus & 0.33 \ci{0.06} & 0.90 \ci{0.04} & \oFourMini & 0.29 \ci{0.06} & 0.80 \ci{0.06} \\
%% \hline
%% \claudeSonnetFour & 0.29 \ci{0.06} & 0.85 \ci{0.05} & \hermes & 0.17 \ci{0.05} & 0.57 \ci{0.07} \\
%% \hline
%% \grokThree & 0.12 \ci{0.05} & 0.74 \ci{0.06} & \dsThree & 0.09 \ci{0.04} & 0.72 \ci{0.06} \\
%% \hline
%% \gptFourOne & 0.07 \ci{0.04} & 0.46 \ci{0.07} & \magistral & 0.02 \ci{0.02} & 0.30 \ci{0.06} \\
%% \hline
%% \nous & 0.01 \ci{0.02} & 0.16 \ci{0.05} & \gptFourO & 0.00 \ci{0.01} & 0.04 \ci{0.03} \\
%% \hline
%% \end{tabular}}
%% \caption{ER performance; \mbf{PLA} = Proof-Level Accuracy, \mbf{SLA} = Step-Level Accuracy.}
%% \label{Tb:ERProofLevelAcc}
%% \end{table}
\renewcommand{\arraystretch}{1.0}

The SLA metric gauges raw ER capability. Top models have basically perfect 
internal mechanisms for solving this problem. PLA, by contrast, measures reliability 
and consistency. Here only two models manage to score higher than 80\%: GPT-5.2 and 
GPT-5.1, with the former getting a nearly perfect score. 
The rest of the models, even elite models like Gemini 3, are not reliable enough 
to get every step right consistently across an entire multi-step proof. 
A single mistake in a 20-step proof is enough to fail the PLA metric. This shows 
that moving from high capability to high reliability in sequential reasoning tasks 
remains a critical frontier for most LLMs. 

PLA's DI (discriminability index) is exceptionally high at 0.36. 

\subsubsection{Error Analysis}
We partition the set of all possible step-level errors into 4 types:
\ben
\item The model's output is ill-formed.
\item The model gives the empty list of supporting equations for a step from $s_i$ to $s_{i+1}$, indicating that the step 
cannot be $\Longrightarrow_{\mathcal{S}}$-justified by any $\mathcal{S} \subseteq \mathcal{E}$, when in fact it can. 
\item The model gives a non-empty list of supporting equations $\mathcal{S}$ for a step when in fact the step cannot be 
$\Longrightarrow_{\mathcal{S}}$-justified by any $\mathcal{S} \subseteq \mathcal{E}$.
\item The model gives a non-empty list of supporting equations $\mathcal{S}$ for a step and the step \emph{is} justifiable, but 
the particular equations given in $\mathcal{S}$ do not justify the step. 
\een
A proof-level error is this: Every element of the list given by the model is properly formatted, but instead of containing 
an element for every step of the proof, the list is cut off prematurely or only contains answers for some of the proof steps, but not all of them. 
We refer to this as a type-5 error. 

\begin{figure}[!htbp]
\centering
%\scalebox{0.5}{\includegraphics[width=\linewidth]{./plots/er_stepwise_error_rates.png}}
%\includegraphics[width=\linewidth]{./plots/er_stepwise_error_rates_smaller.png}
%\includegraphics[width=\linewidth]{./plots/foo.png}
\includegraphics[width=\linewidth]{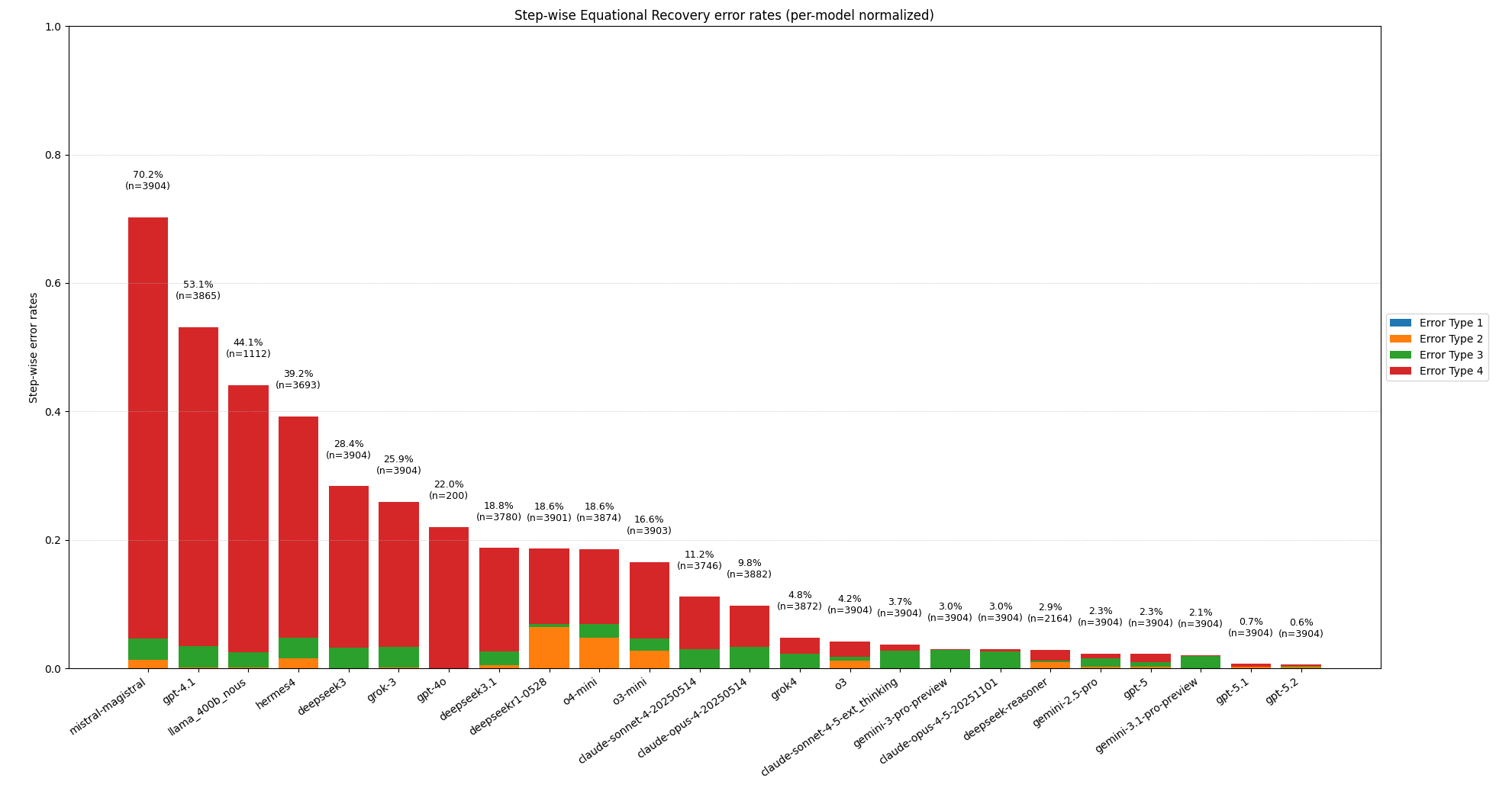}
\caption{Normalized distribution of step-level ER errors.}
\label{Fig:ERStepwiseErrorRates}
\end{figure}

Step-level rates for the 4 possible error types are shown in Figure~\ref{Fig:ERStepwiseErrorRates}. 
Note that the SLAs of Table~\ref{Tb:ERProofLevelAcc} are not equal to 1 minus these error rates. That is because the error rates of Fig.~\ref{Fig:ERStepwiseErrorRates} 
are normalized with respect to the total number of steps that the model has attempted to explain (these are the values \mtt{n=$\,\cdots$} shown above each bar in the figure). 
By contrast, the SLAs of Table~\ref{Tb:ERProofLevelAcc} are 
normalized with respect to the total number of all steps across all proofs, regardless of whether the model has provided a response for a step. 
As an example, consider \gptFourOsp and \nous. These models only provided answers for a small fraction of all steps, particularly \gptFourO, which only handled 
200 out of a total of 3904 steps, a mere 5\%. 
In fact, only 10 models engaged with all possible steps; 
the rest omitted at least a few. The number of steps omitted (the difference between 
3904 and the value of $n$ shown in the figure) is a direct absolute 
measure of a model's type-5 errors. 

No model makes a formatting error. The large majority of ER errors are of type 4. Models correctly identify that the step is valid but they provide the wrong set of justifying equations. 
This is not a failure of pattern recognition in the simple sense. The model is usually ``in the right ballpark'' but lacks precision. It suggests a failure of meticulous verification. 
Models seem to rely on heuristics like ``Axiom $E_5$ looks very similar to the change I see'' and then confidently output $E_5$ without performing the rigorous computational 
step of applying the substitution and ensuring that it produces exactly the target term. 

Type-2 errors (where the model misclassifies the step as unjustifiable by outputting $[]$) are 
%made at considerable rates by mid-tier models 
%like o4-mini, R1, and o3-mini. This is 
a different failure mode that cannot be explained simply by lower reasoning competence,
since there are several models with poor performance but no 
type-2 errors. 
The rate of type-2 errors likely depends on 
whether a model's post-training alignment prioritizes 
helpfulness and engagement versus abstention in the presence of uncertainty. 
Competence still determines how often the model is uncertain, 
but whether the model tries to engage despite the uncertainty is 
largely shaped by its alignment regimen. 

Another finding involves bidirectional rewriting. Our formalization explicitly defines $\Longrightarrow_{\mathcal{S}}$ as left-to-right only, which suffices for this dataset.
%as there is no need for right-to-left rewriting (\iensp, for the symmetric closure of $\Longrightarrow_{\mathcal{S}}$).
Yet several models repeatedly choose sets of equations that can only justify a step if reduction is performed from right to left. 
Strictly speaking, these answers are incorrect because they do not adhere to the prompt instructions, which only allow for left-to-right rewriting. We give full credit to such answers anyway, because they are logically correct, even if not consistent with the prompt. 
While this happens relative rarely,\footnote{o3-mini leads the pack, with bidirectional rewrites in 32 proofs (out of the total 200); followed by GPT-4.1 (17 proofs). Many others use bidirectional rewriting more sporadically.}
it is curious that a model would choose equations that require 
right-to-left rewriting when equally simple equations that work from left to right 
(and are thus fully consistent with the prompt) are available. 
This suggests that many models are following the spirit of equational reasoning (proving two things are the same) while violating the letter
of the specific task instructions (only rewriting from left to right). In other words, their
deeply ingrained semantic understanding of equality 
as a symmetric relation is overriding the specific procedural constraints of the task. 
%% Rather than executing a narrow, syntactic rewriting algorithm, 
%% the models appear to be solving the more general, semantic problem of finding a proof of equivalence. Once their internal search mechanisms identify 
%% a valid logical path, their search terminates without performing the additional compliance check with the directionality constraint.
This underscores how deeply learned priors from pretraining (such as the symmetry of equality) impact a model's operation 
and points to an alignment challenge in getting models to follow instructions that might feel ``unnatural.''

\subsubsection{Equational Gap Filling}

This task is also a subset of \eqonesp with 200 proofs, but each proof has a gap of a randomly chosen size  
inserted at a randomly chosen location. The model is asked to fill that gap with an appropriate sequence of steps. 
The output must be  a dictionary with two main keys, \temv{explanation} and \temv{missingSteps}.
The value of \emph{explanation} is a plain text string describing the model's reasoning strategy. 
The value of \temv{missingSteps} is the main payload: the subproof filling the gap. 
This is expressed as a list of dictionaries representing proof steps, each with 
two keys, \temv{term} and \temv{supportingEquations}. Thus, such a list 
can be viewed as a sequence of pairs $(t_i,N_i)$, $1 \leq i \leq k$  
for some $k \geq 1$, where each $t_i$ is a term (the value of the key \temv{term}) and $N_i$ 
is a list of equation names (the value of the key \temv{supportingEquations}). 
The specified steps correctly fill the gap iff the proof 
\begin{equation}
\begin{aligned}
\rp{s} & = t_1 && \hspace*{-0.08in}\mbox{\kwd{by} $\:N_1$} \\[0.02in]
{} & = t_2 && \hspace*{-0.08in}\mbox{\kwd{by} $\:N_2$}  \\[0.02in]
{} & \vdots && \vdots \\[0.02in]
{} & = t_k  && \hspace*{-0.06in}\mbox{\kwd{by} $\:N_k$} \\[0.025in]
{} & = \rp{t}  && \hspace*{-0.06in}\mbox{\kwd{by} $\:\cdots$} 
\end{aligned}
\label{Eq:GapFillingProof}
\end{equation}

correctly derives the equation $\rp{s} = \rp{t}$, 
where $\rp{s}$ is the term that appears on the step immediately before the gap and $\rp{t}$\mbox{\kwd{ by }}$\cdots$ 
is the step immediately after the gap. 
More precisely, we must have $s' \Longleftrightarrow_{N_1} t_1$; $t_k \Longleftrightarrow_{\cdots} t'$; and 
$t_{i-1} \Longleftrightarrow_{N_{i}} t_{i}$ for every $i = 2,\ldots,k$, where $\Longleftrightarrow_{\mathcal{S}}$ is 
the symmetric closure of $\Longrightarrow_{\mathcal{S}}$. (Thus, unlike ER, this task explicitly allows for 
bidirectional rewriting.) If the model determines that there is no way to fill the gap, it must output the 
empty list as the value of \temv{missingSteps}. We write \mtt{??} to indicate the presence of a gap. 
We explicitly require $\card{N_i} \leq 2$, \iensp, to avoid trivializing the problem, a model may not cite 
more than 2 equations per step. 

For example, given the input: 
\begin{tcAthena}
- Equations:

E1: h(c,V1170,c) = h(V1170,a,f4(f1(c)))
E2: h(V1173,V1174,V1173) = g2(V1173,V1174)
E3: h(a,V1181,f4(f1(c))) = g1(f2(f4(V1181)),f4(e))
E4: f2(f4(V1184)) = g(f3(V1184),f2(e))
E5: g(V1187,f2(e)) = g4(f1(f5(V1187)),d)
E6: f4(V1188) = V1188

- Proof: 

s = g(h(c,a,c),h(a,c,a))                                 
    g(h(a,a,f4(f1(c))),h(a,c,a))          by E1          
    g(h(a,a,f4(f1(c))),g2(a,c))           by E2          
    ??                                    ??
    g(g1(g(f3(a),f2(e)),f4(e)),g2(a,c))   by E4          
    g(g1(g4(f1(f5(f3(a))),d),e),g2(a,c))  by E5, E6      
\end{tcAthena}
a  correct output would be: 
\begin{lstlisting}[language=Python,numbers=none,basicstyle=\scriptsize\ttfamily]
{"missingSteps": [{"term": "g(g1(f2(f4(a)),f4(e)),g2(a,c))", 
                   "supportingEquations": ["E3"]}]}
\end{lstlisting}
since this one step is sufficient to fill the gap. 
As a simple example of a gap that cannot be filled, consider: 
\begin{tcAthena}
- Equations:

E1:  h(V1130,V1131,V1132) = h4(V1130,f(V1132),f2(V1131))
E2:  h(c,c,V1149) = g3(V1149,c)
E3:  g(V1152,V1153) = f2(V1152)
E4:  h4(e,f(V1154),f2(a)) = g3(f2(f1(V1154)),e)
E5:  r(d,c,V1162,V1163) = g(g3(V1163,V1162),f(f4(e)))

- Proof:

s = r(r(d,c,d,b),h(c,c,c),g(d,c),h(e,a,c))                               
    r(r(d,c,d,b),h(c,c,c),g(d,c),h4(e,f(c),f2(a)))        by E1          
    r(r(d,c,d,b),g3(c,c),g(d,c),h4(e,f(c),f2(a)))         by E2          
    ??                                                    ??
    r2(g(g3(b,d),f(f4(e))),g3(c,c),f2(d),g3(f2(f1(c)),e)) by E5        
\end{tcAthena}
Here the correct value for \temv{missingSteps} would be the empty list $[]$, 
%% \setlength{\belowdisplayskip}{0pt}
%% \begin{lstlisting}[language=Python,numbers=none,basicstyle=\scriptsize\ttfamily,belowskip=-0.5em]
%% {"missingSteps": []}
%% \end{lstlisting}
because there are no equations that can introduce an occurrence of $r_2$.
Only a small minority of examples (7 out of 200) have gaps that cannot be filled. 

In addition to \temv{explanation} and \temv{missingSteps}, the top-level JSON output for this task must
contain a third key, \temv{confidence}, whose value is an integer from 1 to 5 representing the model's
confidence in its answer (in the value of \temv{missingSteps}). Here 5 indicates perfect confidence,
1 is zero confidence, 2 is some (low) confidence, 3 is medium confidence, and 4 is strong but not full
confidence.\footnote{In earlier versions we had requested confidence levels to be continuous numbers from 0 to 1,
but in practice models would only output numbers in a small fixed set, such as $\{0.2, 0.4, 0.6, 0.5, 0.8 0.9, 0.95\}$,
so we moved to an inherently discrete formulation.}

\begin{table}[!t]
\centering
\resizebox{0.8\linewidth}{!}{%
\rowcolors{2}{blue!5}{white} % alternate shading, starting with row 2
\begin{tabular}{
  l
  S[table-format=2.2]   % mean
  S[table-format=2.2]   % SD
  S[table-format=2.2]   % median
  S[table-format=2.0]   % min
  S[table-format=2.0]   % max
  l                     % spark
}
\toprule
{Quantity} & {Mean} & {SD} & {Median} & \msp\msp{Min} & \msp{Max}\msp & \multicolumn{1}{c}{Spark} \\
\midrule
Proof length
  & 18.91 & 9.59 & 18.50 & 3 & 35
  & \boxsparkcolored{3}{4}{11}{18.5}{27}{34}{35}{18.91} \\
Gap length
  & 11.72 & 7.07 & 11.5 & 1 & 31
  & \boxsparkcolored{1}{1}{6}{11.5}{17}{23}{31}{11.72} \\
Equations per problem
  & 24.84 & 12.51 & 24.00 & 3 & 53
  & \boxsparkcolored{3}{5.95}{15}{24}{33.25}{46}{53}{24.84} \\
Equations cited per step
  & 1.29 & 0.61 & 1.0 & 1 & 3
   & \boxsparkcolored{1}{0.05}{1.0}{1.0}{2.0}{2.0}{3}{1.29} \\
Individual equation size 
  & 8.59 & 4.02 & 8.00 & 4 & 40
  & \boxsparkcolored{4}{4}{6}{8}{10}{16}{40}{8.590058361843429} \\
\bottomrule
\end{tabular}}
\caption{Data distribution summaries with inline box–sparklines for the gap-filling task slice of \eqone. 
Box = IQR (Interquartile Range); red line = median; dot = mean; whiskers = 5--95th.}
\label{Tb:EqGFDataDistribution}
\end{table}

Table~\ref{Tb:EqGFDataDistribution} shows some elementary statistics about the gap-filling data along 5 dimensions:
the original length of each of the 200 proofs (before the insertion of gaps); the size of each gap (the number of steps removed from the original proof); 
the size of each equational theory (the number of equations in each of the 200 problems); the number of equations
cited per proof step; and the sizes of the equations themselves, viewed as ASTs. As the table shows, 
the average gap is slightly longer than half the size of the proof, so these are quite substantial reasoning holes. 
%% #1 = min, #2 = 5th percentile, #3 = Q1, #4 = median, #5 = Q3, #6 = 95th percentile, #7 = max, #8 = mean.

The main metric we report for this task is accuracy, the ratio \#(correct answers)$\,/\,$\#(all answers). 
When is a gap-filling answer considered correct? If the value of \temv{missingSteps} is the empty list of steps, 
then this is correct iff the gap cannot be filled.\footnote{This is determined 
with the help of an ATP (automated theorem prover).}
If the value of \temv{missingSteps} is a non-empty list of steps specifying a subproof, 
it is considered correct iff the subproof closes the gap correctly
in the sense specified by~(\ref{Eq:GapFillingProof}). We use our own code to decide, for any given $s$, $t$,
and subset $\mathcal{S} \subseteq \mathcal{E}$, whether $s \Longleftrightarrow_{\mathcal{S}} t$.

%% \kanote{For error analysis, LLM errors on gap filling are saved in the files 
%% \[\mtt{code/results/eq/strict\_eq\_gap\_errors.yaml}\]
%% and \mtt{code/results/eq/nonstrict\_eq\_gap\_errors.yaml}.}

%% \begin{table}[h!]
%% \centering
%% \caption{Distribution summaries with inline box–sparklines (per row normalized to that quantity's min–max). Box = IQR; line = median; dot = mean; whiskers = 5--95th.}
%% \begin{tabular}{
%%   l
%%   S[table-format=2.2]   % mean
%%   S[table-format=2.2]   % SD
%%   S[table-format=2.2]   % median
%%   S[table-format=2.0]   % min
%%   S[table-format=2.0]   % max
%%   l                     % spark
%% }
%% \toprule
%% {Quantity} & {Mean} & {SD} & {Median} & {Min} & {Max} & \multicolumn{1}{c}{Spark} \\
%% \midrule
%% Proof length
%%   & 19.91 & 9.59 & 19.50 & 4 & 36
%%   & \boxsparklabeled{4}{5}{12}{19.5}{28}{35}{36}{19.91} \\
%% Gap length
%%   & 2.80 & 3.12 & 1.00 & 1 & 19
%%   & \boxsparklabeled{1}{1}{1}{1}{3.25}{8}{19}{2.795} \\
%% Eqn set size
%%   & 24.85 & 12.51 & 24.00 & 3 & 53
%%   & \boxsparklabeled{3}{5.95}{15}{24}{33.25}{46}{53}{24.845} \\
%% Eqn size
%%   & 8.59 & 4.02 & 8.00 & 4 & 40
%%   & \boxsparklabeled{4}{4}{6}{8}{10}{16}{40}{8.590058361843429} \\
%% \bottomrule
%% \end{tabular}
%% \end{table}

%\clearpage 
Before we present the results, it is interesting to see how many proofs were written by each model (as opposed
to returning an empty list of missing steps, indicating that the gap is unfillable), and how long these proofs were. 
The table below shows the number of proofs written: 
\renewcommand{\arraystretch}{1.2}
\begin{table}[!b]
\centering
\rowcolors{2}{blue!5}{white}
\resizebox{0.8\textwidth}{!}{%
\begin{tabular}{|l|c|l|c|}
\hline
\textbf{Model} & \textbf{Proofs Written} & \textbf{Model} & \textbf{Proofs Written} \\
\hline
\dsThreeOne & 43  & \dsThree & 44 \\
\oFourMini  & 49  & \dsROne & 63 \\
\dsReasoner & 67  & \oThreeMini & 86 \\
\grokFour & 104 & \oThree & 104 \\
\gptFiveOne & 105 & \geminiTwoPointFive & 111 \\
\gptFive & 114 & \claudeOpusFourFiveNov & 129 \\
\claudeSonnetExt & 132 & \opus & 138 \\
\gptFourOne & 140 & \claudeSonnetFour & 146 \\
\gptFiveTwo & 153 & \hermes & 153 \\
\magistral & 154 & \gptFourO & 183 \\ 
\gemThreeOne & 183 & \gemThree & 186  \\ 
\grokThree & 194 & \nous & 199 \\
\hline
\end{tabular}
%% \begin{tabular}{|l|c|l|c|}
%% \hline
%% \textbf{Model} & \textbf{Proofs Written} & \textbf{Model} & \textbf{Proofs Written} \\
%% \hline
%% \dsThreeOne & 43  & \dsThree & 44 \\
%% \oFourMini  & 49  & \dsROne & 63 \\
%% \dsReasoner & 67  & \oThreeMini & 86 \\
%% \grokFour & 104 & \oThree & 104 \\
%% \gptFiveOne & 105 & \geminiTwoPointFive & 111 \\
%% \gptFive & 114 & \claudeSonnetExt & 132 \\
%% \opus & 138 & \gptFourOne & 140 \\
%% \claudeSonnetFour & 146 & \gptFiveTwo & 153 \\
%% \hermes & 153 & \magistral & 154 \\
%% \gptFourO & 183 & \gemThree & 186 \\
%% \grokThree & 194 & \nous & 199 \\
%% \hline
%% \end{tabular}
}
%% \begin{tabular}{|l|c|l|c|}
%% \hline
%% \textbf{Model} & \textbf{Proofs Written} & \textbf{Model} & \textbf{Proofs Written} \\
%% \hline
%% \nous & 199  & \dsThree & 44   \\
%% \gptFive & 114 & \gptFiveOne & 105 \\ 
%% \gptFiveTwo & 153 & \dsReasoner & 67 \\
%% \gemThree & 186 & \claudeSonnetExt & 132 \\
%% \grokThree & 194  & \geminiTwoPointFive & 111  \\
%% \gptFourO & 183  & \grokFour & 104  \\
%% \hermes & 153 & \dsThreeOne & 43 \\ 
%% \magistral & 154  & \oThree & 104  \\
%% \claudeSonnetFour & 146  & \oThreeMini & 86  \\
%% \gptFourOne & 140  & \dsROne & 63  \\
%% \opus & 138  & \oFourMini & 49  \\
%% \hline
%% \end{tabular}}
\caption{Number of nonempty proof completions written for the gap-filling task.}
\label{Tb:NEProofsWritten}
\end{table}
\renewcommand{\arraystretch}{1.0}

\begin{figure}[!htbp]
\centering
%\scalebox{0.5}{\includegraphics[width=\linewidth]{./plots/er_stepwise_error_rates.png}}
%\includegraphics[width=\linewidth]{./plots/er_stepwise_error_rates_smaller.png}
\includegraphics[width=\linewidth]{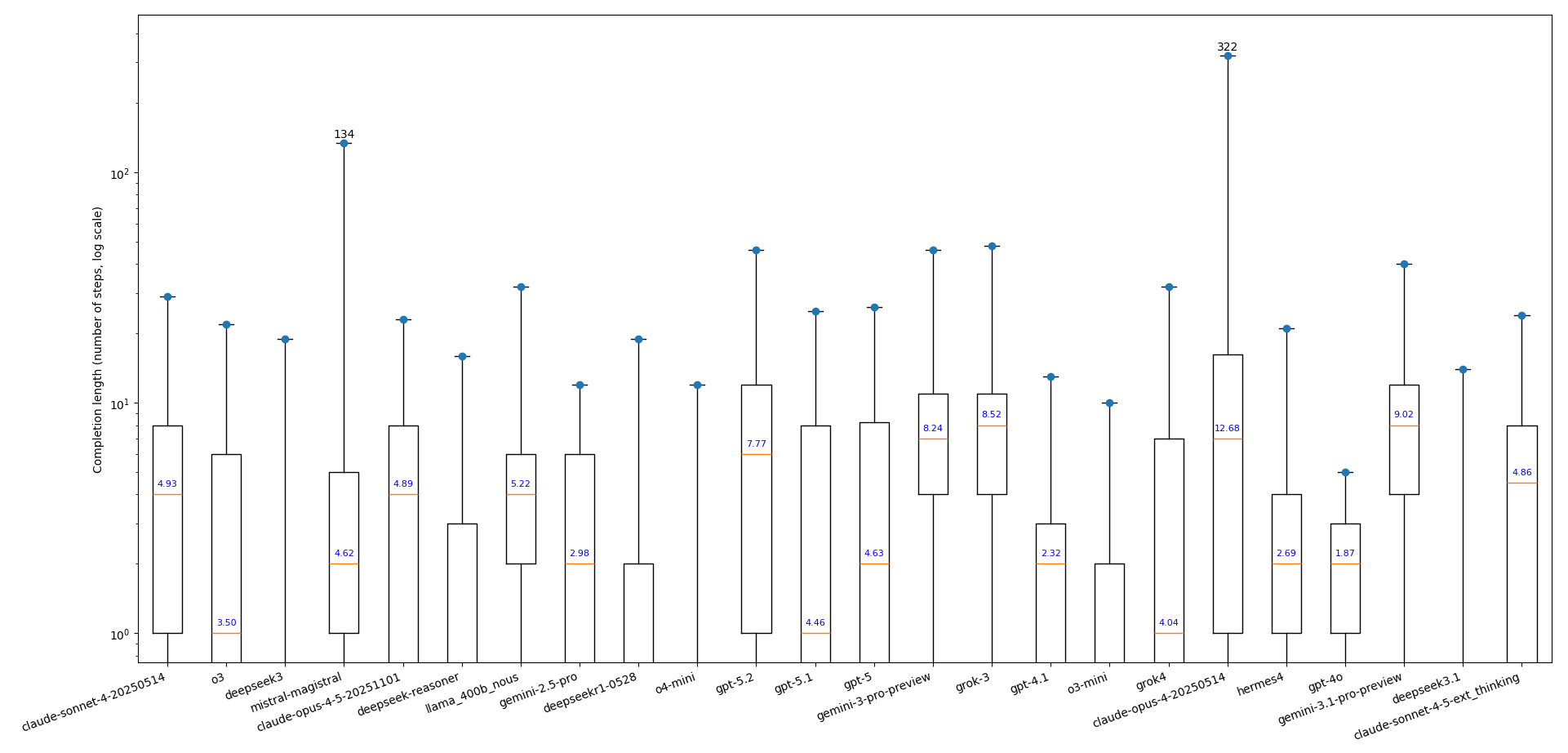}
\caption{Completion length distribution by model (using boxplots and log scale).}
\label{Fig:GFCompLens}
\end{figure}

The length distribution for these proofs
is shown in Fig.~\ref{Fig:GFCompLens}. Recall that the gaps themselves have median
and mean sizes between 11 and 12, with 95\% of all gaps having no more than 23 steps. GPT-5 and Grok-3/Grok-4 are 
the models that come closest to that distribution. Most other models fill the gaps with completions that are either
too short (GPT-4o) or too long (Claude Opus 4.1, which produced a subproof with 322 steps that is not only incorrect 
but also goes in circles, repeatedly reiterating many of the same steps). Claude Opus 4.1 is not the only
model to write proofs with cycles: 64 of the 154 proofs written by Mistral Magistral have cycles, 16/138 for Claude Opus 4.1,
19/194 for Grok-3, 4/140 for GPT-4.1, 4/146 for Claude Sonnet 4, 3/44 for DeepSeek-3, 42/199 for Llama-3.1-405b, 
1/114 for GPT-5 and 1/104 for Grok-4. 
The rest of the models did not have any cycles in their proofs. 
Note, however, that these absolute counts are imperfect measures of duplication given the length distribution in Fig.~\ref{Fig:GFCompLens},
since some models consistently write much shorter proofs and thus have fewer opportunities to 
introduce cycles.\footnote{A more appropriate redundancy measure would the number of non-unique steps 
divided by the total number of steps in the proof.}
%% This is a linear notion, which could be averaged over a set of proofs to give 
%% a dataset-wide redundancy measure for a model. Alternatively, if we care about duplication intensity (how aggressively a proof
%% repeats an inference), we may think of a repetition as a collision between two steps that have the same content.
%% Collisions are pairwise events, so the natural amount of ``opportunity'' for them in a proof of length $n$ is the number of distinct
%% pairs of positions \[\binom{n}{2}=\frac{n(n-1)}{2}.\] If we assume that each pair has a small and fairly stable collision probability $p$,
%% then the expected number of duplicate pairs in a proof of length $n$ would be $p\cdot \binom{n}{2}$. So a quadratic measure of a model's
%% propensity to write proofs with repeated steps would be a micro-averaged pairwise duplicate density metric 
%% along the lines of \[
%% %\frac{\sum_{i=1}^{200} D(i)}{\sum_{i=1}^{200} \binom{n_i}(2}}
%% {\bigsum_{i=1}^{N} \: d_i} \Big/\, {\bigsum_{i=1}^{N} \: \binom{n_i}{2}}
%% \]
%% where $N$ is the total number of proofs written by a model, while $d_i$ and $n_i$ denote the number 
%% of duplicate step pairs in the \nth{i} proof and the length of that proof, respectively. 
%% }

%\Blindtext[2]

%\Blindtext[1]
%\clearpage 

The main results can be found in Table~\ref{Tb:GFAcc}. Note the large performance gap between GPT-5.1 and GPT-5.2,
and the even larger difference between Gemini 2.5 Pro and Gemini 3 Pro Preview. The DI (discriminability index) of these results is 0.26. 
%% In lenient evaluation mode, accuracies increase significantly not because more LLM-generated proof completions are found to be correct, 
%% but because more negative answers ($\temv{missingSteps}\;\smtt{==}\,[]$) are given full credit even though the relevant 
%% gaps are actually fillable. Indeed, if we restrict attention to the half of the task slice that only contains correct 
%% proofs, where the strict/lenient distinction becomes otiose, scores come back down. In particular, there are 96 correct proofs, whose 
%% gaps are therefore fillable by definition. Since none of the 96 proofs have an error, the two evaluation modes coincide. 
%% Performance on this set is shown in Table~\ref{Tb:GFAOC}. It is comparable to performance on the entire slice under 
%% the strict accuracy metric.
\renewcommand{\arraystretch}{1.2}
\begin{table}[!b]
\centering
\rowcolors{2}{blue!5}{white}
\resizebox{0.8\textwidth}{!}{%
\begin{tabular}{|l|c||l|c|}
\hline
\textbf{Model} & \textbf{Accuracy} & \textbf{Model} & \textbf{Accuracy} \\
\hline
\gemThreeOne & \mbf{0.82} \ci{0.05} & \gemThree & 0.62 \ci{0.07} \\
\gptFiveTwo & 0.61 \ci{0.07} & \gptFive & 0.49 \ci{0.07} \\
\gptFiveOne & 0.49 \ci{0.07} & \grokFour & 0.48 \ci{0.07} \\
\claudeOpusFourFiveNov & 0.41 \ci{0.07} & \geminiTwoPointFive & 0.40 \ci{0.07} \\
\oThree & 0.32 \ci{0.06} & \dsReasoner & 0.27 \ci{0.06} \\
\claudeSonnetExt & 0.18 \ci{0.05} & \opus & 0.17 \ci{0.05} \\
\dsROne & 0.15 \ci{0.05} & \oFourMini & 0.15 \ci{0.05} \\
\claudeSonnetFour & 0.07 \ci{0.04} & \dsThreeOne & 0.06 \ci{0.03} \\
\dsThree & 0.06 \ci{0.03} & \oThreeMini & 0.05 \ci{0.03} \\
\magistral & 0.04 \ci{0.03} & \hermes & 0.03 \ci{0.03} \\
\nous & 0.03 \ci{0.03} & \gptFourO & 0.03 \ci{0.03} \\
\grokThree & 0.01 \ci{0.02} & \gptFourOne & 0.01 \ci{0.02} \\
\hline
\end{tabular}
}
\caption{Gap-Filling accuracies.}
\label{Tb:GFAcc}
\end{table}
%% \begin{table}[!bhtp]
%% \centering
%% \rowcolors{2}{blue!5}{white}
%% \resizebox{0.8\textwidth}{!}{%
%% \begin{tabular}{|l|c||l|c|}
%% \hline
%% \textbf{Model} & \textbf{Accuracy} &  \textbf{Model} & \textbf{Accuracy} \\
%% \hline
%% \gemThree & \mbf{0.62} \ci{0.07} & \gptFiveTwo & 0.61 \ci{0.07} \\
%% \gptFive & 0.49 \ci{0.07} & \gptFiveOne & 0.49 \ci{0.07} \\
%% \grokFour & 0.48 \ci{0.07} & \geminiTwoPointFive & 0.40 \ci{0.07} \\
%% \oThree & 0.32 \ci{0.06} & \dsReasoner & 0.27 \ci{0.06} \\
%% \claudeSonnetExt & 0.18 \ci{0.05} & \opus & 0.17 \ci{0.05} \\
%% \dsROne & 0.15 \ci{0.05} & \oFourMini & 0.15 \ci{0.05} \\
%% \claudeSonnetFour & 0.07 \ci{0.04} & \dsThreeOne & 0.06 \ci{0.03} \\
%% \dsThree & 0.06 \ci{0.03} & \oThreeMini & 0.05 \ci{0.03} \\
%% \magistral & 0.04 \ci{0.03} & \hermes & 0.03 \ci{0.03} \\
%% \nous & 0.03 \ci{0.03} & \gptFourO & 0.03 \ci{0.03} \\
%% \grokThree & 0.01 \ci{0.02} & \gptFourOne & 0.01 \ci{0.02} \\
%% \hline
%% \end{tabular}
%% }
%% \caption{Gap-Filling accuracies.}
%% \label{Tb:GFAcc}
%% \end{table}
\renewcommand{\arraystretch}{1.0}

%% \renewcommand{\arraystretch}{1.2}
%% \begin{table}[!htbp]
%% \centering
%% \rowcolors{2}{blue!5}{white}
%% \resizebox{0.6\textwidth}{!}{%
%% \begin{tabular}{|l|c||l|c|}
%% \hline
%% \textbf{Model} & \textbf{AOC} & \textbf{Model} & \textbf{AOC} \\
%% \hline
%% \gptFive & \mbf{0.56} & \claudeSonnetFour & 0.07   \\
%% \grokFour & 0.55  & \dsThree & 0.02   \\
%% \geminiTwoPointFive & 0.43 & \oFourMini & 0.15  \\
%% \oThree & 0.33 & \grokThree & 0.02   \\
%% \oThreeMini & 0.03  & \gptFourO & 0.03   \\
%% \dsROne & 0.17 &   \magistral & 0.01   \\
%% \opus & 0.16 & \nous & 0.01  \\
%%   & & \gptFourO & 0.01   \\
%% \hline
%% \end{tabular}}
%% \caption{Gap-filling accuracies on fillable-only data.}
%% \label{Tb:GFAOC}
%% \end{table}
%% \renewcommand{\arraystretch}{1.0}

%\clearpage

%\begin{tikzpicture}[scale=0.8, transform shape]

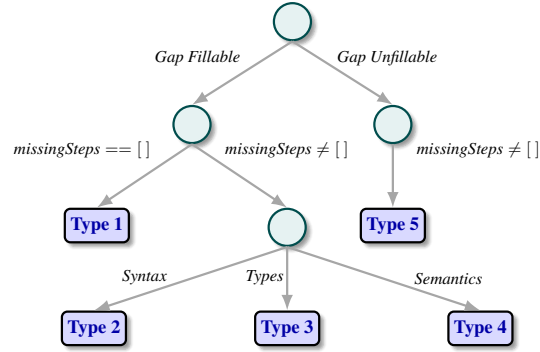
\begin{figure}[!htbp]
\centering
\scalebox{0.7}{
\begin{forest} mytree
  % Root (unlabeled circle)
  [ , decision
    % ---------------- Left branch: Gap Fillable ----------------
    [ , decision,
      edge label={node[condlabel, above, xshift=-24pt, yshift=0pt]{\text{Gap Fillable}}}
      % --> missingSteps == []
      [\textbf{Type 1}, leaf, fill=blue!15, text=blue!65!black,%fill=orange!20!white, text=orange!50!black,
        edge label={node[condlabel, above, pos=.45, xshift=-36pt, yshift=4pt]{$\mbox{\rm\temv{missingSteps}}==[\ ]$}}]
      % --> missingSteps \neq []
      [ , decision,
%%%---->        edge label={node[condlabel, above, pos=.60, xshift=6pt, yshift=4pt]{$\texttt{steps}\neq[\ ]$}}
        edge label={node[condlabel, above, pos=.45, xshift=28pt, yshift=4pt]{$\mbox{\rm\temv{missingSteps}}\neq[\ ]$}}
        % ---- three error kinds ----
        [\textbf{Type 2}, leaf, fill=blue!15, text=blue!65!black,
          edge label={node[condlabel, above, pos=.70, xshift=-4pt, yshift=0pt]{Syntax}}]
        [\textbf{Type 3}, leaf, fill=blue!15, text=blue!65!black, %%fill=purple!15, text=purple!70!black,
          edge label={node[condlabel, above, xshift=-12pt, pos=.70, yshift=0pt]{Types}}]
        [\textbf{Type 4}, leaf, fill=blue!15, text=blue!65!black, % fill=green!18, text=green!50!black,
          edge label={node[condlabel, above, pos=.70, xshift=14pt, yshift=0pt]{Semantics}}]
      ]
    ]
    % ---------------- Right branch: Gap Unfillable ----------------
    [ , decision,
      edge label={node[condlabel, above, xshift=24pt,  yshift=0pt]{\text{Gap Unfillable}}}
      % Only one arc, labeled missingSteps \neq []
      [\textbf{Type 5}, leaf, fill=blue!15, text=blue!65!black, %fill=red!15, text=red!70!black,
        edge label={node[condlabel, above, pos=.45, xshift=46pt, yshift=4pt]{$\mbox{\rm\temv{missingSteps}}\neq[\ ]$}}]
    ]
  ]
\end{forest}
}
\vspace{1em}
\caption{Error typology for the gap-filling task.}
\label{Fig:GFErrors}
\end{figure}

Fig.~\ref{Fig:GFErrors} shows a classification of all possible errors, assuming structurally correct responses by the models. 
There are five main error types: 
\ben
\item {\em Type 1 errors}: the gap is fillable but the model outputs an empty completion, signaling that the gap cannot be filled. 
\item {\em Type 2 errors}: the gap is fillable and the model outputs an non-empty completion, but (at least) one of the terms in the proof has unbalanced parentheses
(a ``syntax error''). 
\item {\em Type 3 errors}: the gap is fillable and the model outputs an non-empty completion, but (at least) one of the terms in the proof has a ``type error,'' meaning
that a function symbol is applied to the wrong number of arguments. 
\item {\em Type 4 errors}: the gap is fillable and the model outputs an non-empty completion, but the proof does not go through because a step is incorrect, in that the 
cited equations do not justify the derivation of the relevant term.
\item {\em Type 5 errors}: the gap is not fillable, but the model outputs a non-empty completion that claims to fill it. 
\een

These can be partitioned into 3 disjoint blocks, $\{1\}$, $\{2, 3, 4\}$, and  $\{5\}$.
Types 2, 3, 4, are not mutually exclusive because a proof can have syntax errors as well as type errors and/or semantic errors. 

\begin{figure}[!b]
\centering
%\scalebox{0.5}{\includegraphics[width=\linewidth]{./plots/er_stepwise_error_rates.png}}
%\includegraphics[width=\linewidth]{./plots/er_stepwise_error_rates_smaller.png}
\scalebox{1.0}{\includegraphics[width=\linewidth]{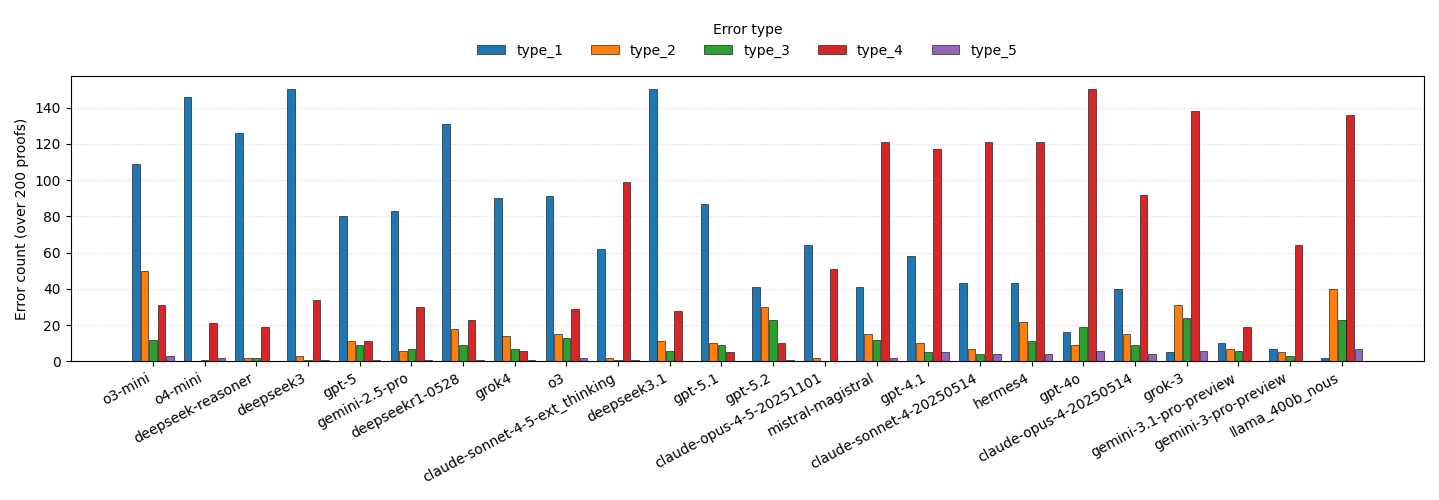}}
\caption{Prevalence of errors of different types in the gap-filling task.}
\label{Fig:GFErrors1}
\end{figure}

Almost all responses by all models were structurally sound (with a sole exception from \nous, one of whose proofs had a step without the required \temv{supportingEquations} key). 
Fig.~\ref{Fig:GFErrors1} shows raw counts for the five error types for all models. 
%\end{tikzpicture}

Some high-level takeaways from these findings: 
\bit
\item The large majority of gaps (193 out of 200) are fillable, so type-5 errors are rare. 

\item As seen in other tasks, when failure-averse models cannot solve a problem, they prefer to fail succinctly rather than generate longer answers,
which means they tend to produce the smallest possible output that adheres to user specifications. DeepSeek models 
lead the pack here. DeepSeek-V3 and DeepSeek-3.1, for example, refuse to fill 156 and 157 out of the 200 gaps, 
respectively, claiming that they cannot be filled. However, this should not be interpreted as the models 
being aware of their limitations.  If the models had such awareness, they would generate 
these answers with low confidence. By contrast, both DeepSeek-V3 and DeepSeek-3.1 (for
example) wrongly claim that more than 150 fillable gaps are unfillable with a very strong 
median confidence of 4. (DeepSeek-V3 does not even once output a ``very low'' confidence of 1.)
In their explanations, these models often admit that they arrived at a negative verdict 
due to the problem's complexity (a favorite phrase is ``Given the time constraints and complexity, 
I concluded that the gap cannot be filled validly'') and yet they claim 
a high confidence in the verdict nevertheless, as if problem complexity by itself was 
evidence of unsolvability. Likewise for R1 and OpenAI's mini models, 
although o4-mini is much more humble in its confidence. 

\item By contrast, the ``compulsively generative'' models with the tall red bars on the right side of the figure 
have few type-1 errors because they tend to plunge ahead and fill
the gaps even if they're not in a position to do so successfully. As a result, the vast majority of errors for these models are of 
type 4, genuinely failing proof steps. 

\item Stronger models also tend to provide better-calibrated confidence levels. 
Here is an explanation provided by Gemini-2.5-Pro: {\em The transformations required, 
particularly for the first and fourth arguments of the main term, are exceptionally complex. Despite significant effort, I could not construct a complete 
and verifiable path from start\_term to t\_k. The search space of possible rewrite sequences is vast due to the large number of equations. 
Given the difficulty in finding a valid path and the high chance of error in such a complex derivation, I am unable to provide a sequence of steps 
with high confidence. It is possible that a key insight or a simpler path was missed, but it is also possible the gap is invalid as stated. 
Therefore, I will indicate that the gap is invalid by providing an empty list, but with low confidence, reflecting the possibility that a 
solution exists but is beyond my current analysis capabilities.} That said, our ECE analysis below shows that \emph{all} 
models are consistently over-confident, and Gemini-2.5-Pro is the worst-calibrated of the elite models.

\item Syntax and type errors (types 2 and 3, respectively) are less common, but they do occur in non-negligible 
numbers across all models, including the very best. Their presence indicates that the models' internal representation 
of these symbolic structures is not quite robust. They  can still lose a parenthesis or misapply a function. 
The high rate of syntax errors by o3-mini is particularly noteworthy---it makes such errors in 
a remarkable 50 proof attempts, or 25\% of all inputs. 

%% \item The top 4 models (GPT-5, Gemini-2.5-Pro, Grok-4, and o3) make relatively few errors of types 2--4, showing that the proofs they write
%% tend to be robust. This highlights a dramatic gap between local and global reasoning competence in different groups of models. Models in the ``compulsive'' group  
%% operate on a greedy, local basis. At each step, they pick an axiom that looks promising and apply it without a clear long-term plan. 
%% This often leads them down dead ends or into loops. By contrast, the success of the elite models in filling some nontrivial gaps suggests a 
%% capacity for global planning. These models are not just taking random walks; they are often constructing coherent paths towards the goal. 

\item Regarding instruction following, recall that the prompt explicitly requires all steps to cite no more than two equations. While virtually all models adhere to that constraint
at the \nth{95} percentile (the only exception is DeepSeek-V3, which 
cites 2.3 equations per step even at the \nth{90} percentile), only a minority of the models (10 out of 24) achieved perfect adherence
to the constraint, without ever citing more than 2 equations: GPT-5.2, GPT-5.1, GPT-5, Gemini-3.1, Claude Opus 4.5, Grok-3, o4-mini, o3-mini, DeepSeek-3.1, and DeepSeek-Reasoner. For all other models, the maximum number of cited equations on a given step ranges from 3 
(for Grok-4) to 10 (for GPT-4.1). 

\item The best models clearly show an ability to compose nontrivial reasoning chains. 
For a problem with 46 equational axioms and a gap size of 23, Gemini 3 Pro Preview 
managed to fill the gap with a subproof of 35 steps, each one citing a single equation. 
However, this is was the longest correct proof written by Gemini 3 and an outlier. 
As we will see shortly, performance drops sharply across the board as the gap 
size increases. 

\eit

%\clearpage

\begin{figure}[!t]
\centering
%\scalebox{0.5}{\includegraphics[width=\linewidth]{./plots/er_stepwise_error_rates.png}}
%\includegraphics[width=\linewidth]{./plots/er_stepwise_error_rates_smaller.png}
\scalebox{0.8}{\includegraphics[width=\linewidth]{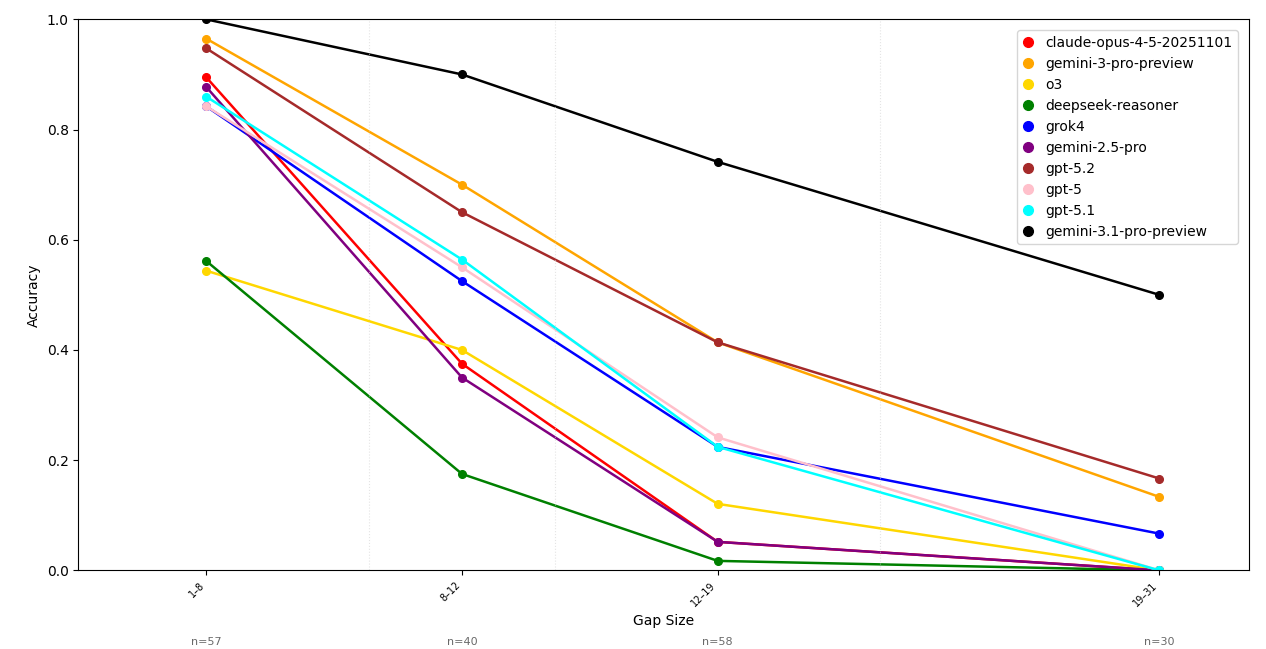}}
\caption{Gap-filling accuracy for the top 10 models as a function of gap size.}
\label{Fig:GFAccVsGapSize}
\end{figure}

\begin{figure}[!b]
\centering
%\scalebox{0.5}{\includegraphics[width=\linewidth]{./plots/er_stepwise_error_rates.png}}
%\includegraphics[width=\linewidth]{./plots/er_stepwise_error_rates_smaller.png}
\scalebox{0.8}{\includegraphics[width=\linewidth]{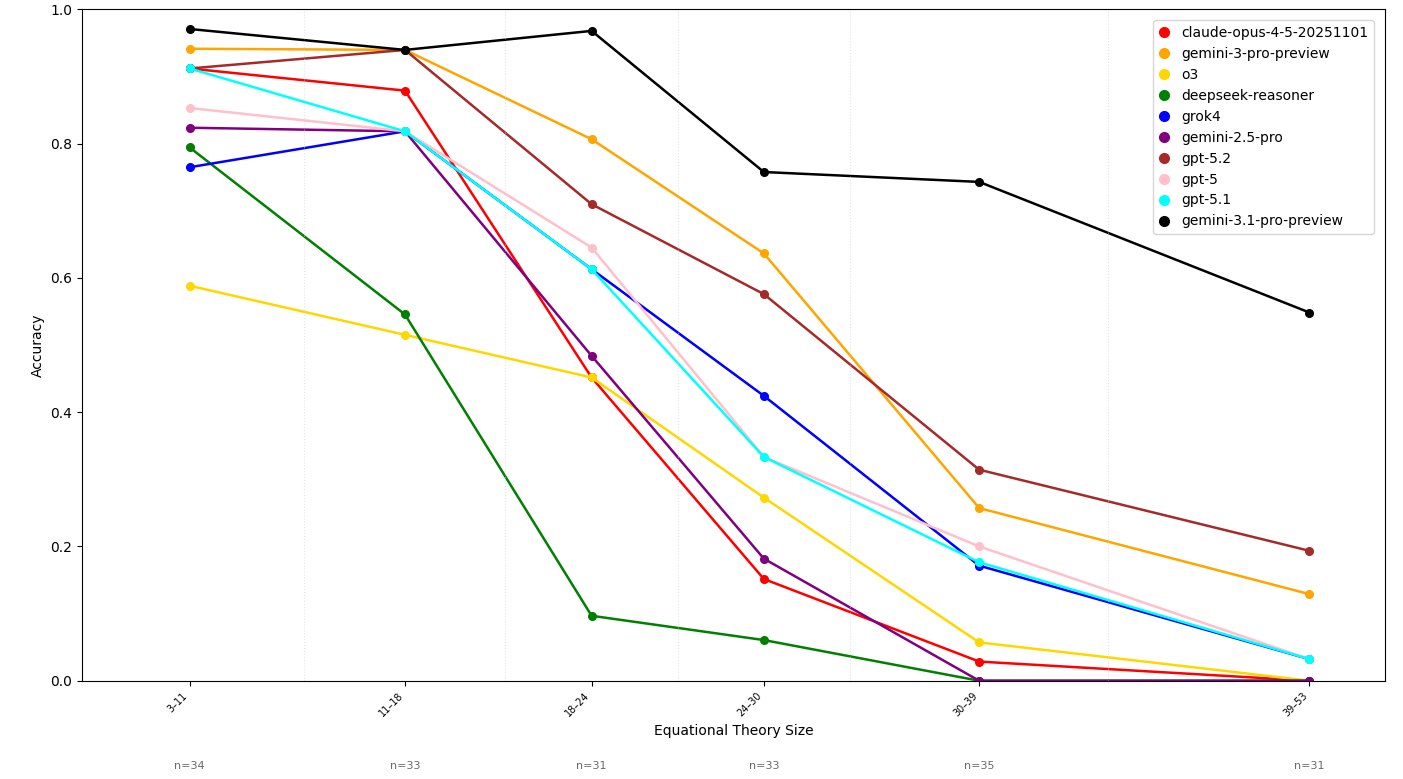}}
\caption{Gap-filling accuracy for the top 10 models as a function of equational theory size.}
\label{Fig:GFAccVsEqTheorySize}
\end{figure}

We now examine how accuracy depends on gap size as well as the size of the equational theory for the top 10 models.
These two relationships are depicted in Fig.~\ref{Fig:GFAccVsGapSize} and 
Fig.~\ref{Fig:GFAccVsEqTheorySize}, respectively. 
The plots show that as the problems become more challenging, performance drops drastically 
even for the strongest models.

In a sense, gap size captures the depth of the required search while theory size captures its breadth. 
The size of the gap is a measure of the number of needed reasoning steps, which taxes the model's ability to maintain a goal, 
avoid getting lost, and chain inferences correctly. By contrast, the number of equational axioms taxes the model's ability to 
select the {\em correct\/} tool at each step from a large set of distractors. It's a measure of its ability to handle a large context 
of rules.

\begin{figure}[!b]
\centering
%\scalebox{0.5}{\includegraphics[width=\linewidth]{./plots/er_stepwise_error_rates.png}}
%\includegraphics[width=\linewidth]{./plots/er_stepwise_error_rates_smaller.png}
\scalebox{0.7}{\includegraphics[width=\linewidth]{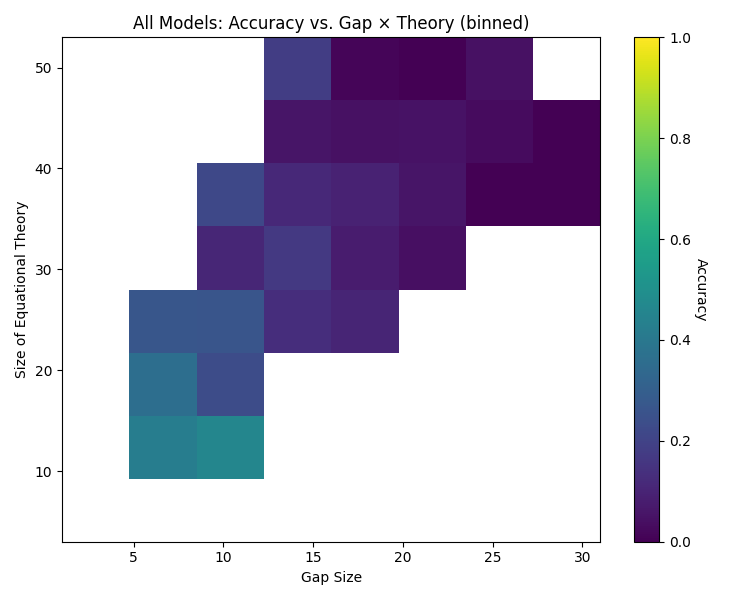}}
\caption{Consolidated accuracy heat map for gap filling, as a function of gap length and equational theory size.}
\label{Fig:GFHeatMap}
\end{figure}

The heat map in Fig.~\ref{Fig:GFHeatMap} shows how model accuracy (across all models) depends jointly on both parameters: gap size on the $x$-axis 
and equational theory size on the $y$-axis. 
Each cell's color represents the fraction of correct proofs among all attempts falling into that combination of gap size and theory size, 
with lighter colors indicating higher accuracy. The pattern reveals a clear gradient: accuracy 
is highest in the bottom-left corner, where gaps are small and theories are simple, and drops sharply as either dimension increases. 
Moving to the right, accuracy decreases with larger gaps; moving upward, it decreases with larger theory sizes, even as gap sizes stay small. 
In the top-right region, where both the gaps and the theories are large, accuracy is near zero, showing that these two sources of difficulty 
combine to severely degrade model performance. Overall, the heat map highlights that while both factors independently hurt success rates, 
their effects compound. Models that can handle a relatively large gap with a small theory, or conversely, break down almost 
completely when both are large. This shows that current models lack the robust search and planning abilities 
needed to navigate complex combinatorial search spaces effectively.

\begin{figure}[!b]
\centering
\scalebox{1.0}{\includegraphics[width=\linewidth]{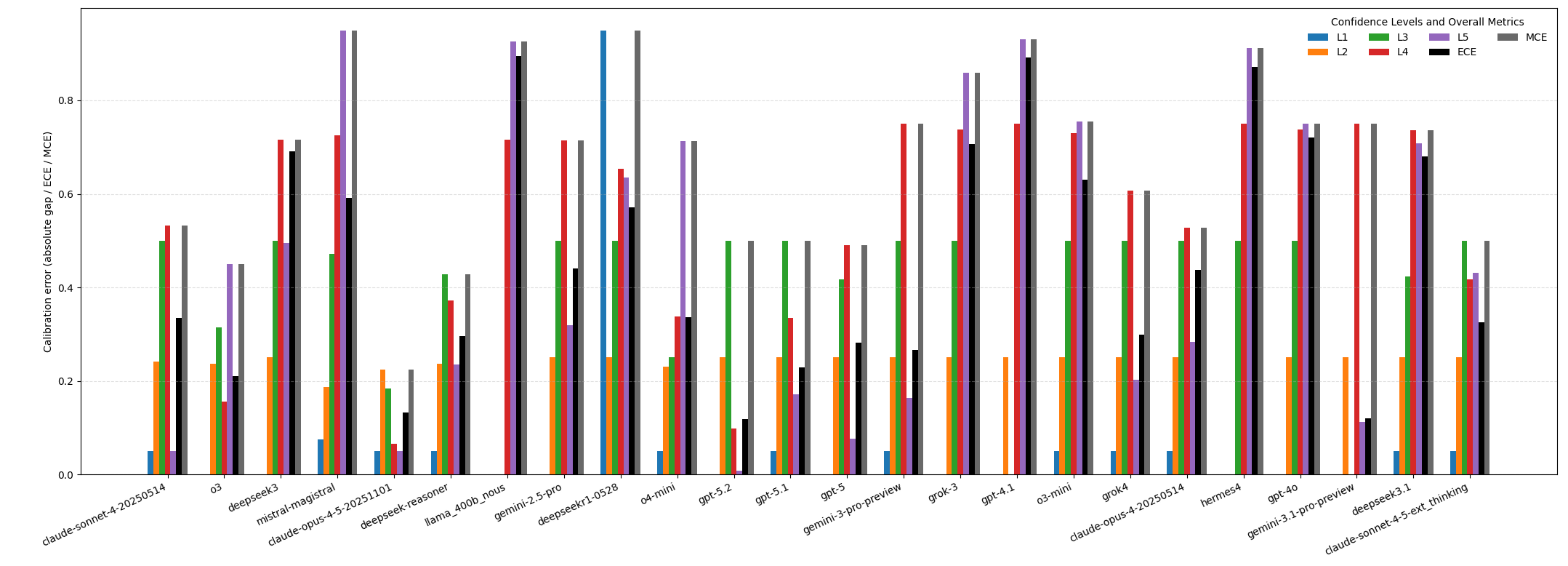}}
\caption{Gap-filling calibration errors across all models (bin-level and overall ECE).}
\label{Fig:GFECE}
\end{figure}

The plot in Fig.~\ref{Fig:GFECE} compares the calibration of the models on the same dataset of 200 proofs. 
For each model, 6 thin bars summarize calibration performance. The first five bars (L1–L5) correspond to absolute 
calibration gaps in each of the model's discrete confidence levels. Recall that we allow for 5 possible confidence levels, 
1 (very low confidence) through 5 (near certainty). We map these levels to probabilities as follows:
\begin{equation}
 \{1 \mapsto 0.05, 2 \mapsto 0.25, 3 \mapsto 0.5, 4 \mapsto 0.75, 5 \mapsto 0.95\}.
\label{Eq:ConfidenceLevelMapping}
\end{equation}
The gap for a given bin is the absolute value of the difference between the empirical accuracy observed in that bin 
and that predicted by the probability assigned to that bin. For instance, the probability assigned to bin 2 (for a confidence
level of 2) is 0.25, so if a model were perfectly calibrated we would expect its outputs to be correct 25\% of the time. If they 
are only accurate 10\% of the time, we have an absolute gap of 15\%---a significant marker of overconfidence. 
In the plot, the height of each bar shows how far the model's confidence in that bin differs from its 
actual empirical accuracy: 
a higher bar means worse calibration. The final, black bar shows the overall Expected Calibration Error (ECE), a weighted average across bins,
with the following standard definition:
\begin{equation*}
\temv{ECE} = \bigsum_{m=1}^{M} \cfrac{\left|B_m\right|}{n} \cdot \left|\temv{acc}(B_m) - \temv{conf}(B_m)\right|,
\end{equation*}
where $M$ is the number of bins (5 in our case); $B_m$ is the set of predictions that fall into bin $m$; $n$ is the
total number of predictions (200 in our case); $\temv{acc}(B_m)$ is the fraction of correct predictions in $B_m$;
and $\temv{conf}(B_m)$ is the average confidence score of the predictions in the bin.\footnote{This is the general definition
of ECE. In our setting, because we don't have continuously varying probabilities,  $\temv{conf}(B_m)$ is fixed for each bin by the mapping~(\ref{Eq:ConfidenceLevelMapping}).} \emph{MCE} is defined as the maximum calibration error across all bins:
\[
\temv{MCE}=\max_{m\in\{1,\ldots,M\}}\left|\temv{acc}(B_m)-\temv{conf}(B_m)\right|.
\]
These are the two canonical calibration metrics that are most commonly reported. 

Looking across all models, none of them are well-calibrated. Only GPT-5.2, Gemini-3.1, 
and Claude Opus 4.5 (with average errors of 0.12, 0.12, and 0.13) 
land in a range that most calibration researchers would consider 
respectable. While not ideal, it is small enough that the models'  
confidence estimates could plausibly carry some actionable signal. 
All other ECE scores are above 0.2, which is generally considered high miscalibration. 
Everything above 0.4 (this includes Gemini 2.5 Pro, o3-mini, and DeepSeek-R1-0528) is 
basically noise, while some models (e.g., \magistral, \nous, and \grokThree) are 
essentially completely uncalibrated, with extremely large per-bin errors ranging
from 0.7 to 0.9. The stated confidence of these models is usually off by 70-90
percentage points, which is not mere miscalibration but complete epistemic collapse. 
The worst bins across 
almost all models are L4 and L5, which correspond to predictions made with high or very high confidence. 
This is classic overconfidence: models severely overestimate their reliability when they are confident. 
In other words, the models are most wrong precisely when they think they're most right, 
which is the opposite of what is needed in mathematical reasoning. 
If confidence levels cannot be trusted specifically when they are high, 
they are of little value as guardrails, \egnsp, thresholding on ``confidence $\geq$ 4'' 
would be unsafe for all of these models.\footnote{To put it in terms of odds, an ECE error of 0.25 in bin 5 means
that when the model says \emph{I'm $\approx$ 95\% sure I'm right}, it's actually right only
about 70\% of the time. Taking the model at its word implies betting odds of 19:1, i.e.,
you expect to win (the model to be right) about 19 times out of 20. But the real 
odds are 7:3, so you'd win only about 7 times out of 10. 
If you are willing to risk \$19 to win \$1 because you believe the 19:1 odds 
(you take the model at its word), then the expected value of your betting strategy per decision
would be $0.7 \cdot 1 + 0.3 \cdot (-19) = -5$, \iensp, on average you would lose \$5 per decision.}
The exception is  GPT-5.2, which has an exceptionally small error in L5, 0.008. Claude 4.5 Opus 
also has a small calibration error in L5, 0.05. 
This suggests that when these two models assign near-maximum confidence, their predictions 
are very likely correct. Gemini-3.1 has a more sizable L5 error, 0.12. 
Note, however, that both GPT-5.2 and Gemini 3.1 still exhibit severe miscalibration in other bins, 
with very high \temv{MCE} values (0.5 and 0.75, respectively), leaving significant room for improvement. 
At 0.22, the MCE of Opus 4.5 is lower but still large. 

%The dramatic ECE gap between GPT-5.2, Gemini 3.1,  and all other models
These results are interesting for a couple of reasons. First, they show that miscalibration is
not inevitable; it can be dramatically reduced. 
Second, combined with the accuracy results, they show that calibration is not a corollary 
of correctness. While accuracy and calibration do tend to increase together, the correlation
is far from perfect. For example, Gemini 3 is the second best-performing model but it is miscalibrated, 
featuring a significant ECE of 0.27---more than twice that of GPT-5.2. 
Conversely, Claude Opus 4.5 is much better calibrated but its accuracy is only 41\%. 
Therefore, calibration is not an emergent property of scale and reasoning competence alone. 
It very likely requires explicit uncertainty supervision and/or architectural or training changes that link 
reasoning depth to confidence. The three best calibrated models almost certainly had an additional 
training signal or mechanism binding confidence to correctness. That could have been
explicit calibration loss, confidence supervision, rejection-aware training, or tighter 
coupling between reasoning depth and confidence output.

\section{Conclusions}

\label{Sec:Conclusions}

This paper introduced \pgrid, a new benchmark designed to assess the reasoning abilities 
of LLMs through tasks that demand rigorous and machine-checkable proof construction and analysis
under minimal formalism. 
Unlike benchmarks that focus solely on answer accuracy, \pgridsp emphasizes
the integrity of a model's inferences and offers a more precise and fine-grained assessment 
of reasoning competence. 

A core theme emerging from our results is one of extraordinary progress in a short time. 
We started this work in late 2024. At that point, the state of the art was GPT-4 and Claude 
Sonnet 3.5, which were essentially incapable of sustained reasoning. We were concerned at the time 
that what are now the easiest tasks in the benchmark, \mbf{PL1-PW} and \mbf{PL1-PC}, 
might be too challenging; performance was well below ceiling and overrun with 
elementary errors. In less than a year, these tasks had 
become easy for the best models and accessible even to mid-tier models. 
Our current results provide many examples of superb fluid reasoning 
on highly challenging abstract problems.
The debate as to whether these systems can reason at all should
be considered settled at this point: they can. 
The evidence is decisive. \pgridsp should provide 
a robust platform for tracking and accelerating these
developments, underscoring the importance of proof-based evaluation as
models move toward deeper integration into scientific and mathematical
workflows.

%% Rapid progress has since pushed top-model performance close 
%% the ceiling on those ``easy'' \pgridsp tasks, which now serve 
%% as foundational or calibration benchmarks rather than stress tests, 
%% as they are approaching saturation for the strongest current models. 
%% to saturation on \plone, reflecting substantial improvements in both formal 
%% judgment and procedural control. At this stage, \plonesp  should be understood 
%% not as a frontier challenge but as a minimal competence test suite for proof-based 
%% reasoning in a formally specified setting. Importantly, however, \plonesp remains 
%% highly informative: it continues to sharply discriminate models across a wide performance 
%% spectrum, exposes systematic failure modes in both generation and auditing, and 
%% provides a diagnostic lens on the interaction between abstract reasoning and formal 
%% execution. In this sense, \plonesp plays a role analogous to foundational benchmarks 
%% in other domains: no longer maximally difficult, but essential for establishing 
%% whether a system has achieved a baseline level of reliable, formally grounded reasoning.
%% Doing well on \plonesp tasks does not demonstrate deep or general reasoning competence, 
%% but failing calls into question a model's ability to reliably reason within even the 
%% simplest deductive systems.

Nevertheless, major challenges remain. Even the best models continue to 
make rudimentary reasoning errors (albeit at a much lower rate than before), 
and all models quickly run up against hard limitations as the difficulty 
of the reasoning increases, especially on structurally rich problems drawn from 
classical families of hard problems. More concerningly, we see high levels of 
what we call \emph{epistemic instability}: the tendency of models to ``forget'' 
what they seem to know well when they are working on a hard problem, 
and to frequently contradict themselves. 
We formalized this phenomenon by introducing a new quantity called ESI 
(epistemic stability index) and we presented the first set of empirical results
on that front. 

In particular, while our results strongly suggest that models possess at least 
partial reasoning competence, they still fall short on procedural grounding and 
cognitive control, which often prevents them from \emph{rigorously} executing 
reasoning tasks without external scaffolding. By and large, they reason in a semantic 
space, not a formal one. They often identify which conclusions follow from which assumptions and 
construct globally coherent proof strategies. However, they fail to reliably 
execute these strategies within a formally constrained system, even a very simple one. 
They struggle to suppress invalid but tempting operations, to adhere 
strictly to syntactic constraints, and to treat inference rules as deterministic 
operators rather than semantic heuristics. In that sense, current models exhibit 
semantic reasoning without procedural grounding. 
%Whether this should count as ``reasoning'' depends on one's definition, but it clearly differs from the kind of rule-governed reasoning exhibited by both humans and symbolic systems.

The issue can perhaps be illuminated through the lens of the manifold hypothesis \citep{bengio2013representation}. 
Neural models perform well when the data distribution lies on a low-dimensional manifold embedded 
in a high-dimensional representation space, as is the case for images or natural language. 
In such domains, nearby points in representation space typically correspond to semantically 
similar objects, and interpolation is meaningful. Formal logical reasoning appears to violate this assumption. 
The space of formal proofs is extremely sparse, and small perturbations to a formula or an inference 
rule often result in radical and disruptive semantic changes rather than smooth variation. Unlike natural 
language, where many paraphrases express essentially the same meaning, 
formal proofs do not allow for much redundancy: a single misplaced connective,
rule application, or even a stray parenthesis can invalidate the entire derivation.

From this perspective, the behavior we observe is unsurprising. Models trained to operate 
in smooth semantic spaces attempt to solve proof tasks by interpolating within a semantic 
manifold that, in this domain, does not meaningfully exist. This results in ``semantic'' 
reasoning that often captures the intended structure of an argument but fails to satisfy 
the discrete, brittle constraints imposed by formal proof systems.
In this sense, current models exhibit reasoning that is well-adapted to continuous 
semantic domains but poorly matched to the sharply discontinuous geometry of rigorous, 
formal proofs. Their failures may therefore reflect a deeper mismatch between 
the inductive biases of deep learning and the structure of symbolic reasoning.
This is not to say that formal reasoning is inherently non-manifold or that deep learning cannot
in principle approximate it well enough; only that, at least in their current training 
and architectural incarnations, the continued symbolic brittleness of models 
may be structurally related to the apparent disconnect between their inductive
bias and the geometry of formal reasoning spaces. 

Of course, the rapid improvement of frontier systems suggests that this 
geometric tension is not insurmountable. But it is worth noting that contemporary 
high-performing models are no longer simple next-token predictors operating in isolation; 
many incorporate inference-time search, reranking, verifier guidance, or other forms of structured deliberation, 
and often distill the results of such search back into their parameters. 
These mechanisms effectively introduce discrete branching, pruning, and constraint 
enforcement steps that differ qualitatively from token-by-token generation
and partially compensate for the smooth, interpolation-driven 
inductive biases of the underlying neural representations. In this sense, the 
gains we observe may not so much reflect  the disappearance of symbolic brittleness, 
but the increasing use of architectural guardrails that approximate modular, 
search-based reasoning within an otherwise continuous semantic substrate.

A related recurrent pattern has been failures of cognitive isolation or 
computational modularity. To put it simplistically,
with transformer-based architectures it is not the case that ``if a model can reliably solve task $S$
and  task $T$  is composed of many $S$ subtasks, the model can also reliably solve $T$."
Reliable problem solving needs reliable context-switching into a cognitively isolated ``system-2 mode'' that can load
a particular subproblem's details into working memory while shutting out everything else as irrelevant noise. 
This is what both humans and conventional symbolic computers do: they use function calling
to encapsulate complexity. When a function is called, its internal logic is 
independent of and isolated from the global state of the program, which can be arbitrarily
complicated without affecting the execution of the function. 
In a  transformer model, by contrast, every token attends to every other token in 
the entire context window. 
In checking \plthreesp proofs, for example, the embeddings for the tokens 
in step 42 are mathematically influenced by the attention patterns from all 
previous steps and the system prompt. So the model's mini-SAT circuits that 
are used to solve the local entailment problems in such a proof are not running 
in isolation; they are running in a big noisy room that subjects the output
to interference from irrelevant sources. Our results underscore that language models 
do not draw upon modular internal reasoning capabilities; 
if they did, their performance would be much more epistemically stable and invariant to context.
Instead, they perform a  context-dependent simulation of reasoning,
essentially approximating system-2 behavior using mechanisms designed for system-1 tasks.
That simulation degrades when the model's architectural and representational resources (such as effective 
attention bandwidth and feature superposition limits, respectively) are stretched thin
by maintaining the global reasoning state. 

%% (Essentially, the previous bullet point (invoking the manifold hypothesis)
%% explains why ``semantic'' or abstract reasoning sort of works, whereas this 
%% bullet point (cognitive isolation/modularity) explains why formal reasoning or 
%% procedural control fails.) 

These findings suggest that proof-based evaluation is not merely a stricter grading scheme, 
but a lens into the structure of current reasoning systems. The gap between semantic fluency 
and formal control reveals where architectural biases still dominate. In that sense, \pgridsp is less 
a leaderboard and more a diagnostic instrument for the evolving nature of machine reasoning.

We close with a brief \mbf{FAQ} that we hope will help to clarify and reinforce some of this paper's key points. 
\bit
\item \mbf{Q1}: Why use formal notation? 
Forcing models to express their reasoning in a formal language like NDL 
introduces confounds, such as artificial syntax and formatting errors. \\[0.03in]
\mbf{A}: From first principles, there are only two main options for a reasoning benchmark: 
models can express their reasoning either in a formal language or in natural language. 

For testing purposes, the core foundational problem with 
%having models express their reasoning in 
natural language is that it doesn't scale. 
%because it makes it impossible to \emph{mechanically} test for correctness. 
%This is simply because 
There is no algorithm for determining the 
correctness of an arbitrary stretch of text that purports 
to reason its way to a conclusion. 
As a result, natural-language reasoning cannot be reliably and 
\emph{automatically} evaluated, and must instead be manually inspected. 
%The only reliable way to do that is to read and analyze the text manually. And 
This difficulty explains why most existing ``reasoning benchmarks'' do not 
actually evaluate the reasoning itself (ignoring, for example, COT traces 
during scoring), because there is no procedure that can take an arbitrary 
natural-language reasoning trace and reliably output \emph{correct} or \emph{incorrect}.
%% This difficulty is why virtually every ``reasoning benchmark'' so far does not actually evaluate the reasoning of the tested models, as expressed in their COT traces for example; because there is no algorithm that can take an arbitrary COT reasoning trace as input and reliably output \emph{correct} or \emph{incorrect}. 
This limitation alone is reason enough for AI benchmarks to move away from natural language 
and toward formal languages, which make reasoning amenable to computational analysis and evaluation. 

Moreover, natural language doesn't remove confounds; it simply 
obscures them. In plain English, we can only determine whether a reasoning trace looks plausible. 
Assumption scope is a simple case in point. In a formal language, this is explicitly
indicated with opening-closing braces (or \mbf{begin}-\mbf{end} pairs, Python-style indentation, etc.),
leaving zero doubt about what hypotheses are active and when. In English, this has to be
guessed from the text by taking into account a wide range of signals, 
including formatting details  like the way in which the text is broken into  paragraphs. 
Experienced authors write in a way that minimizes guesswork, 
but ambiguity often remains: ``Suppose that $p$ holds. Then $\cdots$. Next, consider $q$. We have $\cdots$''
Is $p$ still active inside the second ellipses? What if there is a new paragraph start 
right before ``Next, consider $q$.''? 

It might be argued that natural-language reasoning \emph{can} be mechanically evaluated, using ``LLMs as judges''
\citep{LLMsAsJudges}. But there is an air of circularity in that approach. Who has judged the judges, and 
where can we look for independent ground truth? Using a system of unknown or dubious 
reasoning reliability to judge reasoning correctness seems questionable. 
An NDL proof can be independently evaluated by a proof checker that has a perfectly precise specification,
is written in a few pages of code that can be audited or even formally verified, and does not change
across runs, resulting in an exceptionally small and stable TCB (trusted computing base). An LLM judge, by 
contrast, has billions of parameters, cannot be easily audited, 
has no formal semantics or specification, and its behavior may depend on temperature, prompt, 
and version changes, among other factors. 

Finally, the use of a formal language like NDL provides much higher 
analytical resolution, allowing us to \emph{measure} cognitive control. 
We are able to precisely distinguish parsing errors from 
malformed rule applications, wrong conclusions, purely logical errors,
 unbound identifiers, and so on.  Such fine-grained analysis is simply 
not possible in natural language,  where most issues collapse 
into ``the model's answer seems wrong.'' 
Formal languages, in other words, make deviations from correct reasoning explicit 
and mechanically detectable, whereas natural-language reasoning often obscures 
such failures behind linguistic ambiguity. Models can ``sound right'' while being wrong, 
and logical gaps can easily be hidden in rhetorically compelling prose. 
In this sense, the errors we observe 
when models use NDL are not confounds but diagnostic signals, revealing limitations 
in models' ability to reliably execute rigorous reasoning. A natural-language formulation 
would avoid some of these surface errors, but at the cost of making correctness 
judgments substantially less precise and more subjective, and error analysis much 
more brittle and qualitative.
%% \footnote{To be clear, we are not advocating for wider adoption
%% of formal languages in mathematics in general. While some of these arguments do transfer 
%% over to wider and more general settings, here we are focused on reasoning bechmarks of AI models,
%% not on the much more general (and older) question of whether formal methods should be more
%% widely used in mathematics.}

\item \mbf{Q2}: There is a third, middle-ground alternative: controlled natural language (CNL). 
Why not require models to express their reasoning in a controlled fragment of English that can then 
be formally translated into logic and computationally analyzed? \\[0.03in]
\mbf{A}: CNL sounds like the best of both worlds (natural and formal languages), but in 
this setting it would cause greater difficulties. Crucially, CNL here would not be 
given as \emph{input} to the models (as is widely done, for example, in benchmarks like 
RuleTaker \citep{RuleTaker}, see Section~\ref{Sec:RelatedWork}), but would instead be required 
as the \emph{output} of the models. This immediately presents two problems. 

First, the CNL must be formally defined and its entire spec must be given to the model, so that 
the model knows exactly what it is allowed to say and what sort of sentences are off limits. 
The definition would need to fully specify allowable sentence structures, anaphoras, and tense and voice; 
the interpretation of conjunctions; the expression of implications; scope rules; the introduction and discharge
of assumptions; what  counts as a valid reference; how to disambiguate ambiguous constructions; and so on. 
This would be orders of magnitude harder than specifying NDL. 
The issue is that, for all practical purposes, any CNL is itself a formal language. It must commit to 
a grammar, a vocabulary, scoping and disambiguation rules, an interpretation function, and so on. 
The main difference is that its syntax is less precise, its semantics are harder to specify,
and its failure modes are less transparent. By contrast, a language like NDL has a tiny grammar (it can be written 
in BNF with a handful of productions) and its operational semantics 
can be fully described in a page or two of text. 

Second, CNL would not eliminate issues of procedural control. 
There are no guarantees that models would respect the formal spec
and reliably remain within the bounds of the prescribed fragment, 
just as they may fail to respect the rules of NDL. 
Models would likely be ``almost compliant'' while drifting outside the allowed fragment, 
as a single stray clause, pronoun, or tense change can break parsability. 
With NDL, we can detect and repair or overlook minor deviations, as we have done with 
our instrumented proof checker. Achieving the same level of robustness for a CNL would be much more challenging. 

\item \mbf{Q3}: Why NDL? NDL has hardly ever been seen by models during their pretraining.
Isn't it unfair to make them reason in what is essentially a foreign language to them,
and aren't a lot of the model errors you've encountered due to this unfamiliarity? 
Why not use a language like Lean or Isabelle, which are much more extensively represented 
in a typical pretraining corpus? \\[0.03in]
\mbf{A}: A reasoning benchmark should be universally applicable to any model, 
without requiring prior exposure to a specific formal system. 
In fact, if we are serious about testing fluid reasoning, we should minimize reliance 
on the sort of crystallized knowledge and skills that accrue when a model ingests large 
amounts of pretraining material that it can directly leverage at test time. 
%% In fact, if we are serious about testing \emph{fluid} intelligence, we should minimize reliance 
%% on the sort of crystallized knowledge and skills that accrue when a model ingests large 
%% amounts of directly relevant information during its pretraining. 
We want to measure reasoning ability, not tool familiarity. If a model succeeds in Lean because 
it saw a lot of it in training and has therefore memorized common patterns and tactic heuristics, then 
we are testing recall and imitation just as much as we are testing reasoning. 
%A reasoning benchmark 
We should be testing whether a model can internalize and follow a system 
of rigorous rules from specification, not whether it has memorized one. 
For testing purposes, the absence of strong NDL priors is an advantage, not a liability. 

Another point (already made in Section~\ref{Sec:RelatedWork}, but it bears repeating here)
is that Lean proofs are procedural and tightly coupled with proof search in a sequent calculus.
This is unsurprising, given that Lean is an interactive theorem-proving system and thus naturally 
more focused on tactic orchestration and automation steering. NDL, by contrast, is more focused
on describing the proofs themselves, rather than the search for them. It
mirrors natural deduction, expresses proofs entirely declaratively, makes assumption and logical
scopes explicit, and separates logical structure from proof search. 

Further, as we also already pointed out in Section~\ref{Sec:RelatedWork}, languages 
like Lean are intentionally powerful and thus correspondingly complex, featuring esoteric syntax forms, heavy type theory, 
opaque automation, and a large ecosystem of libraries, modules, and built-in theories. 
This makes Lean very expressive and versatile, but also unsuitable as a fruit fly for rigorous reasoning. 
NDL, by contrast, is intentionally minimal. It fits fully in a short prompt, has no hidden machinery or 
inference automation, no meta-level tactics, and no library dependencies. 
This makes it cleaner, simpler, entirely self-contained, tooling-independent, 
much more diagnostic (because its simplicity exposes real inferential limitations), 
and much more appropriate as a controlled testbed for reasoning. 

Finally, with respect to the fairness concern, we restate our earlier position: the main role 
of a benchmark is to enable meaningful comparison across models, not to extract their maximum achievable performance.
A benchmark is a diagnostic instrument, not an exercise in system optimization. 
Fairness is ensured by clearly specifying the evaluation criteria and applying them 
consistently to all models. 

\item \mbf{Q4}: Why instrument the proof checker? Doesn't that help the models to cheat? \\[0.03in]
\mbf{A}: Instrumentation has been a sharp analytical tool. 
While NDL may be exceptionally simple, it is still a formal language 
and strict checking of NDL proofs would flatline the accuracy of 
many models at exceedingly low levels, often due to minor procedural slips.
This would collapse multiple failure modes into a single low score, 
resulting in poor discrimination and little information about latent reasoning competence.
By contrast, our instrumented proof checker reveals 
a graded competence spectrum, large separation between models, and internally coherent 
error profiles that correlate with other behavioral indicators. 
Far from helping models cheat, instrumentation restores measurement resolution
and disentangles executive brittleness from higher-level reasoning.

\item \mbf{Q5}: 
You claim that using abstract symbolic problems allows you to test fluid reasoning
because the problems don't have any content (other than pure logical content), 
and so the models cannot leverage their huge knowledge banks to retrieve memorized answers. 
However, 
%it is the pretraining that allows models to grasp the logical content of the problems,
%and to write proofs they must clearly understand what a proof is and they must have.
the pretraining of modern models exposes them to a vast range of proof templates, 
patterns of symbolic manipulation, typical lemma structures, and so forth. 
This means that even abstract tasks can partially reduce to pattern reuse. 
Doesn't it follow then that \pgrid-style benchmarks test the ability to generalize reasoning 
strategies absorbed during pre- (and post-) training, rather than tabula-rasa reasoning?  \\[0.03in]
\mbf{A}: That's a valid point, and we agree with it in part. Our claim is not that 
\pgridsp measures tabula-rasa or psychometrically pure fluid reasoning in any strong sense. 
Contemporary models plainly rely on pretraining to acquire the concepts of proof, 
logical entailment, 
% like reasoning by contradiction and case analysis, 
symbolic manipulation, and deductive 
structure. In that respect, \pgridsp tasks (like essentially 
every benchmark for advanced cognition) tests the ability to deploy and generalize previously 
learned reasoning strategies, rather than reasoning ex nihilo. The point of \pgridsp is narrower and 
more concrete: it probes a model's native analytical competence in symbolic reasoning, 
particularly deductive reasoning, and does so in a small, simple, and fully specified 
formalism, on abstract problems whose solutions are extremely unlikely to be directly memorized. 
Because the problems are content-free and procedurally constrained, the model cannot rely 
on world knowledge, standard textbook proofs, or domain-specific theorem memories. 
It must instead instantiate general proof skills in a new setting. So the benchmark is best 
understood not as a measure of pure fluid intelligence, but as a measure of a model's 
ability to compose, verify, analyze, and manipulate proofs under minimal 
semantic scaffolding.

It is also worth reiterating that all nontrivial cognitive benchmarks test the reuse of 
learned strategies. Human IQ tests do too. Nobody reasons from nowhere. Humans bring years 
of schooling, language, symbol use, and learned problem schemas to Raven matrices, 
analogies, and logic puzzles. The relevant question is not whether prior learning matters, 
but what kind of prior learning is useful and how directly task success can be reduced to retrieval.
It may be useful to distinguish three different phenomena: (a) memorization of specific proofs;
(b) reuse of learned proof methods; and (c) the ability to coordinate these methods 
correctly during reasoning when tackling a novel problem. 
The symbolic nature of \pgridsp helps to rule out (a): simple regurgitation of famous contentful 
proofs. This is what we would get if we were just asking for proofs of the infinity of primes or
the irrationality of the square root of 2.  
\pgridsp is not robust against (b), nor should it be. 
If a model has learned useful proof techniques, such as proof by contradiction 
and case analysis, that is part of what mathematical competence consists in. 
And what \pgridsp probes quite effectively is (c): whether a model can compose 
these techniques correctly in a rigorous setting.

\item \mbf{Q6}: Many of the problems in \pgrid, especially in tasks like \mbf{PL3-PW}, look harder 
than what most humans could realistically solve by hand. If humans themselves would struggle to solve 
these problems, why should we regard such tasks as a fair test for models rather than an 
artificially superhuman bar? And if the strongest models are already scoring fairly well on these 
difficult problems (some of which have 50-100 premises), doesn't that suggest that they already 
surpass humans in sustained, hard reasoning?  \\[0.03in]
\mbf{A}: A benchmark is not unfair simply because humans would do poorly on it. 
As already mentioned, fairness in benchmarking requires that the rules are explicit 
and evenly enforced, the scoring is consistent, and the tasks avoid obscure content knowledge, 
highly exotic/specialized skills, or long-context artifacts. A more relevant question is not 
whether a graduate student can solve a task, but whether the task is a clean, interpretable, and discriminating 
probe of the capability we want to measure---rigorous proof-based reasoning in our case. 

There are, broadly speaking, two strands of benchmarking work in AI. The first and older tradition 
has a more psychometric flavor and uses human ability as its reference level, 
%tends to be anchored on human ability levels, typically 
typically for the purpose of comparing human performance against that of computer programs. 
This can make benchmarks conservative and anthropocentric.
Additionally, insisting on human-level difficulty as a fairness criterion risks conflating psychological 
ease with problem-solving essence. Arithmetic with 100-digit numbers is unfair to humans but can be perfectly 
legitimate for models as a measure of their capacity for numerical computation. 
%% Likewise, long rigorous proofs may 
%% be unfair to unaided humans, but they are legitimate probes of whether a system can reliably carry out extended 
%% precise reasoning, something humans themselves typically achieve only with scaffolding. 

More importantly, once models catch up to human baselines, these benchmarks quickly become saturated. 
And if human solvability is treated as a prerequisite, they cannot be expanded as needed 
to keep up with the technology. This explains why benchmarking in games like chess and Go 
continues but has now shifted from questions like ``Can a computer beat the best human?'' to 
engine-vs-engine rating lists and tournaments. More generally, beyond board games, 
this is also why the last two years  have seen the emergence of a new class of benchmarks 
%the older line of benchmarking work has started to become superseded by new tests 
that are not anchored on ordinary human ability. Instead, they target either expert-human 
ability, as in Humanity's Last Exam; superhuman ability, as in FrontierMath’s open-problem track; or 
operational capabilities that we care about regardless of human performance, as in SWE-bench, where 
the key question is whether the issue gets resolved correctly in a reproducible environment. 
\pgridsp falls mainly into this last category. 

Instead of taking humans as the baseline, benchmarks like \pgridsp  can start to locate language 
models with respect to a different and in some ways more relevant and demanding reference class: symbolic 
inference techniques, like theorem provers and automated reasoning procedures in general. On that scale, current language models remain far 
from robust ``hard'' reasoners; the problems in \pgridsp are still trivial by the standards of mature 
methods such as resolution or Knuth-Bendix completion. But part of what makes \pgridsp  useful is precisely 
that it begins to quantify that gap. This makes the benchmark relevant not only to abstract questions 
about reasoning, but also to practical domains such as high-assurance software engineering, where 
the long-term goal is not merely automated generation of code, but automated generation of 
\emph{reliable} code, accompanied by robust justification artifacts, perhaps even including, 
eventually and when appropriate, machine-generated proofs of correctness. 

This brings us to the other relevant question, whether the benchmark is measuring a capability that we should 
care about. The ability to carry out deep chains of inference is an important component of 
highly intelligent systems; it is  crucial in science and (especially) mathematics, but also, as just mentioned, 
in software and other types of engineering. In these domains, correctness is often non-negotiable, and 
small errors can propagate into large failures if not rigorously controlled. Systems capable 
of sustained, internally consistent reasoning could play a transformative role. 

Finally, on the question of whether the current performance of leading models shows that they are 
already better reasoners than humans, the answer depends on how broadly we understand reasoning. 
If we construe it as the general ability to solve inference-intensive problems, then the answer is no. 
Humans can often solve such problems with ease by offloading parts of the work to external tools 
of their own making, such as notes and proof assistants, and by drawing on higher-level abstractions and insight. 
Models are just now approaching the point where they can generate long and rigorously valid 
inference chains under strict rules, chains that are deeper and more exacting than what an unaided 
human could manage.\footnote{They are not quite there yet, as shown by the various reasoning 
pathologies documented in this article.}
On that important but narrow front, humans were already surpassed decades ago by 
automated theorem provers. But humans remain far better at decomposing problems and at socially scaffolded 
forms of problem solving, including the use of tools and abstractions built up by other humans, 
through a process that cultural psychologists call \emph{cumulative cultural evolution}
\citep{tomasello1999cultural}. Language models can ingest the total stock of human knowledge 
up to a fixed snapshot in time, but it remains to be seen whether agentic collections of models
will eventually be able  to simulate this evolutionary process computationally, and achieve on their 
own the same cumulative ratchet effect that powers human problem solving.

\eit

\section{Limitations}

While \pgridsp offers a wide-ranging and deep evaluation of
logical reasoning capabilities in language models, a number of limitations
should be acknowledged.

First, \pgridsp intentionally tests logical reasoning in (minimally) formal languages, 
not natural-language reasoning. 
%Although we did translate all \plonesp proofs to English for one set of experiments, this represents a small portion of our evaluation. 
Thus, \pgrid's results may not fully
generalize to models' reasoning abilities in natural language
contexts.  However, as we argue in the paper, tests of natural-language
reasoning are outcome-based and thus only gauge reasoning competence indirectly. 

Second, our evaluation metrics focus on
accuracy rather than efficiency. Different models may achieve similar
accuracy scores while varying significantly in test-time compute 
or other resource requirements. A more comprehensive evaluation would include these dimensions.  

Third, while propositional and equational logic are rich and fundamental domains, 
they do not exhaust the broader landscape of reasoning abilities that a 
comprehensive benchmark should measure. In particular, the present tasks in \pgridsp
do not evaluate phenomena that arise in settings with quantification, variable binding, 
or more expressive semantic structures. An important direction for future work is 
to broaden the benchmark's coverage.

%% Third, while our test suite is extensive (2.7K problems), it is not exhaustive
%% within the domains of propositional and equational logic. The problems
%% were selected to be challenging yet tractable, and thus may not
%% represent the full spectrum of difficulty in logical reasoning
%% tasks. Additionally, our focus on certain logical formalisms (NDL,
%% NDL+, and equational logic) means that other important logical systems
%% remain unexplored.  

Finally,  given the rapid advancement in AI capabilities, particularly
in reasoning-focused models, the relative performance of various model
families may change significantly in the not-too-distant future.  
In particular, the empirical rankings we report here should be understood 
as a snapshot tied to the models available during the time period of our investigation, 
rather than a stable characterization of model capabilities. 
This kind of temporal fragility is increasingly unavoidable in contemporary 
benchmark studies, especially in fast-moving areas of model development. 
We therefore expect that some of the comparative conclusions in this paper 
will need to be revisited as newer systems become available. 

Despite these limitations, \pgridsp represents a significant advancement in the
evaluation of logical reasoning in AI systems, offering insights into
both the impressive progress and the remaining challenges in this crucial domain.

\section*{Acknowledgments}

We thank Theo Evgeniou for the fruitful discussions that helped to spark this project and contributed to shaping its initial direction.

\section*{Author Contributions}

%% The first author conceived the benchmark, designed the tasks, generated the datasets, performed most 
%% of the analysis of the results, and wrote the prompts and the paper (with the exception of Section~\ref{Sec:PLFour}).
%% Both authors helped to implement the experimental infrastructure. 
%% The second author oversaw and executed large-scale runs 
%% across multiple models, managed the pipeline, and contributed to debugging and validation.
%% The first author also defined and implemented NDL, introduced the Epistemic Stability Index (ESI), 
%% and carried out the IRT implementation and analysis. 

The first author conceived the benchmark, designed the tasks, generated the datasets, 
developed the prompts, led the analysis of the results, and wrote the manuscript (with the exception 
of Section~\ref{Sec:PLFour} and Appendix~\ref{app:technical-details}, which were written jointly). 
Both authors contributed to the experimental infrastructure. 
The second author led the execution of large-scale experiments across multiple models, including 
parallelized API-based evaluation pipelines, and contributed to debugging, validation, analysis, 
experimental reliability, and the provisioning of substantial computational resources. 
The first author additionally defined and implemented NDL and \ndlzero, 
introduced the Epistemic Stability Index (ESI), proposed the use of IRT and Fisher information, 
and carried out the implementation and analysis of the 2PL models. 

\clearpage
\appendix
\section{NDL}
\label{App:NDL}
NDL is a minimalist but rigorously defined formal language for writing natural-deduction proofs. 
It is: 
\ben
\item expressive enough to represent standard natural-deduction proofs; 
\item simple enough to admit a precise interpreter-based semantics; 
\item unforgiving enough to expose reasoning errors rather than hide them behind informal prose.
\een
Below is a sample proof that was already seen earlier, deriving the conclusion $(B \mor D)$ from the three
premises $(A \mif B)$, $(\neg A \mif C)$, and $(C \mif D)$. Comments in NDL start with the hash sign \mtt{\#}. 
\begin{tcAthenaWithNumbers}
assert premise-1 := (A ==> B)
assert premise-2 := (~ A ==> C)
assert premise-3 := (C ==> D)

# We prove (B | D) by a case analysis on (A | ~ A), which holds by the law of the excluded middle.  
{
  (A | ~A) BY ex-middle on A;

  # Case 1: Show that A implies (B | D)
  assume A {
    # Applying modus ponens to premise-1 and the assumption A gives B:  
    B BY mp on premise-1, A;
    # So now (B | D) follows by disjunction introduction: 
    (B | D) by left-either on B, D
  };

  # Case 2: Show that (~ A)  also implies (B | D) 
  assume (~ A) {
    C BY mp on premise-2, (~ A);
    D BY mp on premise-3, C;
    (B | D) BY right-either on B, D 
  };

  # The case analysis is now complete: 
  (B | D) BY cases on (A | ~A), (A ==> B | D), (~A ==> B | D)
}
\end{tcAthenaWithNumbers}
Checking an NDL proof $D$ amounts to evaluating $D$ by running it through the NDL 
interpreter $\mathcal{I}$.\footnote{We use the letter $D$ as a variable ranging
 over NDL proofs ($D$ stands for ``deduction,'' which we use interchangeably with ``proof'').}
The evaluation of a proof can only result in one of two outcomes: either the proof works and it produces 
a unique \emph{conclusion} $p$ as its result, or the proof fails and its evaluation 
generates a relevant error message.  

An NDL proof is always evaluated in the context of---or 
relative to---an {\em assumption base\/} \ab. More precisely, the NDL interpreter $\mathcal{I}$ is a 
function of two arguments: a proof $D$ and an assumption base \ab. Thus, we may only write $\mathcal{I}(D,\ab)$, 
not $\mathcal{I}(D)$.\footnote{Unless we explicitly \emph{curry} $\mathcal{I}$ and give it the signature:
$\mathcal{I}: \mbf{Ded} \rarrls \mbf{AB} \rarrls \mbf{P} \cup \{\temv{error}\}$, 
so that it takes a proof $D \in \mbf{Ded}$ followed by an assumption base $\ab \in \mbf{AB}$ and outputs
either a formula $p \in \mbf{P}$ or the distinguished constant \temv{error}. In that case, $\mathcal{I}(D)$
would be a unary function of an assumption base. Here we stick with the rectangular signature 
$\mathcal{I}: \mbf{Ded} \times \mbf{AB} \rarrls \mbf{P} \cup \{\temv{error}\}$.}
An assumption base is just a finite set of formulas that are taken for granted for the purposes of a proof. 
Initially, the assumption base is empty: $\ab = \eset$. The keyword \skwd{assert} is then used 
to insert---and name---as many premises as we like into the top-level (global) assumption base. 
By convention, premises in \pgridsp are 
always sequentially named: \mtt{premise-1}, \mtt{premise-2}, \etcsp These names can then be used to refer to the 
corresponding premises without having to write out the formulas inline. In this example we have three \skwd{assert} directives on lines
1--3, which insert the corresponding premises into the assumption base. Thus, by 
the time we get to line 4, the top-level $\beta$ contains exactly 3 formulas. 

Following the assertions, we have the main proof (lines 6--26). This is a {\em composite\/} proof of the form $\mtt{\{}D_1; D_2; \cdots; D_m\mtt{\}}$,
where $D_1, \ldots, D_m$ are the \emph{subproofs} being composed. The evaluation of this proof starts on line 6, 
in the assumption base that contains the 3 premises that were asserted earlier. As control flows through the proof, 
the assumption base grows and shrinks to reflect the conclusions derived by the various rule applications, such as 
the application of \mtt{ex-middle} on line 7 or \mtt{mp} (modus ponens) on line 12; or the hypotheses $p$ that 
are conditionally postulated when we enter a subproof of the form $\skwd{assume}\msp{}p\msp{}\rp{D}$ and then automatically 
discharged after $\rp{D}$. 
In this example, the first element of the composition is the application 
of \mtt{ex-middle} to $A$, 
%\mtt{A}, 
which produces the conclusion 
%\mtt{(A$\:\sep\ntildelsft{}$A)}
$(A \mor \neg A)$. The proof then proceeds
by a case analysis of this disjunction. It first shows, on lines 10--15, that \emph{if} we assume $A$, 
we can derive the target conclusion $(B \mor D)$; and then it shows (lines 18-22) that we can do the same 
if we assume the negation of $A$. Since we already know from line 7 that either $A$ 
or its negation holds, case analysis (captured by the inference rule \mtt{cases}) allows us to infer the 
desired $(B \mor D)$ (line 25). 

In general, a core NDL proof $D$ is one of three forms:  
\ben
\item A \emph{rule application}, namely, an application of an inference rule $R$ to a number of arguments $p_1,\ldots,p_k$, written as 
\begin{equation}
R\msp\kwd{on}\msp{}p_1,\ldots,p_k
\label{Eq:NDLRuleApp}
\end{equation}
where $R$ ranges over 24 built-in inference rules, from introduction and elimination rules for the 5 logical connectives to higher-level
rules that are included for convenience, such as \mtt{dsyl} (disjunctive syllogism) 
and \mtt{dm} (De Morgan's).\footnote{A complete list of NDL's inference rules and their operational semantics 
can be found online in the prompts for the tasks that involve NDL, such as \mbf{PL1-PW}.}
Typically, a rule application of the form~(\ref{Eq:NDLRuleApp}) works as follows: if the arguments $p_1,\ldots,p_k$ are in the assumption base, then $R$ derives
some appropriate conclusion $p$, which becomes the result of~(\ref{Eq:NDLRuleApp}). But if some $p_i$ is not in the assumption base,
the application fails with an appropriate error message. 

For instance, consider the application of \mtt{mp}
to \mtt{premise-1} and \mtt{A}, on line 12. At that point in the proof, both of these arguments are in the assumption base: \mtt{premise-1}
is in the assumption base because it was put there by the \kwd{assert} on line 1, while \mtt{A} is in the assumption base because
it was postulated as a hypothesis by the \kwd{assume} on line 10, and at line 12 we are still squarely inside the scope of that hypothesis (meaning
that \mtt{A} is still in the assumption base), as that scope does not close until line 15. 
Thus, because both of its arguments are in the assumption base, \mtt{mp} happily produces the
conclusion \mtt{B}, which becomes the output or result of the step on line 12. 

If one of these arguments were \emph{not} in the assumption base, the application of \mtt{mp}
would cause an error. For instance, suppose that line 12 was instead ``\mtt{D \kwd{BY} mp \kwd{on} (A\psp{}==>\psp{}D), A}''. In that case the first argument, 
\mtt{(A\psp{}==>\psp{}D)}, would not be in the assumption base, and the application would be rejected as unsound.

While this is the typical mode of operation, it is not always the case that every single argument $p_i$ in a rule application~(\ref{Eq:NDLRuleApp})
needs to be in the assumption base at the time of the application. This depends on the rule $R$. For example, the unary rule \mtt{ex-middle} takes
an arbitrary formula $p$ as its sole argument and always produces the conclusion $(p \mor \neg p)$, without caring about whether $p$
is in the assumption base. (Intuitively, this is because the conclusion 
%\mtt{($p$\:$\sep$\ntildelsft$p$)} 
$(p \mor \neg p)$ is always valid; whether $p$ is in the assumption base is immaterial.)

\item A {\em conditional proof\/} of the form 
\begin{equation}
\kwd{assume}\msp{}p\msp\rp{D},
\label{Eq:CondProof}
\end{equation}
where $p$ is the \emph{hypothesis} and $\rp{D}$ is the \emph{body} of the conditional proof, representing the \emph{scope} of the hypothesis $p$. 
The evaluation of such a proof in a given \absp is straightforward: we add the hypothesis $p$ to \absp and evaluate the body $\rp{D}$ in $\ab \cup \{p\}$;
if and when that evaluation produces a conclusion $q$, we produce the conditional 
%\mtt{($p$ ==> $q$)} 
$(p \mif q)$ as the result of~(\ref{Eq:CondProof}).
Otherwise, if the evaluation of  $\rp{D}$ in $\ab \cup \{p\}$ generates an error, that error is returned as the result of~(\ref{Eq:CondProof}). 
\item A {\em composite proof\/} (or simply ``composition'') of the form $\mtt{\big\{}D_1; D_2\mtt{\big\}}$, or more generally, 
$\mtt{\big\{}D_1; D_2; \cdots; D_m\mtt{\big\}}$ for $m > 1$.\footnote{
The composition operator \fmtt{;} associates to the right, so 
$\fmtt{\big\{}D_1; D_2; \cdots; D_m\mtt{\big\}}$ is syntax sugar for $\fmtt{\big\{}D_1; \fmtt{\big\{}D_2; \fmtt{\big\{} \cdots D_m\fmtt{\big\}}\cdots\fmtt{\big\}}\fmtt{\big\}}\fmtt{\big\}}$.}
Composition is the lemma-formation
mechanism in NDL. To evaluate a proof of this form in a given \ab, we start out by evaluating $D_1$ in \ab. If that yields a conclusion $p_1$,
we proceed to evaluate $D_2$ in $\ab \cup \{p_1\}$. Thus, {\em the conclusion of $D_1$ is now available in $D_2$\/} (and indeed in every subsequent $D_i$),
essentially becoming a lemma that downstream proofs can use as needed. If the evaluation of $D_2$ in $\ab \cup \{p_1\}$ produces a conclusion $p_2$,
we go on to evaluate $D_3$ in $\ab \cup \{p_1,p_2\}$, and so on. Continuing this way, the conclusion of the rightmost subproof $D_m$ becomes the conclusion
of the entire composition. 
%$\mtt{\big\{}D_1; D_2; \cdots; D_m\mtt{\big\}}$. 
If any $D_i$ generates an error, then the entire composition
produces that error as its output. 
\een 
Note that rule applications are primitive or atomic proofs, \iensp, they do not contains any subproofs. Conditional 
proofs and compositions are complex, meaning that they (recursively) contain subproofs. 

Any NDL proof $D$ can also be written in conclusion-annotated style, as $p\msp\kwd{BY}\msp{}D$. This can be viewed as an annotation or a promise of sorts:
It says that the evaluation of proof $D$ (in some appropriate \ab) must produce the conclusion $p$. Accordingly, 
it is an error if we evaluate $D$ and obtain a conclusion $\rp{p} \neq p$. 
This construct does not introduce any new inferential power, it only enforces alignment between proof intent and result.

\begin{figure}[b!]
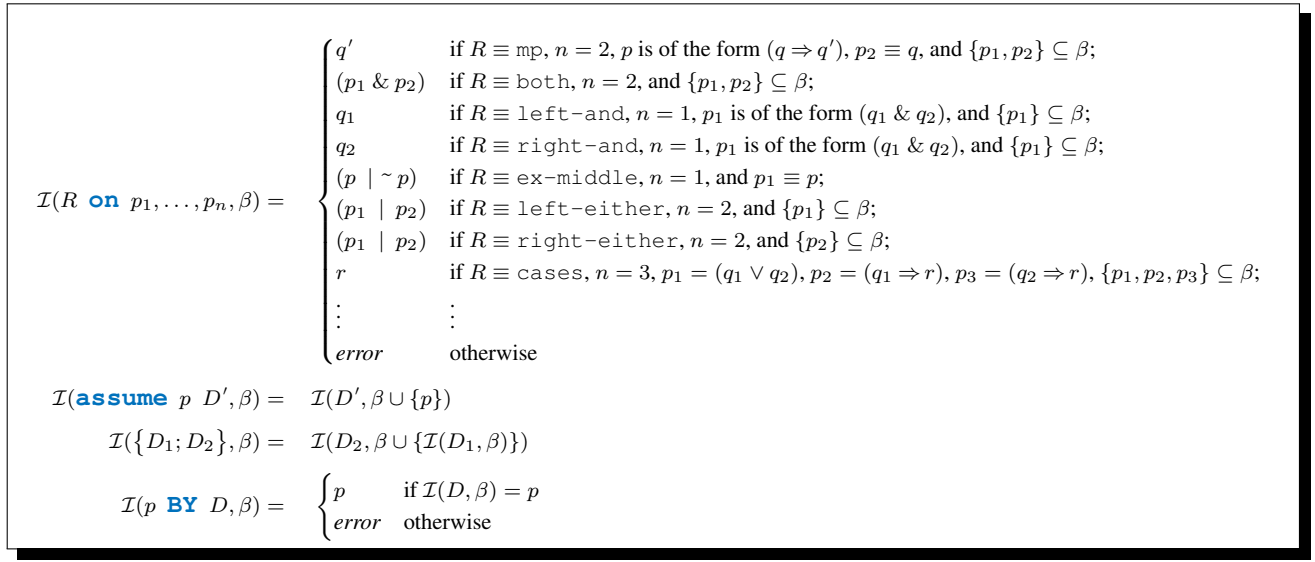

\centering
\shadowbox{%
  \begin{minipage}{1.0\linewidth}
%  optional: 
%\footnotesize 
\inbetweensize
  \[
  \begin{aligned}
\mathcal{I}(R\msp\kwd{on}\msp{}p_1,\ldots,p_n,\ab) & = && 
\begin{cases}
\rp{q} & \mbox{if $R \equiv \mtt{mp}$, $n=2$, $p$ is of the form $(q \mif \rp{q})$, $p_2 \equiv q$, and $\{p_1,p_2\} \subseteq \ab$; } \\
(p_1\:\&\:p_2) & \mbox{if $R \equiv \mtt{both}$, $n=2$, and $\{p_1,p_2\} \subseteq \ab$; } \\
q_1  & \mbox{if $R \equiv \mtt{left-and}$, $n=1$, $p_1$ is of the form $(q_1\:\&\:q_2)$, and $\{p_1\} \subseteq \ab$; } \\
q_2  & \mbox{if $R \equiv \mtt{right-and}$, $n=1$, $p_1$ is of the form $(q_1\:\&\:q_2)$, and $\{p_1\} \subseteq \ab$; } \\
(p\:\sep\ntildelsft{}p)  & \mbox{if $R \equiv \mtt{ex-middle}$, $n=1$, and $p_1 \equiv p$; } \\
(p_1\:\sep\:p_2)  & \mbox{if $R \equiv \mtt{left-either}$, $n=2$, and $\{p_1\} \subseteq \ab$; } \\
(p_1\:\sep\:p_2)  & \mbox{if $R \equiv \mtt{right-either}$, $n=2$, and $\{p_2\} \subseteq \ab$; } \\
r  & \mbox{if $R \equiv \mtt{cases}$, $n=3$, $p_1 = (q_1 \mor q_2)$, $p_2 = (q_1 \mif r)$, $p_3 = (q_2 \mif r)$, $\{p_1, p_2, p_3\} \subseteq \ab$; } \\
\vdots & \vdots  \\
\temv{error} & \mbox{otherwise}
\end{cases}  \\[0.03in]
\mathcal{I}(\kwd{assume}\msp{}p\msp\rp{D},\ab) & = && \mathcal{I}(\rp{D},\ab \cup \{p\}) \\[0.03in]
\mathcal{I}(\big\{D_1; D_2\big\},\ab) & = && \mathcal{I}(D_2,\ab \cup \{\mathcal{I}(D_1,\ab)\}) \\[0.03in]
\mathcal{I}(p\msp\kwd{BY}\msp{}D,\ab) & = && 
\begin{cases} 
p & \mbox{if $\mathcal{I}(D,\ab) = p$} \\
\temv{error} & \mbox{otherwise}
\end{cases}
  \end{aligned}
  \]
  \end{minipage}%
}

\caption{Definition of the NDL interpreter function.}
\label{Fig:NdlInterp}
\end{figure}

The above was a description of both the concrete and abstract syntax of NDL, and indeed of its formal semantics as well. We essentially 
defined the interpreter $\mathcal{I}$ as shown in Figure~\ref{Fig:NdlInterp}.

% usage:

Here is the main soundness and completeness result for NDL. 
\begin{theorem}
If $\mathcal{I}(D,\ab) = p$ then $\ab \models p$. Conversely, if $\ab \models p$ then there is some NDL proof $D$ such that $\mathcal{I}(D,\ab) = p$. 
\end{theorem}
The first (soundness) half says that if a proof successfully derives $p$ from some \ab, then $p$ follows logically from \ab. 
The second half (completeness) says that every logical consequence of \absp is derivable in NDL. We do not prove these results here, but briefly, 
soundness is shown by a straightforward structural induction, while completeness is most easily established by showing that a logic 
system that is already known to be complete can be embedded in NDL.

Along with a full listing of the various inference rules, this constitutes a complete description of NDL. 
The lone bell-and-whistle that NDL allows beyond what we have already covered is this: 
naming the results of intermediate subproofs and/or hypotheses. 
Any element $D_j$ of a composite proof $\mtt{\big\{}D_1; D_2; \cdots; D_m\mtt{\big\}}$ 
can be instead written as $I := D_j$, and 
then going forward (for all subsequent deductions $D_{j'}$, $j' > j$), the name $I$ will be bound to the 
conclusion produced by $D_j$. Likewise, any conditional proof $\kwd{assume}\msp{}p\msp\rp{D}$ can instead be written
as $\kwd{assume}\msp I \mtt{:=} p\msp\rp{D}$, in which case the name $I$ will become bound to the hypothesis $p$
inside (and only inside) the body $\rp{D}$. Thus, for example, we can write: 
\begin{tcAthena}
assume hyp := ((A & B) & (C & D)) {
  l := (A & B) BY left-and on hyp;
  r := (C & D) BY right-and on hyp;
  (r & l) BY both on r, l
} 
\end{tcAthena}
and this will produce the conditional $(((A \mand B) \mand (C \mand D)) \mif ((C \mand D) \mand (A \mand B))))$.\footnote{Since 
NDL is a proper subset of Athena, all proofs in this paper can be evaluated online at 
\href{https://tryathena.org/}{\tt https://tryathena.org/}. But note that Athena does not treat the names 
\fmtt{A}, \fmtt{B}, $\ldots$, as special in any way, so to declare these as atomic sentences, 
the line \fmtt{\fkwd{declare} A, B, C, D, E, F, G, H: Boolean} should be prepended to your input.}
Strictly speaking, this means that the interpreter $\mathcal{I}$ takes 3 arguments: a proof $D$, an 
assumption base \ab, and a {\em lexical environment\/} $\rho$ that maps names to formulas. For those
with a background in programming language theory, extending the denotational-semantics definition of $\mathcal{I}$
given here to account for $\rho$ is a straightforward exercise. 

For simplicity, we assumed that there is a single distinguished value \emph{error}, but in practice
the NDL interpreter produces different types of error messages. As discussed elsewhere, 
most errors\footnote{Some errors do not neatly fit into these categories (for instance, 
references to out-of-scope identifiers).} fall into one of the following three categories:
\bit
\item \emph{Syntax errors}: These could involve either formulas (\egnsp, unbalanced parentheses) 
or  NDL-specific constructs (\egnsp, extra or missing curly braces, hallucinated syntax forms, and so on). 

\item \emph{Structural errors}: These are caused by an incorrect application of an inference rule, \egnsp, to the wrong number of arguments or to 
arguments of the wrong form.

\item \emph{Logic errors}: These occur when a rule application appeals to a formula that does not currently hold (is not in the current assumption base). 

\eit

\clearpage 
\section{Item Response Theory}
\label{Sec:IRT}
\subsection{Introduction}

Item Response Theory (IRT) \citep{IRT3, IRT2} is a family of statistical models for developing 
tests and for analyzing task or benchmark data at the level of individual items/problems. 
Although IRT was not used in the development of \pgrid, we advocate for its use in analyzing 
and understanding various tasks in AI benchmarks, in assessing their discriminating ability, 
and in interpreting benchmark results. In this appendix we introduce IRT and illustrate its 
application to a number of \pgridsp tasks. The theory is valuable for transparency because it 
can reveal  both the strengths and limitations of a benchmark. In our case, IRT indicates 
that most \pgridsp tasks are highly effective at discriminating mid-tier models, and 
several also separate top-tier models well, but none are particularly effective
at distinguishing much weaker models. 

Instead of summarizing performance only with aggregate accuracy, IRT models the probability 
that a respondent answers a particular item correctly as a function of two kinds of latent 
variables: properties of the respondent (typically an unobserved \emph{ability} or proficiency,
usually denoted by $\theta$) and properties of the item (such as its difficulty and, in richer models 
like the ones we build here, its discrimination or susceptibility to guessing). In this way, 
IRT treats observed responses as noisy measurements generated by an underlying interaction 
between respondents and items, rather than as interchangeable Bernoulli trials.

%% In the analyses below, we use standard unidimensional IRT models. 
%% In this setting, the usual assumption is that, within a given task, item responses are driven primarily by a single latent trait and are conditionally independent given that trait. Although multidimensional IRT models also exist, unidimensionality is the conventional assumption in most basic IRT applications. It is rarely exact in practice, but it can still be a useful approximation when one dominant factor accounts for most of the systematic variation in performance.

%\enlargethispage{20pt}

We assume here that a model's responses on a given task  are driven primarily by one underlying latent variable. 
This is known as \emph{unidimensionality}, and it's a standard assumption of IRT.\footnote{Multidimensional 
IRT models are possible, but unidimensionality is the default in most applications.}
This dimension is allowed to vary from task to task. Proof checking 
may tap a different ability than proof writing, and proof masking and gap filling may 
rely on other abilities still. Unidimensionality never quite holds in 
practice, since there are always multiple factors impinging on any test taker's performance, from 
additional abilities (beyond the one that the task aims to measure) to more peripheral 
factors like motivation (in the case of humans) or sampling parameters like temperature (in the case of 
language models). The assumption can be considered sufficiently satisfied as long as there is 
one \emph{dominant} factor responsible for the test taker's performance \citep{hambleton2013item}[p. 17].
Essentially, what matters is that once we know a subject's position on the relevant latent axis, 
there should not be a large amount of systematic variation left over that would call for explanation 
in terms of additional hidden skills. In addition, task items are supposed to be conditionally independent,
an assumption that can be more problematic for some AI benchmarks; we discuss this in greater 
detail on p.~\pageref{Pg:IRTConditionalIndependence}. 

A key benefit of IRT is that it separates benchmark difficulty from model capability. 
Two language models can have the same overall accuracy while differing in the kinds 
of problems they can solve; likewise, two problems solved correctly by the same fraction of 
models need not be equally informative. IRT makes these distinctions explicit by estimating item 
parameters and placing models and items on a shared latent scale. This supports analyses 
such as identifying unusually easy or hard items, finding items that best distinguish between 
stronger and weaker models, and constructing score estimates that are less dependent on 
the particular mix of items in a benchmark task. 

For small to moderately sized benchmark tasks, like those in \pgrid, all of which have fewer than 1K items, 
Wright maps are a particularly useful part of the IRT toolbox. 
A Wright map is a visualization that puts test takers and test items on the same scale. 
On the left, it shows where the respondents fall, from lower-performing to higher-performing. 
On the right, it shows where the problems fall, from easier to harder. The key idea is that 
both are placed on one common line, so one can immediately see whether the questions are well 
matched to the group being tested. If most questions sit far below the test takers, the test 
is too easy; if they sit far above them, it is too hard. A Wright map also shows whether the 
questions cover the full range of ability or bunch up in just one part of the scale. 
In that way, it gives a quick visual summary of how informative and well targeted a task is. 
We publish Wright maps for all \pgridsp tasks for which we have built IRT models; see 
the last four sections of this appendix for the Wright maps of \mbf{PL1-PM}, \mbf{PL3-PC},
\mbf{PL4-PW}, and select \mbf{PL3-PW} subtasks. 

In general, in the context of AI evaluation IRT provides a principled way to move from raw benchmark 
outcomes to a measurement model of competence. This can be useful when comparing systems 
across benchmark versions, studying whether some questions are redundant or poorly calibrated, 
or asking whether improvements reflect broad capability gains versus progress on a narrow subset 
of items. 

\subsection{Modeling and ICCs}

We assume we have $N$ respondents and $M$ task items (problems). In this paper the respondents are language
models and $N = 24$. Taking \mbf{PL1-PW} as an example, we would have $M = 400$ (the number of items
in \mbf{PL2-PW}). We also assume we have test data: a binary response matrix $\mbf{y} \in \{0,1\}^{N\times M}$, 
where $\mbf{y}_{ij}$ indicates whether model $i$ solved problem $j$ correctly. For \mbf{PL2-PW}, $\mbf{y}_{ij} = 1$  
means that model $i$ wrote a proof that derived the expected result for problem $j$, and  $\mbf{y}_{ij} = 0$ 
means that the model's output was incorrect. 

In IRT we begin by specifying a probabilistic response model for each item $j$. 
In the 2PL model,\footnote{In IRT (and psychometrics in general), a ``2PL model'' refers to a 2-parameter logistic model.}
 task item $j$ is associated with an item response function of the following form:\footnote{We write $\sigma$
for the sigmoid function, $\sigma(x) = 1/(1+e^{-x})$.}
\begin{equation}
P(Y_j = 1 \mid \theta, a_j, b_j) = \sigma(a_j (\theta - b_j)) = \frac{1}{1 + e^{-a_j(\theta - b_j)}},
\label{Eq:IRT2PLModel}
\end{equation}
which gives the probability that a hypothetical respondent of ability $\theta$ 
answers item $j$ correctly, given the item's key characteristics, namely its 
\emph{discrimination} $a_j$ and its \emph{difficulty} $b_j$.

It is important to understand the intuition behind the algebraic form $a_j(\theta - b_j)$. Why that 
particular expression and not some other linear combination  $c_1 \theta + c_2 a_j + c_3 b_j$? 
The first thing to note is that $(\theta - b_j)$ represents how much the respondent's ability exceeds
the item's difficulty. If $\theta$ is greater than $b_j$, we should expect the probability of a correct
response to be greater than a random guess, i.e. greater than 1/2. If $\theta$ is less than $b_j$, we 
should expect the probability of a correct response to be less than 1/2. And if $\theta$ is equal to $b_j$,
it should be anybody's guess whether the response will be correct, \iensp,  $\theta = b_j$
should imply $P(Y_j = 1 \mid \theta, a_j, b_j) = 0.5$. And indeed, 
when $\theta = b_j$, the expression on the right-hand side of~(\ref{Eq:IRT2PLModel})
becomes 1/2. So this is the first key idea, that the chance of success depends on the margin $\theta - b_j$. 
The second key idea is that $a_j$ controls how sharply the probability changes as that margin changes.
If the item has high discriminating power, then a small increase in ability makes a big difference in the
odds of a correct response, whereas if $a_j$ is small then the item is fuzzier and the probability of success
changes much more gradually. This is what multiplying $a_j$ by the margin $(\theta - b_j)$ gives us. 
To make this concrete, Table~\ref{tab:irtProbabilities} shows how the probability of success varies for different sample values of 
$\alpha$ and the margin $\theta - b$. 
\renewcommand{\arraystretch}{1.2}
\begin{table}[b!]
\centering
\rowcolors{2}{blue!5}{white}
\resizebox{0.7\textwidth}{!}{%
\begin{tabular}{lccccccccc}
\toprule
$\alpha \backslash (\theta-b_j)$
& $-3$ & $-2$ & $-1$ & $-0.5$ & $0$ & $0.5$ & $1$ & $2$ & $3$ \\
\midrule
$0.2$
& 0.354 & 0.401 & 0.450 & 0.475 & 0.500 & 0.525 & 0.550 & 0.599 & 0.646 \\
$0.5$
& 0.182 & 0.269 & 0.378 & 0.438 & 0.500 & 0.562 & 0.622 & 0.731 & 0.818 \\
$1$
& 0.047 & 0.119 & 0.269 & 0.378 & 0.500 & 0.622 & 0.731 & 0.881 & 0.953 \\
$2$
& 0.002 & 0.018 & 0.119 & 0.269 & 0.500 & 0.731 & 0.881 & 0.982 & 0.998 \\
$3$
& 0.000 & 0.002 & 0.047 & 0.182 & 0.500 & 0.818 & 0.953 & 0.998 & 1.0 \\
\bottomrule
\end{tabular}
}
\caption{Success probabilities under the 2PL model,
\(p = \sigma\!\bigl(\alpha(\theta-b_j)\bigr)\),
for several values of the discrimination parameter \(\alpha\) and the
margin \(\theta-b_j\). As the margin increases, the probability of success
rises; larger values of \(\alpha\) make this rise steeper.}
\label{tab:irtProbabilities}
\end{table}
\renewcommand{\arraystretch}{1.0}

In summary: 
\bit
\item When ability equals difficulty, the item sits at its ``middle point'' and the probability of success is 1/2. 
\item Keeping the difficulty $b_j$ constant, higher ability increases the chances of success. 
\item Keeping the ability $\theta$ constant, higher difficulty decreases the chances of success. 
\item Discrimination controls the sensitivity of the curve, 
\iensp, how sharply the margin matters. 
\eit

\begin{figure}[!tbp]
\scalebox{0.9}{\includegraphics[width=\linewidth]{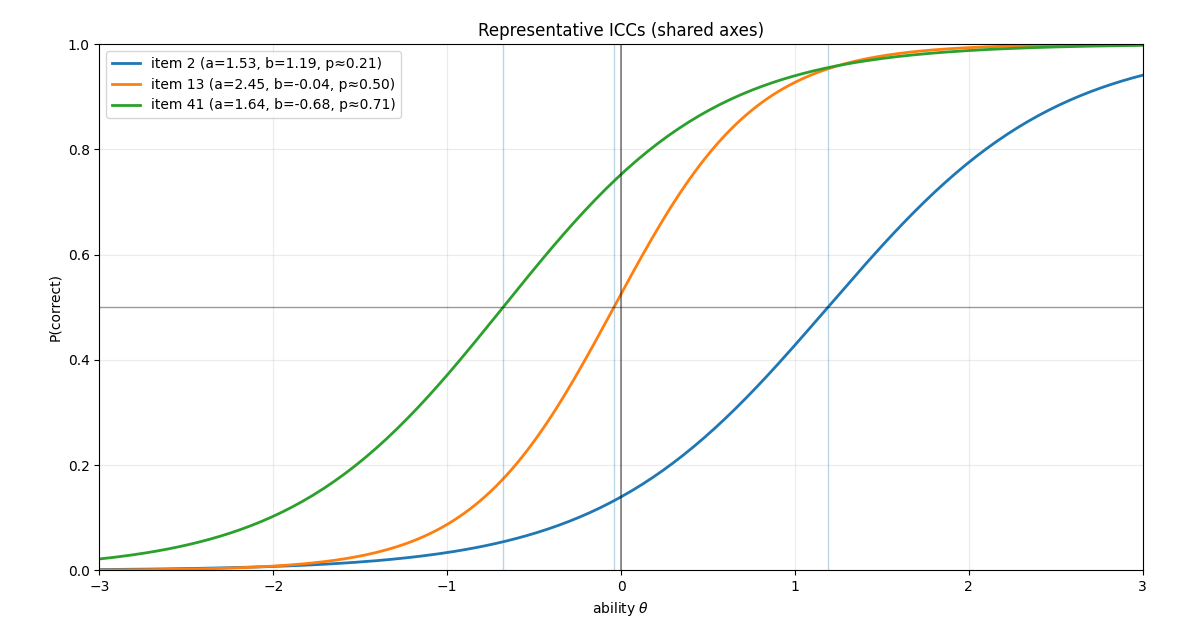}}
\caption{Three ICCs for three sample \mbf{PL2-PW} problems.}
\label{Fig:PL2PWSampleICCs}
\end{figure}

We can visualize a probability model of the form~(\ref{Eq:IRT2PLModel}) by plotting the 
probability on the $y$-axis as a function of ability $\theta$ on the $x$-axis. 
By the definition of~(\ref{Eq:IRT2PLModel}), the result will be a sigmoid curve. 
The item's difficulty $b_j$ shifts the curve to the left or right, while $a_j$
determines its slope. The resulting plot is called an \emph{item characteristic curve} (ICC). 
Figure~\ref{Fig:PL2PWSampleICCs} overlays three sample ICCs on the same axes, for these 
three \mbf{PL2-PW} problems: items 2, 13, and 41. 
Again, the horizontal axis is model ability $\theta$ and the vertical axis 
is the probability of answering an item correctly. Each S-shaped curve 
corresponds to one benchmark item. Where a curve crosses the 50\% line marks that 
item's difficulty $b$: curves shifted to the right (like item 2, $b \approx 1.19$),
are harder---you need higher $\theta$ to reach a 50\% success chance. Curves shifted to 
the left (like item 41, $b \approx -0.68$) are easier, with even lower-ability models having 
a good chance at writing a correct proof. The thin vertical lines (in light blue color) 
at each $b$ visualize this ``where $P=0.5$'' point, and the darker vertical line at $\theta = 0$ plus the horizontal 
line at $P=0.5$ help to identify those crossings.

The steepness of a curve around its midpoint reflects the item’s discrimination $\alpha$---how sharply 
it separates nearby ability levels. A steeper curve (e.g., item 13 with $\alpha \approx 2.45$) makes a big 
probability jump with a small change in $\theta$, so it's excellent for telling slightly weaker
from slightly stronger models near its difficulty. A flatter curve 
changes more gradually, so it's less informative for fine-grained ranking. In short, horizontal position 
reflects difficulty ($b$), while slope reflects discrimination ($a$). Looking at all three together 
shows how this task mixes easier and harder questions, and which ones are better at distinguishing models 
that are close in overall skill.

We can estimate the $N + 2 \cdot M$ parameters $\theta_i, a_j, b_j$ (the ability of each respondent 
plus each item's discrimination and difficulty) by maximizing the likelihood of the observed data \mbf{y}:  
\begin{equation*}
\begin{aligned}
%\theta_i, a_j, b_j 
(\{\hat\theta_i\},\{\hat a_j\},\{\hat b_j\})
%&= \argmax_{\theta_i, a_j, b_j} \, \prod_{i=1}^N \: \prod_{j=1}^M P(Y_j = \mbf{y}_{ij} \mid \theta_i, a_j, b_j)  \\
&= \argmax_{\{\theta_i\},\{a_j\},\{b_j\}} \, \prod_{i=1}^N \: \prod_{j=1}^M P(Y_{ij} = \mbf{y}_{ij} \mid \theta_i, a_j, b_j)  \\
&=  \argmax_{\{\theta_i\},\{a_j\},\{b_j\}} \, \prod_{i=1}^N \: \prod_{j=1}^M \left[P(Y_{ij} = 1 \mid \theta_i, a_j, b_j)^{\mbf{y}_{ij}}\right] 
\cdot \left[1 - P(Y_{ij} = 1 \mid \theta_i, a_j, b_j))^{1- \mbf{y}_{ij}}\right] \\
&= \argmax_{\{\theta_i\},\{a_j\},\{b_j\}} \, \prod_{i=1}^N \: \prod_{j=1}^M 
\left[\frac{1}{1 + e^{-a_j(\theta_i - b_j)}})\right]^{\mbf{y}_{ij}}
\cdot 
\left[
1 - 
\left[\frac{1}{1 + e^{-a_j(\theta_i - b_j)}})\right]^{1- \mbf{y}_{ij}}
\right] \\
&=  \argmax_{\{\theta_i\},\{a_j\},\{b_j\}} \, \log \: \prod_{i=1}^N \: \prod_{j=1}^M 
\left[\frac{1}{1 + e^{-a_j(\theta_i - b_j)}})\right]^{\mbf{y}_{ij}}
\cdot 
\left[
1 - 
\left[\frac{1}{1 + e^{-a_j(\theta_i - b_j)}})\right]^{1- \mbf{y}_{ij}}
\right] \\
&= \argmax_{\{\theta_i\},\{a_j\},\{b_j\}} \,  \bigsum_{i=1}^N \; \bigsum_{j=1}^M \;
{\mbf{y}_{ij}} \cdot \log \:
\left[\frac{1}{1 + e^{-a_j(\theta_i - b_j)}})\right]
+ 
(1- \mbf{y}_{ij}) \cdot \log \:
\left[
1 - 
\left[\frac{1}{1 + e^{-a_j(\theta_i - b_j)}})\right]
\right] 
\end{aligned}
\end{equation*}
where we write $Y_{ij}$ to denote the random variable that records
whether respondent $i$ answers item $j$ correctly, with $Y_{ij} = 1$ for a correct
response and $Y_{ij} = 0$ for an incorrect one. 
We fix the latent scale of $\theta$ by requiring $\theta \sim \mathcal{N}(0,1)$, i.e., that $\theta$ is normally
distributed with a 0 mean and standard deviation 1; and we also impose $a_j > 0$ for all $j = 1,\ldots,M$. 
We describe our algorithm for estimating these parameters in Section~\ref{Sec:IRTFittingAlgo}. 
(Note that we introduce estimation here via the unregularized joint likelihood for conceptual clarity. 
In practice, as described in  Section~\ref{Sec:IRTFittingAlgo}, 
the models reported in this appendix are fit by joint MAP estimation with Gaussian priors, 
which regularizes the item and respondent 
parameters and improves stability when the number of items is large relative to the number of respondents.)

Once these parameters are estimated, we have built the desired model~(\ref{Eq:IRT2PLModel}): we can plug in 
an arbitrary $\theta$ and compute the associated probability. The main use of such a model, however, is 
not prediction (as is usually the case in ML applications), but rather explanation (\egnsp, structured 
explanations of which items are hard/easy, which discriminate, where each model lies on the ability scale, 
and so on). 

Note that the product on the first identity of the above chain 
is justified only under an independence assumption: \label{Pg:IRTConditionalIndependence}
once we condition on the latent trait, item responses should be independent. Informally, this means that
the only reason why two items might co-vary is that stronger respondents do better on both.
This is a fundamental assumption of IRT. Violations are common when items are paraphrases (or duplicates), 
share context, depend on the same passage, or reward the same narrow trick. 
One might think that this is not a concern for stateless language models, but the 
issue is not about inter-item memory. The issue is that even after conditioning on latent ability,
success on item $j_1$ might still tell us something about success on $j_2$, because there may be
additional shared factor(s) on which both items depend. That said, if interpreted strictly
the condition would be impossible to satisfy,\footnote{Consider two very different 
items on a math test. While they may be testing different math skills and content, both may 
depend on the ability to compute binomial roots or apply the Pythagorean theorem, both of which 
have pervasive uses in mathematics. 
%One could consider these as additional factors influencing performance.
} so in practice it is usually understood as an idealization and its satisfaction 
is considered a matter of degree. 

Some tasks in \pgridsp appear to be more at 
odds with the fundamental assumptions of IRT. \mbf{PL3-PW} performance, in particular, 
has a nontrivial dependence on problem type. 
This threatens conditional independence because many items from the same problem family are not just 
``another draw of the same skill.'' They often share the same hidden structure, 
so a model that has the right trick for one item will tend to do well on related items, 
and a model that lacks that trick will tend to fail on many of them. 
Unidimensionality is also suspect here, since different problem families seem to 
demand different types of competence (the ability to perform simple forward inference 
vs counting arguments \etc). There may still be one \emph{dominant} latent factor at play, namely
proof-writing ability, but it is an empirical question whether that 
factor is strong enough to explain most of the covariance among item responses. 
In \mbf{PL3-PC},  by contrast, problem families play a much less prominent causal role. 
The model there is given a list of mini-SAT steps to verify. As long as these steps 
are roughly of the same size and structure across two problem families, then the identities
of these families should not matter much. Proof length and formula size are 
more important factors (and explain why \dbsp proofs, for example, as so hard to verify). 
We build IRT models for most \pgridsp tasks, except that in \mbf{PL3-PW} we build
different models for different problem families. 
In this appendix we present IRT models for \mbf{PL1-PM}, \mbf{PL3-PC}, 
\mbf{PL4-PW}, and subsets of \mbf{PL3-PW} (see Section~\ref{Sec:IRTPLOnePM}---Section~\ref{Sec:IRTPLFourPW}).

\subsection{Introducing Normalized Discrimination Scores}
\label{Sec:RestrictedDiscPower}
%\paragraph{Setup (2PL IRT model).}

Our objective in this section is to introduce and motivate a single quantity that measures how strongly 
discriminating a task is in a given band of ability. The quantity should be normalized to the 
unit interval, with values close to 0 indicating very poor discriminating power and values close
to 1 indicating very high discriminating power. We base our development on the notion of Fisher
information \citep{FisherInfoTutorial}, which we discuss below in detail. 

\begin{figure*}[tbh!]
\centering 
\scalebox{1.0}{\includegraphics[width=\linewidth]{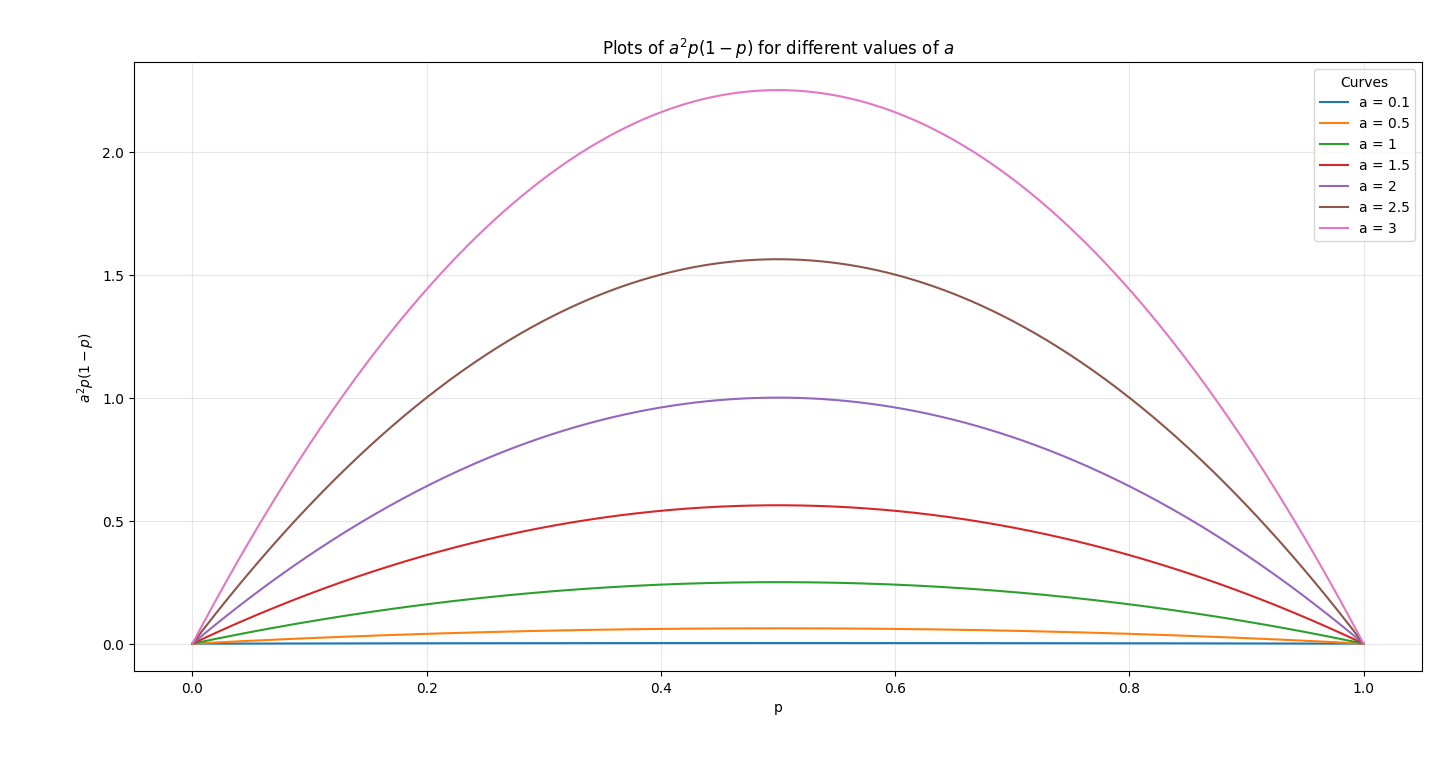}}
\caption{Fisher information of a task item as a function of different $a$ (discrimination) values.}
\label{Fig:FisherInfo}
\end{figure*}

For an item $j$ with discrimination $a_j>0$ and difficulty $b_j\in\mathbb{R}$, equation~(\ref{Eq:IRT2PLModel})
specifies the probability that a language model of latent ability $\theta$ solves the
item as
\begin{equation}
P_j(\theta) = \sigma(a_j (\theta - b_j)) = \frac{1}{1 + e^{-a_j(\theta - b_j)}}\raisedperiod
\label{Eq:ModelEqTwoPL}
\end{equation}
The \emph{Fisher item information} at ability $\theta$ is defined as: 
\begin{equation}
I_j(\theta)\;=\;a_j^2\,P_j(\theta)\,\bigl(1-P_j(\theta)\bigr).
\end{equation} 
To understand the intuition behind this quantity, consider the algebraic behavior of the expression
$a_j^2\,P_j(\theta)\,\bigl(1-P_j(\theta)\bigr)$. First, the bigger the item's discrimination, $a_j$, the bigger 
the $I_j(\theta)$, \iensp, the higher the information conveyed by the item,  and the growth is quadratic: doubling
$a_j$ quadruples $I_j(\theta)$, so this is a strong effect. The second and more important term is the product 
$P_j(\theta)\,\bigl(1-P_j(\theta)\bigr)$ (which, as we note below, is really the variance of a Bernoulli random variable). 
When $P_j(\theta)$ is low, i.e., when a model of ability $\theta$ is unlikely
to solve the item, this product is low. Likewise, the product is low when $P_j(\theta)$ is high. The product is maximized
at 0.5. When we fix a value for $a$, the quantity $a^2\,p\,(1-p)$ can be plotted as a function of $p$, and 
Figure~\ref{Fig:FisherInfo} does this for a range of $a$ values. We thus see that: 
\bit
\item if the item is very likely to be failed by a model of ability $\theta$, the information is low; 
\item ff the item is very likely to be solved by a model of ability $\theta$, the information is also low;  and 
\item the information is highest when the response is most uncertain, i.e., when the response could be correct or incorrect
with roughly equal probability. 
\eit
This holds for any value of $\alpha$, as shown in Figure~\ref{Fig:FisherInfo}, although higher $a$ values result in much higher
information for any fixed $P_j(\theta)$. 

For a more statistical perspective: If we define $X^{\theta}_j$ as a Bernoulli variable representing a hypothetical binary response 
to item $j$ from a respondent of ability $\theta$, then the success-probability parameter of that variable is precisely $P_j(\theta)$, i.e., 
$X^{\theta}_j \sim \mbox{Bernoulli}(P_j(\theta))$, and its variance is $P_j(\theta) \cdot (1 - P_j(\theta))$. 
When $P_j(\theta)$ is low or high, that variance is small because most responses will be either incorrect or correct, respectively. 
The variance increases only when the probability enters middle-ground territory, in which case we can expect more diversity in the
correctness of the responses. So the variance term $P_j(\theta) \cdot (1 - P_j(\theta))$  measures the amount of uncertainty 
regarding responses to item $j$ from test takers of ability $\theta$, and Fisher information is exactly this 
response variability weighted by the square of the item's discrimination ($a_j$). 
For a more analytical perspective, observe that the slope of the ICC of item $j$ at a 
given $\theta$ is the derivative of~(\ref{Eq:ModelEqTwoPL}) with respect to
$\theta$, which is precisely $a_j\,P_j(\theta)\,\bigl(1-P_j(\theta)\bigr)$. 
Therefore, in a 2PL model like this one, $I_j(\theta) = a_j\, \rp{P_j}(\theta)$ and 
item information is proportional to the local slope of the ICC, with $a_j$ as the proportionality factor.

It's worth stressing that this information quantity is not an intrinsic property of an item $j$; it depends crucially on the given $\theta$.
One and the same item could be highly discriminating (informative) for one ability level and quite ineffective for another ability level. 
As a concrete example, consider item 2 in \mbf{PL2-PW}, with $a_2 = 1.53$ and $b_2 = 1.19$. For $\theta = 1.2$ (roughly the ability of Gemini 3), 
we have $P_2(1.2) = 0.504$ and thus $1 - P_2(1.2) = 0.496$, and hence 
\[
I_2(\theta = 1.2) = a_2^2\,P_2(1.2)(1 - P_2(1.2)) = 2.3409 \cdot 0.504 \cdot 0.496 = 0.58. 
\]
However, the information of the same item at the level of $\theta = -1.0$, roughly the ability of Mistral Magistral and Grok-3, is
\[
I_2(-1.0) = a_2^2\,P_2(-1.0)(1 - P_2(-1.0)) = 2.3409 \cdot 0.034 \cdot 0.966 = 0.07. 
\]
This represents an 88\% relative drop in information content between these two ability levels. 

In the 2PL models that we use here, the Fisher information $I_j(\theta)$ of item $j$ is always maximized 
at $\theta = b_j$, since that is precisely the point where $P_j(\theta) = 0.5$ and 
$a_j^2\,P_j(\theta)\,(1-P_j(\theta))$ attains the largest possible value.

Next, given a task $T$, we define the \emph{task information} of $T$ at an ability $\theta$ as the sum of $I_j(\theta)$ across all items $j$ in $T$: 
\begin{equation}
I_{T}(\theta)\;=\;\sum_{j=1}^{\card{T}} I_j(\theta),
\label{Eq:TaskFisherInfoDef}
\end{equation}
where $\card{T}$ is the number of items in task $T$ (only counting 
items contributing to the fit after filtering uninformative all-correct/all-incorrect columns).

In the general theory of Fisher information, the quantity represents how much a random observation $X$ tells us about 
an unknown parameter $\theta$, given a probability model $f(x\mid \theta)$. Its units are $1/(\mbox{units of}\:\theta^2)$,
so if $\theta$ is measured in meters, for example, Fisher information would be measured in $1/m^2$ units. 
In the current IRT context, however, $\theta$ is a latent standardized ability scale, not a physical quantity 
with real-world units. As a result, item information often feels unitless, and raw numbers like 0.4, 1.2 or 2.8
are hard to interpret. The rest of this section aims to turn this scale-dependent quantity that is hard to interpret
in isolation into a [0,1]-normalized quantity representing a relative efficiency score against a clearly defined upper bound.

\paragraph{Restricting to a target ability band.}

As defined by~(\ref{Eq:TaskFisherInfoDef}), the Fisher information of a task $T$ is local in $\theta$: $I_{T}(\theta)$ 
tells us how discriminating the task is at one specific ability level, not overall. 
In practice, we typically care about an entire ability region (say mid-tier models or elite models),
i.e., an interval $\Theta = [\theta_1,\theta_2]$. Therefore, we need to turn the whole curve 
$I_{T}(\theta)$ between $\theta_1$ and $\theta_2$ into a single summary number for that region.
The natural way to do that is to average the task information over the chosen band. 
Conceptually, this is done with an integral, while numerically we implement the averaging over a fine uniform
grid $\{\theta^{(i)}\}_{i=1}^{G}\subset\Theta$: 
\begin{equation}
\overline{I}_{T}(\Theta) 
\;=\;\frac{1}{|\Theta|}\int_{\theta_1}^{\theta_2} I_{T}(\theta)\,d\theta
\;\approx\;\frac{1}{G}\sum_{i=1}^{G} I_{T}(\theta^{(i)}),
\label{Eq:IntegralDefOfTaskTheta}
\end{equation}
where $\theta^{(i)}$ denotes the \nth{i} point in a numerical grid of size $G$ laid across $\Theta$. 
Thus, $\overline{I}_{T}(\Theta)$ is just the average height of the task-information curve over the interval of interest. 
If the curve is high throughout the band, the task discriminates well there. 
If it is high only in a narrow spike but low elsewhere, the average will reflect that. 
So this is just getting a band-specific single-number summary out of a location-dependent quantity,
thereby answering the following question: ``On average, how much local discriminating power 
does this task have over the ability range we care about?''
Intuitively, if many items $j$ have their ICC ``humps'' $b_j$ inside $\Theta$ or at least near it, 
then $\overline{I}_{T}(\Theta)$ will be high, since these are the points that maximize $I_j$. 
But $\overline{I}_{T}(\Theta)$ will be low if many items have their information concentrated 
far from $\Theta$. In that case there will be slim pickings for the 
integral~(\ref{Eq:IntegralDefOfTaskTheta}) to sum over. 

\paragraph{Mean information per item.}
To make the magnitude comparable across tasks of different sizes, we define 
the \emph{mean information per item over $\Theta$} of a task $T$ by dividing by the number of items in $T$: 
\[
\temv{MI}_{T}(\Theta) 
\;=\;\frac{\overline{I}_{T}(\Theta)}{\card{T}}\raisedperiod
\]
This removes the trivial advantage of longer tasks (otherwise, a 400-item task 
will usually have much larger total information than a 50-item task, even if its 
items are not individually better). 

This quantity does not solve the deeper interpretability problem. 
A value like 0.8 for $\temv{MI}_{T}(\Theta)$  still lacks a clean meaning. 
It simply gives us a size-adjusted quantity, not yet a naturally interpretable one. 
By itself, it is mainly useful for comparisons. If, for a given band $\Theta$, 
this quantity is 1.2 for task $T_1$ and 0.8 for task $T_2$, we can say 
that task $T_1$ has more average discrimination per item in $\Theta$
than task $T_2$. This is a valid comparative statement, even if the raw 
numbers have no intuitive absolute benchmark.

\paragraph{Normalized band-targeting score in \([0,1]\)}
As already mentioned, raw Fisher information is useful but its scale is hard to 
interpret directly. To get a clearer score, we compare the task's band-averaged 
information $\bar I_{T}(\Theta)$ to an idealized upper bound defined over $\Theta$. 

We define this idealized upper bound as follows. 
For each item $j$, we keep its estimated discrimination $a_j$ fixed and ask for 
the largest band-averaged information that could be obtained by an imaginary 
item with that same discrimination strength, but with difficulty chosen 
to best match the target band. This gives:\footnote{The maximization operation  \label{Ft:BandTargetingInfoMax}
in~(\ref{Eq:NormalizedBandInfoDef}) ranges over all $b \in \reals$, so one may wonder
if it is well-founded. If we define 
$F_j(b,\Theta)$ as $(1/|\Theta|)\int_{\theta_1}^{\theta_2}a_j^2 \,\sigma\!(a_j(\theta-b)) (1-\sigma\!(a_j(\theta-b)))\, d\theta$,
then the right-hand side of~(\ref{Eq:NormalizedBandInfoDef}) is maximizing $F_j(b,\Theta)$ over all $b \in \reals$. 
In fact, for any finite $\theta_2 > \theta_1$ and fixed $a_j > 0$, the integral 
simplifies to a unary function  of $b$, namely $g(b) = a_j [\sigma(a_j(\theta_2 - b))  - \sigma(a_j(\theta_1 - b))]$
(this follows from elementary calculus, given that 
$
\frac{d}{d \theta} \sigma(a_j(\theta - b)) = a_j \sigma(a_j(\theta - b)) (1 - \sigma(a_j(\theta - b)))
$),
so that $F_j(b,\Theta) = (1/|\Theta|) g(b)$. 
We first note that this quantity is bounded from above, since for all $x \in \reals$ we have 
$0 \leq \sigma(x)(1 - \sigma(x)) \leq 1/4$, and hence for every $b \in \reals$, 
$0 \leq F_j(b,\Theta) \leq a_j^2/4$, where $a_j$ is fixed. So this optimization operation has a finite hard ceiling.
The other question is whether the maximum is actually attained at some finite $b$, and a simple continuity argument
will show that it is. Essentially: $F_j(b,\Theta)$ is a continuous function of $b$, it is bounded (as we just showed), 
and it vanishes at both tails (because, as $b$ tends to either $-\infty$ or $+\infty$, the 
quantity $F_j(b,\Theta)$ tends to 0). This shows that the maximization operation is well defined (and a little
more algebra will show that, assuming $\theta_1 < \theta_2$, a maximum value must occur at the midpoint $(\theta_1 + \theta_2)/2$).}
\begin{equation}
\bar I_j^*(\Theta) 
= 
\max_{b \in \mathbb{R}}
\frac{1}{|\Theta|}
\int_{\theta_1}^{\theta_2}
I_j(\theta) \, d\theta 
=
\max_{b \in \mathbb{R}}
\frac{1}{|\Theta|}
\int_{\theta_1}^{\theta_2}
a_j^2 \,
\sigma\!\bigl(a_j(\theta-b)\bigr)
\Bigl(1-\sigma\!\bigl(a_j(\theta-b)\bigr)\Bigr)
\, d\theta  
%= \frac{\theta_2 - \theta_1}{2}
\raisedperiod
\label{Eq:NormalizedBandInfoDef}
\end{equation}
Summing over items gives the band's attainable upper bound: 
\[
\temv{UB}_{T}(\Theta) 
=
\sum_{j=1}^{\card{T}} \bar I_j^*(\Theta).
\]
We finally define the normalized discrimination score over a band $\Theta$ as follows: 
\begin{equation}
\mc{S}_{T}(\Theta) = \frac{\bar I_{T}(\Theta)}{\temv{UB}_{T}(\Theta)}
\in [0,1].
\end{equation}
This can be read as follows: among all same-length comparison tasks with 
the same per-item discrimination profile $(a_1,\dots,a_{\scriptsize{\card{T}}})$, but with item 
difficulties chosen to fit this band perfectly, what fraction 
of the maximum attainable band-specific information is the actual task $T$ achieving?

What does ``fitting this band perfectly'' mean? It means that the difficulty $b_j$ of each
item $j$ sits right at the midpoint of $\Theta$. (As explained in footnote~\ref{Ft:BandTargetingInfoMax}, 
the maximum value specified by~\ref{Eq:NormalizedBandInfoDef} is always attained exactly at $(\theta_1 + \theta_2)/2$.)
So the normalized band-targeting score 
answers the following question: how informative is $T$ on $\Theta$ compared with the unique 
hypothetical task $\rp{T}$  obtained from $T$ by keeping each item's sharpness $a_j$ fixed
but setting each item’s difficulty $b_j$ exactly to the midpoint of $\Theta$? 
Task $\rp{T}$ is a mathematical fiction, of course: every single item in it has the exact same difficulty $b_j$,
namely, the midpoint of $\Theta$, so all ICC curves are centered at the exact same place and the only
thing distinguishing them is their sharpness (the $a_j$ values). 
This is highly implausible, but $\temv{UB}_{T}(\Theta) $ is not trying to describe a realistic 
benchmark design. It is a deliberately artificial upper bound for one specific band. 
If the only question is ``how much information could these same items deliver on this one target band, 
if one were allowed to place their difficulties optimally for that band?'', then collapsing all $b_j$ 
values to the midpoint is exactly the maximizing move (in the 2PL setting).

$\mc{S}_{T}(\Theta)$ values close to 1 mean that $T$ is very well targeted to the ability band represented by $\Theta$. 
A lower value suggests that, even if the items in $T$ are reasonably sharp, their difficulties 
are not as well aligned to $\Theta$ as they could be in theory. However, this should not be 
over-interpreted: $\mc{S}_{T}(\Theta)$ is a fraction whose denominator sets an exceedingly high bar 
(a very stringent idealized upper bound), so relatively moderate values may already 
indicate strong targeting for practical purposes. 
%% Ultimately, the proper interpretation of $\mc{S}_{T}(\Theta)$ scores 
%% requires some empirical calibration across various tasks and ability bands, 
%% and will become clearer as we present such scores for a number of 
%% \pgridsp tasks and ability regions. 

Finally,  $\mc{S}_{T}(\Theta)$ clearly depends on $\Theta$, so it is not a generic, band-independent measure
of task informativeness. Moreover, its interpretation is most useful for bands of substantive interest.
In one direction, if $\Theta$ is very wide then high values become easier to obtain almost by construction. 
For example, if $\Theta$ is wide enough to cover essentially the entire 
ability range for which the task was designed, then $\mc{S}_{T}(\Theta)$ will often be 
high because many items already have their difficulty levels somewhere inside that 
broad band, so moving them all to the midpoint does not improve things much. 
At the other extreme, if $\Theta$ is very narrow then a high $\mc{S}_{T}(\Theta)$ is 
generally unrealistic, unless the task is unusually concentrated on that narrow slice 
of ability. In that case the score is comparing the real task to an idealized version in 
which every item is perfectly centered on that tiny interval, which is a very demanding 
standard and one that a broadly useful task will usually not meet. 
The measure is most useful for substantively meaningful ability ranges, such as mid-tier, 
top-tier, or lower-tier models, or more systematically defined bands such as quartiles 
or quantiles.

\subsection{Fitting a 2PL IRT Model by Alternating MAP--MLE (EM--style) Updates}
\label{Sec:IRTFittingAlgo}
In standard psychometric uses of IRT, the number of respondents $N$ is typically much
larger than the number of test items $M$. In many recent AI benchmarks, including \pgrid,
the situation is reversed, as we often have $M \gg N$. 
This can introduce certain technical complications, because we have few data points
from which to infer parameter values; with $N = 24$ in a \pgridsp task with $M$ items, 
we have to estimate 2 parameters ($a_j$ and $b_j$) for each $j = 1,\ldots, M$ 
from only 24 binary responses. In this setting, item parameters can become noisy 
(discrimination parameters, in particular, can be especially fragile). In addition, 
estimation in this setting can be subject to separation pathologies and overfitting. 
%Accordingly, IRT should be used with some caution, as a strongly regularized 
%measurement model whose outputs must be interpreted with attention to parameter uncertainty and robustness. 

Here we mitigate the data asymmetry by fitting 2PL models not by gradient ascent
but by using an alternating MAP/Newton procedure that is well suited to the regime of 
many items and relatively few respondents: Gaussian priors regularize the respondent and 
item parameters, optimization in $\log a_j$ enforces positive discriminations and mitigates 
separation pathologies, and blockwise damped Newton updates yield stable low-dimensional 
subproblems for items and respondents. This doesn't quite eliminate the fundamental data asymmetry, 
but it makes estimation substantially more stable and interpretable than unregularized joint 
maximum likelihood in the same regime. The remainder of this section describes our fitting
algorithm in greater detail. 

\paragraph{Goal and data.}
We observe a binary response matrix $Y\in\{0,1\}^{N\times M}$, where $Y_{ij}=1$ if
model (respondent) $i\in\{1,\dots,N\}$ solves item $j\in\{1,\dots,M\}$ and $0$ otherwise.
The 2-parameter logistic (2PL) model posits for each item $j$ a discrimination $a_j>0$
and a difficulty $b_j\in\mathbb{R}$, and for each model $i$ a latent ability $\theta_i\in\mathbb{R}$,
\[
P(Y_{ij}=1\mid \theta_i,a_j,b_j)=\sigma\!\big(a_j(\theta_i-b_j)\big)
\quad\text{with}\quad
\sigma(z)=\frac{1}{1+e^{-z}}.
\]

\paragraph{Objective.}
The algorithm maximizes a regularized (MAP) joint log-posterior over all parameters,
\[
\mathcal{L}(\theta,a,b)
=\sum_{i=1}^N \sum_{j=1}^M \Big[\,Y_{ij}\log P_{ij} + (1-Y_{ij})\log(1-P_{ij})\,\Big]
-\frac{1}{2}\sum_{i=1}^N \theta_i^2
-\frac{1}{2}\sum_{j=1}^M \frac{(\log a_j-\mu_a)^2}{\sigma_a^2}
-\frac{1}{2}\sum_{j=1}^M \frac{b_j^2}{\sigma_b^2},
\]
where $P_{ij}=\sigma(a_j(\theta_i-b_j))$, $\mu_a=0$, and $\sigma_a,\sigma_b>0$ are prior
scales. The Gaussian priors $\,\theta_i\sim\mathcal{N}(0,1)$, $\log a_j\sim\mathcal{N}(\mu_a,\sigma_a^2)$,
$b_j\sim\mathcal{N}(0,\sigma_b^2)$ stabilize estimation (preventing infinite slopes from separation) and enforce $a_j>0$ by optimizing in $\log a_j$.

\enlargethispage{10pt}

\paragraph{Why an EM--style algorithm?}
The likelihood couples language models and items: $P_{ij}$ depends on both $\theta_i$ and $(a_j,b_j)$.
Direct Newton on \emph{all} parameters is brittle on sparse/balanced items and under separation.
Instead, we use an alternating conditional maximization strategy (a close cousin of
ECM/EM): holding $\theta$ fixed we maximize over $(\log a,b)$ item-by-item; then holding $(\log a,b)$ fixed we maximize over $\theta$ model-by-model. Each conditional subproblem is a tiny, well-behaved logistic regression, so Newton steps are reliable; the alternation is a coordinate-ascent on $\mathcal{L}$.

\paragraph{Preprocessing and initialization.}
We  drop non-informative columns (items with all $0$ or all $1$).
We initialize $\theta_i$ from smoothed logits of per-model accuracy and standardize to mean~0, variance~1. We draw $\log a_j\sim\mathcal{N}(0,0.25^2)$ and $b_j\sim\mathcal{N}(0,1)$.
We also enforce a practical box constraint $a_j\in[a_{\min},a_{\max}]$ after each update.

\paragraph{Item (M-step for items) updates.}
Fix $\theta$. For each item $j$ we maximize the \emph{itemwise} objective
\[
\ell_j(\log a_j,b_j)\;=\;\sum_{i=1}^N \Big[\,Y_{ij}\log P_{ij}+(1-Y_{ij})\log(1-P_{ij})\,\Big]
-\frac{(\log a_j-\mu_a)^2}{2\sigma_a^2}
-\frac{b_j^2}{2\sigma_b^2},
\]
where now $P_{ij}=\sigma\!\big(e^{\log a_j}(\theta_i-b_j)\big)$.
We use a single (damped) Newton--Raphson step on the $2\times 2$ Hessian for
$(\log a_j,b_j)$, with gradients/Hessians (letting $a_j=e^{\log a_j}$,
$t_{ij}=\theta_i-b_j$, $p_{ij}=P_{ij}$, $w_{ij}=p_{ij}(1-p_{ij})$):
\allowdisplaybreaks
%% \[
%% \begin{aligned*}
%% \frac{\partial \ell_j}{\partial \log a_j}
%% &=\sum_i (Y_{ij}-p_{ij})\,a_j t_{ij}\;-\;\frac{\log a_j-\mu_a}{\sigma_a^2},\\
%% \frac{\partial \ell_j}{\partial b_j}
%% &=-\sum_i (Y_{ij}-p_{ij})\,a_j \;-\;\frac{b_j}{\sigma_b^2},\\[3pt]
%% \frac{\partial^2 \ell_j}{\partial (\log a_j)^2}
%% &=-\sum_i w_{ij}\,(a_j t_{ij})^2 \;-\;\frac{1}{\sigma_a^2},\\
%% \frac{\partial^2 \ell_j}{\partial b_j^2}
%% &=-\sum_i w_{ij}\,a_j^2 \;-\;\frac{1}{\sigma_b^2},\\
%% \frac{\partial^2 \ell_j}{\partial \log a_j\,\partial b_j}
%% &=-\sum_i w_{ij}\,(a_j t_{ij})\,(-a_j).
%% \end{aligned*}
%% \]
\begin{align*}
\frac{\partial \ell_j}{\partial \log a_j}
&=\sum_i (Y_{ij}-p_{ij})\,a_j t_{ij}\;-\;\frac{\log a_j-\mu_a}{\sigma_a^2},\\
\frac{\partial \ell_j}{\partial b_j}
&=-\sum_i (Y_{ij}-p_{ij})\,a_j \;-\;\frac{b_j}{\sigma_b^2},\\[3pt]
\frac{\partial^2 \ell_j}{\partial (\log a_j)^2}
&=-\sum_i w_{ij}\,(a_j t_{ij})^2 \;-\;\frac{1}{\sigma_a^2},\\
\frac{\partial^2 \ell_j}{\partial b_j^2}
&=-\sum_i w_{ij}\,a_j^2 \;-\;\frac{1}{\sigma_b^2},\\
\frac{\partial^2 \ell_j}{\partial \log a_j\,\partial b_j}
&=-\sum_i w_{ij}\,(a_j t_{ij})\,(-a_j).
%% \frac{\partial \ell_j}{\partial \log a_j}
%% &=\sum_i (Y_{ij}-p_{ij})\,a_j t_{ij}\;-\;\frac{\log a_j-\mu_a}{\sigma_a^2},\\
%% \frac{\partial \ell_j}{\partial b_j}
%% &=-\sum_i (Y_{ij}-p_{ij})\,a_j \;-\;\frac{b_j}{\sigma_b^2},\\
%% \frac{\partial^2 \ell_j}{\partial (\log a_j)^2}
%% &=-\sum_i w_{ij}\,(a_j t_{ij})^2 \;-\;\frac{1}{\sigma_a^2},\\
%% \frac{\partial^2 \ell_j}{\partial b_j^2}
%% &=-\sum_i w_{ij}\,a_j^2 \;-\;\frac{1}{\sigma_b^2},\\
%% \frac{\partial^2 \ell_j}{\partial \log a_j\,\partial b_j}
%% &=-\sum_i w_{ij}\,(a_j t_{ij})\,(-a_j).
\end{align*}
We solve the $2\times 2$ linear system for the Newton step, damp it by a factor
$\in(0,1]$, then clip $a_j$ into $[a_{\min},a_{\max}]$ (by re-adjusting $\log a_j$).
This is done independently for each item $j$.

\paragraph{Model (M-step for models) updates.}
Fix $(\log a,b)$. For each language model $i$ we maximize
\[
\ell_i(\theta_i)\;=\;\sum_{j=1}^M \Big[\,Y_{ij}\log P_{ij}+(1-Y_{ij})\log(1-P_{ij})\,\Big]
-\frac{1}{2}\theta_i^2,
\quad
P_{ij}=\sigma\!\big(a_j(\theta_i-b_j)\big),
\]
by one (damped) Newton step using
\[
\frac{d\ell_i}{d\theta_i}=\sum_j (Y_{ij}-P_{ij})\,a_j\;-\;\theta_i,
\qquad
\frac{d^2\ell_i}{d\theta_i^2}=-\sum_j a_j^2\,P_{ij}(1-P_{ij})\;-\;1.
\]
Again, the update is scalar per model and therefore stable.

\paragraph{Identifiability rescaling.}
The 2PL model is not identifiable up to linear transforms of the latent scale:
$(\theta,a,b)\mapsto\big((\theta-\mu)/s,\;sa,\;(b-\mu)/s\big)$ yields the same probabilities.
After each iteration we fix the scale by standardizing $\theta$ to mean~0 and variance~1:
\[
\theta_i\leftarrow\frac{\theta_i-\bar\theta}{s},\qquad
b_j\leftarrow\frac{b_j-\bar\theta}{s},\qquad
\log a_j\leftarrow \log a_j+\log s,
\]
where $\bar\theta$ and $s$ are the current mean and standard deviation of $\theta$.
(This keeps $a_j(\theta_i-b_j)$ invariant, hence leaves $P_{ij}$ unchanged.)

\paragraph{Monotonicity, damping, and clipping.}
The Newton steps are \emph{damped} to avoid overshooting on items with near-separation,
and $a_j$ is clipped to $[a_{\min},a_{\max}]$ to prevent extreme slopes.
The Gaussian priors play the role of ridge regularizers on $\theta$, $\log a$, and $b$,
ensuring negative-definite Hessians and preventing degeneracy.

\paragraph{Convergence check.}
We track the relative change of the stacked parameter vector
$\big(\theta^\top,\log a^\top,b^\top\big)^\top$ and stop when the $\ell_2$ change
falls below a tolerance. Because each block step increases a local quadratic
approximation of $\mathcal{L}$, the procedure behaves like a block--Newton EM/ECM and
converges rapidly in practice.

\paragraph{Relation to EM / MMLE.}
Classical marginal-maximum-likelihood (MMLE) IRT treats $\theta$ as latent \emph{random}
effects, integrates them out in an E-step (via quadrature), and maximizes the marginal
likelihood over $(a,b)$ in an M-step. Our implementation instead performs \emph{joint}
MAP estimation (sometimes called MAP--JML): we maximize over both $\theta$ and $(a,b)$,
alternating between them. This is an EM-style coordinate-ascent on the
posterior, with the Gaussian priors acting like the penalty that stabilizes each block.
Both approaches are standard; MAP--JML is simple, fast, and well-suited when $M$ is large
and $N$ is modest.

\paragraph{Algorithm summary}
\begin{enumerate}
\item Build $Y$ from the input dictionary; optionally drop all-0/all-1 items.
\item Initialize $\theta$ from logits of accuracy; initialize $\log a$, $b$ from Gaussians.
\item Repeat until convergence:
  \begin{enumerate}
  \item (\emph{Item step}) For each $j$: take one damped Newton step on $(\log a_j,b_j)$
        for $\ell_j$, then clip $a_j$.
  \item (\emph{Language model step}) For each $i$: take one damped Newton step on $\theta_i$ for
        $\ell_i$.
  \item (\emph{Rescale}) Standardize $\theta$ to mean~0, var~1 and adjust $(\log a,b)$.
  \end{enumerate}
\item Return dictionaries for $\theta$, $a$, and $b$ (in original item indices).
\end{enumerate}

\paragraph{Why this works well in practice.}
Each subproblem is low-dimensional with closed-form gradients/Hessians, so one Newton step
per block already yields large ascent. Regularization tames separation and keeps slopes
finite; re-centering fixes the latent scale; and clipping prevents pathological updates.
Empirically, with hundreds of items and a few dozen models, this EM-style
alternating MAP procedure converges quickly and produces stable, interpretable estimates
of abilities $\theta_i$, item difficulties $b_j$, and discriminations $a_j$ suitable
for our downstream analyses (ICCs, Wright maps, information curves, and the normalized discrimination
scores introduced in Section~\ref{Sec:RestrictedDiscPower}).

\subsection{\mbf{PL1-PM} IRT model}
\label{Sec:IRTPLOnePM}

The ability values estimated from the \mbf{PL1-PM} data are shown in Table~\ref{Tb:PLOnePMIRTThetas}.
They induce the same ranking as the accuracies in Table~\ref{Tb:Pl1PMAccuracies}, except 
that GPT-5 and Claude Sonnet 4.5 with extended thinking are now strictly ordered even though 
their accuracies are identical (the latent ability of GPT-5 is estimated to be slightly lower). 

\renewcommand{\arraystretch}{1.2}
\begin{table}[!bt]
\centering
\rowcolors{2}{blue!5}{white}
\resizebox{0.65\textwidth}{!}{%
\begin{tabular}{|l|c|l|c|}
\hline
\textbf{Model} & Ability $\theta$ & \textbf{Model} & Ability $\theta$ \\
\hline
\gemThreeOne & 2.141  & \gemThree & 1.429 \\ \hline
\gptFiveTwo & 1.114  & \claudeOpusFourFiveNov & 1.035 \\ \hline
\gptFiveOne & 1.004  & \grokFour & 0.756 \\ \hline
\geminiTwoPointFive & 0.709  & \claudeSonnetExt & 0.624 \\ \hline
\gptFive & 0.613  & \oThree & 0.506 \\ \hline
\dsReasoner & 0.176  & \dsROne & 0.076 \\ \hline
\oFourMini & -0.033  & \opus & -0.224 \\ \hline
\oThreeMini & -0.434  & \claudeSonnetFour & -0.442 \\ \hline
\dsThreeOne & -0.561  & \grokThree & -0.796 \\ \hline
\dsThree & -0.889  & \hermes & -0.985 \\ \hline
\nous & -1.052  & \gptFourOne & -1.350 \\ \hline
\magistral & -1.637  & \gptFourO & -1.780 \\ \hline
\end{tabular}}
\caption{\mbf{PL1-PM} $\theta$ parameters estimated from the task results.}
\label{Tb:PLOnePMIRTThetas}
\end{table}
\renewcommand{\arraystretch}{1.0}

\begin{figure}[!b]
\centering
\scalebox{0.7}{\includegraphics[width=\linewidth]{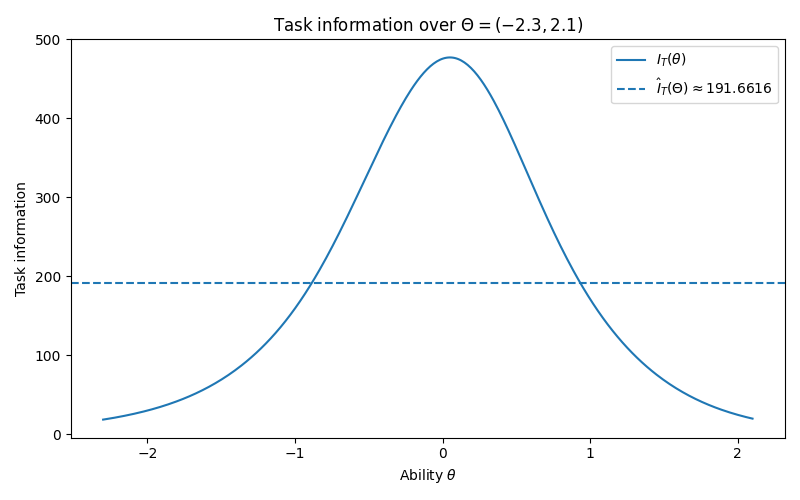}}
\caption{The task-information curve of proof masking over the entire observed ability range.}
\label{Fig:PLOnePMTaskInfoCurve}
\end{figure}

The Wright map for the proof-masking task is shown below, while the task-information curve over the observed ability range
appears in Figure~\ref{Fig:PLOnePMTaskInfoCurve}.
\begin{tcBathena}
PL1-PM MODEL–ITEM MAP: models and items on the same θ scale
θ range: [-2.32, 2.14]
     θ | Models                         | Items (b in bin)
--------------------------------------------------------------------------------
 -2.20 |                                | 41 73 93 107 140 223 243 258
 -1.98 |                                | 119 232
 -1.76 | gpt-4o                         | 18 166
 -1.54 | mistral-magistral              | 164 286
 -1.31 | gpt-4.1                        | 70 82 89 163 219
 -1.09 | hermes4, llama_400b_nous       | 8 29 30 112 129 133 135 174 193 197 221 254
 -0.87 | deepseek3, grok-3              | 48 51 56 58 63 78 98 114 137 168 203 230 244
 -0.64 | deepseek3.1                    | 22 34 88 92 94 118 123 134 144 157 158 192 228 267 281
 -0.42 | o3-mini, 
         claude-sonnet-4-20250514       | 1 19 20 27 37 38 40 44 45 52 54 55 60 65 81 86 101 108 117 121 
                                          127 136 148 149 150 156 160 161 170 182 185 190 207 225 227 249 
                                          250 253 288 289 291 295
 -0.20 | claude-opus-4-20250514         | 11 16 32 39 47 50 57 64 69 71 77 104 122 151 152 154 162 165 
                                          175 180 183 188 199 204 205 206 215 217 224 233 241 257 260 265 
                                          287 292
  0.02 | o4-mini, deepseekr1-0528       | 6 9 23 24 25 31 35 46 53 59 62 67 68 79 95 96 100 103 109 113 
                                          116 120 126 132 139 141 145 153 159 167 169 172 176 177 179 194 
                                          210 213 216 239 240 245 246 252 255 256 264 268 269 271 279 285 
                                          294 297 298
  0.25 | deepseek-reasoner              | 5 17 28 43 66 75 90 99 124 125 128 131 201 208 231 234 247 251 
                                          261 263 278
  0.47 | o3                             | 0 2 4 7 12 21 33 42 49 72 85 87 91 97 105 115 130 155 171 189 
                                          195 196 198 200 202 211 212 214 218 226 235 237 259 272 275 276 
                                          277 280 283
  0.69 | gemini-2.5-pro, grok4, gpt-5, 
         claude-sonnet-4-5-ext_thinking | 10 13 74 80 102 110 111 147 178 181 184 220 242 262 273 274 282 
                                          293 296
  0.92 | gpt-5.1                        | 3 36 61 83 106 138 143 146 236 266 290
  1.14 | claude-opus-4-5-20251101, 
         gpt-5.2                        | 14 26 76 191 222 238 270
  1.36 | gemini-3-pro-preview           | 15 84 186 187 229 299
  1.58 |                                | 
  1.81 |                                | 173 284
  2.03 | gemini-3.1-pro-preview         | 
\end{tcBathena}
A line like this one:
\begin{tcBathena}
 -0.87 | deepseek3, grok-3              | 48 51 56 58 63 78 98 114 137 168 203 230 244
\end{tcBathena}
means that items 48, 51, $\ldots$, 244, have difficulties $b_j$ roughly around -0.87, 
or more precisely, that their estimated difficulties $b_j$ fall in the bin centered 
around -0.87. This means models of ability $\theta \approx -0.87$ (or more precisely, models
whose estimated abilities fall into that same bin), such as DeepSeek-V3 and Grok-3, 
are just as likely to solve these problems as they are to fail them. 

The shape of the Wright map is roughly the discrete counterpart of the task-information curve,
rotated clockwise by 90 degrees. Where the items bunch up on the Wright map, the information curve tends to be high. The peak near 0 means
that many items sit there, so the task becomes especially sensitive in that region. 
This obviously resembles a normal distribution centered at 0, but this is not because
task information curves are inherently normal-shaped, nor is it because $\theta$ is standardized 
to have mean 0 and variance 1. Several tasks in \pgridsp have task-information curves that are skewed 
or irregular; see the task-information curve of \mbf{PL4-PW} as an example, Fig.~\ref{Fig:PLFourPWTableAndCurve}. 
The near-normal appearance observed in Fig.~\ref{Fig:PLOnePMTaskInfoCurve}
tells us something specific about the \mbf{PL1-PM} task items: their difficulties are concentrated 
around the middle of the ability scale, with increasingly fewer items as we move towards either tail end. 
Again, this matches the Wright map: item density is heaviest around the central region and thins out toward both ends.
This means that the items of this task have difficulties that are nicely distributed on the $\theta$ scale, and that the
task's ability to distinguish models is also nicely distributed across that same central region, \iensp, the task 
is most informative near the middle and its informativeness tapers off toward both ends. 
So this is a particularly well-balanced task, with items that are spread broadly across the ability range, 
instead of being piled up in one narrow hard or easy region. In simpler terms, this task has problems 
for weaker, middle, and stronger models, not just for the very strong or weak models. 

This is one of the few \pgridsp tasks that has good discriminating power in the lower half of the ability scale, 
with $\mc{S}_{T}(-2,0.2) = 0.63$.

\subsection{\mbf{PL3-PC} IRT model}
\label{Sec:IRTPLThreePC}

\renewcommand{\arraystretch}{1.2}
\begin{table}[!tb]
\centering
\rowcolors{2}{blue!5}{white}
\resizebox{0.65\textwidth}{!}{%
\begin{tabular}{|l|c|l|c|}
\hline
\textbf{Model} & Ability $\theta$ & \textbf{Model} & Ability $\theta$ \\
\hline
\gemThree & \mbf{1.577} & \gemThreeOne & 1.491 \\ \hline
\claudeOpusFourFiveNov & 1.167 & \gptFiveOne & 0.902 \\ \hline
\gptFive & 0.814 & \claudeSonnetExt & 0.803 \\ \hline
\grokFour & 0.747 & \gptFiveTwo & 0.609 \\ \hline
\oThree & 0.606 & \dsROne & 0.489 \\ \hline
\geminiTwoPointFive & 0.404 & \dsReasoner & 0.358 \\ \hline
\oFourMini & 0.301 & \oThreeMini & -0.214 \\ \hline
\hermes & -0.238 & \dsThreeOne & -0.330 \\ \hline
\nous & -0.356 & \claudeSonnetFour & -0.697 \\ \hline
\magistral & -0.783 & \opus & -0.817 \\ \hline
\gptFourO & -1.516 & \gptFourOne & -1.569 \\ \hline
\grokThree & -1.819 & \dsThree & -1.928 \\ \hline
\end{tabular}}
\caption{Model $\theta$ parameters estimated from \mbf{PL3-PC} results.}
\label{Tb:PLThreePCIRTThetas}
\end{table}
\renewcommand{\arraystretch}{1.0}

Fitting a 2PL IRT model on \mbf{PL3-PC} results yields the latent ability estimates shown in Table~\ref{Tb:PLThreePCIRTThetas}. 
We see that these estimates are quite consistent with the raw accuracies reported in Table~\ref{Tb:PLThreePCAccuracies},
with a couple of minor exceptions.\footnote{Gemini 3 here ranks ahead of Gemini 3.1, and GPT-5 ranks ahead of 
Claude Sonnet 4.5 with extended thinking, almost certainly due to the difficulty profiles (rather than the sheer number)
of the problems solved by these models.}

The following is the Wright map for this task: 
\begin{tcBathena}
PL3-PC MODEL–ITEM MAP: models and items on the same θ scale
θ range: [-2.21, 2.49]
     θ | Models                         | Items (b in bin)
--------------------------------------------------------------------------------
 -2.09 |                                | 215
 -1.85 | deepseek3, grok-3              | 
 -1.62 | gpt-4o, gpt-4.1                | 
 -1.38 |                                | 
 -1.15 |                                | 
 -0.92 | claude-opus-4-20250514         | 7 8
 -0.68 | mistral-magistral, 
         claude-sonnet-4-20250514       | 4 179
 -0.45 | deepseek3.1, llama_400b_nous   | 0 2 3 9 18 98 126 136 140 142 143 149 156 157 158 159 163 169 174 
                                          182 188 189 191 197 199 200 203 207 210 217 219 223 224 225 227 
                                          230 231 232 236 237 239 241 269 281
 -0.21 | o3-mini, hermes4               | 1 6 12 13 14 16 31 32 46 50 52 54 55 59 61 63 64 65 66 68 69 71 
                                          73 74 75 76 78 85 90 92 94 97 101 102 103 105 108 109 110 113 114 
                                          117 118 120 125 133 135 137 139 153 154 160 161 166 167 177 180 
                                          181 186 190 198 201 202 204 206 208 209 212 216 220 221 222 226 
                                          228 233 235 240 242 243 246 247 248 249 253 259 263 265 268 272 
                                          273 275 276 295 296 297 298
  0.02 |                                | 5 10 11 15 19 21 24 25 26 28 29 30 33 34 35 37 38 40 42 43 44 45 
                                          47 49 56 60 62 67 72 77 80 83 84 87 91 93 95 96 100 104 107 111 
                                          115 121 123 124 128 129 131 132 134 138 141 145 146 152 155 162 
                                          164 165 168 170 171 172 173 175 176 178 183 184 194 196 205 218 
                                          229 234 244 250 257 260 261 262 266 270 271 274 277 278 285 299
  0.26 | deepseek-reasoner, o4-mini     | 27 36 39 41 57 58 81 89 99 112 119 122 147 148 150 151 187 193 214 
                                          245 251 252 254 256 258 264 267 292
  0.49 | deepseekr1-0528, o3, 
         gemini-2.5-pro, gpt-5.2        | 48 51 82 86 144 185 211 213 238 255 293 294
  0.73 | claude-sonnet-4-5-ext_thinking, 
         gpt-5, grok4                   | 23 53 70 192 279
  0.96 | gpt-5.1                        | 17 20 22 79 106 127 284
  1.20 | claude-opus-4-5-20251101       | 116 130
  1.43 | gemini-3.1-pro-preview         | 88
  1.66 | gemini-3-pro-preview           | 
  1.90 |                                | 
  2.13 |                                | 290
  2.37 |                                | 282
\end{tcBathena}

We see that the task has items at many nearby difficulty levels across most of the respondent range,
with a particularly dense item mass from about $\theta \approx -0.45$ up to $\theta \approx 0.49$,
plus additional coverage through 0.73 and 0.96. 
The map is clearly strongest in the lower-mid to upper-mid region, and thinner in the extreme tails. 
On the low end, there are only a couple of very easy items below $\approx -0.9$, 
and on the high end there are only a handful of hard items above $\approx 1.2$, with two 
ultra-hard items at 2.13 and 2.37 (items 290 and 282, the only 2 \dbsp problems that are not
failed by all models). So the task covers the middle region beautifully, but the very weakest 
and very strongest models are measured with fewer anchor items. This means the task
should separate the broad population very well, as well as the mid-ability region, 
while the extreme-tail ordering is less stable.

%% This is supported by the normalized discrimination score over the wide ability 
%% band  $\Theta = (-1.9,1.5)$, which is found to be $\mc{S}_{T}(\Theta) = 0.98$, very near 
%% the theoretically perfect score that could be achieved on that band by 
%% any 300-item task with the same per-item discrimination profile. 

The task is also good at separating weaker models (below 0 ability), 
as long as they are not too weak, \egnsp, we have
$\mc{S}_{T}(-0.5,0) = 0.81$. For \temv{very} weak models the discriminating power drops drastically, 
\egnsp, $\mc{S}_{T}(-1.9,-0.6) = 0.21$. 
Finally, as the Wright map suggests, \mbf{PL3-PC} is exceptionally 
good at separating mid-tier models, \egnsp, for $\Theta = (-0.7,0.7)$ we have $\mc{S}_{T}(\Theta) = 0.93$. 

Next, we analyze some tasks that are better at discriminating top-tier models. 

%% A single unitless score in $[0,1]$ is obtained by comparing the achieved band–average
%% $\overline{I}_{\text{band}}$ to a \emph{population-aware, attainable upper bound} over
%% the same band. For each item $j$ we keep its estimated discrimination $a_j$ fixed and
%% optimize the difficulty $b$ to maximize its \emph{band-averaged} information,
%% \[
%% \overline{I}_j^{\star}
%% \;=\;\max_{b\in\mathbb{R}}\;
%% \frac{1}{|\Theta|}\int_{\theta_1}^{\theta_2}
%% a_j^2\,\sigma\!\big(a_j(\theta-b)\big)\bigl(1-\sigma\!\big(a_j(\theta-b)\big)\bigr)\,d\theta,
%% \]
%% which we compute numerically by evaluating a dense grid of trial difficulties $b$ and
%% taking the maximum. Summing over items yields the band’s attainable total,
%% \[
%% \text{UB}_{\text{band}}\;=\;\sum_{j=1}^{M}\overline{I}_j^{\star}.
%% \]
%% The \emph{normalized discrimination score} is then
%% \[
%% \text{normalized\_info\_0to1}
%% \;=\;\frac{\overline{I}_{\text{band}}}{\text{UB}_{\text{band}}}
%% \;\in\;[0,1].
%% \]
%% Intuitively, the denominator asks: if we were free to retarget each item’s difficulty
%% to this specific $\theta$ band (without changing how sharply it discriminates, i.e.,
%% keeping $a_j$ fixed), how much band-average information could we possibly extract?
%% In our case this yields
%% \[
%% \text{normalized\_info\_0to1}\;=\;\frac{313.318}{\text{UB}_{\text{band}}}\;=\;0.681.
%% \]

%% \paragraph{Interpretation.}

\subsection{Type-indexed \mbf{PL3-PW} IRT models}
\label{Sec:IRTPLThreePW}
We illustrate here with 2PL IRT models for two problem families that are useful for sharply 
discriminating the proof writing abilities of top-tier models: Tseitin-formula problems
and relativized-pigeonhole problems. 
The following is the Wright map for \mbf{PL3-PW} Tseitin-formula problems:
\begin{tcBathena}
PL3-PW TSEITIN-FORMULA MODEL–ITEM MAP: models and items on the same θ scale
θ range: [-0.72, 3.12]
     θ | Models                         | Items (b in bin)
--------------------------------------------------------------------------------
 -0.63 | mistral-magistral, deepseek3, 
         gpt-4o, o4-mini, hermes4, 
         deepseekr1-0528, gpt-4.1, o3, 
         deepseek3.1, grok-3, 
         claude-opus-4-20250514, 
         o3-mini, llama_400b_nous       | 
 -0.43 |                                | 
 -0.24 | gemini-2.5-pro, 
         claude-sonnet-4-5-ext_thinking, 
         claude-sonnet-4-20250514       | 
 -0.05 |                                | 
  0.14 |                                | 
  0.34 |                                | 
  0.53 | gpt-5                          | 23
  0.72 | grok4                          | 45
  0.91 | claude-opus-4-5-20251101, 
         deepseek-reasoner              | 
  1.11 | gpt-5.2                        | 36
  1.30 | gpt-5.1                        | 
  1.49 | gemini-3-pro-preview           | 44
  1.68 |                                | 39
  1.87 |                                | 8
  2.07 |                                | 32
  2.26 |                                | 35 38 47
  2.45 |                                | 
  2.64 |                                | 0 7 10 11 12 13 14 15 18 20 21 41 42 43 46 49
  2.84 |                                | 
  3.03 | gemini-3.1-pro-preview         | 1 4 26 27 31 34 37 40 59
\end{tcBathena}
This subtask is very good at discriminating top-tier models, with $\mc{S}_{T}(0.9,3.1) = 0.845$, 
while it is completely ineffective at lower- and mid-tier discrimination. 
The picture for relativized pigeonhole problems is similar, with $\mc{S}_{T}(0.9,3.0) = 0.823$: 
\begin{tcBathena}
PL3-PW RELATIVIZED PIGEONHOLE MODEL–ITEM MAP: models and items on the same θ scale
θ range: [-0.60, 3.36]
     θ | Models                         | Items (b in bin)
--------------------------------------------------------------------------------
 -0.50 | mistral-magistral, deepseek3, 
         gpt-4o, o4-mini, hermes4, 
         deepseekr1-0528, gpt-4.1, o3, 
         claude-sonnet-4-5-ext_thinking, 
         deepseek3.1, grok-3, 
         claude-opus-4-20250514, 
         o3-mini, llama_400b_nous, 
         claude-sonnet-4-20250514       | 
 -0.30 |                                | 
 -0.10 | gemini-2.5-pro                 | 
  0.10 | gpt-5, deepseek-reasoner       | 3
  0.29 |                                | 
  0.49 | gpt-5.1                        | 
  0.69 | gemini-3-pro-preview           | 
  0.89 |                                | 14
  1.08 | claude-opus-4-5-20251101       | 7
  1.28 |                                | 10
  1.48 |                                | 4 8 9 18 22 24
  1.68 | grok4, gpt-5.2                 | 6 12 17 27
  1.88 |                                | 11 16 19 20 23 25
  2.07 |                                | 
  2.27 |                                | 
  2.47 |                                | 5 13 15 21 26 28 29
  2.67 |                                | 
  2.87 |                                | 
  3.06 |                                | 
  3.26 | gemini-3.1-pro-preview         | 
\end{tcBathena}

\subsection{\mbf{PL4-PW} IRT model}
\label{Sec:IRTPLFourPW}
\begin{figure}[!tb]
\centering

\begin{minipage}[t]{0.48\textwidth}
\vspace{0pt}
\centering
\raisebox{-2.8cm}{%
\renewcommand{\arraystretch}{1.2}
\rowcolors{2}{blue!5}{white}
\resizebox{\textwidth}{!}{%
\begin{tabular}{|l|c|l|c|}
\hline
\textbf{Model} & Ability $\theta$ & \textbf{Model} & Ability $\theta$ \\
\hline
\gemThreeOne & 3.268  & \gptFiveTwo & 1.577 \\ \hline
\gemThree & 0.915  & \grokFour & 0.502 \\ \hline
\dsReasoner & 0.425  & \claudeOpusFourFiveNov & 0.424 \\ \hline
\gptFive & 0.372  & \gptFiveOne & 0.358 \\ \hline
\geminiTwoPointFive & 0.307  & \oThree & 0.281 \\ \hline
\claudeSonnetExt & 0.206  & \dsROne & 0.141 \\ \hline
\oFourMini & -0.151  & \oThreeMini & -0.192 \\ \hline
\hermes & -0.388  & \dsThreeOne & -0.501 \\ \hline
\opus & -0.508  & \claudeSonnetFour & -0.508 \\ \hline
\nous & -0.730  & \dsThree & -0.736 \\ \hline
\gptFourO & -0.864  & \magistral & -1.213 \\ \hline
\gptFourOne & -1.213  & \grokThree & -1.772 \\ \hline
\end{tabular}}
\renewcommand{\arraystretch}{1.0}
}
\end{minipage}
\hfill
\begin{minipage}[t]{0.48\textwidth}
\vspace{0pt}
\centering
\includegraphics[width=\textwidth]{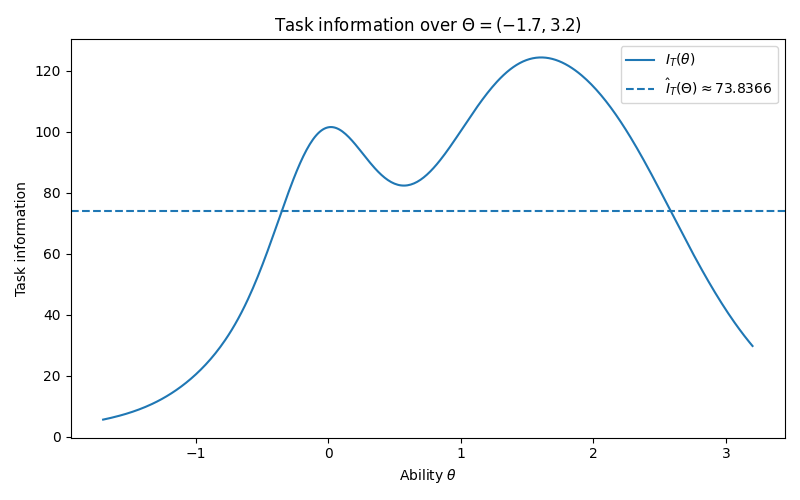}
\end{minipage}

\caption{Ability parameters estimated from \mbf{PL4-PW} results, alongside the task-information curve 
of Hilbert proof-writing over the observed ability range. Information is low for models of ability $\theta < 0$.}
\label{Fig:PLFourPWTableAndCurve}
\end{figure}

The $\theta$ values estimated from \mbf{PL4-PW} data are shown in the table of Fig.~\ref{Fig:PLFourPWTableAndCurve},
next to the task-information curve over the observed ability range. 

The following is the Wright map for \mbf{PL4-PW}:
\begin{tcBathena}
PL4-PW MODEL–ITEM MAP: models and items on the same θ scale
θ range: [-1.77, 3.27]
     θ | Models                         | Items (b in bin)
--------------------------------------------------------------------------------
 -1.65 | grok-3                         | 45 108
 -1.39 |                                | 
 -1.14 | mistral-magistral, gpt-4.1     | 
 -0.89 | gpt-4o                         | 41 172
 -0.64 | deepseek3, llama_400b_nous     | 14 166
 -0.39 | hermes4, deepseek3.1, 
         claude-opus-4-20250514, 
         claude-sonnet-4-20250514       | 10 19 61 62 79
 -0.13 | o4-mini, o3-mini               | 24 75 115 127
  0.12 | deepseekr1-0528, 
         claude-sonnet-4-5-ext_thinking | 6 7 18 56 105 134 187
  0.37 | gemini-2.5-pro, 
         claude-opus-4-5-20251101, 
         gpt-5.1, gpt-5, o3, 
         deepseek-reasoner              | 0 5 17 20 136 185
  0.62 | grok4                          | 2 9
  0.87 | gemini-3-pro-preview           | 42 123 128 189
  1.13 |                                | 15 38 43 60 66 67 97 124 131 157 161
  1.38 |                                | 1 3 4 8 28 29 31 34 39 63 71 72 88 93 95 98 114 120 122 133 138 141 
                                          144 153 171 178 180 183 188 198
  1.63 | gpt-5.2                        | 30 54 74 83
  1.88 |                                | 
  2.13 |                                | 11 12 16 21 25 26 27 33 36 37 44 46 48 51 53 55 57 58 64 65 68 69 73 
                                          77 78 82 87 89 91 94 96 99 103 106 109 111 112 113 116 121 129 132 
                                          137 139 140 146 147 148 149 150 151 152 156 162 164 168 179 182 186 
                                          192 193 196 199
  2.39 |                                | 22 110 197
  2.64 |                                | 163
  2.89 |                                | 
  3.14 | gemini-3.1-pro-preview         | 
\end{tcBathena}

This task too is only effective for discriminating elite models. 
For the ability band $\Theta = (0.5,3.2)$, which corresponds roughly to the models at or above
the 83rd percentile of the observed ability range, the task has a normalized discrimination score of 0.7: 
$\mc{S}_{T}(0.5,3.2) = 0.7$.

\clearpage 
\section{Technical and Operational Details}
\label{app:technical-details}

This appendix summarizes the main operational details of the reported experiments: model identities, 
API routes, prompt formats, decoding and reasoning settings, confidence intervals for our results, 
batching policies, data generation, and evaluation procedures. 
We do not include the prompts here, since they tend to be fairly long; they can be found on 
\href{https://github.com/System-2-Labs/ProofGrid}{\pgrid's Github page}. 

\subsection{Data Generation for \plone}
\label{Sec:PLOneDataGenAlgo}
Data generation for \pltwo, \plthree, and \plfoursp is already discussed in the paper.
For \plone, premises and goals for all problems were generated  randomly as ASTs (abstract syntax trees) 
by choosing a logical operator as the root (or an atom for a leaf) and then recursively  generating subtrees up to a maximum depth.
To minimize degenerate problems, we imposed the following constraints on 
any premises $p_1,\ldots,p_n$ and goal $p$:
\ben
\item all $n+1$ formulas must be distinct;
\item $\{p_1,\ldots,p_n\}$ must logically imply $p$; 
\item every premise $p_i$ must be necessary, meaning that if we remove $p_i$, the goal $p$ is no longer entailed by the remaining premises;
\item the goal does not contain any atoms that do not occur in the premises; and 
\item the premises by themselves are logically consistent (as is the negation of the goal). 
\een
We used a SAT solver to enforce 2, 3, and 5. 

We limit duplicate entries  by normalizing a problem's representation 
as follows: sort the list of all premises, conjoin the resulting elements, form the conditional between that conjunction and the conclusion, and then universally quantify over all atoms in that conditional. We then identify two problems iff these universally quantified normal forms are alpha-equivalent (identical up to renaming of bound variables). Thus, for instance, the argument with $p_1 = (A \mif B)$, $p_2 = (B \mif C)$, $p_3 = \neg C$, 
and goal $\neg A$ is represented by the QBF (Quantified Boolean Formula): 
\[
\forall\: A, B, C \qdot (((A \mif B) \mand (B \mif C) \mand \neg C) \mif \neg A) 
\]
and is thus considered identical to the problem with $p_1 = (C \mif D)$, $p_2 = \neg E$, $p_3 = (D \mif E)$, and goal $\neg C$. 

Terminological note: To improve readability, this paper often uses self-explanatory Latex-friendly 
notation for formulas instead of the plain-ASCII notation actually used in the tasks, 
\egnsp, writing $(A \mand \neg B)$ instead of \mtt{(A\psp\&\psp\ntildels{}B)}, $(p_1 \mif p_2)$ instead 
of \mtt{($p_1$\psp{}==>\psp{}$p_2$)}, $(p_1 \mor p_2)$ instead of \mtt{($p_1$\psp{}|\psp{}$p_2$)}, and so on.

The Athena theorem prover that generated proofs for the \mbf{PL1-PC} problems uses a combination of backward and forward heuristics, 
ordered according to the strategy in Section 4.1.14 of the textbook {\em Fundamental Proof Methods in Computer Science\/} \citep{FPMICS}. 
The theorem prover is written as an Athena {\em method\/}, which is automatically instrumented so that a successful run produces a low-level
{\em certificate}: a fully detailed proof comprising all applications of primitive inference rules that were made during
the run, as well as all conditional (sub)proofs and (sub)proofs by contradiction, properly scoped. This certificate is typically 
a very long proof (many thousands of lines), because it includes all inferences 
made even during paths of the search tree that were ultimately dead ends. These proofs are then simplified by an aggressive 
proof-optimization algorithm \citep{ProofSimp} that removes detours and other inefficiencies 
from natural-deduction proofs, radically simplifying their structure.

The deliberate corruption of a proportion of these proofs used interventions such as altering the logical form of 
an argument to an inference rule (\egnsp, applying conjunction elimination to a disjunction), or the number of arguments, 
removing a step from a chain of subproofs, and so on.

\subsection{Result Archives}

We release result archives for all 15 tasks currently in \pgrid, in the form of one Yaml file per task: \\[0.03in]
{
\centering 
\renewcommand{\arraystretch}{1.2}
\resizebox{0.5\textwidth}{!}{%
\begin{tabular}{|r|l|l|}
\hline
 & Task & Description \\
\hline
1  & \mbf{PL1-PC}   & Proof checking in \plone \\
\hline
2  & \mbf{PL1-PC-c} & Proof checking in \plone, challenge version \\
\hline
3  & \mbf{PL1-PW}   & Proof writing in \plone \\
\hline
4  & \mbf{PL1-PM}   & Proof masking in \plone \\
\hline
5  & \mbf{PL1-PM-c} & Proof masking in \plone, challenge version \\
\hline
6  & \mbf{PL1-GF}   & Proof gap-filling in \plone \\
\hline
7  & \mbf{PL1-GF-c} & Proof gap-filling in \plone, challenge version \\
\hline
8  & \mbf{PL2-PW}   & Proof writing in \pltwo \\
\hline
9  & \mbf{PL3-PC}   & Proof checking in \plthree \\
\hline
10 & \mbf{PL3-PW}   & Proof writing in \plthree \\
\hline
11 & \mbf{PL4-PW}   & Proof writing in \plfour \\
\hline
12 & \mbf{PL4-PW-c} & Proof writing in \plfour, challenge version \\
\hline
13 & \mbf{EQ-PC}    & Proof checking in \eqone \\
\hline
14 & \mbf{EQ-ER}    & Equation recovery in \eqone \\
\hline
15 & \mbf{EQ-GF}    & Gap filling in \eqone \\
\hline
\end{tabular} 
}
}
\renewcommand{\arraystretch}{1.0}
%% \begin{enumerate}
%% \item \mbf{PL1-PC} (proof checking in \plone)
%% \item \mbf{PL1-PC-c} (proof checking in \plone, challenge task)
%% \item \mbf{PL1-PW} (proof writing in \plone)
%% \item \mbf{PL1-PM} (proof masking in \plone)
%% \item \mbf{PL1-PM-c} (proof masking in \plone, challenge task)
%% \item \mbf{PL1-GF} (proof gap-filling in \plone)
%% \item \mbf{PL1-GF-c} (proof gap-filling in \plone, challenge task)
%% \item \mbf{PL2-PW} (proof writing in \pltwo)
%% \item \mbf{PL3-PC} (proof checking in \plthree)
%% \item \mbf{PL3-PW} (proof writing in \plthree)
%% \item \mbf{PL4-PW} (proof writing in \plfour)
%% \item \mbf{PL4-PW-c} (proof writing in \plfour, challenge task)
%% \item \mbf{EQ-PC} (proof checking in \eqone)
%% \item \mbf{EQ-EQ} (equation recovery in \eqone) 
%% \item \mbf{EQ-GF} (gap filling in \eqone) 
%% \end{enumerate}

The released Yaml files contain data documenting our results for all these tasks: 
problem descriptions and metadata, model identifiers, timestamps, and, for most runs, 
raw model outputs, processed model outputs, as well as partial token-usage and/or intermediate 
reasoning information. 

Conceptually, a Yaml result file for task $T$  is a list of 
result dictionaries, one per problem/item in $T$. Each such dictionary has a numeric 
index that serves as an ID, uniquely identifying that particular problem and associated result data 
in the context of $T$.  For example, if our IRT analysis talks about ``item 28'' in \mbf{PL2-PW}, 
it means precisely that record $r$ in the Yaml file for \mbf{PL2-PW} that has \verb|r['index'] == 28|. 

In addition, each result record typically has keys like \smtt{problem}, describing the input problem, 
metadata about the problem (\egnsp, whether or not the input proof to be checked is in
fact correct), and a results dictionary mapping each of the 24 model identifiers $M$ to 
a corresponding result dictionary for $M$. That model-specific dictionary contains a host of information 
about the response of $M$ to the given problem: API call timestamps, the raw output that $M$ generated, 
the processed/evaluated output (as carried out by our pipeline), evaluation results and details, 
and often a metadata key whose value indicates whether $M$'s response was ultimately counted as correct. 
As an example, here is (a truncated version of) item 0 in the results file for \mbf{PL1-PM}: 
\begin{tcAthenaWithNoComments}
  - index: 0
    problem: '(C | A) # ((A ==> B) | (B & (B | A))) # (~C | B) # B'
    conditionalized: true
    proof: |-
      assume h1 := ((C | A) & (((A ==> B) | (B & (B | A))) & (~C | B))) {
        (C | A) BY left-and on h1;
        ... 
      }
    correctProof: true
    complexProof: false
    maskedProof: |-
      assume h1 := ((C | A) & (((A ==> B) | (B & (B | A))) & (~C | B))) {
        MASK38 BY MASK23 on MASK13;
        ...
      }
    masks:
      MASK1: (A ==> B)
      MASK2: C
      MASK3: (B & (B | A))
      ...
    llmMaskFillingResponses:
      claude-opus-4-20250514:
        sentTimestamp: '2025-08-20 14:05:51.817886+00:00'
        receivedTimestamp: '2025-08-20 14:07:08.822000+00:00'
        answer:
          MASK1: (A ==> B)
          MASK2: C
          ...
        unmaskedProof: |-
          assume h1 := ((C | A) & (((A ==> B) | (B & (B | A))) & (~C | B))) {
            (C | A) BY left-and on h1;
            ... 
        correctMasks: false
        instrumentedCheckerResult:
          result: incorrect
          errorType: notInAB
          step: 7
          offendingRule: claim
          args:
            - B
          evaluatedArgs:
            - B
          missingFormula: B
          fixes:
            syntaxCorrections: []
            structuralCorrections:
              -   - step-6
                  - " Overlooking a wrong conclusion that's equivalent up to conj/disj commutativity "
          athenaAcceptedNonInstrumentedProof: false
          athenaResponseToNonInstrumentedProof: |-
            input:9:13: Error: Failed conclusion annotation.
            The expected conclusion was:
            (and C A)
            but the obtained result was:
            (and (or C A)
                 C).
          evaluationTimestamp: '2026-01-07 16:05:49.353478'
\end{tcAthenaWithNoComments}
The \smtt{problem} field indicates the argument/sequent targeted by the relevant \smtt{proof} 
(the proof that was masked). The \smtt{conditionalized} flag indicates whether the problem was conditionalized (whether its 
premises were imported into the conclusion). The \smtt{correctProof} and \smtt{complexProof} flags
are metadata about the starting \smtt{proof}: whether or not it solves the problem correctly and whether or not is has
any complex/nested syntax forms. The value of \smtt{maskedProof} is the main problem-specific input given to the models, 
along with the general prompt for the task; it is obtained
by masking the starting \smtt{proof}. The full mask assignment that recovers \smtt{proof} from \smtt{maskedProof} 
is given by the value of the \smtt{masks} dictionary. Model results are provided under \smtt{llmMaskFillingResponses},
a dictionary mapping our model identifiers to the full corresponding results. One such individual result dictionary, 
for Claude Opus 4, appears between lines 23 and 57 in the above listing (with truncated values for some of the keys). 
The dictionary contains timestamps (lines 23/24), the model's response (lines 25--28), the unmasked proof obtained
by carrying out the assignments specified in the model's response (lines 29--32); a final \smtt{correctMasks} field
indicating whether the model's response was ultimately correct; and detailed evaluation results in the dictionary
\smtt{instrumentedCheckerResult}. The main payload of the latter is the \smtt{result} key, which can take the 3 values \smtt{correct},
\smtt{incorrect}, and (very rarely) \smtt{unknown}. If the result is \mtt{incorrect}, error details provided by the instrumented
proof checked are given, including all syntax and/or semantic repairs. 

Result files for other tasks are similarly structured. 

\subsection{Result Timeline and Interaction Protocol}

\begin{figure}[!tb]
\begin{minipage}{\textwidth}
\centering
\scalebox{0.6}{\includegraphics[width=\linewidth]{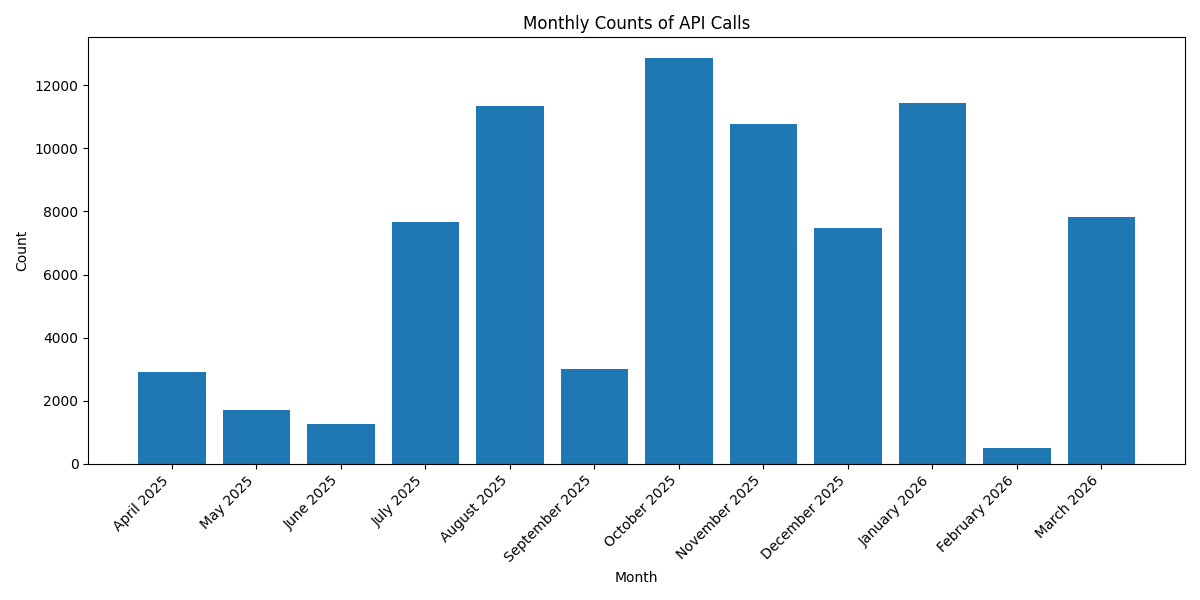}}
\end{minipage}
\caption{API call timeline.}
\label{Fig:APICallTimeline}
\end{figure}

The results were generated from April 2025 through March 2026; the distribution of API calls is shown in
Fig.~\ref{Fig:APICallTimeline}.
All systems were queried as plain language models; we did not use external 
tools, retrieval, web search, or agent tooling. The standard interaction pattern 
was the single-call paradigm used by most classic benchmarks, without additional dialog turns. 
The reported outputs were not obtained by techniques like self-consistency, majority voting, or 
pass@$k$ aggregation.

%Each scored item is derived from a single retained model output, which is then adjudicated deterministically by the benchmark pipeline. 
%All model outputs for all items across all tasks can be found in Yaml files on the benchmark's Github page. 

\subsection{Model Identity and API Routing}

We used several provider APIs. We routed each model request through the model's native API when possible, 
relying on unified/shared API gateways for the rest (mostly OpenRouter, and occasionally Inworld).

\begin{table*}[!b]
\centering
\scriptsize
\setlength{\tabcolsep}{3pt}
\renewcommand{\arraystretch}{1.3}
\begin{tabular}{@{}>{\raggedright\arraybackslash}p{0.15\textwidth} >{\raggedright\arraybackslash}p{0.24\textwidth} >{\raggedright\arraybackslash}p{0.20\textwidth} >{\raggedright\arraybackslash}p{0.12\textwidth} >{\raggedright\arraybackslash}p{0.14\textwidth}@{}}
\toprule
\textbf{Family} & \textbf{Representative models} & \textbf{Primary API path} & \textbf{Batch mode} & \textbf{Default timeout} \\
\midrule
OpenAI reasoning models & \texttt{gpt-5}, \texttt{gpt-5.1}, \texttt{gpt-5.2}, \texttt{o3}, \texttt{o3-mini}, \texttt{o4-mini} & OpenAI Responses API in serial mode; OpenAI batch API available in the codebase & Serial in the reported runs & 30 minutes \\
OpenAI chat models & \texttt{gpt-4o}, \texttt{gpt-4.1} & OpenAI Chat Completions API & Serial in the reported runs & 30 minutes \\
Anthropic Claude models & \texttt{claude-sonnet-4-20250514}, \texttt{claude-sonnet-4.5}, \texttt{claude-opus-4-20250514}, \texttt{claude-opus-4.5} & Anthropic Messages API in serial mode; Anthropic Message Batches in batch mode & Batch used in the reported runs, with serial fill-ins & 30 minutes \\
Google Gemini models & \texttt{gemini-2.5-pro}, \texttt{gemini-3-pro-preview}, \texttt{gemini-3.1-pro-preview} & Google \texttt{genai} client, \texttt{generate\_content} & Serial in the reported runs & 30 minutes \\
DeepSeek models & \texttt{deepseek-v3}, \texttt{deepseek-v3.1}, \texttt{deepseek-r1}, \texttt{deepseek-reasoner} & DeepSeek chat-completions-compatible API & No separate batch wrapper in the main benchmark path & 30 minutes \\
xAI models & \texttt{grok-3}, \texttt{grok-4}, \texttt{grok-4.20} & xAI chat-completions-compatible API & No separate batch wrapper in the main benchmark path & 30 minutes \\
OpenRouter-backed models & \texttt{llama\_400b\_nous}, \texttt{hermes4},\newline \texttt{mistral-magistral},\newline and related fallback models & OpenRouter chat-completions-\newline compatible API & No provider batch path used here & \shortstack[l]{30-second connect timeout;\\30-minute read timeout} \\
\bottomrule
\end{tabular}
\caption{Provider routing used. Model identifiers follow the naming scheme used in this paper; 
our public release artifacts expose the underlying provider identifiers as well.}
\label{tab:tech-routing}
\end{table*}

OpenAI models from the GPT-5 family and from the o-series were not run through the same path 
as GPT-4o and GPT-4.1. The first used the \href{https://developers.openai.com/api/reference/responses/overview}{Responses API}, 
while the second used \href{https://developers.openai.com/api/reference/chat-completions/overview}{Chat Completions}. 
Additionally, while our codebase contains OpenAI batch support and a synchronous 
parallel Gemini wrapper, our reported experiments used batch execution only for Anthropic models.

\begin{table*}[t]
\centering
\scriptsize
\setlength{\tabcolsep}{8pt}
\renewcommand{\arraystretch}{1.2}
\resizebox{0.8\textwidth}{!}{%
\begin{tabular}{@{}lll@{}}
\toprule
\textbf{ID used in this article} & \textbf{Queried model id} & \textbf{Run window} \\
\midrule

\multicolumn{3}{@{}l}{\textit{OpenAI}} \\
\addlinespace[2pt]
\mbox{\gptFive} & \mbox{\texttt{gpt-5}} & Aug.~2025--Mar.~2026 \\
\mbox{\gptFiveOne} & \mbox{\texttt{gpt-5.1}} & Nov.~2025--Mar.~2026 \\
\mbox{\gptFiveTwo} & \mbox{\texttt{gpt-5.2}} & Dec.~2025--Mar.~2026 \\
\mbox{\oThreeMini} & \mbox{\texttt{o3-mini}} & Apr.~2025--Mar.~2026 \\
\mbox{\oFourMini} & \mbox{\texttt{o4-mini}} & Apr.~2025--Mar.~2026 \\
\mbox{\oThree} & \mbox{\texttt{o3}} & Apr.~2025--Mar.~2026 \\
\mbox{\gptFourO} & \mbox{\texttt{gpt-4o}} & Apr.~2025--Mar.~2026 \\
\mbox{\gptFourOne} & \mbox{\texttt{gpt-4.1}} & Jun.~2025--Mar.~2026 \\

\addlinespace[4pt]
\multicolumn{3}{@{}l}{\textit{Google}} \\
\addlinespace[2pt]
\mbox{\texttt{gemini-3.1-pro-preview}} & \mbox{\texttt{gemini-3.1-pro-preview}} & Feb.~2026--Mar.~2026 \\
\mbox{\texttt{gemini-2.5-pro}} & \mbox{\texttt{gemini-2.5-pro}} & Jun.~2025--Mar.~2026 \\
\mbox{\texttt{gemini-3.0-pro-preview}} & \mbox{\texttt{gemini-3-pro-preview}} & Nov.~2025--Mar.~2026 \\

\addlinespace[4pt]
\multicolumn{3}{@{}l}{\textit{Anthropic}} \\
\addlinespace[2pt]
\mbox{\claudeSonnetFour} & \mbox{\texttt{claude-sonnet-4-20250514}} & Jul.~2025--Mar.~2026 \\
%\mbox{\texttt{claude-sonnet-4.5-ext-thinking}}
\mbox{\claudeSonnetExt} & \mbox{\texttt{claude-sonnet-4-5-20250929}} & Oct.~2025--Mar.~2026 \\
\mbox{\opus} & \mbox{\texttt{claude-opus-4-20250514}} & Jul.~2025--Mar.~2026 \\
\mbox{\claudeOpusFourFiveNov} & \mbox{\texttt{claude-opus-4-5-20251101}} & Dec.~2025--Mar.~2026 \\

\addlinespace[4pt]
\multicolumn{3}{@{}l}{\textit{DeepSeek}} \\
\addlinespace[2pt]
\mbox{\dsROne} & \mbox{\texttt{deepseekr1-0528}} & May~2025--Mar.~2026 \\
\mbox{\dsThreeOne} & \mbox{\texttt{deepseek/deepseek-chat-v3.1}} & Nov.~2025--Mar.~2026 \\
\mbox{\dsThree} & \mbox{\texttt{deepseek/deepseek-chat-v3}} & Apr.~2025--Mar.~2026 \\
\mbox{\dsReasoner} & \mbox{\texttt{deepseek-reasoner}} & Dec.~2025--Mar.~2026 \\

\addlinespace[4pt]
\multicolumn{3}{@{}l}{\textit{xAI}} \\
\addlinespace[2pt]
\mbox{\grokFour} & \mbox{\texttt{grok-4-0709}} & Jul.~2025--Mar.~2026 \\
\mbox{\grokThree} & \mbox{\texttt{grok-3-latest}} & Jul.~2025--Mar.~2026 \\

\addlinespace[4pt]
\multicolumn{3}{@{}l}{\textit{Other OpenRouter-backed Models}} \\
\addlinespace[2pt]
\mbox{\texttt{mistral-magistral}} & \mbox{\texttt{mistralai/magistral-medium-2506}} & Jul.~2025--Oct.~2025 \\
\mbox{\texttt{llama-3.1-405b}} & \mbox{\texttt{nousresearch/hermes-3-llama-3.1-405b}} & Apr.~2025--Mar.~2026 \\
\mbox{\hermes} & \mbox{\texttt{nousresearch/hermes-4-405b}} & Nov.~2025--Mar.~2026 \\

\addlinespace[4pt]
\multicolumn{3}{@{}l}{\textit{Inworld-routed models}} \\
\addlinespace[2pt]
\mbox{\texttt{mistral-magistral}} & \mbox{\texttt{mistralai/magistral-medium-latest}} & Feb.~2026--Mar.~2026 \\

\bottomrule
\end{tabular}
}
\caption{The 24-model set reported in the paper, shown under the paper naming scheme together with the exact queried model identifiers and the observed run windows in the released archives.}
\label{tab:tech-model-aliases}
\end{table*}

\subsection{Inputs and Outputs}

Each task used its own prompt, typically in few-shot style with several ICL examples in the prompt. 
The prompts for all tasks can be found on the Github page. 
Most tasks require JSON outputs, the sole exception being gap filling for \plonesp (\mbf{PL1-GF}). 
There, a model generates a plain textual answer organized into labeled \texttt{GAP-\emph{i}:} segments. 
Our evaluation pipeline then reconstructs  a complete proof from that answer and checks it automatically. 
Table~\ref{tab:tech-task-settings} summarizes this information.

%% \begin{table*}[t]
%% \centering
%% \scriptsize
%% \setlength{\tabcolsep}{4pt}
%% \begin{tabular}{@{}p{0.20\textwidth} p{0.31\textwidth} p{0.22\textwidth} p{0.11\textwidth}@{}}
%% \toprule
%% \textbf{Benchmark family} & \textbf{Prompt style} & \textbf{Required output} & \shortstack[c]{\textbf{Task default}\\\textbf{temperature}} \\
%% \midrule
%% Natural-deduction proof writing (PL1/PL2/PL3) & Few-shot formal-proof prompts, in BY style or FROM style depending on the task variant & JSON object containing at least a proof and a short description & 0.6 \\
%% Hilbert proof writing & Hilbert-specific prompts with restricted / expanded variants and a short sequel prompt & JSON object containing a Hilbert proof and a short description & 0.6 \\
%% Proof checking & Few-shot proof-checking prompts; English and formal variants & JSON object containing a Boolean verdict and structured error information & 0.01 \\
%% Mask filling & Few-shot masking prompt with an answer-only sequel & JSON object specifying the model's mask replacements & 0.6 \\
%% Gap filling & Few-shot gap-filling prompt with an answer-only sequel & Free-form text containing one block per gap, labeled \texttt{GAP-\emph{i}:} & 0.6 \\
%% Equational reasoning & Task-specific few-shot prompts for simple checking, equation recovery, and proof-with-gaps & Structured JSON output; the proof-with-gaps task also requested confidence-bearing judgments & 0.6 \\
%% \bottomrule
%% \end{tabular}
%% \caption{Prompt/input and output structure by task family.}
%% \label{tab:tech-task-settings}
%% \end{table*}

\renewcommand{\arraystretch}{1.2}
\begin{table*}[b!]
\centering
\scriptsize
\setlength{\tabcolsep}{4pt}
\begin{tabular}{@{}p{0.28\textwidth} p{0.1\textwidth} p{0.40\textwidth}@{}}
\toprule
\textbf{Task} & \textbf{Prompt style} & \textbf{Required output format} \\
\midrule

NDL/\ndlzerosp proof writing (\plone, \pltwo, \plthree) & 
Few-shot & 
JSON dictionary containing at least a proof and a description/explanation of the proof reasoning \\
Hilbert proof writing (\plfour) & 
Few shot & JSON dictionary containing a Hilbert proof and a short description \\
Proof checking & Few-shot & JSON dictionary containing a Boolean verdict and structured error information \\
Mask filling & Few-shot & JSON dictionary specifying the model's mask assignments \\
Gap filling & Few-shot & Free text containing one block per gap \\
Equational reasoning & Few shot & Structured JSON output (including confidence levels for \mbf{EQ-GF}) \\
\bottomrule
\end{tabular}
\caption{Prompt/input and output structure by task family.}
\label{tab:tech-task-settings}
\end{table*}
\renewcommand{\arraystretch}{1.0}

\subsection{Wilson Confidence Intervals}

The parenthetical numbers in our tables are 95\% confidence intervals for each reported accuracy, 
computed using the Wilson score interval for a binomial proportion. Concretely, for each model 
we treat every task item as a binary outcome: either the model got it right or it did not.
Thus, if a model answers $k$ of $n$ items correctly, its observed accuracy is $\hat p = k/n$. 
Our tables report this point estimate together with the uncertainty due to the finite sample size. 
In the compact notation used in the tables, an entry such as $0.87 \ci{0.02}$ means that the 
estimated accuracy is 0.87 and the Wilson interval has half-width 0.02, so the underlying 
95\% confidence interval is approximately $[0.85, 0.89]$ after rounding to two decimals.

We used Wilson intervals rather than the simpler normal approximation because it behaves better 
for proportions near 0 or 1 and for moderate sample sizes. Our code uses the standard 95\% Wilson 
formula with $z=1.96$: it computes an adjusted center $\tilde p$ and a half-width $h$, then forms 
the interval $[\tilde p-h,\tilde p+h]$. The numbers shown in parentheses are this half-width $h$, 
rounded to two decimal places; the code also computes the full lower and upper bounds explicitly. 
These intervals should be read in the usual frequentist way: if we repeated the same evaluation 
procedure many times on comparable samples, about 95\% of the intervals constructed this way 
would contain the model's true accuracy on that task distribution. 

\subsection{Decoding, Reasoning, and Timeout Settings}

Table~\ref{tab:tech-query-config} summarizes our main decoding and reasoning settings. 
%The table is stated at the level of the benchmark query layer rather than provider marketing labels, since that is what determines the actual runtime behavior. 
Note that temperatures were not uniformly enforced, especially for the more advanced 
reasoning models (such as GPT-5x), which have deemphasized the use of temperature (and sampling parameters in general). 
%In particular, calls to Gemini models via Google used provider defaults, and calls to DeepSeek models used wrapper defaults. 
Unless a setting is explicitly listed in the table, its value was left at the provider default.

\renewcommand{\arraystretch}{1.4}
\begin{table*}[t]
\centering
\scriptsize
\setlength{\tabcolsep}{3pt}
\resizebox{0.9\textwidth}{!}{%
\begin{tabular}{@{}>{\raggedright\arraybackslash}p{0.16\textwidth} >{\raggedright\arraybackslash}p{0.31\textwidth} >{\raggedright\arraybackslash}p{0.29\textwidth} >{\raggedright\arraybackslash}p{0.16\textwidth}@{}}
\toprule
\textbf{Model group} & \textbf{Serial-mode policy} & \textbf{Batch-mode policy} & \textbf{Notes} \\
\midrule
\texttt{gpt-5}, \texttt{gpt-5.1} & Responses API; temperature not set; reasoning effort \texttt{high}; \texttt{max\_output\_tokens}=72K & Not used for the reported results & JSON-constrained tasks use the Responses-API text-format JSON path \\
\texttt{gpt-5.2} & Responses API; temperature not set; reasoning effort \texttt{high}; automatic extra-thinking budget of 64K on tasks that expose it; \texttt{max\_output\_tokens}=112K & Not used for the reported results & Used with the larger GPT-5 token budget path \\
OpenAI o-series (\texttt{o3}, \texttt{o3-mini}, \texttt{o4-mini}) & Responses API; temperature not set; reasoning effort \texttt{high}; \texttt{max\_output\_tokens}=72K & Not used for the reported results & The reported o-series results were effectively serial \\
\texttt{claude-sonnet-}\newline\texttt{4.5-ext-}\newline\texttt{thinking} & Anthropic Messages API; temperature omitted when thinking is enabled; thinking budget 12K; \texttt{max\_tokens}=21{,}333 & Anthropic Message Batches; thinking budget 16K; \texttt{max\_tokens}=64K & Explicit extended thinking; stop reason is recorded when available in serial mode \\
\texttt{claude-opus-}\newline\texttt{4.5} & Anthropic Messages API; temperature omitted when thinking is enabled; 
thinking budget 12K; \texttt{max\_tokens}=21{,}333 
& Anthropic Message Batches; \texttt{max\_tokens}=64K; effective thinking budget 32K after buffer reservation & Explicit extended thinking in both serial and batch paths \\
\texttt{claude-sonnet-}\newline\texttt{420250514} & Anthropic Messages API; task temperature; no explicit thinking; \texttt{max\_tokens}=21{,}333 & Anthropic Message Batches; no explicit thinking; \texttt{max\_tokens}=64K & No extended thinking in the reported runs \\
\texttt{claude-opus-}\newline\texttt{4.1} & Anthropic Messages API; task temperature; no explicit thinking; \texttt{max\_tokens}=21{,}333 & Anthropic Message Batches; no explicit thinking; \texttt{max\_tokens}=32K & No extended thinking in the reported runs \\

Gemini 2.5 / 3 / 3.1 & Google \texttt{genai} client; provider defaults for temperature, output tokens, and reasoning (high ``dynamic thinking level'' by default) & Same provider defaults; ``batch'' runs are synchronous parallel direct calls & The benchmark task temperature is not passed into the Gemini wrapper \\

DeepSeek chat / reasoner & DeepSeek chat uses temperature 0.6 and top-$p=0.9$; DeepSeek reasoner uses the provider reasoner path and enables JSON mode when requested; no explicit token cap & No separate batch wrapper in the main benchmark path & Task-level temperature is not propagated; these models use wrapper defaults \\

xAI Grok & Temperature 0.2; JSON response-format request when needed; no explicit reasoning-effort or output-token cap & No separate batch wrapper in the main benchmark path & Applies to \texttt{grok-3}, \texttt{grok-4}, and \texttt{grok-4.20} \\

OpenRouter-backed models & In the default raw-HTTP path: task temperature, \texttt{max\_tokens}=32K, reasoning enabled with \texttt{high} effort; in the client-library path: task temperature with no explicit reasoning or token cap & No provider batch path used here & The gap-filling benchmark uses the client-library path; most other OpenRouter-backed tasks use the raw-HTTP path \\
\bottomrule
\end{tabular}
}
\caption{Compact summary of the main decoding, reasoning, and token-budget policies.}
\label{tab:tech-query-config}
\end{table*}
\renewcommand{\arraystretch}{1.0}

\paragraph{OpenAI models}
Models from the GPT-5 family and the o-series were queried through the Responses API in serial mode. Temperature were not set explicitly for these models. 
Instead, we specified the values of reasoning parameters. GPT-5, GPT-5.1, GPT-5.2, GPT-5.4,  
and the o-series models used high-effort settings. 
GPT-5.2 was typically given an automatic extra-thinking budget of 64K tokens in tasks that expose this field. 

For \texttt{gpt-4o} and \texttt{gpt-4.1}, we used the Chat Completions API. These models therefore inherited 
our task-level temperature settings (reasoning parameters were inapplicable to these models), which 
are 0.6 for generative tasks like proof writing and 0.01 for discriminative tasks like proof 
checking.\footnote{While 0.6 might seem high for a reasoning task, we have empirically found that
it outperforms lower temperatures for proof writing. In general, proof checking is a low-entropy 
judgment task, while proof writing is a high-entropy search task. Low temperature helps the former 
by reducing noise, wile moderate temperature helps the latter by improving search over possible proof 
trajectories.}

\paragraph{Anthropic Claude models.}
Serial Claude calls used the Anthropic Messages API. When explicit thinking was enabled, 
temperature values were omitted and thinking was requested  directly. For the Claude 4.5 family 
in serial mode, we used a 12K thinking budget with a 21,333-token output cap. (These caps were 
imposed by Anthropic, which only supports higher values for these parameters in streaming mode.) 
Older Claude models were run without explicit thinking controls. 
Serial Claude calls also recorded the provider stop reason when available.

Batch Claude calls used Anthropic Message Batches. We requested provider-side 
thinking for the Claude 4.5 family and for explicit ``extended thinking'' aliases. 
In practice, this meant substantially larger batch-time output budgets than in serial mode. 
For example, the Opus 4.5 batch path used a provider-side 64K token cap together with a 
32K internal thinking budget after reserving buffer space for the visible answer, 
whereas the Sonnet 4.5 batch path used a smaller fixed thinking budget.

\paragraph{Google Gemini models.}
Our Gemini calls used Google's \texttt{genai} client. We did not explicitly 
set temperature, output-token limits, or reasoning controls, instead using provider defaults
(high by default).

\paragraph{DeepSeek models.}
DeepSeek reasoner models were run through the provider's reasoning-enabled path. When JSON 
output was requested, the wrapper also requested JSON formatting and enabled the provider's 
explicit thinking mode. In both cases, the wrapper's own defaults governed these calls rather 
than our task-level temperatures. 

\paragraph{xAI models.}
xAI models were called through an API compatible with Chat Completions. 
We specified temperature values and when JSON output was needed, requested JSON formatting.
We did not set reasoning-effort parameters or output-token caps. 

\paragraph{OpenRouter-backed models.}
For models routed through OpenRouter, we used two related but distinct paths. 
In the default raw-HTTP path, we used the task-level temperature and requested JSON output 
whenever the task required it, capped output at 32K tokens, and enabled OpenRouter's 
reasoning mode with \texttt{high} effort. The wrapper also used a 30-second connect timeout, 
a 30-minute read timeout, and one additional transport-level retry in the case of whitespace-only 
or apparently truncated HTTP bodies. In the client-library path, used only for \mbf{PL1-GF}, 
we did not specify  reasoning or token-cap settings. 

\subsection{Batching, Partial Results, and Failure Policy}

In the reported experiments, the only models run in batch mode were Anthropic models. 
These were run mostly through Anthropic Message Batches, with serial fill-ins used 
in most tasks to complete missing items or rerun failures. Batched outputs were matched 
back to problems through explicit custom identifiers rather than through positional order.

Malformed outputs were generally not discarded silently. Instead, we preserved them 
either as raw outputs or through explicit flags such as \texttt{illFormattedProofOutput} 
or \texttt{illFormattedResponse}, or via task-specific parse-failure markers. 
Proof-checking tasks also recorded explicit API-call failures. 
These malformed or missing outputs remain visible in the released output files.

Our rerun policy was necessarily variable across models. Many models returned valid outputs 
consistently and required little manual intervention. Others, especially some reasoning-oriented 
models on difficult tasks, frequently exhausted their output budget, returned malformed answers, 
or resulted in other API call failures. Grok-4 was especially difficult in this respect, 
and some advanced GPT models also often declined to return an answer on the hardest problems. 
For such models, we used a best-effort fill-in policy: malformed outputs and API-level 
failures were rerun, typically at least three times and sometimes more.

If a problem instance did not yield any response (\egnsp, because of a server-side API-call error), 
that was not counted against the model (the data point was removed from the denominator), but 
if an incomplete or ill-formed response was returned, it was counted as an answer. 
When possible, invalid or partial results were either cleaned with postprocessing or 
the API calls were repeated. 
%% the released archives retain such results only when repeated reruns, usually 
%% at least three attempts, still failed to yield a valid answer. This choice matters 
%% particularly for fragile APIs and long reasoning traces, since omitting such failures 
%% would materially inflate the apparent performance of some models.

\subsection{Evaluation} 

Performance is not computed directly from raw model outputs. 
Each task has a deterministic evaluation pipeline, which forms an 
important part of the experimental design. These are described in our
discussions of the individual tasks and released in full operational detail in 
the Github code base.

\subsection{Reasoning Metadata in the Released Result Files}

%% The released result archives preserve enough information to make the benchmark operationally legible. Across the main benchmark families, the following metadata are commonly preserved:
%% \begin{itemize}
%%     \item model identity;
%%     \item task-level temperature, under either \texttt{temp} or \texttt{temperature};
%%     \item prompt identity;
%%     \item request and completion timestamps;
%%     \item batch identifiers for batch-backed tasks;
%%     \item raw model outputs for many proof-writing, masking, gap-filling, and equational runs.
%% \end{itemize}

Reasoning metadata is present with limited coverage. 
Many GPT- and Claude-based runs preserve one or more of the following:
\begin{itemize}
    \item provider usage summaries;
    \item extracted reasoning summaries;
    \item stored thinking traces;
    \item \texttt{thinkingTokens};
    \item \texttt{outputTokens};
    \item provider stop reasons.
\end{itemize}

However, these fields are not uniformly available across all providers and all tasks. They are common in several proof-checking runs and in subsets of the masking, gap-filling, and equational results, but they are less common for Gemini, xAI, and some OpenRouter-backed models. 

\clearpage
\bibliographystyle{plainnat}
\bibliography{custom}

\end{document}